\documentclass[10pt, a4paper, twoside]{jreport}

\usepackage{color}
\usepackage{bm, helvetic}
\usepackage{times}
\usepackage{calrsfs}
\usepackage{amssymb, amsfonts, amsmath}
\usepackage{fancyhdr, fancyvrb}
\usepackage[framed, thmmarks]{ntheorem}
\usepackage{etoolbox}
\usepackage{graphicx}
\usepackage{bm, bbm, url, natbib, paralist, afterpage, verbatim}
\usepackage{listings}
\usepackage{algorithm2e}
\usepackage{hyperref}

\definecolor{lightgrey}{rgb}{0.9,0.9,0.9}
\definecolor{darkgreen}{rgb}{0,0.6,0}	
\usepackage{xcolor}

\makeatletter
\newcommand{\openbox}{\leavevmode
  \hbox to.77778em{%
  \hfil\vrule
  \vbox to.675em{\hrule width.6em\vfil\hrule}%
  \vrule\hfil}}

\newcommand{\proofname}{Proof.}
\newcounter{proof}%
\newenvironment{proof}[1][\proofname]{
  \th@nonumberplain
  \def\theorem@headerfont{\itshape}%
  \normalfont
  \@thm{proof}{proof}{#1}}%
  {\@endtheorem}
\makeatother

\definecolor{lightgrey}{rgb}{0.9,0.9,0.9}

\newtheoremstyle{example} 
 {\item[\hskip\labelsep \normalfont ##1\ ##2.]}%
 {\item[\hskip\labelsep \normalfont ##1\ ##2, ##3.]}
\theoremstyle{plain}
	\theorembodyfont{\normalfont}
\theoremsymbol{\ensuremath{\square}}
\newtheorem{example}{Example}[chapter]

\theoremseparator{\hspace{-3pt}}
\theorembodyfont{\upshape}
\newtheorem*{contexample}{Example}

\theoremsymbol{}
\newtheorem{proposition}{Proposition}[chapter]
\newtheorem{definition}{Definition}[chapter]
\newtheorem{remark}{Remark}[chapter]
\newtheorem{corollary}{Corollary}[chapter]

\newtheorem{theorem}{Theorem}[chapter]
\newtheorem{lemma}{Lemma}[chapter]
\newtheorem{question}{Question}[chapter]

\newcounter{saveexample}%

\newcounter{saveeqn}%

\newcounter{savefigure}%

\newcounter{savetable}%

\renewcommand{\chaptermark}[1]{\markboth{#1}{}}

\renewcommand{\headwidth}{6.27in}
\fancypagestyle{plain}{%
    \fancyhead{}
   \renewcommand{\headrulewidth}{0pt}}

\bmdefine\ttheta{\theta}
\bmdefine\aalpha{\alpha}
\bmdefine\bbeta{\beta}
\bmdefine\ddelta{\delta}
\bmdefine\eeta{\eta}
\bmdefine\llambda{\lambda}
\bmdefine\ggamma{\gamma}
\bmdefine\nnu{\nu}
\bmdefine\vvarepsilon{\varepsilon}
\bmdefine\mmu{\mu}
\bmdefine\nnu{\nu}
\bmdefine\ttau{\tau}
\bmdefine\SSigma{\Sigma}
\bmdefine\TTheta{\Theta}
\bmdefine\XXi{\Xi}
\bmdefine\PPi{\Pi}
\bmdefine\GGamma{\Gamma}
\bmdefine\DDelta{\Delta}
\bmdefine\ssigma{\sigma}
\bmdefine\UUpsilon{\Upsilon}
\bmdefine\PPsi{\Psi}
\bmdefine\PPhi{\Phi}
\bmdefine\LLambda{\Lambda}
\bmdefine\OOmega{\Omega}

\fancypagestyle{plain}{%
\fancyhf{} 
\fancyfoot[RO,LE]{\small\textsf{\thepage}}
\renewcommand{\headrulewidth}{0pt}}

\DeclareSymbolFont{pazoletters}{OML}{zplm}{m}{it}
\DeclareMathAlphabet{\pazocal}{OML}{zplm}{m}{it}
\DeclareMathAlphabet{\mathcal}{OMS}{zplm}{m}{n}

\newcommand\independent{\protect\mathpalette{\protect\independenT}{\perp}}
\def\independenT#1#2{\mathrel{\rlap{$#1#2$}\mkern2mu{#1#2}}}

\begin{document}
\pagenumbering{roman}
\thispagestyle{empty}
\setcounter{page}{1}

\clearpage
\thispagestyle{empty}
\noindent
{\Huge\textbf{\textsf{Lecture notes on ridge regression
}}}
\\
\\
Version 0.80, August 19, 2026.
\par

\vspace{14.4cm} \noindent {\large\sf\textbf{{Wessel N. van Wieringen}$^{1,2}$}
}
\vspace{.3cm}
\\
\noindent
$^1$ Department of Epidemiology and Data Science, 
\\
$\mbox{ } \, \mbox{}$ Amsterdam Public Health research institute, Amsterdam UMC, location VUmc
\\
$\mbox{ } \, \mbox{}$ P.O. Box 7057, 1007 MB Amsterdam, The Netherlands
\\
$^2$ Department of Mathematics, Vrije Universiteit Amsterdam
\\
$\mbox{ } \, \mbox{}$ De Boelelaan 1111, 1081 HV Amsterdam, The Netherlands
\\
Email: w.vanwieringen@amsterdamumc.nl

\vspace{2.7cm}
\noindent
\textbf{\textsf{License}}
\\
This document is distributed under the Creative Commons Attribution-NonCommercial-ShareAlike license:
\url{http://creativecommons.org/licenses/by-nc-sa/4.0/}

\vspace{-0.5cm}
\begin{figure}[!h]
\begin{tabular}{lcr}
& & \hspace{12cm}
\includegraphics[scale=0.04, angle=0]{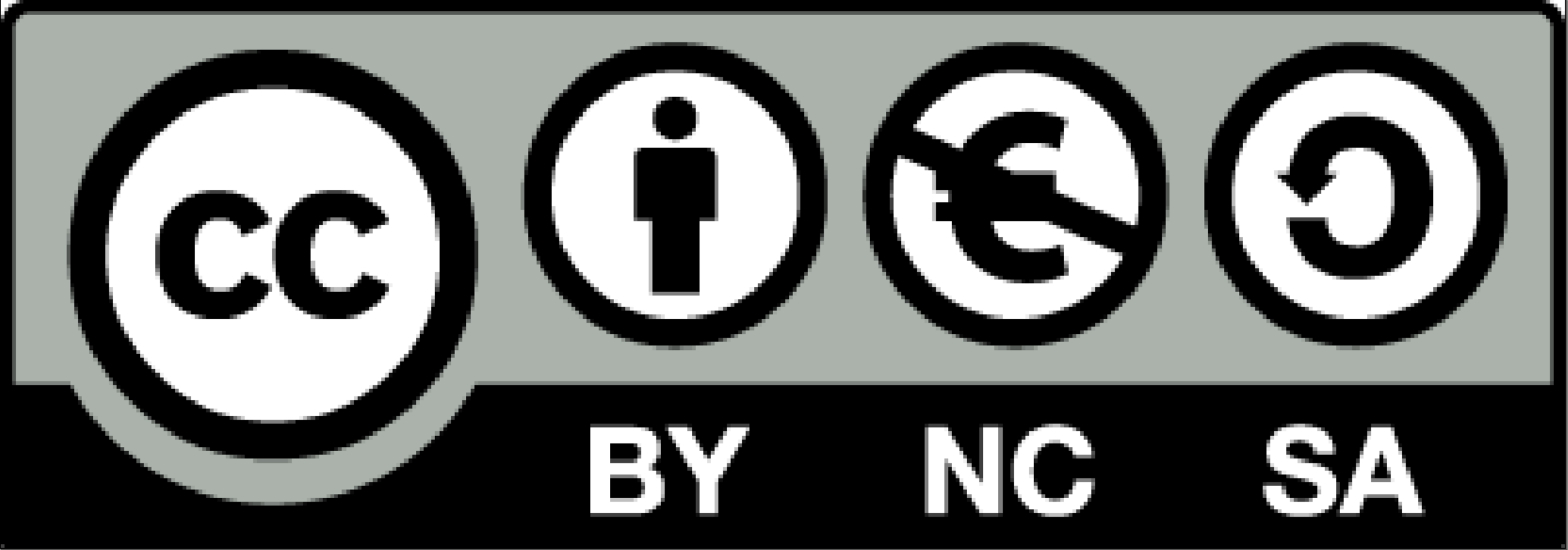}
\end{tabular}
\end{figure}
\afterpage{\clearpage}


\newpage
\thispagestyle{empty}
\[
\]
\[
\]
\[
\]
\[
\]
\[
\]
\[
\]
\[
\]
\[
\]
\[
\]
\[
\]
\[
\]
\[
\]
\[
\]
\[
\]
\[
\]
\[
\]
\[
\]
\[
\]
\[
\]
\[
\]
\[
\]
\[
\]
\[
\]
\[
\]

\[
\]
\[
\]
\[
\]
\[
\]
\[
\]
\[
\]

\noindent
\textbf{\textsf{Disclaimer}}
\\
This document is a collection of many well-known results on ridge regression. The current status of the document is `work-in-progress' as it is incomplete (more results from literature will be included) and it may contain inconsistencies and errors. Hence, reading and believing at own risk. Finally, proper reference to the original source may sometimes be lacking. This is regrettable and these references -- if ever known to the author -- will be included in later versions.

\newpage
\thispagestyle{empty}
\vspace{7cm}
\noindent
\textbf{\textsf{Acknowledgements}}
\\
Many people aided in various ways to the construction of these notes. Mark A. van de Wiel commented on various parts of Chapter \ref{chap:BayesianRegression}. Jelle J. Goeman clarified some matters behind the method described in Section \ref{sect:gradientAscent}. Paul H.C. Eilers pointed to helpful references for Chapter \ref{chap.mixedModel} and provided parts of the code used in Section \ref{sect:Psplines}. Harry van Zanten commented on Chapters \ref{chap:ridgeRegression} and \ref{chap:RidgeAsymptotics}, while St\'{e}phanie van de Pas made suggestions for the improvement of Chapters \ref{chap:RidgeAsymptotics} and \ref{chap:LassoAsymptotics}. Glenn Andrews thoroughly read Chapter \ref{chap:ridgeRegression} filtering out errors, unclarities, and inaccuracies. 

Small typo's or minor errors, that have -- hopefully -- been corrected in the latest version, were pointed out by (among others): Rikkert Hindriks, Micah Blake McCurdy, Jos\'{e} P. Gonz\'{a}lez-Brenes, Hassan Pazira, and numerous students from the \textit{High-dimensional data analysis}- and \textit{Statistics for high-dimensional data}-courses taught at Leiden University and the Vrije Universiteit Amsterdam, respectively.

\newpage
\thispagestyle{empty}
\tableofcontents

\thispagestyle{empty}

\pagenumbering{arabic} \pagestyle{fancy}

\chapter[Ridge regression]{Ridge regression} \label{chap:ridgeRegression}

This chapter provides an expos\'{e} of ridge estimation of the linear regression model. It starts with a brief recap of this model and its maximum likelihood estimator. We then discuss the problems of this estimator when confronted with collinear or high-dimensional study designs. The ridge regression estimator is first presented as a workaround. Subsequently, we discuss its many properties, e.g. moments, mean squared error, loss function, equivalence to constrained estimation, degrees of freedom and so on. In this, we take the maximum likelihood estimator as our point of reference and contrast it to its ridge counterpart in order to enhance the understanding of the latter. Furthermore, we point out connections of the ridge regression estimator with practices from machine learning, discuss the choice of the ridge estimator's parameter, illustrate additional properties of the estimator in various simulations, and close with an application of its use.

\section{Linear regression}
Consider an experiment in which $p$ characteristics of $n$ samples are measured. The data from this experiment are denoted $\mathbf{X}$, with $\mathbf{X}$ as above. The matrix $\mathbf{X}$ is called the \textit{design matrix}. Additional information of the samples is available in the form of $\mathbf{Y}$ (also as above). The variable $\mathbf{Y}$ is generally referred to as the \textit{response variable}. The aim of regression analysis is to explain $\mathbf{Y}$ in terms of $\mathbf{X}$ through a functional relationship like $Y_i = f(\mathbf{X}_{i,\ast})$. Without strong prior knowledge on the form of $f(\cdot)$, it is common to start simple and assume a linear relationship between $\mathbf{X}$ and $\mathbf{Y}$. This assumption gives rise to the \textit{linear regression model}:
\begin{eqnarray}
\label{form.linRegressionModel}
Y_{i} & = & \mathbf{X}_{i,\ast} \, \bbeta + \varepsilon_i 
\, \, \, = \, \, \, \beta_1 \, X_{i,1}  + \ldots   + \beta_{p} \, X_{i, p} +  \varepsilon_i.
\end{eqnarray}
In model (\ref{form.linRegressionModel}), $\bbeta = (\beta_1, \ldots, \beta_p)^{\top}$ is the \textit{regression parameter}. The parameter $\beta_j$, $j=1, \ldots, p$, represents the effect size of covariate $j$ on the response. That is, for each unit change in covariate $j$ (while keeping the other covariates fixed) the observed change in the response is equal to $\beta_j$. The second summand on the right-hand side of the model, $\varepsilon_i$, is referred to as the error. It represents the part of the response not explained by the functional part $\mathbf{X}_{i,\ast} \, \bbeta$ of the model (\ref{form.linRegressionModel}). In contrast to the functional part, which is considered to be systematic (i.e. non-random), the error is assumed to be random. Consequently, $Y_{i}$ need not be equal $Y_{i'}$  for $i \not= i'$, even if $\mathbf{X}_{i,\ast}= \mathbf{X}_{i',\ast}$. To complete the formulation of model (\ref{form.linRegressionModel}), we specify the probability distribution of $\varepsilon_i$. We assume that $\varepsilon_i \sim \mathcal{N}(0, \sigma^2)$ and the $\varepsilon_{i}$ are independent, i.e. $\mbox{Cov}(\varepsilon_{i}, \varepsilon_{i'}) = \sigma^2$ if $i = i'$ and $\mbox{Cov}(\varepsilon_{i}, \varepsilon_{i'}) = 0$ if $i \not= i'$. The randomness of $\varepsilon_i$ implies that $Y_i$ is also a random variable. In particular, $Y_i$ is normally distributed, because $\varepsilon_i \sim \mathcal{N}(0, \sigma^2)$ and $\mathbf{X}_{i,\ast} \, \bbeta$ is a non-random scalar. To specify the parameters of the distribution of $Y_i$ we need to calculate its first two moments. Its expectation equals $\mathbb{E}(Y_i) = \mathbf{X}_{i, \ast} \, \bbeta$,
while its variance $\mbox{Var}(Y_i)$ is $\sigma^2$.
\\
Hence, $Y_i \sim \mathcal{N}( \mathbf{X}_{i, \ast} \, \bbeta, \sigma^2)$. This formulation -- in terms of the normal distribution -- is equivalent to the formulation of model (\ref{form.linRegressionModel}), as both capture the assumptions involved: the linearity of the functional part and the normality of the error.

Model (\ref{form.linRegressionModel}) is often written in a more condensed matrix form:
\begin{eqnarray}
\mathbf{Y} & = & \mathbf{X} \, \bbeta + \vvarepsilon, \label{form.linRegressionModelinMatrix}
\end{eqnarray}
where $\vvarepsilon = (\varepsilon_1, \varepsilon_2, \ldots, \varepsilon_n)^{\top}$ and distributed as $\vvarepsilon \sim \mathcal{N}(\mathbf{0}_{p}, \sigma^2 \mathbf{I}_{nn})$. As above model (\ref{form.linRegressionModelinMatrix}) can be expressed as a multivariate normal distribution: $\mathbf{Y} \sim \mathcal{N}(\mathbf{X} \, \bbeta, \sigma^2 \mathbf{I}_{nn})$.

Model (\ref{form.linRegressionModelinMatrix}) is a so-called hierarchical model (not to be confused with the Bayesian meaning of this term). Here this terminology emphasizes that $\mathbf{X}$ and $\mathbf{Y}$ are not on a par, they play different roles in the model. The former is used to explain the latter. In model (\ref{form.linRegressionModel}) $\mathbf{X}$ is referred as the \textit{explanatory} or \textit{independent} variable, while the variable $\mathbf{Y}$ is generally referred to as the \textit{response} or \textit{dependent} variable.

The covariates may themselves be random. To apply the linear model they are temporarily assumed fixed. The linear regression model is then to be interpreted as $\mathbf{Y} \, | \, \mathbf{X} \sim \mathcal{N}(\mathbf{X} \, \bbeta, \sigma^2 \mathbf{I}_{nn})$

\begin{example} \textit{(Methylation of a tumor-suppressor gene)}
\\
Consider a study where the gene expression levels of a tumor-suppressor genes (TSG) and two methylation markers (MM1 and MM2) have been measured in 67 samples. A methylation marker is a gene that promotes methylation. Methylation refers to attachment of a methyl group to a nucleotide of the DNA. In case this attachment takes place in or close by the promotor region of a gene, this complicates the transcription of the gene. Methylation may down-regulate a gene. This mechanism also works in the reverse direction: removal of methyl groups may up-regulate a gene. A tumor-suppressor gene is a gene that halts the progression of the cell towards a cancerous state.

The medical question associated with these data: do the expression levels methylation markers affect the expression levels of the tumor-suppressor gene? To answer this question we may formulate the following linear regression model:
\begin{eqnarray*}
Y_{i, \texttt{{\tiny tsg}}} & = & \beta_0 + \beta_{\texttt{{\tiny mm1}}} X_{i, \texttt{{\tiny mm1}}} + \beta_{\texttt{{\tiny mm2}}} X_{i, \texttt{{\tiny mm2}}} + \varepsilon_i,
\end{eqnarray*}
with $i = 1, \ldots, 67$ and $\varepsilon_i \sim \mathcal{N}(0, \sigma^2)$. The interest focusses on $\beta_{\texttt{{\footnotesize mm1}}}$ and $\beta_{\texttt{{\footnotesize mm2}}}$. A non-zero value of at least one of these two regression parameters indicates that there is a linear association between the expression levels of the tumor-suppressor gene and that of the methylation markers.

\end{example}

\noindent
The linear regression model (\ref{form.linRegressionModel}) involves the unknown parameters: $\bbeta$ and $\sigma^2$, which need to be learned from the data. The parameters of the regression model, $\bbeta$ and $\sigma^2$ are estimated by means of likelihood maximization. Recall that $Y_i \sim \mathcal{N}( \mathbf{X}_{i,\ast} \, \bbeta, \sigma^2)$ with corresponding density: $ f_{Y_i}(y_i)  =  (2 \, \pi \, \sigma^2)^{-1/2} \, \exp[ - \tfrac{1}{2} \sigma^{-2} (y_i - \mathbf{X}_{i\ast} \bbeta)^2  ]$. The likelihood thus is:
\begin{eqnarray*}
L(\mathbf{Y}, \mathbf{X}; \bbeta, \sigma^2) & = &
\prod_{i=1}^n  (\sqrt{2 \pi} \, \sigma)^{-1} \, \exp[ - \tfrac{1}{2} \sigma^{-2} (Y_i - \mathbf{X}_{i, \ast} \bbeta)^2 ] \, \, \, = \, \, \,   (\sqrt{2 \pi} \, \sigma)^{-n} \, \exp[ - \tfrac{1}{2} \sigma^{-2} \| \mathbf{Y} - \mathbf{X} \bbeta \|_2^2  ],
\end{eqnarray*}
in which the independence of the observations has been used. Because of the strict monotonicity of the logarithm, the maximization of the likelihood coincides with the maximum of the logarithm of the likelihood (called the log-likelihood). Hence, to obtain maximum likelihood estimator of the parameter, it is equivalent to find the maximum of the loglikelihood. 
In order to find the maximum of the log-likelihood, we equate its derivative with respect to $\bbeta$ to zero. 
This gives the estimating equation for $\bbeta$:
\begin{eqnarray} \label{form.normalEquation}
\mathbf{X}^{\top} \mathbf{X} \, \bbeta & = &  \mathbf{X}^{\top} \mathbf{Y}.
\end{eqnarray}
Equation (\ref{form.normalEquation}) is called to the \textit{normal equation}. Pre-multiplication of both sides of the normal equation by $(\mathbf{X}^{\top} \mathbf{X})^{-1}$ yields the maximum likelihood estimator of the regression parameter: $\hat{\bbeta}  =   (\mathbf{X}^{\top} \mathbf{X})^{-1} \, \mathbf{X}^{\top} \mathbf{Y}$, in which it is assumed that $(\mathbf{X}^{\top} \mathbf{X})^{-1}$ is well-defined.

We obtain the maximum likelihood estimator of the residual variance along the same lines as that of the regression parameter. We take the partial derivative of the loglikelihood with respect to $\sigma^2$,
equate it to zero and solve for $\sigma^2$. This yields $\hat{\sigma}^2   =   n^{-1} \| \mathbf{Y} - \mathbf{X} \, \bbeta \|^2_2$. In this expression, $\bbeta$ is unknown and the maximum likelihood estimate of $\bbeta$ is plugged-in.
\\
\\
With explicit expressions of the parameters' maximum likelihood estimator at hand, we can study their properties. The expectation of the maximum likelihood estimator of the regression parameter $\bbeta$ is:
\begin{eqnarray*}
\mathbb{E}(\hat{\bbeta}) & = &  \mathbb{E}[ (\mathbf{X}^{\top} \mathbf{X})^{-1} \, \mathbf{X}^{\top} \mathbf{Y}]
\, \, \, = \, \, \, (\mathbf{X}^{\top} \mathbf{X})^{-1} \, \mathbf{X}^{\top} \mathbb{E}( \mathbf{Y})
\, \, \, = \, \, \, (\mathbf{X}^{\top} \mathbf{X})^{-1} \, \mathbf{X}^{\top} \mathbf{X} \, \bbeta
\, \, \, \, \,  = \, \, \, \bbeta.
\end{eqnarray*}
Hence, the maximum likelihood estimator of the regression coefficients is unbiased.

The variance of the maximum likelihood estimator of $\bbeta$ is:
\begin{eqnarray*}
\mbox{Var}(\hat{\bbeta}) 
& = & \mathbb{E} \{ [(\mathbf{X}^{\top} \mathbf{X})^{-1} \, \mathbf{X}^{\top} \mathbf{Y} - \bbeta] \, [(\mathbf{X}^{\top} \mathbf{X})^{-1} \, \mathbf{X}^{\top} \mathbf{Y} - \bbeta]^{\top} \}
\\
& = & (\mathbf{X}^{\top} \mathbf{X})^{-1} \, \mathbf{X}^{\top}  ( \mathbf{X} \, \bbeta \, \bbeta^{\top} \,  \mathbf{X}^{\top} + \sigma^2 \mathbf{I}_{pp} )  \mathbf{X} \, (\mathbf{X}^{\top} \mathbf{X})^{-1} - \bbeta \, \bbeta^{\top}
\, \, \, = \, \, \, \sigma^2 \, (\mathbf{X}^{\top} \mathbf{X})^{-1},
\end{eqnarray*}
in which we have used that $\mathbb{E} (\mathbf{Y} \mathbf{Y}^{\top}) =   \mathbf{X} \, \bbeta \, \bbeta^{\top} \,  \mathbf{X}^{\top} + \sigma^2 \, \mathbf{I}_{nn}$. From $\mbox{Var}(\hat{\bbeta}) = \sigma^2 \, (\mathbf{X}^{\top} \mathbf{X})^{-1}$, one obtains an estimate of the variance of the estimator of the $j$-th regression coefficient: $\hat{\sigma}^2 (\hat{\beta}_j )  =  \hat{\sigma}^2 [(\mathbf{X}^{\top} \mathbf{X})^{-1}]_{jj}$. This is used to construct a confidence interval for the estimates or test the hypothesis $H_0: \beta_j = 0$. In the latter display, $\hat{\sigma}^2$ should not be the maximum likelihood estimator, but is to be replaced by the residual sum-of-squares divided by $n-p$ rather than $n$. The \textit{residual sum-of-squares} is defined as $\sum_{i=1}^n (Y_i - \mathbf{X}_{i,\ast} \hat{\boldsymbol{\beta}})^2$.
\\
\\
The prediction of $Y_i$, denoted $\widehat{Y}_i$, is the expected value of $Y_i$ according the linear regression model (with its parameters replaced by their estimates). The prediction of $Y_i$ thus equals $\mathbb{E}(Y_i; \hat{\bbeta}, \hat{\sigma}^2) = \mathbf{X}_{i, \ast} \hat{\bbeta}$. In matrix notation the prediction is:
\begin{eqnarray*}
\widehat{\mathbf{Y}} & = & \mathbf{X} \, \hat{\bbeta} \, \, \, = \, \, \,  \mathbf{X} \, (\mathbf{X}^{\top} \mathbf{X})^{-1} \mathbf{X}^{\top} \mathbf{Y} \, \, \, := \, \, \,  \mathbf{H} \mathbf{Y},
\end{eqnarray*}
where $\mathbf{H}$ is the \textit{hat matrix}, as it `puts the hat' on $\mathbf{Y}$. Note that the hat matrix is a projection matrix, i.e. $\mathbf{H}^2 = \mathbf{H}$. 
Thus, the prediction $\widehat{\mathbf{Y}}$ is an orthogonal projection of $\mathbf{Y}$ onto the space spanned by the columns of $\mathbf{X}$.

An estimate of the error $\hat{\varepsilon}_i$, dubbed the \textit{residuals}, is obtained via:
\begin{eqnarray*}
\hat{\vvarepsilon} & = & \mathbf{Y} - \widehat{\mathbf{Y}} \, \, \, = \, \, \, \mathbf{Y} - \mathbf{X} \, \hat{\bbeta} \, \, \, = \, \, \, \mathbf{Y} - \mathbf{X} \, (\mathbf{X}^{\top} \mathbf{X})^{-1} \mathbf{X}^{\top} \mathbf{Y} \, \, \, = \, \, \, ( \mathbf{I} - \mathbf{H} ) \, \mathbf{Y}.
\end{eqnarray*}
Thus, the residuals are a projection of $\mathbf{Y}$ onto the orthogonal complement of the space spanned by the columns of $\mathbf{X}$. The residuals are to be used in diagnostics, e.g. checking of the normality assumption by means of a normal probability plot.
\\
\\
For more on the linear regression model, confer the monograph of \cite{Drap1998}.

\section{Super-collinearity of high-dimensional designs} 

The ridge regression estimator was originally proposed to deal with collinearity. The columns of design matrix $\mathbf{X}$ exhibit then some degree of linear dependence. This hampers the identification of the individual covariates' contributions to the variation of the response. The uncertainty, with respect to the covariate responsible for the variation explained in $\mathbf{Y}$, is reflected in the fit of the linear regression model to data. Collinearity reveals itself in the fit through a large error of the regression parameters' estimates corresponding to the collinear covariates and, consequently, usually accompanied by large values of the estimates.

\begin{example} \textit{(Collinearity)} \label{example.collinearyFlotin}  \\
The flotillins, the FLOT-1 and FLOT-2 genes, have been observed to regulate the proto-oncogene ERBB2 \textit{in vitro} \citep{Pust2013}. One may wish to corroborate this \textit{in vivo}. To this end, we use gene expression data of a breast cancer study, available as a Bioconductor package: {\tt breastCancerVDX}. From this study, the expression levels of probes interrogating the FLOT-1 and ERBB2 genes are retrieved. For clarity of the illustration, the FLOT-2 gene is ignored. After centering, the expression levels of the first ERBB2 probe are regressed on those of the four FLOT-1 probes. The R-code below carries out the data retrieval and analysis.
\begin{lstlisting}
# load packages
library(Biobase)
library(breastCancerVDX)

# load data into memory
data(vdx)

# ids of genes FLOT1
idFLOT1 <- which(fData(vdx)[,5] == 10211)

# ids of ERBB2
idERBB2 <- which(fData(vdx)[,5] == 2064)

# get expression levels of probes mapping to FLOT genes
X <- t(exprs(vdx)[idFLOT1,])
X <- sweep(X, 2, colMeans(X))

# get expression levels of probes mapping to FLOT genes
Y <- t(exprs(vdx)[idERBB2,])
Y <- sweep(Y, 2, colMeans(Y))

# regression analysis
summary(lm(formula = Y[,1] ~ X[,1] + X[,2] + X[,3] + X[,4]))

# correlation among the covariates
cor(X)
\end{lstlisting}
Prior to the regression analysis, we first assess whether there is collinearity among the FLOT-1 probes through evaluation of the correlation matrix. This reveals a strong correlation ($\hat{\rho} = 0.91$) between the second and third probe. All other cross-correlations do not exceed the 0.20 (in absolute sense). Hence, there is strong collinearity among the columns of the design matrix in the to-be-performed regression analysis.
\begin{verbatim}
Coefficients:
           Estimate Std. Error t value Pr(>|t|)
(Intercept)   0.0000     0.0633  0.0000   1.0000
X[, 1]        0.1641     0.0616  2.6637   0.0081 **
X[, 2]        0.3203     0.3773  0.8490   0.3965
X[, 3]        0.0393     0.2974  0.1321   0.8949
X[, 4]        0.1117     0.0773  1.4444   0.1496
---
Signif. codes:  0 ‘***’ 0.001 ‘**’ 0.01 ‘*’ 0.05 ‘.’ 0.1 ‘ ’ 1

Residual standard error: 1.175 on 339 degrees of freedom
Multiple R-squared:  0.04834,	Adjusted R-squared:  0.03711 
F-statistic: 4.305 on 4 and 339 DF,  p-value: 0.002072
\end{verbatim}
The output of the regression analysis above shows the first probe to be significantly associated to the expression levels of ERBB2. The collinearity of the second and third probe reveals itself in the standard errors of the effect size: for these probes the standard error is much larger than those of the other two probes. This reflects the uncertainty in the estimates. Regression analysis has difficulty to decide to which covariate the explained proportion of variation in the response should be attributed. The large standard errors of these effect sizes propagate to the test as the Wald test statistic is the ratio of the estimated effect size and its standard error. Collinear covariates are thus less likely to pass the significance threshold.
\end{example}

\noindent
Ridge regression has seen a renewed interest since the advent of high-dimensional data. This is due to the super-collinearity of a high-dimensional design matrix' covariates. Two (or multiple) covariates are \textit{super-collinear} if they are perfectly linearly dependent. High-dimensionally, a subset of covariates is perfectly linearly related as the rank of the design matrix is maximally equal to $n$ as $\mbox{rank}(\mathbf{X}) \leq \min \{ n, p \}$. This means that the dimension of the subspace spanned by the columns of $\mathbf{X}$ is smaller than or equal to $n$. The columns of $\mathbf{X}$ must thus be linearly dependent.

The super-collinearity of the design matrix prohibits the evaluation of the maximum likelihood estimator of the regression parameter. This estimator, $\hat{\bbeta} = (\mathbf{X}^{\top} \mathbf{X})^{-1} \mathbf{X}^{\top} \mathbf{Y}$, is well-defined if $(\mathbf{X}^{\top} \mathbf{X})^{-1}$ exists. It does if the square, symmetric matrix $\mathbf{X}^{\top} \mathbf{X}$ is non-singular, which requires all its eigenvalues to be nonzero. The invertibility then becomes evident from its spectral decomposition: $\mathbf{X}^{\top} \mathbf{X} = \sum_{j=1}^p \nu_{x,j} \, \mathbf{v}_{x,j} \, \mathbf{v}_{x,j}^{\top}$ with eigenvalue  $\nu_{x,j}$ and corresponding eigenvector $\mathbf{v}_{x,j}$. We can then write $(\mathbf{X}^{\top} \mathbf{X})^{-1} = \sum_{j=1}^p \nu_{x,j}^{-1} \, \mathbf{v}_{x,j} \, \mathbf{v}_{x,j}^{\top}$, which is well-defined if the reciprocals of all eigenvalues are. High-dimensionally, however, $\mbox{rank}(\mathbf{X}^{\top} \mathbf{X}) \leq n < p$. It is not of full rank and at least one of its eigenvalues equals zero, for the rank equals the number of nonzero eigenvalues. The reciprocal to a zero eigenvalue is undefined. Then, so is the inverse of $\mathbf{X}^{\top} \mathbf{X}$. Hence, the regression parameter cannot be learned by the maximum likelihood procedure from high-dimensional experimental designs.
\\
\\
The problem arising from the high-dimensionality of the design is more fundamental (than the fact that the expression $( \mathbf{X}^{\top} \mathbf{X})^{-1} \mathbf{X}^{\top} \mathbf{Y}$ cannot be evaluated numerically): the maximum likelihood estimator is non-unique. To appreciate this, consider the normal equations: $\mathbf{X}^{\top} \mathbf{X} \bbeta  =  \mathbf{X}^{\top} \mathbf{Y}$. The matrix $\mathbf{X}^{\top} \mathbf{X}$ is of rank $n$, while $\bbeta$ is a vector of length $p$. Hence, while there are $p$ unknowns, the system of linear equations from which these are to be solved effectively comprises $n$ degrees of freedom. If $p > n$, the vector $\bbeta$ cannot uniquely be determined from this system of equations. To make this more specific let $\mathcal{U}$ be the $n$-dimensional space spanned by the columns of $\mathbf{X}$ and the $p-n$-dimensional space $\mathcal{V}$ be orthogonal complement of $\mathcal{U}$, i.e. $\mathcal{V} = \mathcal{U}^{\perp}$. Then, $\mathbf{X} \mathbf{v} = \mathbf{0}_{p}$ for all $\mathbf{v} \in \mathcal{V}$. So, $\mathcal{V}$ is the non-trivial null space of $\mathbf{X}$. Consequently, as $\mathbf{X}^{\top} \mathbf{X} \mathbf{v} = \mathbf{X}^{\top} \mathbf{0}_{p} = \mathbf{0}_{n}$, the solution of the normal equations is:
\begin{eqnarray*}
\hat{\bbeta} & = & ( \mathbf{X}^{\top} \mathbf{X})^{+} \mathbf{X}^{\top} \mathbf{Y} + \mathbf{v} \qquad \mbox{for all } \mathbf{v} \in \mathcal{V},
\end{eqnarray*}
where $(\mathbf{X}^{\top} \mathbf{X})^{+}$ denotes the Moore-Penrose inverse of the $\mathbf{X}^{\top} \mathbf{X}$ (adopting the notation of \citealp{Harv2008}). For this square symmetric matrix, the generalized inverse is defined as:
\begin{eqnarray*}
(\mathbf{X}^{\top} \mathbf{X})^{+} & = & \sum\nolimits_{j=1}^p \nu_{x,j}^{-1} \, \mathbbm{1}_{\{ \nu_{x,j} \not= 0 \} } \, \mathbf{v}_{x,j} \, \mathbf{v}_{x,j}^{\top},
\end{eqnarray*}
where $\mathbf{v}_{x,j}$ are the eigenvectors of $\mathbf{X}^{\top} \mathbf{X}$ and are not -- necessarily -- an element of $\mathcal{V}$. The solution of the normal equations is thus determined up to an element from a non-trivial space $\mathcal{V}$, and there is no unique maximum likelihood estimator of the regression parameter.

\section{The ridge regression estimator} \label{sect.ridgeRegression}
The ridge regression estimator \citep{Hoer1970} is a estimator that deals with (super)-collinearity among the columns of the design matrix. It employs an ad-hoc fix to resolve the (almost) singularity of $\mathbf{X}^{\top} \mathbf{X}$. \cite{Hoer1970} simply replace  $\mathbf{X}^{\top} \mathbf{X}$ by $\mathbf{X}^{\top} \mathbf{X} + \lambda \mathbf{I}_{pp}$ with $\lambda \in (0, \infty)$. The scalar $\lambda$ is a tuning parameter, henceforth called the \textit{ridge parameter}, \textit{regularization parameter} or \textit{penalty parameter} for reasons that are evident or will become clear later. The ad-hoc fix solves the singularity as it adds a positive matrix, $\lambda \mathbf{I}_{pp}$, to a positive semi-definite one, $\mathbf{X}^{\top} \mathbf{X}$, making their sum a positive definite matrix (by virtue of Lemma 14.2.4 of \citealp{Harv2008}), rendering it invertible.

\begin{example} \label{example.supercollinearity} \mbox{ }
\\
We evaluate $\mathbf{X}^{\top} \mathbf{X}$ for a particular high-dimensional design matrix: 
\begin{eqnarray*}
\mathbf{X}^{\top} \mathbf{X}  \, \, \, = \, \, \, \left(
\begin{array}{rrr}
5 &  1 &  4
\\
1 &  2 & -1
\\
4 & -1 &  5
\end{array} \right) \qquad \mbox{ for } \qquad 
\mathbf{X} & = & \left(
\begin{array}{rrr}
1 & -1 & 2
\\
2 & 1 & 1
\end{array} \right).
\end{eqnarray*}
The employed design matrix is super-collinear: the first column is the row-wise sum of the other two columns. Consequently, $\mathbf{X}^{\top} \mathbf{X}$ is singular as its eigenvalues are $9, 3, 0$. Those of $\mathbf{X}^{\top} \mathbf{X} + \lambda \mathbf{I}_{pp}$ with $\lambda = 1$ equal 10, 4, and 1. Hence, $\mathbf{X}^{\top} \mathbf{X} + \lambda \mathbf{I}_{pp}$ is non-singular and its inverse is well-defined.
\end{example}

\begin{figure}[!h]
\begin{tabular}{rcl}
\begin{minipage}{5cm}
\mbox{ }
\\
\includegraphics[scale=0.45, angle=0]{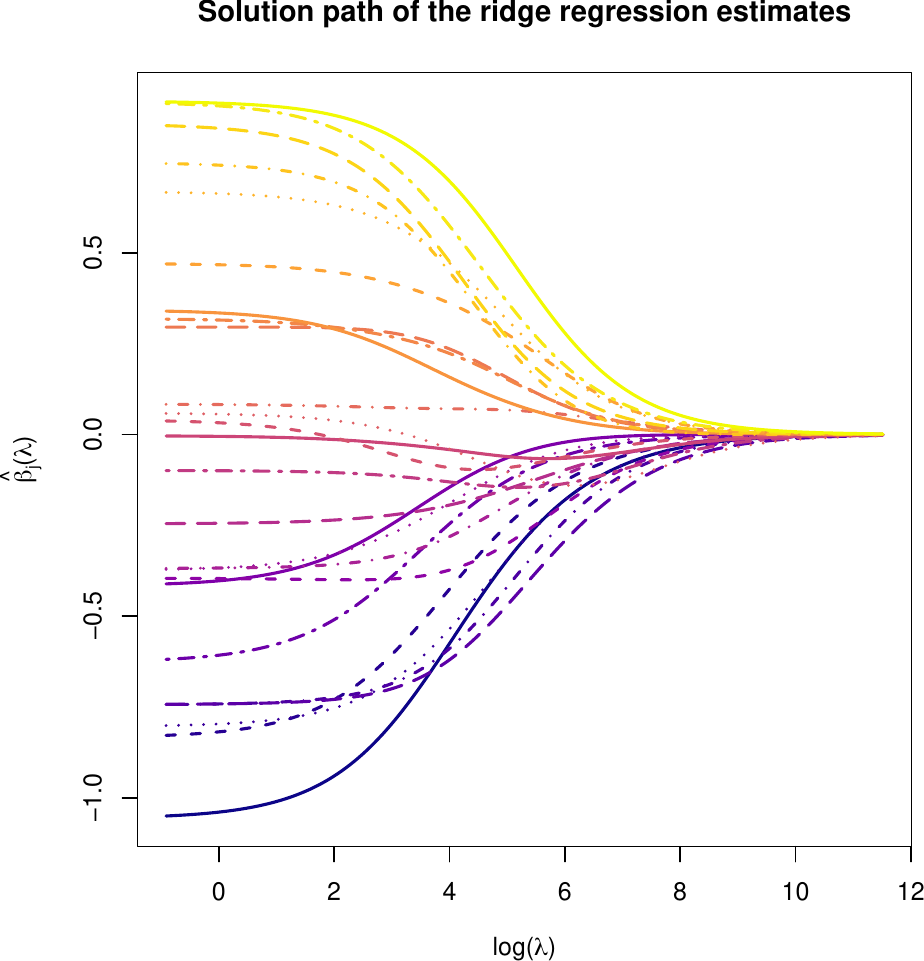}
\end{minipage}
& \begin{minipage}{2cm} \mbox{ } \end{minipage} &
\begin{minipage}{5cm}
\includegraphics[scale=0.26, angle=0]{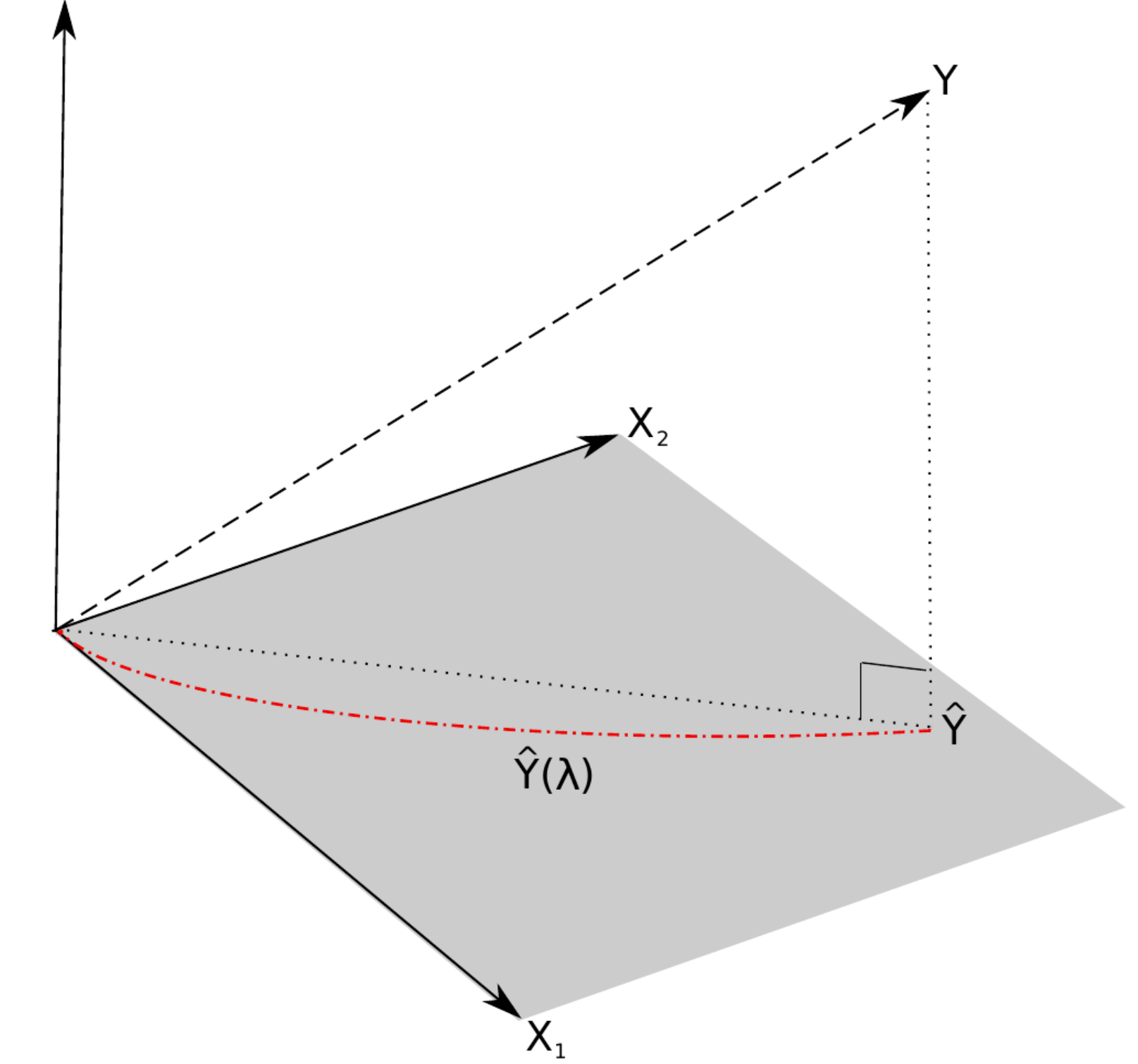}
\end{minipage}
\\
& &
\end{tabular}
\caption{Left-hand side panel: the regularization path of the ridge regression estimator for the artificial data. Right-hand side panel: the `maximum likelihood fit' $\widehat{Y}$ and the `ridge fit' $\widehat{Y}(\lambda)$ (the dashed-dotted red line) to the observation $Y$ in the (hyper)plane spanned by the covariates.
\label{fig.ridgeSolPathPlusVar}}
\end{figure}

\noindent
\cite{Hoer1970} use the ad-hoc fix for the singularity of $\mathbf{X}^{\top} \mathbf{X}$ to define the ridge regression estimator.
\begin{definition} \mbox{ } \\
The \textit{ridge regression estimator} of the regression parameter of the linear regression model is:
\begin{eqnarray} \label{form.ridgeRegressionEstimator}
\hat{\bbeta}(\lambda) & = & (\mathbf{X}^{\top} \mathbf{X} + \lambda \mathbf{I}_{pp})^{-1} \mathbf{X}^{\top} \mathbf{Y},
\end{eqnarray}
for $\lambda \in (0, \infty)$. 
\end{definition}
The ridge regression estimator is a well-defined estimator, even if $\mathbf{X}$ is high-dimensional. However, each choice of the regularization parameter $\lambda$ leads to a different ridge regression estimator. The set of all ridge regression estimates $\{ \hat{\bbeta}(\lambda) \, : \, \lambda \in [0, \infty) \}$ is called the \textit{solution path} or \textit{regularization path} of the ridge regression estimator. An example of a full regularization path of the ridge regression estimator is plotted in the left-hand side panel of Figure \ref{fig.ridgeSolPathPlusVar}. Clearly, the choice of $\lambda$ may lead to substantially different ridge regression estimates. The definition of the ridge regression estimator requires the strict positive $\lambda$. The consequences of a negative regularization parameter are discussed in Exercise \ref{question.negativePenaltyParameter}.

\mbox{ }
\\
\noindent
Low-dimensionally, the ridge regression estimator is linearly related to its maximum likelihood counterpart. To see this, define the linear operator $\mathbf{W}_{\lambda} = (\mathbf{X}^{\top} \mathbf{X} + \lambda \mathbf{I}_{pp})^{-1} \mathbf{X}^{\top} \mathbf{X}$. The ridge regression estimator $\hat{\bbeta}(\lambda)$ can then be expressed as $\mathbf{W}_{\lambda} \hat{\bbeta}$ for:
\begin{eqnarray*}
\mathbf{W}_{\lambda} \hat{\bbeta} & = &  (  \mathbf{X}^{\top} \mathbf{X} + \lambda \mathbf{I}_{pp} )^{-1} \mathbf{X}^{\top} \mathbf{X}  (\mathbf{X}^{\top} \mathbf{X})^{-1} \mathbf{X}^{\top} \mathbf{Y} \, \, \, = \, \, \,  (  \mathbf{X}^{\top} \mathbf{X} + \lambda \mathbf{I}_{pp} )^{-1} \mathbf{X}^{\top} \mathbf{Y} \, \, \, = \, \, \, \hat{\bbeta}(\lambda).
\end{eqnarray*}
The linear operator $\mathbf{W}_{\lambda}$ thus transforms the maximum likelihood estimator of the regression parameter into its ridge regularized counterpart. High-dimensionally, no such linear relation between the ridge and the minimum least squares regression estimators exists.
\\
\\
With an estimate of the regression parameter $\bbeta$ available, we can define the fit. The fit of the ridge regression estimator is defined analogous to the maximum likelihood case:
\begin{eqnarray*}
\widehat{\mathbf{Y}}(\lambda) & = & \mathbf{X} \hat{\bbeta}(\lambda) \, \, \, = \, \, \, \mathbf{X} (\mathbf{X}^{\top} \mathbf{X} + \lambda \mathbf{I}_{pp})^{-1} \mathbf{X}^{\top} \mathbf{Y} \, \, \, := \, \, \, \mathbf{H}(\lambda) \mathbf{Y}.
\end{eqnarray*}
The fit of the maximum likelihood regression estimator could be understood as a projection of $\mathbf{Y}$ onto the subspace spanned by the columns of $\mathbf{X}$. This is depicted in the right panel of Figure \ref{fig.ridgeSolPathPlusVar}, where $\widehat{Y}$ is the projection of the observation $Y$ onto the covariate space. The projected observation $\widehat{Y}$ is orthogonal  to the residual $\varepsilon = Y - \widehat{Y}$. This means the fit is the point in the covariate space closest to the observation. Put differently, the covariate space does not contain a point that is better (in some sense) in explaining the observation. Compare this to the `ridge fit' which is plotted as a dashed-dotted red line  in the right panel of Figure \ref{fig.ridgeSolPathPlusVar}. The `ridge fit' is a line, parameterized by $\{ \lambda : \lambda \in \mathbb{R}_{\geq 0} \}$, where each point on this line matches to the corresponding intersection of the regularization path $\hat{\bbeta}(\lambda)$ and the vertical line $x=\lambda$. The `ridge fit' $\widehat{Y}(\lambda)$ runs from the maximum likelihood fit $\widehat{Y} = \widehat{Y} (0)$ to an intercept-only (if present) fit in which the covariates do not contribute to the explanation of the observation. From the figure it is obvious that for any $\lambda > 0$ the `ridge fit' $\widehat{Y}(\lambda)$ is not orthogonal to the observation $Y$. In other words, the `ridge residuals' $Y - \widehat{Y}(\lambda)$ are not orthogonal to the fit $\widehat{Y}(\lambda)$ (confer Exercise \ref{question.ridgeResidualsProjection} $\!$b).  Hence, the ad-hoc fix of the ridge regression estimator resolves the non-evaluation of the estimator in the face of super-collinearity but yields a `ridge fit' that is not optimal in explaining the observation. Mathematically, this is due to the fact that the fit $\widehat{Y}(\lambda)$ corresponding to the ridge regression estimator is not a projection of $Y$ onto the covariate space (see Exercise \ref{question.ridgeResidualsProjection} $\!$a).

\section{Eigenvalue shrinkage}
We discuss a fundamental difference  between the maximum likelihood and ridge regression estimators. This sheds light on how the ridge regression overcomes the supercollinearity of the design matrix. Moreover, it provides the regularization limits of the ridge regression estimator.

For our purpose, we need the singular value decomposition of the design matrix. This we recap here as it is used at many places in the remainder. The decomposition factorizes the $n \times p$-dimensional design matrix as:
\begin{eqnarray*}
\mathbf{X} & = & \mathbf{U}_x \mathbf{D}_x \mathbf{V}_x^{\top}.
\end{eqnarray*}
In the above, $\mathbf{D}_x$ is an $n \times p$-dimensional block matrix. Its upper left block is a $\mbox{rank}(\mathbf{X})  \times \mbox{rank}(\mathbf{X})$-dimensional diagonal matrix with the singular values on the diagonal. The remaining block of appropriate dimensions comprises zeros only. The matrix $\mathbf{U}_x$ an $n \times n$-dimensional matrix with columns containing the left singular vectors (denoted $\mathbf{u}_{x,i}$ for $i=1, \ldots, n$), and $\mathbf{V}_x$ a $p \times p$-dimensional matrix with columns containing the right singular vectors (denoted $\mathbf{v}_{x,j}$ for $j=1,\ldots, p$).  $\mathbf{U}_x$ and $\mathbf{V}_x$ The matrices $\mathbf{U}_x$ and $\mathbf{V}_x$ are \textit{unitary}, i.e.: $\mathbf{U}_x^{\top} \mathbf{U}_x = \mathbf{I}_{nn} = \mathbf{U}_x\mathbf{U}_x^{\top}$ and $\mathbf{V}_x^{\top} \mathbf{V}_x = \mathbf{I}_{pp} = \mathbf{V}_x\mathbf{V}_x^{\top}$.

Now we rewrite both estimators using the design matrix' singular value decomposition and identify their difference. We start with the maximum likelihood estimator, which is assumed to exist:
\begin{eqnarray*}
\hat{\bbeta} & = & (\mathbf{X}^{\top} \mathbf{X})^{-1} \mathbf{X}^{\top} \mathbf{Y}
\qquad \qquad \qquad \, \, \, \, \, = \, \, \, (\mathbf{V}_x \mathbf{D}_x^{\top} \mathbf{U}_x^{\top} \mathbf{U}_x \mathbf{D}_x \mathbf{V}_x^{\top})^{-1} \mathbf{V}_x \mathbf{D}_x^{\top} \mathbf{U}_x^{\top} \mathbf{Y}
\\
& = & (\mathbf{V}_x \mathbf{D}_x^{\top} \mathbf{D}_x \mathbf{V}_x^{\top})^{-1} \mathbf{V}_x \mathbf{D}_x^{\top} \mathbf{U}_x^{\top} \mathbf{Y} \, \, \, = \, \, \, \mathbf{V}_x (\mathbf{D}_x^{\top} \mathbf{D}_x)^{-1} \mathbf{V}_x^{\top}  \mathbf{V}_x \mathbf{D}_x^{\top} \mathbf{U}_x^{\top} \mathbf{Y}
\\
& = & \mathbf{V}_x (\mathbf{D}_x^{\top} \mathbf{D}_x)^{-1} \mathbf{D}_x^{\top} \mathbf{U}_x^{\top} \mathbf{Y}.
\end{eqnarray*}
The block structure of the design matrix implies that matrix $(\mathbf{D}_x^{\top} \mathbf{D}_x)^{-1}  \mathbf{D}_x$ results in a $p \times n$-dimensional matrix with the reciprocal of the nonzero singular values on the diagonal of the left $p\times p$-dimensional upper left block. Similarly, the ridge regression estimator can be rewritten  as:
\begin{eqnarray}
\nonumber
\hat{\bbeta} (\lambda) & = & (\mathbf{X}^{\top} \mathbf{X} + \lambda \mathbf{I}_{pp})^{-1} \mathbf{X}^{\top} \mathbf{Y}
\\
\nonumber
& = & (\mathbf{V}_x \mathbf{D}_x^{\top} \mathbf{U}_x^{\top} \mathbf{U}_x \mathbf{D}_x \mathbf{V}_x^{\top}  + \lambda \mathbf{I}_{pp})^{-1} \mathbf{V}_x \mathbf{D}_x^{\top} \mathbf{U}_x^{\top} \mathbf{Y}
\\
\nonumber
& = & (\mathbf{V}_x \mathbf{D}_x^{\top} \mathbf{D}_x \mathbf{V}_x^{\top} + \lambda \mathbf{V}_x  \mathbf{V}_x^{\top})^{-1} \mathbf{V}_x \mathbf{D}_x^{\top} \mathbf{U}_x^{\top} \mathbf{Y}
\\
\nonumber
& = & \mathbf{V}_x (\mathbf{D}_x^{\top} \mathbf{D}_x + \lambda \mathbf{I}_{pp})^{-1} \mathbf{V}_x^{\top}  \mathbf{V}_x \mathbf{D}_x^{\top} \mathbf{U}_x^{\top} \mathbf{Y}
\\
\label{form:formRidgeSpectral}
& = & \mathbf{V}_x (\mathbf{D}_x^{\top} \mathbf{D}_x + \lambda \mathbf{I}_{pp})^{-1}  \mathbf{D}_x^{\top} \mathbf{U}_x^{\top} \mathbf{Y}.
\end{eqnarray}
The estimators thus differ only their use of the singular values. We write $(\mathbf{D}_x)_{jj} = d_{x,j}$ for the nonzero singular values on the diagonal of $\mathbf{D}_x$ and compare their use. This reveals that $ d_{x,j}^{-1} \geq d_{x,j} (d_{x,j}^2 + \lambda)^{-1}$ for all $\lambda > 0$. Thus, the ridge regularization parameter shrinks the singular values, while the maximum likelihood estimator leaves them intact.
\\
\\
We return to the problem of the super-collinearity of $\mathbf{X}$ in the high-dimensional setting. The super-collinearity implies the singularity of $\mathbf{X}^{\top} \mathbf{X}$ and prevents the calculation of the maximum likelihood estimator of the regression coefficients. However, $\mathbf{X}^{\top} \mathbf{X} + \lambda \mathbf{I}_{pp}$ is non-singular, with inverse: $(\mathbf{X}^{\top} \mathbf{X} + \lambda \mathbf{I}_{pp})^{-1} = \sum_{j=1}^p (d_{x,j}^2 + \lambda)^{-1} \mathbf{v}_{x,j} \mathbf{v}_{x,j}^{\top}$ where $d_{x,j} = 0$ for $j > \mbox{rank}(\mathbf{X})$. The right-hand side is well-defined for $\lambda > 0$.
\\
\\
The regularization limits of the ridge regression estimator can also be deduced from its `spectral formulation' (\ref{form:formRidgeSpectral}) . The lower regularization limit of the ridge regression estimator $\hat{\bbeta}(0_+) = \lim_{\lambda \downarrow 0} \hat{\bbeta}(\lambda)$ coincides with the minimum least squares estimator. This is immediate when $\mathbf{X}$ is of full rank. In the high-dimensional situation, if the dimension $p$ exceeds the sample size $n$, it follows from the limit:
\begin{eqnarray*}
\lim_{\lambda \downarrow 0} \frac{d_{x,j}}{d_{x,j}^2 + \lambda} & = & \left\{ \begin{array}{lcl} d_{x,j}^{-1} & \mbox{if} & d_{x,j} \not= 0, \\  0 & \mbox{if} & d_{x,j} = 0. \end{array} \right.
\end{eqnarray*}
The upper regularization limit is evident from the fact that $\lim_{\lambda \rightarrow \infty} d_{x,j} (d_{x,j}^2 + \lambda)^{-1} = 0$, which implies $\lim_{\lambda \rightarrow \infty} \hat{\bbeta}(\lambda) = \mathbf{0}_p$. Hence, all coefficients of the estimator are shrunken towards zero as the regularization parameter increases. This also holds for $\mathbf{X}$ with $p > n$. Furthermore, this behaviour is not strictly monotone in $\lambda$: $\lambda_{a} > \lambda_b$ does not necessarily imply  $|\hat{\beta}_j (\lambda_a) | < |\hat{\beta}_j (\lambda_b) |$. Upon close inspection, this can be witnessed from some ridge solution paths in Figure \ref{fig.ridgeSolPathPlusVar}.

The lower regularization limit $\lim_{\lambda \downarrow 0} \hat{\bbeta}(\lambda)$ corresponds to the minimum least squares estimator also known as the \textit{ridgeless} estimator, which is another unique regression estimator for studies with high-dimensional design matrices.
\begin{definition} \citep{Ishwaran2014} \\
The \textit{minimum least squares estimator} of regression parameter minimizes the sum-of-squares criterion and is of minimum length. Formally, $\hat{\bbeta}_{\mbox{{\tiny mls}}} = \arg \min_{\bbeta \in \mathbb{R}^p} \| \mathbf{Y} - \mathbf{X} \bbeta \|_2^2$ such that $\| \hat{\bbeta}_{\mbox{{\tiny mls}}} \|_2^2 < \| \bbeta \|_2^2$ for all $\bbeta$ that minimize $\| \mathbf{Y} - \mathbf{X} \bbeta \|_2^2$.
\end{definition}
The minimum least squares regression and maximum likelihood estimators coincide if $\mathbf{X}$ is of full rank. The latter is a unique minimizer of the sum-of-squares criterion and, thereby, automatically also the minimizer of minimum length. But if $\mathbf{X}$ is rank deficient, they do not coincide as $\hat{\bbeta}_{\mbox{{\tiny mls}}} = (\mathbf{X}^{\top} \mathbf{X})^+ \mathbf{X}^{\top} \mathbf{Y}$. To see this, recall from the above that $\| \mathbf{Y} - \mathbf{X} \bbeta \|_2^2$ is minimized by $(\mathbf{X}^{\top} \mathbf{X})^+ \mathbf{X}^{\top} \mathbf{Y} + \mathbf{v}$ for all $\mathbf{v} \in \mathcal{V}$. The length of these minimizers is:
\begin{eqnarray*}
\| (\mathbf{X}^{\top} \mathbf{X})^+ \mathbf{X}^{\top} \mathbf{Y} + \mathbf{v} \|_2^2 & = & \| (\mathbf{X}^{\top} \mathbf{X})^+ \mathbf{X}^{\top} \mathbf{Y} \|_2^2 + 2  \mathbf{Y}^{\top} \mathbf{X} (\mathbf{X}^{\top} \mathbf{X})^+ \mathbf{v} + \| \mathbf{v} \|_2^2,
\end{eqnarray*}
which, by the orthogonality of $\mathcal{V}$ and the space spanned by the columns of $\mathbf{X}$, equals $\| (\mathbf{X}^{\top} \mathbf{X})^+ \mathbf{X}^{\top} \mathbf{Y} \|_2^2 + \| \mathbf{v} \|_2^2$. Clearly, any nontrivial $\mathbf{v}$, i.e. $\mathbf{v} \not= \mathbf{0}_p$, results in $\| (\mathbf{X}^{\top} \mathbf{X})^+ \mathbf{X}^{\top} \mathbf{Y} \|_2^2 + \| \mathbf{v} \|_2^2 > \| (\mathbf{X}^{\top} \mathbf{X})^+ \mathbf{X}^{\top} \mathbf{Y} \|_2^2$ and, thus, $\hat{\bbeta}_{\mbox{{\tiny mls}}} = (\mathbf{X}^{\top} \mathbf{X})^+ \mathbf{X}^{\top} \mathbf{Y}$.

\subsection{Principal component regression}
Principal component regression is a close relative to ridge regression that too can be applied in high-dimensional settings. Principal component regression explains the response not by the covariates themselves but by linear combinations of the covariates as defined by the principal components of $\mathbf{X}$. Let $\mathbf{U}_x \mathbf{D}_x \mathbf{V}_x^{\top}$ be the singular value decomposition of $\mathbf{X}$. The $k_0$-th principal component of $\mathbf{X}$ is then $\mathbf{X} \mathbf{v}_{k_0}$, henceforth denoted $\mathbf{z}_i$. Let $\mathbf{Z}_k$ be the matrix of the first $k$ principal components, i.e. $\mathbf{Z}_k = \mathbf{X} \mathbf{V}_{x,k}$ where $\mathbf{V}_{x,k}$ contains the first $k$ right singular vectors as columns. Principal component regression then amounts to regressing the response $\mathbf{Y}$ onto $\mathbf{Z}_{k}$, that is, it fits the model $\mathbf{Y} = \mathbf{Z}_k \ggamma + \vvarepsilon$. The least squares estimator of $\ggamma$ then is (with some abuse of notation):
\begin{eqnarray*}
\hat{\ggamma} & = & (\mathbf{Z}_k^{\top} \mathbf{Z}_k)^{-1} \mathbf{Z}_k^{\top} \mathbf{Y}
\\
& = & (\mathbf{V}_{x,k}^{\top} \mathbf{X}^{\top} \mathbf{X} \mathbf{V}_{x,k})^{-1} \mathbf{V}_k^{\top} \mathbf{X}^{\top} \mathbf{Y}
\\
& = & (\mathbf{V}_{x,k}^{\top} \mathbf{V}_x \mathbf{D}_x^{\top} \mathbf{U}_x^{\top} \mathbf{U}_x \mathbf{D}_x \mathbf{V}_x^{\top} \mathbf{V}_{x,k})^{-1} \mathbf{V}_{x,k}^{\top} \mathbf{V}_x \mathbf{D}_x^{\top} \mathbf{U}_x^{\top} \mathbf{Y}
\\
& = & (\mathbf{I}_{kp} \mathbf{D}_x^{\top} \mathbf{D}_x \mathbf{I}_{pk})^{-1} \mathbf{I}_{kp} \mathbf{D}_x^{\top} \mathbf{U}_x^{\top} \mathbf{Y}
\\
& = & (\mathbf{D}_{x,k}^{\top} \mathbf{D}_{x,k})^{-1}  \mathbf{D}_{x,k}^{\top} \mathbf{U}_x^{\top} \mathbf{Y} 
\\
& = &  (\mathbf{D}_{x,k}^{\top} \mathbf{D}_{x,k})^{-1}  \mathbf{D}_{x,k}^{\top} \mathbf{U}_x^{\top} \mathbf{Y},
\end{eqnarray*}
where $\mathbf{D}_{x,k}$ is a submatrix of $\mathbf{D}_x$ formed from $\mathbf{D}_x$ by removal of the last $p-k$ columns. Similarly, $\mathbf{I}_{kp}$ and $\mathbf{I}_{pk}$ are obtained from $\mathbf{I}_{pp}$ by removal of the last $p-k$ rows and columns, respectively. The principal component regression estimator of $\bbeta$ then is $\hat{\bbeta}_{\mbox{{\tiny pcr}}} = \mathbf{V}_{x,k} (\mathbf{D}_{x,k}^{\top} \mathbf{D}_{x,k})^{-1} \mathbf{D}_{x,k}^{\top} \mathbf{U}_x^{\top} \mathbf{Y}$. If $k$ is set equal to the column rank of $\mathbf{X}$ -- and thus to the rank of $\mathbf{X}^{\top} \mathbf{X}$ -- the principal component regression estimator can be written as $\hat{\bbeta}_{\mbox{{\tiny pcr}}} = (\mathbf{X}^{\top} \mathbf{X})^+ \mathbf{X}^{\top} \mathbf{Y}$, where $\mathbf{A}^+$ denotes the Moore-Penrose inverse of matrix $\mathbf{A}$.

The relation between ridge and principal component regression becomes clear when their corresponding estimators are written in terms of the singular value decomposition of $\mathbf{X}$:
\begin{eqnarray*}
\hat{\bbeta}_{\mbox{{\tiny pcr}}} & = & \mathbf{V}_{x,k} (\mathbf{I}_{kp} \mathbf{D}_x^{\top} \mathbf{D}_x \mathbf{I}_{pk})^{-1} \mathbf{I}_{kp} \mathbf{D}_x^{\top} \mathbf{U}_x^{\top} \mathbf{Y},
\\
\hat{\bbeta} (\lambda) & = & \mathbf{V}_x (\mathbf{D}_x^{\top} \mathbf{D}_x + \lambda \mathbf{I}_{pp})^{-1} \mathbf{D}_x^{\top} \mathbf{U}_x^{\top} \mathbf{Y}.
\end{eqnarray*}
Both operate on the singular values of the design matrix. But where principal component regression thresholds the singular values of $\mathbf{X}$, ridge regression shrinks them (depending on their size). Hence, one applies a discrete map on the singular values while the other a continuous one.

\section{Moments} \label{sect:ridgeMoments}
The first two moments of the ridge regression estimator are derived. Next, the performance of the ridge regression estimator is studied in terms of the  mean squared error, which combines the first two moments.

\subsection{Expectation} \label{sect:ridgeExpectation}
The expectation of the ridge regression estimator is:
\begin{eqnarray} 
\nonumber
\mathbb{E} [ \hat{\bbeta}(\lambda) ] & = & \mathbb{E} [ (\mathbf{X}^{\top} \mathbf{X} + \lambda \mathbf{I}_{pp})^{-1} \mathbf{X}^{\top} \mathbf{Y} ] \, \, \, = \, \, \,  (\mathbf{X}^{\top} \mathbf{X} + \lambda \mathbf{I}_{pp})^{-1}  \, \mathbf{X}^{\top} \mathbb{E} ( \mathbf{Y} ) 
\\
\label{form:ridgeExpectation}
& = &  (\mathbf{X}^{\top} \mathbf{X} + \lambda \mathbf{I}_{pp})^{-1} \mathbf{X}^{\top} \mathbf{X} \bbeta \, \, \, \, \, \,  = \, \, \, \bbeta - \lambda  (\mathbf{X}^{\top} \mathbf{X} + \lambda \mathbf{I}_{pp})^{-1}  \bbeta.
\end{eqnarray}
Clearly, $\mathbb{E} \big[ \hat{\bbeta}(\lambda) \big] \not= \bbeta$ for any $\lambda > 0$. Hence, the ridge regression estimator is biased.

\begin{figure}[!h]
\centering
\begin{tabular}{rcl}
\includegraphics[scale=0.45, angle=0]{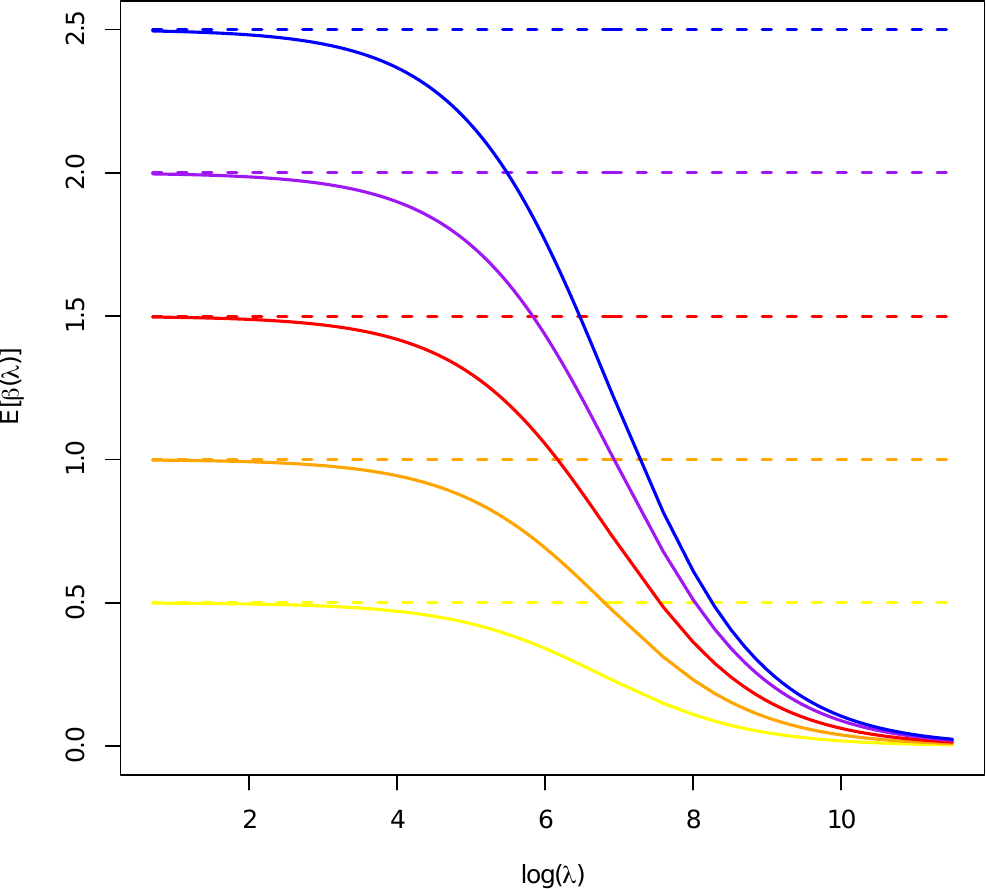}
& 
&
\includegraphics[scale=0.45, angle=0]{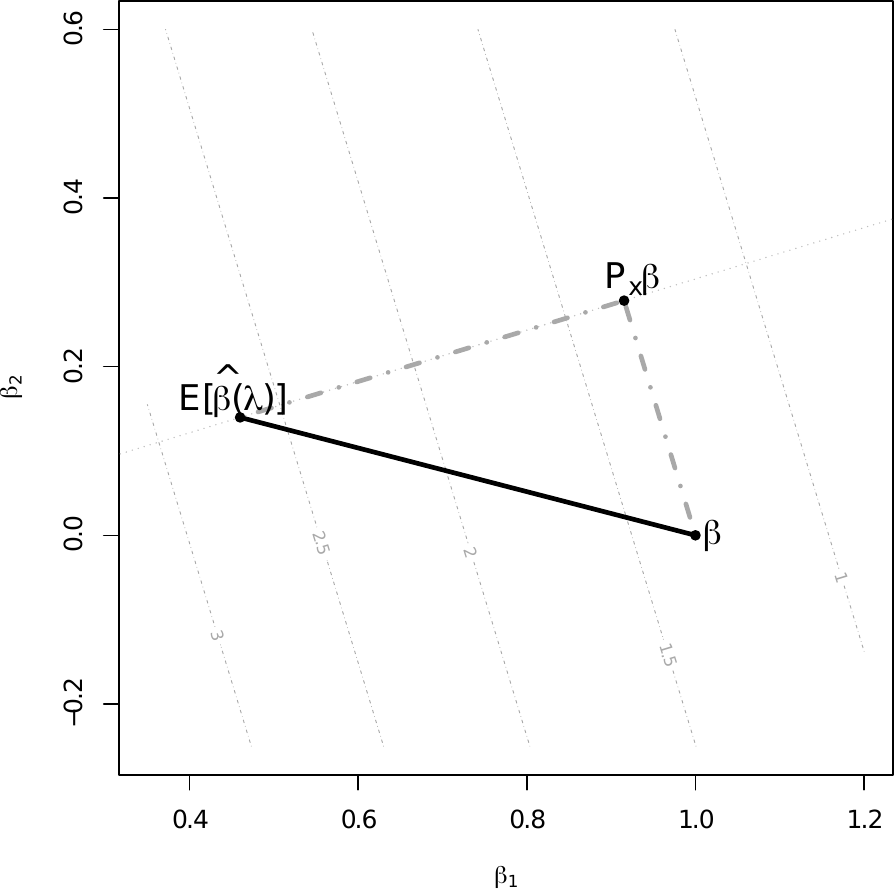}
\end{tabular}
\caption{Left panel: the expectation (solid lines) of the ridge regression estimator vs. the (logarithm of the) regularization parameter. For comparison, the true parameter values has been added (dashed lines).
Right panel: illustration in the $p=2, n=1$-setting of the decomposition of the ridge regression estimator's bias into a high-dimensional and regularization part. The level sets of the sum-of-squares have been added as thin grey lines.  
\label{fig.ridgeExpectationAndBias}}
\end{figure}

\begin{example} \textit{(Orthonormal design matrix)} \label{example.orthoronormalDesign}
\\
Consider an orthonormal design matrix $\mathbf{X}$, i.e.:
$\mathbf{X}^{\top} \mathbf{X}  =  \mathbf{I}_{pp} = (\mathbf{X}^{\top} \mathbf{X})^{-1}$. The relation between the maximum likelihood and ridge regression estimator then is:
\begin{eqnarray*}
\hat{\bbeta}(\lambda) & = & (\mathbf{X}^{\top} \mathbf{X} + \lambda \mathbf{I}_{pp})^{-1} \mathbf{X}^{\top} \mathbf{Y}
\, \, \, = \, \, \, (\mathbf{I}_{pp} + \lambda \mathbf{I}_{pp})^{-1} \mathbf{X}^{\top} \mathbf{Y}
\\
& = & (1 + \lambda)^{-1} \mathbf{I}_{pp} \mathbf{X}^{\top} \mathbf{Y}
\qquad \, \, = \, \, \, (1 + \lambda)^{-1} (\mathbf{X}^{\top} \mathbf{X})^{-1} \mathbf{X}^{\top} \mathbf{Y}
\, \, \, = \, \, \,  (1 + \lambda)^{-1} \hat{\bbeta}.
\end{eqnarray*}
Hence, the ridge regression estimator scales the maximum likelihood estimator by a factor. When taking the expectation on both sides, it is evident that the ridge regression estimator is biased: $\mathbb{E}[ \hat{\bbeta}(\lambda) ] = \mathbb{E}[ (1 + \lambda)^{-1} \hat{\bbeta} ] = (1 + \lambda)^{-1} \mathbb{E}(  \hat{\bbeta} ) = (1 + \lambda)^{-1} \bbeta \not= \bbeta$ if $\lambda > 0$. From this, it also clear that the estimator's expectation vanishes as $\lambda \rightarrow \infty$. This is depicted by the left-hand side panel of Figure \ref{fig.ridgeExpectationAndBias}. The limiting behaviour of the ridge regression estimator does not depend on the specifics of the data set as is evident from the expectation's analytic expression (\ref{form:ridgeExpectation}).
\end{example}

\noindent
The bias of the ridge regression estimator decomposes into two parts, one attributable to the penalization and another to the high-dimensionality of the study design. Prior to that, we show that ridge regression estimator `lives' in the subspace spanned by the rows of the design matrix.

\begin{proposition}  \label{prop:ridgeProjection} \textit{(after \citealp{Shao2012})}
\\
Let $p > n$ and assume $\mbox{rank}(\mathbf{X}) = n$. Define the matrix $\mathbf{P}_x = \mathbf{X}^{\top} (\mathbf{X} \mathbf{X}^{\top})^+ \mathbf{X}$ that projects the parameter space $\mathbb{R}^p$ onto the subspace $\mathcal{R}(\mathbf{X}) \subset \mathbb{R}^p$ spanned by the rows of the design matrix $\mathbf{X}$. Then, $\mathbf{P}_x \hat{\bbeta}(\lambda) = \hat{\bbeta}(\lambda)$,
\end{proposition}

\begin{proof}

We first rewrite the projection matrix. Consider the singular value decomposition $\mathbf{X} = \mathbf{U}_x \mathbf{D}_x \mathbf{V}_x^{\top}$, with matrices defined as before. Substitute this decomposition of $\mathbf{X}$ into the projection matrix and find that:
\begin{eqnarray*}
\mathbf{P}_x & = & \mathbf{X}^{\top} (\mathbf{X} \mathbf{X}^{\top})^+ \mathbf{X}
\qquad \qquad \, \, \, = \, \, \,  \mathbf{V}_x \mathbf{D}_x^{\top} \mathbf{U}_x^{\top}  ( \mathbf{U}_x \mathbf{D}_x \mathbf{V}_x^{\top}  \mathbf{V}_x \mathbf{D}_x^{\top} \mathbf{U}_x^{\top} )^{+} \mathbf{U}_x \mathbf{D}_x \mathbf{V}_x^{\top}
\\
& = &  \mathbf{V}_x \mathbf{D}_x^{\top}  ( \mathbf{D}_x \mathbf{D}_x^{\top}  )^{+}  \mathbf{D}_x \mathbf{V}_x^{\top} \, \, \, = \, \, \, \mathbf{V}_x \mathbf{I}_{pn} \mathbf{I}_{np} \mathbf{V}_x^{\top}.
\end{eqnarray*}
The identity above does not hold if the rows of $\mathbf{X}$ harbor linear dependency. For instance, if the study contains a replicate corresponding to a duplicated row. Hence, the proposition's assumption on the (row) rank of $\mathbf{X}$. This assumption does not hamper the envisioned bias decomposition, for the definition of the projection matrix can be modified. E.g., without one of the instances of the duplicated row, such that the identity in the display above holds and it still projects onto $\mathcal{R}(\mathbf{X})$. 

The claimed identity now follows straightforwardly after substitution of the rewritten projection matrix and the ridge regression estimator's expression (\ref{form:formRidgeSpectral}): 
\begin{eqnarray*}
\mathbf{P}_x \hat{\bbeta}(\lambda) & = & \mathbf{V}_x  \mathbf{I}_{pn} \mathbf{I}_{np} \mathbf{V}_x^{\top} \mathbf{V}_x (\mathbf{D}_x^{\top} \mathbf{D}_x + \lambda \mathbf{I}_{pp})^{-1}  \mathbf{D}_x^{\top} \mathbf{U}_x^{\top} \mathbf{Y} 
\\
& = & \mathbf{V}_x (\mathbf{D}_x^{\top} \mathbf{D}_x + \lambda \mathbf{I}_{pp})^{-1}  \mathbf{I}_{pn} \mathbf{I}_{np} \mathbf{D}_x^{\top} \mathbf{U}_x^{\top} \mathbf{Y}  \qquad \, \, \, \, \, \, = \, \, \,  \hat{\bbeta}(\lambda),
\end{eqnarray*}
where the second identity follows from  \textit{i)} the fact that $\mathbf{V}_x$ is unitary, \textit{ii)} $\mathbf{A} \mathbf{B} = \mathbf{B} \mathbf{A}$  if $\mathbf{A}$ and $\mathbf{B}$ are diagonal matrices and \textit{iii)} $\mathbf{I}_{pn} \mathbf{I}_{np} \mathbf{D}_x^{\top} = \mathbf{D}_x^{\top}$. The ridge regression estimator is thus unaffected by the projection, as $\mathbf{P}_x \hat{\bbeta}(\lambda) = \hat{\bbeta}(\lambda)$, and it must therefore already be an element of the projected subspace $\mathcal{R}(\mathbf{X})$. 
\end{proof}

\begin{corollary}  \label{corol:ridgeBiasDecomposition} \textit{(after \citealp{Shao2012})}
\\
Let $p > n$, assume $\mbox{rank}(\mathbf{X}) = n$, and define $\mathbf{P}_x$ as in Proposition \ref{prop:ridgeProjection}. Then, the bias of the ridge regression estimator decomposes as $\mathbb{E}[ \hat{\bbeta}(\lambda)  - \bbeta] = \mathbf{P}_x  \mathbb{E}[ \hat{\bbeta}(\lambda)  - \bbeta] + (\mathbf{P}_x - \mathbf{I}_{pp}) \bbeta$.
\end{corollary}

\begin{proof}
The bias decomposition follows directly from Proposition \ref{prop:ridgeProjection}:
\begin{eqnarray*}
\mathbb{E}[ \hat{\bbeta}(\lambda)  - \bbeta] & = & \mathbb{E}[ \hat{\bbeta}(\lambda)  - \mathbf{P}_x \bbeta + \mathbf{P}_x \bbeta - \bbeta] \, \, \, = \, \, \, \mathbf{P}_x  \mathbb{E}[ \hat{\bbeta}(\lambda)  - \bbeta] + (\mathbf{P}_x - \mathbf{I}_{pp}) \bbeta,
\end{eqnarray*}
where the second identity is a consequence of the linearity of the expectation operator.
\end{proof}

\noindent
The first summand of Corollary \ref{corol:ridgeBiasDecomposition}'s bias decomposition represents the bias of the ridge regression estimator to the projection of the true parameter value, whereas the second summand is the bias introduced by the high-dimensionality of the study design. Either if \textit{i)} $\mathbf{X}$ is of full row rank (i.e. the study design is low-dimensional and $\mathbf{P}_x = \mathbf{I}_{pp}$) or if \textit{ii)} the true regression parameter $\bbeta$ is an element of the projected subspace (i.e. $\bbeta = \mathbf{P}_x \bbeta \in \mathcal{R}(\mathbf{X})$), the second summand of the bias will vanish. The right-hand side panel of Figure \ref{fig.ridgeExpectationAndBias} illustrates the bias decomposition for the simplest high-dimensional setting with $p=2$ and $n=1$.

\subsection{Variance} \label{sect:ridgeVariance}
The second moment of the ridge regression estimator is straightforwardly obtained as
\begin{eqnarray*}
\mbox{Var}[ \hat{\bbeta}(\lambda) ] & = & \mbox{Var} [ (\mathbf{X}^{\top} \mathbf{X} + \lambda \mathbf{I}_{pp})^{-1} \mathbf{X}^{\top} \mathbf{Y}]  \, \, \, = \, \, \,  \sigma^2 (  \mathbf{X}^{\top} \mathbf{X} + \lambda \mathbf{I}_{pp} )^{-1}  \mathbf{X}^{\top} \mathbf{X}  (  \mathbf{X}^{\top} \mathbf{X} + \lambda \mathbf{I}_{pp} )^{-1},
\end{eqnarray*}
in which we have used $\mbox{Var}(\mathbf{A} \mathbf{Y}) = \mathbf{A} \mbox{Var}( \mathbf{Y}) \mathbf{A}^{\top}$ for a non-random matrix $\mathbf{A}$  and $ \mbox{Var}(\hat{\bbeta} ) = \sigma^2 (\mathbf{X}^{\top} \mathbf{X})^{-1}$. 

We characterize the behavior of the variance of the ridge regression estimator. We first observe that, similar to the expectation, it vanishes as $\lambda$ tends to infinity, i.e. $\lim_{\lambda \rightarrow \infty} \mbox{Var} \big[ \hat{\bbeta}(\lambda) \big] = \mathbf{0}_{pp}$. Hence, the variance of the ridge regression estimator decreases towards zero as the regularization parameter becomes large. This is illustrated in the right panel of Figure \ref{fig.ridgeSolPathPlusVar} for the data of Example \ref{example.supercollinearity}. 
Secondly, the variance of the maximum likelihood estimator is `larger' than that of the ridge regression estimator (Proposition \ref{prop:VarInequalityMLandRidge}). 

\begin{proposition} \label{prop:VarInequalityMLandRidge} \mbox{ } \\
The variance of the maximum likelihood regression estimator exceeds, in the positive definite ordering sense, that of the ridge regression estimator, $\mbox{Var} ( \hat{\bbeta} ) \succeq \mbox{Var}[ \hat{\bbeta}(\lambda) ]$, with the inequality being strict if $\lambda > 0$.
\end{proposition}
\begin{proof}
Use the analytic expression of the variance of the ridge regression estimator to study its difference to that of the maximum likelihood regression estimator:
\begin{eqnarray*}
\mbox{Var}( \hat{\bbeta} ) - \mbox{Var}[ \hat{\bbeta}(\lambda) ] & = & \sigma^2 [(\mathbf{X}^{\top} \mathbf{X})^{-1} - \mathbf{W}_{\lambda}  (\mathbf{X}^{\top} \mathbf{X})^{-1}  \mathbf{W}_{\lambda}^{\top} ]
\\
& = & \sigma^2 \mathbf{W}_{\lambda}  \{ [\mathbf{I} + \lambda (\mathbf{X}^{\top} \mathbf{X})^{-1} ] (\mathbf{X}^{\top} \mathbf{X})^{-1}  [\mathbf{I} + \lambda (\mathbf{X}^{\top} \mathbf{X})^{-1} ]^{\top} - (\mathbf{X}^{\top} \mathbf{X})^{-1} \} \mathbf{W}_{\lambda}^{\top}
\\
& = & \sigma^2 \mathbf{W}_{\lambda}   [ 2 \, \lambda \, (\mathbf{X}^{\top} \mathbf{X})^{-2} + \lambda^2 (\mathbf{X}^{\top} \mathbf{X})^{-3} ] \mathbf{W}_{\lambda}^{\top}
\\
& = & \sigma^2 (  \mathbf{X}^{\top} \mathbf{X} + \lambda \mathbf{I}_{pp} )^{-1}   [ 2 \lambda  \mathbf{I}_{pp} + \lambda^2 (\mathbf{X}^{\top} \mathbf{X})^{-1} ] [ ( \mathbf{X}^{\top} \mathbf{X} + \lambda \mathbf{I}_{pp} )^{-1} ]^{\top}.
\end{eqnarray*}
This difference is non-negative definite as each component in the matrix product is non-negative definite, and even positive definite if either $\lambda > 0$ or $\mbox{rank} (\mathbf{X}^{\top} \mathbf{X}) = p$. 
\end{proof}

\noindent
The variance inequality of Proposition \ref{prop:VarInequalityMLandRidge} can be interpreted in terms of the stochastic behaviour of both involved estimates. This is illustrated by the next example.

\begin{figure}[!h]
\begin{tabular}{crcl}
\hspace{-1.0cm} & \includegraphics[scale=0.48, angle=0]{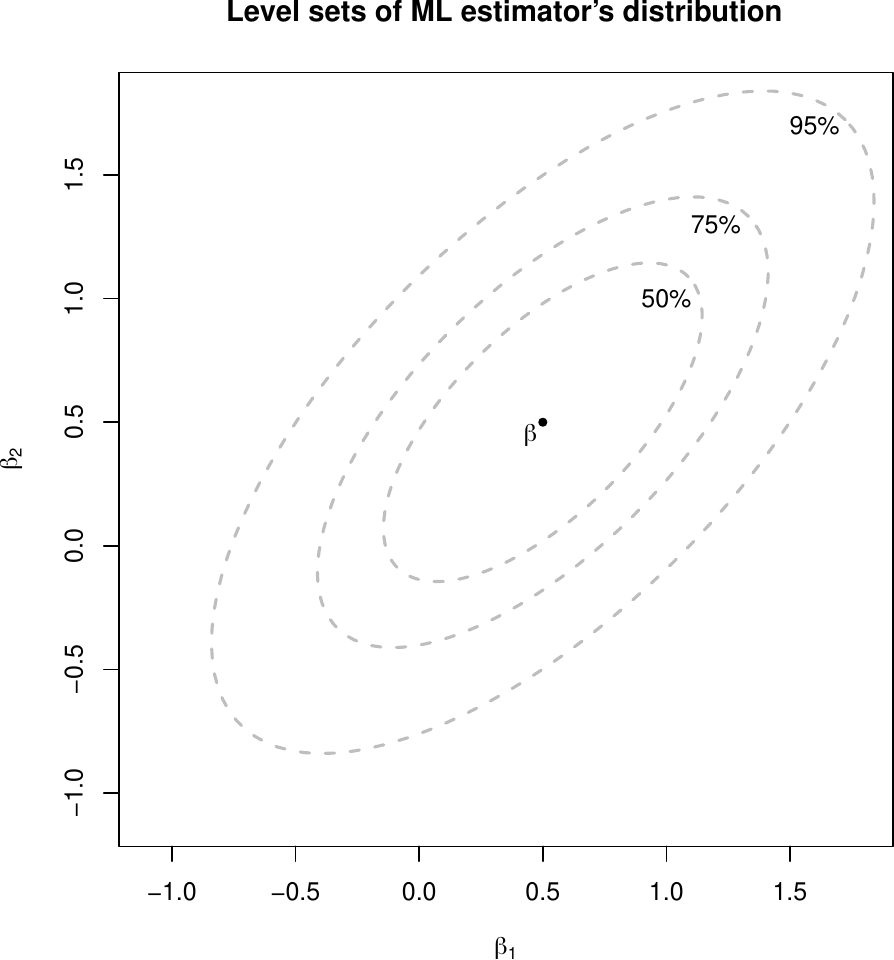}
& \hspace{-0.1cm} &
\includegraphics[scale=0.48, angle=0]{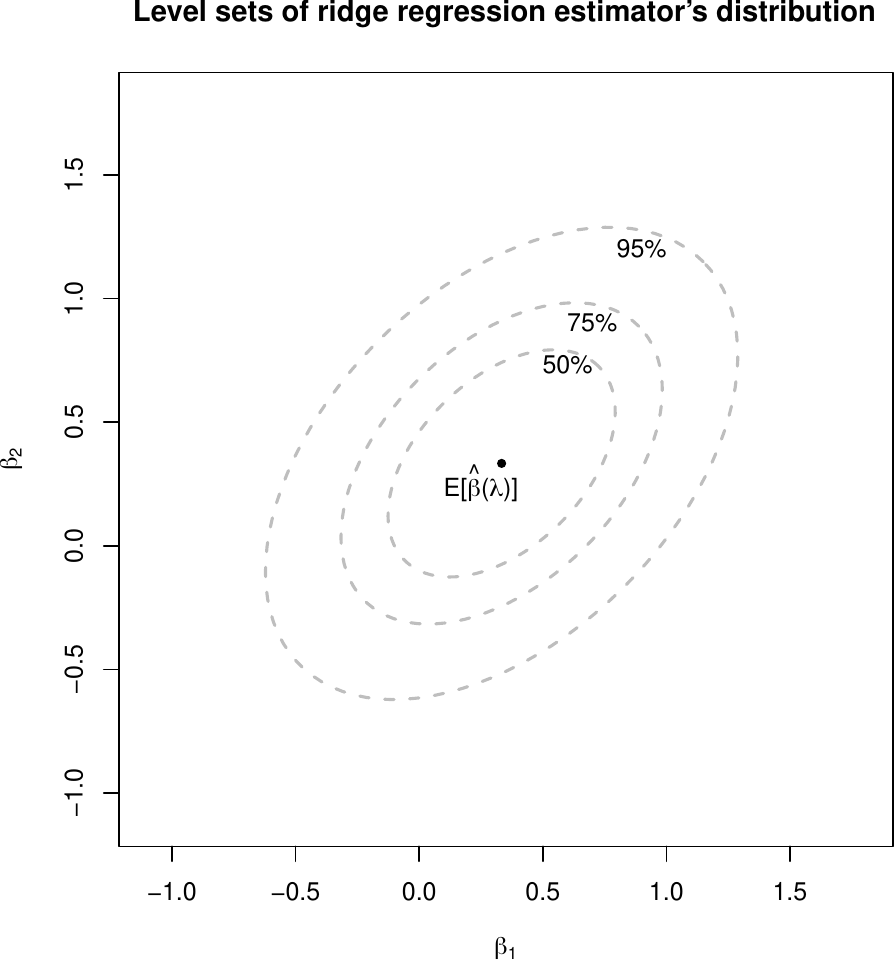}
\end{tabular}
\caption{Level sets of the distribution of the maximum likelihood (left-hand side panel) and ridge (right-hand side panel) regression estimators.} \label{fig.varOLSandRidge}
\end{figure}

\begin{example} \textit{(Variance comparison)}
\\
Consider the design matrix:
\begin{eqnarray*}
\mathbf{X}^{\top} & = & \left(
\begin{array}{rrrr}
-1 & 0 & 2 & 1
\\
2 & 1 & -1 & 0
\end{array} \right).
\end{eqnarray*}
The variances of the maximum likelihood and ridge (with $\lambda=1$) estimates of the regression coefficients then are:
\begin{eqnarray*}
\mbox{Var}(\hat{\bbeta}) & = & \sigma^2 \left(
\begin{array}{rr}
0.3 & 0.2
\\
0.2 & 0.3
\end{array} \right)
\qquad \mbox{and} \qquad
\mbox{Var}[\hat{\bbeta}(\lambda)] \, \, \, = \, \, \, \sigma^2 \left(
\begin{array}{rr}
0.1524 & 0.0698
\\
0.0698 & 0.1524
\end{array} \right).
\end{eqnarray*}
These variances can be used to construct levels sets of the distribution of the estimates. The level sets that contain 50\%, 75\% and 95\% of the distribution of the maximum likelihood and ridge regression estimates are plotted in Figure \ref{fig.varOLSandRidge}. In line with inequality of Proposition \ref{prop:VarInequalityMLandRidge}, the level sets of the ridge regression estimate are smaller than that of the maximum likelihood one. The former thus varies less.
\end{example}

\begin{contexample}\textbf{\ref{example.orthoronormalDesign}} \textit{(Orthonormal design matrix, continued)}
\\
Assume the design matrix $\mathbf{X}$ is orthonormal. Then, $\mbox{Var} ( \hat{\bbeta} ) = \sigma^2 \mathbf{I}_{pp}$ and
\begin{eqnarray*}
\mbox{Var}[ \hat{\bbeta}(\lambda) ] & = & \sigma^2 \mathbf{W}_{\lambda}  (\mathbf{X}^{\top} \mathbf{X})^{-1}  \mathbf{W}_{\lambda}^{\top} \, \, \,  = \, \, \,  \sigma^2 (\mathbf{I}_{pp} + \lambda \mathbf{I}_{pp} )^{-1}   \mathbf{I}_{pp}  [ (\mathbf{I}_{pp} + \lambda \mathbf{I}_{pp} )^{-1} ]^{\top}\, \, \, = \, \, \, \sigma^2 (1 + \lambda )^{-2}   \mathbf{I}_{pp}.
\end{eqnarray*}
As the regularization parameter $\lambda$ is non-negative, the former exceeds the latter. In particular, the  expression after the utmost right equality sign vanishes as $\lambda \rightarrow \infty$.
\end{contexample}

\noindent
The variance of the ridge regression estimator may be decomposed in the same way as its bias (cf. the end of Section \ref{sect:ridgeExpectation}). There is, however, no contribution of the high-dimensionality of the study design as that is non-random and, consequently, exhibits no variation. Hence, the variance only relates to the variation in the projected subspace $\mathcal{R}(\mathbf{X})$ as is obvious from:
\begin{eqnarray*}
\mbox{Var}[ \hat{\bbeta}(\lambda) ] & = & \mbox{Var}[ \mathbf{P}_x \hat{\bbeta}(\lambda) ] \, \, \, = \, \, \, \mathbf{P}_x  \mbox{Var}[ \hat{\bbeta}(\lambda) ] \mathbf{P}_x^{\top} \, \, \, = \, \, \,  \mbox{Var}[ \hat{\bbeta}(\lambda) ].
\end{eqnarray*}
Perhaps this is seen more clearly when writing the variance of the ridge regression estimator in terms of the matrices that constitute the singular value decomposition of $\mathbf{X}$:
\begin{eqnarray*}
\mbox{Var}[ \hat{\bbeta}(\lambda) ] & = & \mathbf{V}_x (\mathbf{D}_x^{\top} \mathbf{D}_x + \lambda \mathbf{I}_{pp})^{-1} \mathbf{D}_x^{\top} \mathbf{D}_x (\mathbf{D}_x^{\top} \mathbf{D}_x + \lambda \mathbf{I}_{pp})^{-1} \mathbf{V}_x^{\top}.
\end{eqnarray*}
High-dimensionally, $(\mathbf{D}_x^{\top} \mathbf{D}_x)_{jj} = 0$  for $j=n+1, \ldots, p$. And if $(\mathbf{D}_x^{\top} \mathbf{D}_x)_{jj} = 0$, so is  $[(\mathbf{D}_x^{\top} \mathbf{D}_x + \lambda \mathbf{I}_{pp})^{-1} \mathbf{D}_x^{\top} \mathbf{D}_x \\ (\mathbf{D}_x^{\top} \mathbf{D}_x + \lambda \mathbf{I}_{pp})^{-1}]_{jj} = 0$. Hence, the variance is determined by the first $n$ columns of $\mathbf{V}_x$. When $n < p$, the variance is then to be interpreted as the spread of the ridge regression estimator (with the same choice of $\lambda$) when the study is repeated with exactly the same design matrix such that the resulting estimator is confined to the same subspace $\mathcal{R}(\mathbf{X})$. The following \texttt{R}-script illustrates this by an arbitrary data example (plot not shown).
\begin{lstlisting}
# set parameters
X      <- matrix(rnorm(2), nrow=1)
betas  <- matrix(c(2, -1), ncol=1)
lambda <- 1

# generate multiple ridge regression estimators with a fixed design matrixs
bHats <- numeric()
for (k in 1:1000){
	Y     <- matrix(X %*% betas + rnorm(1), ncol=1)
	bHats <- rbind(bHats, t(solve(t(X) %*% X + lambda * diag(2)) %*% t(X) %*% Y))
}

# plot the ridge regression estimators
plot(bHats, xlab=expression(paste(hat(beta)[1], "(", lambda, ")", sep="")), 
            ylab=expression(paste(hat(beta)[2], "(", lambda, ")", sep="")), pch=20)





\end{lstlisting}
\mbox{ }
\\
The full distribution of the ridge regression estimator is now known. The estimator, $\hat{\bbeta}(\lambda) = (\mathbf{X}^{\top} \mathbf{X} + \lambda \mathbf{I}_{pp})^{-1} \mathbf{X}^{\top} \mathbf{Y}$  is a linear estimator, linear in $\mathbf{Y}$. As $\mathbf{Y}$ is normally distributed, so is $\hat{\bbeta}(\lambda)$. Moreover, the normal distribution is fully characterized by its  first two moments, which are available. Hence:
\begin{eqnarray*}
\hat{\bbeta}(\lambda) & \sim & \mathcal{N} \{ (\mathbf{X}^{\top} \mathbf{X} + \lambda \mathbf{I}_{pp})^{-1} \mathbf{X}^{\top} \mathbf{X} \, \bbeta, ~\sigma^2 ( \mathbf{X}^{\top} \mathbf{X} + \lambda \mathbf{I}_{pp} )^{-1}  \mathbf{X}^{\top} \mathbf{X} [ ( \mathbf{X}^{\top} \mathbf{X} + \lambda \mathbf{I}_{pp} )^{-1} ]^{\top} \}.
\end{eqnarray*}
Given $\lambda$ and $\mathbf{X}$, the random behavior of the estimator is thus known. High-dimensionally, the variance is semi-positive definite and this $p$-variate normal distribution is degenerate, i.e. there is no probability mass outside $\mathcal{R}(\mathbf{X})$ the subspace of $\mathbb{R}^p$ spanned by the rows of the $\mathbf{X}$.

\subsection{Mean squared error} \label{sect:ridgeMSE}
Previously, we motivated the ridge regression estimator as an ad hoc solution to collinearity. An alternative motivation comes from studying the Mean Squared Error (MSE) of the ridge regression estimator: for a suitable choice of $\lambda$ the ridge regression estimator may outperform the maximum likelihood regression estimator in terms of the MSE. Before we prove this, we first derive the MSE of the ridge regression estimator and quote some auxiliary results. Note that, as the ridge regression estimator is compared to its maximum likelihood counterpart, throughout this subsection $n > p$ is assumed to warrant the uniqueness of the latter.

Recall that (in general) for any estimator of a parameter $\theta$:
\begin{eqnarray*}
\mbox{MSE}( \hat{\theta} ) & = & \mathbb{E} [ ( \hat{\theta} - \theta)^2 ]
\, \, \, = \, \, \,  \mbox{Var}( \hat{ \theta} ) + [\mbox{Bias} ( \hat{\theta} )]^2.
\end{eqnarray*}
This measure of the quality of the estimator thus has a convenient decomposition. The lemma below provides the MSE of the ridge regression estimator.

\begin{lemma} \label{lemma:ridgeMSE} \mbox{ } \\
The mean squared error of the ridge regression estimator is:
\begin{eqnarray}
\mbox{MSE}[\hat{\bbeta}(\lambda)] & = & \sigma^2 \, \mbox{tr} [ \mathbf{W}_{\lambda} \, (\mathbf{X}^{\top} \mathbf{X})^{-1} \, \mathbf{W}_{\lambda}^{\top} ] + \bbeta^{\top} (\mathbf{W}_{\lambda} - \mathbf{I}_{pp})^{\top} (\mathbf{W}_{\lambda} - \mathbf{I}_{pp}) \bbeta, \label{form.ridgeMSE}
\end{eqnarray}
where $\mathbf{W}_{\lambda}$ is as defined in  Section 
\ref{sect.ridgeRegression}.
\end{lemma}

\begin{proof}
Straightforward linear algebra and the expectation calculus yields:
\begin{eqnarray*}
\mbox{MSE}[\hat{\bbeta}(\lambda)] & = & \mathbb{E} [ (\mathbf{W}_{\lambda} \, \hat{\bbeta} - \bbeta)^{\top} \, (\mathbf{W}_{\lambda} \, \hat{\bbeta} - \bbeta) ]  \nonumber
\\
& = & \mathbb{E} ( \hat{\bbeta}^{\top} \mathbf{W}_{\lambda}^{\top} \,\mathbf{W}_{\lambda} \, \hat{\bbeta} ) - \mathbb{E} ( \bbeta^{\top}  \, \mathbf{W}_{\lambda} \, \hat{\bbeta}) - \mathbb{E} ( \hat{\bbeta}^{\top} \mathbf{W}_{\lambda}^{\top} \, \bbeta) + \mathbb{E} ( \bbeta^{\top} \bbeta)  \nonumber
\\
& = & \mathbb{E} ( \hat{\bbeta}^{\top} \mathbf{W}_{\lambda}^{\top} \,\mathbf{W}_{\lambda} \, \hat{\bbeta} ) - \mathbb{E} ( \bbeta^{\top} \, \mathbf{W}_{\lambda}^{\top} \mathbf{W}_{\lambda} \, \hat{\bbeta}) - \mathbb{E} ( \hat{\bbeta}^{\top} \mathbf{W}_{\lambda}^{\top} \, \mathbf{W}_{\lambda} \bbeta) + \mathbb{E} ( \bbeta^{\top} \mathbf{W}_{\lambda}^{\top} \,\mathbf{W}_{\lambda} \, \bbeta )  \nonumber
\\
& & - \mathbb{E} ( \bbeta^{\top} \mathbf{W}_{\lambda}^{\top} \,\mathbf{W}_{\lambda} \, \bbeta ) + \mathbb{E} ( \bbeta^{\top} \, \mathbf{W}_{\lambda}^{\top} \mathbf{W}_{\lambda} \, \hat{\bbeta}) + \mathbb{E} ( \hat{\bbeta}^{\top} \mathbf{W}_{\lambda}^{\top} \, \mathbf{W}_{\lambda} \bbeta)  \nonumber
\\
& &  - \mathbb{E} ( \bbeta^{\top}  \, \mathbf{W}_{\lambda} \, \hat{\bbeta}) - \mathbb{E} ( \hat{\bbeta}^{\top} \mathbf{W}_{\lambda}^{\top} \, \bbeta) + \mathbb{E} ( \bbeta^{\top} \bbeta)  \nonumber
\\
& = & \mathbb{E} [ ( \hat{\bbeta} - \bbeta )^{\top} \mathbf{W}_{\lambda}^{\top} \, \mathbf{W}_{\lambda} \, (\hat{\bbeta} - \bbeta) ]  \nonumber
\\
& & - \bbeta^{\top} \mathbf{W}_{\lambda}^{\top} \,\mathbf{W}_{\lambda} \, \bbeta  + \bbeta^{\top} \, \mathbf{W}_{\lambda}^{\top} \mathbf{W}_{\lambda} \, \bbeta + \bbeta^{\top} \mathbf{W}_{\lambda}^{\top} \, \mathbf{W}_{\lambda} \bbeta  \nonumber
\\
& &  - \bbeta^{\top}  \, \mathbf{W}_{\lambda} \, \bbeta -  \bbeta^{\top} \mathbf{W}_{\lambda}^{\top} \, \bbeta + \bbeta^{\top} \bbeta  \nonumber
\\
& = & \mathbb{E} [ ( \hat{\bbeta} - \bbeta )^{\top} \mathbf{W}_{\lambda}^{\top} \, \mathbf{W}_{\lambda} \, (\hat{\bbeta} - \bbeta) ] +  \bbeta^{\top} (\mathbf{W}_{\lambda} - \mathbf{I}_{pp})^{\top} (\mathbf{W}_{\lambda} - \mathbf{I}_{pp}) \bbeta \nonumber
\\
& = & \sigma^2 \, \mbox{tr} [ \mathbf{W}_{\lambda} \, (\mathbf{X}^{\top} \mathbf{X})^{-1} \, \mathbf{W}_{\lambda}^{\top} ] + \bbeta^{\top} (\mathbf{W}_{\lambda} - \mathbf{I}_{pp})^{\top} (\mathbf{W}_{\lambda} - \mathbf{I}_{pp}) \bbeta. 
\end{eqnarray*}
In the last step, we have used $\hat{\bbeta} \sim  \mathcal{N}[ \bbeta, \sigma^2 \, (\mathbf{X}^{\top} \mathbf{X})^{-1} ]$ and the expectation of the quadratic form of a multivariate random variable $\vvarepsilon \sim \mathcal{N}(\mmu_{\varepsilon}, \SSigma_{\varepsilon})$ that for a nonrandom symmetric positive definite matrix $\LLambda$ is $\mathbb{E} ( \vvarepsilon^{\top}  \LLambda \, \vvarepsilon) = \mbox{tr} ( \LLambda  \SSigma_{\varepsilon}) + \mmu_{\varepsilon}^{\top} \LLambda \mmu_{\varepsilon}$ (cf. \citealt{Math1992}), in which we replace $\vvarepsilon$ by $\hat{\bbeta}$ in this expectation. 
\end{proof}

\noindent
The first summand in the Lemma \ref{lemma:ridgeMSE}'s expression of $\mbox{MSE}[\hat{\bbeta}(\lambda)]$ represents the sum of the variances of the ridge regression estimator, while the second summand can be thought of the ``squared bias'' of the ridge regression estimator. In particular, $\lim_{\lambda \rightarrow \infty} \mbox{MSE}[\hat{\bbeta}(\lambda)] =  \bbeta^{\top} \bbeta$, which is the squared biased for an estimator that equals zero (as does the ridge regression estimator in the limit).

\begin{example} \textit{(Orthonormal design matrix, continued)} \label{example:ridgeMSEorthonormal}
\\
Assume the design matrix $\mathbf{X}$ is orthonormal. Then, $\mbox{MSE}[ \hat{\bbeta} ]  =  p \, \sigma^2$ and
\begin{eqnarray*}
\mbox{MSE}[ \hat{\bbeta}(\lambda) ] & = & p \, \sigma^2 (1+ \lambda)^{-2} +  \lambda^2  (1+ \lambda)^{-2} \bbeta^{\top}  \hspace{-0.03cm} \bbeta.
\end{eqnarray*}
The latter achieves its minimum at: $\lambda = p \hspace{0.02cm}\sigma^2 / \bbeta^{\top} \hspace{-0.03cm} \bbeta = \sigma^2 (p^{-1} \| \bbeta \|_2^2)^{-1}$.
\end{example}

\noindent
The following theorem and proposition are required for the proof of the main result.

\begin{theorem} \textit{(Theorem 1 of \citealp{Theo1974}}) \label{theo.Theobald1}
\\
Let $\hat{\ttheta}_1$ and $\hat{\ttheta}_2$ be (different) estimators of $\ttheta$ with second order moments:
\begin{eqnarray*}
\mathbf{M}_k & = & \mathbb{E} [ (\hat{\ttheta}_k - \ttheta) (\hat{\ttheta}_k - \ttheta)^{\top} ] \qquad \mbox{for } k=1,2,
\end{eqnarray*}
and
\begin{eqnarray*}
\mbox{MSE}(\hat{\ttheta}_k) & = & \mathbb{E} [ (\hat{\ttheta}_k - \ttheta)^{\top} \mathbf{A} (\hat{\ttheta}_k - \ttheta) ] \qquad \mbox{for } k=1,2,
\end{eqnarray*}
where $\mathbf{A} \succeq 0$. Then, $\mathbf{M}_1 - \mathbf{M}_2 \succeq 0$ if and only if $\mbox{MSE}(\hat{\ttheta}_1) - \mbox{MSE}(\hat{\ttheta}_2) \geq 0$ for all $\mathbf{A} \succeq 0$.
\end{theorem}

\begin{proposition} \textit{(\citealp{Fare1976}}) \label{prop.Farebrother}
\\
Let $\mathbf{A}$ be a $p \times p$-dimensional, positive definite matrix, $\mathbf{b}$ be a nonzero $p$ dimensional vector, and $c \in \mathbb{R}_+$. Then, $c \mathbf{A} - \mathbf{b} \mathbf{b}^{\top} \succ 0$ if and only if $\mathbf{b}^{\top} \mathbf{A}^{-1} \mathbf{b} < c$.
\end{proposition}

\noindent
We are now ready to proof the main result, formalized as Theorem \ref{theo.Theobald2}, that, for some value of its regularization parameter, the ridge regression estimator yields a lower MSE than the maximum likelihood regression estimator. Question \ref{question.alternativeSuperiorMSEproof} provides a simpler (?) but more limited proof of this result.

\begin{theorem} \textit{(Theorem 2 of \citealp{Theo1974}}) \label{theo.Theobald2}
\\
There exists $\lambda > 0$ such that $\mbox{MSE}[\hat{\bbeta}(\lambda)] < \mbox{MSE}[\hat{\bbeta}(0)] = \mbox{MSE}(\hat{\bbeta})$.
\end{theorem}

\begin{center}
\begin{figure}[!h]
\hspace{3.3cm}\includegraphics[scale=0.48, angle=0]{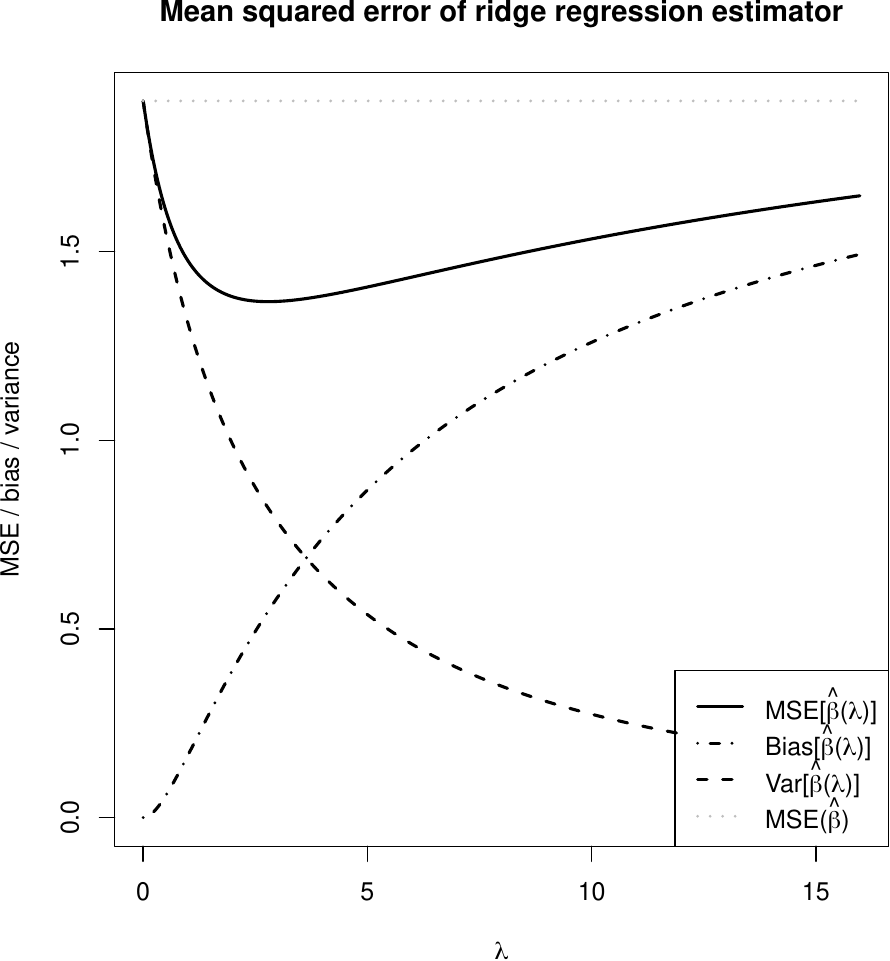}
\caption{The mean squared error, and its `bias' and `variance' parts, of the ridge regression estimator (for artificial data ). For comparison, the mean squared error of the maximum likelihood stimator (for the same artificial data) has been included.} \label{fig.MSEridge}
\end{figure}
\end{center}

\begin{proof}
The second order moment matrix of the ridge regression estimator is:
\begin{eqnarray*}
\mathbf{M} (\lambda) & := & \mathbb{E} [ (\hat{\bbeta}(\lambda) - \bbeta) (\hat{\bbeta} (\lambda) - \bbeta)^{\top} ]
\\
& = & \mathbb{E} \{ \hat{\bbeta}(\lambda) [\hat{\bbeta}(\lambda)]^{\top} \}  - \mathbb{E} [ \hat{\bbeta}(\lambda) ] \{ \mathbb{E} [ \hat{\bbeta}(\lambda) ] \}^{\top} + \mathbb{E} [\hat{\bbeta} (\lambda) - \bbeta)] \{ \mathbb{E} [\hat{\bbeta} (\lambda) - \bbeta)] \}^{\top}
\\
& = & \mbox{Var}[ \hat{\bbeta}(\lambda) ] + \mathbb{E} [\hat{\bbeta} (\lambda) - \bbeta)] \{ \mathbb{E} [\hat{\bbeta} (\lambda) - \bbeta)] \}^{\top}.
\end{eqnarray*}
Then:
\begin{eqnarray*}
\mathbf{M} ( 0 ) - \mathbf{M}(\lambda) & = & \sigma^2 (\mathbf{X}^{\top} \mathbf{X})^{-1} - \sigma^2 \mathbf{W}_{\lambda}  (\mathbf{X}^{\top} \mathbf{X})^{-1}  \mathbf{W}_{\lambda}^{\top} - (\mathbf{W}_{\lambda} - \mathbf{I}_{pp}) \bbeta \bbeta^{\top} (\mathbf{W}_{\lambda} -\mathbf{I}_{pp})^{\top}
\\
& = & \sigma^2 \mathbf{W}_{\lambda}  (\mathbf{X}^{\top} \mathbf{X})^{-1} (\mathbf{X}^{\top} \mathbf{X}+ \lambda \mathbf{I}_{pp}) (\mathbf{X}^{\top} \mathbf{X})^{-1}  (\mathbf{X}^{\top} \mathbf{X}+ \lambda \mathbf{I}_{pp}) (\mathbf{X}^{\top} \mathbf{X})^{-1} \mathbf{W}_{\lambda}^{\top}
\\
& & - \sigma^2 \mathbf{W}_{\lambda}  (\mathbf{X}^{\top} \mathbf{X})^{-1}  \mathbf{W}_{\lambda}^{\top} - (\mathbf{W}_{\lambda} - \mathbf{I}_{pp}) \bbeta \bbeta^{\top} (\mathbf{W}_{\lambda} -\mathbf{I}_{pp})^{\top}
\\
& = & \sigma^2 \mathbf{W}_{\lambda}   [ 2 \, \lambda \, (\mathbf{X}^{\top} \mathbf{X})^{-2} + \lambda^2 (\mathbf{X}^{\top} \mathbf{X})^{-3} ] \mathbf{W}_{\lambda}^{\top}
\\
& &  - \lambda^2 (\mathbf{X}^{\top} \mathbf{X} + \lambda \mathbf{I}_{pp})^{-1} \bbeta \bbeta^{\top} [ (\mathbf{X}^{\top} \mathbf{X} + \lambda \mathbf{I}_{pp})^{-1} ]^{\top}
\\
& = & \sigma^2  ( \mathbf{X}^{\top} \mathbf{X} + \lambda \mathbf{I}_{pp} )^{-1}  [ 2 \, \lambda \, \mathbf{I}_{pp} + \lambda^2 (\mathbf{X}^{\top} \mathbf{X})^{-1} ] [ (\mathbf{X}^{\top} \mathbf{X} + \lambda \mathbf{I}_{pp})^{-1} ]^{\top}
\\
& & - \lambda^2 ( \mathbf{X}^{\top} \mathbf{X} + \lambda \mathbf{I}_{pp} )^{-1} \bbeta \bbeta^{\top} [ ( \mathbf{X}^{\top} \mathbf{X} + \lambda \mathbf{I}_{pp})^{-1} ]^{\top}
\\
& = &  \lambda ( \mathbf{X}^{\top} \mathbf{X} + \lambda \mathbf{I}_{pp})^{-1}  [ 2 \, \sigma^2  \, \mathbf{I}_{pp} + \lambda \sigma^2 (\mathbf{X}^{\top} \mathbf{X})^{-1}  - \lambda \bbeta \bbeta^{\top}  ] [ (\mathbf{X}^{\top} \mathbf{X} + \lambda \mathbf{I}_{pp} )^{-1} ]^{\top}.
\end{eqnarray*}
This is positive definite if and only if $ 2 \, \sigma^2  \, \mathbf{I}_{pp} + \lambda \sigma^2 (\mathbf{X}^{\top} \mathbf{X})^{-1}  - \lambda \bbeta \bbeta^{\top}  \succ 0$. Hereto it suffices to show that $2 \, \sigma^2  \, \mathbf{I}_{pp} - \lambda \bbeta \bbeta^{\top} \succ 0$. By Proposition \ref{prop.Farebrother}, this holds for $\lambda$ such that $2 \sigma^2 (\bbeta^{\top} \bbeta)^{-1} > \lambda$. For these values of $\lambda$, we thus have $\mathbf{M} ( 0 ) - \mathbf{M}(\lambda)$. Application of Theorem \ref{theo.Theobald1} concludes the proof.
\end{proof}
This result of \cite{Theo1974} is generalized by \cite{Fare1976} to the class of design matrices $\mathbf{X}$ with $\mbox{rank}(\mathbf{X}) < p$.
\\
\\
Theorem \ref{theo.Theobald2} can be used to illustrate that the ridge regression estimator strikes a balance between the bias and variance. This is illustrated in the left panel of Figure \ref{fig.MSEridge}. For small $\lambda$, the variance of the ridge regression estimator dominates the MSE. This may be understood when realizing that in this domain of $\lambda$ the ridge regression estimator is close to the unbiased maximum likelihood regression estimator. For large $\lambda$, the variance vanishes and the bias dominates the MSE. For small enough values of $\lambda$, the decrease in variance of the ridge regression estimator exceeds the increase in its bias.   As the MSE is the sum of these two, the MSE first decreases as $\lambda$ moves away from zero. In particular, as $\lambda = 0$ corresponds to the maximum likelihood regression estimator, the ridge regression estimator yields a lower MSE for these values of $\lambda$. In the right panel of Figure \ref{fig.MSEridge}, $\mbox{MSE}[ \hat{\bbeta}(\lambda)] < \mbox{MSE}[ \hat{\bbeta}(0)]$ for $\lambda < 7$ (roughly) and the ridge regression estimator outperforms its maximum likelihood counterpart.
\mbox{ }
\\
\noindent
Besides another motivation behind the ridge regression estimator, the use of Theorem \ref{theo.Theobald2} is limited. The optimal choice of $\lambda$ depends on the quantities $\bbeta$ and $\sigma^2$. These are unknown in practice. Then, the penalty parameter is chosen in data-driven fashion (see e.g. Section \ref{subsect.crossvalidation}  and various other places).
\\
\\
\noindent
Theorem \ref{theo.Theobald2} may be of limited practical use, it does give insight in when the ridge regression estimator may be preferred over its ML counterpart. Ideally, the range of regularization parameters for which the ridge regression estimator outperforms -- in the MSE sense -- the ML regression estimator is as large as possible. The factors that influence the size of this range may be deduced from the optimal penalty $\lambda_{\mbox{{\tiny opt}}} = \sigma^2  (\bbeta^{\top} \bbeta / p)^{-1}$ found under the assumption of an orthonormal $\mathbf{X}$ (see Example \ref{example:ridgeMSEorthonormal}). But also from the bound on the penalty parameter, $\lambda_{\mbox{{\tiny max}}} = 2 \sigma^2  (\bbeta^{\top} \bbeta )^{-1}$ such that $\mbox{MSE}[\hat{\bbeta}(\lambda)] < \mbox{MSE}[\hat{\bbeta}(0)]$ for all $\lambda \in (0, \lambda_{\mbox{{\tiny max}}})$, derived in the proof of Theorem \ref{theo.Theobald2}. Firstly, an increase of the error variance $\sigma^2$ yields a larger $\lambda_{\mbox{{\tiny opt}}}$ and $\lambda_{\mbox{{\tiny max}}}$. Put differently, more noisy data benefit the ridge regression estimator. Secondly, $\lambda_{\mbox{{\tiny opt}}}$ and $\lambda_{\mbox{{\tiny max}}}$ also become larger when their denominators decreases. The denominator $p^{-1} \| \bbeta^{\top} \|_2^2$ may be viewed as an estimator of the `signal' variance `$\sigma^2_{\beta}$'. A quick conclusion would be that ridge regression profits from less signal. But more can be learned from the denominator. Contrast the two regression parameters $\bbeta_{\mbox{{\tiny unif}}} = \mathbf{1}_p$ and $\bbeta_{\mbox{{\tiny sparse}}}$ which comprises of only zeros except the first element which equals $p$, i.e. $\bbeta_{\mbox{{\tiny sparse}}} = (p, 0, \ldots, 0)^{\top}$. Then, the $\bbeta_{\mbox{{\tiny unif}}}$ and $\bbeta_{\mbox{{\tiny sparse}}}$ have comparable signal in the sense that $\sum_{j=1}^p | \beta_j | = p$. The denominator of $\lambda_{\mbox{{\tiny opt}}}$ corresponding both parameters equals $1$ and $p$, respectively. This suggests that ridge regression will perform better in the former case where the regression parameter is not dominated by a few elements but rather all contribute comparably to the explanation of the variation in the response. Of course, more factors contribute. For instance, collinearity among the columns of $\mathbf{X}$, which gave rise to ridge regression in the first place.
\\
\\
The choice of the penalty parameter on the basis of the mean squared error strikes a balance between the bias and variance, a so-called \textit{bias-variance trade-off}. In the extremes, the model has either a high variance but low bias and will overfit the data. Or, the model exhibits little variance but has a high bias and as a result underfits the data. Ideally, the balance between bias and variance results in a model that neither over- nor underfits. Apart from the regularization parameter, the bias-variance trade-off is affected by other means. For instance, the addition of more covariates (or transformations thereof) to the model is likely to result in a lower bias but also in a high variance, while the employment of a simpler model has the opposite effect. Alternative, a larger sample size typically reduces the variance. 
\\
\begin{remark} \mbox{ }
\\
Theorem \ref{theo.Theobald2} can also be used to conclude on the biasedness of the ridge regression estimator. The Gauss-Markov theorem \citep{Rao1973} states (under some assumptions) that the ML regression estimator is the best linear unbiased estimator (BLUE) with the smallest MSE. As the ridge regression estimator is a linear estimator and outperforms (in terms of MSE) this maximum likelihood estimator, it must be biased (for it would otherwise refute the Gauss-Markov theorem).
\end{remark}

\subsection{Debiasing}
Despite a potential superior performance in the MSE sense of the ridge regression estimator, it remains biased. This bias hampers the direct application of most machinery presented in the statistical literature as that is geared towards unbiased estimators. Unbiasedness facilitates proper inference and confidence interval construction. For instance, an estimator with a large bias and small variance may yield a confidence interval that does not contain the true parameter value. Hence, several proposals to de-bias the ridge regression estimator have been presented. 

A straightforward approach would be to correct for the bias of the estimator (\ref{form:ridgeExpectation}), using the `non-debiased' ridge regression estimator as an estimator for the regression parameter. This debiased ridge regression estimator is exactly what \cite{zhang2022ridge} propose:
\begin{eqnarray*}
\hat{\bbeta}_d (\lambda) & = & \hat{\bbeta} (\lambda) + \lambda (\mathbf{X}^{\top} \mathbf{X} + \lambda \mathbf{I}_{pp})^{-1} \hat{\bbeta} (\lambda).
\end{eqnarray*}
To study the effect of this debiasing, note that 
\begin{eqnarray*}
\hat{\bbeta}_d (\lambda) - \bbeta & = & - \lambda^2 (\mathbf{X}^{\top} \mathbf{X} + \lambda \mathbf{I}_{pp})^{-2}  \bbeta + (\mathbf{X}^{\top} \mathbf{X} + \lambda \mathbf{I}_{pp})^{-1}
[ \mathbf{I}_{pp} + \lambda (\mathbf{X}^{\top} \mathbf{X} + \lambda \mathbf{I}_{pp})^{-1} ] \mathbf{X}^{\top} \vvarepsilon,
\end{eqnarray*}
where we have \textit{i)} substituted the analytic expression of the non-debiased ridge regression estimator, \textit{ii)} substituted $\mathbf{X} \bbeta + \vvarepsilon$ for $\mathbf{Y}$ by virtue of the linear model, and \textit{iii)} manipulated the resulting expression using straightforward linear algebra. From the expression of the preceeding display, we directly obtain the bias, $\mathbb{E}[\hat{\bbeta}_d (\lambda)] - \bbeta = - \lambda^2 (\mathbf{X}^{\top} \mathbf{X} + \lambda \mathbf{I}_{pp})^{-2}  \bbeta$, and variance,
\begin{eqnarray*}
\mbox{Var}[\hat{\bbeta}_d (\lambda)]  & = & \sigma^2 [ (\mathbf{X}^{\top} \mathbf{X} + \lambda \mathbf{I}_{pp})^{-1} + \lambda (\mathbf{X}^{\top} \mathbf{X} + \lambda \mathbf{I}_{pp})^{-2} ] 
\\
& & \, \, \, \, \, \times \, \mathbf{X}^{\top} \mathbf{X} \, [ (\mathbf{X}^{\top} \mathbf{X} + \lambda \mathbf{I}_{pp})^{-1} + \lambda (\mathbf{X}^{\top} \mathbf{X} + \lambda \mathbf{I}_{pp})^{-2} ].
\end{eqnarray*}
From these expressions, it is -- as  \cite{zhang2022ridge} point out -- clear that, if $\lambda d_{x,n}^{-2}$ tends to zero as $n \rightarrow \infty$, the bias of debiased estimator is much smaller than that of its non-debiased counterpart, while the difference in their variances becomes negligible. It is not immediate how this condition on $\lambda d_{x,n}^{-2}$ translates practically to finite sample sizes in terms of a class of design matrices and a domain of the regularization parameter. 

An alternative debiased ridge regression estimator is proposed in \cite{Buhlmann2013statistical}. It starts from the bias decomposition into a part attributable to the regularization and one to the high-dimensionality (see 
Proposition \ref{prop:ridgeProjection}). \cite{Buhlmann2013statistical} assumes a relatively small penalty parameter such that effectively the regularization bias, $\mathbf{P}_x \{ \mathbb{E}[\hat{\bbeta}(\lambda)] - \bbeta\}$, is smaller than the standard error and, thereby, not of primary concern. We are then left to reduce the bias introduced by the high-dimensionality, $(\mathbf{P}_x - \mathbf{I}_{pp}) \bbeta$, called \textit{projection bias} in \cite{Buhlmann2013statistical}. Hereto \cite{Buhlmann2013statistical} assumes the existence of an alternative but accurate estimator $\hat{\bbeta}^{\mbox{{\tiny (acc)}}}$. Under (strong?) assumptions, the lasso regression estimator (Chapter \ref{chap:lassoRegression}) may serve as such. With this alternative, accurate estimator at hand, the projection bias is eliminated by replacing $\bbeta$ by $\hat{\bbeta}^{\mbox{{\tiny (acc)}}}$ and substracting the projection bias term from the non-debiased ridge regression estimator.

\section{Constrained estimation} \label{sect.constrainedEstimation}
The ad-hoc fix of \cite{Hoer1970} to super-collinearity of the design matrix can been motivated post-hoc through the use of a loss function. The ridge regression estimator minimizes the \textit{ridge loss function}, which is defined as:
\begin{eqnarray}
\mathcal{L}_{\mbox{{\tiny ridge}}}(\bbeta; \lambda) & = & \| \mathbf{Y} - \mathbf{X}  \bbeta \|^2_2 + \lambda \| \bbeta \|^2_2 \, \, \,  = \, \, \, \sum\nolimits_{i=1}^n (Y_i - \mathbf{X}_{i\ast}  \bbeta)^2 + \lambda \sum\nolimits_{j=1}^p \beta_j^2. \label{form.ridgeLossFunction}
\end{eqnarray}
This loss function is the traditional sum-of-squares augmented with a \textit{penalty}. The particular form of the penalty, $\lambda \| \bbeta \|^2_2$ is referred to as the \textit{ridge penalty} or \textit{ridge regularization term} and $\lambda$ as the \textit{penalty parameter} or \textit{regularization parameter}. For $\lambda=0$, minimization of the ridge loss function yields the ML estimator (if it exists). For any $\lambda > 0$, the ridge penalty contributes to the loss function, affecting its minimum and its location. The minimum of the sum-of-squares is well-known. The minimum of the ridge penalty is attained at $\bbeta = \mathbf{0}_{p}$ whenever $\lambda > 0$. The $\bbeta$ that minimizes $\mathcal{L}_{\mbox{{\tiny ridge}}}(\bbeta; \lambda)$ then balances the sum-of-squares and the penalty. The effect of the penalty in this balancing act is to shrink the regression coefficients towards zero, its minimum. In particular, the larger $\lambda$, the larger the contribution of the penalty to the loss function, the stronger the tendency to shrink non-zero regression coefficients to zero (and decrease the contribution of the penalty to the loss function). This motivates the name `penalty' as non-zero elements of $\bbeta$ increase (or penalize) the loss function.

To verify that the ridge regression estimator indeed minimizes the ridge loss function, proceed as usual. Take the derivative with respect to $\bbeta$:
\begin{eqnarray*}
\frac{\partial}{\partial \bbeta} \mathcal{L}_{\mbox{{\tiny ridge}}}(\bbeta; \lambda) & = & -2  \mathbf{X}^{\top} (\mathbf{Y} - \mathbf{X} \bbeta)  + 2 \lambda  \mathbf{I}_{pp}  \bbeta \, \, \, = \, \, \, -2  \mathbf{X}^{\top} \mathbf{Y} + 2 ( \mathbf{X}^{\top} \mathbf{X} + \lambda  \mathbf{I}_{pp}) \bbeta.
\end{eqnarray*}
Equate the derivative to zero and solve for $\bbeta$. This yields the ridge regression estimator.

The ridge regression estimator is thus a stationary point of the ridge loss function. A stationary point corresponds to a minimum if the Hessian matrix with second order partial derivatives is positive definite. The Hessian matrix of the ridge loss function is
\begin{eqnarray*}
\frac{\partial^2}{\partial \bbeta \, \partial \bbeta^{\top}} \mathcal{L}_{\mbox{{\tiny ridge}}}(\bbeta; \lambda) & = & 2  ( \mathbf{X}^{\top} \mathbf{X} + \lambda  \mathbf{I}_{pp}).
\end{eqnarray*}
This Hessian matrix is the sum of the (semi-)positive definite matrix $\mathbf{X}^{\top} \mathbf{X}$ and the positive definite matrix $\lambda \mathbf{I}_{pp}$. Lemma 14.2.4 of \cite{Harv2008} then states that the sum of these matrices is itself a positive definite matrix. Hence, the Hessian matrix is positive definite and the ridge regression loss function has a stationary point at the ridge regression estimator, which is a minimum.

\begin{figure}[!h]
\begin{tabular}{rcl}
\includegraphics[scale=0.38, angle=0]{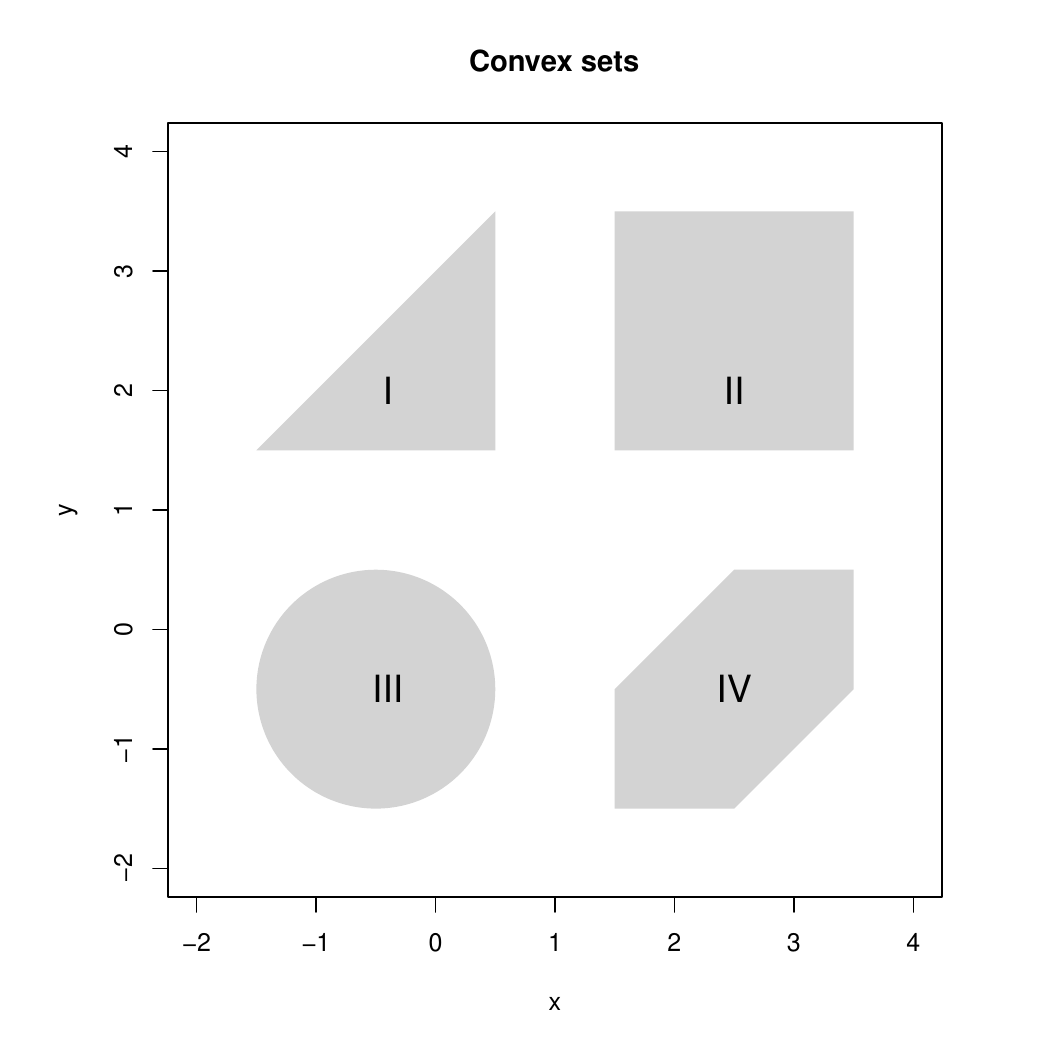}
& &
\includegraphics[scale=0.38, angle=0]{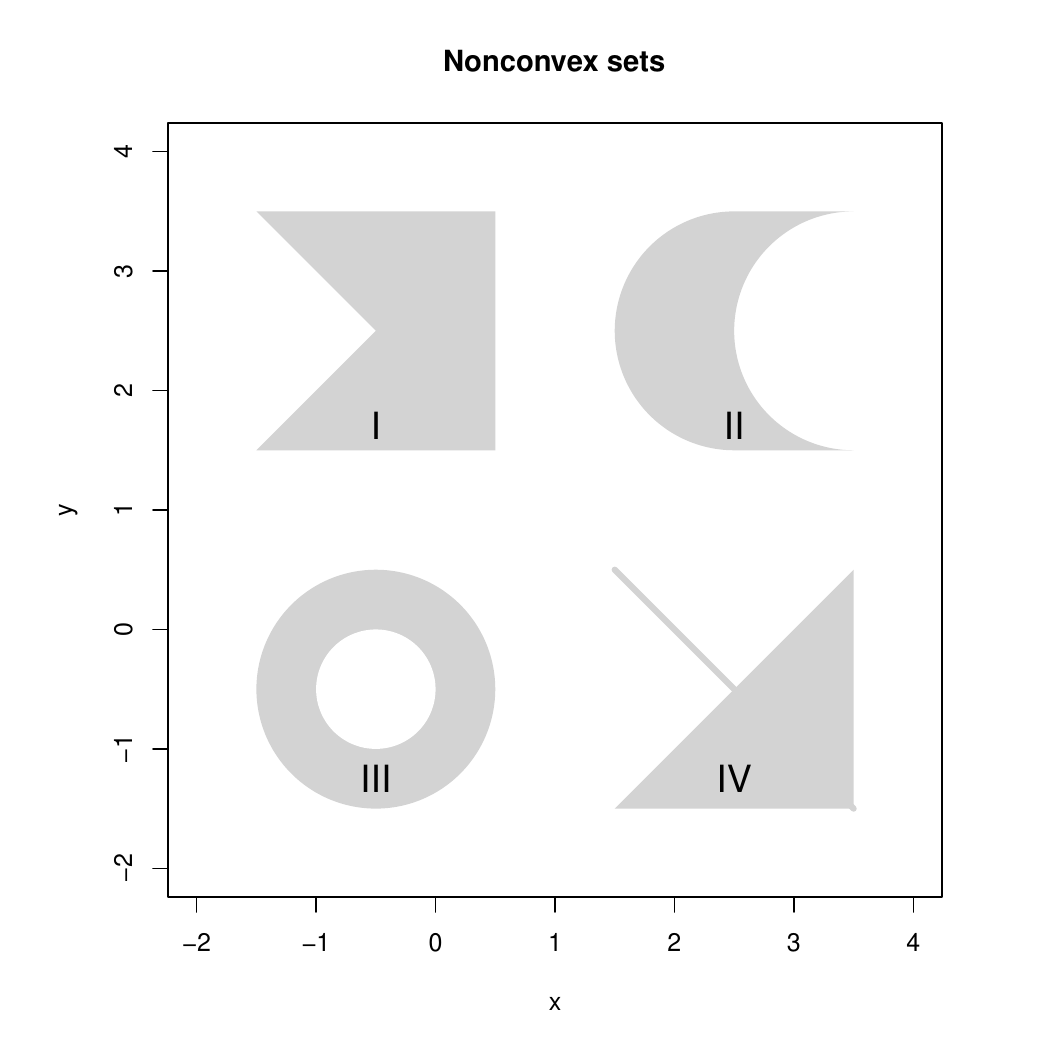}
\\
\includegraphics[scale=0.38, angle=0]{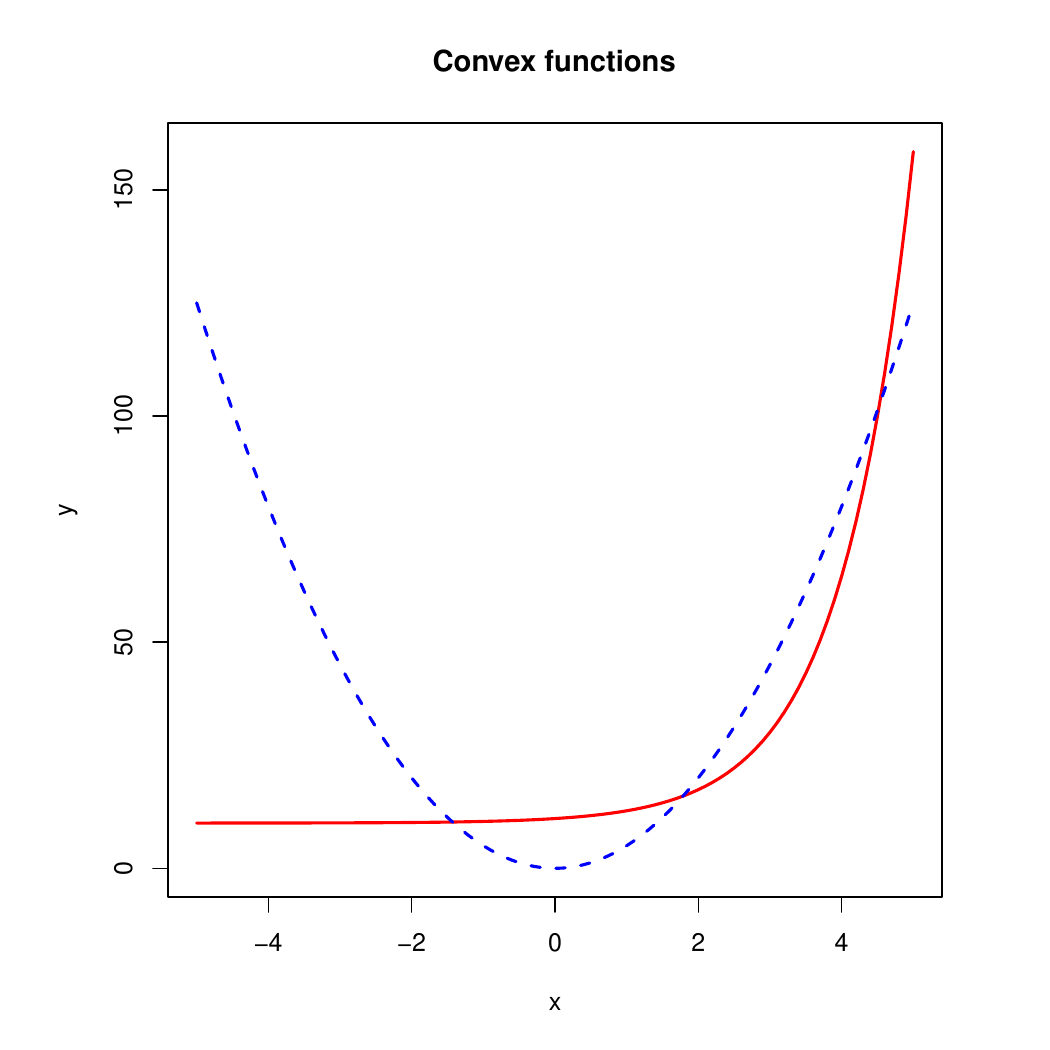}
& &
\includegraphics[scale=0.38, angle=0]{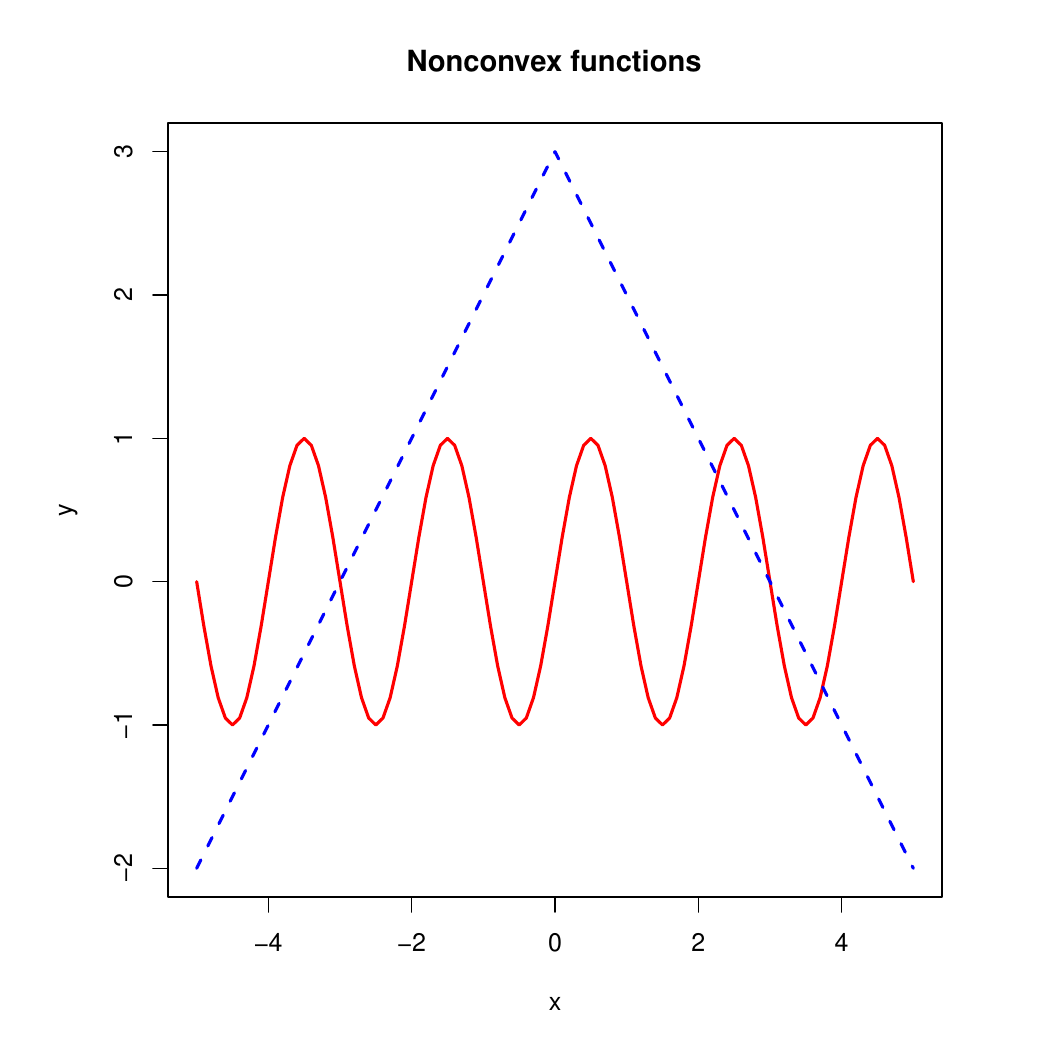}
\\
\includegraphics[scale=0.38, angle=0]{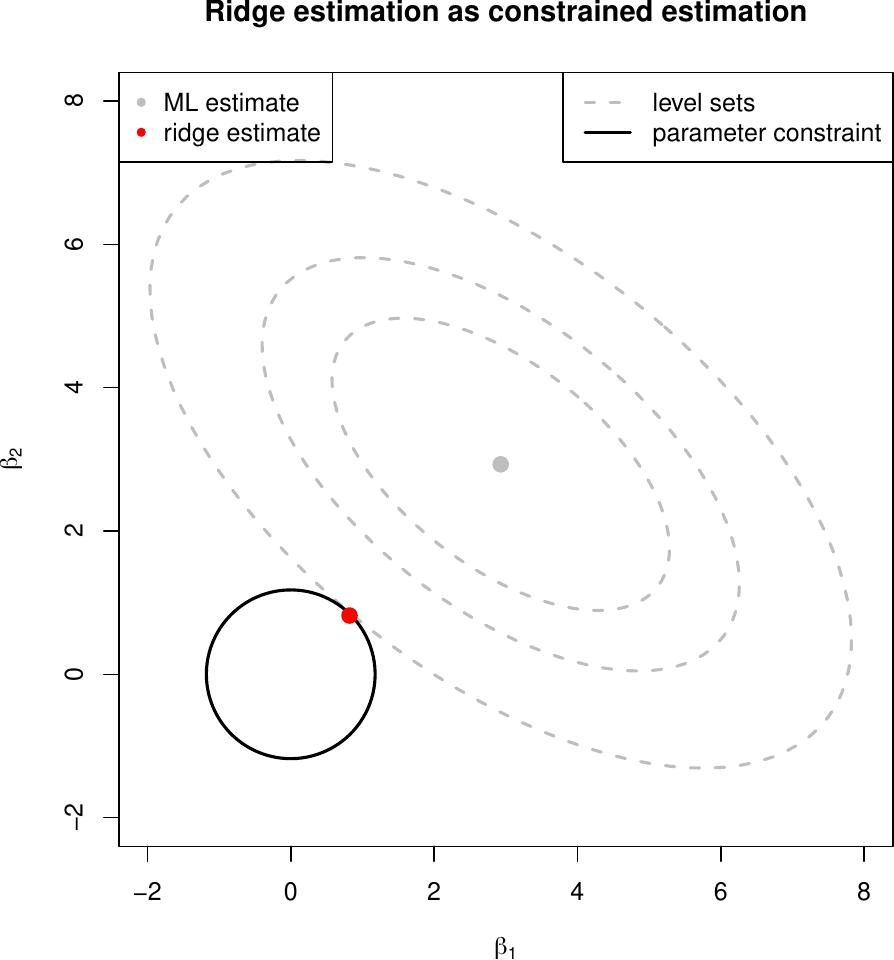}
& &
\includegraphics[scale=0.38, angle=0]{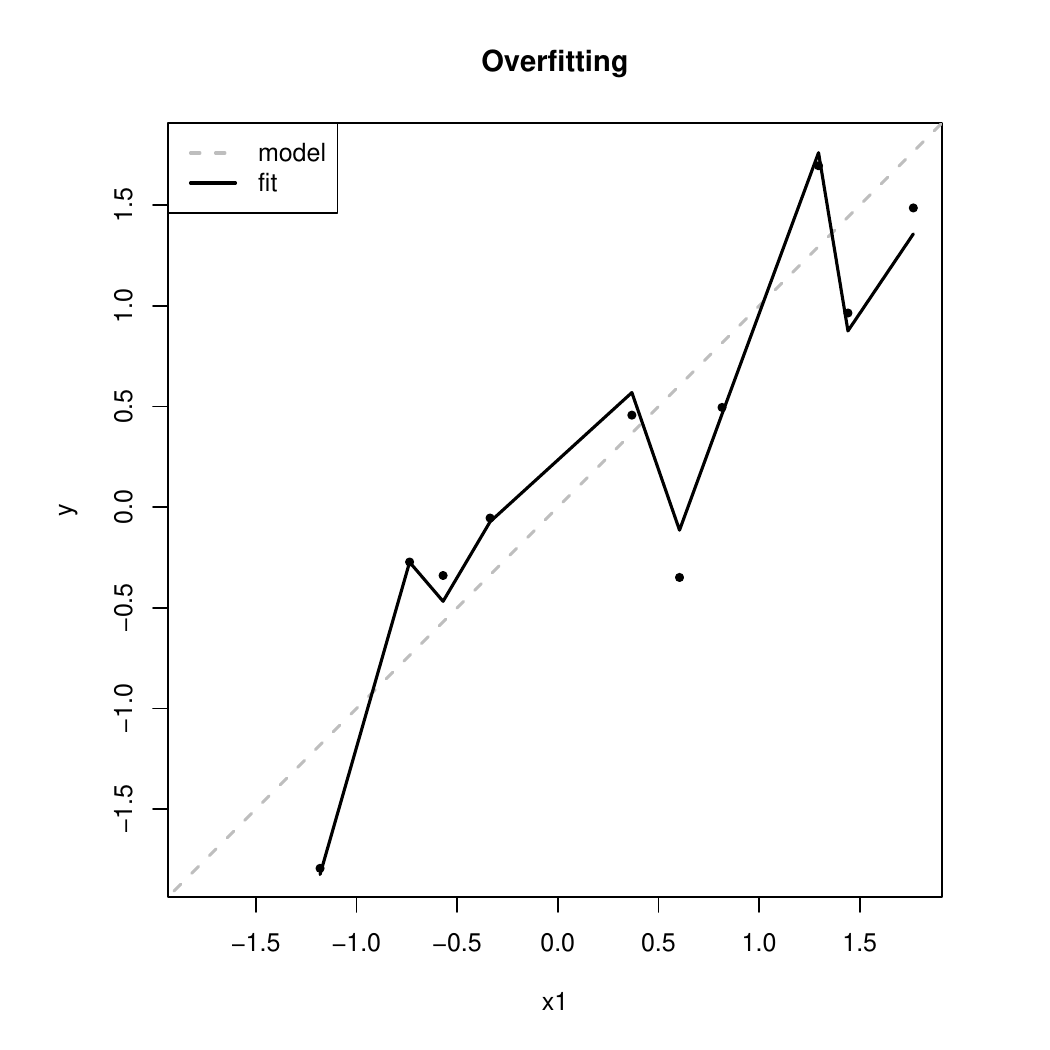}
\end{tabular}
\caption{Top panels show examples of convex (left) and nonconvex (right) sets. Middle panels show examples of convex (left) and nonconvex (right) functions.
The left bottom panel illustrates the ridge estimation as a constrained estimation problem. The ellipses represent the contours of the ML loss function, with the blue dot at the center the ML estimate. The circle is the ridge parameter constraint. The red dot is the ridge estimate. It is at the intersection of the ridge constraint and the smallest contour with a non-empty intersection with the constraint. The right bottom panel shows the data corresponding to Example \ref{exam.overfitting}. The grey line represents the `true' relationship, while the black line the fitted one.
} \label{fig.ridgeAsConstrainedEst}
\end{figure}

The ridge regression estimator minimizes the ridge loss function. It remains to verify that it is a global minimum. To this end, we introduce the concept of a convex function. As a prerequisite, a set $\mathcal{S} \subset \mathbb{R}^p$ is called \textit{convex} if for all $\bbeta_1, \bbeta_2 \in \mathcal{S}$ their weighted average $\bbeta_{\theta}  =  (1 - \theta) \bbeta_1 + \theta \bbeta_2$ for all $\theta \in [0, 1]$ is itself an element of $\mathcal{S}$, thus $\bbeta_{\theta} \in \mathcal{S}$. If for all $\theta \in (0, 1)$, the weighted average $\bbeta_{\theta}$ is inside $\mathcal{S}$ and not on its boundary, the set is called \textit{strictly convex}. Examples of (strictly) convex and nonconvex sets are depicted in Figure \ref{fig.ridgeAsConstrainedEst}. A function $f(\cdot)$ is \textit{(strictly) convex} if the set $\{ y \, : \, y \geq f(\bbeta) \mbox{ for all } \bbeta \in \mathcal{S} \mbox{ for any convex } \mathcal{S} \}$, called the epigraph of $f(\cdot)$, is (strictly) convex. Examples of (strictly) convex and nonconvex functions are depicted in Figure \ref{fig.ridgeAsConstrainedEst}. The ridge loss function is the sum of two parabola's: one is at least convex and the other strictly convex in $\bbeta$. The sum of a convex and strictly convex function is itself strictly convex (see Lemma 9.4.2 of \citealt{Flet2008}). The ridge loss function is thus strictly convex. Theorem 9.4.1 of \citealt{Flet2008} then warrants, by the strict convexity of the ridge loss function, that the ridge regression estimator is a global minimum.
\\
\\
The limiting behavior of the variance of the ridge regression estimator can also be understood from the ridge loss function. The ridge penalty with its minimum $\bbeta = \mathbf{0}_{p}$ does not involve data and, consequently, the variance of its minimum equals zero. With the ridge regression estimator being a compromise between the maximum likelihood estimator and the minimum of the penalty, so is its variance a compromise of their variances. As $\lambda$ tends to infinity, the ridge regression estimator and its variance converge to the minimizer of the loss function and the variance of the minimizer, respectively. Hence, in the limit (large $\lambda$) the variance of the ridge regression estimator vanishes. Understandably, as the penalty now fully dominates the loss function and, consequently, it does no longer involve data (i.e. randomness).
\afterpage{\clearpage}
\\
\\
The loss function of the ridge regression estimator facilitates another view on the estimator. Hereto now define the ridge regression estimator  as:
\begin{eqnarray} \label{form.ridgeEstViaPenEst}
\hat{\bbeta}(\lambda) & = & \arg \min\nolimits_{\bbeta \in \mathbb{R}^p} \| \mathbf{Y} - \mathbf{X} \, \bbeta \|^2_2 + \lambda \| \bbeta \|^2_2.
\end{eqnarray}
This minimization problem can be reformulated into the following constrained optimization problem:
\begin{eqnarray} \label{form.constrEstProblemRidge}
\hat{\bbeta}(\lambda) & = &  \arg \min\nolimits_{\{ \bbeta \in \mathbb{R}^p \, : \, \| \bbeta \|_2^2 \leq c\}} \| \mathbf{Y} - \mathbf{X} \, \bbeta \|^2_2,
\end{eqnarray}
for some suitable $c > 0$. The constrained optimization problem  (\ref{form.constrEstProblemRidge}) can be solved by means of the Karush-Kuhn-Tucker (KKT) multiplier method \citep{Flet2008}, which minimizes a function subject to inequality constraints. The KKT multiplier method states that, under some regularity conditions (all met here), there exists a constant $\nu \geq 0$, called the \textit{multiplier}, such that the solution $\hat{\bbeta}(\nu)$ of the constrained minimization problem (\ref{form.constrEstProblemRidge}) satisfies the so-called KKT conditions. The first KKT condition (referred to as the stationarity condition) demands that the gradient (with respect to $\bbeta$) of the Lagrangian associated with the minimization problem equals zero at the solution $\hat{\bbeta}(\nu)$. The Lagrangian for problem (\ref{form.constrEstProblemRidge}) is:
\begin{eqnarray*}
\| \mathbf{Y} - \mathbf{X} \, \bbeta \|^2_2 + \nu ( \| \bbeta \|^2_2 - c).
\end{eqnarray*}
The second KKT condition, also known as the complementarity condition, requires that $\nu (\| \hat{\bbeta}(\nu) \|_2^2 - c) = 0$. If $\nu = \lambda$ and $c = \| \hat{\bbeta}(\lambda) \|_2^2$, the ridge regression estimator $\hat{\bbeta} (\lambda)$ satisfies both KKT conditions. Hence, both problems have the same solution if  $c = \| \hat{\bbeta}(\lambda) \|_2^2$.

The constrained estimation interpretation of the ridge regression estimator is illustrated in the left bottom panel of Figure \ref{fig.ridgeAsConstrainedEst}. It shows the level sets of the sum-of-squares criterion and centered around zero the circular ridge parameter constraint, parametrized by $\{ \bbeta \, : \, \| \bbeta \|_2^2 = \beta_1^2 + \beta_2^2 \leq c \}$ for some $c > 0$. The ridge regression estimator is then the point where the  sum-of-squares' smallest level set hits the constraint. Exactly at that point, the sum-of-squares is minimized over those $\bbeta$'s that live on or inside the constraint. In the high-dimensional setting, the ellipsoidal level sets are degenerated. For instance, in the 2-dimensional case of the left bottom panel of Figure \ref{fig.ridgeAsConstrainedEst}, the ellipsoids would then be lines but the geometric interpretation is unaltered.

The ridge regression estimator is always to be found on the boundary of the ridge parameter constraint and is never an interior point. To see this, we need the following technical result that translates the eigenvalue shrinkage to the size of the estimator.
\begin{lemma} \label{lemma.estimatorSize} \mbox{ } \\
The ridge regression estimator is smaller or equal than its maximum likelihood counterpart, i.e. $\| \hat{\bbeta} ( \lambda) \|_2^2 \leq \| \hat{\bbeta}^{\mbox{{\tiny ml}}} \|_2^2$ for all $\lambda \geq 0$.
\end{lemma}
\begin{proof}
The radius of the ridge parameter constraint can be bounded as follows
\begin{eqnarray*}
\| \hat{\bbeta} ( \lambda) \|_2^2 & = & \mathbf{Y}^{\top} \mathbf{X} (\mathbf{X}^{\top} \mathbf{X} + \lambda \mathbf{I}_{pp})^{-2} \mathbf{X}^{\top} \mathbf{Y} \, \, \,\leq \, \, \, \mathbf{Y}^{\top} \mathbf{X} [(\mathbf{X}^{\top} \mathbf{X})^{+}]^2 \mathbf{X}^{\top} \mathbf{Y} \, \, \, \leq \, \, \, \| \hat{\bbeta}^{\mbox{{\tiny ml}}} \|_2^2.
\end{eqnarray*}
The inequality in the display above follows from \textit{i)} $\mathbf{X}^{\top} \mathbf{X} \succ 0$ and $\lambda \mathbf{I}_{pp} \succeq 0$, \textit{ii)} $\mathbf{X}^{\top} \mathbf{X} + \lambda \mathbf{I}_{pp} \succ \mathbf{X}^{\top} \mathbf{X}$ (due to Lemma 14.2.4 of \citealp{Harv2008}), and \textit{iii)} $[(\mathbf{X}^{\top} \mathbf{X})^{+}]^2 \succeq (\mathbf{X}^{\top} \mathbf{X} + \lambda \mathbf{I}_{pp})^{-2}$ (inferring Corollary 7.7.4 of \citealp{Horn2012matrix}). 
\end{proof}
We now distinguish between two cases. 
\begin{compactitem}
\item[\textit{i)}] The ridge parameter constraint encompasses the maximum likelihood estimator. The constraint is then not active and $\hat{\bbeta} ( \lambda) =  \hat{\bbeta}^{\mbox{{\tiny ml}}}$. All constraints under which the two estimators coincide are equivalent to one with radius $\| \hat{\bbeta}^{\mbox{{\tiny ml}}} \|_2$ and have $\hat{\bbeta} ( \lambda)$ on the boundary of the constraint $\{ \bbeta \in \mathbb{R}^p \, : \, \| \bbeta \|_2^2 = \| \hat{\bbeta}^{\mbox{{\tiny ml}}} \|_2^2 \}$.
\item[\textit{ii)}] The ridge parameter constraint does not contain the maximum likelihood estimator. This corresponds to the situation displayed in the left bottom panel of Figure \ref{fig.ridgeAsConstrainedEst}: the constraint is centered around the origin, the ridge regression estimator is confined to this constraint, and location at which the loss is miminized (i.e., the maximum likelihood estimator) is -- by Lemma \ref{lemma.estimatorSize} -- situated further away from the origin. Now consider a sequence of ever wider level sets of the loss function that radiate out from the maximum likelihood estimator. The loss increases with the width of these level sets. Among this sequence of levels sets, our interest is in those that intersect with the ridge parameter constraint.  From these intersecting level sets, we single out the one that corresponds to the lowest loss. This singled out level set has smallest width and intersects with the ridge parameter constraint only at the boundary. Furthermore, any level set that goes through the interior of the constraint is wider and corresponds to a poorer loss. By construction, the ridge regression estimator must then be on the constraint's boundary.
\end{compactitem}
Both cases reveal that $\hat{\bbeta} ( \lambda)$ is always on the boundary of the ridge parameter constraint.

The size of the spherical ridge parameter constraint shrinks monotonously as $\lambda$ increases, and eventually, in the $\lambda \rightarrow \infty$-limit collapses to zero (as is formalized by Proposition \ref{prop:constraintBehavior}). 
\begin{proposition} \label{prop:constraintBehavior} \mbox{ } \\
The squared norm of the ridge regression estimator satisfies:
\begin{compactitem}
\item[\textit{i)}]
$d \| \hat{\bbeta} ( \lambda) \|_2^2 / d\lambda < 0$ for $\lambda > 0$,

\item[\textit{ii)}] $\lim_{\lambda \rightarrow \infty} \| \hat{\bbeta} ( \lambda) \|_2^2 = 0$.
\end{compactitem}
\end{proposition}

\begin{proof}
For part \textit{i)}, we need to verify that $\| \hat{\bbeta}(\lambda) \|_2^2 > \| \hat{\bbeta}(\lambda + h) \|_2^2$ for $h > 0$. Hereto substitute the singular value decomposition of the design matrix, $\mathbf{X} = \mathbf{U}_x \mathbf{D}_x \mathbf{V}_x^{\top}$, into the display above. Then, after a little algebra, 
take the derivative with respect to $\lambda$, and obtain:
\begin{eqnarray*}
\frac{d \, \, }{d\, \lambda} \| \hat{\bbeta}(\lambda) \|_2^2 & = & - 2\sum\nolimits_{i=1}^n 
 d_{x,i}^2 (d_{x,i}^2 + \lambda)^{-3} (\mathbf{Y}^{\top} \mathbf{U}_{x,\ast,i})^2.
\end{eqnarray*}
This is negative for all $\lambda > 0$. Indeed, the parameter constraint thus becomes smaller and smaller as $\lambda$ increases, and so does the size of the estimator. 

Part \textit{ii)} follows from $\lim_{\lambda \rightarrow \infty} \hat{\bbeta}(\lambda) = \mathbf{0}_p$, which has been concluded previously by other means. 
\end{proof}
The relevance of viewing the ridge regression estimator as the solution to a constrained estimation problem becomes obvious when considering a typical threat to high-dimensional data analysis: overfitting. \textit{Overfitting} refers to the phenomenon of modelling the noise rather than the signal. In case the true model is parsimonious (i.e., few covariates driving the response) and data on many covariates are available, it is likely that a linear combination of all covariates yields a higher likelihood than a combination of the few that are actually related to the response. As only the few covariates related to the response contain the signal, the model involving all covariates then cannot but explain more than the signal alone: it also models the error. Hence, it overfits the data. In high-dimensional settings overfitting is a real threat. The number of explanatory variables exceeds the number of observations. It is thus possible to form a linear combination of the covariates that perfectly explains the response, including the noise.

Large estimates of regression coefficients are often an indication of overfitting. Augmentation of the estimation procedure with a constraint on the regression coefficients is a simple remedy to large parameter estimates. As a consequence, it decreases the probability of overfitting. Overfitting is illustrated in the next example.

\begin{example} \textit{(Overfitting)} \label{exam.overfitting}
\\
Consider an artificial data set comprising of ten observations on a response $Y_i$ and nine covariates $X_{i,j}$. All covariate data are sampled from the standard normal distribution: $X_{i,j} \sim \mathcal{N}(0, 1)$. The response is generated by $Y_i = X_{i,1} + \varepsilon_i$ with $\varepsilon_{i} \sim \mathcal{N}(0, \tfrac{1}{4})$. Hence, only the first covariate contributes to the response.

The regression model $Y_i = \sum_{j=1}^9  X_{i,j} \beta_j+ \varepsilon_i$ is fitted to the artificial data using \texttt{R}. This yields the regression parameter estimates:
\begin{eqnarray*}
\hat{\bbeta}^{\top} & = & (0.048, -2.386, -5.528, 6.243, -4.819, 0.760, -3.345, -4.748, 2.136).
\end{eqnarray*}
As $\bbeta^{\top} = (1, 0, \ldots, 0)$, many regression coefficient are clearly over-estimated.

The fitted values $\widehat{Y}_i = \mathbf{X}_i \hat{\bbeta}$ are plotted against the values of the first covariates in the right bottom panel of Figure \ref{fig.ridgeAsConstrainedEst}. As a reference, the line $x=y$, which represents the `true' model,  is added. The fitted model follows the `true' relationship. But it also captures the deviations from this line that represent the errors.
\end{example}


\section{Degrees of freedom} \label{sect.ridgeDOF}
The degrees of freedom consumed by the ridge regression estimator, which may aid in the choice of the value of the penalty parameter, is derived. A formal definition of the degrees of freedom, which can among others be found in \cite{Efron1986biased}, is
\begin{eqnarray} \label{form.DOFdef}
\mbox{df}(\lambda) & = &  \sum\nolimits_{i=1}^n [\mbox{Var}(Y_i)]^{-1} \mbox{Cov}(\widehat{Y}_i, Y_i).
\end{eqnarray}
It represents the effective number of parameters used by the estimator. It can also be viewed as the amount of self-explanation by the observations of the fit. Recall from ordinary regression that $\widehat{\mathbf{Y}}  =  \mathbf{X} (\mathbf{X}^{\top} \mathbf{X})^{-1} \mathbf{X}^{\top} \mathbf{Y} = \mathbf{H} \mathbf{Y}$ where $\mathbf{H}$ is the hat matrix. Application of the definition (\ref{form.DOFdef}) then yields the degrees of freedom used by the maximum likelihood regression estimator and are equal to $\mbox{tr}(\mathbf{H})$, the trace of $\mathbf{H}$. In particular, if $ \mathbf{X}$ is of full rank, i.e. $\mbox{rank}(\mathbf{X}) = p$, then $\mbox{tr}(\mathbf{H}) = p$.

We adopt the degrees of freedom definition (\ref{form.DOFdef}) for the ridge regression estimator. We then find  
\begin{eqnarray*}
\mbox{df}(\lambda) & = &  \sum\nolimits_{i=1}^n [\mbox{Var}(Y_i)]^{-1} \mbox{Cov}(\widehat{Y}_i, Y_i)
\\
& = & \sigma^{-2}  \sum\nolimits_{i=1}^n \mbox{Cov}[\mathbf{X}_{i, \ast} \hat{\bbeta} (\lambda), Y_i]
\\
& = & \sigma^{-2}  \sum\nolimits_{i=1}^n \mbox{Cov}[\mathbf{X}_{i, \ast} (\mathbf{X}^{\top} \mathbf{X} + \lambda \mathbf{I}_{pp})^{-1} \mathbf{X}^{\top} \mathbf{Y}, Y_i]
\\
& = & \sigma^{-2}  \sum\nolimits_{i=1}^n \mathbf{X}_{i, \ast} (\mathbf{X}^{\top} \mathbf{X} + \lambda \mathbf{I}_{pp})^{-1} \mathbf{X}^{\top} \mbox{Cov} (\mathbf{Y}, Y_i)
\\
& = & \mbox{tr} [ \mathbf{X} ( \mathbf{X}^{\top}   \mathbf{X}^{\top} + \lambda \mathbf{I}_{pp})^{-1} \mathbf{X}^{\top}]
\\
& = &  \sum\nolimits_{j=1}^p  (\mathbf{D}_x^{\top} \mathbf{D}_x)_{jj} [(\mathbf{D}_x^{\top} \mathbf{D}_x)_{jj} + \lambda]^{-1},
\end{eqnarray*}
where we have used the independence among the observations. High-dimensionally, the sum on the right-hand side of the last line of the display above may be limited to $n$. The degrees of freedom consumed by the ridge regression estimator is monotone decreasing in $\lambda$. In particular, $\lim_{\lambda \rightarrow \infty} \mbox{df}(\lambda) = 0$. That is, in the limit no information from $\mathbf{X}$ is used. Indeed, $\hat{\bbeta}(\lambda)$ is forced to equal $\mathbf{0}_{p}$ which is not derived from data. Finally, from the derivation in the display above, we may deduce a definition of the `ridge hat matrix': $\mathbf{H}(\lambda) := \mathbf{X} (\mathbf{X}^{\top} \mathbf{X} + \lambda \mathbf{I}_{pp})^{-1} \mathbf{X}^{\top}$. We can then write, in analogy to the ordinary regression case, $\mbox{df}(\lambda) = \mbox{tr}[ \mathbf{H}(\lambda)]$.



\section{Computationally efficient evaluation} \label{sect.ridgeEfficientCalculation}
In the high-dimensional setting, the number of covariates $p$ is large compared to the number of samples $n$. For instance, in a high-throughput experiment, it is not uncommon to find $p = 40000$ and $n= 100$. To perform ridge regression in this context, the following expression needs to be evaluated numerically: $(\mathbf{X}^{\top} \mathbf{X} + \lambda \mathbf{I}_{pp})^{-1} \mathbf{X}^{\top} \mathbf{Y}$. For $p=40000$, this requires the inversion of a $40000 \times 40000$ dimensional matrix. This is not feasible on most desktop computers. 

There is a workaround to the computational burden. 
Hereto consider the so-called `thin' singular value decomposition of $\mathbf{X} = \mathbf{U}_x \tilde{\mathbf{D}}_x \tilde{\mathbf{V}}_x^{\top}$, where $\tilde{\mathbf{D}}_x$ and $\tilde{\mathbf{V}}_x$ are formed by dropping from $\mathbf{D}_x$ and $\mathbf{V}_x$, respectively, the columns that correspond to the zero singular values. The resulting $\tilde{\mathbf{D}}_x$ and $\tilde{\mathbf{V}}_x$ are $n \times n$ and $p \times n$-dimensional, respectively. Dropping these columns has no effect on the matrix factorization of $\mathbf{X}$, i.e. $\mathbf{X} = \mathbf{U}_x \mathbf{D}_x \mathbf{V}_x^{\top} = \mathbf{U}_x \tilde{\mathbf{D}}_x \tilde{\mathbf{V}}_x^{\top}$. The ridge regression estimator can be rewritten in terms of the thin singular value decomposition matrices, which follows from the ridge regression estimator's formulation (\ref{form:formRidgeSpectral}):
\begin{eqnarray*}
\hat{\bbeta}(\lambda) & = & \mathbf{V}_x (\mathbf{D}_x^{\top} \mathbf{D}_x  + \lambda \mathbf{I}_{pp})^{-1} \mathbf{D}_x^{\top} \mathbf{U}_x^{\top} \mathbf{Y} \, \, \, = \, \, \, \tilde{\mathbf{V}}_x (\tilde{\mathbf{D}}_x^{\top} \tilde{\mathbf{D}}_x  + \lambda \mathbf{I}_{nn})^{-1} \tilde{\mathbf{D}}_x^{\top} \mathbf{U}_x^{\top} \mathbf{Y}.
\end{eqnarray*}
Hence, the reformulated ridge regression estimator involves the inversion of an $n \times n$-dimensional matrix. With a sample size in the order of hundreds, this is feasible on most standard computers.

\begin{figure}[!h]
\centering
\includegraphics[scale=0.40, angle=90]{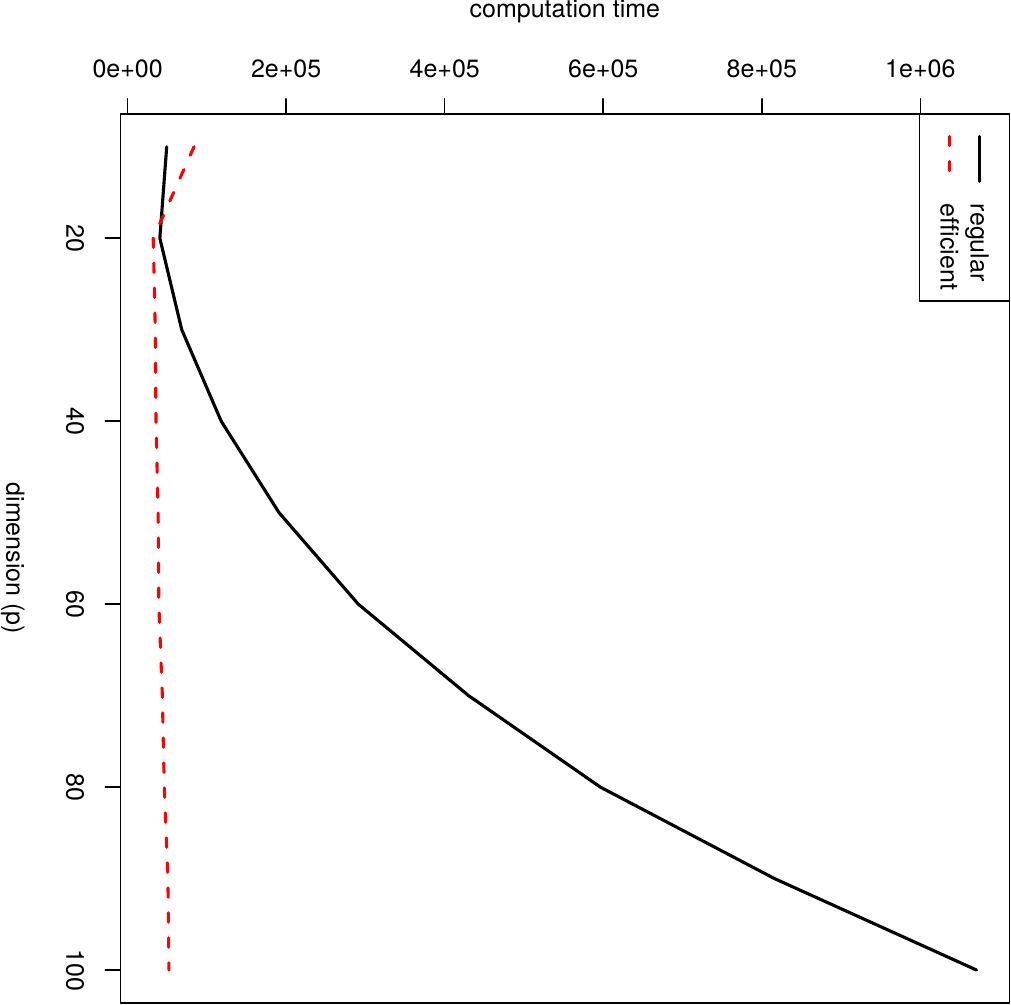}
\caption{Computation time of the evaluation of the ridge regression estimator using the naive and efficient implementation against the dimension $p$. 
\label{fig.compTimeEffRidge}}
\end{figure}

\cite{Hast2004} point out that, with the SVD-trick above, the number of computation operations reduces from $\mathcal{O}(p^3)$ to $\mathcal{O}(p n^2)$. In addition, they point out that this computational short-cut can be used in combination with other loss functions, e.g. that of standard generalized linear models (see Chapter \ref{chap.ridgeLogistic}). This computation is illustrated in Figure \ref{fig.compTimeEffRidge}, which shows the substantial gain in computation time of the evaluation of the ridge regression estimator using the efficient over the naive implementation against the dimension $p$. Details of this figure are provided in Question \ref{question:efficientEvaluation}.

The inversion of the $p \times p$-dimensional matrix can be avoided in an other way. Hereto one needs the Woodbury identity. Let $\mathbf{A}$, $\mathbf{U}$ and $\mathbf{V}$ be $p \times p$-, $p \times n$- and $n \times p$-dimensional matrices, respectively.
The (simplified form of the) Woodbury identity then is:
\begin{eqnarray*}
(\mathbf{A} + \mathbf{U} \mathbf{V})^{-1} & = & \mathbf{A}^{-1} - \mathbf{A}^{-1} \mathbf{U} (\mathbf{I}_{nn} + \mathbf{V} \mathbf{A}^{-1} \mathbf{U})^{-1} \mathbf{V} \mathbf{A}^{-1}.
\end{eqnarray*}
Application of the Woodbury identity to the matrix inverse in the ridge estimator of the regression parameter gives:
\begin{eqnarray} 
\nonumber
\hat{\bbeta}(\lambda) & = & (\mathbf{X}^{\top} \mathbf{X} + \lambda \mathbf{I}_{pp})^{-1} \mathbf{X}^{\top} \mathbf{Y} \hspace{1.2cm} = \, \, \,  [\lambda^{-1} \mathbf{I}_{pp} - \lambda^{-2} \mathbf{X}^{\top} (\mathbf{I}_{nn} + \lambda^{-1} \mathbf{X} \mathbf{X}^{\top})^{-1} \mathbf{X}] \mathbf{X}^{\top} \mathbf{Y} \qquad \mbox{ }
\\
\label{form.effRidgeEstimator}
& = & 
\lambda^{-1} \mathbf{X}^{\top} (\mathbf{I}_{nn} + \lambda^{-1} \mathbf{X} \mathbf{X}^{\top})^{-1} \mathbf{Y}.
\end{eqnarray}
The inversion of the $p \times p$-dimensional matrix $\mathbf{X}^{\top} \mathbf{X} + \lambda \mathbf{I}_{pp}$ is thus replaced by that of the $n \times n$-dimensional matrix $\mathbf{I}_{nn} + \lambda^{-1} \mathbf{X} \mathbf{X}^{\top}$. In addition, this expression of the ridge regression estimator avoids the singular value decomposition of $\mathbf{X}$, which may in some cases introduce additional numerical errors (e.g. at the level of machine precision).

\section{Penalty parameter selection} \label{sect:penaltyParameterSelectionL2}
Throughout the introduction of the ridge regression estimator and the subsequent discussion of its properties, we considered the penalty parameter to be known or `given'. In practice, it is unknown and the user needs to make an informed decision on its value. We present several strategies to facilitate such a decision. Prior to that, we discuss some sanity requirements one may wish to impose on the ridge regression estimator. Those requirements do not yield a specific choice of the penalty parameter but they specify a domain of sensible penalty parameter values. 

The evaluation of the ridge regression estimator may be subject to numerical inaccuracy, which ideally is to be avoided. This numerical inaccuracy results from the ill-conditionedness of $\mathbf{X}^{\top} \mathbf{X} + \lambda \mathbf{I}_{pp}$. A matrix is \textit{ill-conditioned} if its condition number is high. The \textit{condition number} of a square positive definite matrix $\mathbf{A}$ is the ratio of its largest and smallest eigenvalue. If the smallest eigenvalue is zero, the conditional number is undefined and so is $\mathbf{A}^{-1}$. Furthermore, a high condition number is indicative of the loss (on a log-scale) in numerical accuracy of the evaluation of $\mathbf{A}^{-1}$. To ensure the numerical accurate evaluation of the ridge regression estimator, the choice of the penalty parameter is thus to be restricted to a subset of the positive real numbers such that it yields a well-conditioned  matrix $\mathbf{X}^{\top} \mathbf{X} + \lambda \mathbf{I}_{pp}$. Clearly, the penalty parameter $\lambda$ should not be too close to zero. There is however no consensus on the criteria on the condition number for a matrix to be well-defined. This depends among others on how much numerical inaccuracy is tolerated by the context. Of course, as pointed out in Section \ref{sect.ridgeEfficientCalculation}, inversion of the $\mathbf{X}^{\top} \mathbf{X} + \lambda \mathbf{I}_{pp}$ matrix can be circumvented. One then still needs to ensure the well-conditionedness of $\lambda^{-1} \mathbf{X} \mathbf{X}^{\top} + \mathbf{I}_{nn}$, which too results in a lower bound for the penalty parameter. Practically, following \cite{Peeters2019}, who do so in a different context, we suggest to generate a conditional number plot. It plots (on some convenient scale) the condition number of the matrix $\mathbf{X}^{\top} \mathbf{X} + \lambda \mathbf{I}_{pp}$ against the penalty parameter $\lambda$. From this plot, we identify the domain of the penalty parameter value associated with well-conditionedness. To guide in this choice, \cite{Peeters2019} overlay this plot with a curve indicative of the numerical inaccurary.

Traditional text books on regression analysis suggest, in order to prevent over-fitting, to limit the number of covariates in the model. This ought to leave enough degrees of freedom to estimate the error and facilitate proper inference. While ridge regression is commonly used for prediction purposes and inference need not be the objective, over-fitting is certainly to be avoided. This can be achieved by limiting the degrees of freedom spent on the estimation of the regression parameter \citep{harrell2001regression}. We thus follow \cite{Saleh2019theory}, who illustrate this in the ridge regression context, and use the degrees of freedom to bound the search domain of the penalty parameter. This requires the specification of a maximum degrees of freedom, denoted by $\nu$ with $0 \leq \nu \leq n$, one wishes to spend on the construction of the ridge regression estimator. Now choose $\lambda$ such that $\mbox{df}[ \hat{\bbeta}(\lambda) ] \leq \nu$. To find the bound on the penalty parameter, note that 
\begin{eqnarray*}
\mbox{df}[ \hat{\bbeta}(\lambda) ] \, \, \, = \, \, \, \mbox{tr}[ \mathbf{H}(\lambda)] & = & \sum\nolimits_{j=1}^{\min\{n,p\}}  (\mathbf{D}_x^{\top} \mathbf{D}_x)_{jj} [(\mathbf{D}_x^{\top} \mathbf{D}_x)_{jj} + \lambda]^{-1} 
\\
& \leq & \min\{n,p\} \, [ d_1 (\mathbf{X})]^2 \{ [d_1 (\mathbf{X})]^2 + \lambda \}^{-1}.
\end{eqnarray*}
Then, if $\lambda \geq \max\{ 0, \nu^{-1} [d_1 (\mathbf{X})]^2 (\min\{n,p\} - \nu)\}$, the degrees of freedom consumed by the ridge regression estmator is smaller than $\nu$. But how to choose the maximum degrees of freedom $\nu$ to be spend on the estimator? This may be determined by the context. If that does not resolve the matter, \cite{harrell2001regression} suggests as a rule of thumb to choose $\nu$ as a fraction of the sample size but provides no recommendation on the size of this fraction. As a last resort, we can base this choice of $\nu$ on the degrees of freedom necessary to obtain a reliable estimate of the error variance, which suggests a conservative upperbound of $\max\{0, n- 30\}$ on $\nu$.

\subsection{Analytic expressions}
Ideally, we would have an analytic prescription how to choose the penalty parameter. Typically, these are somewhat ad-hoc as they are not motivated from a general principle. For instance, in their original paper on ridge regression, \cite{Hoer1970} provide one. They base their choice on the observation that if $\mathbf{X}$ is orthonormal, the $\mbox{MSE}[ \hat{\boldsymbol{\beta}} ( \lambda) ]$ is minimized for $\lambda = p \sigma^2 \| \boldsymbol{\beta} \|_2^{-2}$ (see Example \ref{example:ridgeMSEorthonormal}), and propose to use:
\begin{eqnarray*}
\hat{\lambda} & = & p \hat{\sigma}^2 \| \hat{\boldsymbol{\beta}}_{\mbox{{\tiny ml}}} \|_2^{-2},
\end{eqnarray*}
with $\hat{\sigma}^2  = n^{-1} \| \mathbf{Y} - \mathbf{X} \hat{\boldsymbol{\beta}}_{\mbox{{\tiny ml}}} \|_2^2$. In the high-dimensional context, this proposal is of little use as it depends on the existence of the maximum likelihood estimator. This may be circumvented by replacing the maximum likelihood estimator of $\bbeta$ by its minimum least squared counterpart. In this vain, many proposals have been made in the literature: \cite{mermi2024new} have gathered 366 (!) such proposals from literature, and even proposed an additional one. The behavior of these proposals has been studied extensively by simulation (see the references in \citealp{mermi2024new}). These are limited by their very nature. For instance, none of the analytic choices of $\lambda$ in the overview of \cite{mermi2024new} comes with the guarantee of a superior MSE of $\hat{\boldsymbol{\beta}}(\hat{\lambda})$ over $\hat{\boldsymbol{\beta}}_{\mbox{{\tiny ml}}}$. That said, they may still be sensible choices for the case at hand.

\subsection{Information criterion}
A popular strategy is to choose a penalty parameter that yields a good but parsimonious model. Information criteria measure the balance between model fit and model complexity. Here we present the Akaike's information criterion (AIC, \citealp{Akai1974}), but many other criteria have been presented in the literature (see, e.g. \citealp{Schw1978}). The AIC measures model fit by the loglikelihood and model complexity as measured by the number of parameters used by the model. The number of model parameters in regular regression simply corresponds to the number of covariates in the model. Or, by the degrees of freedom consumed by the model, which is equivalent to the trace of the hat matrix. For ridge regression, it thus seems natural to define model complexity analogously by the trace of the ridge hat matrix. This yields the AIC for the linear regression model with ridge regression estimates:
\begin{eqnarray*}
\mbox{AIC}(\lambda) & = & 2 \, (\mbox{\# estimated model parameters}) - 2 \log\{L[\mathbf{Y}, \mathbf{X}; \hat{\bbeta}(\lambda), \hat{\sigma}^2(\lambda)]\}
\\
& = & 2 \, \mbox{tr} [\mathbf{H}(\lambda)] - 2 \log\{L[\mathbf{Y}, \mathbf{X}; \hat{\bbeta}(\lambda), \hat{\sigma}^2(\lambda)]\}
\\
& = & 2 \, \sum\nolimits_{j=1}^p (\mathbf{D}_x^{\top} \mathbf{D}_x)_{jj} [(\mathbf{D}_x^{\top} \mathbf{D}_x)_{jj} + \lambda]^{-1} 
\\
& & + 2 n \, \log[\sqrt{2 \, \pi} \, \hat{\sigma}(\lambda)] +  \hat{\sigma}^{-2}(\lambda)  \sum\nolimits_{i=1}^n [y_i - \mathbf{X}_{i, \ast} \, \hat{\bbeta}(\lambda)]^2.
\\
& = & 2 \, \sum\nolimits_{j=1}^p (\mathbf{D}_x^{\top} \mathbf{D}_x)_{jj} [(\mathbf{D}_x^{\top} \mathbf{D}_x)_{jj} + \lambda]^{-1} + 2 n \, \log[\sqrt{2 \, \pi} \, \hat{\sigma}(\lambda)] +  n,
\end{eqnarray*}
as $\hat{\sigma}^{-2}(\lambda) = n^{-1} \| \mathbf{Y} - \mathbf{X} \hat{\bbeta}(\lambda) \|_2^2$. The value of $\lambda$ which minimizes $\mbox{AIC}(\lambda)$ corresponds to the `optimal' balance of model complexity and overfitting.

One may question the use of information criteria to guide the choice of the penalty parameter within the context of ridge regression. Information criteria guide the decision process when having to decide among various different models. Different models use different sets of explanatory variables to explain the behaviour of the response variable. In that sense, the use of information criteria for the deciding on the ridge penalty parameter may be considered inappropriate: ridge regression uses the same set of explanatory variables irrespective of the value of the penalty parameter.  Moreover, often ridge regression is employed to predict a response and not to provide an insightful explanatory model. The latter need not yield the best predictions. 

\subsection{Cross-validation} \label{subsect.crossvalidation}
Cross-validation requires the regularization parameter to yield a model with good predictive performance. Commonly, this performance is evaluated on novel data. Novel data need not be easy to come by and one has to make do with the data at hand. The setting of `original' and novel data is then mimicked by sample splitting: the data set is divided into two (groups of) samples. One of these two data sets, called the \textit{training set}, plays the role of `original' data on which the model is built. The second of these data sets, called the \textit{test set}, plays the role of the `novel' data and is used to evaluate the prediction performance (often operationalized as the loglikelihood or the prediction error) of the model built on the training data set. This procedure (model building and prediction evaluation on training and test set, respectively) is done for a collection of possible penalty parameter choices. The penalty parameter that yields the model with the best prediction performance is to be preferred. The thus obtained performance evaluation depends on the actual split of the data set. To remove this dependence, the data set is split many times into a training and test set. Per split, the model parameters are estimated for all choices of $\lambda$ using the training data and estimated parameters are evaluated on the corresponding test set. The penalty parameter, that on average over the test sets performs best (in some sense), is then selected.

When the repetitive splitting of the data set is done randomly,  samples may accidently end up in a vast majority of the splits in either training or test set. Such samples may have an unbalanced influence on either model building or prediction evaluation. To avoid this $k$-fold cross-validation structures the data splitting. The samples are divided into $k$ more or less equally sized exhaustive and mutually exclusive subsets. In turn (at each split) one of these subsets plays the role of the test set while the union of the remaining subsets constitutes the training set. Such a splitting warrants a balanced representation of each sample in both training and test set over the splits. Still the division into the $k$ subsets involves a degree of randomness. This may be fully excluded when choosing $k=n$. This particular case is referred to as leave-one-out cross-validation (LOOCV). For illustration purposes the LOOCV procedure is detailed fully below:
\begin{compactitem}
\item[0)] Define a range of interest for the penalty parameter.

\item[1)] Divide the data set into training and test set comprising samples $\{1, \ldots, n\} \setminus i$ and $\{ i \}$, respectively.

\item[2)] Fit the linear regression model by means of ridge estimation  for each $\lambda$ in the grid using the training set. This yields:
\begin{eqnarray*}
\hat{\bbeta}_{-i}(\lambda) & = & ( \mathbf{X}_{-i, \ast}^{\top}
\mathbf{X}_{-i, \ast} + \lambda \mathbf{I}_{pp})^{-1}
\mathbf{X}_{-i, \ast}^{\top} \mathbf{Y}_{-i},
\end{eqnarray*}
where $\mathbf{X}_{-i, \ast}$ and $\mathbf{Y}_{-i}$ are the design matrix and response vector with the $i$-th row and element, respectively, excluded. The corresponding estimate of the error variance $\hat{\sigma}_{-i}^2(\lambda)$. 

\item[3)] Evaluate the prediction performance of these models on the test set by $\log\{L[Y_i, \mathbf{X}_{i, \ast}; \hat{\bbeta}_{-i}(\lambda), \hat{\sigma}_{-i}^2(\lambda)]\}$. Or, by the prediction error $|Y_i - \mathbf{X}_{i, \ast} \hat{\bbeta}_{-i}(\lambda)|$, possibly squared.

\item[4)] Repeat steps 1) to 3) such that each sample plays the role of the test set once.

\item[5)] Average the prediction performances of the test sets at each grid point of the penalty parameter:
\begin{eqnarray*}
n^{-1} \sum\nolimits_{i = 1}^n \log\{L[Y_i, \mathbf{X}_{i, \ast}; \hat{\bbeta}_{-i}(\lambda), \hat{\sigma}_{-i}^2(\lambda)]\}.
\end{eqnarray*}
The quantity above is called the \textit{cross-validated loglikelihood}. It is an estimate of the prediction performance of the model corresponding to this value of the penalty parameter on novel data.

\item[6)] The value of the penalty parameter that maximizes the cross-validated loglikelihood is the value of choice.
\end{compactitem}
The procedure is straightforwardly adopted to $k$-fold cross-validation, a different criterion, and different estimators.

In the LOOCV procedure above, resampling can be avoided when the predictive performance is measured by Allen's PRESS (Predicted Residual Error Sum of Squares) statistic \citep{Alle1974}. For then, the LOOCV predictive performance can be expressed analytically in terms of the known quantities derived from the design matrix and response (as pointed out but not detailed in \citealt{Golu1979}). Define the optimal penalty parameter to minimize Allen's PRESS statistic:
\begin{eqnarray*}
\lambda_{\mbox{{\tiny opt}}} & = & \arg \min\nolimits_{\lambda > 0} n^{-1} \sum\nolimits_{i=1}^n [Y_i - \mathbf{X}_{i, \ast} \hat{\bbeta}_{-i}(\lambda)]^2 \, \, \, = \, \, \, \arg \min\nolimits_{\lambda > 0}  n^{-1} \| \mathbf{B}(\lambda) [\mathbf{I}_{nn} - \mathbf{H}(\lambda)] \mathbf{Y} \|_ 2^2,
\end{eqnarray*}
where $\mathbf{B}(\lambda)$ is diagonal with $[\mathbf{B}(\lambda)]_{ii} = [ 1 - \mathbf{H}_{ii}(\lambda)]^{-1}$. The second equality in the preceding display is elaborated in Exercise \ref{question:efficientLOOCV}. Its main takeaway is that the predictive performance for a given $\lambda$ can be assessed directly from the ridge hat matrix and the response vector without the recalculation of the $n$ leave-one-out ridge regression estimators. Computationally, this is a considerable gain.

No such analytic expression of the cross-validated loss as above exists for general $K$-fold cross-validation, but considerable computational gain can nonetheless be achieved \citep{vdWiel2020fast}. This exploits the fact that the ridge regression estimator appears in Allen's PRESS statistics -- and the likelihood -- only in combination with the design matrix, together forming the linear predictor. There is thus no need to evaluate the estimator itself when interest is only in its predictive performance. Then, if $\mathcal{G}_1, \ldots, \mathcal{G}_K \subset \{1, \ldots, n\}$ are the mutually exclusive and exhaustive $K$-fold sample index sets, the linear predictor for the $k$-th fold can be expressed as:
\begin{eqnarray} \label{form.efficientCVlinearpredictor}
\mathbf{X}_{\mathcal{G}_k, \ast} \hat{\bbeta}_{-\mathcal{G}_k}(\lambda) & = & \lambda^{-1} \mathbf{X}_{\mathcal{G}_k, \ast} \mathbf{X}_{-\mathcal{G}_k, \ast}^{\top} 
(\lambda^{-1} \mathbf{X}_{-\mathcal{G}_k, \ast} \mathbf{X}_{-\mathcal{G}_k, \ast}^{\top} + \mathbf{I}_{n-|\mathcal{G}_k|, n-|\mathcal{G}_k|})^{-1} \mathbf{Y}_{-\mathcal{G}_k, \ast},
\end{eqnarray}
where we have used the Woodbury identity again. Finally, for each fold the computationally most demanding matrices of this expression, $\mathbf{X}_{\mathcal{G}_k, \ast} \mathbf{X}_{-\mathcal{G}_k, \ast}^{\top}$ and $\mathbf{X}_{-\mathcal{G}_k, \ast} \mathbf{X}_{-\mathcal{G}_k, \ast}^{\top}$, are both submatrices of $\mathbf{X} \mathbf{X}^{\top}$. If the latter matrix is evaluated prior to the cross-validation loop, all calculations inside the loop involve only matrices of dimensions $n \times n$, maximally, and obtained from $\mathbf{X} \mathbf{X}^{\top}$ by subsetting.

\subsection{Generalized cross-validation}
Generalized cross-validation is another method to guide the choice of the penalty parameter. It is like cross-validation but with a different criterion to evaluate the performance of the ridge regression estimator on novel data. This criterion, denoted $\mbox{GCV}(\lambda)$ (where GCV is an acronym of Generalized Cross-Validation), is an approximation to Allen's PRESS statistic. In the previous subsection, this statistic was reformulated as:
\begin{eqnarray*}
n^{-1} \sum\nolimits_{i=1}^n [Y_i - \mathbf{X}_{i, \ast} \hat{\bbeta}_{-i}(\lambda)]^2 & = & n^{-1} \sum\nolimits_{i=1}^n  [ 1 - \mathbf{H}_{ii}(\lambda)]^{-2} [ Y_i - \mathbf{X}_{i, \ast}^{\top} \hat{\bbeta}(\lambda) ]^2.
\end{eqnarray*}
The identity $\mbox{tr}[\mathbf{H}(\lambda)] = \sum_{i=1}^n [\mathbf{H}(\lambda)]_{ii}$ suggests $[\mathbf{H}(\lambda)]_{ii} \approx n^{-1} \mbox{tr}[\mathbf{H}(\lambda)]$. The approximation thus proposed by \cite{Golu1979}, which they endow with a `weighted version of Allen's PRESS statistic'-interpretation, is:
\begin{eqnarray*}
\mbox{GCV}(\lambda) & = & n^{-1} \sum\nolimits_{i=1}^n  \{ 1 - n^{-1} \mbox{tr}[ \mathbf{H}(\lambda)]\}^{-2} [ Y_i - \mathbf{X}_{i, \ast}^{\top} \hat{\bbeta}(\lambda) ]^2
\\
& = & n^{-1}  \{ 1 - n^{-1} \mbox{tr}[ \mathbf{H}(\lambda)]\}^{-2} \| [ \mathbf{I}_{nn} - \mathbf{H}(\lambda) ] \mathbf{Y} \|_2^2.
\end{eqnarray*}
The need for this approximation is pointed out by \cite{Golu1979} by example. They present an `extreme' case where the minimization of Allen's PRESS statistic fails to produce a well-defined choice of the penalty parameter $\lambda$. This `extreme' case requires a (unit) diagonal design matrix $\mathbf{X}$. Straightforward (linear) algebraic manipulations of Allen's PRESS statistic then yield:
\begin{eqnarray*}
n^{-1} \sum\nolimits_{i=1}^n  [ 1 - \mathbf{H}_{ii}(\lambda)]^{-2} [ Y_i - \mathbf{X}_{i, \ast}^{\top} \hat{\bbeta}(\lambda) ]^2 & = & n^{-1} \sum\nolimits_{i=1}^n Y_i^2,
\end{eqnarray*}
which indeed has no unique minimizer in $\lambda$. Additionally, the $\mbox{GCV}(\lambda)$ criterion may be preferred in cases where it is computationally easier to evaluate $\mbox{tr}[ \mathbf{H}(\lambda)]$ (e.g. from the singular values) than the individual diagonal elements of  $\mathbf{H}(\lambda)$. 

\subsection{Other methods}
Other methods to choose the penalty parameter in informative ways have been proposed. Some will be encountered in later chapters of this book, e.g. empirical Bayes (Section \ref{sect.empBayes}), via mixed model estimation (Section \ref{sect:mixedModel2ridgeRegression}) , stability selection (Section \ref{sect:stabilitySelection}). But also more implicitly approaches of choosing the penalty parameter such as drop out (Section  \ref{sect:dropout}), early stopping (Subsection \ref{sect:earlyStopping}), data augmentation and adding noise (Subsection \ref{sect:dataAugmentation}) will be discussed.

\subsection{Randomness}
The discussed procedures for penalty parameter selection all depend on the data at hand. As a result, so does the selected penalty parameter. It should therefore be considered a statistic, a quantity calculated from data. In case of $K$-fold cross-validation, the random formation of the splits adds another layer of randomness. Irrespectively, a statistic exhibits randomness.   This randomness propagates into the ridge regression estimator. The distributional properties of the ridge regression estimator derived in Section \ref{sect:ridgeMoments} are thus conditional on the penalty parameter. 
\begin{figure}[!h]
\begin{tabular}{rcl}
\includegraphics[scale=0.22, angle=0]{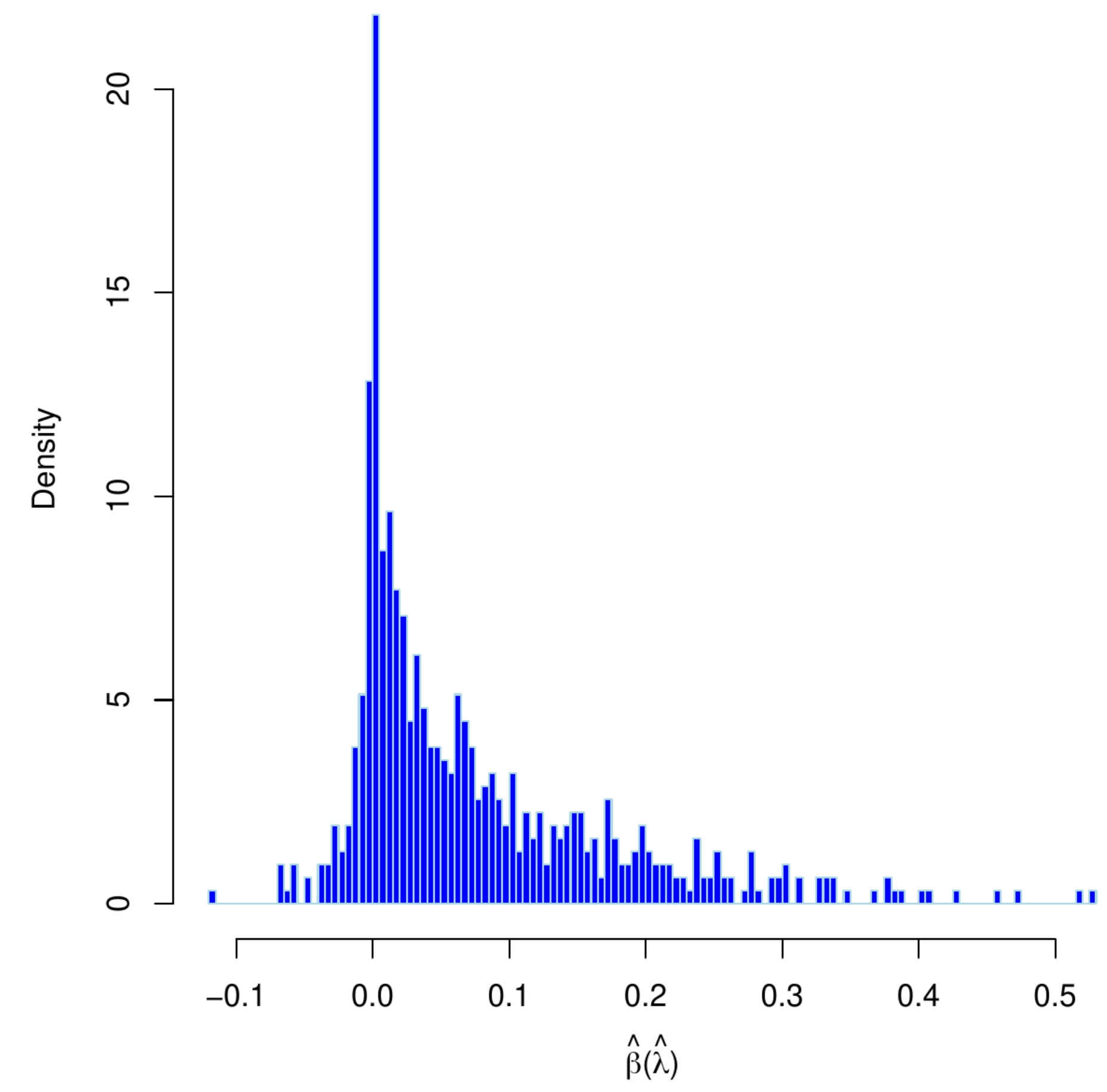}
&
\includegraphics[scale=0.22, angle=0]{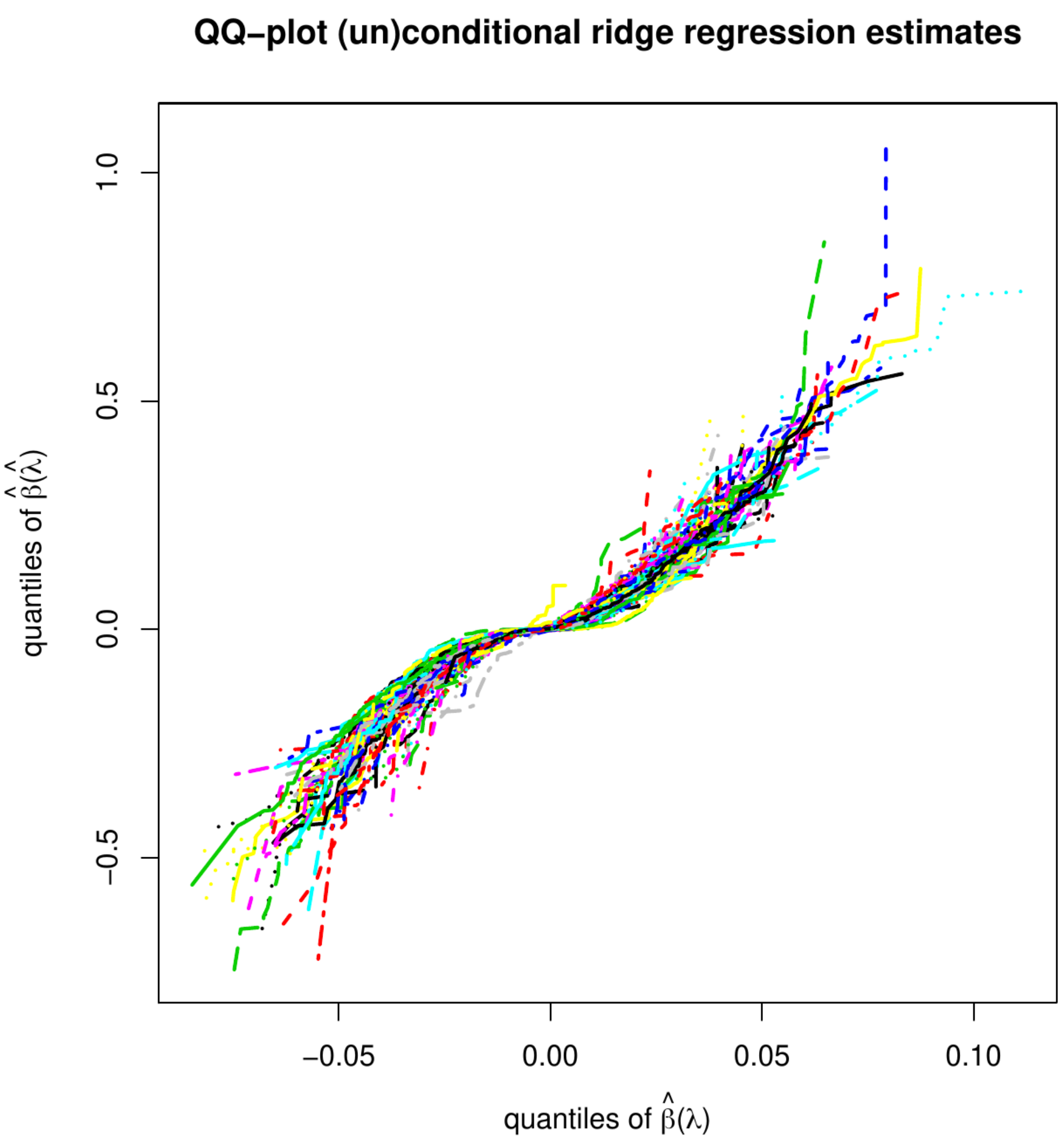}
\end{tabular}
\caption{Left panel: histogram of estimates of an element of $\hat{\bbeta}(\hat{\lambda})$ obtained from thousand bootstrap samples with re-tuned penalty parameters. Right penal: qq-plot of the estimates estimates of an element of $\hat{\bbeta}(\hat{\lambda})$ obtained from thousand bootstrap samples with re-tuned vs. fixed penalty parameters.}
\label{fig.unconditionalBetas}
\end{figure}
Those of the unconditional distribution of the ridge regression estimator, now denoted $\hat{\bbeta}(\hat{\lambda})$ with $\hat{\lambda}$ indicating that the penalty depends on the data (and possibly the particulars of the splits), may be rather different. Analytic finite sample results appear to be unavailable. An impression of the distribution of $\hat{\bbeta}(\hat{\lambda})$ can be obtained through simulation. Hereto we have first drawn a data set from the linear regression model with dimension $p=100$, sample size $n=25$, a standard normally distribution error, the rows of the design matrix sampled from a zero-centered multivariate normal distribution with a uniformly correlated but equivariant covariance matrix, and a regression parameter with elements equidistantly distributed over the interval $[-2\tfrac{1}{2}, 2 \tfrac{1}{2}]$. From this data set, we then generate thousand nonparametric bootstrapped data sets. For each bootstrapped data set, we select the penalty parameter by means of $K=10$-fold cross-validation and evaluate the ridge regression estimator using the selected penalty parameter. The left panel of Figure \ref{fig.unconditionalBetas} shows the histogram of the thus acquired estimates of an arbitrary element of the regression parameter. The shape of the distribution clearly deviates from the normal one of the conditional ridge regression estimator. In the right panel of Figure \ref{fig.unconditionalBetas}, we have plotted element-wise quantiles of the conditional vs. unconditional ridge regression estimators. It reveals that not only the shape of the distribution, but also its moments are affected by the randomness of the penalty parameter.

\subsection{Behavior around $n\approx p$}
How is the cross-validated regularization parameter affected by an increase in the number of covariates? This we investigate using the following simulation set-up. Data are sampled from the linear regression model. Throughout the sample size is fixed at $n=100$, while the elements of the design matrix and the error are drawn from the standard normal distribution. We vary the dimension $p$ over $10, 20, \ldots, 200$ and let the elements of the regression parameter be equidistantly distributed from $-2 \tfrac{1}{2} p^{-1/2}$ to $-2 \tfrac{1}{2} p^{-1/2}$. The scaling by the square root $p$ ensures the signal is of comparable size among the dimensions. For all $(n,p)$-combinations, we  generate 200 data sets, from which we form a $1000$ nonparametrically bootstrapped versions. With each data set, we choose the regularization parameter to optimize the 10-fold cross-validated loss. The bootstrappped data sets capture the within-data-set uncertainty of the chosen regularization parameter, which is then averaged over the 200 Monte Carlo draws. With the optimal regularization parameter at hand, we evaluate the ridge regression estimator and its mean squared error.

\begin{figure*}[b!]
\centering
\begin{tabular}{ll}
\includegraphics[angle=0, scale=0.41]{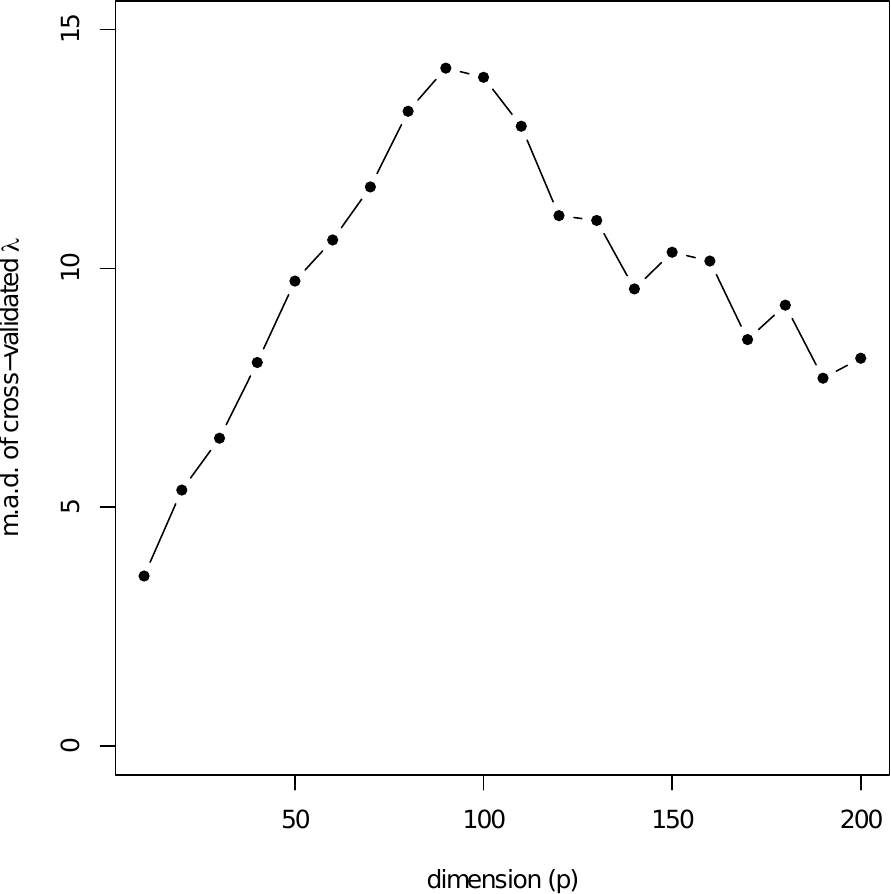}
&
\includegraphics[angle=0, scale=0.41]{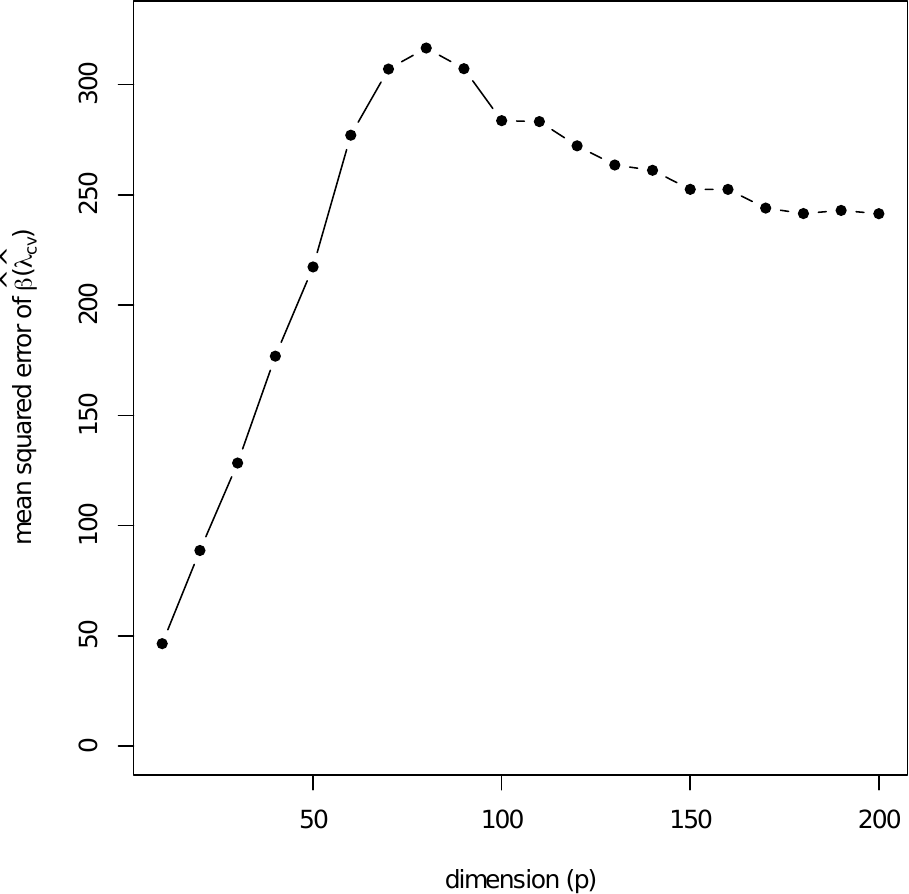}
\end{tabular}
\caption{Left panel: the median  absolute deviation of the cross-validated regularization parameter against the dimension. Right panel: the estimated mean squared error of the ridge regression estimator at the optimal regularization parameter.} \label{fig.lambdaAndMse2dimBehavior}
\hfill{}
\end{figure*}

Figure \ref{fig.lambdaAndMse2dimBehavior} depicts the simulation results. The left panel reveals the uncertainty -- operationalized by the median absolute deviation -- of the optimal regularization parameter over the dimensions. It first goes up with the dimension, until it reaches a peak around $n \approx p$ and then slowly goes down. We explain the peak behavior below. We first note that this uncertainty behavior propagates into the estimator. This can be witnessed from the right-hand side plot, which depicts the estimator's mean squared error against the dimension. It has a peak around the same location. This detoriation in performance around $n \approx p$ has been observed previously in other settings (see, e.g. \citealp{duin2000classifiers}).

Why do we see the largest in uncertainty in the cross-validated regularization parameter around $n \approx p$? The reason is to be found in the inversion of the matrix $\mathbf{X}^{\top} \mathbf{X} + \lambda \mathbf{I}_{pp}$. 
\begin{compactitem}
\item[$\circ$] In the low-dimensional regime, $n \gg p$, the eigenvalues of the $\mathbf{X}^{\top} \mathbf{X}$ matrix are all clearly distinct from zero. This is not (much) affected if we remove a few rows from the design matrix, as we do in cross-validation. The inverse of $\mathbf{X}^{\top} \mathbf{X} + \lambda \mathbf{I}_{pp}$ is then well-defined, irrespective of the choice of regularization parameter $\lambda$.
Consequently, the predictions of a left-out sample do not differ dramatically for close but different choices of $\lambda$. In turn, the cross-validated loss does not vary dramatically with small changes in $\lambda$. This loss function is (relatively) smooth in $\lambda$ and has a stable optimum.

\item[$\circ$] On the end of the spectrum, in the high-dimensional regime, where $p \gg n$, many eigenvalues of $\mathbf{X}^{\top} \mathbf{X}$ are zero. The folds in cross-validation are more high-dimensional, making this issue more pronounced. Irrespective, zero eigenvalues are a given. To ensure a well-defined inverse of $\mathbf{X}^{\top} \mathbf{X} + \lambda \mathbf{I}_{pp}$, the regularization parameter has to be large. For then, then the conditional number of this matrix, $\lambda^{-1} (d_{x,1}^2 + \lambda)$, is not too large and the inverse well-defined. A sensible prediction thus requires a substantially large regularization parameter. If $\lambda$ is clearly bounded away from zero, the predictions for a left-out sample with a large and a large-plus-a-little-bit $\lambda$ tend to be close. For large $\lambda$, the cross-validated loss is well-behaved function in $\lambda$ again quite stable.

\item[$\circ$] At the boundary of those two regimes, $n\approx p$, life is different. There, the eigenvalues of $\mathbf{X}^{\top} \mathbf{X}$ tend to be close, if not equal, to zero. Suppose that the sample size and dimension of the row subsetted design matrices in the cross-validation folds are approximately equal. The corresponding $\mathbf{X}^{\top} \mathbf{X}$ matrix product of these design matrix then have eigenvalues close or equal to zero. The exact values of the smallest eigenvalues vary over the folds. The differences in these smallest eigenvalues have large consequences, as the contribution of $\lambda$ on the condition number is most notable if the smallest eigenvalue is close to zero. The smallest eigenvalues thus determine the amount of regularization that warrants a well-defined inverse of $\mathbf{X}^{\top} \mathbf{X} + \lambda \mathbf{I}_{pp}$. For instance, some covariates maybe collinear in one fold but not in another fold. In the first instance, a large regularization parameter is needed. While in the other fold, the eigenvalues may fall conveniently somewhat away from zero. There, less, and maybe even little, regularization is needed to yield a well-defined inverse and a reasonable prediction. This fold-specific variation in regularization propagates in the left-out predictions and the cross-validated loss and yields more variation in the optimum. 
\end{compactitem}
The behavior of the regularization parameter around $n=p$ is not specific for cross-validation. Other methods to choose the hyperparameter, like empirical Bayes (discussed in Section \ref{sect.empBayes}), have been shown to exhibit similar behavior around $n=p$ \citep{vandeWiel2019learning}.

\section{Connections}
The ridge regression estimator is closely related to various other statistical methods and machine learning techniques. In this section, we discuss its connection to other pragmatic ways of regularization. In subsequent chapters, we elaborate on other machine learning procedures that are also closely connected, if not one-to-one, to ridge regression.

\subsection{Early stopping} \label{sect:earlyStopping}
\textit{Early stopping} is a regularization technique from the `old days' with limited computing power, when it was not feasible to run an iterative optimizer till convergence. Early stopping results in a form of regularization. This exploits the equivalence between regularized and constrained estimation. Intuitively, it should be clear that, when iteratively optimizing and starting from some arbitrary initial value, premature stopping restricts the space of the final solution. It is to be found in a subset around the initial value. Effectively, the parameter space is constrained (and the estimator thus regularized). And we accept that we are not at the optimum but have nonetheless come closer to that optimum. 

We illustrate early stopping for the linear regression model. We still aim to optimize the sum of squares $\| \mathbf{Y} - \mathbf{X} \boldsymbol{\beta} \|_2^2$ with respect to  $\boldsymbol{\beta}$. We temporarily forget that we have an analytic expression for the optimum available and use \textit{gradient descent} to minimize the loss. Gradient descent finds a function's minimum in an iterative manner. The procedure starts with an initial guess $\boldsymbol{\beta}^{(0)}$ of the regression parameter. It evaluates the direction of steepest descent at the current parameter value $\boldsymbol{\beta}^{(t)}$, which is the gradient of the sum-of-squares: $2\mathbf{X}^{\top} \mathbf{X} \boldsymbol{\beta} - 2 \mathbf{X}^{\top} \mathbf{Y}$. We arrive at an update $\boldsymbol{\beta}^{(t+1)}$ of the regression parameter by proceeding in this direction for a step size, denoted $\delta$ and called the \textit{learning rate}. The update amounts to: $\boldsymbol{\beta}^{(t+1)}  =  \boldsymbol{\beta}^{(t)} - 2 \delta ( \mathbf{X}^{\top} \mathbf{X} \boldsymbol{\beta}^{(t)} - \mathbf{X}^{\top} \mathbf{Y})$. By construction, each update yields a lower loss and is closer to the optimum. Traditionally, i.e. low-dimensionally, we continue this updating until convergence. High-dimensionally, however, we terminate the updating prematurely: we stop early. 

To show that early stopping constrains the estimator, we use the series expansion of the generalized inverse: $(\mathbf{I}_{pp}-\mathbf{A})^{+} = \mathbf{I}_{pp} + \mathbf{A} + \mathbf{A}^2 + \ldots$ for symmetric, singular matrix $\mathbf{A}$ with largest (in absolute sense) eigenvalue smaller than one. We then rewrite the gradient descent update as
\begin{eqnarray*}
\boldsymbol{\beta}^{(t+1)} & = & \boldsymbol{\beta}^{(t)} - 2 \delta ( \mathbf{X}^{\top} \mathbf{X} \boldsymbol{\beta}^{(t)} - \mathbf{X}^{\top} \mathbf{Y})
\\
& = & \ldots 
\\
& = & (\mathbf{I}_{pp} - 2 \delta  \mathbf{X}^{\top} \mathbf{X})^{t+1} \boldsymbol{\beta}^{(0)} + 2 \delta \big[ \sum\nolimits_{t'=0}^t
(\mathbf{I}_{pp} - 2 \delta  \mathbf{X}^{\top} \mathbf{X})^{t'} \big]  \mathbf{X}^{\top} \mathbf{Y}
\\
& = & (\mathbf{I}_{pp} - 2 \delta  \mathbf{X}^{\top} \mathbf{X})^{t+1} \boldsymbol{\beta}^{(0)} +  [ \mathbf{I}_{pp} - (\mathbf{I}_{pp} - 2 \delta  \mathbf{X}^{\top} \mathbf{X})^{t+1}] ( \mathbf{X}^{\top} \mathbf{X})^{+} \mathbf{X}^{\top} \mathbf{Y},
\end{eqnarray*}
where $\delta$ is such that the largest eigenvalue assumption holds. To simplify the illustration, assume \textit{i)} $\mathbf{X}^{\top} \mathbf{X}$ is nonsingular and $(\mathbf{X}^{\top} \mathbf{X})^+ = (\mathbf{X}^{\top} \mathbf{X})^{-1}$ and \textit{ii)} $\boldsymbol{\beta}^{(0)} = \mathbf{0}_{p}$. Then,
\begin{eqnarray*}
\| \boldsymbol{\beta}^{(t+1)} - \hat{\boldsymbol{\beta}}^{\mbox{{\tiny ml}}} \|_2^2 & = & \mathbf{Y}^{\top} \mathbf{X} ( \mathbf{X}^{\top} \mathbf{X})^{-1}  (\mathbf{I}_{pp} - 2 \delta  \mathbf{X}^{\top} \mathbf{X})^{2t+2} ( \mathbf{X}^{\top} \mathbf{X})^{-1} \mathbf{X}^{\top} \mathbf{Y}.
\end{eqnarray*}
The distance between the update and the optimum is controlled by $t$ (and $\delta$). Hence, if we fix $\delta$ and $t$, we know how close our estimator is to the optimum.  

Early stopping and ridge regression are one-to-one related if \textit{i)} all nonzero eigenvalues of $\mathbf{X}^{\top} \mathbf{X}$ are identical and \textit{ii)} $\boldsymbol{\beta}^{(0)} = \mathbf{0}_p$. We can then, using the eigendecomposition of $\mathbf{X}^{\top} \mathbf{X}$, express $\lambda$ in terms of $t$. This is left as an exercise.

\subsection{Data augmentation and adding noise} \label{sect:dataAugmentation}
The ridge regression estimator can be evaluated with standard software designed for the evaluation of the maximum likelihood estimator. Of course, this is an obsolete issue. But the way, in which it is done, has an interesting interpretation. Moreover, it is analogous to a common practice in machine learning to prevent overfitting.  

To evaluate the ridge regression estimator as a maximum likelihood one, requires \textit{data augmentation}. The design matrix $\mathbf{X}$ and response vector $\mathbf{Y}$ are augmented with $p$ additional rows $\sqrt{\lambda} \mathbf{I}_{pp}$ and $p$ zeros, respectively. That is, define $\tilde{\mathbf{X}} = (\mathbf{X}^{\top}  \, \, \sqrt{\lambda} \mathbf{I}_{pp})^{\top}$ and $\tilde{\mathbf{Y}} = (\mathbf{Y}^{\top} \, \, \mathbf{0}_{p}^{\top})^{\top}$. The maximum likelihood estimator of the regression parameter from the experiment with augmented data then is:
\begin{eqnarray*}
\hat{\boldsymbol{\beta}}(\lambda)  & = & (\tilde{\mathbf{X}}^{\top} \tilde{\mathbf{X}})^{-1} \tilde{\mathbf{X}}^{\top} \tilde{\mathbf{Y}}
\hspace{1.3cm} = \, \, \, \left[ (\mathbf{X}^{\top}  \, \sqrt{\lambda} \mathbf{I}_{pp})
\left( \begin{array}{c}
\mathbf{X}
\\
\sqrt{\lambda} \mathbf{I}_{pp}
\end{array}
\right)
\right]^{-1}
(\mathbf{X}^{\top}  \, \sqrt{\lambda} \mathbf{I}_{pp})
\left( \begin{array}{c}
\mathbf{Y}
\\
\mathbf{0}_{p}
\end{array}
\right)
\\
&  = & (\mathbf{X}^{\top} \mathbf{X}  +  \lambda \mathbf{I}_{pp})^{-1} \mathbf{X}^{\top} \mathbf{Y}.
\end{eqnarray*}
This is identical to the ridge regression estimator.

Another and similar but approximate way to evaluate the ridge regression estimator would be to replace the augmented part of the design matrix by rows with random instead of fixed values for the covariates. That is, augment the original design matrix $\mathbf{X}$ with $p$ rows. The $i$-th row, $i > n$, harbors a covariate information vector $\tilde{\mathbf{X}}_{i,\ast}^{\top}$ drawn from $\mathcal{N}( \mathbf{0}_p, \lambda \mathbf{I}_{pp})$. The thus augmented design matrix, again denoted $\tilde{\mathbf{X}}$, satisfies $\tilde{\mathbf{X}}^{\top} \tilde{\mathbf{X}} \approx \mathbf{X}^{\top} \mathbf{X} + p \lambda \mathbf{I}_{pp}$ as $\tilde{\mathbf{X}}_{i,\ast} \tilde{\mathbf{X}}_{i',\ast}^{\top} \approx 0$ for $i, i' > n$ and $i \not= i'$. As before, the response vector is augmented with zero's.

There is an alternative but analogous approach employed in machine learning that prevents overfitting. It amounts to the duplication of an observation, both covariate information vector and the response, say, $\mathbf{X}_{i, \ast}^{\top}$ and $Y_i$. The covariate information vector is replaced by a draw from $\mathcal{N}(\mathbf{X}_{i,\ast}^{\top}, \lambda \mathbf{I}_{pp})$. This is done for each observation repeatedly. As such, the data set is enlarged by perturbed versions of the observations. These duplicated and perturbed versions are different from the original ones, but -- should the perturbation be small -- still close. This too prevents overfitting, as $\tilde{\mathbf{X}}^{\top} \tilde{\mathbf{X}} \approx q \mathbf{X}^{\top} \mathbf{X} + q \lambda \mathbf{I}_{pp}$ and $\tilde{\mathbf{X}}^{\top} \tilde{\mathbf{Y}} \approx q \mathbf{X}^{\top} \mathbf{Y}$ and the maximum likelihood estimator of the perturbed data approximates the ridge regression one.

\section{Simulations}
Simulations are presented that illustrate properties of the ridge regression estimator not discussed explicitly in the previous sections of this chapter.

\subsection{Role of the variance of the covariates} \label{ridge:covariateVariances}
In many applications of high-dimensional data, the covariates are standardized prior to the execution of the ridge regression. Before we discuss whether this is appropriate, we first illustrate the effect of ridge regularization on covariates with distinct variances using simulated data.

The simulation involves one response to be (ridge) regressed on fifty covariates. Data (with $n=1000$) for the covariates, denoted $\mathbf{X}$, are drawn from a multivariate normal distribution: $\mathbf{X} \sim \mathcal{N}(\mathbf{0}_{50}, \mathbf{\Sigma})$ with $\mathbf{\Sigma}$ diagonal and $(\mathbf{\Sigma})_{jj} = j / 10$. From this the response is generated through $\mathbf{Y} = \mathbf{X} \bbeta + \vvarepsilon$ with $\bbeta = \mathbf{1}_{50}$ and $\vvarepsilon \sim \mathcal{N}(\mathbf{0}_{100}, \mathbf{I}_{100, 100})$.

\begin{figure}[!h]
\begin{tabular}{rcl}
\multicolumn{3}{c}{\hspace{4cm} \includegraphics[scale=0.45, angle=0]{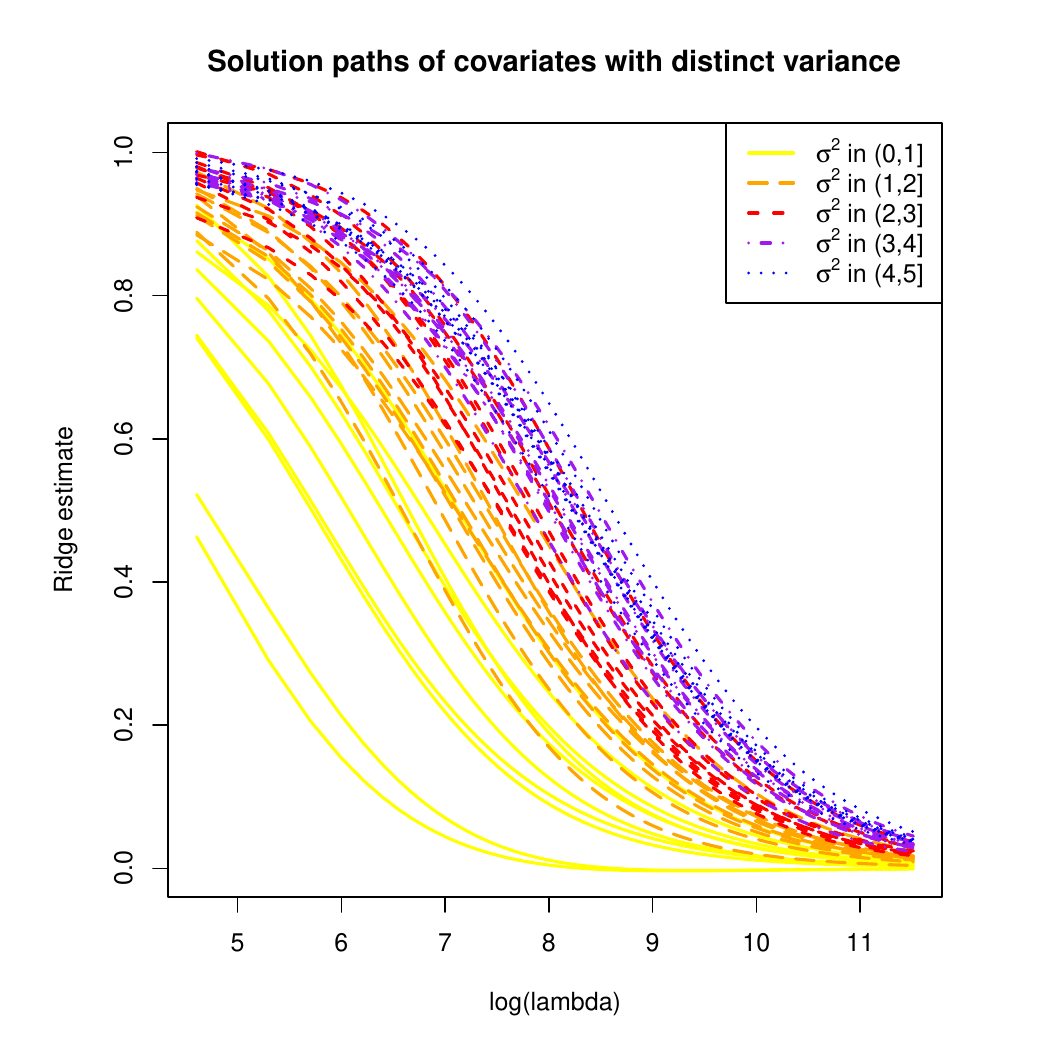}}
\\
\includegraphics[scale=0.45, angle=0]{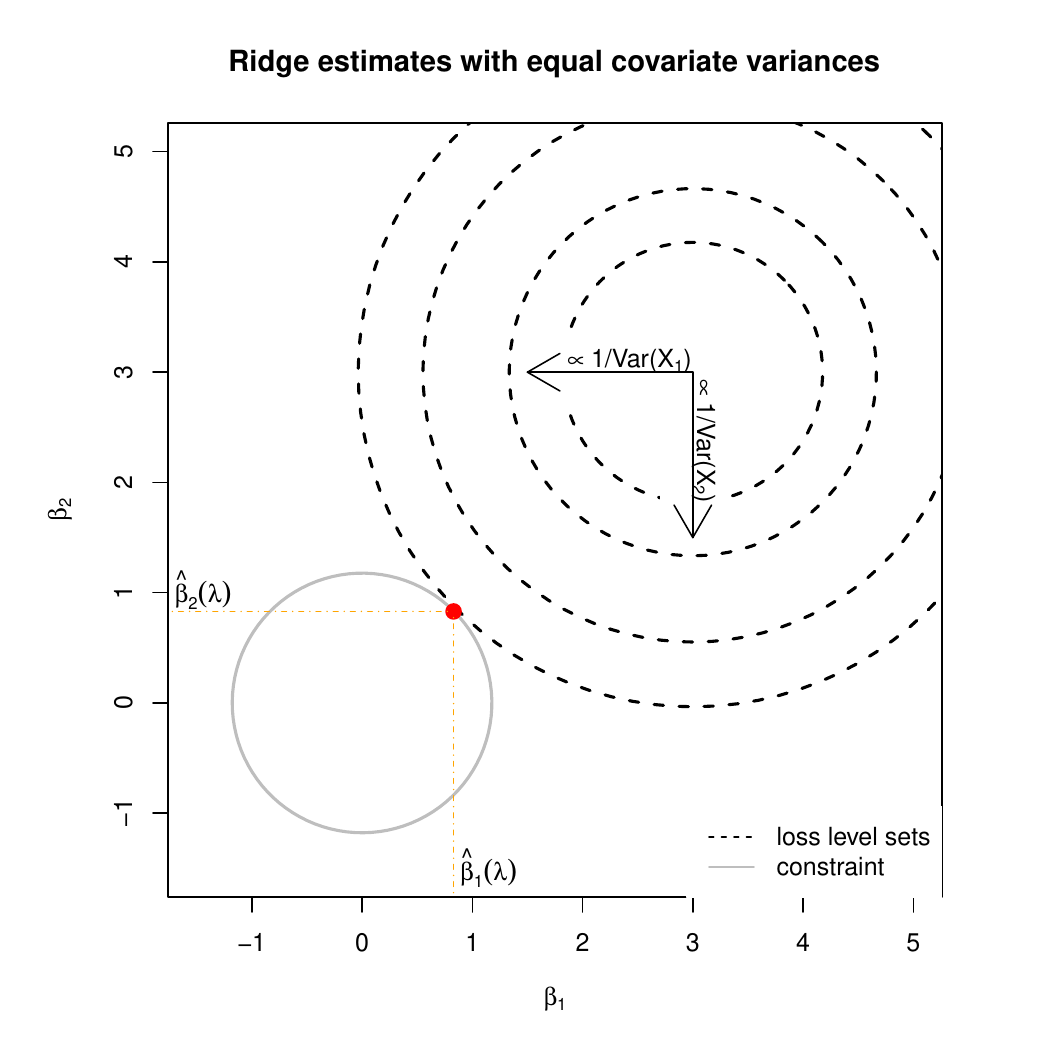}
&
\includegraphics[scale=0.45, angle=0]{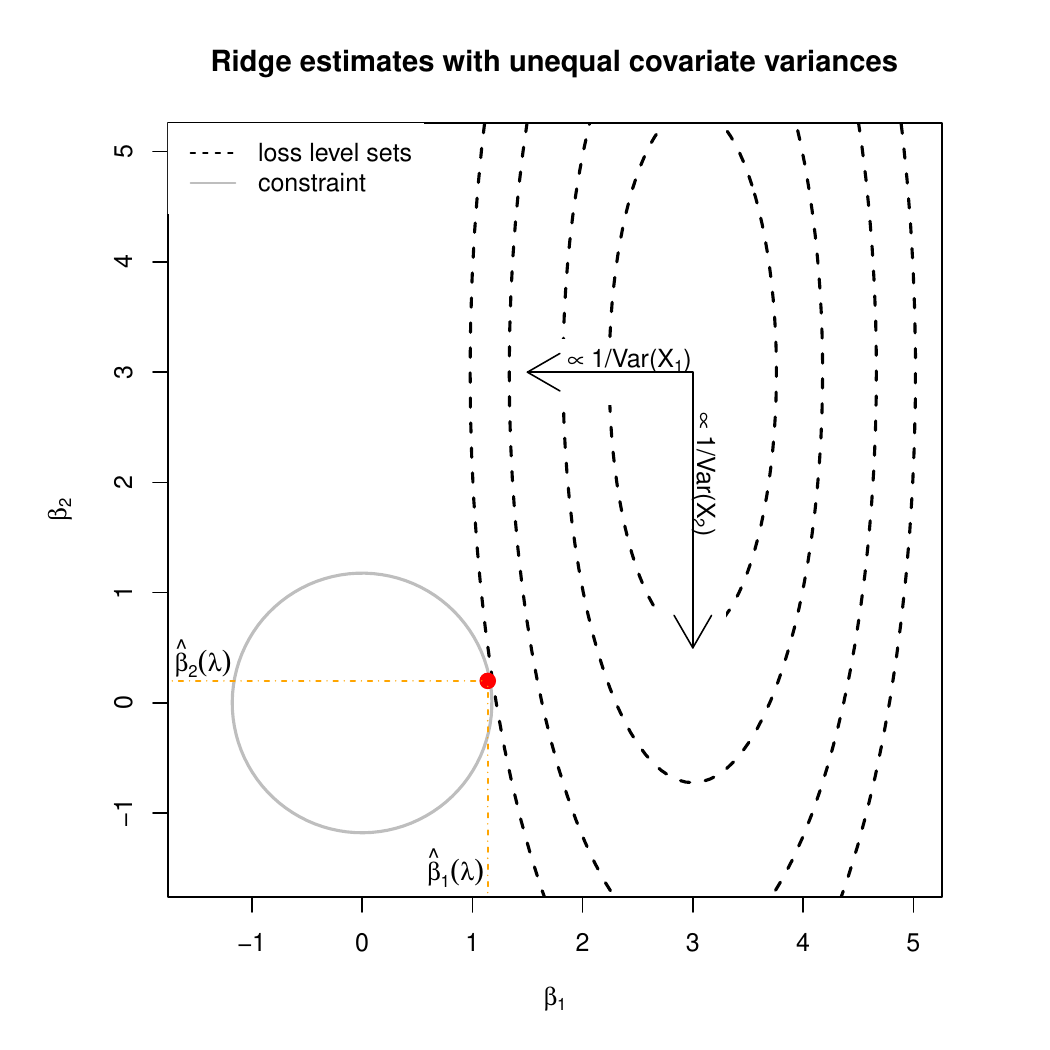}
\end{tabular}
\caption{Top panel: Ridge regularization paths for coefficients of the 50 uncorrelated covariates with distinct variances. Color and line type indicated the grouping of the covariates by their variance. Bottom panels:
Graphical illustration of the effect of a covariate's variance on the ridge regression estimator. The grey circle depicts the ridge parameter constraint. The dashed black ellipsoids are the level sets of the least squares loss function. The red dot is the ridge regression estimate. Left and right panels represent the cases with equal and unequal, respectively, variances of the covariates.}
\label{fig.effectOfRidge_effectOfVariances}
\end{figure}

With the simulated data at hand the ridge regression estimators of $\bbeta$ are evaluated for a large grid of the penalty parameter $\lambda$. The resulting ridge regularization paths of the regression coefficients are plotted (Figure \ref{fig.effectOfRidge_effectOfVariances}). All paths start  close to one and vanish as $\lambda \rightarrow \infty$. However, ridge regularization paths of regression coefficients corresponding to covariates with a large variance dominate those with a low variance.

Ridge regression's preference of covariates with a large variance can intuitively be understood as follows. First note that the ridge regression estimator now can be written as:
\begin{eqnarray*}
\bbeta (\lambda) & = & [ \mbox{Var}(\mathbf{X}) + \lambda \mathbf{I}_{50, 50}]^{-1} \mbox{Cov}(\mathbf{X}, \mathbf{Y})
\\
& = & ( \SSigma + \lambda \mathbf{I}_{50, 50})^{-1} \SSigma [ \mbox{Var}(\mathbf{X}) ]^{-1} \mbox{Cov}(\mathbf{X}, \mathbf{Y})
\\
& = & ( \SSigma + \lambda \mathbf{I}_{50, 50})^{-1} \SSigma \bbeta.
\end{eqnarray*}
Plug in the employed parametrization of $\mathbf{\Sigma}$, which gives: $[\bbeta (\lambda)]_j = j (j + 50 \lambda)^{-1} (\bbeta)_j$. Hence, the larger the covariate's variance (corresponding to the larger $j$), the larger its ridge regression coefficient estimate. Ridge regression thus prefers, among a set of covariates with comparable effect sizes, those with larger variances.

The reformulation of ridge penalized estimation as a constrained estimation problem offers a geometrical interpretation of this phenomenon. Let $p=2$ and the design matrix $\mathbf{X}$ be orthogonal, while both covariates contribute equally to the response. Contrast the cases with $\mbox{Var}(X_1) \approx \mbox{Var}(X_2)$ and $\mbox{Var}(X_1) \gg \mbox{Var}(X_2)$. The level sets of the least squares loss function associated with the former case are circular, while that of the latter are strongly ellipsoidal (see Figure \ref{fig.effectOfRidge_effectOfVariances}). The diameters along the principal axes (that -- due to the orthogonality of $\mathbf{X}$ -- are parallel to that of the $\beta_1$- and $\beta_2$-axes) of both circle and ellipsoid are reciprocals of the variance of the covariates. When the variances of both covariates are equal, the level sets of the loss function expand equally fast along both axes. With the two covariates having the same regression coefficient, the point of these level sets closest to the parameter constraint is to be found on the line $\beta_1 = \beta_2$ (Figure \ref{fig.effectOfRidge_effectOfVariances}, left panel). Consequently, the ridge regression estimator satisfies $\hat{\beta}_1 (\lambda) \approx \hat{\beta}_2(\lambda)$. With unequal variances between the covariates, the ellipsoidal level sets of the loss function have diameters of rather different sizes. In particular, along the $\beta_1$-axis it is narrow (as $\mbox{Var}(X_1)$ is large), and -- vice versa -- wide along the $\beta_2$-axis. Consequently, the point of these level sets closest to the circular parameter constraint will be closer to the $\beta_1$- than to the $\beta_2$-axis (Figure \ref{fig.effectOfRidge_effectOfVariances}, left panel). For the ridge estimator of the regression parameter this implies $0 \ll \hat{\beta}_1 (\lambda) < 1$ and $0 < \hat{\beta}_2 (\lambda) \ll 1$. Hence, the covariate with a larger variance yields the larger ridge regression estimator.

Should one thus standardize the covariates prior to ridge regression analysis? That question is context dependent. Often, the data have been subjected to a series of pre-processing steps. For instance, when dealing with gene expression data, preprocessing comprises quality control, background correction, within- and between-normalization \citep{nguyen2002dna}. The purpose of these steps is to make the data comparable between samples. In the example of gene expression data, the expression levels of genes are made comparable both within and between hybridizations. The preprocessing should thus be considered an inherent part of the measurement. As such, it is to be done independently of whatever down-stream analysis is to follow and further tinkering with the data is preferably to be avoided (as it may mess up the `comparable-ness' of the expression levels as achieved by the preprocessing). For other data, types different considerations may apply.

Among the considerations to decide on standardization of the covariates, one should also include the fact that ridge regression estimates prior and posterior to scaling do not simply differ by a factor. To see this, assume that the covariates have been centered. Scaling of the covariates amounts to post-multiplication of the design matrix by a $p \times p$-dimensional diagonal matrix $\mathbf{A}$ with the reciprocals of the covariates' scale estimates on its diagonal \citep{Sard2008}. Hence, the ridge regression estimator (for the rescaled data) is then given by:
\begin{eqnarray*}
\min\nolimits_{\bbeta \in \mathbb{R}^p} \| \mathbf{Y} - \mathbf{X} \mathbf{A} \bbeta \|_2^2 + \lambda \| \bbeta \|_ 2^2.
\end{eqnarray*}
Apply the change-of-variable $\ggamma = \mathbf{A} \bbeta$ and obtain:
\begin{eqnarray*}
\min_{\ggamma} \| \mathbf{Y} - \mathbf{X} \gamma \|_2^2 + \lambda \| \mathbf{A}^{-1} \ggamma \|_ 2^2 & = & \min_{\ggamma} \| \mathbf{Y} - \mathbf{X} \gamma \|_2^2 + \sum\nolimits_{j=1}^p \lambda [(\mathbf{A})_{jj}]^{-2} \gamma_j^2.
\end{eqnarray*}
Effectively, the scaling is equivalent to covariate-wise penalization (see Chapter \ref{chap:genRidge} for more on this). The `scaled' ridge regression estimator may then be derived along the same lines as before in Section \ref{sect.constrainedEstimation}:
\begin{eqnarray*}
\hat{\bbeta}^{\mbox{{\tiny (scaled)}}} (\lambda) & = & \mathbf{A}^{-1} \hat{\ggamma} (\lambda)
\, \, \,  = \, \, \, \mathbf{A}^{-1} (\mathbf{X}^{\top} \mathbf{X} + \lambda \mathbf{A}^{-2})^{-1} \mathbf{X}^{\top} \mathbf{Y}.
\end{eqnarray*}
In general, this is unequal to the ridge regression estimator without the rescaling of the columns of the design matrix. Moreover, it should be clear that $\hat{\bbeta}^{\mbox{{\tiny (scaled)}}} (\lambda) \not= \mathbf{A} \hat{\bbeta}(\lambda)$.

\subsection{Ridge regression and collinearity} \label{sect:collinearCovariates}
Initially, ridge regression was motivated as an ad-hoc fix of (super)-collinear covariates in order to obtain a well-defined estimator. We now study the effect of this ad-hoc fix on the regression coefficient estimates of collinear covariates. In particular, their ridge regularization paths are contrasted to those of `non-collinear' covariates.

To this end, we consider a simulation in which one response is regressed on 50 covariates. The data of these covariates, stored in a design matrix denoted $\mathbf{X}$, are sampled from a multivariate normal distribution, with mean zero and a $5 \times 5$ block covariance matrix:
\begin{eqnarray*}
\mathbf{\Sigma} & = & \left(
\begin{array}{ccccc}
\mathbf{\Sigma}_{11} & \mathbf{0}_{10, 10} & \mathbf{0}_{10, 10}
& \mathbf{0}_{10, 10} & \mathbf{0}_{10, 10}
\\
\mathbf{0}_{10, 10} & \mathbf{\Sigma}_{22} & \mathbf{0}_{10, 10}
& \mathbf{0}_{10, 10} & \mathbf{0}_{10, 10}
\\
\mathbf{0}_{10, 10} & \mathbf{0}_{10, 10}
& \mathbf{\Sigma}_{33} & \mathbf{0}_{10, 10} & \mathbf{0}_{10, 10}
\\
\mathbf{0}_{10, 10} & \mathbf{0}_{10, 10}
& \mathbf{0}_{10, 10} & \mathbf{\Sigma}_{44} & \mathbf{0}_{10, 10}
\\
\mathbf{0}_{10, 10} & \mathbf{0}_{10, 10}
& \mathbf{0}_{10, 10} & \mathbf{0}_{10, 10} & \mathbf{\Sigma}_{55}
\end{array}
\right)
\end{eqnarray*}
with
\begin{eqnarray*}
\mathbf{\Sigma}_{kk} & = & \tfrac{1}{5}(k-1) \, \mathbf{1}_{10, 10} + \tfrac{1}{5} (6-k) \, \mathbf{I}_{10, 10}.
\end{eqnarray*}
The data of the response variable $\mathbf{Y}$ are then obtained through: $\mathbf{Y}  = \mathbf{X} \bbeta + \vvarepsilon$, with $\vvarepsilon \sim \mathcal{N}( \mathbf{0}_{n}, \mathbf{I}_{nn})$ and $\bbeta = \mathbf{1}_{50}$. Hence, all covariates contribute equally to the response. Would the columns of $\mathbf{X}$ be orthogonal, little difference in the ridge estimates of the regression coefficients is expected.

\begin{figure}[!h]
\begin{tabular}{rcl}
\includegraphics[scale=0.45, angle=0]{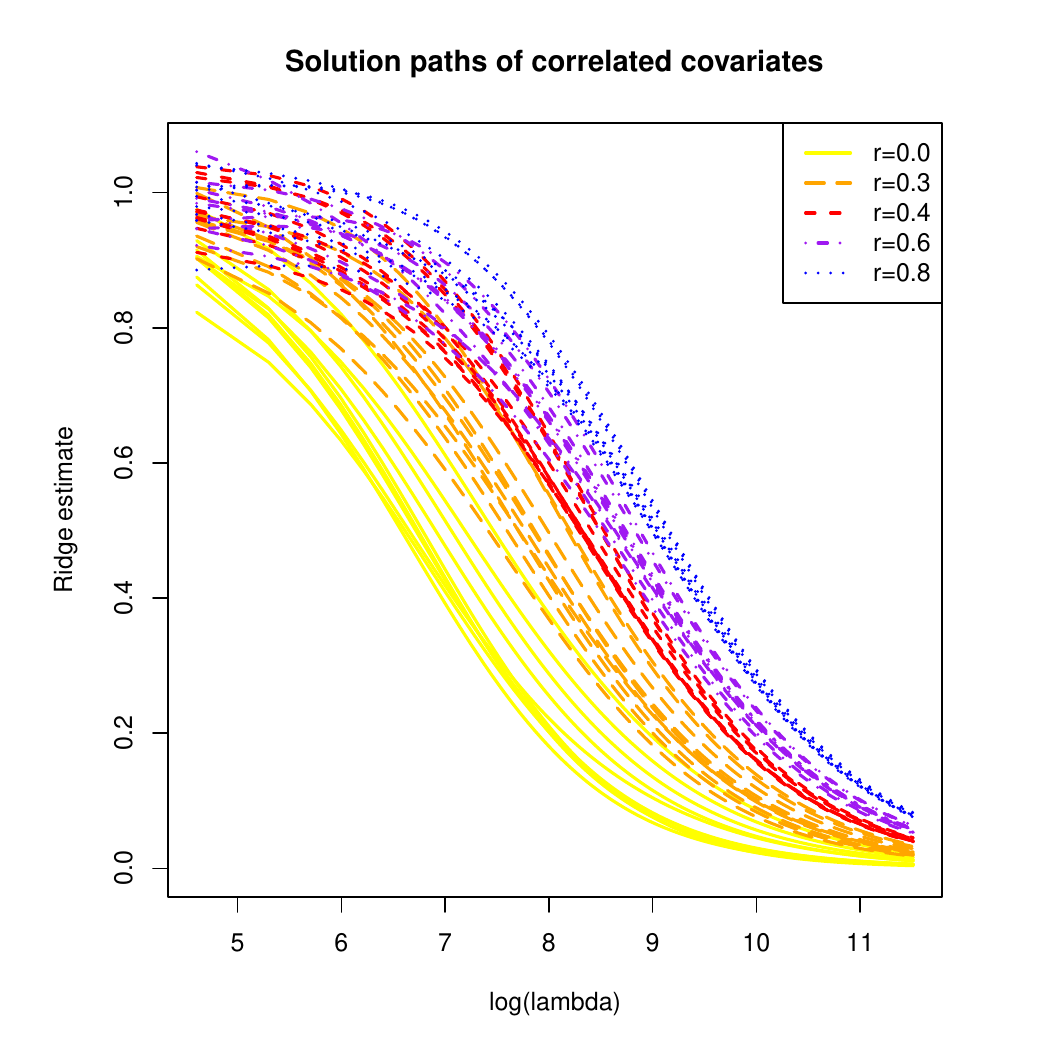}
& &
\includegraphics[scale=0.45, angle=0]{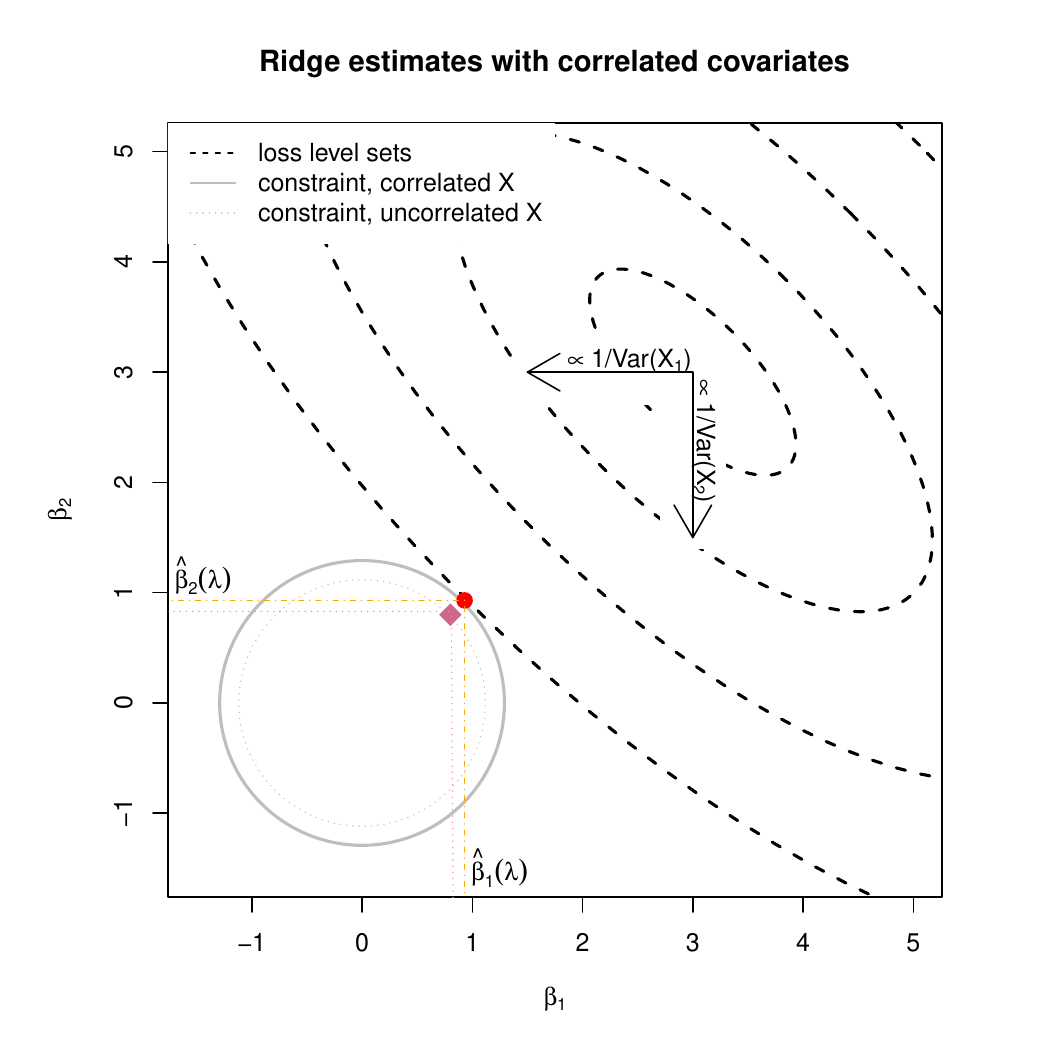}
\end{tabular}
\caption{
Left panel: Ridge regularization paths for coefficients of the 50 covariates, with various degree of collinearity but equal variance. Color and line type correspond to the five blocks of the covariate matrix $\mathbf{\Sigma}$. Right panel: Graphical illustration of the effect of the collinearity among  covariates on the ridge regression estimator. The solid and dotted grey circles depict the ridge parameter constraint for the collinear and orthogonal cases, respectively. The dashed black ellipsoids are the level sets of the sum-of-squares squares loss function. The red dot and violet diamond are the ridge regression for the positive collinear and orthogonal case, respectively.}
\label{fig.ridgeEstimates_correlatedCovariates}
\end{figure}

The results of this simulation study with sample size $n=1000$ are presented in Figure \ref{fig.ridgeEstimates_correlatedCovariates}. All 50 regularization paths start close to one as $\lambda$ is small and converge to zero as $\lambda \rightarrow \infty$. But the paths of covariates of the same block of the covariance matrix $\mathbf{\Sigma}$ quickly group, with those corresponding to a block with larger off-diagonal elements above those with smaller ones. Thus, ridge regression prefers (i.e. shrinks less) coefficient estimates of strongly positively correlated covariates.

Intuitive understanding of the observed behaviour may be obtained from the $p=2$ case. Let $U$, $V$ and $\varepsilon$ be independent random variables with zero mean. Define $X_1 = U + V$, $X_2 = U - V$, and $Y = \beta_1 X_1 + \beta_2 X_2 + \varepsilon$ with  $\beta_1$ and $\beta_2$ constants. Hence, $\mathbb{E}(Y) = 0$. Then:
\begin{eqnarray*}
Y & = & (\beta_1 + \beta_2) U  + (\beta_1 - \beta_2) V + \varepsilon \, \, \, = \, \, \, \gamma_u U + \gamma_v V + \varepsilon
\end{eqnarray*}
and $\mbox{Cor}(X_{1}, X_{2}) = [\mbox{Var}(U) - \mbox{Var}(V)] / [ \mbox{Var}(U) + \mbox{Var}(V) ]$.
The random variables $X_1$ and $X_2$ are strongly positively correlated if  $\mbox{Var}(U) \gg \mbox{Var}(V)$. The ridge regression estimator associated with regression of $Y$ on $U$ and $V$ is:
\begin{eqnarray*}
\ggamma(\lambda) & = & \left(
\begin{array}{rr}
\mbox{Var}(U) + \lambda & 0
\\
0 &  \mbox{Var}(V) + \lambda
\end{array}
\right)^{-1}
\left(
\begin{array}{r}
\mbox{Cov}(U, Y)
\\
\mbox{Cov}(V, Y)
\end{array}
\right).
\end{eqnarray*}
For large enough $\lambda$
\begin{eqnarray*}
\ggamma(\lambda)
& \approx & \frac{1}{\lambda}
\left(
\begin{array}{rr}
\mbox{Var}(U)  & 0
\\
0 &  \mbox{Var}(V)
\end{array}
\right)
\left(
\begin{array}{r}
\beta_1 + \beta_2
\\
\beta_1 - \beta_2
\end{array}
\right).
\end{eqnarray*}
If $\mbox{Var}(U) \gg \mbox{Var}(V)$ and $\beta_1 \approx \beta_2$, the ridge estimate of $\gamma_v$ vanishes for large $\lambda$. Hence, ridge regression prefers positively covariates with similar effect sizes.

This phenomenon can also be explained geometrically. For the illustration consider ridge estimation with $\lambda=1$ of the linear model $\mathbf{Y} = \mathbf{X} \bbeta + \vvarepsilon$ with $\bbeta = (3, 3)^{\top}$, $\vvarepsilon \sim \mathcal{N}(\mathbf{0}_2, \mathbf{I}_{22})$ and the columns of $\mathbf{X}$ strongly and positively collinear. The level sets of the sum-of-squares loss, $\| \mathbf{Y} - \mathbf{X} \bbeta \|_2^2$, are plotted in the right panel of Figure \ref{fig.ridgeEstimates_correlatedCovariates}. Recall that the ridge regression estimate is found by looking for the smallest loss level set that hits the ridge contraint. The sought-for estimate is then the point of intersection between this level set and the constraint, and -- for the case at hand -- is on the $x=y$-line. This is no different from the case with orthogonal $\mathbf{X}$ columns. Yet their estimates differ, even though the same $\lambda$ is applied. The difference is to due to fact that the radius of the ridge constraint depends on $\lambda$, $\mathbf{X}$ and $\mathbf{Y}$. This is immediate from the fact that the radius of the constraint equals $\| \hat{\bbeta}(\lambda) \|_2^2$ (see Section \ref{sect.constrainedEstimation}). To study the effect of $\mathbf{X}$ on the radius, we remove its dependence on $\mathbf{Y}$ by considering its expectation, which is:
\begin{eqnarray*}
\mathbb{E}[ \| \hat{\bbeta}(\lambda) \|_2^2 ] & = & \mathbb{E} \{ [(\mathbf{X}^{\top} \mathbf{X} + \lambda \mathbf{I}_{pp})^{-1} (\mathbf{X}^{\top} \mathbf{X}) \, \hat{\bbeta}]^{\top} \, (\mathbf{X}^{\top} \mathbf{X} + \lambda \mathbf{I}_{pp})^{-1} (\mathbf{X}^{\top} \mathbf{X}) \, \hat{\bbeta} \} \nonumber 
\\
& = & \mathbb{E} [ \mathbf{Y}^{\top} \mathbf{X} (\mathbf{X}^{\top} \mathbf{X} + \lambda \mathbf{I}_{pp})^{-2} \mathbf{X}^{\top} \mathbf{Y} ] \nonumber
\\
& = & \sigma^2 \, \mbox{tr}\big\{ \mathbf{X} (\mathbf{X}^{\top} \mathbf{X} + \lambda \mathbf{I}_{pp})^{-2} \mathbf{X}^{\top} \big\} + \bbeta^{\top} \mathbf{X}^{ \top} \mathbf{X} (\mathbf{X}^{\top} \mathbf{X} + \lambda \mathbf{I}_{pp})^{-2} \mathbf{X}^{\top} \, \mathbf{X} \bbeta. 
\end{eqnarray*}
In the last step, we have used $\mathbf{Y} \sim  \mathcal{N}( \mathbf{X} \bbeta, \sigma^2 \mathbf{I}_{nn})$ and the expectation of the quadratic form of a multivariate random variable $\vvarepsilon \sim \mathcal{N}(\mmu_{\varepsilon}, \SSigma_{\varepsilon})$ is $\mathbb{E} ( \vvarepsilon^{\top} \LLambda \, \vvarepsilon)  = \mbox{tr} ( \LLambda \, \SSigma_{\varepsilon}) + \mmu_{\varepsilon}^{\top} \LLambda  \mmu_{\varepsilon}$ (cf. \citealp{Math1992}). The expression for the expectation of the radius of the ridge constraint can now be evaluated for the orthogonal $\mathbf{X}$ and the strongly, positively collinear $\mathbf{X}$. It turns out that the latter is larger than the former. This results in a larger ridge constraint. For the larger ridge constraint, there is a smaller level set that hits it first. The point of intersection, still on the $x=y$-line, is now thus closer to $\bbeta$ and further from the origin (cf. right panel of Figure \ref{fig.ridgeEstimates_correlatedCovariates}). The resulting estimate is thus larger than that from the orthogonal case.

The above needs some attenuation. Among others, it depends on: {\it i)} the number of covariates in each block, {\it ii)} the size of the effects, i.e. regression coefficients of each covariate, and {\it iii)} the degree of collinearity. Possibly, there are more factors influencing the behaviour of the ridge regression estimator presented in this subsection.

This behaviour of ridge regression is to be understood if a certain (say) clinical outcome is to be predicted from (say) gene expression data. Genes work in concert to fulfil a certain function in the cell. Consequently, one expects their expression levels to be correlated. Indeed, gene expression studies exhibit many co-expressed genes, that is, genes with correlating transcript levels. But also in most other applications with many explanatory variables, collinearity will be omnipresent and similar issues are to be considered.

\subsection{Variance inflation factor}
The ridge regression estimator was introduced to resolve the undefinedness of its maximum likelihood counterpart in the face of (super-)collinearity among the explanatory variables. The effect of collinearity on the uncertainty of estimates is often quantified by the Variance Inflation Factor (VIF). The VIF measures the change in the variance of the estimate due to the collinearity. Here we investigate how penalization affects the VIF. This requires a definition of the VIF of the ridge regression estimator.

The VIF of the maximum likelihood estimator of the $j$-th element of the regression parameter is defined as a factor in the following factorization of the variance of $\hat{\beta}_j$:
\begin{eqnarray*}
\mbox{Var}(\hat{\beta}_j) & = & n^{-1} \sigma^2 [(\mathbf{X}^{\top} \mathbf{X})^{-1}]_{jj}
\\
& & n^{-1} \sigma^2 [\mbox{Var}(X_{i,j} \, | \, X_{i,1}, \ldots, X_{i,j-1}, X_{i,j+1}, \ldots, X_{i,p})]^{-1}
\\
& = & \frac{n^{-1} \sigma^2}{ \mbox{Var}(X_{i,j}) } \cdot \frac{ \mbox{Var}(X_{i,j}) }{ \mbox{Var}(X_{i,j} \, | \, X_{i,1}, \ldots, X_{i,j-1}, X_{i,j+1}, \ldots, X_{i,p})}
\\
& := & n^{-1} \sigma^2 [\mbox{Var}(X_{i,j})]^{-1}  \mbox{VIF}(\hat{\beta}_j),
\end{eqnarray*}
in which it assumed that the $X_{i,j}$'s are random and -- using the column-wise zero `centered-ness'of $\mathbf{X}$ -- that $n^{-1} \mathbf{X}^{\top} \mathbf{X}$ is an estimator of their covariance matrix. Moreover, the identity used to arrive at the second line of the display, $[(\mathbf{X}^{\top} \mathbf{X})^{-1}]_{jj} = [\mbox{Var}(X_{i,j} \, | \, X_{i,1}, \ldots, X_{i,j-1}, X_{i,j+1}, \ldots, X_{i,p})]^{-1}$, originates from Corollary 5.8.1 of \cite{Whit1990}. Thus, $\mbox{Var}(\hat{\beta}_j)$ factorizes into $\sigma^{2} [\mbox{Var}(X_{i,j})]^{-1}$ and the variance inflation factor $\mbox{VIF}(\hat{\beta}_j)$. When the $j$-th covariate is orthogonal to the other, i.e. there is no collinearity, then the VIF's denominator, $\mbox{Var}(X_{i,j} \, | \, X_{i,1}, \ldots, X_{i,j-1}, X_{i,j+1}, \ldots, X_{i,p})$, equals $\mbox{Var}(X_{i,j})$. Consequently, $\mbox{VIF}(\hat{\beta}_j) = 1$. When there is collinearity among the covariates $\mbox{Var}(X_{i,j} \, | \, X_{i,1}, \ldots, X_{i,j-1}, X_{i,j+1}, \ldots, X_{i,p}) < \mbox{Var}(X_{i,j})$ and $\mbox{VIF}(\hat{\beta}_j) > 1$. The VIF then inflates the variance of the estimator of $\beta_j$ under orthogonality -- hence, the name -- by a factor attributable to the collinearity.

The definition of the VIF needs modification to be applicable to the ridge regression estimator. In \cite{Marq1970} the `ridge VIF' is defined analogously to the above definition of the VIF of the maximum likelihood regression estimator as:
\begin{eqnarray*}
\mbox{Var}[\hat{\beta}_j (\lambda)] & = & \sigma^2 [(\mathbf{X}^{\top} \mathbf{X} + \lambda \mathbf{I}_{pp})^{-1} \mathbf{X}^{\top} \mathbf{X} (\mathbf{X}^{\top} \mathbf{X} + \lambda \mathbf{I}_{pp})^{-1}]_{jj}
\\
& = & \frac{n^{-1} \sigma^2}{ \mbox{Var}(X_{i,j}) } \cdot n \mbox{Var}(X_{i,j}) [(\mathbf{X}^{\top} \mathbf{X} + \lambda \mathbf{I}_{pp})^{-1} \mathbf{X}^{\top} \mathbf{X} (\mathbf{X}^{\top} \mathbf{X} + \lambda \mathbf{I}_{pp})^{-1}]_{jj}
\\
& := & n^{-1} \sigma^2 [\mbox{Var}(X_{i,j})]^{-1}  \mbox{VIF}[ \hat{\beta}_j (\lambda) ],
\end{eqnarray*}
where the factorization is forced in line with that of the `maximum likelihood VIF' but lacks a similar interpretation. When $\mathbf{X}$ is orthogonal, $\mbox{VIF}[ \hat{\beta}_j (\lambda) ] = [\mbox{Var}(X_{i,j})]^{2} [\mbox{Var}(X_{i,j}) + \lambda]^{-2} < 1$ for  $\lambda > 0$. Penalization then deflates the VIF.

\begin{figure}[!b]
\begin{tabular}{rcl}
\includegraphics[scale=0.23, angle=0]{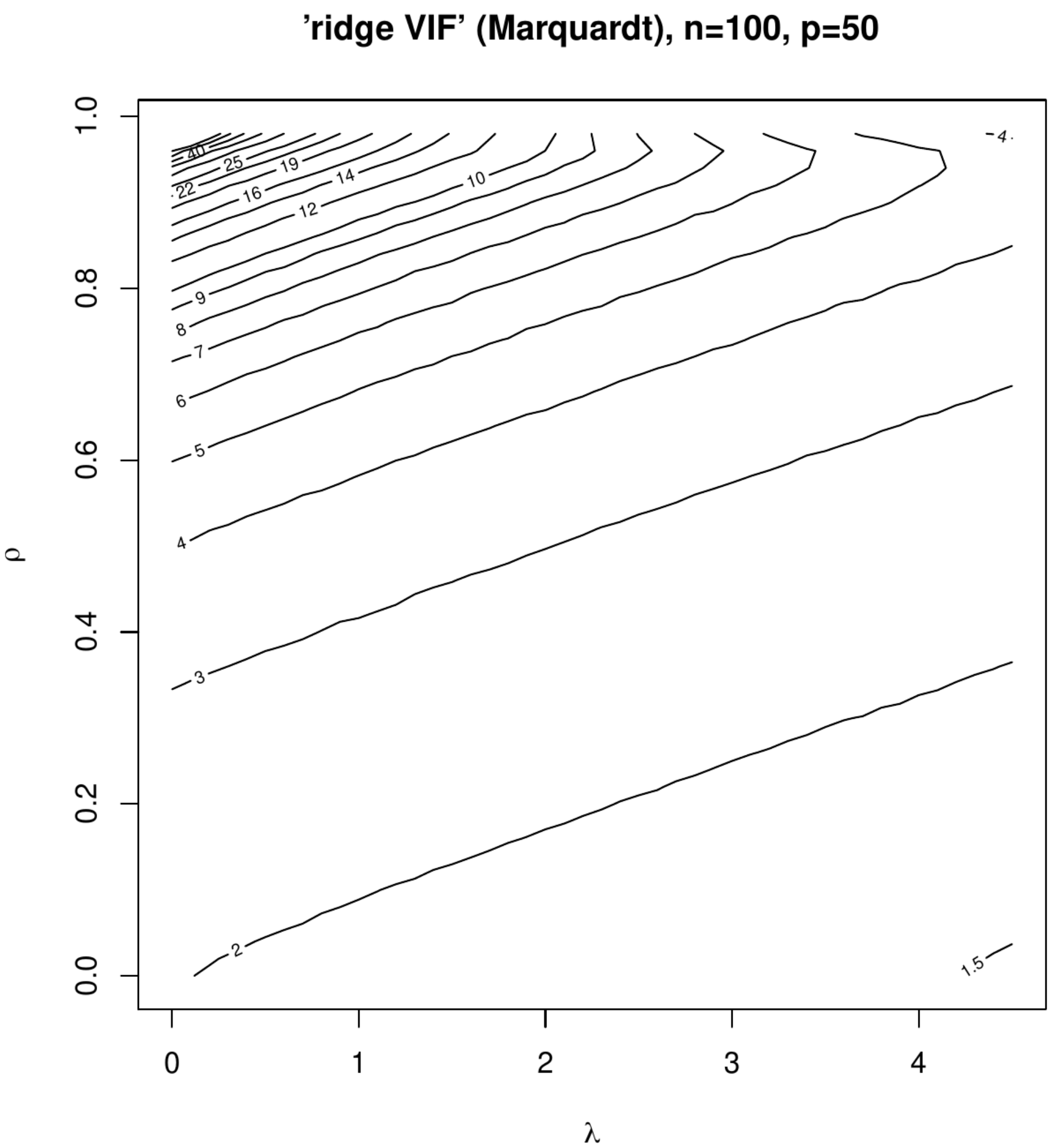}
& &
\includegraphics[scale=0.23, angle=0]{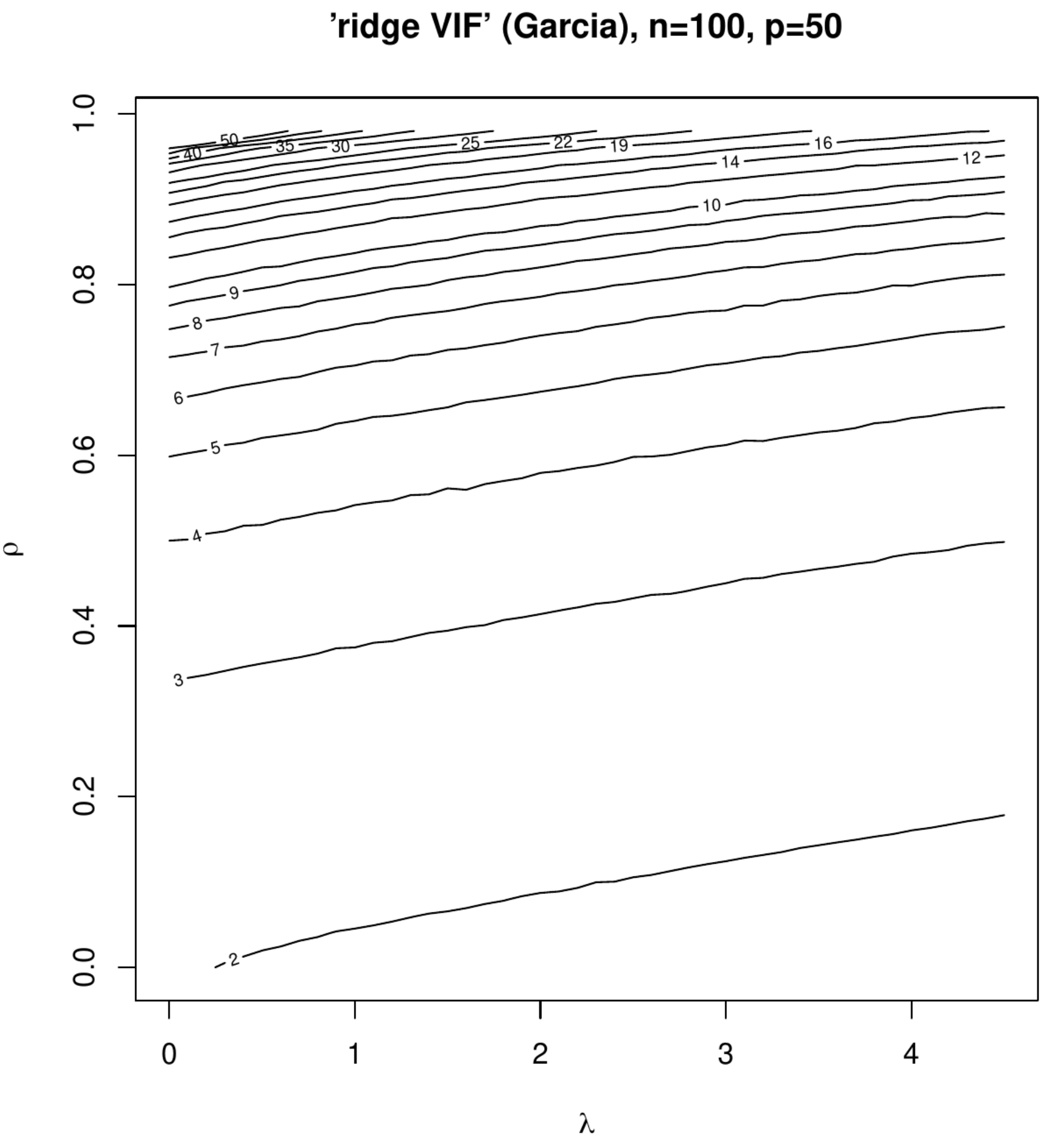}
\\
\includegraphics[scale=0.23, angle=0]{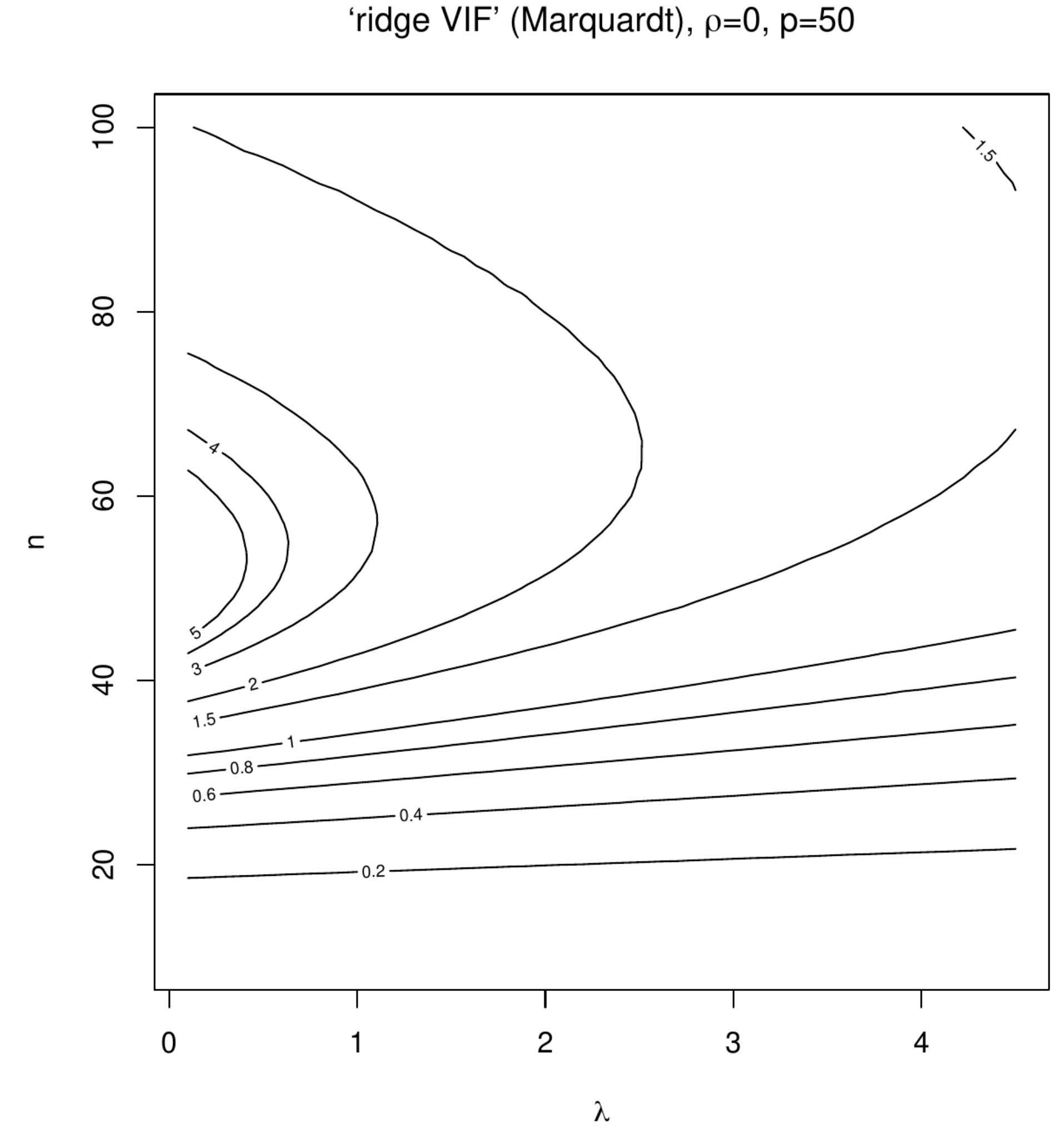}
& &
\includegraphics[scale=0.23, angle=0]{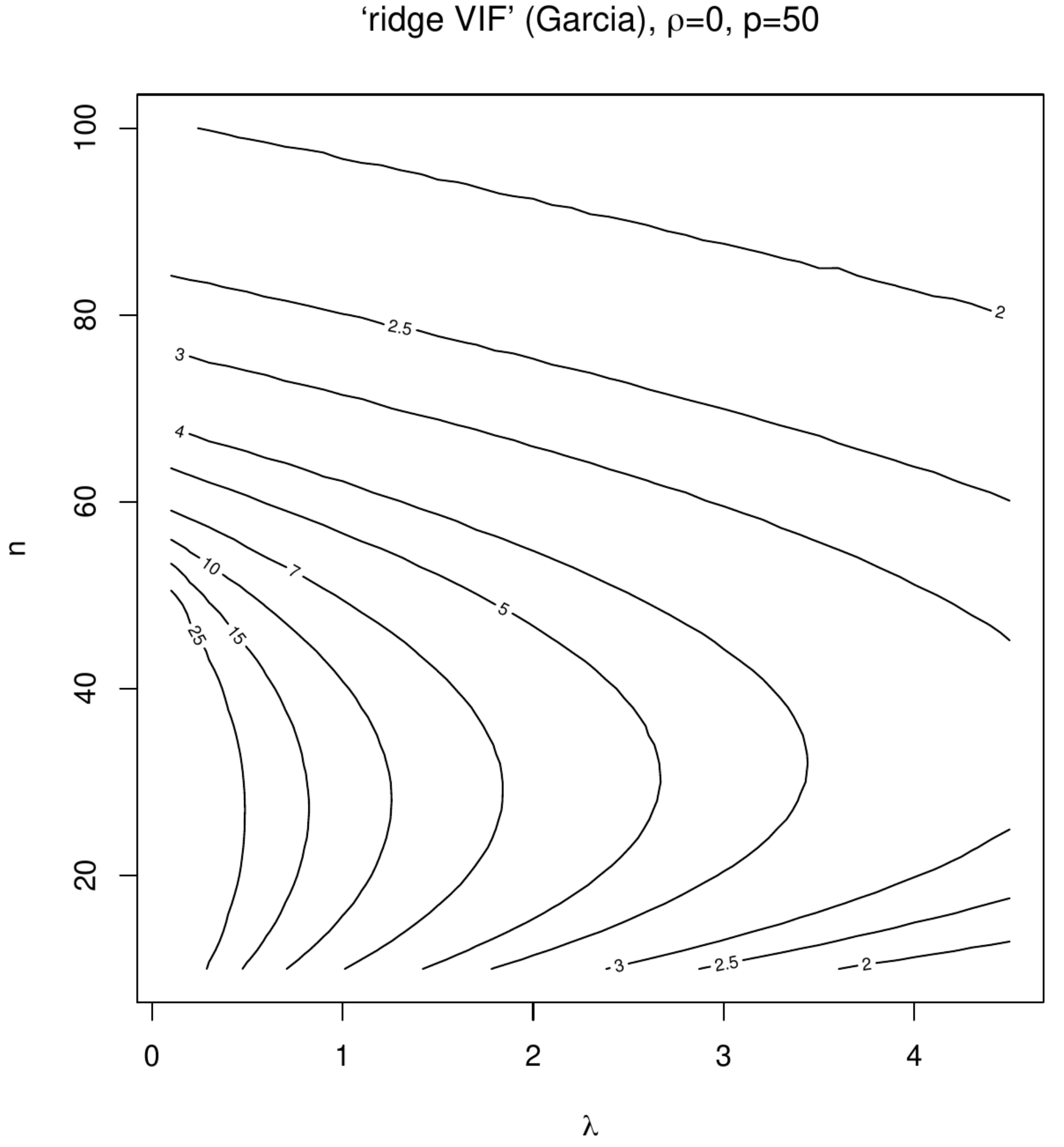}
\end{tabular}
\caption{Contour plots of `Marquardt VIFs' and `Garcia VIFs', left and right columns,  respectively. The top panels show these VIFs against the degree of penalization ($x$-axis) and collinearity ($y$-axis) for a fixed sample size ($n=100$) and dimension ($p=50$). The bottom panels show these VIFs against the degree of penalization ($x$-axis) and the sample size ($y$-axis) for a fixed dimension ($p=50$).} \label{fig.contourVIFs}
\end{figure}

An alternative definition of the `ridge VIF' presented by \cite{Garc2015} for the `$p=2$'-case, which they motivate from counterintuitive behaviour observed in the `ridge VIF' defined by \cite{Marq1970}, adopts the `maximum likelihood VIF' definition but derives the ridge regression estimator from augmented data to comply with the maximum likelihood approach. This requires to augment the response vector with $p$ zeros, i.e. $\mathbf{Y}_{\mbox{{\tiny aug}}} = (\mathbf{Y}^{\top}, \mathbf{0}_p^{\top})^{\top}$ and the design matrix with $p$ rows as $\mathbf{X}_{\mbox{{\tiny aug}}} = (\mathbf{X}^{\top}, \sqrt{\lambda} \mathbf{I}_{pp})^{\top}$. The ridge regression estimator can then be written as  $\hat{\bbeta} (\lambda) = (\mathbf{X}_{\mbox{{\tiny aug}}}^{\top} \mathbf{X}_{\mbox{{\tiny aug}}})^{-1} \mathbf{X}_{\mbox{{\tiny aug}}}^{\top} \mathbf{Y}_{\mbox{{\tiny aug}}}$ (see Section \ref{sect:dataAugmentation}).  This reformulation of the ridge regression estimator in the form of its maximum likelihood counterpart suggests the adoption of latter's VIF definition for the former. However, in the `maximum likelihood VIF' the design matrix is assumed to be zero centered column-wise. Within the augmented data formulation
(Section \ref{sect:dataAugmentation}), this may be achieved by the inclusion of a column of ones in $\mathbf{X}_{\mbox{{\tiny aug}}}$  representing the intercept. The inclusion of an intercept, however, requires a modification of the estimators of $\mbox{Var}(X_{i,j})$ and $\mbox{Var}(X_{i,j} \, | \, X_{i,1}, \ldots, X_{i,j-1}, X_{i,j+1}, \ldots, X_{i,p})$. The former is readily obtained, 
while the latter is given by the reciprocals of the $2^{\mbox{{\tiny nd}}}$ to the $(p+1)$-th diagonal elements of the inverse of:
\begin{eqnarray*}
\left(
\begin{array}{rc}
\mathbf{1}_{n} & \mathbf{X}
\\
\mathbf{1}_{p} & \sqrt{\lambda} \mathbf{I}_{pp}
\end{array} \right)^{\top}
\left(
\begin{array}{rc}
\mathbf{1}_{n} & \mathbf{X}
\\
\mathbf{1}_{p} & \sqrt{\lambda} \mathbf{I}_{pp}
\end{array} \right)
&  = &  \left( \begin{array}{rr} n+p  & \sqrt{\lambda} \mathbf{1}_{p}^{\top}
\\
\sqrt{\lambda} \mathbf{1}_{p} & \mathbf{X}^{\top} \mathbf{X} + \lambda \mathbf{I}_{pp}
\end{array}
\right).
\end{eqnarray*}
The lower right block of this inverse is obtained using well-known linear algebra results \textit{i)} the analytic expression of the inverse of a $2\times 2$-partitioned block matrix and \textit{ii)} the inverse of the sum of an invertible and a rank one matrix (given by the Sherman-Morrison formula, see Corollary 18.2.10, \citealp{Harv2008}). It equals:
\begin{eqnarray*}
( \mathbf{X}^{\top} \mathbf{X} + \lambda \mathbf{I}_{pp} )^{-1} - (1+c)^{-1} (n+p)^{-1} \lambda ( \mathbf{X}^{\top} \mathbf{X} + \lambda \mathbf{I}_{pp} )^{-1} \mathbf{1}_p \mathbf{1}_p^{\top} ( \mathbf{X}^{\top} \mathbf{X} + \lambda \mathbf{I}_{pp} )^{-1},
\end{eqnarray*}
where $c = (n+p)^{-1} \lambda \,  \mathbf{1}_p^{\top} ( \mathbf{X}^{\top} \mathbf{X} + \lambda \mathbf{I}_{pp} )^{-1}  \mathbf{1}_p = (n+p)^{-1} \lambda \sum_{j, j'=1}^p [( \mathbf{X}^{\top} \mathbf{X} + \lambda \mathbf{I}_{pp} )^{-1} ]_{j,j'}$. Substitution of these expressions in the ratio of $\mbox{Var}(X_{i,j})$ and $\mbox{Var}(X_{i,j} \, | \, X_{i,1}, \ldots, X_{i,j-1}, X_{i,j+1}, \ldots, X_{i,p})$ in the above yields the alternative VIF of \cite{Garc2015}.


The effect of collinearity on the variance of the ridge regression estimator and the influence of penalization on this effect is studied in two small simulation studies. In the first study, the rows of the design matrix $\mathbf{X}_{i,\ast}^{\top}$ are sampled from  $\mathcal{N}(\mathbf{0}_p, \mathbf{\Sigma})$ with $\mathbf{\Sigma} = (1-\rho) \mathbf{I}_{pp} + \rho \mathbf{1}_{pp}$ fixing the dimension at $p=50$ and the sample size at $n=100$. The correlation coefficient $\rho$ is varied over the unit interval $[0,1)$, representing various levels of collinearity. The columns of the thus drawn $\mathbf{X}$ are zero centered. The response is then formed in accordance with the linear regression model $\mathbf{Y} = \mathbf{X} \bbeta + \vvarepsilon$ with $\bbeta = \mathbf{1}_{p}$ and the error drawn from $\mathcal{N}(\mathbf{0}_n, \tfrac{1}{4} \mathbf{I}_{nn})$. The effect of penalization is studied by varying $\lambda$. For each $(\rho, \lambda)$-combination data, $\mathbf{X}$ and $\mathbf{Y}$, are sampled thousand times and both `ridge VIFs', for discriminative purposes referred as the Marquardt and Garcia VIFs (after the first author of the proposing papers), are calculated and averaged. Figure \ref{fig.contourVIFs} shows the contour plots of the averaged Marquardt and Garcia VIF against $\rho$ and $\lambda$. At $(\rho, \lambda) \approx (0, 0)$ both VIFs are close to two, which -- although the set-up is not high-dimensional in the strict `$p>n$'-sense -- is due to the streneous $p/n$-ratio. Nonetheless, as expected an increase in collinearity results in an increase in the VIF (irrespective of the type). Moreover, indeed some counterintuitive behaviour is observed in the Marquardt VIF: at (say) a value of $\lambda=3$ the VIF reaches a maximum for $\rho \approx 0.95$ and declines for larger degrees of collinearity. This is left without further interpretation as interest is not in deciding on the most appropriate operationalization among both VIFs of the ridge regression estimator, but is primarily in the effect of penalization on the VIF. In this respect, both VIFs exhibit the same behaviour: a monotone decrease of both VIFs in $\lambda$.

In the second simulation, the effect of `spurious collinearity' introduced by the varying the sample size on the ridge regression estimator is studied. The settings are identical to that of the first simulation but with $\rho=0$ and a sample size $n$ that varies from ten to hundred. For $p=50$ in particular, the lower sample sizes will exhibit high degrees of collinearity as the sample correlation coefficient has been seen to inflate then. 
The resulting contour plots (Figure \ref{fig.contourVIFs}) now present the VIFs against $n$ and $\lambda$. Although some counterintuitive behaviour can be seen around $n=p$ and small $\lambda$, the point to be noted and relevant here -- as interest is in the VIF's behaviour for data sets with a fixed sample size and covariates drawn without collinearity -- is the monotone decrease of both VIFs in $\lambda$ at any sample size.

In summary, the penalization does not remove collinearity but, irrespective of the choice of VIF, it reduces the effect of collinearity on the variance of the ridge regression estimator (as measured by the VIFs above). This led \cite{Garc2015} -- although admittingly their focus appears to be low-dimensionally -- to suggest that the VIFs may guide the choice the penalty parameter: choose $\lambda$ such that the variance of the estimator is increased at most by a user-specified factor.

\section{Illustration} \label{sect.ridgeRegressionDataIllustration}
The application of ridge regression to actual data aims to illustrate its use in practice.

\subsection{MCM7 expression regulation by microRNAs}
Recently, a new class of RNA was discovered, referred to as microRNA. MicroRNAs are non-coding, single stranded RNAs of approximately 22 nucleotides. Like mRNAs, microRNAs are encoded in and transcribed from the DNA. MicroRNAs play an important role in the regulatory mechanism of the cell. MicroRNAs down-regulate gene expression by either of two post-transcriptional mechanisms: mRNA cleavage or transcriptional repression. This depends on the degree of complementarity between the microRNA and the target. Perfect or nearly perfect complementarity of the mRNA to the microRNA will lead to cleavage and degradation of the target mRNA. Imperfect complementarity will repress the productive translation and reduction in protein levels without affecting the mRNA levels.
A single microRNA can bind to and regulate many different mRNA targets. Conversely, several microRNAs can bind to and cooperatively control a single mRNA target (\citealp{Bart2004}; \citealp{Esqu2006}; \citealp{Kim2006}).

In this illustration, we wish to confirm the regulation of mRNA expression by microRNAs in an independent data set. We cherry pick an arbitrary finding from literature reported in \cite{Ambs2008}, which focusses on the microRNA regulation of the MCM7 gene in prostate cancer. The MCM7 gene is involved in DNA replication \citep{Tye1999}, a cellular process often derailed in cancer. Furthermore, MCM7 interacts with the tumor-suppressor gene RB1 \citep{Ster1998}. Several studies indeed confirm the involvement of MCM7 in prostate cancer \citep{Padm2004}. And recently, it has been reported that in prostate cancer MCM7 may be regulated by microRNAs \citep{Ambs2008}.

We here assess whether the MCM7 down-regulation by microRNAs can be observed in a data set other than the one upon which the microRNA-regulation of MCM7 claim has been based. To this end, we download from the Gene Expression Omnibus (GEO) a prostate cancer data set (presented by \citealp{Wang2009GeneNetworks}). This data set (with GEO identifier: GSE20161) has both mRNA and microRNA profiles for all samples available. The preprocessed (as detailed in \citealp{Wang2009GeneNetworks}) data are downloaded and require only minor further manipulations to suit our purpose. These manipulations comprise \textit{i)} averaging of duplicated profiles of several samples, \textit{ii)} gene- and mir-wise zero-centering of the expression data, \textit{iii)} averaging the expression levels of the probes that interrogate MCM7. Eventually, this leaves 90 profiles each comprising of 735 microRNA expression measurements.

\begin{lstlisting}
# load libraries
library(GEOquery)
library(RmiR.hsa)
library(penalized)

# extract data
slh     <- getGEO("GSE20161", GSEMatrix=TRUE)
GEdata  <- slh[1][[1]]
MIRdata <- slh[2][[1]]

# average duplicate profiles
Yge  <- numeric()
Xmir <- numeric()
for (sName in 1:90){
	Yge  <- cbind(Yge,  apply(exprs(GEdata)[,sName,drop=FALSE],  1, mean))
	Xmir <- cbind(Xmir, apply(exprs(MIRdata)[,sName,drop=FALSE], 1, mean))
}
colnames(Yge)  <- paste("S", 1:90, sep="")
colnames(Xmir) <- paste("S", 1:90, sep="")

# extact mRNA expression of the MCM7N tumor suppressor gene
entrezID <- c("4176")
geneName <- "MCM7"
Y        <- Yge[which(fData(GEdata)[,10] == entrezID),]

# average gene expression levels over probes
Y <- apply(Y, 2, mean)

# mir-wise centering mir expression data
X <- t(sweep(Xmir, 1, rowMeans(Xmir)))

# generate cross-validated likelihood profile
profL2(Y, penalized=X, minlambda2=1, maxlambda2=20000, plot=TRUE)

# decide on the optimal penalty value directly
optLambda <- optL2(Y, penalized=X)$lambda

# obtain the ridge regression estimages
ridgeFit <- penalized(Y, penalized=X, lambda2=optLambda)

# plot them as histogram
hist(coef(ridgeFit, "penalized"), n=50, col="blue", border="lightblue", 
     xlab="ridge regression estimates with optimal lambda", 
     main="Histogram of ridge estimates")

#  linear prediction from ridge
Yhat <- predict(ridgeFit, X)[,1]
plot(Y ~ Yhat, pch=20, xlab="pred. MCM7 expression", 
                       ylab="obs. MCM7 expression")
\end{lstlisting}

\begin{figure}[!h]
\begin{tabular}{rcl}
\includegraphics[scale=0.45, angle=0]{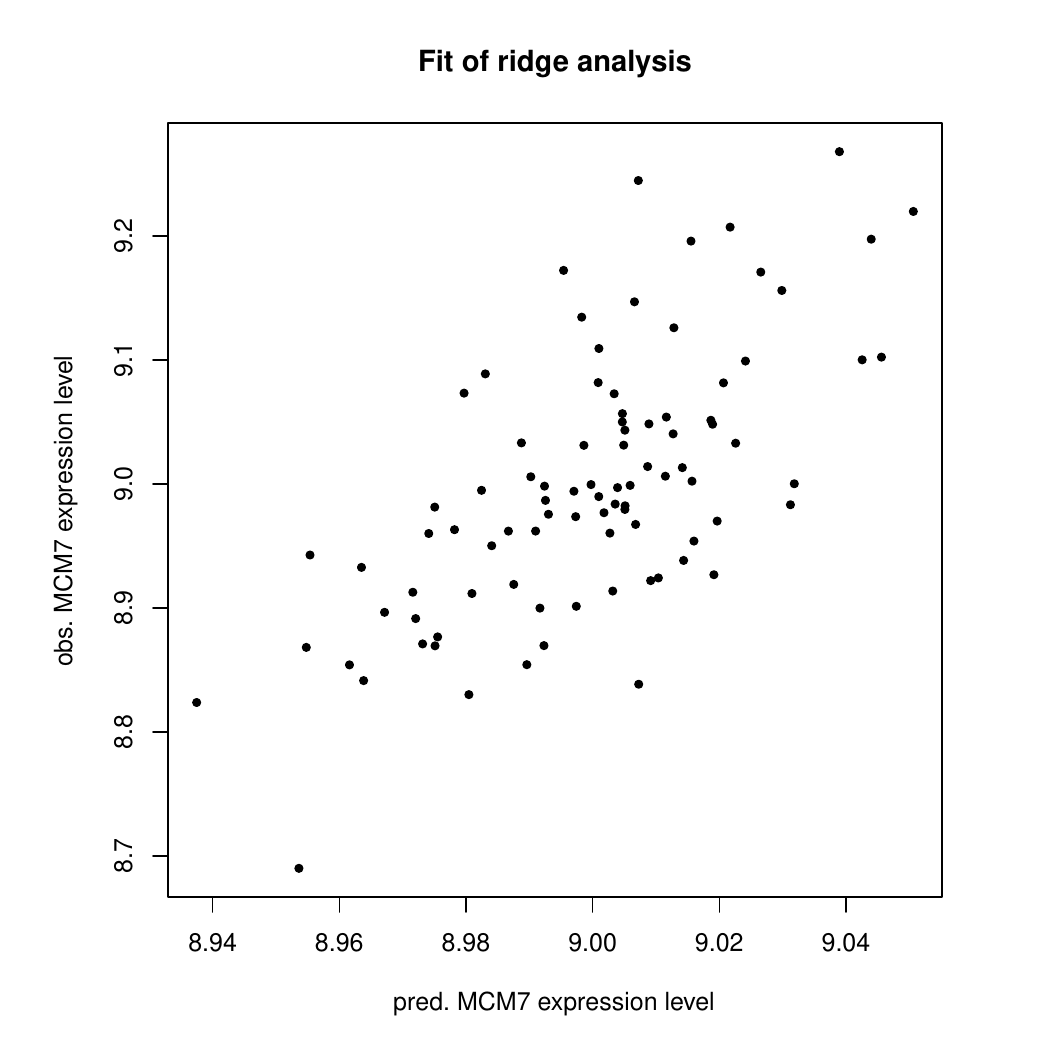}
& &
\includegraphics[scale=0.45, angle=0]{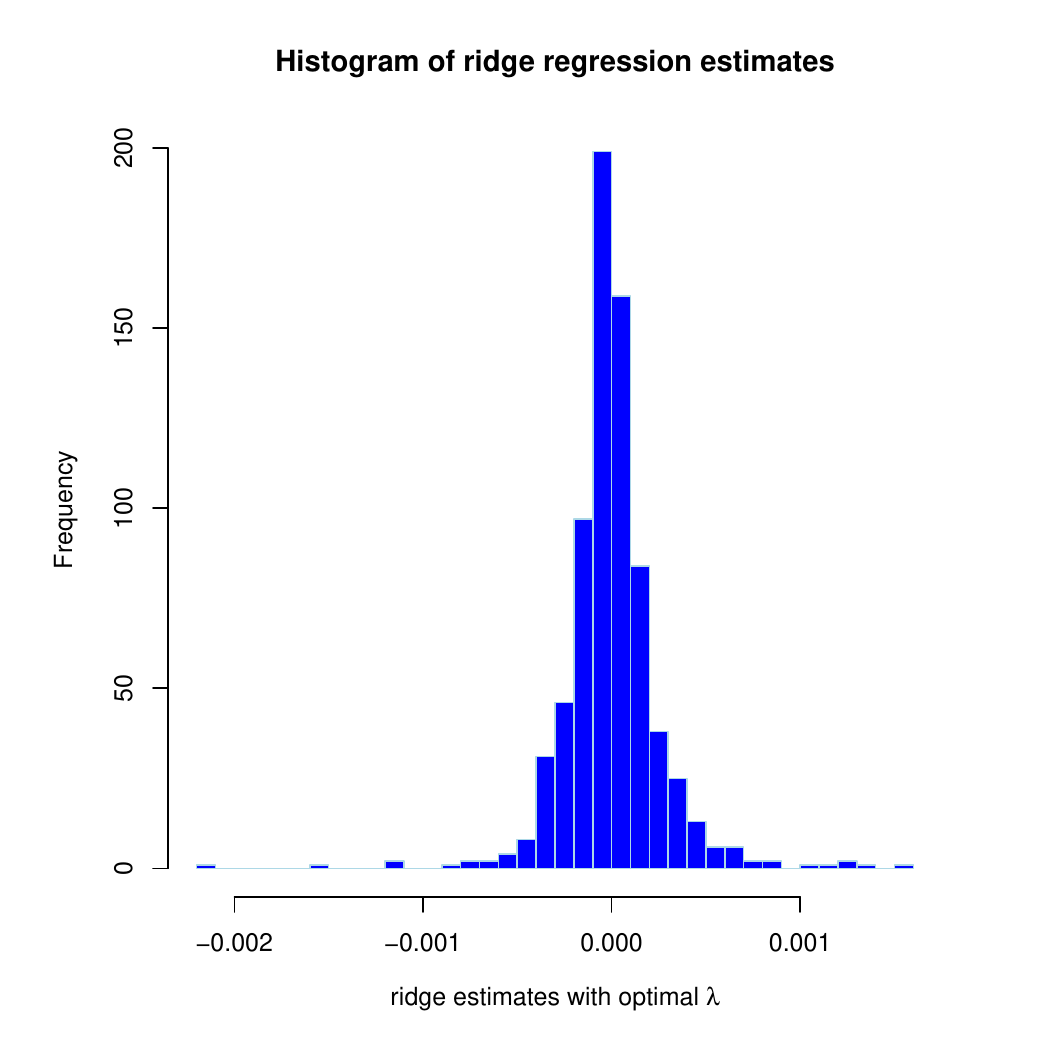}
\end{tabular}
\caption{Left panel: Observed vs. (ridge) fitted MCM7 expression values. Right panel: Histogram of the ridge regression coefficient estimates.} \label{fig.RidgeProstateExample}
\end{figure}

With this prostate data set at hand we now investigate whether MCM7 is regulated by microRNAs. Hereto we fit a linear regression model regressing the expression levels of MCM7 onto those of the microRNAs. As the number of microRNAs exceeds the number of samples, ordinary least squares fails and we resort to the ridge estimator of the regression coefficients. First, an informed choice of the penalty parameter is made through maximization of the LOOCV log-likelihood, resulting in $\lambda_{\mbox{{\tiny opt}}} = 1812.826$. Having decided on the value of the to-be-employed penalty parameter, the ridge regression estimator can now readily be evaluated. The thus fitted model allows for the evaluation of microRNA-regulation of MCM7. E.g., by the proportion of variation of the MCM7 expression levels by the microRNAs as expressed in coefficient of determination: $R^2 = 0.4492$. Alternatively, but closely related, observed expression levels may be related to the linear predictor of the MCM7 expression levels: $\hat{\mathbf{Y}}(\lambda_{\mbox{{\tiny opt}}}) = \mathbf{X} \hat{\bbeta} (\lambda_{\mbox{{\tiny opt}}})$. The Spearman correlation of response and predictor equals 0.6295. A visual inspection is provided by the left panel of Figure \ref{fig.RidgeProstateExample}. Note the difference in scale of the $x$- and $y$-axes. This is due to the fact that the regression coefficients have been estimated in penalized fashion, consequently shrinking estimates of the regression coefficients towards zero leading to small estimates and in turn compressing the range of the linear prediction. The above suggests there is indeed association between the microRNA expression levels and those of MCM7.

The overall aim of this illustration was to assess whether microRNA-regulation of MCM7 could also be observed in this prostate cancer data set. In this endeavour, the dogma (stating this regulation should be negative) has nowhere been used. A first simple assessment of the validity of this dogma studies the signs of the estimated regression coefficients. The ridge regression estimate has 394 out of the 735 microRNA probes with a negative coefficient. Hence, a small majority has a sign in line with the `microRNA $\downarrow$ mRNA' dogma. When, in addition, taking the size of these coefficients into account (Figure \ref{fig.RidgeProstateExample}, right panel), the negative regression coefficient estimates do not substantially differ from their positive counterparts (as can be witnessed from their almost symmetrical distribution around zero). Hence, the value of the `microRNA $\downarrow$ mRNA' dogma is not confirmed by this ridge regression analysis of the MCM7-regulation by microRNAs. Nor is it refuted.

The implementation of ridge regression in the {\tt penalized}-package offers the possibility to fully obey the dogma on negative regulation of mRNA expression by microRNAs. This requires all regression coefficients to be negative. Incorporation of the requirement into the ridge estimation augments the constrained estimation problem with an additional constraint:
\begin{eqnarray*}  
\hat{\bbeta}(\lambda) & = &  \arg \min\nolimits_{ \{ \| \bbeta \|_2^2 \leq c (\lambda); \beta_j \leq 0 \, \mbox{{\footnotesize  for all $j$}} \}} \, \, \| \mathbf{Y} - \mathbf{X} \, \bbeta \|^2_2.
\end{eqnarray*}
With the additional non-positivity constraint on the parameters, there is no explicit solution for the estimator. The ridge estimate of the regression parameters is then found by numerical optimization using e.g. the Newton-Raphson algorithm or a gradient descent approach. The next listing gives the R-code for ridge estimation with the non-positivity constraint of the linear regression model.

\begin{lstlisting}
# decide on the optimal penalty value with sign constraint on paramers
optLambda <- optL2(Y, penalized=-X, positive=rep(TRUE, ncol(X)))$lambda

# obtain the ridge regression estimages
ridgeFit <- penalized(Y, penalized=-X, lambda2=optLambda, 
                      positive=rep(TRUE, ncol(X)))

#  linear prediction from ridge
Yhat <- predict(ridgeFit, -X)[,1]
plot(Y ~ Yhat, pch=20, xlab="predicted MCM7 expression level", 
                       ylab="observed MCM7 expression level")
cor(Y, Yhat, m="s")
summary(lm(Y ~ Yhat))[8]

\end{lstlisting}

\begin{figure}[!h]
\begin{tabular}{rcl}
\includegraphics[scale=0.45, angle=0]{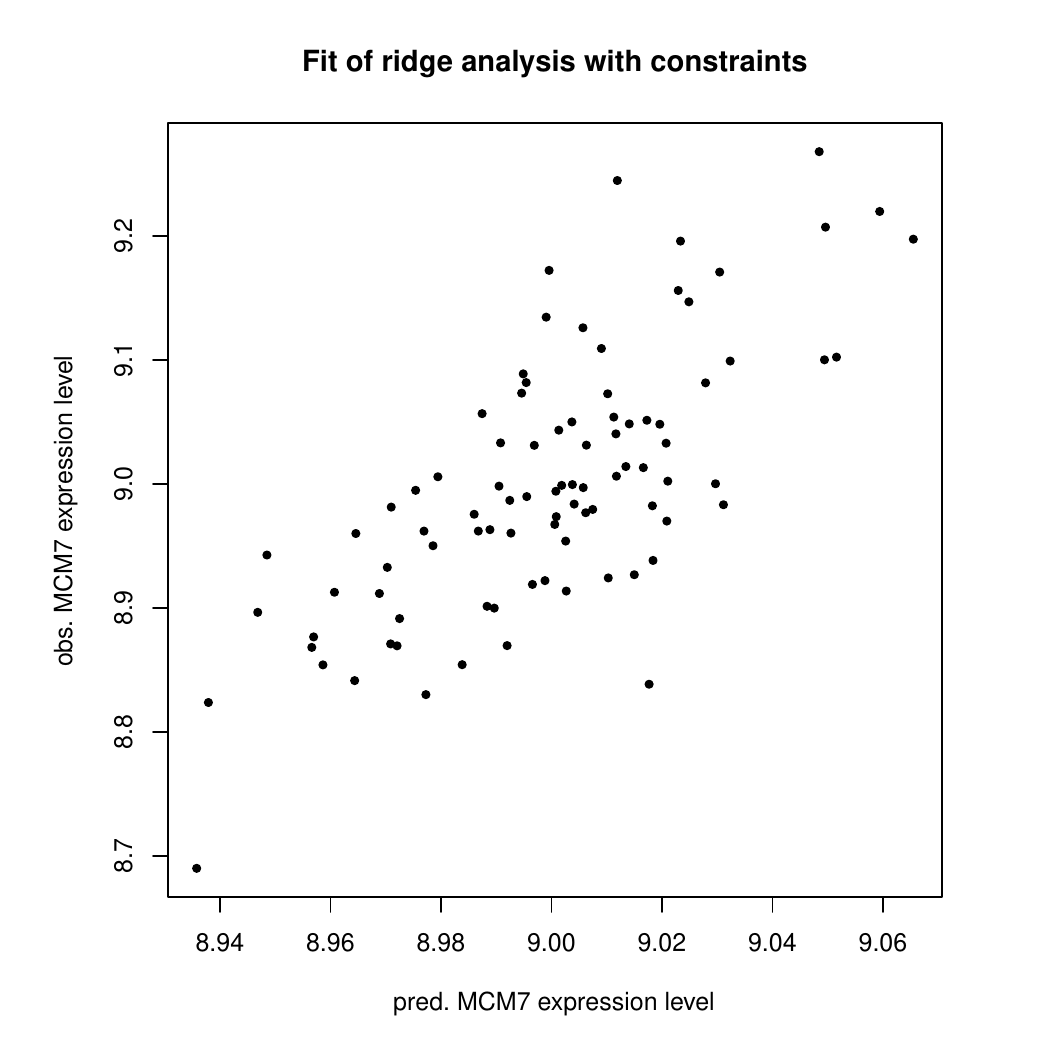}
& &
\includegraphics[scale=0.45, angle=0]{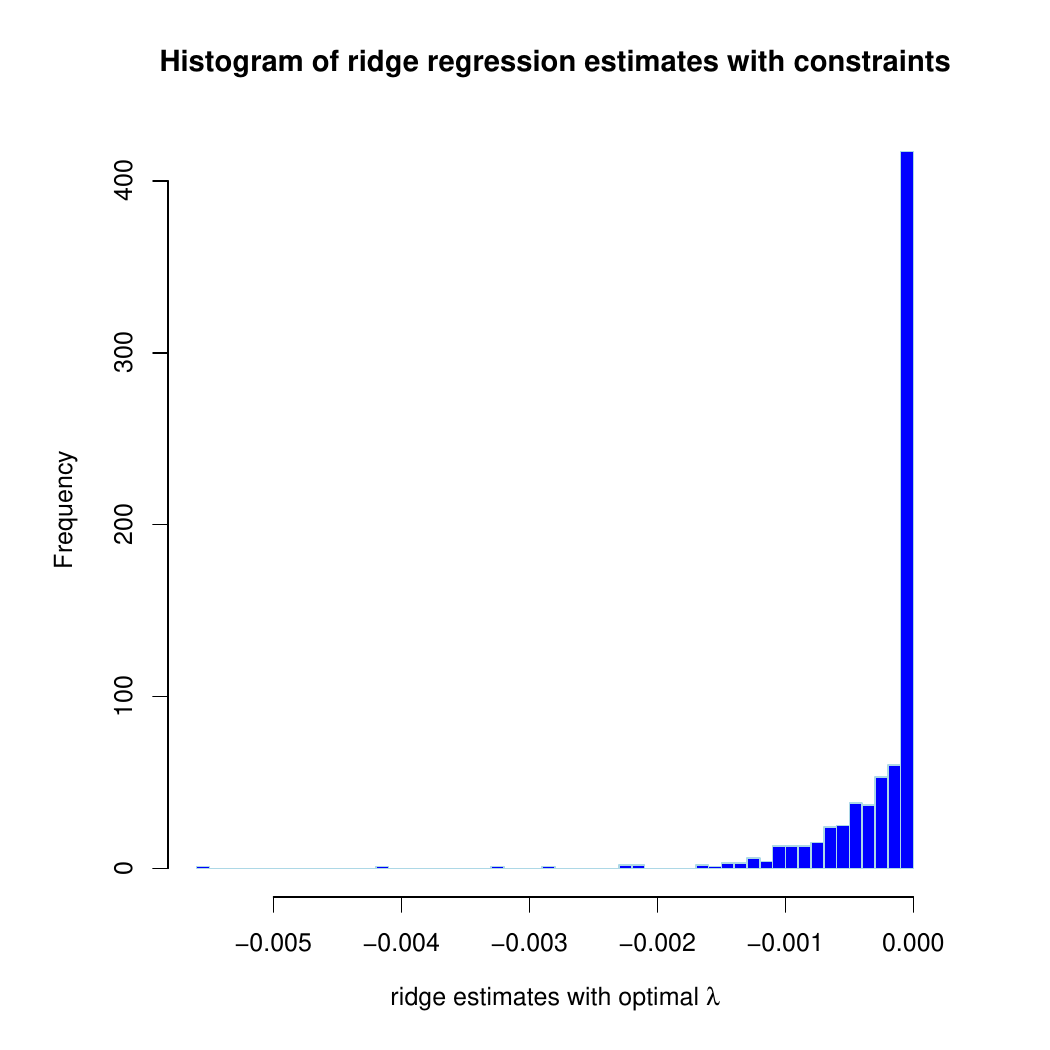}
\end{tabular}
\caption{Left panel: Observed vs. (ridge) fitted MCM7 expression values (with the non-positive constraint on the parameters in place). Right panel: Histogram of the ridge regression coefficient estimates (from the non-positivity constrained analysis).} \label{fig.RidgeProstateExample_constrainedAnalysis}
\end{figure}

The linear regression model linking MCM7 expression to that of the microRNAs is fitted by ridge regression while simultaneously obeying the `negative regulation of mRNA by microRNA'-dogma to the prostate cancer data. In the resulting model 401 out of 735 microRNA probes have a nonzero (and negative) coefficient. There is a large overlap in microRNAs with a negative coefficient between those from this and the previous fit. The models are also compared in terms of their fit to the data. The Spearman rank correlation coefficient between response and predictor for the model without positive regression coefficients equals 0.679 and its coefficient of determination 0.524 (confer the left panel of \ref{fig.RidgeProstateExample_constrainedAnalysis} for a visualization). This is a slight improvement upon the unconstrained ridge estimated model. The improvement may be small but it should be kept in mind that the number of parameters used by both models is 401 (for the model without positive regression coefficients) vs. 735. Hence, with close to half the number of parameters the dogma-obeying model gives a somewhat better description of the data. This may suggest that there is some value in the dogma as inclusion of this prior information leads to a more parsimonious model without any loss in fit.

The dogma-obeying model selects 401 microRNAs that aid in the explanation of the variation in the gene expression levels of MCM7. There is an active field of research, called \textit{target prediction}, trying to identify which microRNAs target the mRNA of which genes. Within {\tt R} there is a collection of packages that provide the target prediction of known microRNAs. The packages differ on the method (e.g. experimental or sequence comparison) that has been used to arrive at the prediction. These target predictions may be used to evaluate the value of the found 401 microRNAs. Ideally, there would be a substantial amount of overlap. The {\tt R}-script that loads the target predictions and does the comparison is below.

\begin{lstlisting}
# extract mir names and their (hypothesized) mrna target
mir2target     <- numeric()
mirPredProgram <- c("targetscan", "miranda", "mirbase", "pictar", "mirtarget2") 
for (program in mirPredProgram){    
    slh           <- dbReadTable(RmiR.hsa_dbconn(), program)
    slh           <- cbind(program, slh[,1:2])
    colnames(slh) <- c("method", "mir", "target")
    mir2target    <- rbind(mir2target, slh)
}
mir2target <- unique(mir2target)
mir2target <- mir2target[which(mir2target[,3] == entrezID),]
uniqMirs   <- tolower(unique(mir2target[,2]))

# extract names of mir-probe on array
arrayMirs <- tolower(levels(fData(MIRdata)[,3])[fData(MIRdata)[,3]])

# which mir-probes are predicted to down-regulate MCM7
selMirs <- intersect(arrayMirs, uniqMirs)
ids     <- which(arrayMirs %in% selMirs)

# which ridge estimates are non-zero
nonzeroBetas <- (coef(ridgeFit, "penalized") != 0)

# which mirs are predicted to 
nonzeroPred      <- 0 * betas
nonzeroPred[ids] <- 1

# contingency table and chi-square test
table(nonzeroBetas, nonzeroPred)
chisq.test(table(nonzeroBetas, nonzeroPred))


\end{lstlisting}

\begin{table}[h!]
\centering
\begin{tabular}{rrr}
\hline \hline \vspace{-7pt} & & \\
\vspace{3pt} &         $\hat{\beta}_j = 0$ &        $\hat{\beta}_j < 0$ \\
\hline \vspace{-4pt} & & \\
microRNA not target &  323 &  390 \\
microRNA target &  11 &  11 \\
\vspace{-9pt} & & \\
\hline \hline
\end{tabular}
\caption{Cross-tabulation of the microRNAs being a potential target of MCM7 vs. the value of its regression coefficient in the dogma-obeying model.} \label{table.mir2mcm7prediction}
\end{table}

With knowledge available on each microRNA whether it is predicted (by at least one target prediction package) to be a potential target of MCM7, it may be cross-tabulated against its corresponding regression coefficient estimate in the dogma-obeying model being equal to zero or not. Table \ref{table.mir2mcm7prediction} contains the result.  Somewhat superfluous considering the data, we may test whether the targets of MCM7 are overrepresented in the group of strictly negatively estimated regression coefficients. The corresponding chi-squared test (with Yates' continuity correction) yields the test statistic $\chi^2 = 0.0478$ with a $p$-value equal to 0.827. Hence, there is no enrichment among the 401 microRNAS of those that have been predicted to target MCM7. This may seem worrisome. However, the microRNAs have been selected for their predictive power of the expression levels of MCM7. Variable selection has not been a criterion (although the sign constraint implies selection). Moreover, criticism on the value of the microRNA target prediction has been accumulating in recent years (REF).

\section{Conclusion}
We discussed ridge regression as a modification of linear regression to  overcome the empirical non-identifiability of the latter when confronted with high-dimensional data. The means to this end was the addition of a (ridge) penalty to the sum-of-squares loss function of the linear regression model, which turned out to be equivalent to constraining the parameter domain. This warranted the identification of the regression coefficients, but came at the cost of introducing bias in the estimates. Several properties of ridge regression like moments and its MSE have been reviewed. Finally, its behaviour and use have been illustrated in simulation and omics data.

\section{Exercises}
\begin{question} \mbox{ } \\
Consider the linear regression model $\mathbf{Y} = \mathbf{X} \bbeta + \vvarepsilon$ with $\vvarepsilon \sim \mathcal{N} ( \mathbf{0}_n, \sigma_{\varepsilon}^2 \mathbf{I}_{nn})$. This model (without intercept) is fitted to data using the ridge regression estimator $\hat{\bbeta}(\lambda) = (\mathbf{X}^{\top} \mathbf{X} + \lambda \mathbf{I}_{pp})^{-1} \mathbf{X}^{\top} \mathbf{Y}$ with $\lambda > 0$. The data  are:
\begin{eqnarray*}
\mathbf{X}^{\top} = \left( \begin{array}{rrrr} 
-1 & 1 &  1 &  -1 \\
\end{array} \right) \, \mbox{ and } \,
\mathbf{Y}^{\top} 
= \left( \begin{array}{rrrr} 
-1.5 &  2.9 & -3.5 & 0.7 
\end{array} \right).
\end{eqnarray*}
\begin{compactitem}
\item[\textit{a)}] Evaluate the maximum likelihood estimator of the regression parameter, i.e. $\hat{\beta}(\lambda)$ for $\lambda=0$.

\item[\textit{b)}] Evaluate the ridge regression estimator for $\lambda=1$.  

\item[\textit{c)}] Verify that the ridge regression estimator $\hat{\beta}(\lambda)$ shrinks as $\lambda$ increases. Hereto combine your answer to parts \textit{a)} and \textit{b)} and the evaluation of the ridge regression estimator for $\lambda=10$ and $\lambda=1000$. Is the order of the employed choices of $\lambda$ and the absolute value of the corresponding estimates concordant?  
\end{compactitem}
\end{question}

\begin{question}\footnote{This exercise is inspired by one from \cite{Drap1998}}
\label{question:ridgeNumericalExerciseFindEstimatorGivenLambda}
\\
Consider the simple linear regression model $Y_i = \beta_0 + \beta_1 X_i + \varepsilon_i$ with $\varepsilon_i \sim \mathcal{N}(0, \sigma^2)$. The data on the covariate and response are: $(X_1, X_2, \ldots, X_{8})^{\top} = (-2, -1, -1, -1, 0, 1, 2, 2)^{\top}$ and $(Y_1, Y_2, \ldots, Y_{8})^{\top} = (35, 40, 36, 38, 40, 43, 45, 43)^{\top}$, with corresponding elements in the same order.
\begin{compactitem}
\item[\textit{a)}] Find the ridge regression estimator for the data above for a general value of $\lambda$. \textit{Hint:} do not forget to include the intercept in the design matrix!
\item[\textit{b)}] Evaluate the fit, i.e. $\widehat{Y}_i(\lambda)$ for $\lambda=10$. Would you judge the fit as good? If not, what is the most striking feature that you find unsatisfactory?
\item[\textit{c)}] Now zero center the covariate and response data, denote it by $\tilde{X}_i$ and $\tilde{Y}_i$, and evaluate the ridge regression estimator of $\tilde{Y}_i = \beta_1 \tilde{X}_i + \varepsilon_i$ at $\lambda=4$. Verify that in terms of original data the resulting predictor now is: $\widehat{Y}_i(\lambda) = 40 + 1.75 X$. 
\end{compactitem}
\textit{Note}: the employed estimate in the predictor found in part \textit{c)} is effectively a combination of a maximum likelihood and ridge regression one for intercept and slope, respectively. Put differently, only the slope has been shrunken. More on this in Chapter \ref{chap:genRidge}.
\end{question}


\begin{question}  \mbox{ }
\\
Plot the regularization path of the ridge regression estimator over the range $\lambda \in (0, 20.000]$ using the data of Example \ref{example.collinearyFlotin}. 
\end{question}



\begin{question} \label{question.ridgeResidualsProjection} \mbox{ }
\\
The coefficients $\bbeta$ of a linear regression model, $\mathbf{Y} = \mathbf{X} \bbeta + \vvarepsilon$, are estimated by $\hat{\bbeta} = (\mathbf{X}^{\top} \mathbf{X})^{-1} \mathbf{X}^{\top} \mathbf{Y}$. The associated fitted values then given by $\widehat{\mathbf{Y}} = \mathbf{X} \, \hat{\bbeta} = \mathbf{X} (\mathbf{X}^{\top} \mathbf{X})^{-1} \mathbf{X}^{\top} \mathbf{Y} = \mathbf{H} \mathbf{Y}$, where $\mathbf{H} =\mathbf{X} (\mathbf{X}^{\top} \mathbf{X})^{-1} \mathbf{X}^{\top}$ referred to as the hat matrix. The hat matrix $\mathbf{H}$ is a projection matrix as it satisfies $\mathbf{H} = \mathbf{H}^2$. Hence, linear regression projects the response $\mathbf{Y}$ onto the vector space spanned by the columns of $\mathbf{Y}$. Consequently, the residuals $\hat{\vvarepsilon}$ and $\hat{\mathbf{Y}}$ are orthogonal. Now consider the ridge estimator of the regression coefficients: $\hat{\bbeta}(\lambda) = (\mathbf{X}^{\top} \mathbf{X} + \lambda \mathbf{I}_{pp})^{-1} \mathbf{X}^{\top} \mathbf{Y}$. Let $\hat{\mathbf{Y}}(\lambda) =  \mathbf{X} \hat{\bbeta}(\lambda)$ be the vector of associated fitted values.

\begin{compactitem}
\item[\textit{a)}] Show that the ridge hat matrix $\mathbf{H}(\lambda) = \mathbf{X} (\mathbf{X}^{\top} \mathbf{X} + \lambda \mathbf{I}_{pp})^{-1} \mathbf{X}^{\top}$, associated with ridge regression, is not a projection matrix (for any $\lambda > 0$), i.e. $\mathbf{H}(\lambda) \not= [\mathbf{H}(\lambda)]^2$.

\item[\textit{b)}] Show that for any $\lambda > 0$ the `ridge fit' $\widehat{\mathbf{Y}}(\lambda)$ is  not orthogonal to the associated `ridge residuals' $\hat{\vvarepsilon}(\lambda)$, defined as $\vvarepsilon(\lambda) = \mathbf{Y} - \mathbf{X} \hat{\bbeta}(\lambda)$.
\end{compactitem}
\end{question}

\begin{question} \mbox{ }
\\
Consider the standard linear regression model $Y_i = \mathbf{X}_{i,\ast} \bbeta + \varepsilon_i$ for $i=1, \ldots, n$ and with the $\varepsilon_i \sim_{i.i.d} \mathcal{N}(0, \sigma^2)$. Suppose the parameter $\bbeta$ is estimated by the ridge regression estimator $\hat{\bbeta}(\lambda) = (\mathbf{X}^{\top} \mathbf{X} + \lambda \mathbf{I}_{pp})^{-1} \mathbf{X}^{\top} \mathbf{Y}$.

\begin{compactitem}
\item[\textit{a)}] The vector of `ridge residuals', defined as $\vvarepsilon(\lambda) = \mathbf{Y} - \mathbf{X} \hat{\bbeta}(\lambda)$, are normally distributed. Why?

\item[\textit{b)}] Show that $\mathbb{E}[\vvarepsilon(\lambda)] =  [\mathbf{I}_{nn} - \mathbf{X} (\mathbf{X}^{\top} \mathbf{X} + \lambda \mathbf{I}_{pp})^{-1} \mathbf{X}^{\top}] \mathbf{X} \bbeta$.

\item[\textit{c)}] Show that $\mbox{Var}[\vvarepsilon(\lambda)] =  \sigma^2 [\mathbf{I}_{nn} - \mathbf{X} (\mathbf{X}^{\top} \mathbf{X} + \lambda \mathbf{I}_{pp})^{-1} \mathbf{X}^{\top}]^2$.

\item[\textit{d)}] Could the normal probability plot, i.e. a qq-plot with the quantiles of standard normal distribution plotted against those of the ridge residuals, be used to assess the normality of the latter? Motivate.
\end{compactitem}
\end{question}

\begin{question} \label{question.BiasOfTheRidgeEstimator} \mbox{ }
\\
Consider the linear regression model $\mathbf{Y} = \mathbf{X} \bbeta + \vvarepsilon$ with $\vvarepsilon \sim \mathcal{N} ( \mathbf{0}_n, \sigma^2 \mathbf{I}_{nn})$. This model is fitted to data, $\mathbf{X}_{1,\ast} = (4, -2)$ and $Y_1 =(10)$, using the ridge regression estimator $\hat{\bbeta}(\lambda) = (\mathbf{X}^{\top}_{1,\ast} \mathbf{X}_{1,\ast} + \lambda \mathbf{I}_{22})^{-1} \mathbf{X}_{1,\ast}^{\top} Y_1$. Throughout use $\lambda=5$
\begin{compactitem}
\item[\textit{a)}] Evaluate the ridge regression estimator.

\item[\textit{b)}] Suppose $\bbeta = (1,-1)^{\top}$. Evaluate the bias of the ridge regression estimator.

\item[\textit{c)}] Decompose the bias into a component due to the regularization and one attributable to the high-dimensionality of the study.

\item[\textit{d)}] Would $\bbeta$ have equalled $(2,-1)^{\top}$, the bias' component due to the high-dimensionality vanishes. Explain why.
\end{compactitem}
\end{question}

\begin{question} \textit{(Numerical inaccuracy)} \\
The linear regression model, $\mathbf{Y} =\mathbf{X} \bbeta + \vvarepsilon$ with $\vvarepsilon \sim \mathcal{N}(\mathbf{0}_n, \sigma^2 \mathbf{I}_{nn})$, is fitted by to the data with the following response, design matrix, and relevant summary statistics:
\begin{eqnarray*}
\mathbf{X}  = \left( \begin{array}{rr} 0.3 & -0.7 \end{array} \right), \, 
\mathbf{Y}  = \left( \begin{array}{r} 0.2  \end{array} \right), \,
\mathbf{X}^{\top} \mathbf{X} = \left( \begin{array}{rr} 0.09 & -0.21 \\ -0.21 & 0.49 \end{array} \right), \mbox{ and } \, \mathbf{X}^{\top} \mathbf{Y} = \left( \begin{array}{r} 0.06 \\ -0.14 \end{array} \right).
\end{eqnarray*}
Hence, $p=2$ and $n=1$. The fitting uses the ridge regression estimator.

\begin{compactitem}
\item[\textit{a)}] Section \ref{sect:ridgeExpectation} states that the regularization path of the ridge regression estimator, i.e. $\{ \hat{\bbeta}(\lambda) : \lambda > 0\}$, is confined to a line in $\mathbb{R}^2$. Draw this line in the $(\beta_1, \beta_2)$-plane.

\item[\textit{b)}] Verify numerically, for a set of penalty parameter values, whether the corresponding estimates $\hat{\bbeta}(\lambda)$ are indeed confined to the line found in part \textit{a)}. Do this by plotting the estimates in the $(\beta_1, \beta_2)$-plane (along with the line found in part \textit{a)}. In this use the following set of $\lambda$'s:
\begin{lstlisting}
lambdas <- exp(seq(log(10^(-15)), log(1), length.out=100))
\end{lstlisting}

\item[\textit{c)}] Part \textit{b)} reveals that, for small values of $\lambda$, the estimates fall outside the line found in part \textit{a)}. Using the theory outlined in Section \ref{sect:ridgeExpectation}, the estimates can be decomposed into a part that falls on this line and a part that is orthogonal to it. The latter is given by $(\mathbf{I}_{22} - \mathbf{P}_x) \hat{\bbeta}(\lambda)$ where $\mathbf{P}_x$ is the projection matrix onto the space spanned by the columns of $\mathbf{X}$. Evaluate the projection matrix $\mathbf{P}_x$.

\item[\textit{d)}] Numerical inaccuracy, resulting from the ill-conditionedness of $\mathbf{X}^{\top} \mathbf{X} + \lambda \mathbf{I}_{22}$, causes $(\mathbf{I}_{22} - \mathbf{P}_x) \hat{\bbeta}(\lambda) \not= \mathbf{0}_2$. Verify that the observed non-null $(\mathbf{I}_{22} - \mathbf{P}_x) \hat{\bbeta}(\lambda)$ are indeed due to numerical inaccuracy. Hereto generate a log-log plot of the condition number of $\mathbf{X}^{\top} \mathbf{X} + \lambda \mathbf{I}_{22}$ vs. the $\| (\mathbf{I}_{22} - \mathbf{P}_x) \hat{\bbeta}(\lambda) \|_2$ for the provided set of $\lambda$'s. \textit{Recall}: the condition number of the ratio of the matrix' largest and the smallest eigenvalue. 
\end{compactitem}
\end{question}

\begin{question}\hspace{-0.05cm}$^*$ \mbox{ } \label{question.alternativeSuperiorMSEproof} \\
Provide an alternative proof of Theorem \ref{theo.Theobald2} that states the existence of a positive value of the penalty parameter for which the ridge regression estimator has a superior MSE compared to that of its maximum likelihood counterpart. 

\begin{compactitem}
\item[\textit{a)}] Show that the derivative of the MSE with respect to the penalty parameter is negative at zero. In this use the following results from matrix calculus:
\begin{eqnarray*}
\frac{d}{d  \lambda}  \mbox{tr} [ \mathbf{A} (\lambda ) ]  & = &  \mbox{tr} \Big[ \frac{d}{d  \lambda} \mathbf{A} (\lambda ) \Big], \qquad  \frac{d}{d  \lambda}  (\mathbf{A} + \lambda \mathbf{B})^{-1} \, \, \,  = \, \, \, - (\mathbf{A} + \lambda \mathbf{B})^{-1}  \mathbf{B} (\mathbf{A} + \lambda \mathbf{B})^{-1},
\end{eqnarray*}
and the chain rule
\begin{eqnarray*}
\frac{d}{d \lambda}  \mathbf{A} (\lambda ) \, \mathbf{B} (\lambda ) & = & 
\Big[ \frac{d}{d  \lambda} \mathbf{A} (\lambda ) \Big] \, \mathbf{B} (\lambda ) + \mathbf{A} (\lambda ) \, \Big[ \frac{d}{d  \lambda} \mathbf{B} (\lambda ) \Big],
\end{eqnarray*}
where $\mathbf{A} (\lambda)$ and $\mathbf{B} (\lambda)$ are square, symmetric matrices parameterized by the scalar $\lambda$.

\item[\textit{b)}] Use Von Neumann's trace inequality to show that $\mbox{MSE}[ \hat{\bbeta} (\lambda) ] < \mbox{MSE}[ \hat{\bbeta} (0) ]$ for $0 < \lambda < \sigma^2 \| \bbeta \|_2^{-2}$. Von Neumann's trace inequality states that, for $p \times p$-dimensional matrices $\mathbf{A}$ and $\mathbf{B}$ with singular values $d_{a,1} \geq d_{a,2} \geq \ldots \geq d_{a,p}$  and $d_{b,1} \geq d_{b,2} \geq \ldots \geq d_{b,p}$ respectively,  $| \mbox{tr} (\mathbf{A} \mathbf{B}) | \leq \sum_{j=1}^p d_{a,j} d_{b,j}$.  
\end{compactitem}
\textit{Note:} the proof in the lecture notes is a stronger one, as it provides a (larger) interval on the penalty parameter where the MSE of the ridge regression estimator is superior to that of the maximum likelihood one.
\begin{compactitem}
\item[\textit{c)}] Recall that there exists $\lambda >  0$ such  that $\mbox{MSE}(\hat{\bbeta}) > \mbox{MSE}[\hat{\bbeta}(\lambda)]$. Then, show that, for that $\lambda$, the mean squared error of the linear predictor satisfies $\mbox{MSE}(\widehat{\mathbf{Y}}) = \mbox{MSE}(\mathbf{X} \hat{\bbeta}) \geq \mbox{MSE}[\mathbf{X}\hat{\bbeta}(\lambda)]$. \textit{Hint:} If $\mathbf{A}$ and $\mathbf{B}$ are nonnegative definite square matrices, then $\mbox{tr} (\mathbf{A} \mathbf{B}) \geq 0$. 
\end{compactitem}
\end{question}

\begin{question} \textit{(Constrained estimation)} \label{question.constraindEstExercise} 
\\
The ridge regression estimator can be viewed as the solution of a constraint estimation problem. Consider the two panels of Figure \ref{fig.constrainedEstExercise}. Both show the ridge parameter constraint  represented by the grey circle around the origin. The left panel also contains an ellopsoid representing a level set of the sum-of-squares centered around the maximum likelihood regression estimate, while the right panel contains a line formed by the solutions of the normal equations. Both panels have four red dots on the boundary of the ridge parameter constraint. Identify for both cases, which of these four dots is (closest to) the ridge regression estimate?
\begin{figure}[h!]
\begin{center}
\centering
\begin{tabular}{ll}
\includegraphics[angle=0, scale=0.23]{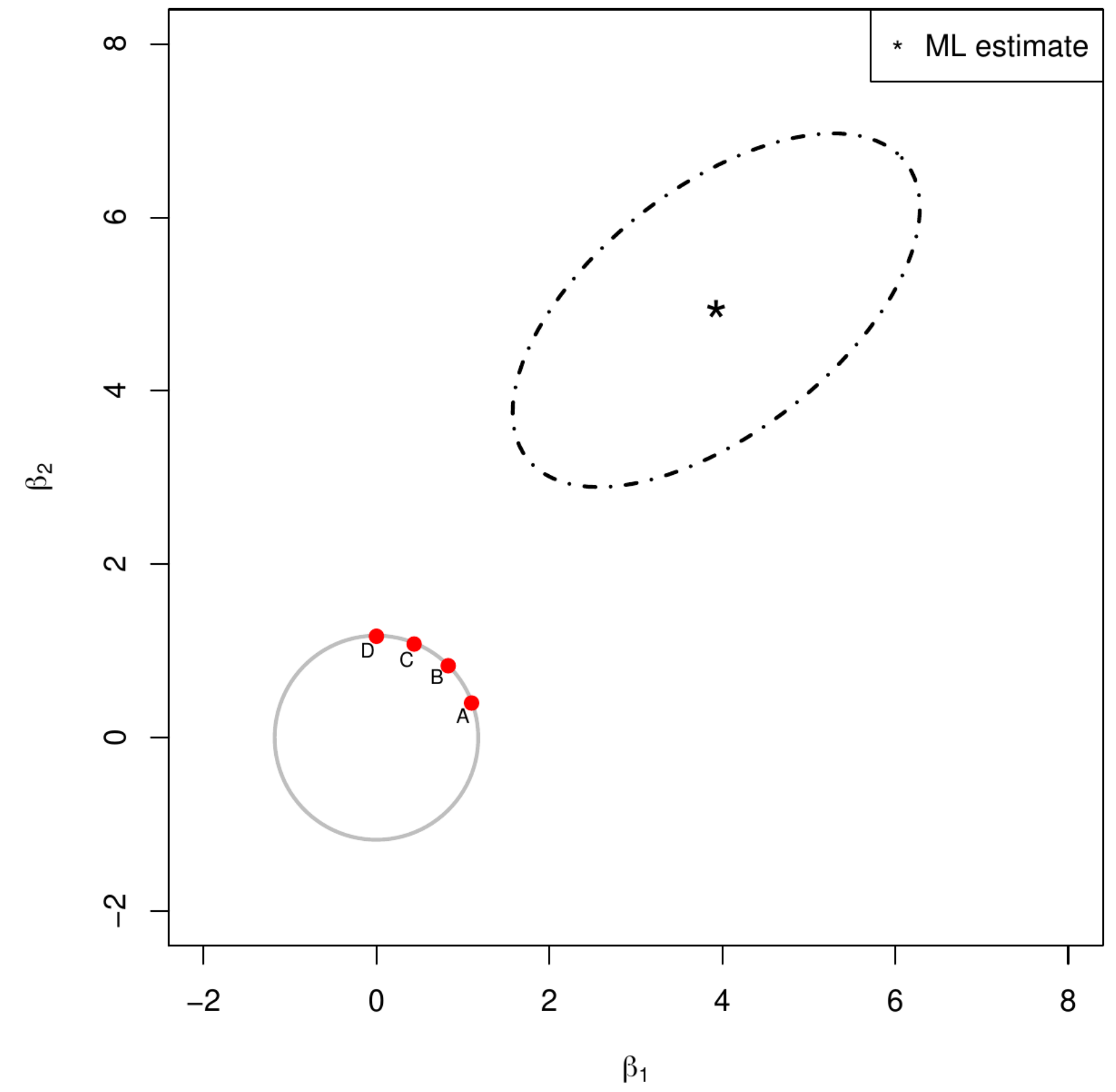}
&
\includegraphics[angle=0, scale=0.23]{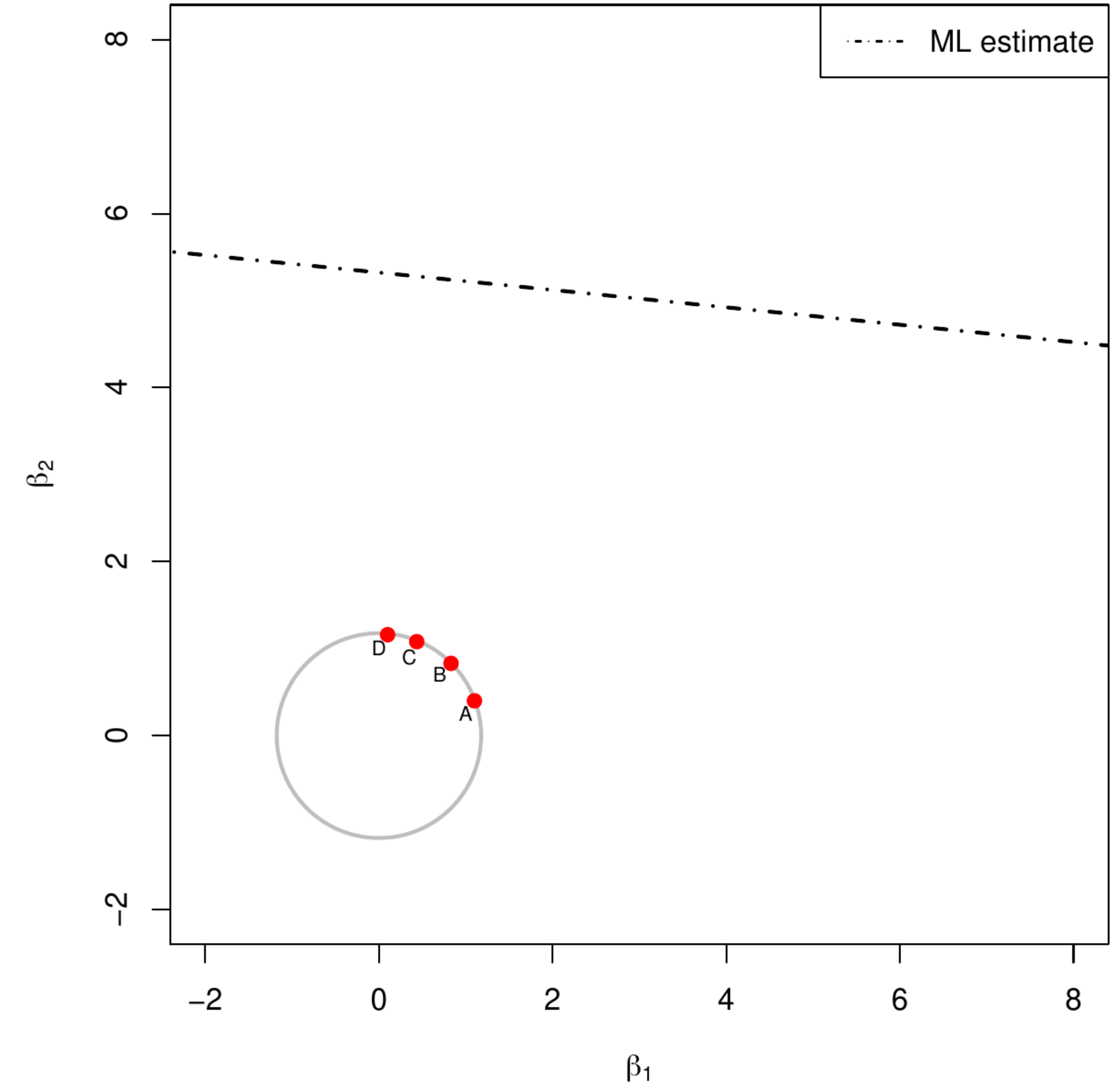}
\end{tabular}
\end{center}
\caption{Figure related to Exercise \ref{question.constraindEstExercise}. \label{fig.constrainedEstExercise}}
\end{figure}
\end{question}

\begin{question} \mbox{ } \\
Consider the regularization paths of the elements of the ridge regression estimator of the linear regression model with two covariates. Could it be that, for $\lambda' > \lambda > 0$, we find the estimates
\begin{compactitem}
\item[\textit{a)}] $\hat{\bbeta} (\lambda) = (2,1)^{\top}$ and $\hat{\bbeta} (\lambda') = (1\tfrac{1}{2},1\tfrac{1}{10})^{\top}$?

\item[\textit{b)}] $\hat{\bbeta} (\lambda') = (2,1)^{\top}$ and $\hat{\bbeta} (\lambda) = (1\tfrac{1}{2},1\tfrac{1}{10})^{\top}$?

\item[\textit{c)}] $\hat{\bbeta} (\lambda) = (2,0)^{\top}$ and $\hat{\bbeta} (\lambda') = (1\tfrac{1}{2},0)^{\top}$?
\end{compactitem}
\end{question}

\begin{question}  \textit{(Negative penalty parameter)} \label{question.negativePenaltyParameter}
\\
\noindent
Consider fitting the linear regression model, $\mathbf{Y} = \mathbf{X} \bbeta + \vvarepsilon$ with $\varepsilon \sim \mathcal{N}(\mathbf{0}_n, \sigma^2 \mathbf{I}_{nn})$, to data by means of the ridge regression estimator. This estimator involves the penalty parameter which is said to be positive. It has been suggested, by among others \cite{Hua1983generalized}, to extend the range of the penalty parameter to the whole set of real numbers. That is, also tolerating negative values. Let's investigate the consequences of allowing negative values of the penalty parameter. Hereto use in the remainder the following numerical values for the design matrix, response, and corresponding summary statistics:
\begin{eqnarray*}
\mathbf{X} \, \, \, = \, \, \, \left( \begin{array}{rr} 5 & 4 \\ -3 & -4 \end{array} \right), \quad \mathbf{Y} \, \, \, = \, \, \, \left( \begin{array}{r} 3 \\ -4 \end{array} \right), \quad \mathbf{X}^{\top} \mathbf{X}  \, \, \, = \, \, \,  \left( \begin{array}{rr} 34 & 32 \\ 32 & 32 \end{array} \right), \quad \mbox{and} \quad \mathbf{X}^{\top} \mathbf{Y}  \, \, \, = \, \, \,  \left( \begin{array}{r} 27  \\ 28  \end{array} \right). 
\end{eqnarray*}

\begin{compactitem}
\item[\textit{a)}] For which $\lambda < 0$ is the ridge regression estimator $\hat{\bbeta}(\lambda) = (\mathbf{X}^{\top} \mathbf{X} + \lambda \mathbf{I}_{22})^{-1} \mathbf{X}^{\top} \mathbf{Y}$ well-defined? 

\item[\textit{b)}] Now consider the ridge regression estimator to be defined via the ridge loss function, i.e. 
\begin{eqnarray*}
\hat{\bbeta} ( \lambda) & = & \arg \min\nolimits_{\bbeta  \in \mathbb{R}^2} \| \mathbf{Y} - \mathbf{X} \bbeta \|_2^2 + \lambda \| \bbeta \|_2^2.
\end{eqnarray*}
Let $\lambda = -20$. Plot the level sets of this loss function, and add a point with the corresponding ridge regression estimate $\hat{\bbeta}(-20)$.

\item[\textit{c)}] Verify that the ridge regression estimate $\hat{\bbeta}(-20)$ is a saddle point of the ridge loss function, as can also be seen from the contour plot generated in part \textit{b)}. Hereto study the eigenvalues of its Hessian matrix. Moreover, specify the range of negative penalty parameters for which the ridge loss function is convex (and does have a unique well-defined minimum).

\item[\textit{d)}] Find the minimum of the ridge loss function.
\end{compactitem}
\end{question}


\begin{question} \mbox{ }
\\
Consider the standard linear regression model $Y_i = \mathbf{X}_{i,\ast} \bbeta + \varepsilon_i$ with $\varepsilon_i \sim_{\mbox{{\tiny i.i.d}}} \mathcal{N}(0, \sigma^2)$ for $i=1, \ldots, n$. The model involves two covariates, neither represents the intercept.

\begin{compactitem}
\item[\textit{a)}] Temporarily assume $\mathbf{X}_{\ast, 1} = \mathbf{X}_{\ast, 2}$ and consider the ridge regression estimator $\hat{\bbeta}(\lambda)$. Show that $[\hat{\bbeta}(\lambda)]_1 = [\hat{\bbeta}(\lambda)]_2$ for all $\lambda > 0$.

\item[\textit{b)}] The covariates are now, and in parts \textit{c)} and \textit{d)}, related as  $\mathbf{X}_{\ast, 1} = - 2 \mathbf{X}_{\ast, 2}$. Data on the response and the covariates are:
\begin{eqnarray*}
\{(y_i, x_{i,1}, x_{i,2})\}_{i=1}^6 & = & \{ (1.5,  1.0, -0.5),   (1.9, -2.0,  1.0), (-1.6,  1.0, -0.5),
\\
& & \, \, \, (0.8,  4.0, -2.0), (0.9,  2.0, -1.0), (\textcolor{white}{-} 0.5,  4.0, -2.0) \}.
\end{eqnarray*}
Evaluate the ridge regression estimator for these data with $\lambda = 1$.

\item[\textit{c)}] The data are as in part \textit{b)}. Show $\hat{\bbeta}(\lambda+\delta) = (52.5 + \lambda) (52.5 + \lambda + \delta)^{-1}  \hat{\bbeta}(\lambda)$ for a fixed $\lambda$ and any $\delta > 0$. That is, given the ridge regression estimator evaluated for a particular value of the penalty parameter $\lambda$, the remaining regularization path $\{ \hat{\bbeta}(\lambda + \delta) \}_{\delta \geq 0}$ is known analytically. \textit{Hint:} Use the singular value decomposition of the design matrix $\mathbf{X}$ and the fact that its largest singular value equals $\sqrt{52.5}$.

\item[\textit{d)}] The data are as in part \textit{b)}. Consider the model $Y_i = X_{i,1} \gamma + \varepsilon_i$. The parameter $\gamma$ is estimated through minimization of $\sum_{i=1}^6 (Y_i - X_{i,1} \gamma)^2 + \lambda_{\gamma} \gamma^2$. The perfectly linear relation of the covariates suggests that the regularization paths of the linear predictors $X_{i,1} \hat{\gamma}(\lambda_{\gamma})$ and $\mathbf{X}_{i,\ast} \hat{\bbeta}(\lambda)$ overlap. Find the functional relationship $\lambda_{\gamma} = f(\lambda)$ such that the resulting linear predictor $X_{i,1} \hat{\gamma}(\lambda_{\gamma})$ indeed coincides with that obtained from the estimate evaluated in part \textit{b)} of this exercise, i.e. $\mathbf{X} \hat{\bbeta}(\lambda)$.
\end{compactitem}
\end{question}

\begin{question}\hspace{-0.05cm}$^*$ \label{question.RidgeEstimatorWithIdenticalCovariates} \mbox{ }
\\
Consider the standard linear regression model $\mathbf{Y} = \mathbf{X} \bbeta + \vvarepsilon$ with $\vvarepsilon \sim \mathcal{N} (\mathbf{0}_n, \sigma^2 \mathbf{I}_{nn})$. The design matrix comprises identical standardized covariates only, i.e. $\mathbf{X}_{\ast,j} = \mathbf{X}_{\ast,j'}$ for all $j, j'=1, \ldots, p$ and $\| \mathbf{X}_{\ast,j} \|_2 = 1$. 

\begin{compactitem}
\item[\textit{a)}] Show that the effect size estimate of a single covariate is spread out uniformly over all elements of the ridge regression parameter estimate. Mathematically, 
show that the ridge regression estimator equals:
\begin{eqnarray*}
\hat{\bbeta}(\lambda) & = & b  (\lambda + p)^{-1}  \mathbf{1}_p,
\end{eqnarray*}
where $b = \mathbf{X}_{\ast,1}^{\top} \mathbf{Y}$. \textit{Hint}: use the Sherman-Morrison formula. Let $\mathbf{A}$ and $\mathbf{B}$ be symmetric matrices of the same dimension, with $\mathbf{A}$ invertible and $\mathbf{B}$ of rank one. Moreover, define $g = \mbox{tr}( \mathbf{A}^{-1} \mathbf{B})$. Then: $(\mathbf{A} + \mathbf{B})^{-1} = \mathbf{A}^{-1} - (1+g)^{-1} \mathbf{A}^{-1} \mathbf{B} \mathbf{A}^{-1}$.

\item[\textit{b)}] Write $\bar{\beta} = p^{-1} \mathbf{1}_p^{\top} \bbeta$ and show that $\mbox{MSE}  [\hat{\bbeta} (\lambda)]  = \| \bar{\beta} \mathbf{1}_p - \bbeta \|_2^2 + p  \lambda^2 (\lambda + p)^{-2} \bar{\beta}^2 +  p (\lambda + p)^{-2} \sigma^2$.
\end{compactitem}
In the remainder, let $\bbeta^{(p)} = p^{-1} \beta_0 \mathbf{1}_p$ with $\beta_0 \in \mathbb{R}$ for all $p \in \mathbb{N}$ and denote the corresponding ridge regression estimator by $\hat{\bbeta}^{(p)} (\lambda)$.
\begin{compactitem}
\item[\textit{c)}]  Verify that $\mathbf{Y} = \sum_{j=1}^p \mathbf{X}_{\ast, j} \beta_j^{(p)} + \vvarepsilon = \sum_{j=1}^{p'} \mathbf{X}_{\ast, j} \beta_j^{(p')} + \vvarepsilon$ for $p \not= p_{j'}$.

\item[\textit{d)}]  Verify that $\lim_{p \rightarrow \infty} \mbox{MSE}  [\hat{\bbeta}^{(p)} (\lambda)] = 0$.

\item[\textit{e)}] Parts \textit{c)} and \textit{d)} suggest we best explain the variation in $\mathbf{Y}$ by as many as possible identical covariates. Would you agree with this suggestion? Motivate.
\end{compactitem}
\end{question}

\begin{question} \mbox{ } \\
Consider the standard linear regression model $Y_i = \mathbf{X}_{i,\ast} \bbeta + \varepsilon_i$ for $i=1, \ldots, n$ and with the $\varepsilon_i$ i.i.d. normally distributed with zero mean and a common but unknown variance. Information on the response, design matrix and relevant summary statistics are:
\begin{eqnarray*}
\mathbf{X}^{\top}  = \left( \begin{array}{rrr} 2 & 1 & -2 \end{array} \right), \, 
\mathbf{Y}^{\top}  = \left( \begin{array}{rrr} -1 & -1 & 1 \end{array} \right), \,
\mathbf{X}^{\top} \mathbf{X} = \left( \begin{array}{r} 9 \end{array} \right), \mbox{ and } \, \mathbf{X}^{\top} \mathbf{Y} = \left( \begin{array}{r} -5 \end{array} \right),
\end{eqnarray*}
from which the sample size and dimension of the covariate space are immediate.
\begin{compactitem}
\item[\textit{a)}] Evaluate the ridge regression estimator $\hat{\bbeta}(\lambda)$ with $\lambda=1$.

\item[\textit{b)}] Evaluate the variance of the ridge regression estimator, i.e.$\widehat{\mbox{Var}}[\hat{\bbeta}(\lambda)]$, for $\lambda = 1$. In this the error variance $\sigma^2$ is estimated by $n^{-1} \| \mathbf{Y} - \mathbf{X} \hat{\bbeta}(\lambda) \|_2^2$.

\item[\textit{c)}] Recall that the ridge regression estimator $\hat{\bbeta}(\lambda)$ is normally distributed. Consider the interval 
\begin{eqnarray*}
\mathcal{C} & = & \big(\hat{\bbeta}(\lambda) - 2 \{ \widehat{\mbox{Var}}[\hat{\bbeta}(\lambda)] \}^{1/2}, \, \hat{\bbeta}(\lambda) + 2 \{ \widehat{\mbox{Var}}[\hat{\bbeta}(\lambda)] \}^{1/2} \big).
\end{eqnarray*}
Is this a genuine (approximate) $95\%$ confidence interval for $\bbeta$? If so, motivate. If not, what is the interpretation of this interval?

\item[\textit{d)}] Suppose the design matrix is augmented with an extra column identical to the first one. Moreover, assume $\lambda$ to be fixed. Is the estimate of the error variance unaffected, or not? Motivate.

\end{compactitem}
\end{question}

\begin{question}\hspace{-0.05cm}$^*$ \mbox{ } \\
Fit the linear regression model, $\mathbf{Y} = \mathbf{X} \bbeta + \vvarepsilon$ with the usual assumptions, by means of the ridge regression estimator $\hat{\bbeta}(\lambda) = (\mathbf{X}^{\top} \mathbf{X} + \lambda \mathbf{I}_{pp})^{-1} \mathbf{X}^{\top} \mathbf{Y}$.
\begin{compactitem}
\item[\textit{a)}] Consider the ridge regression estimator  Show that, for $\lambda$ large enough, $\mbox{sign} [ \hat{\bbeta}(\lambda) ] = \mbox{sign} ( \mathbf{X}^{\top} \mathbf{Y} )$. \textit{Hint:} If $\mathbf{A}$ is a nonsingular matrix and  the largest (in absolute sense) singular value of $\mathbf{B} \mathbf{A}^{-1}$ is smaller than one, then$(\mathbf{A} + \mathbf{B})^{-1} = \mathbf{A}^{-1} +  \mathbf{A}^{-1} \sum_{k=1}^{\infty} (-1)^k (\mathbf{B} \mathbf{A}^{-1})^k$.

\item[\textit{b)}] Let $\mbox{df}(\lambda)$ denote the degrees of freedom consumed by the ridge regression estimator. Show that $\lim_{\lambda \rightarrow \infty} [n- \mbox{\emph{df}}(\lambda)]^{-1} \|\mathbf{Y} - \mathbf{X} \hat{\bbeta}(\lambda) \|_2^2 = n^{-1} \| \mathbf{Y} \|_2^2$, the maximum likelihood estimator of the variance in the response.
\end{compactitem}
\end{question}

\begin{question} \mbox{ } \\
The linear regression model, $\mathbf{Y} =\mathbf{X} \bbeta + \vvarepsilon$ with $\vvarepsilon \sim \mathcal{N}(\mathbf{0}_n, \sigma^2 \mathbf{I}_{nn})$, is fitted to the data with the following design matrix, response and relevant summary statistics:
\begin{eqnarray*}
\mathbf{X}  = \left( \begin{array}{rr} 2 & 1 \end{array} \right), \, 
\mathbf{Y}  = \left( \begin{array}{r} 0.5  \end{array} \right), \, \mathbf{X}^{\top} \mathbf{X} = \left( \begin{array}{rr} 4 & 2 \\ 2 & 1 \end{array} \right), \mbox{ and } \, \mathbf{X}^{\top} \mathbf{Y} = \left( \begin{array}{r} 1 \\ 0.5 \end{array} \right).
\end{eqnarray*}
Hence, $p=2$ and $n=1$. 

\begin{compactitem}
\item[\textit{a)}] Verify by matrix multiplication that the singular value decomposition of $\mathbf{X}$ is as given below:
\begin{eqnarray*}
\mathbf{X} & = & \mathbf{U}_x \mathbf{D}_x \mathbf{V}_x^{\top} \, \, \, := \, \, \, 
 \left( \begin{array}{r} -1  \end{array} \right) \left( \begin{array}{rr} \sqrt{5} & 0 \end{array} \right) \left( \begin{array}{rr} -2 / \sqrt{5} &  - 1/\sqrt{5}  \\
 1 / \sqrt{5} &  - 2/\sqrt{5} 
 \end{array} \right).
\end{eqnarray*}

\item[\textit{b)}] Evaluate the minimum least squares estimator of the regression parameter. 

\item[\textit{c)}] Consider the ridge regression estimator. Suppose the parameter constraint induced by its ridge penalty has a radius equalling $\tfrac{1}{4} \sqrt{1/5}$. What is the value of its penalty parameter? 

\item[\textit{d)}] How many degrees of freedom are consumed by the ridge regression estimator for the penalty parameter found in part \textit{c)}? 
\end{compactitem}
\end{question}

\begin{question} \textit{(Computationally efficient evaluation)} \label{question:efficientEvaluation}
\\
Consider the linear regression model $\mathbf{Y} =  \mathbf{X} \bbeta + \vvarepsilon$, without intercept and $\vvarepsilon \sim \mathcal{N} ( \mathbf{0}_n, \sigma^2 \mathbf{I}_{nn})$, to explain the variation in the response $\mathbf{Y}$ by a linear combination of the columns of the design matrix $\mathbf{X}$. The linear regression model is fitted by means of ridge estimation. The estimator is evaluated directly from its regular expression and a computationally efficient one:
\begin{eqnarray*}
\hat{\bbeta}_{\mbox{{\tiny reg}}} (\lambda) & = &   (\lambda \mathbf{I}_{pp} + \mathbf{X}^{\top} \mathbf{X})^{-1} \mathbf{X}^{\top} \mathbf{Y},
\\
\hat{\bbeta}_{\mbox{{\tiny eff}}} (\lambda) & = & \lambda^{-1} \mathbf{X}^{\top} \mathbf{Y} - \lambda^{-2} \mathbf{X}^{\top} (\mathbf{I}_{nn} + \lambda^{-1} \mathbf{X} \mathbf{X}^{\top})^{-1} \mathbf{X} \mathbf{X}^{\top}  \mathbf{Y},
\end{eqnarray*}
respectively. In the remainder study the computational gain of the latter. Hereto carry out the following instructions:

\begin{compactitem}
\item[\textit{a)}] Load the \texttt{R}-package \texttt{microbenchmark} \citep{Mers2014}.

\item[\textit{b)}] Generate data. In this fix the sample size at $n=10$, and let the dimension range from $p=10, 20, 30, \ldots, 100$. Sample the elements of the response and the ten design matrices from the standard normal distribution. 

\item[\textit{c)}] Verify the superior computation time of the latter by means of the \texttt{microbenchmark}-function with default settings. Throughout use the first design matrix and $\lambda=1$. Write the output of the \texttt{microbenchmark}-function to an \texttt{R}-object. It will be a \texttt{data.frame} with two slots \texttt{expr} and \texttt{time} that contain the function calls and the corresponding computation times, respectively. Each call has by default been evaluated a hundred times in random order. Summary statistics of these individual computation times are given when printing the object on the screen.

\item[\textit{d)}] Use the \texttt{crossprod}- and \texttt{tcrossprod}-functions to improve the computation times of the evaluation of both $\hat{\bbeta}_{\mbox{{\tiny reg}}} (\lambda)$ and $\hat{\bbeta}_{\mbox{{\tiny eff}}} (\lambda)$ as much as possible. 

\item[\textit{e)}] Use the \texttt{microbenchmark}-function to evaluate the (average) computation time both $\hat{\bbeta}_{\mbox{{\tiny reg}}} (\lambda)$ and $\hat{\bbeta}_{\mbox{{\tiny eff}}} (\lambda)$ on all ten data sets, i.e. defined by the ten design matrices with different dimensions. 

\item[\textit{f)}] Plot, for both $\hat{\bbeta}_{\mbox{{\tiny reg}}} (\lambda)$ and $\hat{\bbeta}_{\mbox{{\tiny eff}}} (\lambda)$, the (average) computation time (on the $y$-axis) against the dimension of the ten data sets. Conclude on the computation gain from the plot. 
\end{compactitem}
\end{question}

\begin{question}\hspace{-0.05cm}$^*$ \textit{(Computationally efficient LOOCV)} \label{question:efficientLOOCV}
\\
Consider the ridge regression estimator $\hat{\bbeta}(\lambda)$ of the linear regression model parameter $\bbeta$. Its penalty parameter $\lambda$ may be chosen as the minimizer of 
Allen's PRESS statistic, i.e.:
$\lambda_{\mbox{{\tiny opt}}} =  \arg \min_{\lambda > 0}  n^{-1} \sum\nolimits_{i=1}^n [Y_i - \mathbf{X}_{i, \ast} \hat{\bbeta}_{-i}(\lambda)]^2$,
with \vspace{-2pt} the LOOCV ridge regression estimator $\hat{\bbeta}_{-i}(\lambda) = (\mathbf{X}_{- i, \ast}^{\top} \mathbf{X}_{- i, \ast} + \lambda \mathbf{I}_{pp})^{-1} \mathbf{X}_{- i, \ast}^{\top} \mathbf{Y}_{- i}$. This is computationally demanding as it involves $n$ evaluations of $\hat{\bbeta}_{-i}(\lambda)$, which can be circumvented by rewriting Allen's PRESS statistics. Hereto:
\begin{compactitem}
\item[\textit{a)}] Use the Woodbury matrix identity to verify:
\begin{eqnarray*}
(\mathbf{X}_{- i, \ast}^{\top} \mathbf{X}_{- i, \ast} + \lambda \mathbf{I}_{pp})^{-1} & = & (\mathbf{X}^{\top} \mathbf{X} + \lambda \mathbf{I}_{pp})^{-1} 
\\
& &  + (\mathbf{X}^{\top} \mathbf{X} + \lambda \mathbf{I}_{pp})^{-1} \mathbf{X}_{i, \ast}^{\top}  [ 1 - \mathbf{H}_{ii}(\lambda)]^{-1} \mathbf{X}_{i, \ast} (\mathbf{X}^{\top} \mathbf{X} + \lambda \mathbf{I}_{pp})^{-1},
\end{eqnarray*}
in which $\mathbf{H}_{ii}(\lambda) = \mathbf{X}_{i, \ast} (\mathbf{X}^{\top} \mathbf{X} + \lambda \mathbf{I}_{pp})^{-1}  \mathbf{X}_{i, \ast}^{\top}$.

\item[\textit{b)}] Rewrite the LOOCV ridge regression estimator to:
\begin{eqnarray*}
\hat{\bbeta}_{- i}(\lambda) & = & \hat{\bbeta}(\lambda) - (\mathbf{X}^{\top} \mathbf{X} + \lambda \mathbf{I}_{pp})^{-1} \mathbf{X}_{i, \ast}^{\top} [ 1 - \mathbf{H}_{ii}(\lambda)]^{-1} [ Y_i - \mathbf{X}_{i, \ast} \hat{\bbeta}(\lambda) ].
\end{eqnarray*}
In this use part \textit{a)} and the identity $\mathbf{X}_{-i}^{\top} \mathbf{Y}_{-i} = \mathbf{X}^{\top} \mathbf{Y} - \mathbf{X}_{i, \ast}^{\top} Y_i$.

\item[\textit{c)}] Reformulate, using part \textit{b)}, the prediction error as 
$Y_i - \mathbf{X}_{i, \ast} \hat{\bbeta}_{-i}(\lambda)  = [ 1 - \mathbf{H}_{ii}(\lambda)]^{-1} [ Y_i - \mathbf{X}_{i, \ast}^{\top} \hat{\bbeta}(\lambda) ]$ and express Allen's PRESS statistic as:
\begin{eqnarray*}
n^{-1} \sum\nolimits_{i=1}^n [Y_i - \mathbf{X}_{i, \ast} \hat{\bbeta}_{-i}(\lambda)]^2 & = &   n^{-1} \| \mathbf{B}(\lambda) [\mathbf{I}_{nn} - \mathbf{H}(\lambda)] \mathbf{Y} \|_ F^2,
\end{eqnarray*}
where $\mathbf{B}(\lambda)$ is diagonal with $[\mathbf{B}(\lambda)]_{ii} = [ 1 - \mathbf{H}_{ii}(\lambda)]^{-1}$. 
\end{compactitem}
\end{question}


\begin{question} \textit{(Optimal penalty parameter)} \\
Consider the linear regression model $\mathbf{Y} = \mathbf{X} \bbeta + \vvarepsilon$ with $\bbeta \in \mathbb{R}^p$, $\vvarepsilon \sim \mathcal{N}(\mathbf{0}_n, \sigma^2 \mathbf{I}_{nn})$,
\begin{eqnarray*}
\mathbf{X} & = & \left( \begin{array}{rr} 1 & 1 \\ -1 & -1 \end{array} \right), \quad \mbox{ and } \, \, \,  \mathbf{Y} \, \, \, = \, \, \, \left( \begin{array}{r} -3 \\ 5 \end{array} \right).
\end{eqnarray*}
Find the penalty parameter:
\begin{compactitem}
\item[\textit{a)}] that minimizes the Mean Squared Error (MSE) of the ridge regression estimator for this data set. 
\item[\textit{b)}] by means of leave-one-out cross-validation, minimizing Allen's PRESS statistic. 
\item[\textit{c)}] that minimizes the Akaike's information criterion. 
\end{compactitem}
\end{question}

\begin{question}  \textit{(LOOCV)}
\\
The linear regression model, $\mathbf{Y} = \mathbf{X} \bbeta + \vvarepsilon$ with $\varepsilon \sim \mathcal{N}(\mathbf{0}_n, \sigma^2 \mathbf{I}_{nn})$ is fitted by means of the ridge regression estimator. The design matrix and response are:
\begin{eqnarray*}
\mathbf{X} = \left( \begin{array}{rr} 2 & -1 \\ 0 & 1 \end{array} \right) \quad \mbox{ and } \quad \mathbf{Y} = \left( \begin{array}{r} 1 \\ \tfrac{1}{2} \end{array} \right).
\end{eqnarray*}
The penalty parameter is chosen as the minimizer of the leave-one-out cross-validated squared error of the prediction (i.e. Allen's PRESS statistic). Show that $\lambda = \infty$.
\end{question}

\begin{question} \mbox{ } \\
Consider fitting the linear regression model, $\mathbf{Y} = \mathbf{X} \bbeta + \vvarepsilon$ with design matrix $\mathbf{X}$, $\bbeta \in \mathbb{R}^p$, and $\vvarepsilon \sim \mathcal{N}(\mathbf{0}_n, \sigma^2 \mathbf{I}_{nn})$, by means of the ridge regression estimator, $\hat{\bbeta} (\lambda) = (\mathbf{X}^{\top} \mathbf{X} + \lambda \mathbf{I}_{pp})^{-1} \mathbf{X}^{\top} \mathbf{Y}$, for the below provided choices of $\mathbf{X}$ and $\mathbf{Y}$: 
\begin{eqnarray*}
\mathbf{X} & = & \left( \begin{array}{rr} 1 & 0 \\ 0 & 1 \\ -1 & -1 \end{array} \right), \quad \mathbf{H} (\lambda) \, \, \, = \, \, \, \frac{1}{(\lambda+1) (\lambda + 3)}  \left( \begin{array}{rrr}  2 + \lambda  & -1 & -(1+\lambda) \\  -1 & 2 + \lambda & -(1+\lambda) \\  -(1+\lambda) & -(1+\lambda) & 2 (1 + \lambda) \end{array} \right),
\end{eqnarray*}
and $\mathbf{Y} = ( -2, 1,  2)^{\top}$. The above display also contains the ridge hat matrix, which comes in handy for the answer. Prior to fitting, choose the penalty parameter through generalized cross-validation. Show that then $\lambda_{\mbox{{\tiny opt}}} \in (0, \frac{6}{53})$, and that this interval indeed contains a single minimum, which is indeed the global minimum of $\mbox{GCV}(\lambda)$ on the positive real line. 
\end{question}

\begin{question}  \textit{(Penalty parameter selection)}
\\
PSA (Prostate-Specific Antigen) is a prognostic indicator of prostate cancer. Low and high PSA values indicate low and high risk, respectively. PSA interacts with the VEGF pathway. In cancer the VEGF pathway aids in the process of angiogenesis, i.e. the formation of blood vessels in solid tumors. Assume the aforementioned interaction can -- at least partially -- be captured by a linear relationship between PSA and the constituents of the VEGF pathway. Use the prostate cancer data of \cite{RossAdams2015} to estimate this linear relationship using the ridge regression estimator. The following \texttt{R}-script downloads and prepares the data. 

\begin{lstlisting}
# load the necessary libraries
library(Biobase)
library(prostateCancerStockholm)
library(penalized)
library(KEGG.db)

# load data
data(stockholm)

# prepare psa data
psa <- pData(stockholm)[,9]
psa <- log(as.numeric(levels(psa)[psa]))

# prepare VEGF pathway data
X <- exprs(stockholm)
kegg2vegf    <- as.list(KEGGPATHID2EXTID)
entrezIDvegf <- as.numeric(unlist(kegg2vegf[names(kegg2vegf) == "hsa04370"]))
entrezIDx    <- as.numeric(fData(stockholm)[,10])
idX2VEGF     <- match(entrezIDvegf, entrezIDx)
entrezIDvegf <- entrezIDvegf[-which(is.na(idX2VEGF))]
idX2VEGF     <- match(entrezIDvegf, entrezIDx)
X            <- t(X[idX2VEGF,])
gSymbols     <- fData(stockholm)[idX2VEGF,13]
gSymbols     <- levels(gSymbols)[gSymbols]
colnames(X)  <- gSymbols

# remove samples with missing outcome
X   <- X[-which(is.na(psa)), ]
X   <- sweep(X, 2, apply(X, 2, mean))
X   <- sweep(X, 2, apply(X, 2, sd), "/")
psa <- psa[-which(is.na(psa))]
psa <- psa - mean(psa)
\end{lstlisting}
\mbox{ }
\begin{compactitem}
\item[\textit{a})] Find the ridge penalty parameter by means of AIC minization. \textit{Hint:} the likelihood can be obtained from the \texttt{penFit}-object that is created by the \texttt{penalized}-function of the \texttt{R}-package \texttt{penalized}.
\item[\textit{b})] Find the ridge penalty parameter by means of leave-one-out cross-validation, as implemented by the \texttt{optL2}-function provided by the \texttt{R}-package \texttt{penalized}. 
\item[\textit{c})] Find the ridge penalty parameter by means of leave-one-out cross-validation using Allen's PRESS statistic as performance measure (see Section \ref{subsect.crossvalidation}). 
\item[\textit{d})] Discuss the reasons for the different values of the ridge penalty parameter obtained in parts \textit{a)}, \textit{b)}, and \textit{c)}. Also investigate the consequences of these values on the corresponding regression estimates.
\end{compactitem}
\end{question}

\begin{question}  \textit{(Covariate rescaling)}
\\
The variance of the covariates influences the amount of shrinkage of the regression estimator induced by ridge regularization. Some deal with this through rescaling of the covariates to have a common unit variance. This is discussed at the end of Section  \ref{ridge:covariateVariances}. Investigate this numerically using the data of the microRNA-mRNA illustration discussed in Section \ref{sect.ridgeRegressionDataIllustration}. 
\begin{compactitem}
\item[\textit{a)}] Load the data by running the first \texttt{R}-script of Section \ref{sect.ridgeRegressionDataIllustration}.

\item[\textit{b)}] Fit the linear regression model by means of the ridge regression estimator with $\lambda=1$ using both the scaled and unscaled covariates. Compare the order of the coefficients between the both estimates as well as their corresponding linear predictors. Does the top 50 largest (in absolute size) coefficients differ much between the two estimates?

\item[\textit{c)}] Repeat part \textit{b)}, now with the penalty parameter chosen by means of LOOCV. 
\end{compactitem}
\end{question}

\begin{question} \mbox{ } \\
Consider the linear regression model $\mathbf{Y} = \mathbf{X} \bbeta + \vvarepsilon$ with $\vvarepsilon \sim \mathcal{N} ( \mathbf{0}_n, \sigma_{\varepsilon}^2 \mathbf{I}_{nn})$. The model is fitted to the following (summaries of the) data:
\begin{eqnarray*}
\mathbf{X} = \left( \begin{array}{rr} -2 & 1 \\ 1 & -2 \end{array} \right), 
\mathbf{X}^{\top} \mathbf{X} = \left( \begin{array}{rr} 5 & -4 \\ -4 & 5 \end{array} \right), \mathbf{Y} = \left( \begin{array}{r} 1 \\ -3 \end{array} \right), \, \mbox{ and } \, \mathbf{X}^{\top} \mathbf{Y} = 
\left( \begin{array}{r} -5 \\ 7 \end{array} \right).
\end{eqnarray*}
This is done using the ridge regression estimator, i.e. $\hat{\bbeta}(\lambda) = \arg \min\nolimits_{\bbeta \in \mathbb{R}^2} \, \| \mathbf{Y} - \mathbf{X} \bbeta \|_2^2 + \lambda \| \bbeta\|_2^2$.  
\begin{compactitem}
\item[\textit{a)}] Evaluate the ridge regression estimator for $\lambda = 1$.
\item[\textit{b)}] Suppose $\bbeta = (0, 1)^{\top}$. Evaluate the part of the ridge regression estimator's bias attributable to the high-dimensionality for $\lambda = 1$.

\item[\textit{c)}] For which $\lambda$ does the ridge regression estimator consume a single degree of freedom? \textit{Hint}: the eigenvalues of $\mathbf{X}^{\top} \mathbf{X}$ equal 9 and 1.

\item[\textit{d)}] Still take $\bbeta = (0, 1)^{\top}$. Suppose additionally we believe $\sigma_{\varepsilon}^2 = 1$. We then define $\lambda_{\sigma_{\varepsilon}^2=1}^{\mbox{{\tiny (opt)}}} = \arg \min_{\lambda >0} \mbox{MSE}[\hat{\bbeta}(\lambda)]$, where MSE is the acronym for Mean Squared Error. Would our believe of the error variance be off and in fact  $\sigma_{\varepsilon}^2 = 2$, does $\mbox{MSE}[\hat{\bbeta}(0)] > \mbox{MSE}[\hat{\bbeta}(\lambda_{\sigma_{\varepsilon}^2=1}^{\mbox{{\tiny (opt)}}})]$ still hold? Motivate your answer (not necessarily mathematically but certainly intuitively, possibly using an analogous setting).

\item[\textit{e)}]  Evaluate the `ridge linear predictor', i.e. $\mathbf{X}
\hat{\bbeta}(\lambda)$ for $\lambda = 1$.

\item[\textit{f)}] Evaluate the variance of the ridge linear predictor for $\lambda = 1$ and estimate $\sigma^2$ by the maximum likelihood method.

\item[\textit{g)}] The variance of the ridge linear predictor decreases monotonically with $\lambda$. Show that the estimator of the error variance  
$\hat{\sigma}^2_{\varepsilon}(\lambda) = n^{-1} \| \mathbf{Y} - \mathbf{X} \hat{\bbeta}(\lambda) \|_2^2$  increases monotonically with $\lambda$. 

\item[\textit{h)}] Suppose the study would have been high-dimensional, i.e. $p \gg n$. Can we make a (good) prediction at every point of the design space? Motivate your answer.
\end{compactitem}
\end{question}

\begin{question} \mbox{ } \\
Consider the linear regression model $\mathbf{Y} = \mathbf{X} \bbeta + \vvarepsilon$ with $\vvarepsilon \sim \mathcal{N} ( \mathbf{0}_n, \sigma_{\varepsilon}^2 \mathbf{I}_{nn})$. The model is fitted to the following (summaries of the) data:
\begin{eqnarray*}
\mathbf{X} = \left( \begin{array}{rr} -2 & 4 \\ 1 & -2 \end{array} \right), 
\mathbf{X}^{\top} \mathbf{X} = \left( \begin{array}{rr} 5 & -10 \\ -10 & 20 \end{array} \right), \mathbf{Y} = \left( \begin{array}{r} 1 \\ -2 \end{array} \right), \, \mbox{ and } \, \mathbf{X}^{\top} \mathbf{Y} = 
\left( \begin{array}{r} -4 \\ 8 \end{array} \right),
\end{eqnarray*}
using the regular ridge regression estimator $\hat{\bbeta}(\lambda_2)  =  \arg \min\nolimits_{\bbeta \in \mathbb{R}^2} \| \mathbf{Y} - \mathbf{X} \bbeta \|_2^2 + \lambda \| \bbeta \|_2^2$.
\begin{compactitem}
\item[\textit{a)}] Evaluate the ridge regression estimator for $\lambda = 15$. 

\item[\textit{b)}] Could $\hat{\bbeta} (\lambda) = (1, 2)^{\top}$ for some $\lambda > 0$ and $\mathbf{Y} \in \mathbb{R}^2$? Motivate.

\item[\textit{c)}] Verify that $\mbox{AIC}(\lambda) \propto 50 ( \lambda + 25)^{-1} + 2 \log (\lambda^2 + 18 \lambda + 225) - 4 \log(\lambda + 25)$. 

\item[\textit{d)}] Find the optimal $\lambda$, based on the $\mbox{AIC}(\lambda)$.  

\item[\textit{e)}] Typically, `more is better' when it comes to sample size. However, the inclusion of this extra observation does not reduce the bias attributable to the high-dimensionality of the design. Argue or show why. 

\item[\textit{f)}] Is the regularization attributable bias affected by the inclusion of the third observation? Motivate.

\item[\textit{g)}] Does the ridge regression estimator's variance changes with the inclusion of the third observation? Motivate. 
\end{compactitem}
\end{question}

\pagestyle{fancy}

\chapter[Bayesian regression]{Bayesian regression} \label{chap:BayesianRegression}
The ridge regression estimator is equivalent to a Bayesian regression estimator. On one hand this equivalence provides another interpretation of the ridge regression estimator. But it also shows that within the Bayesian framework, the high-dimensionality of the data need not frustrate the numerical evaluation of the estimator. In addition, the framework provides ways to quantify the consequences of high-dimensionality on the uncertainty of the estimator. Within this chapter, focus is on the equivalence of Bayesian and ridge regression. In this particular case, the connection is immediate from the analytic formulation of the Bayesian regression estimator. After this has been presented, it is shown how this estimator may also be obtained by means of sampling. The relevance of the sampling for the evaluation of the estimator and its uncertainty becomes apparent in subsequent chapters, where we discuss other regularized estimators for which the connection cannot be captured in analytic form.

\section{A minimum of prior knowledge on Bayesian statistics}
The schism between Bayesian and frequentist statistics centers on the interpretation of the concept of probability. A frequentist views the probability of an event as the limiting frequency of observing the event among a large number of trials. In contrast, a Bayesian considers it to be a measure of believe (in the occurence) of the event. The difference between the two interpretations becomes clear when considering events that can occur only once (or a small number of times). A Bayesian would happily discuss the probability of this (possibly hypothetical) event happening, which would be meaningless to a frequentist.

What are the consequences of this schism for the estimating a regression model? Exponents from both paradigms assume a statistical model, e.g. here the linear regression model, to describe the data generating mechanism. But a frequentist treats the parameters as platonic quantities, for which a true value exists that is to be estimated from the data. A Bayesian, however, first formalizes his/her current believe/knowledge on the parameters in the form of probability distributions. This is referred to as the prior -- to the experiment -- distribution. The parameters are thus random variables and their distributions are to be interpreted as reflecting the (relative) likelihood of the parameters' values. From the Bayesian point it then makes sense to talk about the random behaviour of the parameter, e.g., what is the probability of the parameter being larger than a particular value. In the frequentist context, one may ask this of the estimator, but not of the parameter. The parameter either \textit{is} or \textit{is not} larger than this value as a platonic quantity is not governed by a chance process. Then, having specified her/his believe on the parameters, a Bayesian uses the data to update this believe of the parameters of the statistical model, which again is a distribution and called the posterior -- to the experiment -- distribution.

To illustrate this process of updating, assume that the data $Y_1, \ldots, Y_n \sim \{0, 1 \}$ are independently and identically drawn from a Bernouilli distribution with parameter $\theta = P(Y_i = 1)$. The likelihood of these data for a given choice of the parameter $\theta$ then is: $L(\mathbf{Y} = \mathbf{y}; \theta) = P(\mathbf{Y} = \mathbf{y} \, | \, \theta) = P(Y_1 = y_1, \ldots, Y_n = y_n \, | \, \theta) = \prod_{i=1}^n P(Y_i = y_i \, | \, \theta) = \prod_{i=1}^n \theta^{y_i} (1-\theta)^{1-y_i}$. A Bayesian now needs to specificy the prior distribution, denoted $\pi_{\theta}(\cdot)$, on the parameter $\theta$. A common choice in this case would be a beta-distribution: $\theta \sim \mathcal{B}(\alpha, \beta)$. The posterior distribution is now arrived at by the use of Bayes' rule:
\begin{eqnarray*}
\pi ( \theta \, | \, \mathbf{Y} = \mathbf{y}) & = & \frac{P(\mathbf{Y} = \mathbf{y} \, | \, \theta) \, \pi (\theta) }{ P(\mathbf{Y} = \mathbf{y} ) }
\, \, \, = \, \, \, \frac{P(\mathbf{Y} = \mathbf{y} \, | \, \theta) \, \pi (\theta) }{ \int_{0,1} P(\mathbf{Y} = \mathbf{y} \, | \, \theta) \, \pi (\theta) \, d\theta }.
\end{eqnarray*}
The posterior distribution thus contains all knowledge on the parameter. As the denomenator -- referred to as the distribution's normalizing constant -- in the expression of the posterior distribution does not involve the parameter $\theta$, one often writes $\pi ( \theta \, | \, \mathbf{Y} = \mathbf{y}) \propto P(\mathbf{Y} = \mathbf{y} \, | \, \theta) \, \pi (\theta)$, thus specifying the (relative) density of the posterior distribution of $\theta$ as `likelihood $\times$ prior'.

While the posterior distribution is all one wishes to know, often a point estimate is required. A point estimate of $\theta$ can be obtained from the posterior by taking the mean or the mode. Formally, the Bayesian point estimator of a parameter $\ttheta$ is defined as the estimator that minimizes the Bayes risk over a prior distribution of the parameter $\theta$. The Bayes risk is defined as $\int_{\theta} \mathbb{E} [(\hat{\theta} - \theta)^2] \pi_{\theta}(\theta; \alpha) d\theta$, where $\pi_{\theta}(\theta; \alpha)$ is the prior distribution of $\theta$ with hyperparameter $\alpha$. It is thus a weighted average of the Mean Squared Error, with weights specified through the prior. The Bayes risk is minimized by the posterior mean:
\begin{eqnarray*}
\mathbb{E}_{\theta}(\theta \, | \,  \mbox{data}) & = & \frac{ \int_{\theta} \theta \,  P(\mathbf{Y} = \mathbf{y} \, | \, \theta) \, \pi (\theta) \, d \theta }{ \int_{\theta} P(\mathbf{Y} = \mathbf{y} \, | \, \theta) \, \pi (\theta) \, d\theta }.
\end{eqnarray*}
(cf., e.g., \citealp{Bijm2017}). The Bayesian point estimator of $\theta$ yields the smallest possible expected MSE, under the assumption of the employed prior. This estimator  thus depends on the likelihood \textit{and} the prior, and a different prior yields a different estimator.

A point estimator is preferably accompanied by a quantification of its uncertainty. Within the Bayesian context, this is done  by so-called credible intervals or regions, the Bayesian equivalent of confidence intervals or regions. A Bayesian credible interval encompasses a certain percentage of the probability mass of the posterior. For instance, a subset of the parameter space $\mathcal{C}$ forms an $100 (1-\alpha) \%$ credible interval if $\int_{\mathcal{C}} \pi(\theta \, | \, \mbox{data}) \, d \theta = 1- \alpha$.

In the example above, it is important to note the role of the prior in the updating of the knowledge on $\theta$. First, when the prior distribution is uniform on the unit interval, the mode of the posterior coincides with the maximum likelihood estimator. Furthermore, when $n$ is large the influence of the prior in the posterior is negligible. It is therefore that for large sample sizes frequentist and Bayesian analyses tend to produce similar results. However, when $n$ is small, the prior's contribution to the posterior distribution cannot be neglected. The choice of the prior is then crucial as it determines the shape of the posterior distribution and, consequently, the posterior estimates. It is this strong dependence of the posterior on the arbitrary (?) choice of the prior that causes unease with a frequentist, leading some to accuse Bayesians of subjectivity. In the high-dimensional setting, the sample size is usually small, especially in relation to the parameter dimension, and the choice of the prior distribution then matters. Interest here is not in resolving the frequentist's unease, but in identifying or illustrating the effect of the choice of the prior distribution on the parameter estimates.

\section{Connection to ridge regression} \label{sect.Bayes2ridge}
\begin{figure}[!h]
\begin{tabular}{rc}
\mbox{ } \qquad \qquad \qquad \qquad \mbox{ } &
\includegraphics[scale=0.45, angle=0]{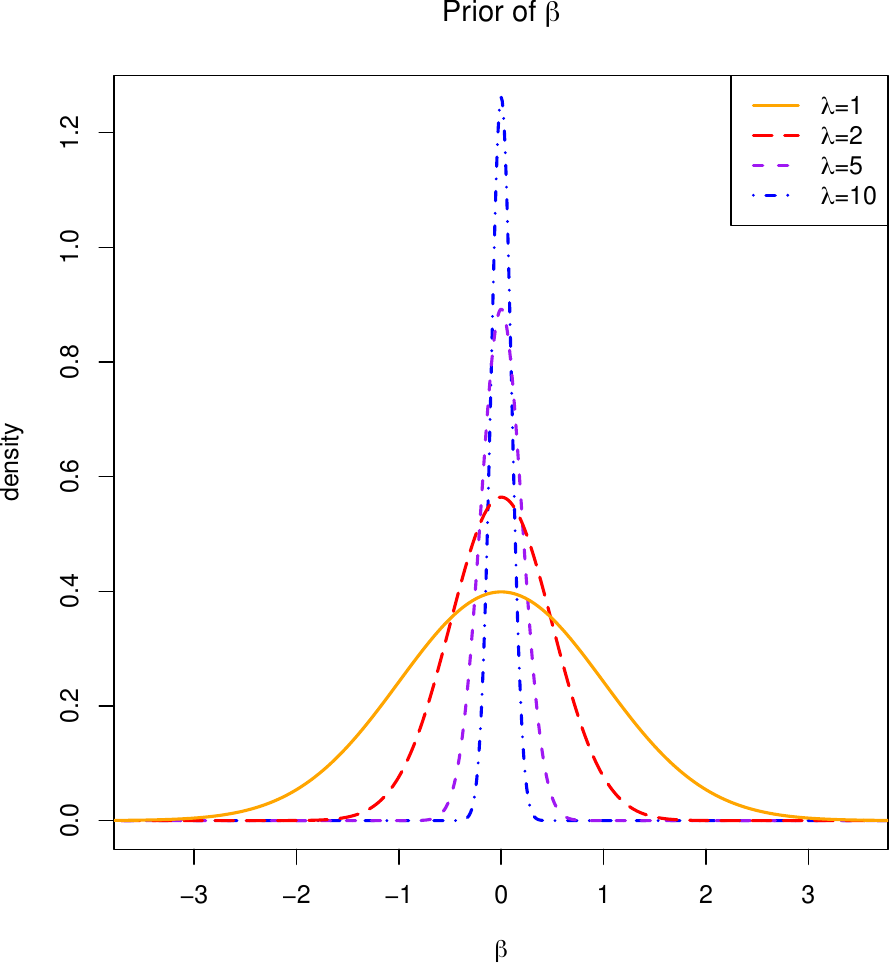}
\end{tabular}
\caption{Conjugate prior of the regression parameter $\bbeta$ for various choices of $\lambda$, the penalty parameters c.q. precision.} \label{fig.ridgePriorOfBeta}
\end{figure}

Ridge regression has a close connection to Bayesian linear regression. Bayesian linear regression assumes the parameters $\bbeta$ and $\sigma^2$ to be the random variables, while at the same time considering $\mathbf{X}$ and $\mathbf{Y}$ as fixed. Within the regression context, the commonly chosen priors of $\bbeta$ and $\sigma^2$ are $\bbeta \, | \, \sigma^2 \sim \mathcal{N}(\mathbf{0}_p,  \sigma^2 \lambda^{-1} \mathbf{I}_{pp})$ and $\sigma^2 \sim \mathcal{IG}(a_0, b_0)$, where $\mathcal{IG}$ denotes the inverse Gamma distribution with shape parameter $a_0$ and scale parameter $b_0$. The penalty parameter can be interpreted as the (scaled) precision of the prior of $\bbeta$, determining how informative the prior should be. A smaller penalty (i.e. precision) corresponds to a wider prior, and a larger penalty to a more informative, concentrated prior (Figure \ref{fig.ridgePriorOfBeta}).

The joint posterior distribution of $\bbeta$ and $\sigma^2$ is, under the assumptions of the (likelihood of) the linear regression model and the priors above:
\begin{eqnarray*}
f_{\bbeta, \sigma^2} (\bbeta, \sigma^2 \, | \, \mathbf{Y}, \mathbf{X}) & \propto & f_Y (\mathbf{Y} \, | \, \mathbf{X}, \bbeta, \sigma^2) \, f_{\beta}(\bbeta | \sigma^2) \, f_{\sigma}(\sigma^2)
\\
& \propto & \sigma^{-n} \exp [ - \tfrac{1}{2}\sigma^{-2} ( \mathbf{Y} - \mathbf{X} \bbeta)^{\top} ( \mathbf{Y} - \mathbf{X} \bbeta) ]
\\
&  & \times \, \, \sigma^{-p} \exp ( - \tfrac{1}{2} \sigma^{-2} \lambda \bbeta^{\top} \bbeta ) \, \times \, \, (\sigma^2)^{-a_0-1} \exp ( -  b_0 \sigma^{-2} ).
\end{eqnarray*}
The posterior distribution can be expressed as a multivariate normal distribution. Hereto group the arguments of the exponential functions that involve $\bbeta$ and manipulate as follows:
\begin{eqnarray*}
& & \hspace{-1.5cm} ( \mathbf{Y} - \mathbf{X} \bbeta)^{\top} ( \mathbf{Y} - \mathbf{X} \bbeta) + \lambda \bbeta^{\top} \bbeta
\\
& = & \mathbf{Y}^{\top} \mathbf{Y} - 2 \bbeta^{\top} \mathbf{X}^{\top} \mathbf{Y} + \bbeta^{\top} \mathbf{X}^{\top} \mathbf{X} \bbeta + \lambda \bbeta^{\top} \bbeta
\\
& = & \mathbf{Y}^{\top} \mathbf{Y} - 2 \bbeta^{\top} (\mathbf{X}^{\top} \mathbf{X} + \lambda \mathbf{I}_{pp}) (\mathbf{X}^{\top} \mathbf{X} + \lambda \mathbf{I}_{pp})^{-1} \mathbf{X}^{\top} \mathbf{Y} + \bbeta^{\top} (\mathbf{X}^{\top} \mathbf{X} + \lambda \mathbf{I}_{pp}) \bbeta
\\
& = & \mathbf{Y}^{\top} \mathbf{Y} - \bbeta^{\top} (\mathbf{X}^{\top} \mathbf{X} + \lambda \mathbf{I}_{pp})
\hat{\bbeta} (\lambda) - [ \hat{\bbeta} (\lambda) ]^{\top} (\mathbf{X}^{\top} \mathbf{X} + \lambda \mathbf{I}_{pp}) \bbeta + \bbeta^{\top} (\mathbf{X}^{\top} \mathbf{X} + \lambda \mathbf{I}_{pp}) \bbeta
\\
& = & \mathbf{Y}^{\top} \mathbf{Y} - \mathbf{Y}^{\top} \mathbf{X} (\mathbf{X}^{\top} \mathbf{X} + \lambda \mathbf{I}_{pp})^{-1} \mathbf{X}^{\top} \mathbf{Y} + [ \bbeta - \hat{\bbeta}(\lambda) ]^{\top} (\mathbf{X}^{\top} \mathbf{X} + \lambda \mathbf{I}_{pp}) [ \bbeta - \hat{\bbeta}(\lambda) ].
\end{eqnarray*}
Using this result, the posterior distribution can be rewritten to:
\begin{eqnarray*}
f_{\bbeta, \sigma^2} (\bbeta, \sigma^2 \, | \, \mathbf{Y}, \mathbf{X}) & \propto & g_{\bbeta | \sigma^2} (\bbeta \, | \, \sigma^2, \mathbf{Y}, \mathbf{X}) \, g_{\sigma^2} (\sigma^2 \, | \, \mathbf{Y}, \mathbf{X})
\end{eqnarray*}
with
\begin{eqnarray*}
g_{\bbeta | \sigma^2} (\bbeta \, | \, \sigma^2, \mathbf{Y}, \mathbf{X}) & \propto & \exp \big\{ - \tfrac{1}{2} \sigma^{-2} \big[ \bbeta - \hat{\bbeta}(\lambda) \big]^{\top} (\mathbf{X}^{\top} \mathbf{X} + \lambda \mathbf{I}_{pp}) \big[ \bbeta - \hat{\bbeta}(\lambda) \big] \big\}.
\end{eqnarray*}
In the display above, we recognize the form of a multivariate normal density and conclude that $\bbeta \, | \, \sigma^2, \mathbf{Y}, \mathbf{X} \sim \mathcal{N}  ( \cdot, \cdot)$ with the conditional posterior mean of $\bbeta$ is $\mathbb{E}(\bbeta \, | \, \sigma^2, \mathbf{Y}, \mathbf{X}) = \hat{\bbeta}(\lambda)$. Hence, the ridge regression estimator can be viewed as the Bayesian posterior mean estimator of $\bbeta$ if we impose a Gaussian prior on the regression parameter.

The correspondence between Bayesian and ridge regression continues beyond the estimators. The level sets of the prior distribution are the parameter constraints implied by the regularization. This becomes obvious after a simple reformulation of the multivariate normal distribution's level sets:
\begin{eqnarray*}
\{ \boldsymbol{\beta} \in \mathbb{R}^p \, : \, (2 \pi \lambda^{-1} \sigma^2)^{-1/2} \exp[ - (2\sigma^2)^{-1} \lambda \boldsymbol{\beta}^{\top} \boldsymbol{\beta}] = c \}
& = & \{ \boldsymbol{\beta} \in \mathbb{R}^p \, : \,   \| \boldsymbol{\beta} \|_2^2 = - 2 \lambda^{-1} \sigma^2 \log[ (2 \pi \lambda^{-1} \sigma^2)^{1/2} c ]  \},
\end{eqnarray*}
with constant $c$ such that $0 \leq c \leq (2 \pi \lambda^{-1} \sigma^2)^{-1/2}$. The right-hand side of the preceeding display is the parameter constraint induced by ridge regularization. This correspondence endows the parameter constraint with an probabilistic interpretation: the $100\alpha$\%  of the prior's probability mass, i.e. the $100\alpha$\% most likely values of the parameter according to the prior. The percentage $100\alpha\%$ relates directly to the regularization parameter.
\\
\\
In general, any penalized estimator has a Bayesian equivalent.  A penalized estimator always coincides with the Bayesian MAP (Maximum A Posteriori) estimator. Then, the MAP estimates a parameter $\ttheta$ by the mode, i.e. the maximum, of the posterior density. To see the equivalence of both estimators, we derive the MAP estimator. The latter is defined as: 
\begin{eqnarray*}
\hat{\ttheta}_{\mbox{{\tiny map}}} \, \, \, = \, \, \, \arg \max_{\ttheta} \, \pi ( \ttheta \, | \, \mathbf{Y} = \mathbf{y}) & = & \arg \max_{\ttheta} \frac{P(\mathbf{Y} = \mathbf{y} \, | \, \ttheta) \, \pi (\theta) }{ \int P(\mathbf{Y} = \mathbf{y} \, | \, \ttheta) \, \pi (\ttheta) \, d\ttheta }.
\end{eqnarray*}
To find the maximum of the posterior density take the logarithm (which does not change the location of the maximum) and drop terms that do not involve $\ttheta$. The MAP estimator then is:
\begin{eqnarray*}
\hat{\ttheta}_{\mbox{{\tiny map}}} & = & \arg \max_{\ttheta} \, \log[ P(\mathbf{Y} = \mathbf{y} \, | \, \ttheta) ] + \log[ \pi (\ttheta) ].
\end{eqnarray*}
The first summand on the right-hand side is the loglikelihood, while the second is the logarithm of the prior density. The latter, in case of a normal prior, is proportional to $\sigma^{-2} \lambda \| \bbeta \|_2^2$. This is -- up to some factors -- exactly the loss function of the ridge regression estimator. If the quadratic ridge penalty is replaced by different one, it is easy to see what form the prior density should have in order for both estimators -- the penalized and MAP -- to coincide.

Finally, the ridge regression estimator also equals the Bayesian regression MAP estimator. In this case, the posterior mean and MAP estimator coincide due to the normal form of its likelihood and conjugate prior.
\\
\\
For the joint posterior distribution of $\bbeta$ and $\sigma$, we still need the marginal posterior of $\sigma^2$. After a little more rewriting of $f_{\bbeta, \sigma}$, we find it is given by:
\begin{eqnarray*}
g_{\sigma^2} (\sigma^2 \, | \, \mathbf{Y}, \mathbf{X}) & \propto & (\sigma^2)^{-(n/2 + a_0)}
\exp [ - \tfrac{1}{2} \sigma^{-2} ( \| \mathbf{Y} \|_2^2  - \lambda\| \hat{\bbeta}(\lambda) \|_2^2 - \| \mathbf{X} \hat{\bbeta}(\lambda) \|_2^2  + 2 b_0)  ],
\end{eqnarray*}
in which we recognize the shape of an inverse gamma distribution. This posterior does not involve the regression parameter $\bbeta$, which has effectively been integrated out. We can now study the joint distribution of $\bbeta$ and $\sigma^2$ through sampling. The pseudo-code below (Algorithm \ref{alg.directSamplerRidgeRegression}) describes how to draw from $f_{\bbeta, \sigma}$. 
\\
\mbox{ }
\\
\begin{minipage}[h]{\textwidth}
\begin{center}
\fbox{\parbox{14cm}{
\begin{algorithm}[H] 
\SetKwInOut{Input}{input}
\SetKwInOut{Output}{output}
\RestyleAlgo{boxed}
\LinesNumbered
\SetKwFor{inpar}{for}{do in parallel}{\text{ end synchronize}}
\Input{sample size $T$;
\\
\, data $\{\mathbf{x}_i, \mathbf{y}_i \}_{i=1}^n$;
\\
\, conditional distributions $g_{\sigma^2} (\sigma^2; \{\mathbf{x}_i, \mathbf{y}_i \}_{i=1}^n)$ and $g_{\bbeta \hspace{1pt} | \hspace{1pt} \sigma^2} (\bbeta, \sigma^2; \{\mathbf{x}_i, \mathbf{y}_i \}_{i=1}^n)$.
}
\Output{draws from the joint posterior $\pi(\bbeta, \sigma^2 \, | \, \{\mathbf{x}_i, \mathbf{y}_i \}_{i=1}^n)$.}
\vspace{0.25cm}
\For{ $t = 1$ to $T$  }{
\BlankLine
draw $\sigma_{t}^2$ from conditional distribution $g_{\sigma^2} (\sigma^2; \{\mathbf{x}_i, \mathbf{y}_i \}_{i=1}^n)$,
\\
draw $\bbeta_{t}$ from conditional distribution $g_{\bbeta | \sigma^2} (\bbeta ; \sigma_t^2, \{\mathbf{x}_i, \mathbf{y}_i \}_{i=1}^n)$.
\\
\BlankLine
}
\BlankLine
\vspace{0.25cm}
\caption{{\small Pseudocode of the sampler of the joint posterior of the Bayesian regression parameters.}} \label{alg.directSamplerRidgeRegression}
\end{algorithm}
}
}
\end{center}
\end{minipage}
\mbox{ }
\\
At each iteration, Algorithm \ref{alg.directSamplerRidgeRegression} samples a $\{\bbeta_t, \sigma_t^2\}$-pair directly from the joint posterior distribution. Direct sampling from the posterior is, however, not always possible (as marginalization need not be analytically trackable). More intricate sampling schemes are then needed. These require the conditional posterior of $\sigma^2$:
\begin{eqnarray*}
g_{\sigma^2 | \bbeta} (\sigma^2 \, | \, \bbeta, \mathbf{Y}, \mathbf{X}) & \propto & (\sigma^2)^{-[(n+p)/2 + a_0 + 1]} \exp [ - \tfrac{1}{2} \sigma^{-2} ( \| \mathbf{Y} - \mathbf{X} \bbeta \|_2^2  + 2 b_0)  ].
\end{eqnarray*}
The use of this conditional distribution will become clear in Section \ref{sect.MCMCsampling}.


\section{High-dimensionality}
\begin{figure}[!h]
\begin{tabular}{lcr}
\includegraphics[scale=0.45, angle=0]{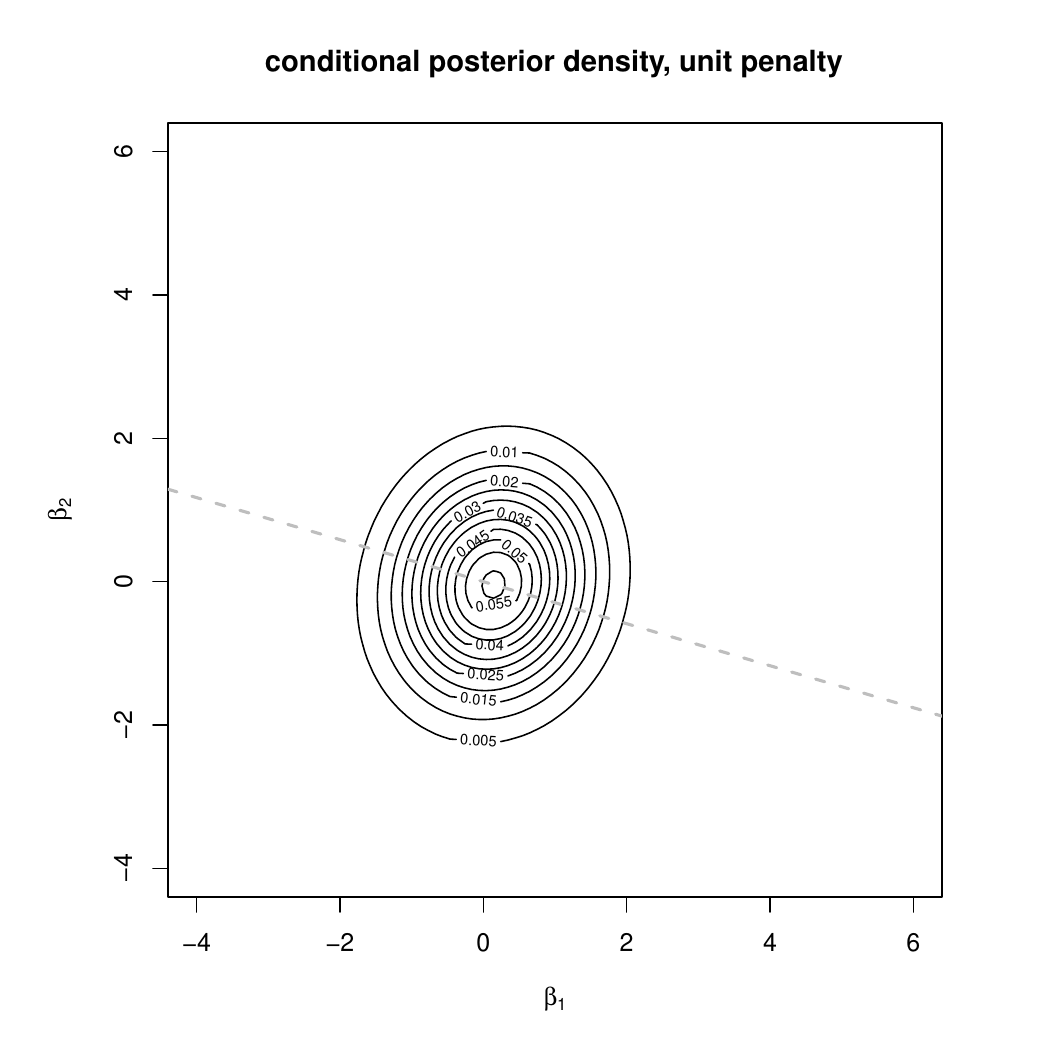}
 &
\includegraphics[scale=0.45, angle=0]{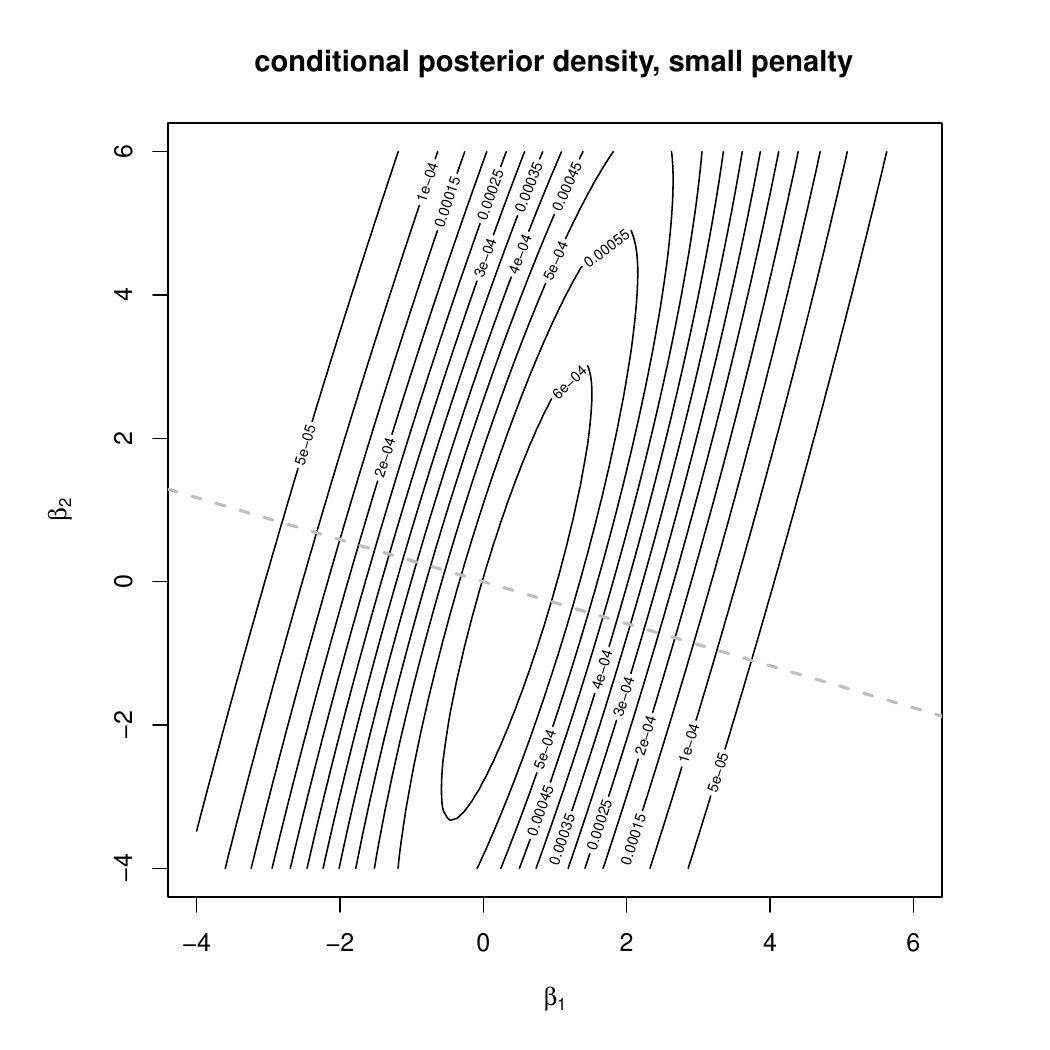}
\end{tabular}
\caption{Level sets of the (unnormalized) conditional posterior of the regression parameter $\bbeta$. The grey dashed line depicts the support of the degenerated normal distribution of the frequentist's ridge regression estimate.} \label{fig.conditionalPosteriorOfBeta}
\end{figure}

It is often claimed that Bayesian methods have no issue with high-dimensionality. This is true in the sense that typically the posterior, and thereby its mean or mode, of Bayesian regression is well-defined. However, Bayesian methods too cannot learn if the experiment provides no information on (parts of) the parameter. That is, there are domains of the parameter space where the believe on the parameter is unaffected by the experiment. Translated to the estimation of the regression parameter from high-dimensional data, the likelihood provides no information in the direction orthogonal to the rows of design matrix. In that subspace of the parameter space, the Bayesian updating of the believe has no affect and prior and posterior coincide there. Put differently, any information in that direction extracted from the posterior is due to the prior. Consequently, uncertainty quantification (by credible intervals) from the posterior, one of the strengths of Bayesian methods, is not meaningful. In the sense, that the uncertainty is a reflection of the prior believe instead of the information from the experiment. We illustrate this by contrasting the posterior distributions after updating from a high-dimensional experiment with an informative and uninformative prior. 
These conditional posterior distributions of $\bbeta$, $g_{\bbeta} (\bbeta \, | \, \sigma^2, \mathbf{Y}, \mathbf{X})$, related to a high-dimensional situation ($n=1$, $p=2$) are depicted in Figure \ref{fig.conditionalPosteriorOfBeta} for two choices of the penalty parameter $\lambda$. Put differently, for two different priors. In one, the left panel, $\lambda$ is arbitrarily set equal to one. The choice of the employed $\lambda$, i.e. $\lambda=1$, is irrelevant, the resulting posterior distribution only serves as a reference. This choice results in a posterior distribution that is concentrated around the Bayesian point estimate, which coincides with the ridge regression estimate. The almost spherical level sets around the point estimate may be interpreted as credible intervals for $\bbeta$. The grey dashed line, spanned by the row of the design matrix, represents the support of the degenerated normal distribution of the frequentist's ridge regression estimate (cf. end of Section  \ref{sect:ridgeVariance}). The contrast between the degenerated frequentist and well-defined Bayesian normal distribution illustrates that -- for a suitable choice of the prior -- within the Bayesian context high-dimensionality need not be an issue with respect to the evaluation of the posterior. In the right panel of Figure \ref{fig.conditionalPosteriorOfBeta}, the penalty parameter $\lambda$ is set equal to a very small value, i.e. $0 < \lambda \ll 1$. This represents a very imprecise or uninformative prior. It results in a posterior distribution with level sets that are far from spherical and are very stretched in the direction orthogonal to the subspace spanned by the row of the design matrix. This indicates in which direction there is most uncertainty with respect to $\bbeta$. In particular, when $\lambda \downarrow 0$, the level sets loose their ellipsoid form and the posterior collapses to the degenerated normal distribution of the frequentist's ridge regression estimate. Hence, also in the Bayesian context, we do not learn in the direction orthogonal to the rows of the design matrix. Finally, in combination with the left-hand plot, this illustrates the large effect of the prior in the high-dimensional context.

\subsection{Credible set}
Uncertainty quantification through credible intervals may even be useless in high-dimensions. We illustrate this by example from \cite{stein1962confidence}. The example considers the normal means model, i.e. the linear regression model with a design matrix equal to the identity matrix. That is, there is only a single observation per element of the regression parameter available for its estimation. We write this formally as $\mathbf{Y} \sim \mathcal{N}(\boldsymbol{\mu}, \sigma^2 \mathbf{I}_{pp})$ with $\boldsymbol{\mu} = (\mu_1, \ldots, \mu_p)^{\top}$. For simplicity, we assume $\sigma^2 = 1$. Moreover, we have only one draw from this distribution available. 

We complete the Bayesian regression model by adopting a flat prior on $\boldsymbol{\mu}$, i.e. $\pi(\boldsymbol{\mu}) \propto 1$. This is an improper prior, in the sense that it is not a valid distribution if the parameter domain is infinite. But the corresponding posterior is a valid distribution. It is a multivariate normal distribution with $\mathbb{E} ( \boldsymbol{\mu} \, | \, \mathbf{Y}) = \mathbf{Y}$ and $\mbox{Var} ( \boldsymbol{\mu} \, | \, \mathbf{Y}) = \mathbf{I}_{pp}$.  

We aim to construct a credible interval for the parameter $\theta = \sum_{j=1}^p \mu_j^2$ and assess its coverage probability in high dimensions. The posterior for $\theta$ is a noncentral $\chi^2$ distribution with $p$ degrees of freedom and a noncentrality parameter equal to $\sum_{j=1}^p Y_j^2$. From the moments of this $\chi^2$ distribution, we directly obtain the posterior mean and variance: $\mathbb{E} ( \theta \, | \, \mathbf{Y}) = p + \sum_{j=1}^p Y_j^2$ and $\mbox{Var} ( \theta \, | \, \mathbf{Y}) = 2 p + 4\sum_{j=1}^p Y_j^2$.

The credible interval of $\theta$ is of the form $[c_{\ell, \alpha}, \infty)$ with the lower bound $c_{\ell, \alpha}$ such that $P \{ \theta \in [c_{\ell, \alpha}, \infty) \, | \, \mathbf{Y} \} = 1-\alpha$ for $\alpha \in [0,1]$. To find the lower bound, we note that, for large $p$, this noncentral $\chi^2$ is well approximated by the normal distribution.  Conveniently, we can thus derive the lower bound $c_{\ell, \alpha}$ analogously to that of a confidence interval for the mean of the normal distribution.  Hence, for sufficiently large $p$, 
\begin{eqnarray*}
c_{\ell, \alpha} & \approx & p + \sum\nolimits_{j=1}^p Y_j^2 - z_{1-\alpha} \big[  2 p + 4\sum\nolimits_{j=1}^p Y_j^2 \big]^{1/2}.
\end{eqnarray*}
where $z_{1-\alpha}$ is the $1-\alpha$-th percentage point of the normal distribution. 

Before we proceed, we derive an approximation of the first two moments of the lower bound. The expectation is:
\begin{eqnarray*}
\mathbb{E}( c_{\ell, \alpha} ) & = & p + \sum\nolimits_{j=1}^p \mathbb{E} ( Y_j^2 ) - z_{1-\alpha} \mathbb{E} \big\{ \big[  2 p + 4\sum\nolimits_{j=1}^p Y_j^2 \big]^{1/2} \big\}
\\
& \approx & \theta + 2 p - \sqrt{ 4 \theta + 6 p }-  \tfrac{1}{8} z_{1-\alpha}  (4 \theta + 2p) ( 4 \theta + 6 p)^{-3/2},
\end{eqnarray*}
where we have used the moments of the noncentral $\chi^2$ distribution. The approximation in the last step hinges upon that of $\mathbb{E} (\sqrt{X})$. The Delta method provides: $\mathbb{E} (\sqrt{X})  \approx \sqrt{ \mathbb{E}(X) } - \tfrac{1}{8} \mbox{Var}(X)  [ \mathbb{E}(X) ]^{-3/2}$, which holds for large values of $X$ and a relative low variance. We now turn to the variance of the lower bound: 
\begin{eqnarray*}
\mbox{Var}(c_{\ell, \alpha} ) & \approx & \big\{ 1 - 2 z_{1-\alpha} \big[ 4 \mathbb{E} \big( \sum\nolimits_{j=1}^p Y_j^2 \big)  + 2p \big]^{-1/2} \big\}^2 \mbox{Var} \big( \sum\nolimits_{j=1}^p Y_j^2 \big)
\\
& = & [ 1 - 2 z_{1-\alpha} ( 4 \theta + 6 p )^{-1/2} ]^2 (4 \theta + 2p),
\end{eqnarray*}
in which we again rely on the moments of the noncentral $\chi^2$ distribution. The approximation in the first line of the above display follows from the Delta method: $\mbox{Var}[g(X)] \approx \{ g'[\mathbb{E}(X)]\}^2 \, \mbox{Var}(X)$ for random variable $X$ and smooth function $g(\cdot)$.

We move forward with our original aim to assess the coverage of the credible interval for $\theta$, i.e. $P( c_{\ell, \alpha} \leq \theta ) = 1 - \alpha$? We first use set theory to include the event in another more convenient one:
\begin{eqnarray*}
\{\theta \, : \, c_{\ell, \alpha} \leq \theta  \} &  = &  \{ \theta :  c_{\ell, \alpha} - \mathbb{E}( c_{\ell, \alpha} ) \leq \theta - \mathbb{E}( c_{\ell, \alpha} ) \} 
\hspace{0.4cm}  = \, \, \,  \{\theta \, : \,  - [ c_{\ell, \alpha} - \mathbb{E}( c_{\ell, \alpha} )] \geq  \mathbb{E}( c_{\ell, \alpha} ) - \theta \}
\\
& \subseteq & \{ \theta \, : \,   | c_{\ell, \alpha} - \mathbb{E}( c_{\ell, \alpha} ) |  \geq  \mathbb{E}( c_{\ell, \alpha} ) - \theta \}.
\end{eqnarray*} 
Then, using that $\mathbb{E}( c_{\ell, \alpha} ) \approx \theta + 2p$ for large enough $p$, we bound the coverage probability as:
\begin{eqnarray*}
P( c_{\ell, \alpha} \leq \theta  ) & \leq & P [  | c_{\ell, \alpha} - \mathbb{E}( c_{\ell, \alpha} ) |  \geq  \mathbb{E}( c_{\ell, \alpha} ) - \theta ] \, \, \, \approx \, \, \, P [  | c_{\ell, \alpha} - \mathbb{E}( c_{\ell, \alpha} ) |  \geq  2p ].
\end{eqnarray*} 
Finally, apply Chebyshev's inequality and the fact that $\mbox{Var}(c_{\ell, \alpha} ) \approx 4\theta + 2p$ for large $p$, to bound even further:
\begin{eqnarray*}
P( c_{\ell, \alpha} \leq \theta  ) & \leq &  P [  | c_{\ell, \alpha} - \mathbb{E}( c_{\ell, \alpha} ) |  \geq  2p ] \, \, \, \leq \, \, \, (2p)^{-2} \, \mbox{Var}(c_{\ell, \alpha} ) \, \, \, \approx \, \, \, (2p)^{-1},
\end{eqnarray*} 
for finite $\theta$. If $p \rightarrow \infty$, the right-hand side vanishes. High-dimensionally, the coverage probability of the credible interval of $\theta$ is thus zero, far from the nominal $1-\alpha$. 

The intuition behind the poor coverage of the credible interval comes from the curse of dimensionality. A flat prior implies that all possible parameter values are believed to be equally likely. This is unproblematic if we have only a few dimenions, but high dimensionally it is not. Then, most probability mass of the distribution concentrates in the `crust', i.e. outer limits of the parameter domain. Consequently, by adopting a flat prior in a high dimensional setting, we essentially assume that the true parameter is most likely to be found in the crust. And our flat prior forces the probability mass of the posterior distribution to the crust, away from the data. This phenomenon can be resolved by adopting a subjective prior. 

\subsection{Efficient sampling}
High-dimensionally, Bayesian regression also encounters practical issues. Sampling from the conditional posterior of $\bbeta$ becomes horrorifically slow in high dimensions. This is overcome by the computationally efficient sampling scheme proposed by \cite{bhattacharya2016fast}, which draws a sample from the multivariate normal distribution with a structured mean vector and covariance matrix,
\begin{eqnarray} \label{form:sampleFromNormalDist}
\mathcal{N} [ (\mathbf{X}^{\top} \mathbf{X} + \lambda \mathbf{I}_{pp})^{-1} \mathbf{X}^{\top} \mathbf{Y}, \sigma^2 (\mathbf{X}^{\top} \mathbf{X} + \lambda \mathbf{I}_{pp})^{-1}].
\end{eqnarray}
The pseudo-code of the algorithm comprises
\\
\\
\vspace{-0.54cm}
\\
\begin{minipage}[h]{\textwidth}
\begin{center}
\fbox{\parbox{14cm}{
\begin{algorithm}[H]
\SetKwInOut{Input}{input}
\SetKwInOut{Output}{output}
\RestyleAlgo{boxed}
\LinesNumbered
\Input{design matrix $\mathbf{X}$, response vector $\mathbf{Y}$, variance $\sigma^2$, and penalty parameter $\lambda$.
}
\Output{a draw from distribution (\ref{form:sampleFromNormalDist}).}
\vspace{0.25cm}
{Sample $\mathbf{u} \sim \mathcal{N}( \mathbf{0}_p, \sigma^2 \lambda^{-1} \mathbf{I}_{pp})$ and $\ddelta \sim \mathcal{N} ( \mathbf{0}_n, \mathbf{I}_{nn})$ independently}.
\\
{Set $\mathbf{v} =  \sigma^{-1} \mathbf{X} \mathbf{u} + \ddelta$}.
\\
{Solve $(\lambda^{-1} \mathbf{X} \mathbf{X}^{\top} + \mathbf{I}_{nn}) \mathbf{w} = \sigma^{-1} \mathbf{Y} - \mathbf{v}$ for $\mathbf{w}$}.
\\
{Set $\bbeta = \mathbf{u} + \sigma \lambda^{-1} \mathbf{X}^{\top} \mathbf{w}$}.
\vspace{0.25cm}
\caption{{\small High-dimensional efficient sampling of distribution (\ref{form:sampleFromNormalDist}).}} \label{alg.highdimNormal}
\end{algorithm}
}
}
\end{center}
\end{minipage}
\mbox{ }
\\
\mbox{ }
\vspace{-0.2cm}
\\
The next proposition warrants that the Algorithm \ref{alg.highdimNormal} indeed samples from distribution (\ref{form:sampleFromNormalDist}).
\begin{proposition} \label{prop:highdimNormSampling} \textit{(Proposition 1 of \citealp{bhattacharya2016fast})} \mbox{ } \\
If $\bbeta$ is sampled in accordance with Algorithm \ref{alg.highdimNormal}, then 
$\bbeta$ is distributed as (\ref{form:sampleFromNormalDist}). 
\end{proposition}
\begin{proof}
It is immediate that $\mathbf{v} \sim \mathcal{N} ( \mathbf{0}_n, \lambda^{-1} \mathbf{X} \mathbf{X}^{\top}  + \mathbf{I}_{nn})$. Furthermore,  as
\begin{eqnarray*}
\bbeta & = & \mathbf{u} + \sigma \lambda^{-1} \mathbf{X}^{\top} (\lambda^{-1} \mathbf{X} \mathbf{X}^{\top} + \mathbf{I}_{nn})^{-1} ( \sigma^{-1} \mathbf{Y} - \mathbf{v}),
\end{eqnarray*}
$\bbeta$ is normally distributed. Its mean is
$\mathbb{E} (\bbeta)  =  \lambda^{-1}  \mathbf{X}^{\top} ( \lambda^{-1} \mathbf{X} \mathbf{X}^{\top} + \mathbf{I}_{nn})^{-1} \mathbf{Y}$, which is the expression of the ridge regression estimator  (\ref{form.effRidgeEstimator}). The variance is 
\begin{eqnarray*}
\mbox{Var} (\bbeta) & = & \mbox{Var}[ \mathbf{u} - \sigma \lambda^{-1} \mathbf{X}^{\top} (\lambda^{-1} \mathbf{X} \mathbf{X}^{\top} + \mathbf{I}_{nn})^{-1} \mathbf{v}]
\\
& = & \sigma^{2} \lambda^{-1} \mathbf{I}_{pp} - \sigma^2 \lambda^{-2} \mathbf{X}^{\top} (\lambda^{-1} \mathbf{X} \mathbf{X}^{\top} + \mathbf{I}_{nn})^{-1}  \mathbf{X}
\\
& = & \sigma^{2} (  \mathbf{X}^{\top} \mathbf{X} +  \lambda \mathbf{I}_{pp})^{-1}, 
\end{eqnarray*}
where we have used the rules of variance calculus, the independence between $\mathbf{u}$ and $\ddelta$, and the Woodbury matrix identity. 
\end{proof}
Algorithm \ref{alg.highdimNormal} is valid for any $(n,p)$-combination, but most computational speed is gained if $p \gg n$. 

Even with an efficient sampler at hand, sampling is not trivial in high dimensional settings. A single draw from a high dimensional distribution is easily sampled. But a representative sample is difficult to acquire, for it should be representative of the whole parameter space. A representative sample is needed to warrant accurate quantification of its uncertainty through a credible region. Typically, this means a large sample is to be drawn in order to cover all relevant features of the distribution over the whole parameter space. Of course, if interest is only in a summary statistic like the posterior mean, one can do with much smaller sample sizes.

\section{Markov chain Monte Carlo} \label{sect.MCMCsampling}
The ridge regression estimator was shown to coincide with the posterior mean of the regression parameter if it is endowed with a normal prior. This particular choice of the prior yielded a closed form expression of the posterior, and its mean. Prior distributions that result in well-characterized posterior ones are referred to as \textit{conjugate}. E.g., for the standard linear regression model a normal prior is conjugate. Conjugate priors, however, are the exception rather than the rule. In general, an arbitrary prior distribution will not result in a posterior distribution that is familiar (amongst others due to the fact that its normalizing constant is not analytically evaluable). For instance, would a wildly different prior for the regression parameter be chosen, its posterior is unlikely to be analytically known. Then, although analytically unknown, such posteriors can be investigated \textit{in silico} by means of the Markov chain Monte Carlo (MCMC) method. In this section the MCMC method is explained and illustrated -- despite its conjugancy -- on the linear regression model with a normal prior.

Markov chain Monte Carlo is a general purpose technique for sampling from complex probabilistic models/distributions. It draws a sample from a Markov chain that has been constructed such that its stationary distribution equals the desired distribution (read: the analytically untractable posterior distribution). The chain is, after an initial number of iterations (called the \textit{burn-in} period), believed to have converged and reached stationarity. A sample from the stationary chain is then representative of the desired distribution. Statistics derived from this sample can be used to characterize the desired distribution. Hence, estimation is reduced to setting up and running a Markov process on a computer. For more on the MCMC, see \cite{chib1995understanding} and \cite{geyer2011introduction} and the references therein. 

Recall Monte Carlo integration. Assume the analytical evaluation of $\mathbb{E}[h(\mathbf{Y})] = \int h(\mathbf{y}) \pi(\mathbf{y}) d \mathbf{y}$ is impossible. This quantity can nonetheless be estimated when samples from $\pi(\cdot)$ can be drawn. For, if $\mathbf{Y}^{(1)}, \ldots, \mathbf{Y}^{(T)} \sim \pi(\mathbf{y})$, the integral may be estimated by: $\widehat{\mathbb{E}}[h(\mathbf{Y})] = T^{-1} \sum_{t=1}^T h(\mathbf{Y}^{(t)})$. By the law of large numbers this is a consistent estimator, i.e. if $T \rightarrow \infty$ the estimator converges (in probability) to its true value. Thus, irrespective of the specifics of the posterior distribution $\pi(\cdot)$, if one is able to sample from it, its moments (or other quantities) can be estimated. Or, by the same principle it may be used in the Bayesian regression framework of Section \ref{sect.Bayes2ridge} to sample from the marginal posterior of $\bbeta$,
\begin{eqnarray*}
f_{\bbeta}( \bbeta \, | \, \mathbf{Y}, \mathbf{X}) & = & \int_{0}^{\infty}
f_{\bbeta, \sigma^2}( \bbeta, \sigma^2 \, | \, \mathbf{Y}, \mathbf{X}) \, d \sigma^2 \, \, \, = \, \, \, \int_{0}^{\infty} \, g_{\bbeta}( \bbeta \, | \, \sigma^2,  \mathbf{Y}, \mathbf{X}) \, g_{\sigma^2}(\sigma^2 \, | \, \mathbf{Y}, \mathbf{X}) \, d \sigma^2,
\end{eqnarray*}
given that it of $\sigma^2$ is known. By drawing a sample $\sigma^{2,(1)}, \ldots, \sigma^{2,(T)}$ of $g_{\sigma^2}(\sigma^2 \, | \, \mathbf{Y}, \mathbf{X}) $ and, subsequently, a sample $\{ \bbeta^{(t)} \}_{t=1}^T$ from $g_{\bbeta}( \bbeta \, | \, \sigma^{2,(t)},  \mathbf{Y}, \mathbf{X})$ for $t=1, \ldots, T$, the parameter $\sigma^2$ has effectively been integrated out and the resulting sample of $\bbeta^{(1)}, \bbeta^{(2)}, \ldots, \bbeta^{(T)}$ is from  $f_{\bbeta}( \bbeta \, | \, \mathbf{Y}, \mathbf{X})$.

It is clear from the above that the ability to sample from the posterior distribution is key to the estimation of the parameters. But this sampling is hampered by knowledge of the normalizing constant of the posterior. This is overcome by the \textit{acceptance-rejection sampler}. This sampler generates independent draws from a target density $\pi(\mathbf{y}) = f(\mathbf{y}) / K$, where $f(\mathbf{y})$ is an unnormalized density and $K$ its (unknown) normalizing constant. The sampler relies on the existence of a density $h(\mathbf{y})$ that dominates $f(\mathbf{y})$, i.e., $f(\mathbf{y}) \leq c \, h(\mathbf{y})$ for some known $c \in \mathbb{R}_{>0}$, from which it is possible to simulate. Then, in order to draw a $\mathbf{Y}$ from the posterior $\pi(\cdot)$ run Algorithm \ref{alg.accrejSampler}:
\\
\\
\vspace{0.25cm}
\begin{minipage}[h]{\textwidth}
\begin{center}
\fbox{\parbox{14cm}{
\begin{algorithm}[H]
\SetKwInOut{Input}{input}
\SetKwInOut{Output}{output}
\RestyleAlgo{boxed}
\LinesNumbered
\SetKwFor{inpar}{for}{do in parallel}{\text{ end synchronize}}
\Input{densities $f(\cdot)$ and $h(\cdot)$;
\\
constant $c >0$;
}
\Output{a draw from $\pi(\cdot)$.}
\vspace{0.25cm}
{Generate $\mathbf{Y}$ from $h(\cdot)$}.
\\
{Generate $U$ from the uniform distribution $\mathcal{U}[0,1]$}.
\\
\uIf{ $U \leq f(\mathbf{Y}) \, / \,[ c \times h(\mathbf{Y})]$  }{Return $\mathbf{Y}$.}\Else{Go back to line 1.}
\vspace{0.25cm}
\caption{{\small Acceptance-rejection sampling.}} \label{alg.accrejSampler}
\end{algorithm}
}
}
\end{center}
\end{minipage}
A proof that this indeed yields a random sample from $\pi(\cdot)$ is given in \cite{Flury1990}. For this method to be computationally efficient, $c$ is best set equal to $\sup_{\mathbf{y}} \{ f(\mathbf{y})/h(\mathbf{y}) \}$. The \texttt{R}-script below illustrates this sampler for the unnormalized density $f(y) = \cos^2(2 \pi y)$ on the unit interval. As $\max_{y \in [0,1]} \cos^2(2 \pi y) = 1$, the density $f(y)$ is dominated by the density function of the uniform distribution $\mathcal{U}[0,1]$ and, hence, the script uses $h(y) = 1$ for all $y \in [0, 1]$.
\begin{lstlisting}
# define unnormalized target density
cos2density <- function(x){ (cos(2*pi*x))^2 }

# define acceptance-rejection sampler
acSampler <- function(n){
	draw <- numeric()
	for(i in 1:n){
		accept <- FALSE
		while (!accept){
			Y <- runif(1)
			if (runif(1) <= cos2density(Y)){
				accept <- TRUE
				draw   <- c(draw, Y)
			}
		}
	}
	return(draw)
}

# verify by eye-balling
hist(acSampler(100000), n=100)

\end{lstlisting}
In practice, it may not be possible to find a density $h(x)$ that dominates the unnormalized target density $f(\mathbf{y})$ over the whole domain. Or, the constant $c$ is too large, yielding a rather small acceptance probability, which would make the acceptance-rejection sampler impractically slow. A different sampler is then required.

The Metropolis-Hastings sampler overcomes the problems of the acceptance-rejection sampler and generates a sample from a target distribution that is known up to its normalizing constant. Hereto it constructs a Markov chain that converges to the target distribution. A Markov chain is a sequence of random variables $\{ \mathbf{Y}_t \}_{t=1}^{\infty}$ with $\mathbf{Y}_t \in \mathbb{R}^p$ for all $t$ that exihibit a simple dependence relationship among subsequent random variables in the sequence. Here that simple relationship refers to the fact/assumption that the distribution of $\mathbf{Y}_{t+1}$ only depends on $\mathbf{Y}_t$ and not on the part of the sequence preceeding it. The Markov chain's random walk is usually initiated by a single draw from some distribution, resulting in $\mathbf{Y}_1$. From then on the evolution of the Markov chain is described by the \textit{transition kernel}. The transition kernel is a the conditional density $g_{\mathbf{Y}_{t+1} \, | \, \mathbf{Y}_t} (\mathbf{Y}_{t+1} = \mathbf{y}_a \, | \, \mathbf{Y}_t =\mathbf{y}_b)$ that specifies the distribution of the random variable at the next instance given the realization of the current one. Under some conditions (akin to aperiodicity and irreducibility for Markov chains in discrete time and with a discrete state space), the influence of the initiation washes out and the random walk converges. Not to a specific value, but to a distribution. This is called the stationary distribution, denoted by $\varphi (\mathbf{y})$, and $\mathbf{Y}_t \sim \varphi (\mathbf{y})$ for large enough $t$.  The stationary distribution satisfies:
\begin{eqnarray} \label{form.stationary.dist.markov.chain}
\varphi (\mathbf{y}_a) & = & \int_{\mathbb{R}^p} g_{\mathbf{Y}_{t+1} \, | \, \mathbf{Y}_t} (\mathbf{Y}_{t+1} = \mathbf{y}_a \, | \, \mathbf{Y}_t =\mathbf{y}_b) \, \varphi(\mathbf{y}_b) \, d\mathbf{y}_b.
\end{eqnarray}
That is, the distribution of $\mathbf{Y}_{t+1}$, obtained by marginalization over $\mathbf{Y}_{t}$, coincides with that of $\mathbf{Y}_{t}$. Put differently, the mixing by the transition kernel does not affect the distribution of individual random variables of the chain. To verify that a particular distribution $\varphi(\mathbf{y})$ is the stationary one, it is sufficient to verify that it satisfies the detailed balance equations:
\begin{eqnarray*}
g_{\mathbf{Y}_{t+1} \, | \, \mathbf{Y}_t} (\mathbf{y}_b \, | \, \mathbf{y}_a) \, \varphi(\mathbf{y}_a)  & = & g_{\mathbf{Y}_{t+1} \, | \, \mathbf{Y}_t} (\mathbf{y}_a \, | \, \mathbf{y}_b) \, \varphi(\mathbf{y}_b),
\end{eqnarray*}
for all choices of $\mathbf{y}_a, \mathbf{y}_b \in \mathbb{R}^p$. If a Markov chain satisfies this detailed balance equations, it is said to be \textit{reversible}. Reversibility means so much as that, from the realizations, the direction of the chain cannot be discerned as: $f_{\mathbf{Y}_t,\mathbf{Y}_{t+1}} (\mathbf{y}_a, \mathbf{y}_b ) = f_{\mathbf{Y}_t,\mathbf{Y}_{t+1}} (\mathbf{y}_b, \mathbf{y}_a )$, that is, the probability of starting in state $\mathbf{y}_a$ and finishing in state $\mathbf{y}_b$ equals that of starting in $\mathbf{y}_b$ and finishing in $\mathbf{y}_a$. The sufficiency of this condition is evident after its integration on both sides with respect to $\mathbf{y}_b$, from which condition (\ref{form.stationary.dist.markov.chain}) follows.

MCMC assumes the stationary distribution of the Markov chain to be known up to a scalar -- this is the target density from which is to be sampled -- but the transition kernel is unknown. This poses a problem as the transition kernel is required to produce a sample from the target density. An arbitrary kernel is unlikely to satisfy the detailed balance equations with the desired distribution as its stationary one. If, indeed, it does not, then:
\begin{eqnarray*}
g_{\mathbf{Y}_{t+1} \, | \, \mathbf{Y}_t} (\mathbf{y}_b \, | \, \mathbf{y}_a) \, \varphi(\mathbf{y}_a)  & > & g_{\mathbf{Y}_{t+1} \, | \, \mathbf{Y}_t} (\mathbf{y}_a \, | \, \mathbf{y}_b) \, \varphi(\mathbf{y}_b),
\end{eqnarray*}
for some $\mathbf{y}_a$ and $\mathbf{y}_b$. This may (loosely) be interpreted as that the process moves from $\mathbf{y}_a$ to $\mathbf{y}_b$ too often and from $\mathbf{y}_b$ to $\mathbf{y}_a$ too rarely. To correct this, the probability of moving from $\mathbf{y}_b$ to $\mathbf{y}_a$ is introduced to reduce the number of moves from $\mathbf{y}_a$ to $\mathbf{y}_b$. As a consequence not always a new value is generated, the algorithm may decide to stay in $\mathbf{y}_a$ (as opposed to acceptance-rejection sampling). To this end a kernel is constructed and comprises compound transitions that propose a possible next state which is simultaneously judged to be acceptable or not. Hereto take an arbitrary \textit{candidate-generating} density function $g_{\mathbf{Y}_{t+1} \, | \, \mathbf{Y}_t} (\mathbf{y}_{t+1} \, | \, \mathbf{y}_{t})$. Given the current state $\mathbf{Y}_{t} = \mathbf{y}_{t}$, this kernel produces a suggestion $\mathbf{y}_{t+1}$ for the next point in the Markov chain. This suggestion may take the random walk too far afield from the desired and known stationary distribution. If so, the point is to be rejected in favour of the current state, until a new suggestion is produced that is satisfactory close to or representative of the stationary distribution. Let the probability of an acceptable suggestion $\mathbf{y}_{t+1}$, given the current state $\mathbf{Y}_t = \mathbf{y}_t$, be denoted by $\mathcal{A}(\mathbf{y}_{t+1} \, | \,  \mathbf{y}_t)$. The thus constructed transition kernel then formalizes to:
\begin{eqnarray*}
h_{\mathbf{Y}_{t+1} \, | \, \mathbf{Y}_{t}}(\mathbf{y}_{a} \, | \, \mathbf{y}_b) & = & g_{\mathbf{Y}_{t+1} \, | \, \mathbf{Y}_t} (\mathbf{y}_a \, | \, \mathbf{y}_b) \, \mathcal{A} (\mathbf{y}_{a} \, | \,  \mathbf{y_b}) + r(\mathbf{y}_a) \, \delta_{\mathbf{y_b}} ( \mathbf{y}_a ) ,
\end{eqnarray*}
where $\mathcal{A} (\mathbf{y}_{a} \, | \,  \mathbf{y_b})$ is the acceptance probability of the suggestion $\mathbf{y}_a$ given the current state $\mathbf{y}_b$ and is defined as:
\begin{eqnarray*}
\mathcal{A} & = & \min \Big\{ 1, \frac{\varphi(\mathbf{y}_{a})}{\varphi(\mathbf{y}_{b})} \frac{g_{\mathbf{Y}_{t+1} \, | \, \mathbf{Y}_t} (\mathbf{y}_{b} \, | \, \mathbf{y}_a)}{g_{\mathbf{Y}_{t+1} \, | \, \mathbf{Y}_t} (\mathbf{y}_a \, | \, \mathbf{y}_b)} \Big\}.
\end{eqnarray*}
Furthermore, $\delta_{\mathbf{y_b}} (\cdot)$ is the Dirac delta function and $r(\mathbf{y}_a) = 1 - \int_{\mathbb{R}^p} g_{\mathbf{Y}_{t+1} \, | \, \mathbf{Y}_t} (\mathbf{y}_a \, | \, \mathbf{y}_b)  \mathcal{A} (\mathbf{y}_{a} \, | \,  \mathbf{y_b}) \, d\mathbf{y}_a$, which is associated with the rejection. The transition kernel $h_{\mathbf{Y}_{t+1} \, | \, \mathbf{Y}_{t}}(\cdot \, | \, \cdot)$ equipped with the above specified acceptance probability is referred to as the Metropolis-Hastings kernel. Note that, although the normalizing constant of the stationary distribution may be unknown, the acceptance probability can be evaluated as this constant cancels out in the  $\varphi(\mathbf{y}_{a}) \, [\varphi(\mathbf{y}_{b})]^{-1}$ term.

It rests now to verify that $\varphi(\cdot)$ is the stationary distribution of the thus defined Markov process. Hereto we first assess the reversibility of the proposed kernel. The contribution of the second summand of the kernel to the detailed balance equations, $r(\mathbf{y}_a) \, \delta_{\mathbf{y_b}} ( \mathbf{y}_a ) \varphi ( \mathbf{y}_b)$, exists only  if $\mathbf{y}_a = \mathbf{y}_b$. Thus, if it contributes to the detailed balance equations, it does so equally on both sides of the equation. It is now left to assess whether:
\begin{eqnarray*}
g_{\mathbf{Y}_{t+1} \, | \, \mathbf{Y}_t} (\mathbf{y}_a \, | \, \mathbf{y}_b) \, \mathcal{A} (\mathbf{y}_{a} \, | \,  \mathbf{y_b}) \varphi(\mathbf{y}_b) & = & g_{\mathbf{Y}_{t+1} \, | \, \mathbf{Y}_t} (\mathbf{y}_b \, | \, \mathbf{y}_a) \, \mathcal{A} (\mathbf{y}_{b} \, | \,  \mathbf{y_a}) \varphi(\mathbf{y}_a).
\end{eqnarray*}
To verify this equality use the definition of the acceptance probability from which we note that if $\mathcal{A} (\mathbf{y}_{a} \, | \,  \mathbf{y_b}) \leq 1$ (and thus $\mathcal{A} (\mathbf{y}_{a} \, | \,  \mathbf{y_b}) = [ \varphi(\mathbf{y}_{a})   g_{\mathbf{Y}_{t+1} \, | \, \mathbf{Y}_t} (\mathbf{y}_{b} \, | \, \mathbf{y}_a) ] [ \varphi(\mathbf{y}_{b}) g_{\mathbf{Y}_{t+1} \, | \, \mathbf{Y}_t} (\mathbf{y}_a \, | \, \mathbf{y}_b) ]^{-1}$), then $\mathcal{A} (\mathbf{y}_{b} \, | \,  \mathbf{y}_a) = 1$ (and vice versa). In either case, the equality in the preceeding display holds, and thereby, together with the observation regarding the second summand, so do the detailed balance equations for Metropolis-Hastings kernel. Then, $\varphi(\cdot)$ is indeed the stationary distribution of the constructed transition kernel $h_{\mathbf{Y}_{t+1} \, | \, \mathbf{Y}_t} (\mathbf{y}_a \, | \,  \mathbf{y}_b)$ as can be verified from Condition (\ref{form.stationary.dist.markov.chain}):
\begin{eqnarray*}
& & \hspace{-1.5cm}  \int_{\mathbb{R}^p} h_{\mathbf{Y}_{t+1} \, | \, \mathbf{Y}_t} (\mathbf{y}_a \, | \,  \mathbf{y}_b) \, \varphi(\mathbf{y}_b) \, d\mathbf{y}_b \, \, \, = \, \, \, \int_{\mathbb{R}^p} h_{\mathbf{Y}_{t+1} \, | \, \mathbf{Y}_t} (\mathbf{y}_b \, | \,  \mathbf{y}_a) \, \varphi(\mathbf{y}_a) \, d\mathbf{y}_b
\\
& = & \int_{\mathbb{R}^p} g_{\mathbf{Y}_{t+1} \, | \, \mathbf{Y}_t} (\mathbf{y}_b \, | \,  \mathbf{y}_a) \, \mathcal{A} (\mathbf{y}_b \, | \,  \mathbf{y}_a) \, \varphi(\mathbf{y}_a) \, d \mathbf{y}_b  +  \int_{\mathbb{R}^p}  r(\mathbf{y}_b) \delta_{\mathbf{y}_a} (\mathbf{y}_b) \, \varphi(\mathbf{y}_a) \, d \mathbf{y}_b
\\
& = &  \varphi(\mathbf{y}_a) \int_{\mathbb{R}^p} g_{\mathbf{Y}_{t+1} \, | \, \mathbf{Y}_t} (\mathbf{y}_b \, | \, \mathbf{y}_a) \,  \mathcal{A} (\mathbf{y}_b \, | \,  \mathbf{y}_a) \, d \mathbf{y}_b  + \varphi(\mathbf{y}_a) \int_{\mathbb{R}^p}  r(\mathbf{y}_b) \delta_{\mathbf{y}_a} (\mathbf{y}_b) \,  d \mathbf{y}_b
\\
& = &  \varphi(\mathbf{y}_a) [ 1- r(\mathbf{y}_a)]  + r(\mathbf{y}_a)  \, \varphi(\mathbf{y}_a) \, \, \,  = \, \, \,  \varphi(\mathbf{y}_a),
\end{eqnarray*}
in which the reversibility of $h_{\mathbf{Y}_{t+1} \, | \, \mathbf{Y}_t} (\cdot, \cdot)$, the definition of $r(\cdot)$, and the properties of the Dirac delta function have been used.
\\
\\
The Metropolis-Hastings algorithm is now described by:
\\
\\
\vspace{0.25cm}
\begin{minipage}[h]{\textwidth}
\begin{center}
\fbox{\parbox{14cm}{
\begin{algorithm}[H]
\SetKwInOut{Input}{input}
\SetKwInOut{Output}{output}
\RestyleAlgo{boxed}
\LinesNumbered
\SetKwFor{inpar}{for}{do in parallel}{\text{ end synchronize}}
\Input{densities $f(\cdot)$ and $h(\cdot)$;
\\
constant $c >0$;
}
\Output{A draw from $\pi(\cdot)$.}
\vspace{0.25cm}
{Choose starting value $x^{(0)}$.}.
\\
{Generate $y$ from $q(x^{(j)}, \cdot)$ and $U$ from $\mathcal{U}[0,1]$}.
\\
\uIf{ $U \leq \alpha(X^{(j)}, y)$  }{set $x^{(j+1)} = y$.}\Else{set $x^{(j+1)} = x^{(j)}$.}
Return the values $\{ x^{(1)}, x^{(2)}, x^{(3)}, \ldots, \}$
\vspace{0.25cm}
\caption{{\small Metropolis-Hastings algorithm.}} \label{alg.MetropHastings}
\end{algorithm}
}
}
\end{center}
\end{minipage}
The convergence rate of this sequence is subject of on-going research.
\\
\\
The following \texttt{R}-script shows how the Metropolis-Hastings sampler can be used to from a mixture of two normals $f(y) = \theta \phi_{\mu=2, \sigma^2=0.5}(y) + (1-\theta) \phi_{\mu=-2, \sigma^2=2}(y)$ with $\theta = {1}{4}$. The sampler assumes this distribution is known up to its normalizing constant and uses its unnormalized density $\exp[-(y-2)^2] + 3\exp[-(y+2)^2/4]$. The sampler employs the Cauchy distribution centered at the current state as transition kernel from a candidate for the next state is drawn. The chain is initiated arbitrarily with $Y^{(1)} = 1$.
\begin{lstlisting}
# define unnormalized mixture of two normals 
mixDensity <- function(x){
	# unnormalized mixture of two normals 
	exp(-(x-2)^2) + 3*exp(-(x+2)^2/4)
}

# define the MCMC sampler
mcmcSampler <- function(n, initDraw){
	draw <- initDraw
	for (i in 1:(n-1)){
		# sample a candidate
		candidate <- rcauchy(1, draw[i])

		# calculate acceptance probability:
		alpha <- min(1, mixDensity(candidate) / mixDensity(draw[i]) * 
		                dcauchy(draw[i], candidate) / dcauchy(candidate, draw[i]))

		# accept/reject candidate
		if (runif(1) < alpha){ 
			draw <- c(draw, candidate)
		} else {
			draw <- c(draw, draw[i])
		}
	}
	return(draw)
}

# verify by eye-balling
Y <- mcmcSampler(100000, 1)
hist(Y[-c(1:1000)], n=100)






\end{lstlisting} \label{Rcode.MCMCmixtureOfnormals}
The histogram (not shown) shows that the sampler, although using a unimodal kernel, yields a sample from the bimodal mixture distribution.
\\
\\
The \textit{Gibbs sampler} is a particular version of the MCMC algorithm. The Gibbs sampler enhances convergence to the stationary distribution (i.e. the posterior distribution) of the Markov chain. It requires, however, the full conditionals of all random variables (here: model parameters), i.e. the conditional distributions of one random variable given all others, to be known analytically. Let the random vector $\mathbf{Y}$ for simplicity -- more refined partitions possible -- be partioned as $(\mathbf{Y}_{a}, \mathbf{Y}_{b})$ with subscripts $a$ and $b$ now refering to index sets that partition the random vector $\mathbf{Y}$ (instead of the previously employed meaning of referring to two -- possibly different -- elements from the state space). The Gibbs sampler thus requires that both $f_{\mathbf{Y}_{a} \, | \, \mathbf{Y}_b} (\mathbf{y}_{a}, \mathbf{y}_{b})$ and $f_{\mathbf{Y}_b \, | \, \mathbf{Y}_a} (\mathbf{y}_a, \mathbf{y}_b)$ are known. Being a specific form of the MCMC algorithm the Gibbs sampler seeks to draw $\mathbf{Y}_{t+1} = (\mathbf{Y}_{a,t+1}, \mathbf{Y}_{b,t+1})$ given the current state $\mathbf{Y}_{t} = (\mathbf{Y}_{a,t}, \mathbf{Y}_{b,t})$. It draws, however, only a new instance for a single element of the partition (e.g. $\mathbf{Y}_{a,t+1}$) keeping the remainder of the partition (e.g. $\mathbf{Y}_{b,t}$) temporarily fixed. Hereto define the transition kernel:
\begin{eqnarray*}
g_{\mathbf{Y}_{t+1} \, | \, \mathbf{Y}_{t} } (\mathbf{Y}_{t+1} = \mathbf{y}_{t+1}, \mathbf{Y}_{t} = \mathbf{y}_{t}) & = & 
\left\{ \begin{array}{ll} 
f_{\mathbf{Y}_{a,t+1} \, | \, \mathbf{Y}_{b,t+1}} (\mathbf{y}_{a,t+1}, \mathbf{y}_{b,t+1}) & \mbox{if }  \mathbf{y}_{b,t+1} = \mathbf{y}_{b,t},
\\
0 & \mbox{otherwise.}
\end{array}  
\right.
\end{eqnarray*}
Using the definition of the conditional density, the acceptance probability for the $t+1$-th proposal of the subvector $\mathbf{Y}_a$ then is:
\begin{eqnarray*}
\mathcal{A}(\mathbf{y}_{t+1} \, | \, \mathbf{y_t}) & = & \min \Big\{ 1, \frac{\varphi(\mathbf{y}_{t+1})}{\varphi(\mathbf{y}_{t})} \frac{g_{\mathbf{Y}_{t+1} \, | \, \mathbf{Y}_t} (\mathbf{y}_{t}, \mathbf{y}_{t+1})}{g_{\mathbf{Y}_{t+1} \, | \, \mathbf{Y}_t} (\mathbf{y}_{t+1}, \mathbf{y}_t)} \Big\}
\\
& = &  \min \Big\{ 1, \frac{\varphi(\mathbf{y}_{t+1})}{\varphi(\mathbf{y}_{t})} \frac{f_{\mathbf{Y}_{a,t+1} \, | \, \mathbf{Y}_{b,t+1}} (\mathbf{y}_{a,t}, \mathbf{y}_{b,t}) }{f_{\mathbf{Y}_{a,t+1} \, | \, \mathbf{Y}_{b,t+1}} (\mathbf{y}_{a,t+1}, \mathbf{y}_{b,t+1}) } \Big\}
\\
& = & \min \Big\{ 1, \frac{\varphi(\mathbf{y}_{t+1})}{\varphi(\mathbf{y}_{t})} \frac{f_{\mathbf{Y}_{a,t+1}, \mathbf{Y}_{b,t+1}} (\mathbf{y}_{a,t}, \mathbf{y}_{b,t}) /  f_{\mathbf{Y}_{b,t+1}} (\mathbf{y}_{b,t}) }{f_{\mathbf{Y}_{a,t+1}, \mathbf{Y}_{b,t+1}} (\mathbf{y}_{a,t+1}, \mathbf{y}_{b,t+1}) / f_{\mathbf{Y}_{b,t+1}} (\mathbf{y}_{b,t+1}) } \Big\} 
\\
& = &  \min \Big\{ 1,  \frac{f_{\mathbf{Y}_{b,t+1}} (\mathbf{y}_{b,t+1}) }{f_{\mathbf{Y}_{b,t+1}} (\mathbf{y}_{b,t}) } \Big\}  \, \, \, = \, \, \, 1,
\end{eqnarray*}
where the fourth equality uses the reversibility of the kernel. The acceptance probability of each proposal is thus one (which contributes to the enhanced convergence of the Gibbs sampler to the joint posterior). Having drawn an acceptable proposal for this element of the partition, the Gibbs sampler then draws a new instance for the next element of the partition, i.e. now $\mathbf{Y}_b$, keeping $\mathbf{Y}_a$ fixed. This process of subsequently sampling each partition element is repeated until enough samples have been drawn.

To illustrate the Gibbs sampler revisit Bayesian regression. In Section \ref{sect.Bayes2ridge} the full conditional distributions of $\bbeta$ and $\sigma^2$ were derived. The Gibbs sampler now draws in alternating fashion from these conditional distributions (see Algorithm
\ref{alg.gibbsSamplerRidgeRegression} for its pseudo-code).
\\
\mbox{ }
\\
\begin{minipage}[h]{\textwidth}
\begin{center}
\fbox{\parbox{14cm}{
\begin{algorithm}[H]
\SetKwInOut{Input}{input}
\SetKwInOut{Output}{output}
\RestyleAlgo{boxed}
\LinesNumbered
\SetKwFor{inpar}{for}{do in parallel}{\text{ end synchronize}}
\Input{
sample size $T$
\\
length of burn in period $T_{\mbox{{\tiny burn-in}}}$
\\
thinning factor $f_{{\mbox{{\tiny thinning}}}}$
\\
data $\{\mathbf{x}_i, \mathbf{y}_i \}_{i=1}^n$;
\\
conditional distributions $f_{\sigma^2 \, | \, \bbeta} (\sigma^2, \bbeta; \{\mathbf{x}_i, \mathbf{y}_i \}_{i=1}^n)$ and $f_{\beta \, | \, \sigma^2} (\bbeta, \sigma^2; \{\mathbf{x}_i, \mathbf{y}_i \}_{i=1}^n)$.
}
\Output{draws from the joint posterior $\pi(\bbeta, \sigma^2 \, | \, \{\mathbf{x}_i, \mathbf{y}_i \}_{i=1}^n)$.}
\vspace{0.25cm}
{\bf initialize} {$(\bbeta_0, \sigma^2_{0})$}.
\\
\vspace{0.25cm}
\For{ $t = 1$ to $T_{\mbox{{\tiny burn-in}}} + T f_{{\mbox{{\tiny thinning}}}}$  }{
\BlankLine
draw $\bbeta_{t}$ from conditional distribution $f_{\bbeta \, | \, \sigma^2} (\bbeta, \sigma_{t-1}^2; \{\mathbf{x}_i, \mathbf{y}_i \}_{i=1}^n)$,
\\
draw $\sigma_{t}^2$ from conditional distribution $f_{\sigma^2 \, | \, \bbeta} (\sigma^2, \bbeta_t; \, \{\mathbf{x}_i, \mathbf{y}_i \}_{i=1}^n)$.
\\
\BlankLine
}
\BlankLine
Remove the first $T_{\mbox{{\tiny burn-in}}}$ draws (representing the burn-in phase).
\\
Select every $f_{\mbox{{\tiny thinning}}}$-th sample (thinning).
\vspace{0.25cm}
\caption{{\small Pseudocode of the Gibbs sampler of the joint posterior of the Bayesian regression parameters.}} \label{alg.gibbsSamplerRidgeRegression}
\end{algorithm}
}
}
\end{center}
\end{minipage}

\section{Empirical Bayes} \label{sect.empBayes}
Empirical Bayes (EB) is a branch of Bayesian statistics that meets the subjectivity criticism of frequentists. Instead of fully specifying the prior distribution empirical Bayesians identify only its form. The hyper parameters of this prior distribution are left unspecified and are to be found empirically. In practice, these hyper parameters are estimated from the data at hand. However, the thus estimated hyper parameters are used to obtain the Bayes estimator of the model parameters. As such the data are then used multiple times. This is usually considered an inappropriate practice but is deemed acceptable when the number of model parameters is large in comparison to the number of hyper parameters. Then, the data are not used twice but `$1+\varepsilon$'-times (i.e. once-and-a-little-bit) and only little information from the data is spent on the estimation of the hyper parameters. Having obtained an estimate of the hyper parameters, the prior is fully known, and the posterior distribution (and summary statistics thereof) are readily obtained by Bayes' formula.

The most commonly used procedure for the estimation of the hyper parameters is marginal likelihood maximization, which is a maximum likelihood-type procedure. But the likelihood cannot directly be maximized with respect to the hyper parameters as it contains the model parameters that are assumed to be random within the Bayesian framework. This may be circumvented by choosing a specific value for the model parameter but that would render the resulting hyper parameter estimates dependent on this choice. Instead of maximization with the model parameter set to a particular value, one would preferably maximize over all possible realizations. The latter is achieved by marginalization with respect to the random model parameter, in which the (prior) distribution of the model parameter is taken into account. This amounts to integrating out the model parameter from the posterior distribution, i.e. $\int P(\mathbf{Y} = \mathbf{y} \, | \, \theta) \, \pi (\theta) \, d \theta$, resulting in the so-called marginal posterior. After marginalization the specifics of the model parameter have been discarded and the marginal posterior is a function of the observed data and the hyper parameters. The estimator of the hyper parameter is now defined as the maximizer of this marginal posterior.

To illustrate the estimation of hyper parameters of the Bayesian linear regression model through marginal likelihood maximization assume the regression parameter $\bbeta$ and and the error variance $\sigma^2$ to be endowed with conjugate priors: $\bbeta \, | \, \sigma^2 \sim \mathcal{N}(\mathbf{0}_p,  \sigma^2 \lambda^{-1} \mathbf{I}_{pp})$ and $\sigma^2 \sim \mathcal{IG}(a_0, b_0)$. Three hyper parameters are thus to be estimated: the shape and scale parameters, $\alpha_0$ and $\beta_0$, of the inverse gamma distribution and the $\lambda$ parameter related to the variance of the regression coefficients. Straightforward application of the outlined marginal likelihood principle does, however, not work here. The joint prior, $\pi(\bbeta \, | \, \sigma^2) \pi (\sigma^2)$, is too flexible and does not yield sensible estimates of the hyper parameters. As interest is primarily in $\lambda$, this is resolved by setting the hyper parameters of $\pi (\sigma^2)$ such that the resulting prior is uninformative, i.e. as objectively as possible. This is operationalized as a very flat distribution. Then, with the particular choices of $a_0$ and $b_0$ that produce an uninformative prior for $\sigma^2$, the empirical Bayes estimate of $\lambda$ is:
\begin{eqnarray*}
\hat{\lambda}_{eb} & = & \arg \max_{\lambda} \int_{0}^{\infty} \int_{\mathbb{R}^p}  f_{\bbeta, \sigma^2} (\bbeta, \sigma^2 \, | \, \mathbf{Y}, \mathbf{X}) \, d \bbeta \, d \sigma^2
\\
& = & \arg \max_{\lambda} \int_{0}^{\infty} \int_{\mathbb{R}^p}  \sigma^{-n} \exp [ - \tfrac{1}{2}\sigma^{-2} ( \mathbf{Y} - \mathbf{X} \bbeta)^{\top} ( \mathbf{Y} - \mathbf{X} \bbeta) ]
\\
&  & \qquad \qquad \qquad \qquad \times \, \, \sigma^{-p} \exp ( - \tfrac{1}{2}\sigma^{-2} \lambda \bbeta^{\top} \bbeta ) \, \times \, \, (\sigma^2)^{-a_0 - 1} \exp ( - b_0 \sigma^{-2} )  \, d \bbeta \, d \sigma^2
\\
& = & \arg \max_{\lambda} \int_{0}^{\infty} \int_{\mathbb{R}^p} \sigma^{-n} \exp \{ - \tfrac{1}{2} \sigma^{-2} [ \mathbf{Y}^{\top} \mathbf{Y} - \mathbf{Y}^{\top} \mathbf{X} (\mathbf{X}^{\top} \mathbf{X} + \lambda \mathbf{I}_{pp})^{-1} \mathbf{X}^{\top} \mathbf{Y} \}
\\
& & \qquad \qquad \qquad \qquad \times \, \, \sigma^{-p} \exp \{ - \tfrac{1}{2} \sigma^{-2}  \, [ \bbeta - \hat{\bbeta}(\lambda) ]^{\top} (\mathbf{X}^{\top} \mathbf{X} + \lambda \mathbf{I}_{pp}) [ \bbeta - \hat{\bbeta}(\lambda) ] \}
\\
&  & \qquad \qquad \qquad \qquad \times \, \, (\sigma^2)^{-a_0-1} \exp ( - b_0 \sigma^{-2} )  \, d \bbeta \, d \sigma^2
\\
& = & \arg \max_{\lambda} \int_{0}^{\infty} \sigma^{-n} \exp \{ - \tfrac{1}{2} \sigma^{-2} [ \mathbf{Y}^{\top} \mathbf{Y} - \mathbf{Y}^{\top} \mathbf{X} (\mathbf{X}^{\top} \mathbf{X} + \lambda \mathbf{I}_{pp})^{-1} \mathbf{X}^{\top} \mathbf{Y} \}
\\
&  & \qquad \qquad \qquad \qquad \times \, \, | \mathbf{X}^{\top} \mathbf{X} + \lambda \mathbf{I}_{pp}|^{-1/2}  \, (\sigma^2)^{-a_0-1} \exp ( - b_0\sigma^{-2} ) \, d \sigma^2
\\
& = & \arg \max_{\lambda} \, | \mathbf{X}^{\top} \mathbf{X} + \lambda \mathbf{I}_{pp}|^{-1/2}  \int_{0}^{\infty} \exp \big( - \sigma^{-2} \{ b_0 +  \tfrac{1}{2} [ \mathbf{Y}^{\top} \mathbf{Y} - \mathbf{Y}^{\top} \mathbf{X} (\mathbf{X}^{\top} \mathbf{X} + \lambda \mathbf{I}_{pp})^{-1} \mathbf{X}^{\top} \mathbf{Y} ] \} \big)
\\
&  & \qquad \qquad \qquad \qquad \qquad \qquad \times \, \,  \, (\sigma^2)^{-a_0-n/2 - 1} \, d \sigma^2
\\
& = & \arg \max_{\lambda}  \, |\mathbf{X}^{\top} \mathbf{X} + \lambda \mathbf{I}_{pp}|^{-1/2} \,  \{ b_0 + \tfrac{1}{2} [ \mathbf{Y}^{\top} \mathbf{Y} -  \mathbf{Y}^{\top} \mathbf{X} (\mathbf{X}^{\top} \mathbf{X} + \lambda \mathbf{I}_{pp} )^{-1}  \mathbf{X}^{\top} \mathbf{Y} ]\}^{-a_0 - n/2}, 
\end{eqnarray*}
where the factors not involving $\lambda$ have been dropped throughout. Prior to the maximization of the marginal likelihood the logarithm is taken. That changes the maximum, but not its location, and yields an expression that is simpler to maximize. With the empirical Bayes estimate $\hat{\lambda}_{eb}$ at hand, the Bayesian estimate of the regression parameter $\bbeta$ is $\hat{\bbeta}_{eb} = (\mathbf{X}^{\top} \mathbf{X} + \hat{\lambda}_{eb} \mathbf{I}_{pp})^{-1} \mathbf{X}^{\top} \mathbf{Y}$. Finally, the particular choice of the hyper parameters of the prior on $\sigma^2$ is not too relevant. Most values of $a_0$ and $b_0$ that correspond to a rather flat inverse gamma distribution yield resulting point estimates $\hat{\bbeta}_{eb}$ that do not differ too much numerically.

\[
\]

\section{Conclusion}
Bayesian regression was introduced and shown to be closely connected to ridge regression. Under a conjugate Gaussian prior on the regression parameter the Bayesian regression estimator coincides with the ridge regression estimator, which endows the ridge penalty with the interpretation of this prior. While an analytic expression of these estimators is available, a substantial part of this chapter was dedicated to evaluation of the estimator through resampling. The use of this resampling will be evident when other penalties and non-conjugate priors will be studied (cf. Excercise \ref{exercise.mixturePrior} and Sections \ref{sect.BayesianLogistic} and \ref{sect.BayesianLasso}). Finally, another informative procedure, empirical Bayes, to choose the hyper parameter was presented.

\section{Exercises}
\begin{question} \mbox{ }
\\
Consider the linear regression model $Y_i = X_i \beta + \varepsilon_i$ with the $\varepsilon_i$ i.i.d. following a standard normal law $\mathcal{N}(0, 1)$. Data on the response and covariate are available: $\{(y_i, x_i)\}_{i=1}^8 = \{ (-5, -2),  (0,  -1), \\ (-4, -1), (-2, -1), (0, 0), (3,1), (5,2), (3,2) \}$.
\begin{compactitem}
\item[\textit{a)}] Assume a zero-centered normal prior on $\beta$. What variance, i.e. which $\sigma_{\beta}^2 \in \mathbb{R}_{>0}$, of this prior yields a mean posterior $\mathbb{E}(\beta \, | \, \{(y_i, x_i)\}_{i=1}^8, \sigma_{\beta}^2)$ equal to $1.4$?
\item[{b)}]   Assume a non-zero centered normal prior. What (mean, variance)-combinations for the prior will yield a mean posterior estimate $\hat{\beta} = 2$?
\end{compactitem}
\end{question}

\begin{question} \mbox{ }
\\
Consider the Bayesian linear regression model $\mathbf{Y} = \mathbf{X} \bbeta + \vvarepsilon$ with $\vvarepsilon \sim \mathcal{N} ( \mathbf{0}_n, \sigma^2 \mathbf{I}_{nn})$ and priors
$\bbeta \, | \, \sigma^2 \sim \mathcal{N} ( \mathbf{0}_p, \sigma_{\beta}^{2} \mathbf{I}_{pp})$ and $\sigma^2 \sim \mathcal{IG}(a_0, b_0)$ where $ \sigma_{\beta}^{2} =  c \sigma^{2}$ for some $c > 0$ and $a_0$ and $b_0$ are the shape and scale parameters, respectively, of the inverse Gamma distribution. This model is fitted to data from a study where the response is explained by a single covariate, and henceforth $\bbeta$ is replaced by $\beta$, with the following relevant summary statistics: $\mathbf{X}^{\top} \mathbf{X} = 2$ and $\mathbf{X}^{\top} \mathbf{Y} = 5$.
\begin{compactitem}
\item[\textit{a)}] Suppose $\mathbb{E}( \beta \, | \, \sigma^2=1, c, \mathbf{X}, \mathbf{Y}) = 2$. What amount of regularization should be used such that the ridge regression estimate $\hat{\beta}(\lambda)$ coincides with the aforementioned posterior (conditional) mean?

\item[\textit{b)}] Give the (posterior) distribution of $\beta \, | \, \{ \sigma^2=2, c=2, \mathbf{X}, \mathbf{Y} \}$.

\item[\textit{c)}] Discuss how a different prior of $\sigma^2$ affects the correspondence between $\mathbb{E} (\beta \, | \, \sigma^2, c, \mathbf{X}, \mathbf{Y})$ and the ridge regression estimator.
\end{compactitem}
\end{question}

\begin{question} \mbox{ }
\\
Revisit the microRNA data example of Section \ref{sect.ridgeRegressionDataIllustration}. Use the empirical Bayes procedure outlined in Section \ref{sect.empBayes} to estimate the penalty parameter. Compare this to the one obtained via cross-validation. In particular, compare the resulting point estimates of the regression parameter.
\end{question}

\begin{question} \mbox{ }
\\
Revisit question \ref{question.negativePenaltyParameter}. From a Bayesian perspective, is the suggestion of a negative ridge penalty parameter sensible?
\end{question}

\begin{question} \mbox{ } \\
Consider the linear regression model $\mathbf{Y} = \mathbf{X} \bbeta + \vvarepsilon$ with $\mathbf{X} \in \mathcal{M}^{n,p}$, $\bbeta \in \mathbb{R}^p$, and $\vvarepsilon_i \sim \mathcal{N}(0, \sigma^2 \mathbf{I}_{nn})$. Assume the $\beta_j$ are independently and identically distributed with a generalized normal prior distribution. The latter has density function $f(x; \mu, \alpha_1, \alpha_2) = \alpha_2 [2 \alpha_1 \Gamma(\alpha_2^{-1})]^{-1} \exp[ - (|x-\mu|/\alpha_1)^{\alpha_2}]$ with location parameter $\mu$, scale parameter $\alpha_1$ and shape parameter $\alpha_2$. For which choice of these hyperparameter does the MAP estimator coincide with the bridge regression one? The \textit{bridge regression estimator} of $\bbeta$ is defined as:
\begin{eqnarray*}
\hat{\bbeta}(\lambda_\gamma) & = & \arg \min\nolimits_{\bbeta \in \mathbb{R}} \, \| \mathbf{Y} - \mathbf{X} \bbeta \|_2^2 + \lambda_\gamma \sum\nolimits_{j=1}^p | \beta_j |^\gamma,
\end{eqnarray*}
with penalty parameter $\lambda_\gamma$ and shrinkage parameter $\gamma > 0$. 
\end{question}

\begin{question} \mbox{ } \\
Consider the linear regression model $\mathbf{Y} =  \mathbf{X} \bbeta + \vvarepsilon$ without intercept and $\vvarepsilon \sim \mathcal{N} ( \mathbf{0}_n, \sigma^2 \mathbf{I}_{nn})$, to explain the variation in the response $\mathbf{Y}$ by a linear combination of the columns of the design matrix $\mathbf{X}$. Execute the \texttt{R}-code below to sample data from this model.
\begin{lstlisting}
# set dimension and sample size
n <-  50
p <- 100

# create covariates
X <- matrix(rnorm(n*p),     ncol=p)

#  create coefficients and draw response
betas <- matrix(seq(-2, 2, length.out=p), ncol=1)
betas <- (betas - min(betas)+0.001)
betas <- qnorm(betas / (max(betas)+0.001)) / 5
Y     <- X %*% betas + rnorm(n, sd=1/4)

\end{lstlisting}
With these sampled data, find a Bayesian estimate, the posterior mean, of the regression parameter. Hereto adopt the following priors for the parameters: $\bbeta \, | \, \sigma^2 \sim \mathcal{N}(\mathbf{0}_p, \sigma^2 \lambda \mathbf{I}_{pp})$ with $\lambda=1$ and $\sigma^2 \sim \mathcal{IG}(a_0, b_0)$ with $a_0 = 1 = b_0$. 

\begin{compactitem}
\item[\textit{a)}] Build a Gibbs sampler (Algorithm 3 in the lecture notes). Initiate the sampler with $\bbeta_0 = \mathbf{0}_p$ and $\sigma^2_0 = 1$. The sampler then iterates between drawing new $\bbeta$ and $\sigma^2$, respectively, from their conditional posteriors: 
\begin{eqnarray*}
\bbeta \, | \, \sigma^2_t, \mathbf{Y}, \mathbf{X} & \sim & \mathcal{N} [ (\mathbf{X}^{\top} \mathbf{X} + \lambda \mathbf{I}_{pp})^{-1} \mathbf{X}^{\top} \mathbf{Y}, \sigma_t^2  (\mathbf{X}^{\top} \mathbf{X}  + \lambda \mathbf{I}_{pp})^{-1} ],
\\ 
\sigma^2 \, | \, \bbeta_t, \mathbf{Y}, \mathbf{X} & \sim & \mathcal{IG}(a, b),
\end{eqnarray*}
with shape parameter $a=a_0 + \tfrac{1}{2} (n+p)$ and rate parameter $b = b_0 + \tfrac{1}{2} \| \mathbf{Y} \|_2^2  - \tfrac{1}{2} \| \mathbf{X} \hat{\bbeta} (\lambda) \|_2^2  - \tfrac{1}{2} \lambda \| \hat{\bbeta}(\lambda) \|_2^2$ and in which $t$ represents the index for the iteration. To sample from both distributions, use the \texttt{rmvnorm}- and the \texttt{rinvgamma}-functions in \texttt{R}.
  
\item[\textit{b)}] Use the Gibbs sampler to draw $10100$ times from both conditional posteriors in alternating fashion. To remove the dependency on the choice of the initiation $\sigma^2_0$ throw away the first $100$ (i.e. the burn-in period) draws for both parameters. To reduce the dependency between subsequent draws, apply thinning: keep each $10$-th draw after the burn-in period. After both operations (removal of the burning-in period and the thinning) only those draws corresponding to iterations $110, 120, \ldots, 10100$ are preserved. Those are considered a representative sample from the joint posterior of $\bbeta$ and $\sigma^2$.

\item[\textit{c)}] Calculate the lower bound of the 95\% credible interval, containing the central $(100-\alpha)$\% with $\alpha =0.05$ of the posterior probability mass, of the second element of the regression parameter. Use the \texttt{quantile}-function for the credible interval construction. 

\item[\textit{d)}] Investigate the dependency among subsequent draws, i.e. without the thinning, of $\sigma^2$ from the Gibbs sampler. In this contrast the case with $\lambda=1$ to that with $\lambda = 10^{6}$. Is there a difference in their $1^{\mbox{{\tiny st}}}$ order dependencies? If so, explain this difference. If not, explain the absence thereof. 
\end{compactitem}
\end{question}

\pagestyle{fancy}

\chapter[Generalizing ridge regression]{Generalizing ridge regression} \label{chap:genRidge}
The expos\'{e} on ridge regression may be generalized in many ways. Among others, different generalized linear models may be considered (confer Section \ref{sect.ridgeLogistic}). In this section we stick to the linear regression model with the usual assumptions. But we  now fit it in weighted fashion -- to accommodate the ridge estimation of the logistic regression model (see Chapter  \ref{chap.ridgeLogistic}) -- and generalize the common, spherical penalty. 

\section{Generalized ridge regression}
Our generalization of the ridge loss function, a weighted least squares criterion augmented with a generalized ridge penalty, is:
\begin{eqnarray} \label{form:generalizedRidgeLoss}
(\mathbf{Y} - \mathbf{X} \bbeta)^{\top} \mathbf{W} (\mathbf{Y} - \mathbf{X} \bbeta) + (\bbeta - \bbeta_0)^{\top} \mathbf{\Delta}
(\bbeta - \bbeta_0).
\end{eqnarray}
In the above display $\mathbf{W}$ is a $n \times n$-dimensional, diagonal matrix with $(\mathbf{W})_{ii} \in [0,1]$ representing the weight of the $i$-th observation. The minimizer of loss function (\ref{form:generalizedRidgeLoss}) is the generalized ridge regression estimator.

The generalized ridge penalty in loss function (\ref{form:generalizedRidgeLoss}) is now a quadratic form with penalty parameter $\mathbf{\Delta}$, a $p \times p$-dimensional, positive definite, symmetric matrix. If $\mathbf{\Delta} = \lambda \mathbf{I}_{pp}$, one regains the spherical penalty of `regular ridge regression'. This penalty shrinks each element of the regression parameter $\bbeta$ equally along the unit vectors $\mathbf{e}_j$. Generalizing $\mathbf{\Delta}$ to the class of symmetric, positive semi-definite matrices $\mathcal{S}_{+}^p$ allows for \textit{i)} different penalization per regression parameter, and \textit{ii)} joint (or correlated) shrinkage among the elements of $\bbeta$. The penalty parameter $\mathbf{\Delta}$ determines the speed and direction of shrinkage.  The $p$-dimensional column vector $\bbeta_0$ is a user-specified, non-random target towards which $\bbeta$ is shrunken as the penalty parameter increases. We illustrate this in Example \ref{example:generalizedRidgeRegularizationPaths}.

The addition of the generalized ridge penalty to the sum-of-squares typically ensures the existence of a unique regression estimator in the face of super-collinearity. The generalized penalty is a non-degenerated quadratic form in $\bbeta$ if the penalty matrix $\mathbf{\Delta}$ is positive definite. The generalized ridge penalty is then strictly convex. Consequently, the generalized ridge regression loss function (\ref{form:generalizedRidgeLoss}), being the sum of a convex and strictly convex function, is also strictly convex. This warrants the existence of a unique global minimum and, thereby, a unique estimator.

There is an analytic expression for the optimum of the generalized ridge loss function (\ref{form:generalizedRidgeLoss}). To see this, obtain the estimating equation of $\bbeta$ through equating its derivative with respect to $\bbeta$ to zero:
\begin{eqnarray*}
2 \mathbf{X}^{\top} \mathbf{W} \mathbf{Y} - 2 \mathbf{X}^{\top} \mathbf{W} \mathbf{X} \bbeta - 2 \mathbf{\Delta} \bbeta  + 2 \mathbf{\Delta} \bbeta_0 & = & \mathbf{0}_{p}.
\end{eqnarray*}
This is solved by:
\begin{eqnarray} \label{form:generalizedRidgeEstimator}
\hat{\bbeta}(\mathbf{\Delta}) & = & (\mathbf{X}^{\top} \mathbf{W} \mathbf{X} + \mathbf{\Delta})^{-1} (\mathbf{X}^{\top} \mathbf{W} \mathbf{Y} + \mathbf{\Delta} \bbeta_0).
\end{eqnarray}
Clearly, this reduces to the `regular' ridge regression estimator by setting $\mathbf{W} = \mathbf{I}_{nn}$,  $\bbeta_0 = \mathbf{0}_{p}$, and $\mathbf{\Delta} = \lambda \mathbf{I}_{pp}$.

\begin{example} \label{example:generalizedRidgeRegularizationPaths} \mbox{ } \\
The effects of the generalized ridge penalty on the coefficients of corresponding estimator are reflected in the regularization paths of the estimator's coefficients. This is illustrated on data from the linear  regression model $\mathbf{Y} = \mathbf{X} \bbeta + \vvarepsilon$ with $X_{i,j} \sim \mathcal{N}(0,1)$ for all $i,j$, $\bbeta = (2, 0, 1, 3)^{\top}$, and $\vvarepsilon \sim \mathcal{N}(\mathbf{0}_n, \mathbf{I}_{nn})$. We fit the  model with the generalized ridge regression estimator with $\mathbf{W} = \mathbf{I}_{nn}$, target $\bbeta_0 = (0, 0, 2, 2)^{\top}$, and penalty matrix 
\begin{eqnarray*}
\mathbf{\Delta} = \lambda \left( \begin{array}{rrrr} 1 & 0 & 0 & 0 \\ 0 & 1 & -\tfrac{1}{2} & 0 \\ 0 & -\tfrac{1}{2} & 1 & 0 \\ 0 & 0 & 0 & 1 \end{array} \right).
\end{eqnarray*}
We evaluate the estimator for a grid of $\lambda$'s. Figure  \ref{fig:generalizedRidgeIllustration} shows the regularization paths for the coefficients of the four explanatory variables of our linear regression model.
\begin{figure}[!h]
\centering
\begin{tabular}{c}
\includegraphics[scale=0.40, angle=0]{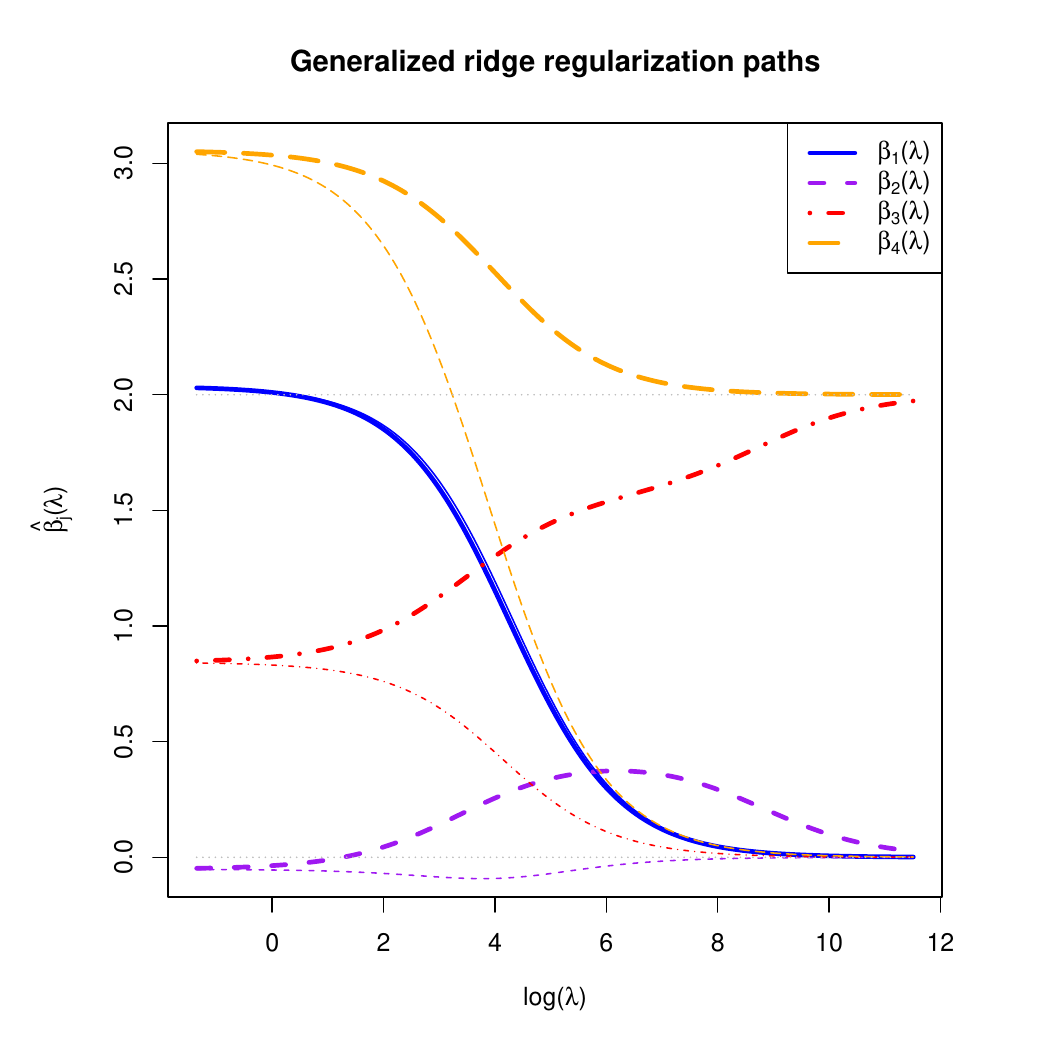}
\end{tabular}
\caption{Generalized (fat coloured lines) and `regular' (thin coloured lines) regularization paths  of four regression coefficients. The dotted grey (straight) lines indicated the targets towards the generalized ridge penalty shrinks regression coefficient estimates.} \label{fig:generalizedRidgeIllustration}
\end{figure}
Most striking is the limiting behaviour of the estimates of $\beta_3$ and $\beta_4$ for large values of the penalty parameter $\lambda$: they convergence to a non-zero value (as was specified by the nonzero elements of $\bbeta_0$). More subtle is the (temporary) convergence of the regularization paths of the estimates of $\beta_2$ and $\beta_3$. That of $\beta_2$ is pulled away from zero (its true value and approximately its unpenalized estimate) towards the estimate of $\beta_3$. In the regularization path of $\beta_3$, this can be observed in a delayed convergence to its nonzero target value (for comparison consider that of $\beta_4$). For reference, the corresponding regularization paths of the `regular' ridge estimates (as thinner lines of the same colour) are included in Figure \ref{fig:generalizedRidgeIllustration}.
\end{example}

\noindent
The generalized ridge regression estimator is an affine transformation of its regular counterpart. This can be seen from the following: 
\begin{eqnarray*}
\hat{\boldsymbol{\beta}} (\mathbf{\Delta}, \boldsymbol{\beta}_0) & = &  (\mathbf{X}^{\top} \mathbf{W} \mathbf{X} + \mathbf{\Delta})^{-1} ( \mathbf{X}^{\top} \mathbf{W} \mathbf{Y} + \mathbf{\Delta} \boldsymbol{\beta}_0)
\\
& = &  (\mathbf{X}^{\top} \mathbf{W} \mathbf{X} + \mathbf{\Delta})^{-1} (\mathbf{X}^{\top} \mathbf{W} \mathbf{X} + \lambda \mathbf{I}_{pp}) (\mathbf{X}^{\top} \mathbf{W} \mathbf{X} + \lambda \mathbf{I}_{pp})^{-1} \mathbf{X}^{\top} \mathbf{Y}  + (\mathbf{X}^{\top} \mathbf{W} \mathbf{X} + \mathbf{\Delta})^{-1}  \mathbf{\Delta} \boldsymbol{\beta}_0
\\
& := & \mathbf{A}(\mathbf{\Delta}, \lambda) \hat{\boldsymbol{\beta}}(\lambda) + \mathbf{a}_0,
\end{eqnarray*}
where the translation matrix $\mathbf{A}(\mathbf{\Delta}, \lambda)$ and the offset vector $\mathbf{a}_0$ are appropriately defined. 
\\
\\
\noindent
One particular version of the generalized ridge penalty deserves special attention. It distinghuises between penalized and unpenalized covariates. The latter comprises explanatory variables that ought to be in the model, and are not to be shrunken. This aims to keep the bias of their estimates to a minimum and make them somewhat comparable to these reported in the literature. For instance, in the statistical analysis of clinical trials factors like `age' and `sex' are known to associated with the response, and any model without them would not be acceptable to the field. Additionally, the patients of the clinical trial may have been molecularly characterized at baseline. The resulting high-dimensional molecular information are to be included in the model, alongside the aforementioned factors. The effect of the high-dimensional molecular covariates can then only be estimated in penalized fashion, while that of `age' and `gender' are preferably estimated in an unpenalized manner.

To present our estimator with penalized and unpenalized covariates, we introduce separate notation for them. This modifies the model to $\mathbf{Y} = \mathbf{U} \ggamma + \mathbf{X} \bbeta + \varepsilon$, where $\mathbf{U}$ is a full rank $n \times q$-dimensional design matrix of the $q$ unpenalized covariates and $\ggamma$ the associated regression parameter.  The ridge regression estimator is then generalized to:
\begin{eqnarray*}
\hat{\ggamma}(\mathbf{\Delta}, \bbeta_0), \hat{\bbeta}(\mathbf{\Delta}, \bbeta_0) & = & \arg \min_{\ggamma \in \mathbb{R}^q, \bbeta \in \mathbb{R}^p} \| \mathbf{Y} - \mathbf{U} \ggamma - \mathbf{X} \bbeta \|_2^2 + (\bbeta - \bbeta_0)^{\top} \mathbf{\Delta} (\bbeta - \bbeta_0).
\end{eqnarray*}
Note that the estimator of regression parameter $\ggamma$ of the unpenalized covariates $\mathbf{U}$ too depends on the penalty parameters. This is due to the fact that it is jointly estimated with the regression parameter $\bbeta$ of their penalized counterparts $\mathbf{X}$. This bears consequence for the bias of the estimate of $\ggamma$ (see Corollary \ref{corr:biasGenRidgeEstimator}). Again an analytic expression of the estimator exists:
\begin{eqnarray} 
\label{form:genRidgeEstimatorWithUnAndPenCovariates}
\left(
\begin{array}{l}
\hat{\ggamma}(\mathbf{\Delta}, \bbeta_0)
\\
\hat{\bbeta}(\mathbf{\Delta}, \bbeta_0)
\end{array}
\right)
& = & 
\left(
\begin{array}{ll} \mathbf{U}^{\top} \mathbf{U} & \mathbf{U}^{\top} \mathbf{X}
\\
\mathbf{X}^{\top} \mathbf{U} & \mathbf{X}^{\top} \mathbf{X}  + \mathbf{\Delta}
\end{array}
\right)^{-1}
\left(
\begin{array}{l} \mathbf{U}^{\top} \mathbf{Y} 
\\
\mathbf{X}^{\top} \mathbf{Y}  + \mathbf{\Delta} \bbeta_0
\end{array}
\right).
\end{eqnarray}
This expression can be `simplified', as is done in \citep{Lettink2023twoDimFusedRidge}, to:
\begin{eqnarray*}
\hat{\ggamma}(\mathbf{\Delta}, \bbeta_0) & = &  \{\mathbf{U}^{\top}  [\mathbf{X} \mathbf{\Delta}^{-1} \mathbf{X}^{\top} + \mathbf{I}_{nn}]^{-1} \mathbf{U}\}^{-1} \mathbf{U}^{\top} [\mathbf{X} \mathbf{\Delta}^{-1} \mathbf{X}^{\top} + \mathbf{I}_{nn}]^{-1} ( \mathbf{Y} - \mathbf{X} \bbeta_0),
\\
\hat{\bbeta}(\mathbf{\Delta}, \bbeta_0) & = & \bbeta_0 + (\mathbf{X}^{\top} \mathbf{X}  + \mathbf{\Delta})^{-1} \mathbf{X}^{\top} [ \mathbf{Y}  - \mathbf{U} \hat{\ggamma}(\mathbf{\Delta}, \bbeta_0) - \mathbf{X} \bbeta_0],
\end{eqnarray*}
which is obtained by means of the Woodbury matrix identity and the analytic expression of the inverse of a $2 \times 2$ block matrix.

\section{Constrained estimation}
Generalized ridge estimation can be recast as a constrained estimation problem. In the latter framework, the generalized ridge regression estimator is defined as
\begin{eqnarray*}
\hat{\bbeta} (\mathbf{\Delta}, \bbeta_0) & = & \arg \, \, \min_{\{ \bbeta \in \mathbb{R}^p \, : \, (\bbeta - \bbeta_0)^{\top} \mathbf{\Delta} (\bbeta - \bbeta_0) \leq c(\mathbf{\Delta}, \bbeta) \} }\, \, \| \mathbf{Y} - \mathbf{X} \bbeta \|_2^2.
\end{eqnarray*}
The constrained estimator $\hat{\bbeta} (\mathbf{\Delta}, \bbeta_0)$ is located at the boundary of the generalized ridge parameter constraint. The two formulations, penalized and constrained estimation, of the generalized ridge regression estimator coincide if  $c(\mathbf{\Delta}, \bbeta) = [\hat{\bbeta} (\mathbf{\Delta}, \bbeta_0)  - \bbeta_0]^{\top} \mathbf{\Delta} [\hat{\bbeta} (\mathbf{\Delta}, \bbeta_0)  - \bbeta_0]$. 

\begin{figure}[!h]
\begin{tabular}{rcl}
\includegraphics[scale=0.45, angle=0]{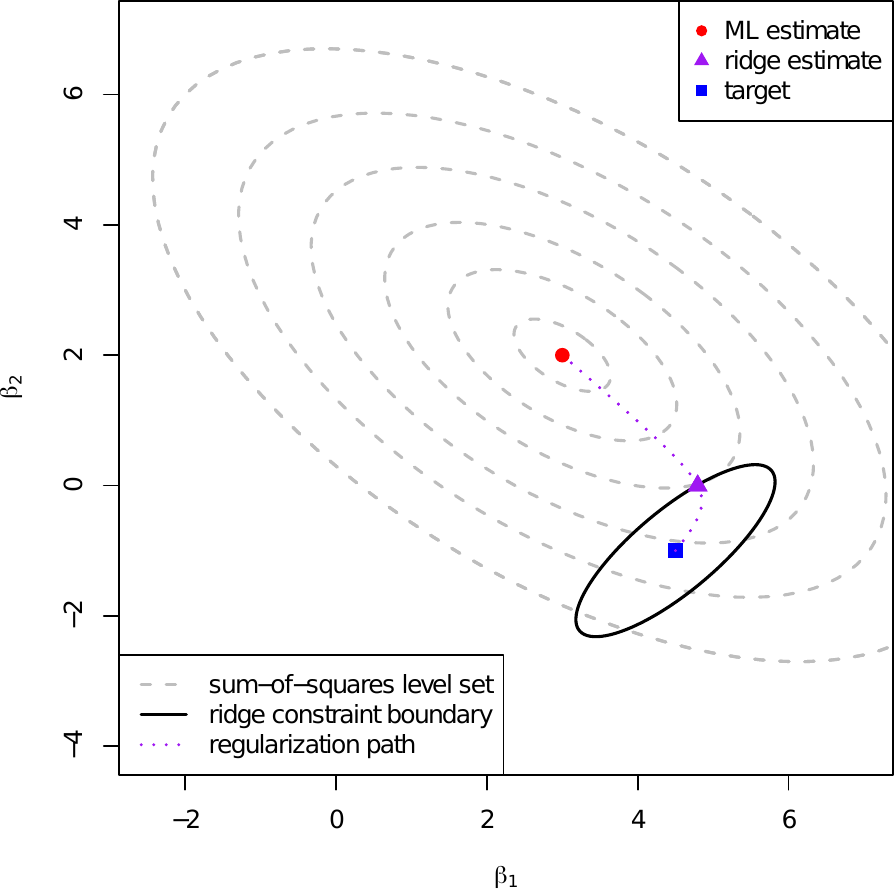}
& &
\includegraphics[angle=0, scale=0.45]{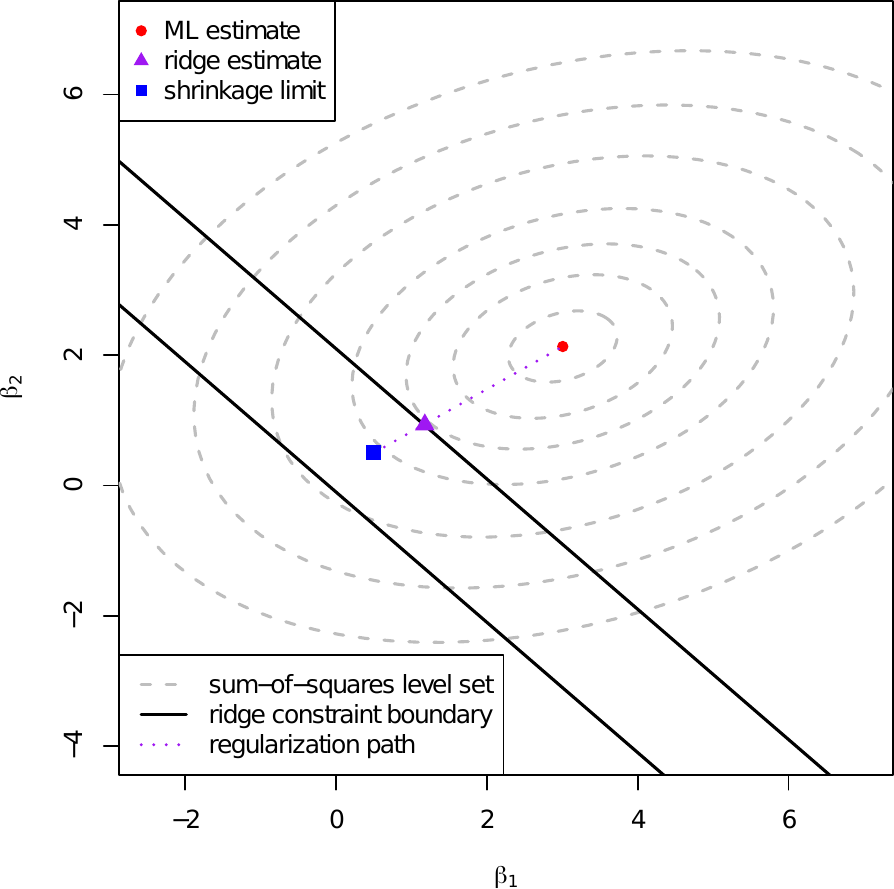}
\end{tabular}
\caption{Illustrations of the generalized ridge regression estimators as constrained estimators with a positive definite (left-hand side panel) and nonnegative definite (right-hand side panel). } \label{fig:constrainedEstimationGeneralizedRidge}
\end{figure}

The implications of the generalized ridge penalty can be visualized. The left panel of Figure \ref{fig:constrainedEstimationGeneralizedRidge} depicts the level sets of the sum-of-squares as dashed grey lines and the generalized ridge parameter constraint. The parameter constraint is not centered at zero and ellipsoidal (and not spherical) as the generalized ridge penalty is a quadratic form with 
\begin{eqnarray*}
\bbeta_0 \, \, \, = \, \, \, \left( \begin{array}{rr} 0 \\ 1 \end{array} \right) & \mbox{ and } & \mathbf{\Delta} \, \, \, = \, \, \, \left( \begin{array}{rr} 5 & 4 \\ 4 & 5 \end{array} \right).
\end{eqnarray*}
The shape of the penalty can be a degenerated ellipsoid. This happens if  $\mathbf{\Delta}$ is positive semi-definite, rendering the quadratic form of the generalized ridge penalty degenerate. The right panel of Figure \ref{fig:constrainedEstimationGeneralizedRidge} shows an example with $2 \times 2$-dimensional penalty matrix $\mathbf{\Delta}$ comprising only $2$'s, which is positive semi-definite as its eigenvalues are $4$ and $0$. The parameter constraint is the area inbetween the two black lines.

The study of the parameter constraint reveals the regularization limit of the generalized ridge regression estimator. Hereto we suppose that, for large $\lambda >0$, we can write $\mathbf{\Delta} = \lambda \mathbf{\Delta}_{\infty}$ and $\mathbf{\Delta}_{\infty}$ is independent of $\lambda$. Then, if $\mathbf{\Delta}_{\infty}$ is positive definite, $\lim_{\lambda \rightarrow \infty} \hat{\bbeta} (\lambda \mathbf{\Delta}_{\infty}, \bbeta_0) = \bbeta_0$ as only then the penalty is minimized to zero, neutralizing the penalization. The purple dotted line  in Figure \ref{fig:constrainedEstimationGeneralizedRidge} is the regularization path that connects the minimizers of the sum-of-squares on ever smaller ellipsoidal parameter constraints that eventually collapse onto the shrinkage target $\bbeta_0$. If $\mathbf{\Delta}$ is positive semi-definite, as is the case in the right-hand side panel of Figure \ref{fig:constrainedEstimationGeneralizedRidge}, the $\lambda \rightarrow \infty$-regularization limit is not directly obvious. In this limit, the parameter constraint collapses to the $\beta_2 = 1 - \beta_1$-line. The estimator is shrunken to a point on this line that is determined by the sum-of-squares (should it not be degenerated itself with level sets parallel to that of the constraint). An analytic expression of this point (the estimator) can be derived. Hereto consider the eigendecomposition $\mathbf{V}_{\Delta} \mathbf{D}_{\Delta} \mathbf{V}_{\Delta}^{\top}$ of $\mathbf{\Delta}$, where $\mathbf{V}_{\Delta}$ comprises the eigenvectors as columns and diagonal $\mathbf{D}_{\Delta}$ contains the eigenvalues. Separate columns of $\mathbf{V}_{\Delta}$ and diagonal elements of $\mathbf{D}_{\Delta}$ by the non- and zero eigenvalues and write $ \mathbf{V}_{\Delta} = (\mathbf{V}_{\Delta,=0} \, \,  \mathbf{V}_{\Delta,\not=0})$ and $\mbox{diag}(\mathbf{D}_{\Delta}) =  [\mbox{diag} (\mathbf{D}_{\Delta, \not=0}), \mathbf{0}_q]$. The ridge regression estimator can then be expressed as $\hat{\bbeta} (\mathbf{\Delta}, \bbeta_0)  = \bbeta_0 + \mathbf{V}_{\Delta} [ \hat{\ddelta}_{\not=0}^{\top} (\mathbf{\Delta}, \bbeta_0), \hat{\ddelta}_{=0}^{\top} (\mathbf{\Delta}, \bbeta_0) ]^{\top}$ with
\begin{eqnarray*}
\hat{\ddelta}_{\neg0} (\mathbf{\Delta}, \bbeta_0) & = & ( \mathbf{V}_{\Delta,\neg0}^{\top} \mathbf{X}^{\top} \mathbf{X} \mathbf{V}_{\Delta,\neg0}   + \mathbf{D}_{\Delta, \neg 0})^{-1} \mathbf{V}_{\Delta,\neg0}^{\top} [ (\mathbf{Y}   - \mathbf{X} \bbeta_0) -  \mathbf{X} \mathbf{V}_{\Delta,=0}  \hat{\ddelta}_{=0} (\mathbf{\Delta}, \bbeta_0)],
\\
\hat{\ddelta}_{=0} (\mathbf{\Delta}, \bbeta_0) & = &  \{ \mathbf{V}_{\Delta,=0}^{\top} \mathbf{X}^{\top} [\mathbf{X} \mathbf{V}_{\Delta,\neg0} \mathbf{D}_{\Delta, \neg 0}^{-1} \mathbf{V}_{\Delta,\neg0}^{\top} \mathbf{X}^{\top} + \mathbf{I}_{nn}]^{-1}  \mathbf{X} \mathbf{V}_{\Delta,=0}  \}^{-1} 
\\
& &  \, \, \, \mathbf{V}_{\Delta,=0}^{\top} \mathbf{X}^{\top} [\mathbf{X} \mathbf{V}_{\Delta,\neg0} \mathbf{D}_{\Delta, \neg 0}^{-1}  \mathbf{V}_{\Delta,\neg0}^{\top} \mathbf{X}^{\top} + \mathbf{I}_{nn}]^{-1} ( \mathbf{Y} - \mathbf{X} \bbeta_0).
\end{eqnarray*}
The derivation of the display above uses the analytic expression of the ridge regression estimator with un- and penalized covariates. Now exploit the assumption that $\mathbf{\Delta} = \lambda \mathbf{\Delta}_{\infty}$ for large $\lambda $. We can then write $\mathbf{D}_{\Delta, \neg 0} = \lambda \tilde{\mathbf{D}}_{\Delta, \neg 0}$ and let $\lambda$ tend to infinity to find:
\begin{eqnarray*}
\hat{\bbeta} (\lambda \mathbf{\Delta}_{\infty}, \bbeta_0) & = & \bbeta_0 + \mathbf{V}_{\Delta} \left( \begin{array}{r} \mathbf{0}_q
\\
\{ \mathbf{V}_{\Delta,=0}^{\top} \mathbf{X}^{\top} \mathbf{X} \mathbf{V}_{\Delta,=0}  \}^{-1}  \mathbf{V}_{\Delta,=0}^{\top} \mathbf{X}^{\top} ( \mathbf{Y} - \mathbf{X} \bbeta_0)
\end{array} \right).
\end{eqnarray*}
We leave the verification -- by substitution -- that this solution satisfies the generalized ridge parameter constraint as an exercise.

\section{Moments}
The distribution, expectation and variance of $\hat{\bbeta}(\bbeta_0, \mathbf{\Delta})$ are obtained through application of the same arguments, matrix algebra, and expectation and covariance rules used in the derivation of their counterparts of the `regular' ridge regression estimator. This gives 
\begin{eqnarray} 
\label{form:genRidgeEstimatorExpectation}
\mathbb{E}[\hat{\bbeta}(\mathbf{\Delta}, \bbeta_0) \, | \, \bbeta_0, \mathbf{\Delta}] & = & \textcolor{white}{\sigma^2 } (\mathbf{X}^{\top} \mathbf{W} \mathbf{X} + \mathbf{\Delta})^{-1} (\mathbf{X}^{\top} \mathbf{W} \mathbf{X} \bbeta + \mathbf{\Delta} \bbeta_0),
\\
\label{form:genRidgeEstimatorVariance}
\mbox{Var}[\hat{\bbeta}(\mathbf{\Delta}, \bbeta_0) \, | \, \bbeta_0, \mathbf{\Delta}] & = & \sigma^2 (\mathbf{X}^{\top} \mathbf{W} \mathbf{X} + \mathbf{\Delta})^{-1} \mathbf{X}^{\top} \mathbf{W}^2 \mathbf{X} (\mathbf{X}^{\top} \mathbf{W} \mathbf{X} + \mathbf{\Delta})^{-1},
\end{eqnarray}
and error variance $\sigma^2$. The distribution of the generalized ridge regression estimator, by its linearity, is now also known:
\begin{eqnarray*}
\hat{\bbeta} (\mathbf{\Delta}, \bbeta_0) \, | \, \bbeta_0, \mathbf{\Delta} & \sim &  \mathcal{N} \{ \mathbb{E}[\hat{\bbeta}(\mathbf{\Delta}, \bbeta_0) \, | \, \bbeta_0, \mathbf{\Delta}], \mbox{Var}[\hat{\bbeta}(\mathbf{\Delta}, \bbeta_0) \, | \, \bbeta_0, \mathbf{\Delta}]\} 
\end{eqnarray*}
where we condition on $\bbeta_0$ and $\mathbf{\Delta}$ to emphasize that aspects of $\bbeta_0$ and $\mathbf{\Delta}$ are typically
chosen in data-driven fashion. In the remainder of this chapter, we reduce notation clutter and write $\mathbb{E}[\hat{\bbeta}(\mathbf{\Delta}, \bbeta_0) ]$ and $\mbox{Var}[\hat{\bbeta}(\mathbf{\Delta}, \bbeta_0)]$ for 
$\mathbb{E}[\hat{\bbeta}(\mathbf{\Delta}, \bbeta_0) \, | \, \bbeta_0, \mathbf{\Delta}]$ and $\mbox{Var}[\hat{\bbeta}(\mathbf{\Delta}, \bbeta_0) \, | \, \bbeta_0, \mathbf{\Delta}]$, respectively.
\\
\\
In Chapter \ref{chap:ridgeRegression}, we have derived a decomposition of the bias of `regular' ridge regression estimator. A similar decomposition can be obtained for the generalized ridge estimator. To that end, we first need the following technical result.  
\begin{proposition} \label{prop:projectionGenRidgeEstimator} \mbox{ }  \\
Consider the projection matrix $\mathbf{P}_x = \mathbf{X}^{\top} (\mathbf{X} \mathbf{X})^+ \mathbf{X}$. Then, $\mathbf{P}_{x, \Delta} = (\mathbf{X}^{\top} \mathbf{X} + \mathbf{\Delta})^{-1}\mathbf{P}_x (\mathbf{X}^{\top} \mathbf{X} + \mathbf{\Delta})$ is also a projection matrix. Moreover, $\mathbf{P}_{x, \Delta} [ \hat{\boldsymbol{\beta}} (\mathbf{\Delta}, \boldsymbol{\beta}_0) -  \boldsymbol{\beta}_0 ] = \hat{\boldsymbol{\beta}} (\mathbf{\Delta}, \boldsymbol{\beta}_0) - \boldsymbol{\beta}_0$.
\end{proposition}
\begin{proof} We first verify that $\mathbf{P}_{x, \Delta}$ is a projection matrix:
\begin{eqnarray*}
\mathbf{P}_{x, \Delta}^2 & = & (\mathbf{X}^{\top} \mathbf{X} + \mathbf{\Delta})^{-1}\mathbf{P}_x (\mathbf{X}^{\top} \mathbf{X} + \mathbf{\Delta}) (\mathbf{X}^{\top} \mathbf{X} + \mathbf{\Delta})^{-1}\mathbf{P}_x (\mathbf{X}^{\top} \mathbf{X} + \mathbf{\Delta})
\, \, \, = \, \, \, \mathbf{P}_{x, \Delta},
\end{eqnarray*}
where we have used that $\mathbf{P}_x^2 = \mathbf{P}_x$.

We now prove the proposition's identity:
\begin{eqnarray*}
\mathbf{P}_{x, \Delta}  \hat{\boldsymbol{\beta}} (\mathbf{\Delta}, \boldsymbol{\beta}_0) & = & \mathbf{P}_{x, \Delta} (\mathbf{X}^{\top} \mathbf{X} + \mathbf{\Delta})^{-1} ( \mathbf{X}^{\top} \mathbf{Y} + \mathbf{\Delta} \boldsymbol{\beta}_0)
\\
& = & \mathbf{P}_{x, \Delta} (\mathbf{X}^{\top} \mathbf{X} + \mathbf{\Delta})^{-1} \mathbf{X}^{\top} \mathbf{Y} +  \mathbf{P}_{x, \Delta} (\mathbf{X}^{\top} \mathbf{X} + \mathbf{\Delta})^{-1} ( \mathbf{X}^{\top} \mathbf{X} + \mathbf{\Delta} - \mathbf{X}^{\top} \mathbf{X}) \boldsymbol{\beta}_0
\\
& = & \mathbf{P}_{x, \Delta} (\mathbf{X}^{\top} \mathbf{X} + \mathbf{\Delta})^{-1} \mathbf{X}^{\top} ( \mathbf{Y} -  \mathbf{X} \boldsymbol{\beta}_0)   + \mathbf{P}_{x, \Delta} \boldsymbol{\beta}_0
\\
& = & (\mathbf{X}^{\top} \mathbf{X} + \mathbf{\Delta})^{-1}\mathbf{P}_x  \mathbf{X}^{\top} ( \mathbf{Y} -  \mathbf{X} \boldsymbol{\beta}_0)   + \mathbf{P}_{x, \Delta} \boldsymbol{\beta}_0
\\
& = & (\mathbf{X}^{\top} \mathbf{X} + \mathbf{\Delta})^{-1}  \mathbf{X}^{\top} ( \mathbf{Y} -  \mathbf{X} \boldsymbol{\beta}_0)   + \mathbf{P}_{x, \Delta} \boldsymbol{\beta}_0
\\
& = & (\mathbf{X}^{\top} \mathbf{X} + \mathbf{\Delta})^{-1}  \mathbf{X}^{\top}  \mathbf{Y} - (\mathbf{X}^{\top} \mathbf{X} + \mathbf{\Delta})^{-1}  (\mathbf{X}^{\top}  \mathbf{X} + \mathbf{\Delta} - \mathbf{\Delta})\boldsymbol{\beta}_0   + \mathbf{P}_{x, \Delta} \boldsymbol{\beta}_0
\\
& = & (\mathbf{X}^{\top} \mathbf{X} + \mathbf{\Delta})^{-1}  (\mathbf{X}^{\top}  \mathbf{Y} + \mathbf{\Delta}\boldsymbol{\beta}_0) - \boldsymbol{\beta}_0   + \mathbf{P}_{x, \Delta} \boldsymbol{\beta}_0
\\
& = & \hat{\boldsymbol{\beta}} (\mathbf{\Delta}, \boldsymbol{\beta}_0)  - \boldsymbol{\beta}_0   + \mathbf{P}_{x, \Delta} \boldsymbol{\beta}_0,
\end{eqnarray*}
where the fourth equality uses the definition of $\mathbf{P}_{x,\Delta}$ and the fifth uses the identity $\mathbf{P}_x \mathbf{X}^{\top} = \mathbf{X}^{\top}$.
\end{proof}
With this proposition at hand, we formulate a generalization of Corollary \ref{corol:ridgeBiasDecomposition}, which decomposes the bias in a part attributable to the regularization of the estimator and to the high-dimensionality of the study. 
\begin{corollary} \mbox{ } \\
The bias can be decomposed as: $\mathbb{E} [ \hat{\boldsymbol{\beta}}(\mathbf{\Delta}, \boldsymbol{\beta}_0)] - \boldsymbol{\beta} = \mathbb{E} [ \mathbf{P}_{x, \Delta} \hat{\boldsymbol{\beta}}(\mathbf{\Delta}, \boldsymbol{\beta}_0) - \mathbf{P}_{x, \Delta} \boldsymbol{\beta}]  + (\mathbf{P}_{x, \Delta} - \mathbf{I}_{pp} )  (\boldsymbol{\beta} - \boldsymbol{\beta}_0)$.
\end{corollary}
\begin{proof}
The bias decomposition follows from
\begin{eqnarray*}
\mathbb{E} [ \hat{\boldsymbol{\beta}}(\mathbf{\Delta}, \boldsymbol{\beta}_0)] - \boldsymbol{\beta} & = &
\mathbb{E} [ \hat{\boldsymbol{\beta}}(\mathbf{\Delta}, \boldsymbol{\beta}_0) - \boldsymbol{\beta}_0] + \boldsymbol{\beta}_0 - \boldsymbol{\beta} + \mathbf{P}_{x, \Delta}  \boldsymbol{\beta} - \mathbf{P}_{x, \Delta} \boldsymbol{\beta}
\\
& = & \mathbb{E} [ \mathbf{P}_{x, \Delta} \hat{\boldsymbol{\beta}}(\mathbf{\Delta}, \boldsymbol{\beta}_0) - \mathbf{P}_{x, \Delta} \boldsymbol{\beta}_0] - \mathbf{P}_{x, \Delta} \boldsymbol{\beta} + \boldsymbol{\beta}_0 - (\mathbf{I}_{pp} - \mathbf{P}_{x, \Delta})  \boldsymbol{\beta} 
\\
& = & \mathbb{E} [ \mathbf{P}_{x, \Delta} \hat{\boldsymbol{\beta}}(\mathbf{\Delta}, \boldsymbol{\beta}_0) - \mathbf{P}_{x, \Delta} \boldsymbol{\beta}]  + (\mathbf{P}_{x, \Delta} - \mathbf{I}_{pp} )  (\boldsymbol{\beta} - \boldsymbol{\beta}_0),
\end{eqnarray*}
where the second equality follows from Proposition \ref{prop:projectionGenRidgeEstimator}. 
\end{proof}
The bias of the generalized ridge regression estimator can thus be decomposed in a contribution attributable to the regularization and the high-dimensionality of the study. The latter vanishes if either  $\mathbf{P}_{x, \Delta} = \mathbf{I}_{pp}$ or $\mathbf{P}_{x,\Delta} (\boldsymbol{\beta} - \boldsymbol{\beta}_0) = \boldsymbol{\beta} - \boldsymbol{\beta}_0$. This requires either \textit{i)} the study to have a low dimensional design as $\mbox{rank} ( \mathbf{P}_{x, \Delta} ) \leq \min \{  \mbox{rank} ( \mathbf{X}^\top \mathbf{X} + \mathbf{\Delta} ),  \mbox{rank} ( \mathbf{P}_{x} ) \}$ or \textit{ii)} $\boldsymbol{\beta}$ and $\boldsymbol{\beta}_0$ both to be in the subspace onto which $\mathbf{P}_{x, \Delta}$ projects.
\\
\\
Intuitively, we expect the generalized ridge regression estimator to behave poorly if $\bbeta_0$ is way off. Strong penalization then drags the estimator away from the true parameter $\bbeta$. This is reflected in the bias. The squared bias of the generalized ridge regression estimator is (after some linear algebraic manipulations):
\begin{eqnarray*}
\{ \mbox{Bias}[\hat{\bbeta}(\mathbf{\Delta}, \bbeta_0)] \}^{\top} \mbox{Bias}[\hat{\bbeta}(\mathbf{\Delta}, \bbeta_0)] & = & \{ \mathbb{E}[\hat{\bbeta}(\mathbf{\Delta}, \bbeta_0)] - \bbeta \}^{\top} \{ \mathbb{E}[\hat{\bbeta}(\mathbf{\Delta}, \bbeta_0)] - \bbeta \}
\\
& = & ( \bbeta_0 - \bbeta)^{\top} \mathbf{\Delta} (\mathbf{X}^{\top} \mathbf{W} \mathbf{X} + \mathbf{\Delta})^{-2}  \mathbf{\Delta} ( \bbeta_0 - \bbeta).
\end{eqnarray*}
This can be bounded from below and above as
\begin{eqnarray*}
d_{p} \| \bbeta_0 - \bbeta \|_2^2 \, \, \, \leq \, \, \, \{ \mbox{Bias}[\hat{\bbeta}(\mathbf{\Delta}, \bbeta_0)] \}^{\top} \mbox{Bias}[\hat{\bbeta}(\mathbf{\Delta}, \bbeta_0)]  & \leq & d_{1} \| \bbeta_0 - \bbeta \|_2^2,
\end{eqnarray*}
where $d_{p}$ and $d_{1}$ denote the smallest and the largest eigenvalues of the symmetric, positive definite matrix $\mathbf{\Delta} (\mathbf{X}^{\top} \mathbf{W} \mathbf{X} + \mathbf{\Delta})^{-2}  \mathbf{\Delta}$. Hence, the further away $\bbeta_0$ from $\bbeta$, the larger the bias of the generalized ridge regression estimator. On the other hand, should $\bbeta_0$ be close to $\bbeta$, the opposite happens and it will improve the estimator's bias.

The limiting penalization behaviour of the generalized ridge regression estimator can be deduced from these expressions. Hereto let $\mathbf{V}_{\Delta} \mathbf{D}_{\Delta} \mathbf{V}_{\Delta}^{\top}$ be the eigendecomposition of $\mathbf{\Delta}$ and $d_{\Delta,j} = (\mathbf{D}_{\Delta})_{jj}$. Furthermore, define (with a horrible abuse of notation) $\lim_{\mathbf{\Delta} \rightarrow \infty}$ as the limit of all $d_{\Delta,j}$ simultaneously tending to infinity. It is then straightforwardly seen that $\lim_{\mathbf{\Delta} \rightarrow \infty} \mathbb{E}[\hat{\bbeta}(\mathbf{\Delta}, \bbeta_0)] = \bbeta_0$ and $\lim_{\mathbf{\Delta} \rightarrow \infty} \mbox{Var}[\hat{\bbeta}(\mathbf{\Delta}, \bbeta_0)] = \mathbf{0}_{pp}$. Hence, this behavior mimicks that of the `regular' ridge regression case, in the sense \textit{i)} the bias increases (should the target $\bbeta_0$ be off) and \textit{ii)} the variance vanishes. 
\\
\\
The `regular' and generalized ridge regression estimators can now be compared in terms of their moments. Which estimator is to be preferred, depends on the particulars of the case at hand (as can be witnessed from the next example).

\begin{example} \mbox{ } \label{example.gRidge2orthonormalDesign}
\\
Let $\mathbf{X}$ be an $n \times p$-dimensional, orthonormal design matrix with $p \geq 2$. Contrast the regular and generalized ridge regression estimator, the latter with $\mathbf{W} = \mathbf{I}_{pp}$, $\bbeta_0 = \mathbf{0}_p$ and $\mathbf{\Delta} = \lambda [(1-\rho) \mathbf{I}_{pp} + \rho \mathbf{1}_{pp}]$ for $\rho \in (-(p-1)^{-1}, 1)$. For $\rho =0$, the two estimators coincide. The variance of the generalized ridge regression estimator then is $\mbox{Var}[ \hat{\bbeta}(\mathbf{\Delta}, \bbeta_0 = \mathbf{0}_p)] = (\mathbf{I}_{pp} + \mathbf{\Delta})^{-2}$. The efficiency of this estimator, measured by its generalized variance, is:
\begin{eqnarray*}
\det \{ \mbox{Var}[ \hat{\bbeta}(\mathbf{\Delta}, \bbeta_0 = \mathbf{0}_p)] \} & = & \{ [1 + \lambda + (p-1) \rho]  (1 + \lambda-\rho)^{p-1} \}^{-2}.
\end{eqnarray*}
This efficiency attains its minimum at $\rho = 0$. In the present case, the regular ridge regression estimator is thus more efficient than its generalized counterpart.
\end{example}

\noindent
In the particular case of the generalized ridge regression estimator with un- and penalized covariates, the moments are as in the next lemma.
\begin{lemma}  \label{lemma:momentsGenRidgeEstimator} \mbox{ } \\
The generalized ridge regression estimator (\ref{form:genRidgeEstimatorWithUnAndPenCovariates}) of the linear regression model $\mathbf{Y} = \mathbf{U} \ggamma + \mathbf{X} \bbeta + \vvarepsilon$ with $\vvarepsilon \sim \mathcal{N}(\mathbf{0}_{n}, \sigma^2 \mathbf{I}_{nn})$ is normally distributed with mean:
\begin{eqnarray*}
\mathbb{E}  \left(
\hspace{-0.1cm}
\begin{array}{l}
\hat{\ggamma}(\cdot)
\\
\hat{\bbeta}(\cdot)
\end{array} \hspace{-0.1cm}
\right) 
& = & \left(
\begin{array}{l}
\ggamma + \{\mathbf{U}^{\top} [ \mathbf{I}_{nn} -  \mathbf{X} (\mathbf{X}^{\top} \mathbf{X} + \mathbf{\Delta} )^{-1} \mathbf{X}^{\top}  ] \mathbf{U}\}^{-1} \mathbf{U}^{\top} \mathbf{X}  (\mathbf{X}^{\top} \mathbf{X} +  \mathbf{\Delta} )^{-1} \mathbf{\Delta}  (\bbeta - \bbeta_0)
\\
\bbeta -   \{\mathbf{X}^{\top} [\mathbf{I}_{nn} - \mathbf{U} (\mathbf{U}^{\top}  \mathbf{U})^{-1} \mathbf{U}^{\top}] \mathbf{X} +  \mathbf{\Delta} \}^{-1}   \mathbf{\Delta}  (\bbeta - \bbeta_0)
\end{array} 
\right)
\end{eqnarray*}
and (co)variances:
\begin{eqnarray*}
\mbox{Var}[\hat{\ggamma}(\cdot)] & = &  \sigma^2 \mathbf{E}^{-1} \mathbf{U}^{\top}  ( \mathbf{X} \mathbf{\Delta}^{-1} \mathbf{X}^{\top} + \mathbf{I}_{nn})^{-2}  \mathbf{U} \mathbf{E}^{-1},
\\
\mbox{Cov}[\hat{\ggamma}(\cdot), \hat{\bbeta}(\cdot)] & = & \sigma^2 \mathbf{E}^{-1} \mathbf{U}^{\top} (\mathbf{X} \mathbf{\Delta}^{-1} \mathbf{X}^{\top} + \mathbf{I}_{nn})^{-1} \mathbf{X} \mathbf{D}^{-1}  - \mbox{Var}[\hat{\ggamma}(\cdot)] \mathbf{U}^{\top}  \mathbf{X} \mathbf{D}^{-1},
\\
\mbox{Var}[\hat{\bbeta}(\cdot)] & = & \sigma^2 \mathbf{D}^{-1}  \mathbf{X}^{\top} \mathbf{X} \mathbf{D}^{-1}  - \mbox{Cov}[\hat{\bbeta}(\cdot), \hat{\ggamma}(\cdot)] \mathbf{U}^{\top} \mathbf{X} \mathbf{D}^{-1}
\\
& & \, \, -  \mathbf{D}^{-1}  \mathbf{X}^{\top}  \mathbf{U} \mbox{Cov}[\hat{\ggamma}(\cdot), \hat{\bbeta}(\cdot)]  -  \mathbf{D}^{-1}  \mathbf{X}^{\top}  \mathbf{U} \mbox{Var}[ \hat{\ggamma}(\cdot)] \mathbf{U}^{\top}  \mathbf{X} \mathbf{D}^{-1},
\end{eqnarray*}
where  $\mathbf{E} =  \mathbf{U}^{\top} (  \mathbf{I}_{nn} +  \mathbf{X} \mathbf{\Delta}^{-1} \mathbf{X}^{\top} )^{-1}  \mathbf{U}$ and $\mathbf{D} = \mathbf{X}^{\top} \mathbf{X} + \mathbf{\Delta}$.
\end{lemma}
\begin{proof} The normality follows from the linearity of the estimator. The expectation and covariances can be derived  straightforwardly using linear algebra, the rules of the expectation and variance calculus, the fact that $ \mathbf{Y} \sim \mathcal{N}( \mathbf{U} \ggamma + \mathbf{X} \bbeta, \sigma^2 \mathbf{I}_{nn})$, the Woodbury matrix identity, and the analytic expression of the inverse of a $2 \times 2$ block matrix.
\end{proof}

\noindent
Should there be no covariates left unpenalized, the expectation and variance of the generalized ridge regression estimator provided in Lemma \ref{lemma:momentsGenRidgeEstimator} reduce to those given in Displays (\ref{form:genRidgeEstimatorExpectation}) and (\ref{form:genRidgeEstimatorVariance}).

Lemma \ref{lemma:momentsGenRidgeEstimator} reveals that the moments of the estimators of the regression parameters $\ggamma$ and $\bbeta$, associated with the un- and penalized covariates, respectively, are related. This is to be expected as both contribute to the explanation of the response. Their moments are communicating vessels. For instance, if the estimator of $\bbeta$ is off, the other one will try to compensate. The relation between the bias of these estimators is detailed in Corollary \ref{corr:biasGenRidgeEstimator}

\begin{corollary} \label{corr:biasGenRidgeEstimator} \mbox{ }
\\
Consider the generalized ridge regression estimator (\ref{form:genRidgeEstimatorWithUnAndPenCovariates}) of the linear regression model $\mathbf{Y} = \mathbf{U} \ggamma + \mathbf{X} \bbeta + \vvarepsilon$ with $\vvarepsilon \sim \mathcal{N}(\mathbf{0}_{n}, \sigma^2 \mathbf{I}_{nn})$. The bias of the individual regression parameters' estimators and linear predictors are related as:
\begin{eqnarray*}
\mathbb{E}[\hat{\ggamma}(\cdot)] - \ggamma  & = & - (\mathbf{U}^{\top}  \mathbf{U})^{-1} \mathbf{U}^{\top} \mathbf{X} \{ \mathbb{E}[ \hat{\bbeta}(\cdot) ] - \bbeta \}, 
\\
\mathbb{E}[\mathbf{U} \hat{\ggamma}(\cdot)] - \mathbf{U}  \ggamma  & = &  - \mathbf{P}_u \{ \mathbb{E}[ \mathbf{X} \hat{\bbeta}(\cdot) ] - \mathbf{X}\bbeta \},
\end{eqnarray*}  
respectively, where $\mathbf{P}_u = \mathbf{U} (\mathbf{U}^{\top}  \mathbf{U})^{-1} \mathbf{U}^{\top}$. 
\end{corollary}

\begin{proof}
Reformulate the expectation of $\hat{\bbeta} (\cdot)$ to:
\begin{eqnarray} 
\label{form.expBeta2Gamma}
(\mathbf{X}^{\top} \mathbf{X}  + \mathbf{\Delta})^{-1}  \mathbf{\Delta}  (\bbeta - \bbeta_0) & = & \bbeta - \mathbb{E}[ \hat{\bbeta}(\cdot) ]  +  (\mathbf{X}^{\top} \mathbf{X} +  \mathbf{\Delta})^{-1} \mathbf{X}^{\top} \mathbf{U} \{ \ggamma - \mathbb{E}[ \hat{\ggamma}(\cdot) ] \}. 
\end{eqnarray}
Substitute this into the expectation of $\hat{\ggamma} (\cdot)$, pre-multiply the both sides of the preceeding display by   $\mathbf{U}^{\top}  [ \mathbf{I}_{nn} - \mathbf{X} (\mathbf{X}^{\top} \mathbf{X} +  \mathbf{\Delta} )^{-1} \mathbf{X}^{\top} ] \mathbf{U}$, and arrive at 
\begin{eqnarray*}
& & \hspace{-1.5cm} \{\mathbf{U}^{\top}  [ \mathbf{I}_{nn} - \mathbf{X} (\mathbf{X}^{\top} \mathbf{X} +  \mathbf{\Delta} )^{-1} \mathbf{X}^{\top} ] \mathbf{U}\} \{ \mathbb{E}[\hat{\ggamma}(\cdot)] - \ggamma \} 
\\
& = &  \mathbf{U}^{\top} \mathbf{X} \{ \bbeta - \mathbb{E}[ \hat{\bbeta}(\cdot) ] \} + \mathbf{U}^{\top} \mathbf{X}(\mathbf{X}^{\top} \mathbf{X} + \mathbf{\Delta})^{-1} \mathbf{X}^{\top} \mathbf{U} \{ \ggamma - \mathbb{E}[ \hat{\ggamma}(\cdot) ] \}.
\end{eqnarray*}
Or, after simplication: $\mathbb{E}[\hat{\ggamma}(\cdot)] = \ggamma - (\mathbf{U}^{\top}  \mathbf{U})^{-1} \mathbf{U}^{\top} \mathbf{X} \{ \mathbb{E}[ \hat{\bbeta}(\cdot) ] - \bbeta \}$. The above thus gives the relation between the bias of both estimators. In particular, if $\mathbf{X} = \mathbf{U}$, we have $\mathbb{E}[ \hat{\ggamma}(\cdot)] - \ggamma = - \{ \mathbb{E}[ \hat{\bbeta}(\cdot)] -   \bbeta \}$. Furthermore, if both sides are pre-multiplied by $\mathbf{U}$, we get the bias of the linear predictor $\mathbb{E}[\mathbf{U} \hat{\ggamma}(\cdot)] - \mathbf{U}  \ggamma =  -\mathbf{U} (\mathbf{U}^{\top}  \mathbf{U})^{-1} \mathbf{U}^{\top}  \{ \mathbb{E}[ \mathbf{X} \hat{\bbeta}(\cdot) ] - \mathbf{X}\bbeta \} =   - \mathbf{P}_u \{ \mathbb{E}[ \mathbf{X} \hat{\bbeta}(\cdot) ] - \mathbf{X}\bbeta \}$.
\end{proof}

\noindent
The relationships among the bias and covariances between $\hat{\ggamma}(\cdot)$ and $\hat{\bbeta}(\cdot)$, as stated in Lemma \ref{lemma:momentsGenRidgeEstimator} and Corollary \ref{corr:biasGenRidgeEstimator}, vanish if the columns of $\mathbf{U}$ are orthogonal to those of $\mathbf{X}$, i.e. $\mathbf{U}^{\top} \mathbf{X} = \mathbf{0}_{qp}$. Put differently, the bias in one does no longer affect the other. In particular, $\hat{\ggamma} (\cdot)$ is an unbiased estimator of $\ggamma$. Furthermore, the covariance between the two estimators vanishes. And, for instance, the variance of $\hat{\bbeta}(\cdot)$ equals that specified in Equation \ref{form:genRidgeEstimatorVariance}. 

\subsection{Mean squared error}
The mean squared error of the generalized ridge regression estimator is now readily obtained from its moments provided in Displays \ref{form:genRidgeEstimatorExpectation} and \ref{form:genRidgeEstimatorVariance}. We may then hope for a result in a similar vain like Theorem \ref{chap:genRidge}. The next proposition indeed shows that -- under mild conditions -- the generalized ridge regression estimator, like its regular counterpart, outperforms the maximum likelihood one in terms of the mean squared error.

\begin{proposition} \label{prop:MSEofGeneralizedRidge} \mbox{ } \\
Assume the ridge penalty matrix is parametrized by a single parameter $\lambda$, denoted by $\mathbf{\Delta}(\lambda)$, such that \textit{i)} $\mathbf{\Delta}(\lambda) = \mathbf{0}_{pp}$ for $\lambda=0$ and \textit{ii)} $ \mathbf{\Delta}(\lambda) = \lambda \mathbf{\Delta}_0  + \mathcal{O}(\lambda^2)$ for $0 < \lambda \ll 1$ with non-trivial $\mathbf{\Delta}_0 \succeq 0$. There exists a $\lambda > 0$ for which $\mbox{MSE}\{ \hat{\bbeta}[\mathbf{\Delta}(\lambda), \bbeta_0]\} < \mbox{MSE}(\hat{\bbeta}^{\mbox{{\tiny ml}}})$.
\end{proposition}
\begin{proof} We prove that $\tfrac{d}{d \lambda}  \mbox{MSE}\{ \hat{\bbeta}[\mathbf{\Delta}(\lambda), \bbeta_0]\} \big|_{\lambda=0} < 0$. To evaluate the derivative we use the following results from matrix calculus:
\begin{eqnarray*}
\tfrac{d}{d\lambda} \mbox{tr} [\mathbf{A}(\lambda)] \, \, \, = \, \, \, \mbox{tr} \big[ \tfrac{d}{d\lambda} \mathbf{A} (\lambda) \big] \quad \mbox{and} \quad \tfrac{d}{d\lambda} [\mathbf{B}^{\top} + \mathbf{A} (\lambda)]^{-1} \, \, \, = \, \, \, - [\mathbf{B} + \mathbf{A} (\lambda)]^{-1}  \big[ \tfrac{d}{d\lambda} \mathbf{A} (\lambda) \big]  [\mathbf{B} + \mathbf{A} (\lambda)]^{-1}
\end{eqnarray*}
for symmetric matrices $\mathbf{A}(\lambda)$ and $\mathbf{B}$ of equal dimensions such that $\mathbf{B} + \mathbf{A}(\lambda)$ is invertible. Now take the derivative of the generalized ridge regression estimator's bias and variance with respect to the penalty parameter and evaluate these derivatives at zero:
\begin{eqnarray*} 
\tfrac{d}{d \lambda} \mbox{Bias}[\hat{\bbeta}(\mathbf{\Delta}, \bbeta_0)] \big|_{\lambda=0} & = & 2 ( \bbeta_0 - \bbeta)^{\top} \{ \mathbf{I}_{pp} - \mathbf{\Delta} (\lambda)  [\mathbf{X}^{\top} \mathbf{W} \mathbf{X} +  \mathbf{\Delta}(\lambda)]^{-1} \}
\\
& & \textcolor{white}{2} \times \big[ \tfrac{d}{d \lambda}  \mathbf{\Delta} (\lambda) \big] [\mathbf{X}^{\top} \mathbf{W} \mathbf{X} + \mathbf{\Delta} (\lambda) ]^{-2} \mathbf{\Delta} (\lambda) ( \bbeta_0 - \bbeta) \big|_{\lambda=0}
\, \, \, = \, \, \, 0
\end{eqnarray*}
and
\begin{eqnarray*} 
\tfrac{d}{d \lambda} \mbox{tr} \big( \mbox{Var}\{ \hat{\bbeta}[\mathbf{\Delta}(\lambda), \bbeta_0]\} \big) \big|_{\lambda=0}  & = &  - 2 \sigma^2 \mbox{tr} \big\{ \mathbf{X}^{\top} \mathbf{W}^2 \mathbf{X}  [\mathbf{X}^{\top} \mathbf{W} \mathbf{X} +  \mathbf{\Delta}(\lambda)]^{-2}  
\\
& & \textcolor{white}{-2} \, \, \qquad \times \big[ \tfrac{d}{d \lambda}  \mathbf{\Delta} (\lambda) \big] [\mathbf{X}^{\top} \mathbf{W} \mathbf{X} + \mathbf{\Delta} (\lambda) ]^{-1}  \big\}  \big|_{\lambda=0}
\\
& = &  - 2 \sigma^2 \mbox{tr} [  \mathbf{X}^{\top} \mathbf{W}^2 \mathbf{X}  (\mathbf{X}^{\top} \mathbf{W} \mathbf{X} )^{-2} \mathbf{\Delta}_0 (\mathbf{X}^{\top} \mathbf{W} \mathbf{X} )^{-1}  ], 
\end{eqnarray*}
where we have used that $\mbox{tr}(\mathbf{A} \mathbf{B} \mathbf{C}) = \mbox{tr}(\mathbf{A} \mathbf{C} \mathbf{B}) = \mbox{tr}(\mathbf{C} \mathbf{A} \mathbf{B}) = \mbox{tr}(\mathbf{C} \mathbf{B} \mathbf{A}) = \mbox{tr}(\mathbf{B} \mathbf{C} \mathbf{A}) = \mbox{tr}(\mathbf{B} \mathbf{A} \mathbf{C})$ if $\mathbf{A}$, $\mathbf{B}$, $\mathbf{C}$ are all symmetric. Put together and use the properties of the involved matrices to conclude that the derivative of the $\mbox{MSE}\{ \hat{\bbeta}[\mathbf{\Delta}(\lambda), \bbeta_0]\}$ with respect to $\lambda$ at zero is negative. This indicates that there is a positive penalty parameter value that yields a lower MSE of the generalized ridge regression estimator than that of its maximum likelihood counterpart.
\end{proof}

\noindent
Proposition \ref{prop:MSEofGeneralizedRidge} holds independently of the choice of the target $\bbeta_0$.  This is not at odds with the previously observed increase in bias for targets further away from the true parameter value. The proof of Proposition \ref{prop:MSEofGeneralizedRidge} reveals that the derivative of the $\hat{\bbeta}[\mathbf{\Delta}(\lambda), \bbeta_0]$'s mean squared error with respect to $\lambda$ evaluated at zero does not depend on the bias. Nonetheless, the further $\bbeta_0$ is off, the larger the bias. That results in a smaller interval of the penalty parameter where the superior mean squared error of the generalized ridge regression estimator is achieved. In contrast, if $\bbeta_0$ is close to $\bbeta$ a superior mean squared error performance of the generalized ridge regression estimator over its maximum likelihood counterpart is achieved irrespectively of the choice of the penalty parameter (see the next example). 
	
\begin{example} \textit{(MSE with perfect target)}
\\ 
Set $\bbeta_0 = \bbeta$, i.e. the target is equal to the true value of the regression parameter. Then:
\begin{eqnarray*}
\mathbb{E}[\hat{\bbeta}(\mathbf{\Delta}, \bbeta_0 = \bbeta)] & = & (\mathbf{X}^{\top} \mathbf{W} \mathbf{X} + \mathbf{\Delta})^{-1} (\mathbf{X}^{\top} \mathbf{W} \mathbf{X} \bbeta + \mathbf{\Delta} \bbeta) \, \, \, = \, \, \, \bbeta.
\end{eqnarray*}
Hence, irrespective of the choice of $\mathbf{\Delta}$, the generalized ridge estimator is then unbiased. Thus:
\begin{eqnarray*}
\mbox{MSE}[\hat{\bbeta}(\mathbf{\Delta}, \bbeta_0 = \bbeta)] & = & \mbox{tr} \{
\mbox{Var}[\hat{\bbeta}(\mathbf{\Delta}, \bbeta_0 = \bbeta)] \}
\\
& = & \mbox{tr}[ \sigma^{2} (\mathbf{X}^{\top} \mathbf{W} \mathbf{X} + \mathbf{\Delta})^{-1} \mathbf{X}^{\top} \mathbf{W}^2 \mathbf{X} (\mathbf{X}^{\top} \mathbf{W} \mathbf{X} + \mathbf{\Delta})^{-1}]
\\
& = & \sigma^2 \mbox{tr}[ \mathbf{X}^{\top} \mathbf{W}^2 \mathbf{X} (\mathbf{X}^{\top} \mathbf{W} \mathbf{X} + \mathbf{\Delta})^{-2}].
\end{eqnarray*}
When $\mathbf{\Delta} = \lambda \mathbf{I}_{pp}$, this MSE is smaller than that of the ML regression estimator, irrespective of the choice of $\lambda$. 
\end{example}

\section{Degrees of freedom}
The degrees of freedom consumed by the generalized ridge regression estimator is 
\begin{eqnarray*}
\mbox{df}[  \hat{\bbeta} (\mathbf{\Delta}, \bbeta_0)] & = & \sum\nolimits_{i=1}^n [ \mbox{Var} (Y_i )]^{-1} \mbox{Cov} (Y_i, \hat{Y}_i) \, \, \, = \, \, \, \mbox{tr} [(\mathbf{X}^{\top} \mathbf{W} \mathbf{X} + \mathbf{\Delta})^{-1} \mathbf{X}^{\top} \mathbf{W} \mathbf{X}]. 
\end{eqnarray*}
This coincides with the degrees of freedom of the regular ridge regression with $\mathbf{W}= \mathbf{I}_{nn}$ and $\mathbf{\Delta} = \lambda \mathbf{I}_{pp}$. The degrees of freedom decreases as the penalization increases. In particular,  if $\mathbf{\Delta} = \lambda \mathbf{\Delta}_0$ for large $\lambda$ and $\mathbf{\Delta}_0$ positive definite, then $\lim_{\lambda \rightarrow \infty} \mbox{df}[  \hat{\bbeta} (\mathbf{\Delta}, \bbeta_0)] = 0$, which reflects the fact that the estimator's shrinkage limit $\bbeta_0$ is nonrandom. In contrast, if $\mathbf{\Delta}_0$ is nonnegative definite, the regularization limit still involves $\mathbf{Y}$ and
\begin{eqnarray*}
\lim_{\lambda \rightarrow \infty} \mbox{df}[  \hat{\bbeta} (\lambda \mathbf{\Delta}_0, \bbeta_0)] & = & \lim_{\lambda \rightarrow \infty} 
\sum\nolimits_{i=1}^n [ \mbox{Var} (Y_i) ]^{-1} \mbox{Cov} [Y_i, \mathbf{X}_{i,\ast} \hat{\bbeta} (\lambda \mathbf{\Delta}_0, \bbeta_0)]
\\
& = & \sum\nolimits_{i=1}^n [ \mbox{Var} (Y_i) ]^{-1} \mbox{Cov} \left[ Y_i, \mathbf{X}_{i,\ast} \mathbf{V}_{\Delta,=0} \{ \mathbf{V}_{\Delta,=0}^{\top} \mathbf{X}^{\top} \mathbf{X} \mathbf{V}_{\Delta,=0}  \}^{-1}  \mathbf{V}_{\Delta,=0}^{\top} \mathbf{X}^{\top}  \mathbf{Y} 
\right]
\\
& = & \sum\nolimits_{i=1}^n  \mathbf{X}_{i,\ast} \mathbf{V}_{\Delta,=0} \{ \mathbf{V}_{\Delta,=0}^{\top} \mathbf{X}^{\top} \mathbf{X} \mathbf{V}_{\Delta,=0}  \}^{-1}  \mathbf{V}_{\Delta,=0}^{\top} \mathbf{X}_{i,\ast}^{\top}  
\\
& = & \max \{0, n - p + \mbox{rank}(\mathbf{\Delta})\}.
\end{eqnarray*}
Hence, the degrees of freedom of the generalized ridge regression estimator does not necessarily vanish if $\mathbf{\Delta}_0$ is nonnegative definite.

\section{The Bayesian connection} \label{sect:genRidgeBayes}
This generalized ridge regression estimator can, like the regular ridge regression estimator, be viewed as a Bayesian estimator. It requires to replace the conjugate prior on $\bbeta$ by a more general normal law, $\bbeta \sim \mathcal{N}(\bbeta_0, \sigma^2 \mathbf{\Delta}^{-1})$, but retains the inverse gamma prior on $\sigma^2$. The joint posterior distribution of $\bbeta$ and $\sigma^2$ is then obtained analogously to the derivation of posterior with a standard normal prior on $\bbeta$ as presented in Chapter \ref{chap:BayesianRegression} (the details are left as Exercise \ref{question.generalizedRidgeAndBayes}):
\begin{eqnarray*}
f_{\bbeta, \sigma^2} (\bbeta, \sigma^2 \, | \, \mathbf{Y}, \mathbf{X}) & = & f_Y (\mathbf{Y} \, | \, \mathbf{X}, \bbeta, \sigma^2) \, f_{\beta | \sigma^2}(\bbeta \, | \, \sigma^2) \, f_{\sigma}(\sigma^2)
\\
& \propto & g_{\bbeta | \sigma^2, \mathbf{Y}, \mathbf{X} } (\bbeta \, | \, \sigma^2, \mathbf{Y}, \mathbf{X}) \,
g_{\sigma^2 |  \mathbf{Y}, \mathbf{X}} (\sigma^2 \, | \, \mathbf{Y}, \mathbf{X})
\end{eqnarray*}
with
\begin{eqnarray*}
g_{\bbeta | \sigma^2, \mathbf{Y}, \mathbf{X}} (\bbeta \, | \, \sigma^2, \mathbf{Y}, \mathbf{X})
& \propto & \exp \{ - \tfrac{1}{2} \sigma^{-2} \big[ \bbeta - \hat{\bbeta}(\mathbf{\Delta}) \big]^{\top} (\mathbf{X}^{\top} \mathbf{X} +  \mathbf{\Delta}) \big[ \bbeta - \hat{\bbeta}(\mathbf{\Delta}) \big] \}.
\end{eqnarray*}
This implies $\mathbb{E}(\bbeta  \, | \, \sigma^2, \mathbf{Y}, \mathbf{X}) = \hat{\bbeta}(\mathbf{\Delta}, \bbeta_0)$. Hence, the generalized ridge regression estimator too can be viewed as the Bayesian posterior mean estimator of $\bbeta$ when imposing a multivariate Gaussian prior on the regression parameter.

The Bayesian formulation of our generalization of ridge regression provides additional intuition of the penalty parameters $\bbeta_0$ and $\mathbf{\Delta}$. For instance, a better initial guess, i.e. $\bbeta_0$, yields a better posterior mean and mode. Or, less uncertainty in the prior, i.e. a smaller (in the positive definite ordering sense) variance $\mathbf{\Delta}$, yields a more concentrated posterior. These claims are underpinned in Question \ref{question.consequencesOfHyperparameters}. The intuition they provide guides in the choice of penalty parameters $\bbeta_0$ and $\mathbf{\Delta}$.

We can construct a Gibbs sampler to draw from the generalized ridge regression estimator's posterior distribution. High-dimensionally, the main obstacle is then to draw from the conditional distribution of $\bbeta$: 
\begin{eqnarray} \label{form:sampleFromGenRidgeDist}
\bbeta \, | \, \sigma^2, \mathbf{Y}, \mathbf{X} & \sim & \mathcal{N} [ (\mathbf{X}^{\top} \mathbf{X} + \mathbf{\Delta})^{-1} (\mathbf{X}^{\top} \mathbf{Y} + \mathbf{\Delta} \bbeta_0), \sigma^2 (\mathbf{X}^{\top} \mathbf{X} + \mathbf{\Delta})^{-1}].
\end{eqnarray}
This can be done computational efficiently. Hereto let $\mathbf{L}$ be the lower triangular matrix of the Cholesky decomposition of $\mathbf{\Delta}$ such that $\mathbf{\Delta} = \mathbf{L} \mathbf{L}^{\top}$. The pseudo-code of the efficient algorithm is provided in the next algorithm. 
\\
\\
\vspace{-0.54cm}
\\
\begin{minipage}[h]{\textwidth}
\begin{center}
\fbox{\parbox{14cm}{
\begin{algorithm}[H]
\SetKwInOut{Input}{input}
\SetKwInOut{Output}{output}
\RestyleAlgo{boxed} 
\LinesNumbered
\Input{design matrix $\mathbf{X}$, response vector $\mathbf{Y}$, variance $\sigma^2$, and \\ penalty matrix $\mathbf{\Delta}$ and target $\bbeta_0$.
}
\Output{a draw from distribution (\ref{form:sampleFromGenRidgeDist}).}
\vspace{0.25cm}
{Sample $\mathbf{u} \sim \mathcal{N}(\mathbf{L}^{\top} \bbeta_0, \sigma^2  \mathbf{I}_{pp})$ and $\ddelta \sim \mathcal{N} ( \mathbf{0}_n, \mathbf{I}_{nn})$ independently}.
\\
{Solve $\mathbf{L}^{\top} \tilde{\mathbf{u}} =  \mathbf{u}$ for $\tilde{\mathbf{u}}$ by backward substitution}.
\\
{Set $\mathbf{v} =  \sigma^{-1} \mathbf{X} \tilde{\mathbf{u}} + \ddelta$}.
\\
{Solve $(\mathbf{X} \mathbf{\Delta}^{-1} \mathbf{X}^{\top} + \mathbf{I}_{nn}) \mathbf{w} = \sigma^{-1} \mathbf{Y} - \mathbf{v}$ for $\mathbf{w}$}.
\\
{Set $\bbeta = \tilde{\mathbf{u}} + \sigma \mathbf{\Delta}^{-1} \mathbf{X}^{\top} \mathbf{w}$}.
\vspace{0.25cm}
\caption{{\small High-dimensional efficient sampling of distribution (\ref{form:sampleFromGenRidgeDist}).}} \label{alg.highdimNormalGenRidge}
\end{algorithm}
} 
}
\end{center}
\end{minipage}
\mbox{ }
\\
\mbox{ }
\vspace{-0.2cm}
\\
The proof that $\bbeta$, if sampled in accordance with Algorithm \ref{alg.highdimNormal}, is distributed as in accordance with the law in Display (\ref{form:sampleFromGenRidgeDist}) is analogous to that of Proposition \ref{prop:highdimNormSampling} and left as an exercise.

\section{Variants}
We illustrate the generalized ridge regression estimator through the presentation of its various usages that have appeared in the literature. They all boil down to a particular choice of the target $\bbeta_0$ and the ridge penalty matrix $\mathbf{\Delta}$. These choices are motivated from the context.

\subsection{Eigenvalue-wise regularization}
What is historically referred to as `generalized ridge regression' (cf. \citealp{Hoer1970, Hemm1975}) is the particular case of loss function (\ref{form:generalizedRidgeLoss}) in which $\mathbf{W} = \mathbf{I}_{nn}$,  $\bbeta_0 = \mathbf{0}_{p}$, and $\mathbf{\Delta} = \mathbf{V}_{x} \mathbf{\Lambda} \mathbf{V}_x^{\top}$, where $\mathbf{V}_x$ is obtained from the singular value decomposition of $\mathbf{X}$ (i.e., $\mathbf{X} = \mathbf{U}_{x} \mathbf{D}_x \mathbf{V}_x^{\top}$ with its constituents endowed with the usual interpretation) and $\mathbf{\Lambda}$ a positive definite diagonal matrix. This gives the estimator:
\begin{eqnarray*}
\hat{\bbeta}(\mathbf{\Lambda}) & = & (\mathbf{X}^{\top} \mathbf{X} + \mathbf{\Delta})^{-1} \mathbf{X}^{\top} \mathbf{Y}
\\
& = &  (\mathbf{V}_x \mathbf{D}_x^{\top} \mathbf{U}_x^{\top} \mathbf{U}_x \mathbf{D}_x \mathbf{V}_x^{\top} + \mathbf{V}_x \mathbf{\Lambda} \mathbf{V}_x^{\top})^{-1} \mathbf{V}_x \mathbf{D}_x^{\top} \mathbf{U}_x^{\top} \mathbf{Y}
\\
& = & \mathbf{V}_x (\mathbf{D}_x^{\top} \mathbf{D}_x + \mathbf{\Lambda})^{-1} \mathbf{D}_x^{\top} \mathbf{U}_x^{\top} \mathbf{Y}.
\end{eqnarray*}
From this last expression it becomes clear how this estimator generalizes the `regular ridge estimator'. The latter shrinks all eigenvalues, irrespectively of their size, in the same manner through a common penalty parameter. The `generalized ridge estimator', through differing penalty parameters (i.e. the diagonal elements of $\mathbf{\Lambda}$), shrinks them individually.

The generalized ridge estimator coincides with the Bayesian linear regression estimator with the normal prior $\mathcal{N}[\mathbf{0}_p, (\mathbf{V}_x \mathbf{\Lambda} \mathbf{V}_x^{\top})^{-1}]$ on the regression parameter $\bbeta$ (and preserving the inverse gamma prior on the error variance). Assume $\mathbf{X}$ to be of full column rank and choose $\mathbf{\Lambda} = g^{-1} \mathbf{D}_x^{\top} \mathbf{D}_x$ with $g$ a positive scalar. The prior on $\bbeta$ then -- assuming $(\mathbf{X}^{\top} \mathbf{X})^{-1}$ exists -- reduces to Zellner's $g$-prior: $\bbeta \sim \mathcal{N}[\mathbf{0}_p, g (\mathbf{X}^{\top} \mathbf{X})^{-1}]$ \citep{Zell1986}. The corresponding posterior mean estimator of the regression coefficient is: $\hat{\bbeta}(g)  = g (1+g)^{-1} (\mathbf{X}^{\top} \mathbf{X})^{-1} \mathbf{X}^{\top} \mathbf{Y}$, which is proportional to the unpenalized ordinary least squares estimator of $\bbeta$.

For convenience of notation in the analysis of the generalized ridge estimator the linear regression model is usually rewritten as:
\begin{eqnarray*}
\mathbf{Y} & = & \mathbf{X} \bbeta + \vvarepsilon \, \, \, = \, \, \,
\mathbf{X} \mathbf{V}_x  \mathbf{V}_x^{\top} \bbeta + \vvarepsilon \, \, \, = \, \, \, \tilde{\mathbf{X}} \aalpha + \vvarepsilon,
\end{eqnarray*}
with $\tilde{\mathbf{X}} = \mathbf{X} \mathbf{V}_x = \mathbf{U}_x
\mathbf{D}_x$ (and thus $\tilde{\mathbf{X}}^{\top} \tilde{\mathbf{X}} = \mathbf{D}_x^{\top} \mathbf{D}_x$) and $\aalpha = \mathbf{V}_x^{\top} \bbeta$ with loss function $(\mathbf{Y} - \tilde{\mathbf{X}} \aalpha)^{\top} (\mathbf{Y} - \tilde{\mathbf{X}} \aalpha) + \aalpha^{\top} \mathbf{\Lambda} \aalpha$. In the notation above the generalized ridge estimator is then:
\begin{eqnarray*}
\hat{\aalpha}(\mathbf{\Lambda}) & = & (\tilde{\mathbf{X}}^{\top} \tilde{\mathbf{X}} + \mathbf{\Lambda})^ {-1} \tilde{\mathbf{X}}^{\top} \mathbf{Y} \, \, \, = \, \, \, (\mathbf{D}_x^{\top} \mathbf{D}_x + \mathbf{\Lambda})^{-1} \tilde{\mathbf{X}}^{\top} \mathbf{Y},
\end{eqnarray*}
from which one obtains $\hat{\bbeta}(\mathbf{\Lambda}) = \mathbf{V}_x \hat{\aalpha}(\mathbf{\Lambda})$. Using $\mathbb{E}[\hat{\aalpha}(\mathbf{\Lambda})] = (\mathbf{D}_x^{\top} \mathbf{D}_x + \mathbf{\Lambda})^{-1} \mathbf{D}_x^{\top} \mathbf{D}_x \aalpha$ and $\mbox{Var}[\hat{\aalpha}(\mathbf{\Lambda})] = \sigma^2 (\mathbf{D}_x^{\top} \mathbf{D}_x + \mathbf{\Lambda})^{-1} \mathbf{D}_x^{\top} \mathbf{D}_x (\mathbf{D}_x^{\top} \mathbf{D}_x + \mathbf{\Lambda})^{-1}$, the MSE for the generalized ridge estimator can be written as:
\begin{eqnarray*}
\mbox{MSE}[\hat{\aalpha}(\mathbf{\Lambda})] & = & \sum_{j=1}^p ( \sigma^2 d_{x,j}^2 + \alpha_j^2 \lambda_{j}^2 ) (d_{x,j}^2 +  \lambda_{j} )^{-2},
\end{eqnarray*}
where $d_{x,j} = (\mathbf{D}_x)_{jj}$ and $\lambda_j = (\mathbf{\Lambda})_{jj}$. Having $\aalpha$ and $\sigma^ 2$ available, it is easily seen (equate the derivative w.r.t. $\lambda_j$ to zero and solve) that the MSE of $\hat{\aalpha}(\mathbf{\Lambda})$ is minimized by $\lambda_j = \sigma^2 / \alpha_j^2$ for all $j$. With $\aalpha$ and $\sigma^2$ unknown, \cite{Hoer1970} suggest an iterative procedure to estimate the $\lambda_j$'s. Initiate the procedure with the OLS estimates of $\aalpha$ and $\sigma^2$, followed by sequentially updating the $\lambda_j$'s and the estimates of $\aalpha$ and $\sigma^2$. An analytic expression of the limit of this procedure exists (\citealp{Hemm1975}). This limit, however, still depends on the observed $\mathbf{Y}$ and as such it does not necessarily yield the minimal attainable value of the MSE. This limit may nonetheless still yield a potential gain in MSE. This is investigated in \cite{Lawl1981}. Under a variety of cases it seems to indeed outperform the OLS estimator, but there are exceptions.

A variation on this theme is presented by \cite{Guilkey1975} and dubbed ``directed'' ridge regression. Directed ridge regression only applies the above `generalized shrinkage' in those eigenvector directions that have a corresponding small(er) -- than some user-defined cut-off -- eigenvalue. This intends to keep the bias low and yield good (or supposedly better) performance.

\subsection{Fused ridge estimation} \label{sect:fusedRidgeEstimation}
An example of a generalized ridge penalty is the \textit{fused ridge penalty} (as introduced by \citealp{Goem2008}). Consider the standard linear model $\mathbf{Y} = \mathbf{X} \bbeta + \vvarepsilon$. The fused ridge  regression estimator of $\bbeta$ then minimizes:
\begin{eqnarray} \label{form:fusedRidgeLoss}
\| \mathbf{Y} - \mathbf{X} \bbeta \|_2^2 + \lambda \sum\nolimits_{j=2}^p ( \beta_{j} - \beta_{j-1} )^2.
\end{eqnarray}
The penalty in the loss function above can be written as a generalized ridge penalty:
\begin{eqnarray*}
\lambda \sum\nolimits_{j=2}^p ( \beta_{j} - \beta_{j-1} )^2 & = &  \bbeta^{\top}
\left(
\begin{array}{rrrrrr}
\lambda & -\lambda & 0 & \ldots & \ldots &  0
\vspace{-4pt}
\\
-\lambda & 2 \lambda & -\lambda & \ddots & & \vdots
\vspace{-4pt}
\\
0 & -\lambda & 2 \lambda & \ddots & \ddots   & \vdots
\vspace{-4pt}
\\
\vdots & \ddots & \ddots & \ddots & \ddots & 0
\vspace{-4pt}
\\
\vdots &  & \ddots & \ddots & \ddots & -\lambda
\\
0 & \ldots & \ldots & 0 & -\lambda &  \lambda
\end{array}\right)
\bbeta.
\end{eqnarray*}
The matrix $\mathbf{\Delta}$  employed above is semi-positive definite and therefore the loss function (\ref{form:fusedRidgeLoss}) need not be strictly convex. Hence, often a regular ridge penalty $\| \bbeta \|_2^2$ is added (with its own penalty parameter).

\begin{figure}[!h]
\begin{tabular}{rcl}
\includegraphics[scale=0.40, angle=0]{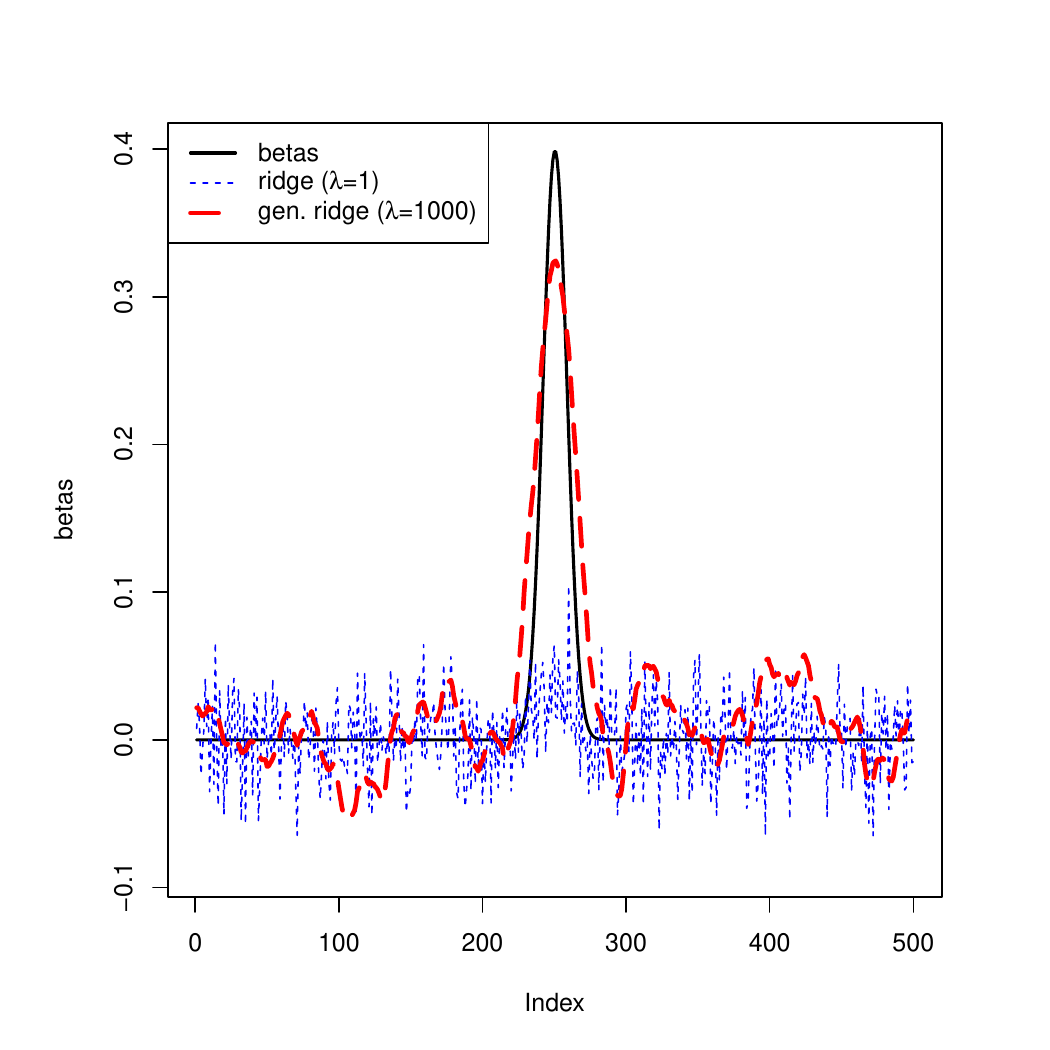}
&
\includegraphics[angle=0, scale=0.40]{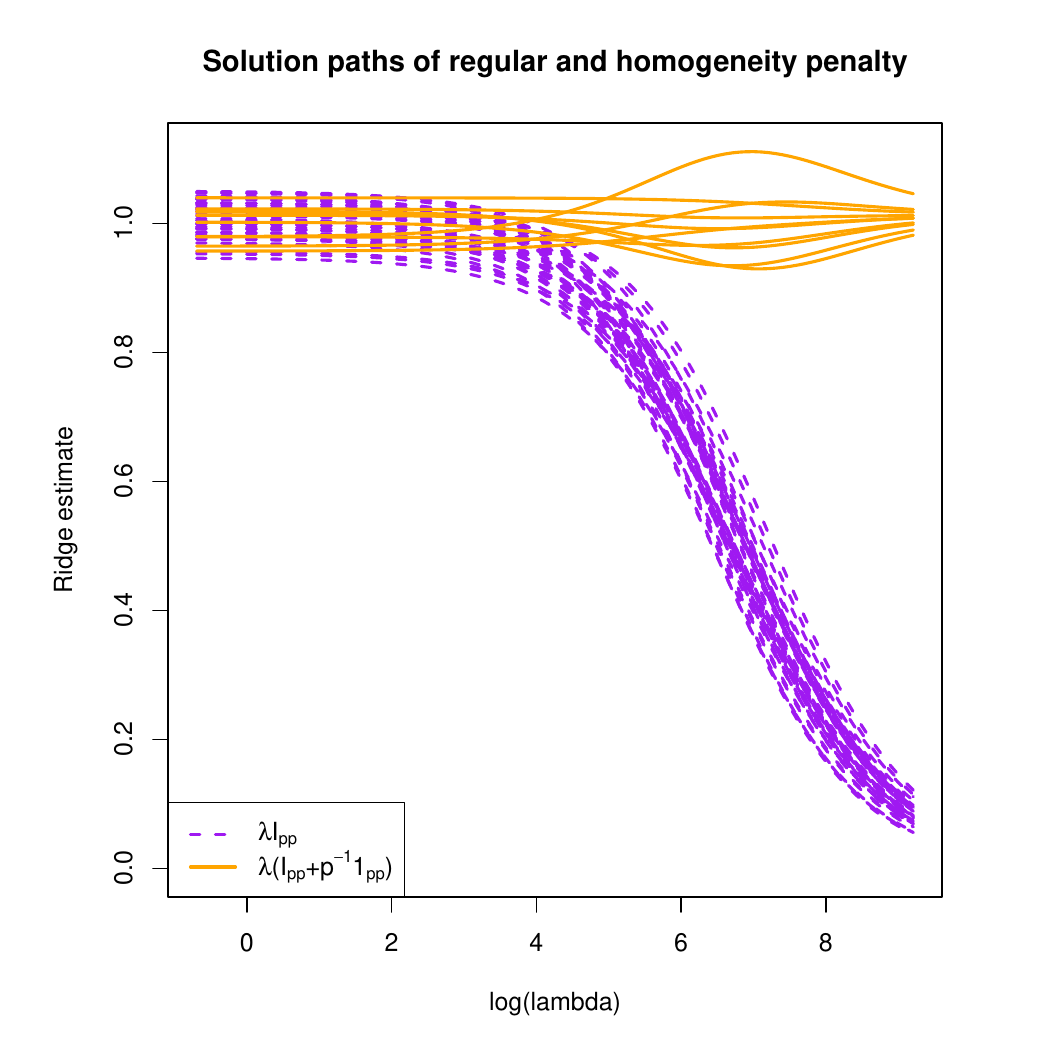}
\\
\includegraphics[angle=0, scale=0.40]{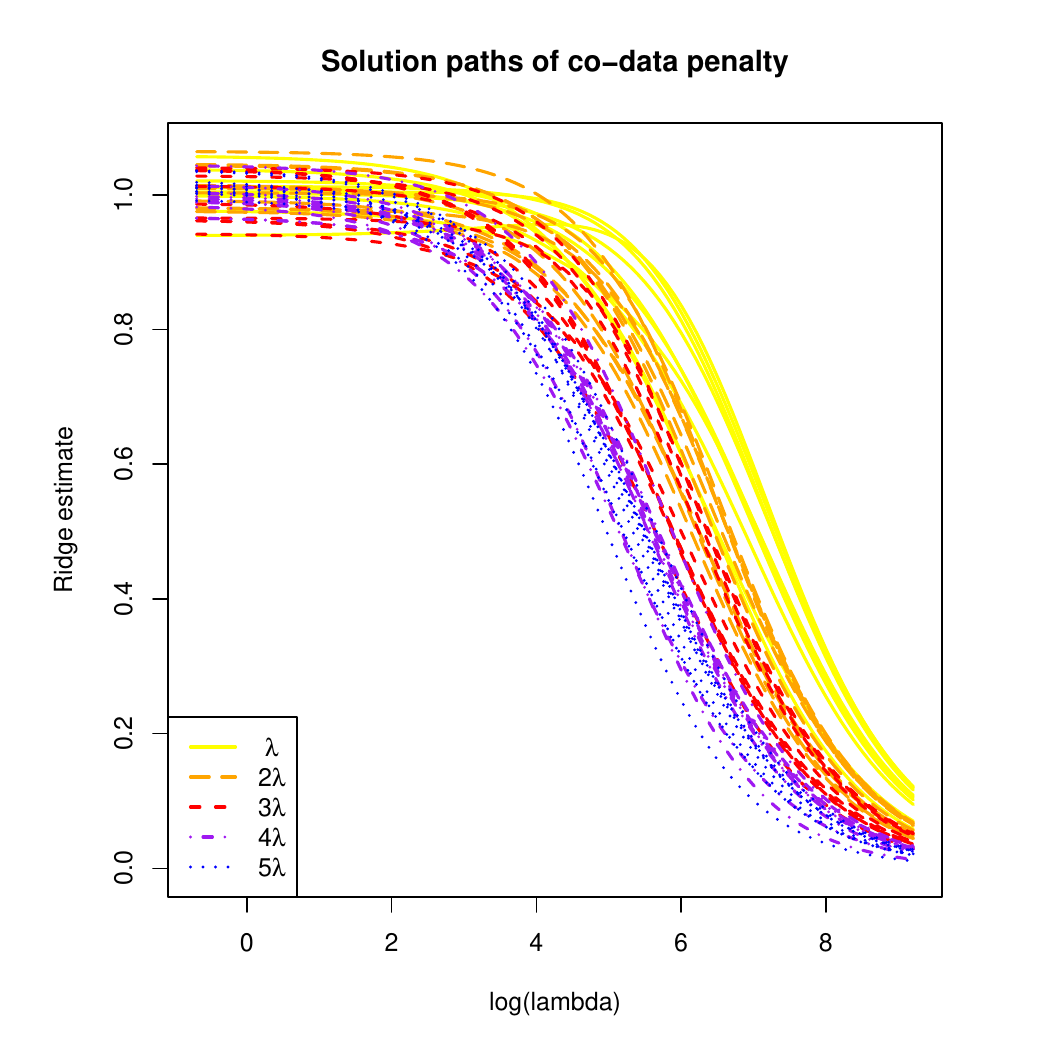}
& 
\includegraphics[scale=0.19, angle=0]{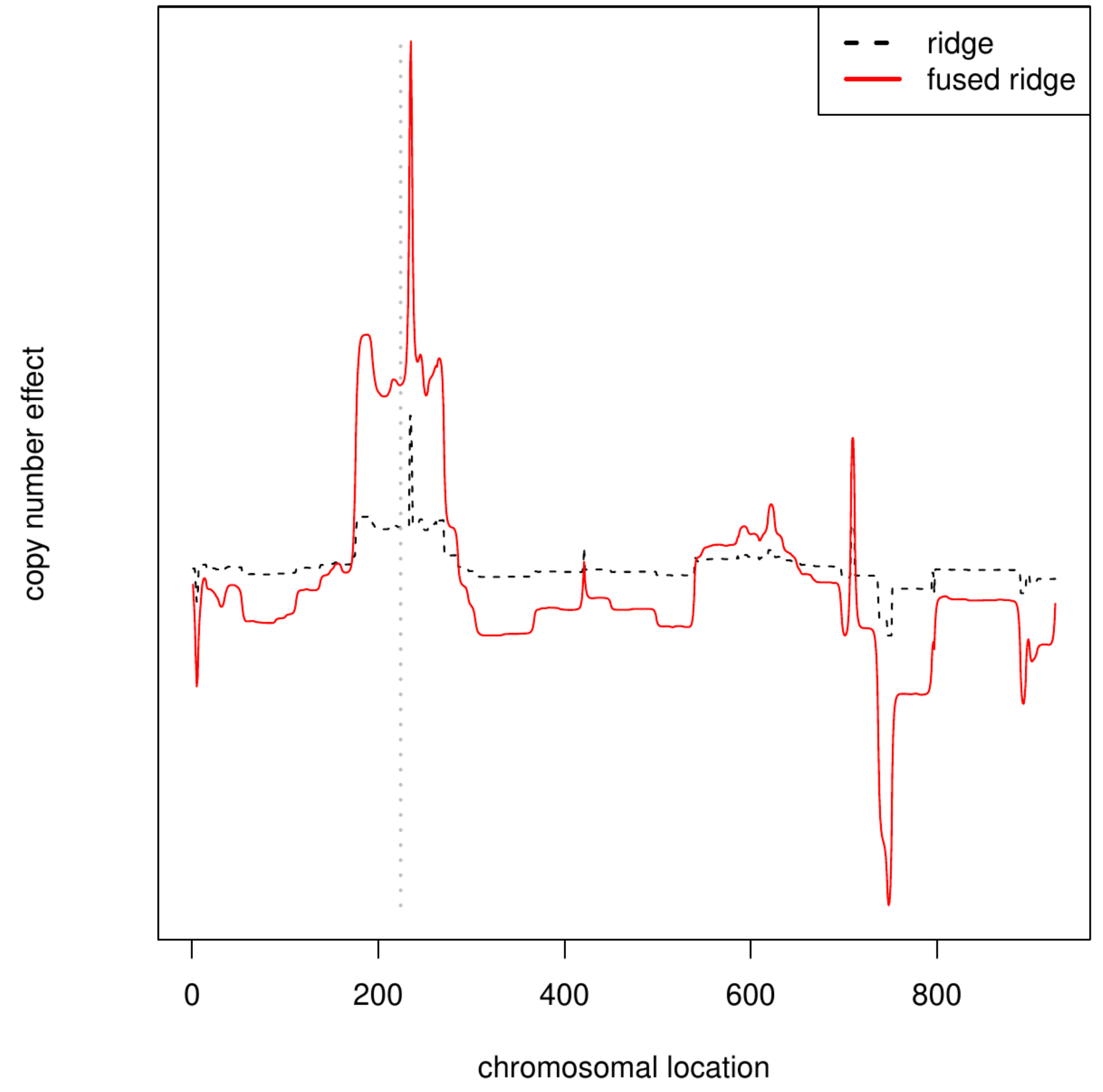}
\end{tabular}
\caption{Top left panel: illustration of the fused ridge estimator (in simulation). The true parameter $\bbeta$ and its ridge and fused ridge estimates against their spatial order. Top right panel: regularization path of the ridge regression estimator with a `homogeneity' penalty matrix. Bottom left panel: regularization path of the ridge regression estimator with a penalty matrix which group-wise regularization. Bottom right panel: Ridge vs. fused ridge estimates of the DNA copy effect on KRAS expression levels. The dashed, grey vertical bar indicates the location of the KRAS gene. } \label{fig:fusedRidgeIllustration}
\end{figure}

To illustrate the effect of the fused ridge penalty on the estimation of the linear regression model $\mathbf{Y} = \mathbf{X} \bbeta + \vvarepsilon$, let $\beta_j = \phi_{0,1}(z_j)$ with $z_j =-30 + \tfrac{6}{50} j$ for $j=1, \ldots, 500$. Sample the elements of the design matrix $\mathbf{X}$ and those of the error vector $\vvarepsilon$ from the standard normal distribution, then form the response $\mathbf{Y}$ from the linear model. The regression parameter is estimated through fused ridge loss minimization with $\lambda=1000$. The estimate is shown in the left panel of Figure \ref{fig:fusedRidgeIllustration} (red line). For reference the figure includes the true $\bbeta$ (black line) and the `regular ridge' estimate with $\lambda=1$ (blue line). Clearly, the fused ridge regression estimate yields a nice smooth vector of $\bbeta$ estimates.

The fused ridge penalty employed in the fused ridge loss function (\ref{form:fusedRidgeLoss}) shrinks one-dimensionally. It shrinks,  when viewing the $\beta_j$ as equidistantly distributed on a line ordered by their position in the regression parameter, contiguous elements towards one other. Should we expect the covariates' effects to be similar, i.e. of equal size and sign, depending on their spatial proximity on a two-dimensional grid, this may be incorporated in the two-dimensional fused ridge regression estimator \citep{Lettink2023twoDimFusedRidge}.

\subsection{A common but unknown target} \label{sect:homogeneityRidge}
A different form of fusion proposed by \cite{mohler2018penalized} and \cite{anatolyev2020ridge} shrinks (a subset of) the regression parameters to a common value. In the work of \cite{mohler2018penalized} and \cite{anatolyev2020ridge}, this common value is the mean of these regression parameters. This mean is also learned from data, alongside the individual elements of regression parameter. Hereto the sum-of-squares is augmented by the following penalty:
\begin{eqnarray} \label{form.homogeneityPenalty}
\lambda \bbeta^{\top} ( \mathbf{I}_{pp} - p^{-1} \mathbf{1}_{pp}) \bbeta & = & \lambda \sum\nolimits_{j=1}^p \big( \beta_j -
p^{-1} \sum\nolimits_{j'=1}^p \beta_{j'} \big)^2 \, \, \, = \, \, \, \lambda p^{-1} \sum\nolimits_{j < j'} (\beta_j - \beta_{j'})^2.
\end{eqnarray}
In \cite{mohler2018penalized}, this is called the \textit{fairness penalty}. The penalty matrix corresponding to penalty (\ref{form.homogeneityPenalty}), which we write as $\lambda \mathbf{\Delta} := \lambda ( \mathbf{I}_{pp} - p^{-1} \mathbf{1}_{pp})$, is nonnegative definite. Hence, in the face of (super-) collinearity, an additional penalty term is required for a well-defined estimator. 

The ridge regression estimator with penalty (\ref{form.homogeneityPenalty}) can  be reformulated, but this is left as an exercise, using straightforward linear algebra as
\begin{eqnarray*}
\hat{\bbeta} (\lambda) & = & a \mathbf{1}_p + (\mathbf{\Delta} \mathbf{X}^{\top} \mathbf{X} \mathbf{\Delta} + \lambda \mathbf{\Delta} )^+ \mathbf{X}^{\top} ( \mathbf{Y} -  a \mathbf{X}  \mathbf{1}_p),
\end{eqnarray*}
with
$a =  [ \mathbf{Y}^{\top} ( \lambda \mathbf{I}_{nn} +    \mathbf{X} \mathbf{\Delta} \mathbf{X}^{\top})^{-1}  \mathbf{X}  \mathbf{1}_p ] [ \mathbf{1}_p^{\top} \mathbf{X}^{\top} ( \lambda \mathbf{I}_{nn} +    \mathbf{X} \mathbf{\Delta} \mathbf{X}^{\top})^{-1} \mathbf{X} \mathbf{1}_p]^{-1}$.
It follows that $\lim_{\lambda \rightarrow \infty} \hat{\bbeta} (\lambda) = a \mathbf{1}_p$. That is, all elements are shrunken to the same value as the amount of regularization increases. 

A closely related approach has been proposed by \cite{wang2022average}. They augment the least squares loss with a targeted ridge penalty of the form $\lambda \| \bbeta - \hat{a} \mathbf{1}_p \|_2^2$. In this, $\hat{a}$ is chosen to equal $\arg \min_a \| \mathbf{Y} - a \mathbf{X} \mathbf{1}_p \|_2^2$. The resulting targeted ridge regression estimator shrinks to this common value. It is thereby reminiscent of the estimators of  \cite{mohler2018penalized} and \cite{anatolyev2020ridge}.

The Bayesian counterpart of the ridge regression estimator with penalty (\ref{form.homogeneityPenalty}) would be a multivariate normal prior with common variance of and uniform correlation between the variates. That is, $\boldsymbol{\beta} \sim \mathcal{N}(\mathbf{0}_p, \mathbf{\Sigma}_u)$, where $\mathbf{\Sigma}_u = \sigma^2 [(1-\rho) \mathbf{I}_{pp} + \rho \mathbf{1}_{pp}]$. But penalty (\ref{form.homogeneityPenalty}) uses the particular choice $\rho = -p^{-1}$, which renders $\mathbf{\Sigma}_u$ to be nonnegative definite and the corresponding prior degenerate.

We illustrate the effect of penalty (\ref{form.homogeneityPenalty}) on the regression estimates. We sample from the linear regression model $\mathbf{Y} = \mathbf{X} \bbeta + \vvarepsilon$ with $\bbeta = \mathbf{1}_{50}$ and both the elements of the design matrix as well as the error sampled from the standard normal distribution. We then evaluate the regularization path of the estimator (\ref{form:generalizedRidgeEstimator}) with $\mathbf{\Delta}$ a diagonal $2 \times 2$ block matrix. The first diagonal block $\mathbf{\Delta}_{11} = \lambda (\mathbf{I}_{10, 10} - \tfrac{1}{10} \mathbf{1}_{10,10})$ and the second $\mathbf{\Delta}_{22} = \lambda \mathbf{I}_{40, 40}$. The resulting regularization paths are shown in the top right panel  of Figure \ref{fig:fusedRidgeIllustration}.

Straightforward application of penalty (\ref{form.homogeneityPenalty}) within the context of the linear regression model may seem farfetched. That is, why would a common effect of all covariates be desirable? Indeed, the motivation of \cite{mohler2018penalized} and \cite{anatolyev2020ridge} stems from a different context. Translated to the present standard linear regression model context, that motivation could be thought of -- loosely -- as an application of the penalty (\ref{form.homogeneityPenalty}) to subsets of the regression parameter. Situations arise where many covariates are only slightly different operationalizations of the same trait. For instance, in brain image data the image itself is often summarized in many statistics virtually measuring the same thing. In extremis, those could be the mean, median and trimmed mean of the intensity of a certain region of the image. It would be ridiculous to assume these three summary statistics to have a wildly different effect. Shrinkage to a common value then seems sensible practice.

\subsection{Group-wise penalization} \label{sect:codataRidge}
Another utilization of the generalized ridge regression estimator (\ref{form:generalizedRidgeEstimator}) can be found in applications where groups of covariates are deemed to be differentially important for the explanation of the response. A shared group membership of covariates is indicative of an equal (implicit) relevance for model fitting. The possible important differences among the covariate groups are accommodated through the augmentation of the loss by a sum of group-wise regular ridge penalties. That is, the elements of the regression parameter that correspond to the covariates of the same group are penalized by the sum of the square of these elements multiplied by a common, group-specific penalty parameter. To formalize this, let $\mathcal{J}_1, \ldots, \mathcal{J}_K$ be $K$ mutually exclusive and exhaustive subsets of the covariate index set $\{1, \ldots, p\}$, i.e. $\mathcal{J}_k \cap \mathcal{J}_{k'} = \emptyset$ for all $k \not= k'$ and $\cup_{k=1}^K \mathcal{J}_k = \{1, \ldots, p\}$. The estimator then is:
\begin{eqnarray} \label{form.codataEstimator}
\hat{\bbeta}(\lambda_1, \ldots, \lambda_K) & = &  \arg \min_{\bbeta \in \mathbb{R}^p} \| \mathbf{Y} - \mathbf{X} \bbeta \|_2^2 + \sum\nolimits_{k=1}^K \lambda_k \sum\nolimits_{j \in \mathcal{J}_k} \beta_j^2,
\end{eqnarray}
To express the above in terms the estimator (\ref{form:generalizedRidgeEstimator}), it involves $\mathbf{W} = \mathbf{I}_{nn}$, $\bbeta_0 = \mathbf{0}_p$, and a diagonal $\mathbf{\Delta}$. A group's importance is reflected in the size of the penalty parameter $\lambda_k$, relative to the other penalty parameters. The larger a group's penalty parameter, the smaller the ridge constraint of the corresponding elements of the regression parameter. And a small constraint allows these elements little room to vary, and the corresponding covariates little flexibility to accommodate the variation in the response.  For instance, suppose the group index is concordant with the hypothesized importance of the covariates for the linear model. Ideally, the $\lambda_k$ are then reciprocally concordant to the group index. Of course, the penalty parameters need not adhere to this concordance, as they are selected by means of data. For more on their selection, see \cite{vdWiel2016}. Clearly, the first regression coefficients of the first ten covariates are shrunken towards a common value, while the others are all shrunken towards zero.

We illustrate the effect of the group-wise penalization on the same data as that used to illustrate the effect of the `homogeneity' penalty (\ref{form.homogeneityPenalty}). Now we employ a diagonal $5 \times 5$ block penalty matrix $\mathbf{\Delta}$, with blocks  
$\mathbf{\Delta}_{11} = \lambda \mathbf{I}_{10,10}$,  $\mathbf{\Delta}_{22} = 2 \lambda \mathbf{I}_{10,10}$, $\mathbf{\Delta}_{33} = 3 \lambda \mathbf{I}_{10,10}$, $\mathbf{\Delta}_{33} = 4 \lambda \mathbf{I}_{10,10}$, and $\mathbf{\Delta}_{55} = 5 \lambda \mathbf{I}_{10,10}$. The regularization paths of the estimator (\ref{form.codataEstimator}) are shown in the bottom left panel of Figure \ref{fig:fusedRidgeIllustration}. Clearly, the regression coefficient estimates are shrunken less if they belong to a group with high important (i.e. lower penalty).

Covariate groups are formed in various ways:
\begin{compactitem}
\item[$\circ$] \cite{vdWiel2016} suggests to form the covariate groups on the basis of \textit{co-data}. The co-data concept refers to auxillary data on the covariates not necessary directly related to but possibly implicitly informative for the to-be-estimated model. For instance, for each covariate a marginal $p$-value or effect estimate from a different study may be available. The covariate groups may then be formed on the basis of the size of this statistic. 

\item[$\circ$] \cite{nunez2019voxelwise} present \textit{banded ridge regression} that too bases the covariate grouping on external information. In their area of application, brian image analysis, the groups may be defined by brain halfs/regions. Similarly, within molecular biology, genes may be grouped by pathway membership.

\item[$\circ$] The \textit{graphical group ridge} presented by \cite{aldahmani2020graphical} reconstructs -- prior to the evaluation of the estimator -- the groups from the design matrix itself. They identify groups of covariates that are strongly linearly related conditionally on the remaining covariates (=collinear) by fitting a sparse graphical model to the design matrix' columns. Strongly collinear covariates form the groups that share a common penalty parameter in the ridge regression estimator (\ref{form.codataEstimator}). 

\item[$\circ$] \cite{yang2017bayesian} form the covariate groups on the basis of a scaled version of their marginal estimates. If this scaled estimate exceeds a threshold, the corresponding element of regression parameter is penalized by $\lambda$ and by $\tfrac{1}{2} \lambda$ otherwise.
\end{compactitem}


\subsection{Updating via targeted ridge regression}
An example that illustrates the use of the target of the generalized ridge penalty can be found in sequential learning (as is pointed out in \citealp{vanWieringen2022transfer}). Sequential learning takes the lessons learned, i.e. estimates from previous studies of the same phenomenon, into account
when updating these estimates from a data of a novel such study. The setting may be formalized by assuming  
\begin{compactitem}
\item[\textit{A1)}] an infinite sequence of data sets $\{ \mathbf{Y}^{(t)}, \mathbf{X}^{(t)} \}_{t=1}^\infty$ sampled from the linear regression model $\mathbf{Y}^{(t)} = \mathbf{X}^{(t)} \bbeta + \vvarepsilon^{(t)}$ with $\vvarepsilon^{(t)} \sim \mathcal{N}(\mathbf{0}_{n_t}, \sigma^2 \mathbf{I}_{n_t, n_t})$ for all $t = 1, \ldots, T$ and $\vvarepsilon^{(t)} \independent \vvarepsilon^{(t')}$ for $t \not= t'$.  
\end{compactitem}
Effectively, the linear regression model is the same for all data sets as its parameters $\bbeta$ and $\sigma^2$ are common to all. The number of covariates is fixed across data sets but the covariates' settings, i.e. their values, may differ among the data sets. Similarly, the sample sizes $n_t \in \mathbb{N}$ need not be common to all data sets.

\begin{figure*}[h!]
\centering
\begin{tabular}{ll}
\includegraphics[angle=0, scale=0.41]{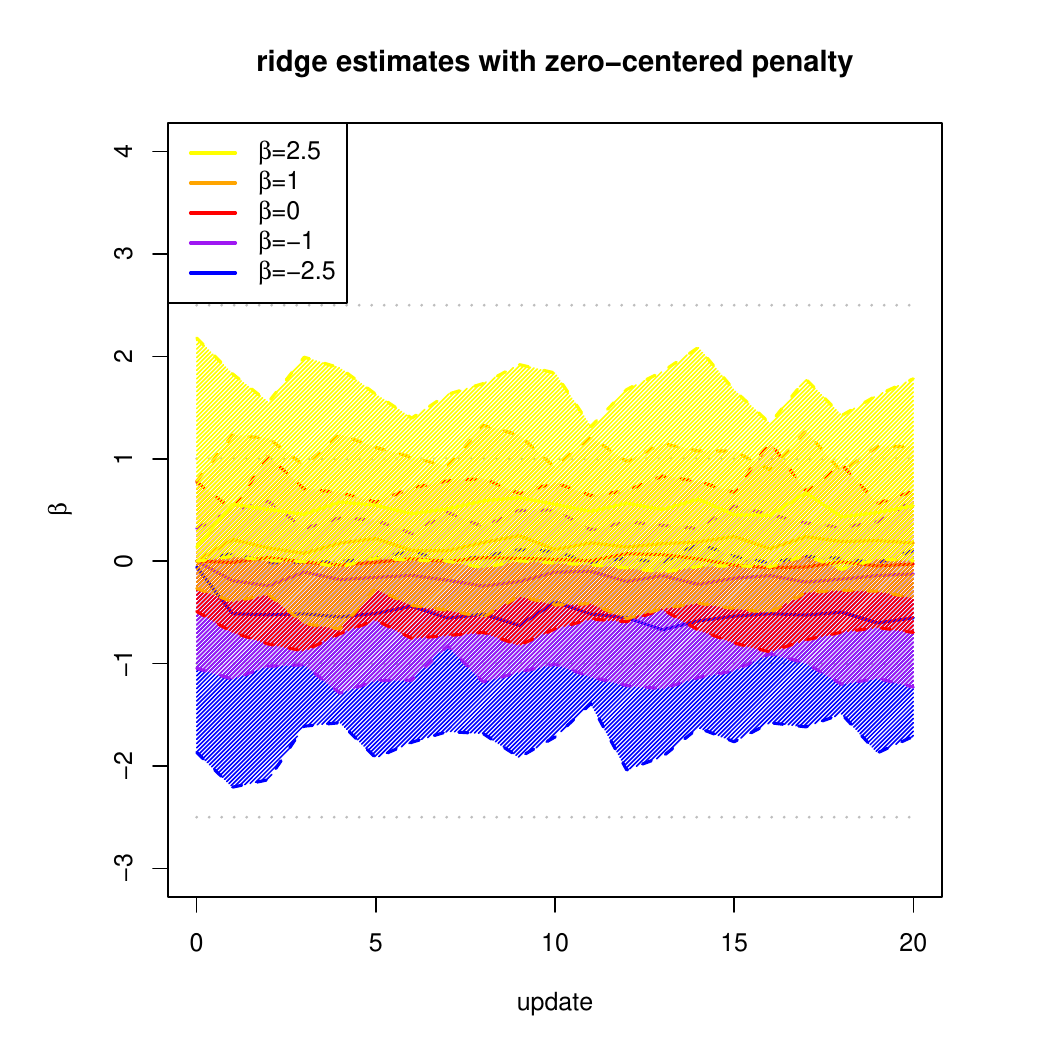}
&
\includegraphics[angle=0, scale=0.41]{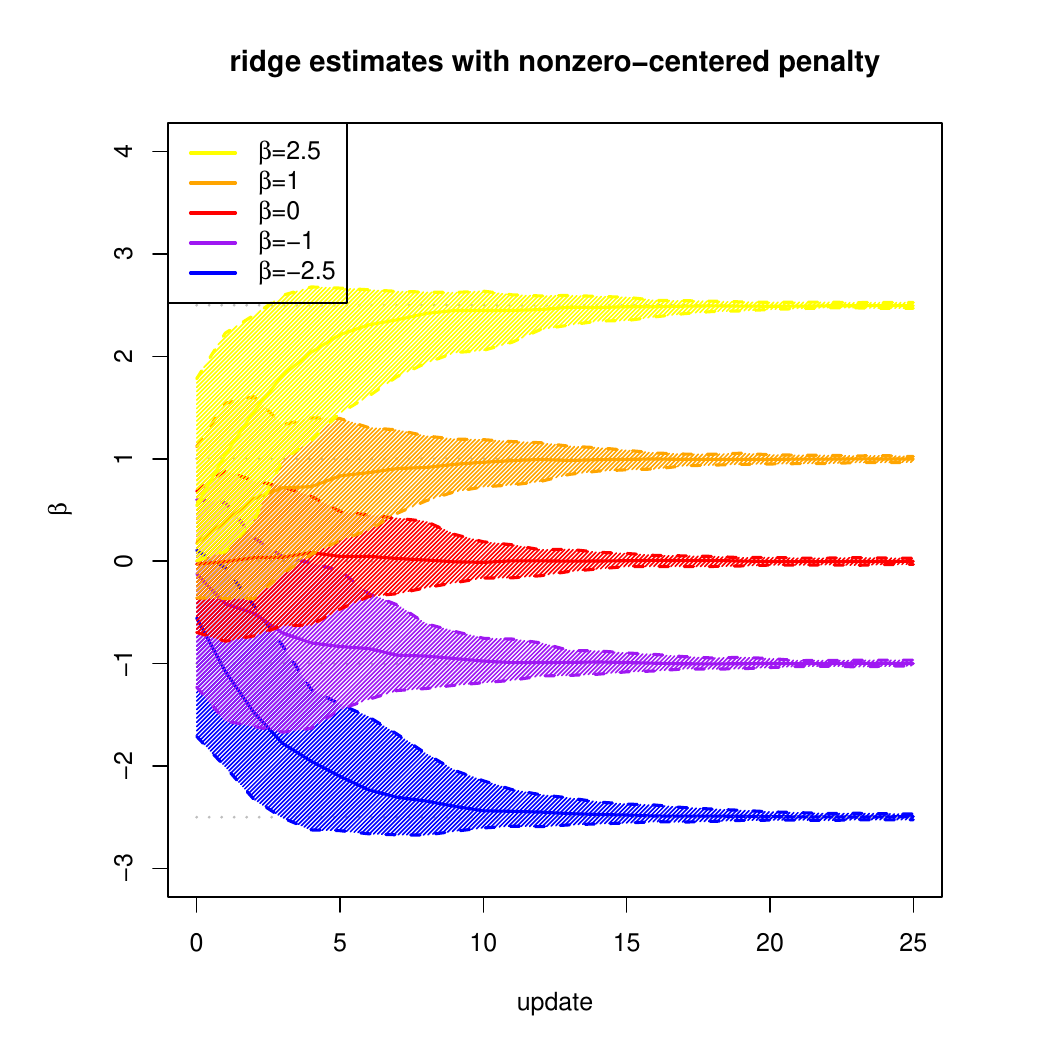}
\end{tabular}
\caption{The top panels show the $(5\%,  95\%$-quantile intervals of the  traditional (left) and updated (right) ridge estimates of $\beta_{j}$ with $j \in \{ 0, 30, 50, 70, 100 \}$ plotted against $t$. The solid, colored line inside these intervals is the corresponding $50\%$ quantile. The dotted, grey lines are the true values of the $\beta_j$'s. } \label{fig.regular2updateRidge}
\hfill{}
\end{figure*}

We now estimate the regression parameter from each data set by the generalized ridge regression estimator using the estimate of the preceeding data set as the target. Formally,
\begin{eqnarray} \label{form:updatedRidgeEstimator}
\hat{\bbeta}^{(t)} (\lambda_t) & = & \arg \min\nolimits_{\bbeta \in \mathbb{R}^p} \, \| \mathbf{Y}^{(t)} - \mathbf{X}^{(t)} \bbeta \|_2^2 + \lambda_t \| \bbeta - \hat{\bbeta}^{(t-1)} (\lambda_{t-1}) \|_2^2
\end{eqnarray}
for $t=1, \ldots$. Effectively, with the arrival of novel data we update our current estimator of the regression parameter  $\hat{\bbeta}^{(t-1)} (\lambda_{t-1})$, resulting in the updated one $ \hat{\bbeta}^{(t)} (\lambda_{t})$. We thus obtain a sequence of estimators $\{ \hat{\bbeta}^{(t)} (\lambda_t) \}_{t=1}^{\infty}$, which is initiated by  any nonrandom target, e.g. the null vector. Over time we expect this sequence of estimators to incorporate the lessons from the past.

Let us investigate by simulation whether the updated ridge regression estimator (\ref{form:updatedRidgeEstimator}) indeed accumulates knowledge on the regression parameter. Hereto we draw a sequence of data sets from the linear regression model $\mathbf{Y}_t = \mathbf{X}_t \bbeta + \vvarepsilon_t$ with $t=1,\ldots, 25$. The elements of the design matrices are sampled from the standard normal distribution, the elements of $\bbeta$ chosen as $\beta_j = (j-50)/20$ for $j=0, 1, 2, \ldots, 100$, while $\vvarepsilon_t \sim \mathcal{N}(\mathbf{0}_n, 0.04 \mathbf{I}_{nn})$. The regression parameter is estimated from each data set using both the regular and updated ridge estimators, in which the latter uses its previous estimate as target. The penalty parameter of both estimators are determined by cross-validation. We repeat all this a hundred times. The results, by the  $5\%$, $50\%$ and $95\%$ quantiles, of the regular and updated ridge regression estimates of $\beta_{j}$ with $j \in \{ 0, 30, 50, 70, 100 \}$ are plotted against $t$ (Figure \ref{fig.regular2updateRidge}). The quantiles of the regular ridge regression estimates of the selected elements of $\bbeta$ are virtually constant over times (left panel of Figure \ref{fig.regular2updateRidge}). In contrast, the quantiles of the update ridge regression estimates (right panel of Figure \ref{fig.regular2updateRidge}) clearly improve as $t$ increases. The improvement is two-fold: \textit{i)} they become less biased, and \textit{ii)} the distance between the $5\%$ and $95\%$ quantiles vanishes.
The quantiles' behavior indicates that the updating does lead to an accumulation of knowledge

In Figure \ref{fig.regular2updateRidge} the $5\%$, $50\%$ and $95\%$ quantiles of the traditional and updated ridge estimates of $\beta_{j}$ with $j \in \{ 0, 30, 50, 70, 100 \}$ are plotted against $t$. These quantiles of the traditional ridge estimates of these elements of $\bbeta$ are constant over $t$ (left panel of Figure \ref{fig.regular2updateRidge}). Those of the update ridge estimates (right panel of Figure \ref{fig.regular2updateRidge}) clearly improve as $t$ increases. The improvement is two-fold: \textit{i)} they become less biased, and \textit{ii)} the distance between the $5\%$ and $95\%$ quantiles vanishes.

The simulation results can be underpinned theoretically. Theorem \ref{theorem:consistency} states (under conditions) the consistency in $t$ of the sequence for the estimation of $\bbeta$ and the associated linear predictor.
\begin{theorem} \label{theorem:consistency} \mbox{(Theorem 2 of \citealp{vanWieringen2022transfer})}
\\
Adopt assumption \textit{A1} and let $\{ \hat{\bbeta}_{t} (\lambda_t) \}_{t=1}^{\infty}$ the corresponding sequence of update ridge regression estimators (\ref{form:updatedRidgeEstimator}). Furthermore, 
\begin{compactitem}
\item[\textit{i)}] initiate the estimator sequence by any nonrandom target.
\item[\textit{ii)}]  let  $\cap_{t=T}^{\infty} \mbox{null}(\mathbf{X}_t) = \mathbf{0}_p$ for sufficiently large $T \in \mathbb{N}$.
\item[\textit{iii)}] choose the regularization scheme $\{ \lambda_t \}_{t=1}^{\infty}$ such that $\lim_{t\rightarrow \infty} \sigma_{\varepsilon}^2 \, p \, d^{2}_{1}(\mathbf{X}_{t}) \, \lambda_{t}^{-2} = 0$ with $d_1 (\mathbf{X}_t)$ the largest singular value of $\mathbf{X}_t$. 
\end{compactitem} 
Then, for every $c>0$:
\begin{eqnarray*}
\lim_{t \rightarrow \infty} P [ \| \hat{\bbeta}_{t}(\lambda_{t}) - \bbeta \| \geq c \, | \, \{ \lambda_{\tau} \}_{\tau=1}^{t} ]  \rightarrow 0,
\\
\lim_{t \rightarrow \infty} P [ \| \mathbf{X}_{\mbox{{\tiny new}}} \hat{\bbeta}_{t}(\lambda_{t}) - \mathbf{X}_{\mbox{{\tiny new}}} \bbeta \| \geq c \, | \, \{ \lambda_{\tau} \}_{\tau=1}^{t} ]  \rightarrow 0,
\end{eqnarray*}
where $\mathbf{X}_{\mbox{{\tiny new}}}$ is the design matrix of a novel study.
\end{theorem}

\begin{proof} See \cite{vanWieringen2022transfer}.
\end{proof}

\noindent
The updating procedure above is a frequentist analogue of Bayesian updating \citep{berger2013stat}. From the Bayesian perspective, the ridge penalty corresponds to a normal prior on the regression parameter (see Chapter \ref{chap:BayesianRegression}). With this normal prior, the posterior of $\bbeta$ is also normal. In turn, this normal posterior serves as (normal) prior in the next update of $\bbeta$. The prior for the $t+1$-th update then becomes $\bbeta_{t+1} \, | \, \sigma^2 \sim \mathcal{N}[  \bbeta_{t}(\lambda_{t}, \bbeta_{t-1}), \sigma^2 \lambda_{t+1}^{-1} \mathbf{I}_{pp}]$, which yields a posterior mean $\mathbb{E}[ \bbeta_{t+1} \, | \, \mathbf{Y}_{t+1}, \mathbf{X}_{t+1}, \sigma^2, \bbeta_{t}(\lambda_{t}, \beta_{t-1})]$  coinciding with the frequentist estimator $\hat{\bbeta}_{t+1}(\lambda_{t+1}, \bbeta_{t})$.

\subsection{Hierarchical ridge}
\cite{kawaguchi2022hierarchical} present the hierarchical ridge regression estimator shrinks the regression parameter to a data-driven target. To limit the burden of this additional layer of complexity, the target is assumed to have a low-dimensional form that can easily be learned from data. For instance, mutations in the same gene may have a similar effect size on the disease outcome. If this hypothesis is deemed to be informative context knowledge, genes then form a natural grouping on the covariates. The average of the effect sizes of a gene's mutations could be a reasonable shrinkage target that can straightforwardly be estimated from data. \cite{kawaguchi2022hierarchical} assume such information, on which effect sizes contribute to what element of the target, to be available and encoded in a $p \times q$-dimensional matrix $\mathbf{Z}$ with $q$ small, i.e. $q < n$ and preferably much smaller than the sample size. The low-dimensional shrinkage target is then $\bbeta_0 = \mathbf{Z} \ggamma$ with $\ggamma \in \mathbb{R}^q$, where $\ggamma$ is to be learned alongside $\bbeta$ from the data. Hereto \cite{kawaguchi2022hierarchical} propose the hierarchical ridge regression estimator:
\begin{eqnarray} \label{form:hierarchicalRidgeLoss}
\hat{\bbeta} (\lambda_\beta, \lambda_\gamma), \hat{\ggamma} (\lambda_\beta, \lambda_\gamma) & = & \arg \min_{\bbeta \in \mathbb{R}^p, \ggamma \in \mathbb{R}^q} \| \mathbf{Y} - \mathbf{X} \bbeta \|_2^2 + \lambda_\beta \| \bbeta - \mathbf{Z} \ggamma \|_2^2 + \lambda_\gamma \| \ggamma \|_2^2,
\end{eqnarray}
with penalty parameters $\lambda_\beta$ and $\lambda_\gamma$. The first regulates the amount of shrinkage to the target, while the latter controls the shrinkage of the target estimator. 

To evaluate the hierarchical ridge regression estimator the loss function (\ref{form:hierarchicalRidgeLoss}) is rewritten using the change-of-variable $\eeta = \bbeta - \mathbf{Z} \ggamma$ to a generalized ridge regression one:
\begin{eqnarray*}
\left\| \mathbf{Y} - \left( \begin{array}{rr}  \mathbf{X} & \mathbf{X} \mathbf{Z} \end{array} \right) \left( \begin{array}{c}  \eeta \\ \ggamma \end{array} \right)  \right\|_2^2 + \left( \begin{array}{cc}  \eeta^{\top} & \ggamma^{\top} \end{array} \right)  \left( \begin{array}{rr}  \lambda_\beta \mathbf{I}_{pp} &  \mathbf{0}_{pq} \\ \mathbf{0}_{qp} & \lambda_\gamma \mathbf{I}_{pp}  \end{array} \right)  \left( \begin{array}{c}  \eeta \\ \ggamma \end{array} \right).
\end{eqnarray*} 
An analytical expression of the optimizer of the latter loss can be derived in similar fashion as for the generalized ridge regression estimator with un- and penalized covariates. Hence, equate the gradient with respect to
both parameters $\bbeta$ and $\ggamma$ to zero, apply the Woodbury matrix identity and the analytic expression of the inverse of the $2 \times 2$ block matrix, and obtain:
\begin{eqnarray*}
\hat{\bbeta} (\lambda_\beta, \lambda_\gamma)  & = & \lambda_\beta^{-1} \mathbf{X}^{\top} (\lambda_\beta^{-1} \mathbf{X} \mathbf{X}^{\top} + \mathbf{I}_{nn})^{-1}  [ \mathbf{Y} -  \mathbf{X}  \mathbf{Z} \hat{\ggamma} (\lambda_\beta, \lambda_\gamma) ]  + \mathbf{Z} \hat{\ggamma} (\lambda_\beta, \lambda_\gamma),
\\
\hat{\ggamma} (\lambda_\beta, \lambda_\gamma) & = & \big[(\mathbf{X} \mathbf{Z})^{\top} (\lambda_\beta^{-1} \mathbf{X} \mathbf{X}^{\top} + \mathbf{I}_{nn})^{-1} \mathbf{X} \mathbf{Z}  + \lambda_\gamma \mathbf{I}_{qq} \big]^{-1} (\mathbf{X} \mathbf{Z})^{\top} ( \lambda_\beta^{-1} \mathbf{X} \mathbf{X}^{\top} + \mathbf{I}_{nn})^{-1} \mathbf{Y}.
\end{eqnarray*}
For practical purposes, we have provided an analytic formulation of the hierachical ridge regression estimator that is computationally most efficiently evaluated if $p \gg n > q$.

The analytic expression of the hierarchical ridge regression estimator enables the study of the limits of the regularization paths. The estimators of both parameters reduce to their regular ridge regression estimator as:
\begin{eqnarray*}
\lim\nolimits_{\lambda_\gamma \rightarrow \infty} \hat{\bbeta}(\lambda_\beta, \lambda_\gamma) & = & (\mathbf{X}^{\top} \mathbf{X} + \lambda_\beta  \mathbf{I}_{pp})^{-1}  \mathbf{X}^{\top} \mathbf{Y},
\\
\lim\nolimits_{\lambda_\beta \rightarrow \infty} \hat{\ggamma}(\lambda_\beta, \lambda_\gamma) & = & \big[ (\mathbf{X} \mathbf{Z})^{\top}  \mathbf{X} \mathbf{Z} + \lambda_\gamma \mathbf{I}_{qq} \big]^{-1} (\mathbf{X} \mathbf{Z})^{\top}  \mathbf{Y}.
\end{eqnarray*}
While not directly obvious from the above provided analytic expression of the hierarchical ridge regression estimators, an the other end of the regularization path we have $\lim_{\lambda_{\beta} \downarrow 0} \hat{\bbeta}(\lambda_\beta, \lambda_\gamma) = \hat{\bbeta}^{\mbox{{\tiny mls}}}$, the minimum least squares estimator. 

The hierarchical ridge regression estimator of $\bbeta$ is effectively a targeted ridge regression estimator. This can be seen from the rewritten hierarchical ridge regression estimator of $\bbeta$:
\begin{eqnarray*}
\hat{\bbeta}(\lambda_\beta, \lambda_\gamma)  & = &  (\mathbf{X}^{\top} \mathbf{X} + \lambda_\beta  \mathbf{I}_{pp})^{-1} [ \mathbf{X}^{\top} \mathbf{Y} +  \lambda_{\beta} \mathbf{Z} \hat{\ggamma} (\lambda_\beta, \lambda_\gamma)  ].
\end{eqnarray*}
This formulation reveals that the low rank parametrization $\mathbf{Z} \ggamma$ of $\bbeta$ plays the same role as the target $\bbeta_0$. The key difference is that the target is now estimated in penalized fashion from data.

The Bayesian equivalent of the hierarchical ridge regression estimator uses -- surprise, surprise -- a hierarchical prior: $\bbeta \, | \, \ggamma \sim \mathcal{N} ( \mathbf{Z} \ggamma, \sigma^2 \lambda_{\beta}^{-1} \mathbf{I}_{pp})$ with $\ggamma \sim \mathcal{N}( \mathbf{0}_q, \sigma^2 \lambda_{\gamma}^{-1} \mathbf{I}_{qq})$. Hence, $\ggamma$ controls the mean of $\bbeta$, and unconditionally results in a structured covariance matrix of $\bbeta$.

The hierarchical ridge regression estimator is closely related to other instances of the generalized ridge regression estimator discussed in this section. For instance, suppose the covariates can be separated into exhaustive and mutually exclusive sets. This can be encoded by a $\mathbf{Z}$ with indicator columns, one for each set. The hierarchical ridge regression estimator then shrinks the effects of covariates from the same set to a common effect. So does the homogeneity ridge regression estimator of Subsection \ref{sect:homogeneityRidge}. But in contrast to the former estimator, the latter estimates this common effect unpenalizedly. The co-data ridge regression estimator of Subsection \ref{sect:codataRidge} can also incorporate the above set structure of the covariates. The co-data ridge regression estimator then shrinks all effects to zero but allows for different penalization among the sets. Effectively, taking the Bayesian perspective, the co-data ridge regression estimator employs a prior with a common zero mean but with set-wise variances, while the hierarchical ridge estimator adopts a (conditional) prior with set-wise means but a common variance.

\section{Other connections}
The generalized ridge regression estimator appears in other contexts where it has not been proposed as a penalized estimator. Here we discuss some of them.

\subsection{Dropout} \label{sect:dropout}
The generalized ridge regression estimator corresponds to the dropout estimator  Dropout \citep{srivastava2014dropout} is an iterative estimation procedure for models with many parameters. Dropout estimates a large model with many parameters through the estimation of a long sequence of reduced, i.e. with substantially fewer parameters, models that are nested within the original, full model (but not in each other). At each iteration, only those parameters of the full model are updated that have been included in the reduced model at hand. Eventually, the updated parameter estimate converges (under conditions) to an estimate of the full model. 

Dropout tackles two problems. Firstly, it allows to learn the full model if its direct estimation is computationally too intensive. Secondly, dropout counters a phenomenon that in field of machine learning is called \textit{co-adaptation}. Co-adaptation refers to the phenomenon of strong dependence among covariates regarding their contribution to the prediction. This may remind us of collinearity. But collinearity refers to a strong linear relationship among the covariates, while co-adaptation refers to a strong dependence in the learning of the coefficient of covariates. Collinearity and co-adaptation are closely related. For instance, in a linear regression model with three identical covariates only the sum of their regression coefficients, $\beta_1 + \beta_2 + \beta_3 = c$, can identified. In nonlinear models, e.g. deep learners (see Section \ref{sect:deepLearning}), co-adaptation need not be rooted in collinearity. Dropout thus performs a form of regularization. 

Dropout effectively regulates the parameter estimation in generalized $\ell_2^2$-fashion.  It achieves this through random selection of the parameters to be included in the reduced models. Then, each time a reduced model is fitted to the data, an included parameter is estimated jointly with a random subset of other ones. The randomness in subset creation breaks any reliance of a parameter's estimation on the presence of other parameters in the model. In fact, it can only rely on the average behavior of the other parameters. Within the context of the linear regression model, an included covariate in the reduced model competes with a random set of other covariates to explain the variation in the response. The estimation of its effect can then not anticipate on the contribution of other covariates but must adapt to the precence of a random collection of competitors. Hereby dropout prevents very large parameter estimates and yields ones that are more uniform in size \citep{hinton2012improving}. Put differently, dropout prevents overfitting. 

Let us make things precise. Suppose we learn the linear regression model's parameter by means of gradient descent. We would then start with an initial parameter estimate $\tilde{\boldsymbol{\beta}}^{(0)}$ and update this at a fixed learning rate $\alpha > 0$ by:
\begin{eqnarray*}
\tilde{\boldsymbol{\beta}}^{(t+1)} & = &  \tilde{\boldsymbol{\beta}}^{(t)} + \alpha \mathbf{X}^{\top} (\mathbf{Y} - \mathbf{X} \tilde{\boldsymbol{\beta}}^{(t)}),
\end{eqnarray*}
until convergence or until the maximum number of iterations has been reached.
Dropout replaces the gradient in the update rule by the gradient of a randomly reduced model. At each iteration, the reduced model is formed by randomly subsetting the set of covariates that is included in the model. More precisely, each covariate has a probability $\kappa$ to be selected and enter the model. The succes and failures, denoted by $1$ and $0$, respectively, of this random selection procedure are stored on diagonal of the $p \times p$ diagonal dropout matrix $\mathbf{D}$. On average the drop matrix has $p \kappa$ and $p (1-\kappa)$ unit and zero diagonal entries, respectively. Then, gradient descent with dropout amounts to the update rule:
\begin{eqnarray*}
\tilde{\boldsymbol{\beta}}^{(t+1)} & = &  \tilde{\boldsymbol{\beta}}^{(t)} + \alpha \mathbf{D}_{t+1} \mathbf{X}^{\top} (\mathbf{Y} - \mathbf{X} \tilde{\boldsymbol{\beta}}^{(t)}),
\end{eqnarray*}
where the dropout matrices $\mathbf{D}_{t+1}$ have generated independently and identically. The post-multiplication of the design matrix $\mathbf{X}$ by the dropout matrix effectively `deletes' the columns with a zero diagonal element of the dropout matrix.  

Does the dropout updating of the regression parameter converge and, if so, to what? Intuitively, it appears that dropout still minimizes the sum-of-squares but now averaged over all possible randomly reduced models (with selection probability $\kappa$). Translated mathematically, that would amount to the dropout updates converging to the minimum of $\mathbb{E}_{\mathbf{D}} [ \| \mathbf{Y} - \mathbf{X} \mathbf{D}  \boldsymbol{\beta} \|_2^2 \, | \, \mathbf{Y} ]$ \, where the subscript of the expectation operator indicates that it is taken with respect to all possible dropout matrices. This is indeed the case, as we show below. 

To study the converge of the dropout updates, we first look at the minimizer  of its hypothesized loss. We rewrite this loss as follows:
\begin{eqnarray*}
\mathbb{E}_{\mathbf{D}} [ \| \mathbf{Y} - \mathbf{X} \mathbf{D}  \boldsymbol{\beta} \|_2^2 \, | \, \mathbf{Y} ] & = & \mathbb{E}_{\mathbf{D}} ( \mathbf{Y}^{\top} \mathbf{Y} - 2 \mathbf{Y}^{\top} \mathbf{X} \mathbf{D}  \boldsymbol{\beta} + \boldsymbol{\beta}^{\top}  \mathbf{D} \mathbf{X}^{\top} \mathbf{X} \mathbf{D}  \boldsymbol{\beta})
\\
& = & \| \mathbf{Y} - \kappa \mathbf{X} \boldsymbol{\beta} \|_2^2 + \kappa (1-\kappa) \boldsymbol{\beta}^{\top} \mathbf{A}_x \boldsymbol{\beta},
\end{eqnarray*}
where the last step hinges upon the linearity of the expectation operation and  Lemma C.2 of \cite{Clara2023dropout} that states that $\mathbb{E}_{\mathbf{D}} (\mathbf{D}) = \kappa \mathbf{I}_{pp}$ and $\mathbb{E}_{\mathbf{D}} (\mathbf{D} \mathbf{X}^{\top} \mathbf{X} \mathbf{D}) = \kappa^2 \mathbf{X}^{\top} \mathbf{X} + \kappa (1-\kappa) \mathbf{A}_x$ with diagonal matrix $\mathbf{A}_x$ and $(\mathbf{A}_x)_{jj} = ( \mathbf{X}^{\top} \mathbf{X} )_{jj}$. In the last line of the preceeding display we recognize a generalized ridge regression loss function. This is minimized by
\begin{eqnarray*}
\tilde{\boldsymbol{\beta}} & = & \arg \min\nolimits_{\boldsymbol{\beta} \in \mathbb{R}^p} \mathbb{E}_{\mathbf{D}} [ \| \mathbf{Y} - \mathbf{X} \mathbf{D}  \boldsymbol{\beta} \|_2^2 \, | \, \mathbf{Y} ] \, \, \, = \, \, \,
 [\kappa \mathbf{X}^{\top} \mathbf{X} +  (1-\kappa) \mathbf{A}_x]^{-1} \mathbf{X}^{\top} \mathbf{Y}.
\end{eqnarray*}
Hence, $\tilde{\boldsymbol{\beta}}$ is a generalized ridge regression estimator. Finally, if the columns of $\mathbf{X}$ are orthogonal, $\mathbf{X}^{\top} \mathbf{X} = \mathbf{A}_x$ and $\tilde{\boldsymbol{\beta}}$ coincides with the maximum likelihood estimator of the linear regression model.

The next lemma states that the sequence of dropout updates converges in expectation exponentially fast to the generalized ridge regression estimator $\tilde{\boldsymbol{\beta}}$. 

\begin{lemma} \textit{(Lemma 4.1 of \citealp{Clara2023dropout})} \mbox{ } \\
Choose the learning rate $\alpha$ and dropout probability $\kappa$ such that $\alpha \kappa  \| \mathbf{X}^{\top} \mathbf{X} \|_2 < 1$. Let the randomness of the initialization $\tilde{\boldsymbol{\beta}}_0$ be independent of that of $\mathbf{Y}$ and the dropout matrices. Finally, assume the design matrix $\mathbf{X}$ has no zero columns. Then,
\begin{eqnarray*}
\| \mathbb{E}_{(\mathbf{Y}, \mathbf{D})} (\tilde{\boldsymbol{\beta}}_t - \tilde{\boldsymbol{\beta}}) \|_2 & \leq & \| \mathbf{I}_{pp} - \alpha [ \kappa^2 \mathbf{X}^{\top} \mathbf{X} + \kappa (1-\kappa) \mathbf{A}_x] \|^t \, \| \mathbb{E}_{\mathbf{Y}} (\tilde{\boldsymbol{\beta}}_0 - \tilde{\boldsymbol{\beta}}) \|_2
\end{eqnarray*}
for $t=0, 1, 2, \ldots$.
\end{lemma} 

\begin{proof}
See \cite{Clara2023dropout}.
\end{proof}
While the sequence of dropout updates converges in expectation to the generalized ridge regression estimator $\tilde{\boldsymbol{\beta}}$, it exhibits larger variance due to the additional randomness introduced by the dropout matrix. This can be circumvented by tinkering with the update rule \citep{Clara2023dropout}.

\section{Application}
\begin{figure}[!h]
\centering
\includegraphics[scale=0.20, angle=0]{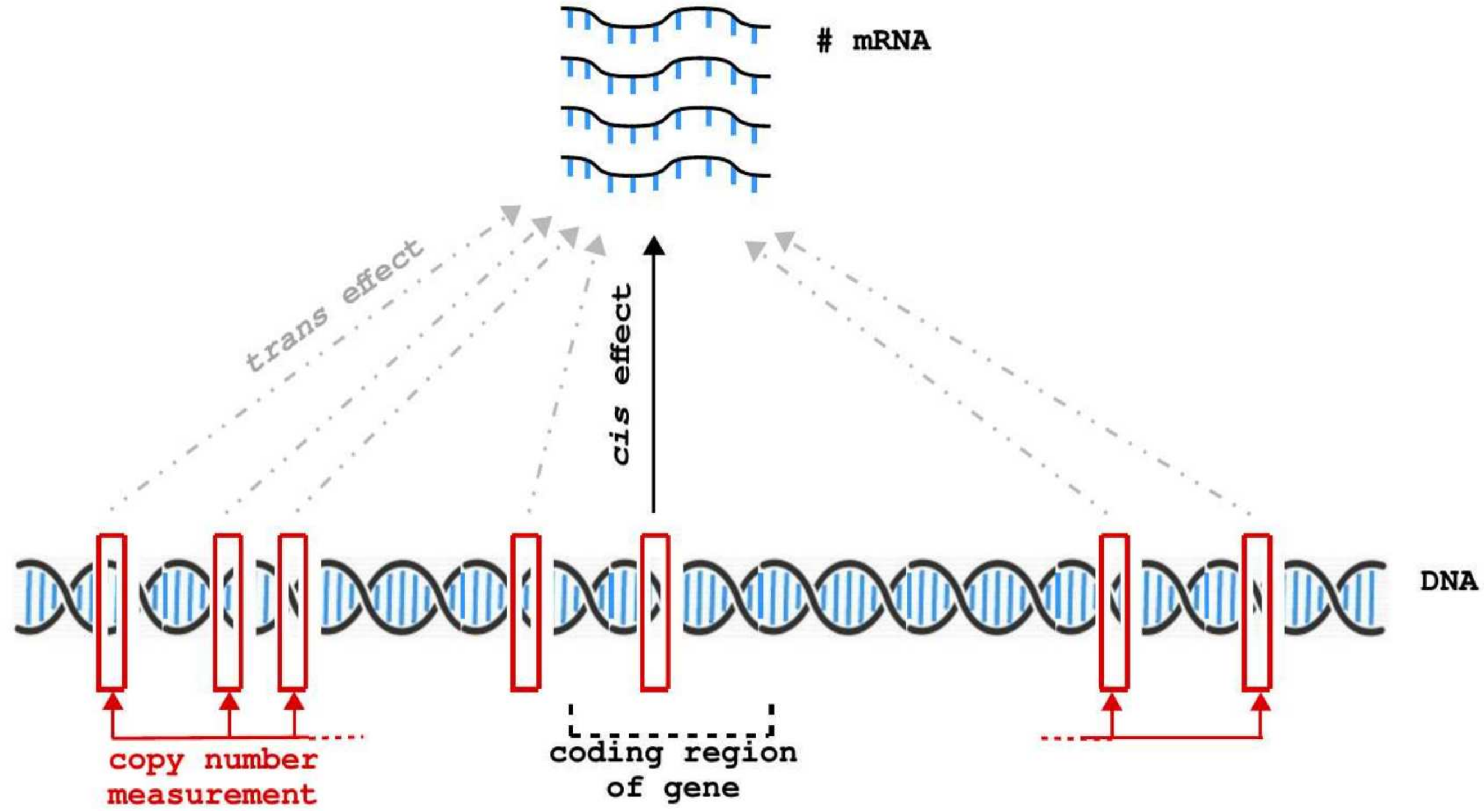}
\caption{Illustration of the \textit{cis}- and \textit{trans}-effect of DNA copy number on gene expression levels.} \label{fig:cis2transEffect}
\end{figure}

An illustration involving omics data can be found in the explanation of a gene's expression levels in terms of its DNA copy number. The latter is simply the number of gene copies encoded in the DNA. For instance, for most genes on the autosomal chromosomes the DNA copy number is two, as there is a single gene copy on each chromosome and autosomal chromosomes come in pairs.  Alternatively, in males the copy number is one for genes that map to the X or Y chromosome, while in females it is zero for genes on the Y chromosome. In cancer the DNA replication process has often been compromised leading to a (partially) reshuffled and aberrated DNA. Consequently, the cancer cell may exhibit gene copy numbers well over a hundred for classic oncogenes. A faulted replication process does -- of course -- not nicely follow the boundaries of gene encoding regions. This causes contiguous genes to commonly share aberrated copy numbers. With genes being transcribed from the DNA and a higher DNA copy number implying an enlarged availability of the gene's template, the latter is expected to lead to elevated expression levels. Intuitively, one expects this effect to be localized (a so-called \textit{cis}-effect), but some suggest that aberrations elsewhere in the DNA may directly affect the expression levels of distant genes (referred to as a \textit{trans}-effect). Figure \ref{fig:cis2transEffect} shows a cartoon of the \textit{cis}- and \textit{trans}-effect of DNA copy number on transcription.

The \textit{cis}- and \textit{trans}-effects of DNA copy aberrations on the expression levels of the KRAS oncogene in colorectal cancer are investigated. Data of both molecular levels from the TCGA (The Cancer Genome Atlas) repository are downloaded \citep{TCGA2012colon}. The gene expression data are limited to that of KRAS, while for  the DNA copy number data only that of chromosome 12, which harbors KRAS, is retained. This leaves genomic profiles of 195 samples comprising 927 aberrations. Both molecular data types are zero centered feature-wise. Moreover, the data are limited to ten -- conveniently chosen? -- samples. The KRAS expression levels are explained by the DNA copy number aberrations through the linear regression model. The model is fitted by means of ridge regression, with $\lambda \mathbf{\Delta}$ and $\mathbf{\Delta} = \mathbf{I}_{pp}$ and a single-banded $\mathbf{\Delta}$ with unit diagonal and the elements of the first off-diagonal equal to the arbitrary value of $-0.4$. The latter choice appeals to the spatial structure of the genome and encourages similar regression estimates for contiguous DNA copy numbers. The penalty parameter is chosen by means of leave-one-out cross-validation using the squared error loss. 

\begin{lstlisting}
# load libraries
library(cgdsr)
library(biomaRt)
library(Matrix)

# get list of human genes
ensembl  <- useMart("ensembl", dataset="hsapiens_gene_ensembl")
geneList <- getBM(attributes=c("ensembl_gene_id", "external_gene_name", 
                               "entrezgene_trans_name", "chromosome_name", 
                               "start_position", "end_position"), mart=ensembl)

# remove all gene without entrezID
geneList <- geneList[!is.na(geneList[,3]),]

# remove all genes not mapping to chr 12
geneList <- geneList[which(geneList[,4] %in% c(12)),]
geneList <- geneList[,-1]
geneList <- unique(geneList)
geneList <- geneList[order(geneList[,3], geneList[,4], geneList[,5]),]

# create CGDS object
mycgds   <- CGDS("http://www.cbioportal.org/public-portal/")

# get available case lists (collection of samples) for a given cancer study
mycancerstudy <- getCancerStudies(mycgds)[37,1]
mycaselist    <- getCaseLists(mycgds,mycancerstudy)[1,1]

# get available genetic profiles
mrnaProf      <- getGeneticProfiles(mycgds,mycancerstudy)[c(4),1]
cnProf        <- getGeneticProfiles(mycgds,mycancerstudy)[c(6),1]

# get data slices for a specified list of genes, genetic profile and case list
cnData   <- numeric()
geData   <- numeric()
geneInfo <- numeric()
for (j in 1:nrow(geneList)){
	geneName <- as.character(geneList[j,1])
	geneData <- getProfileData(mycgds, geneName, c(cnProf, mrnaProf), mycaselist)
	if (dim(geneData)[2] == 2 & dim(geneData)[1] > 0){
		cnData <- cbind(cnData, geneData[,1])
		geData <- cbind(geData, geneData[,2])
		geneInfo <- rbind(geneInfo, geneList[j,])
	}
}
colnames(cnData) <- rownames(geneData)
colnames(geData) <- rownames(geneData)

# preprocess data
Y  <- geData[, match("KRAS", geneInfo[,1]), drop=FALSE]
Y  <- Y - mean(Y)
X  <- sweep(cnData, 2, apply(cnData, 2, mean))

# subset data
idSample <- c(50, 58, 61, 75, 66, 22, 67, 69, 44, 20)
Y        <- Y[idSample]
X        <- X[idSample,]

# generate banded penalty matrix
diags <- list(rep(1, ncol(X)), rep(-0.4, ncol(X)-1))
Delta <- as.matrix(bandSparse(ncol(X), k=-c(0:1), diag=c(diags), symm=TRUE))

# define loss function
CVloss <- function(lambda, X, Y, Delta){
    loss <- 0
    for (i in 1:nrow(X)){
        betasLoo <- solve(crossprod(X[-i,]) + lambda * Delta) %*% 
                          crossprod(X[-i,], Y[-i])
        loss <- loss + as.numeric((Y[i] - X[i,,drop=FALSE] %*% betasLoo)^2)
    }  
    return(loss)
}

# optimize penalty parameter
limitsL <- c(10^(-10), 10^(10))
optLr  <- optimize(CVloss, limitsL, X=X, Y=Y, Delta=diag(ncol(X)))$minimum 
optLgr <- optimize(CVloss, limitsL, X=X, Y=Y, Delta=Delta)$minimum 

# evaluate (generalized) ridge estimators  
betasGr <- solve(crossprod(X) + optLgr * Delta)         %*% crossprod(X, Y)
betasR  <- solve(crossprod(X) + optLr  * diag(ncol(X))) %*% crossprod(X, Y)

# plot estimates vs. location
ylims <- c(min(betasR, betasGr), max(betasR, betasGr))
plot(betasR, type="l", ylim=ylims, ylab="copy number effect", 
             lty=2,    yaxt="n",   xlab="chromosomal location")
lines(betasGr, lty=1, col="red")
lines(seq(ylims[1], ylims[2], length.out=50) ~ 
      rep(match("KRAS", geneInfo[,1]), 50), col="grey", lwd=2, lty=3)
legend("topright", c("ridge", "fused ridge"), lwd=2,  
                   col=c("black", "red"),     lty=c(2, 1))



\end{lstlisting}

The right panel of Figure \ref{fig:fusedRidgeIllustration} shows the ridge regression estimate with both choices of $\mathbf{\Delta}$ and optimal penalty parameters plotted against the chromosomal order. The location of KRAS is indicated by a vertical dashed bar. The ordinary ridge regression estimates show a minor peak at the location of KRAS but is otherwise more or less flat. In the generalized ridge estimates the peak at KRAS is emphasized. Moreover, the region close to KRAS exhibits clearly elevated estimates, suggesting co-abberated DNA. For the remainder the generalized ridge estimates portray a flat surface, with the exception of a single downward spike away from KRAS. Such negative effects are biologically nonsensible (more gene templates leading to reduced expression levels?). On the whole the generalized ridge estimates point towards the \textit{cis}-effect as the dominant genomic regulation mechanism of KRAS expression. The isolated spike may suggest the presence of a \textit{trans}-effect, but its sign is biological nonsensible and the spike is fully absent in the ordinary ridge estimates. This leads us to ignore the possibility of a genomic \textit{trans}-effect on KRAS expression levels in colorectal cancer.

The sample selection demands justification. It yields a clear illustrate-able difference between the ordinary and fused ridge estimates. When all samples are left in, the \textit{cis}-effect is clearly present, discernable from both estimates that yield a virtually similar profile.

\section{Conclusion}
We close with a note of caution. The generalized ridge penalty is extremely flexible. It can incorporate any prior knowledge on the parameter values (through specification of $\bbeta_0$) and the relations among these parameters (via $\mathbf{\Delta}$). While a pilot study or literature may provide a suggestion for $\bbeta_0$, it is less obvious how to choose an informative $\mathbf{\Delta}$, although a spatial structure is a nice exception. In general, exact knowledge on the parameters should not be incorporated implicitly via the penalty (read: prior) but preferably be used explicitly in the model -- the likelihood -- itself. Though this may be the viewpoint of a prudent frequentist and a subjective Bayesian might disagree.

\section{Exercises}
\begin{question} \mbox{ }
\\
Suppose a researcher has hundred realizations, $i=1, \ldots, 100$, of a response $Y_{i}$ and covariates $X_{i,1}$, and $X_{i,2}$ available. In order to assess how the effect of the two covariates on the response the researcher fits the linear regression model $Y_{i} = \beta_1 X_{i,1} + \beta_2 X_{i,2} + \varepsilon_{i}$ with $\varepsilon_i \sim \mathcal{N}(0, \sigma^2)$ by means of ridge regression, but with a separate penalty parameter, $\lambda_{1}$ and $\lambda_{2}$, for the two regression coefficients, $\beta_1$ and $\beta_2$, respectively.

\begin{compactitem}
\item[\textit{a)}] Write down the ridge penalized loss function employed by the researcher.

\item[\textit{b)}] Does a different choice of penalty parameter for the second regression coefficient affect the estimation of the first regression coefficient? Motivate your answer.

\item[\textit{c)}] The researcher decides that the second covariate $X_{i,2}$ is irrelevant. Instead of removing the covariate from model, the researcher decides to set $\lambda_{2} = \infty$. Show that this results in the same ridge estimate for $\beta_1$ as when fitting (again by means of ridge regression) the model without the second covariate.
\end{compactitem}
\end{question}

\begin{question} \mbox{} \\
Consider the linear regression model $Y_i = \beta_1 X_{i,1} + \beta_2 X_{i,2} + \varepsilon_i$ for $i=1, 2, 3$. Information on the response, design matrix and relevant summary statistics are:
\begin{eqnarray*}
\mathbf{X}^{\top}  = \left( \begin{array}{rrr} -1 & 0 & 2 \\ 1 & -2 & 1 \end{array} \right), \, 
\mathbf{Y}^{\top}  = \left( \begin{array}{rrr} -1 & -1 & 2 \end{array} \right), \,
\mathbf{X}^{\top} \mathbf{X} = \left( \begin{array}{rr} 5 & 1 \\ 1 & 6 \end{array} \right), \mbox{ and } \, \mathbf{X}^{\top} \mathbf{Y} = \left( \begin{array}{r} 5 \\ 3 \end{array} \right).
\end{eqnarray*}
\begin{compactitem}
\item[\textit{a)}] Evaluate the fused ridge regression estimator with $\lambda = 2$. 

\item[\textit{b)}] Draw the contour of the parameter constraint induced by the fused ridge penalty. 

\item[\textit{c)}] Verify that for $p=2$ the ridge homogeneity penalty is equivalent to the fused ridge penalty.
\end{compactitem} 
\end{question}

\begin{figure}[!h]
\begin{tabular}{rcl}
\hspace{-0.1cm}
\includegraphics[scale=0.31, angle=0]{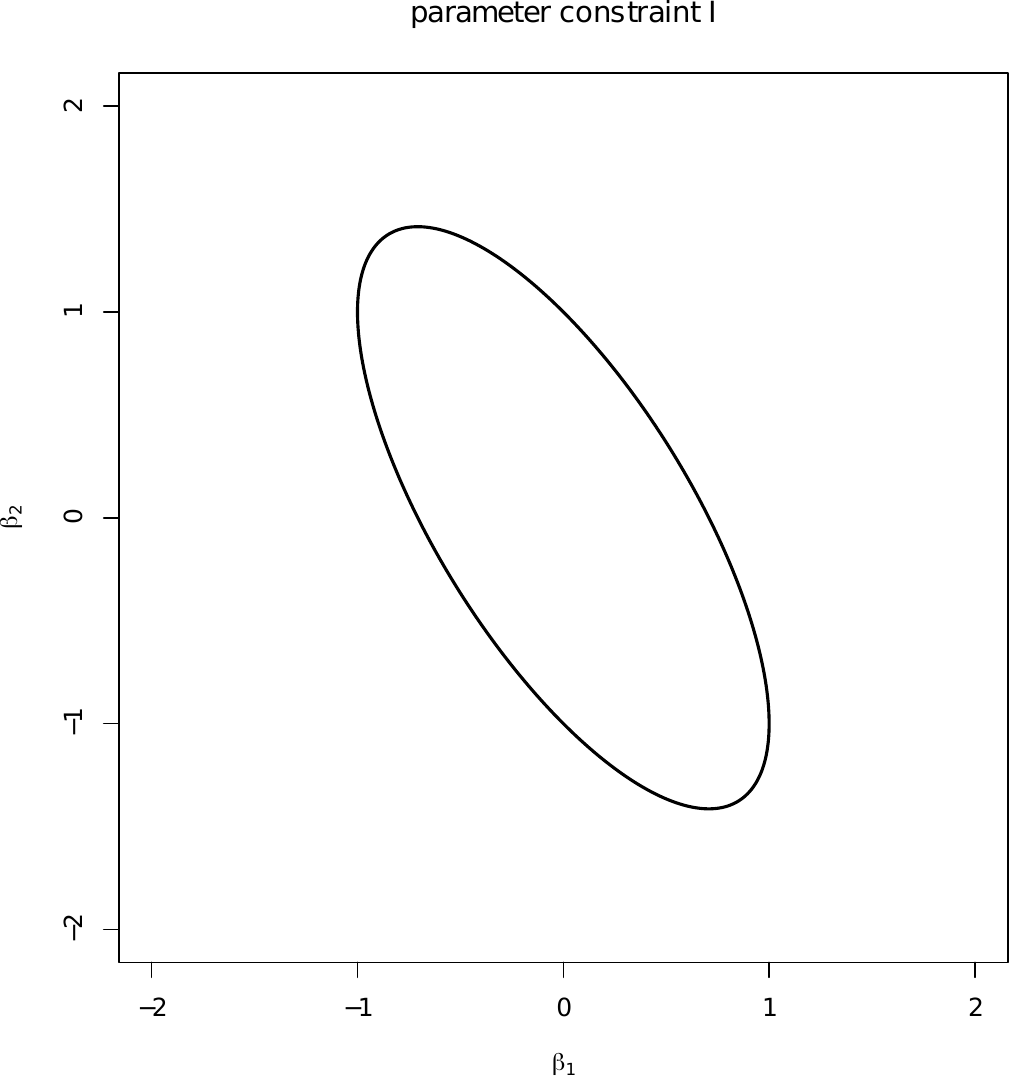}
& 
\hspace{-0.7cm}
\includegraphics[scale=0.31, angle=0]{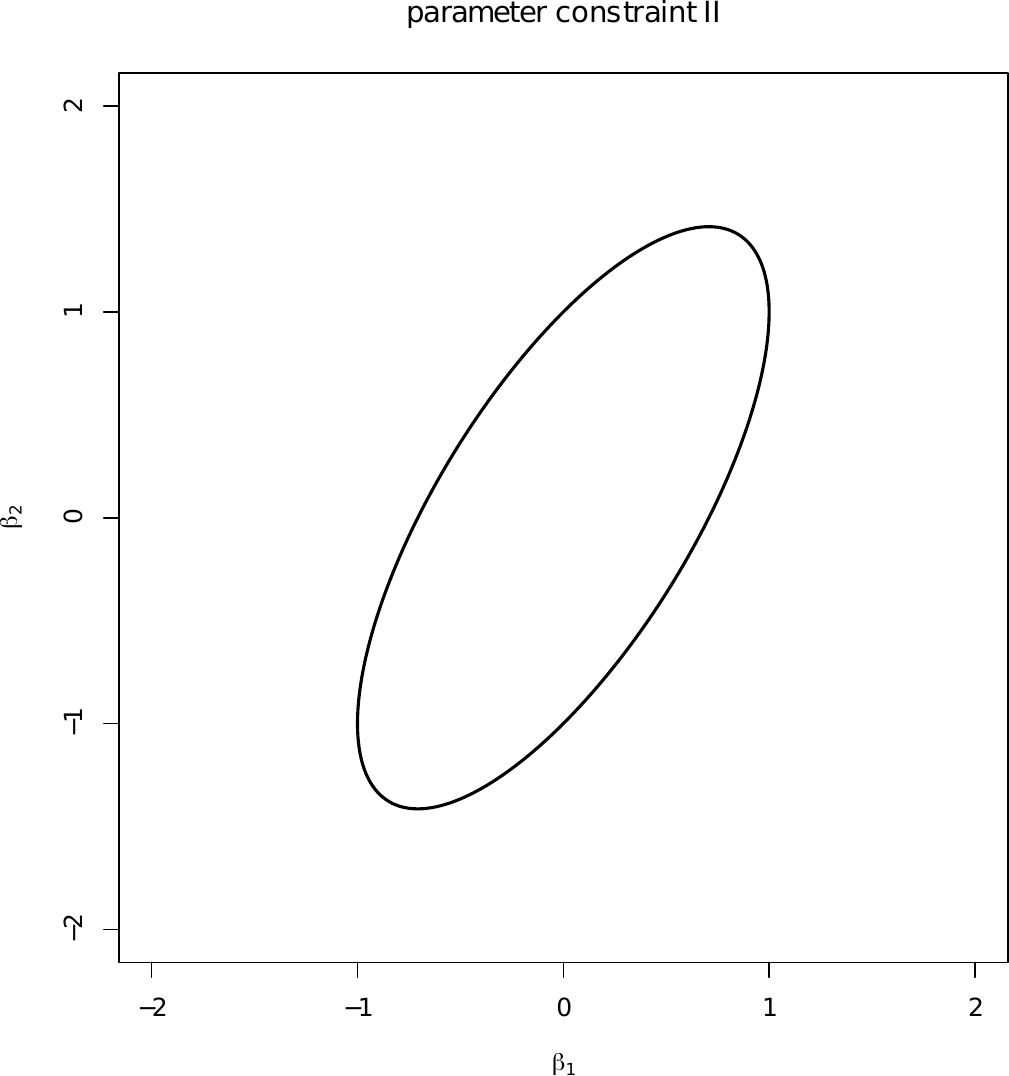}
&
\hspace{-0.7cm}
\includegraphics[scale=0.31, angle=0]{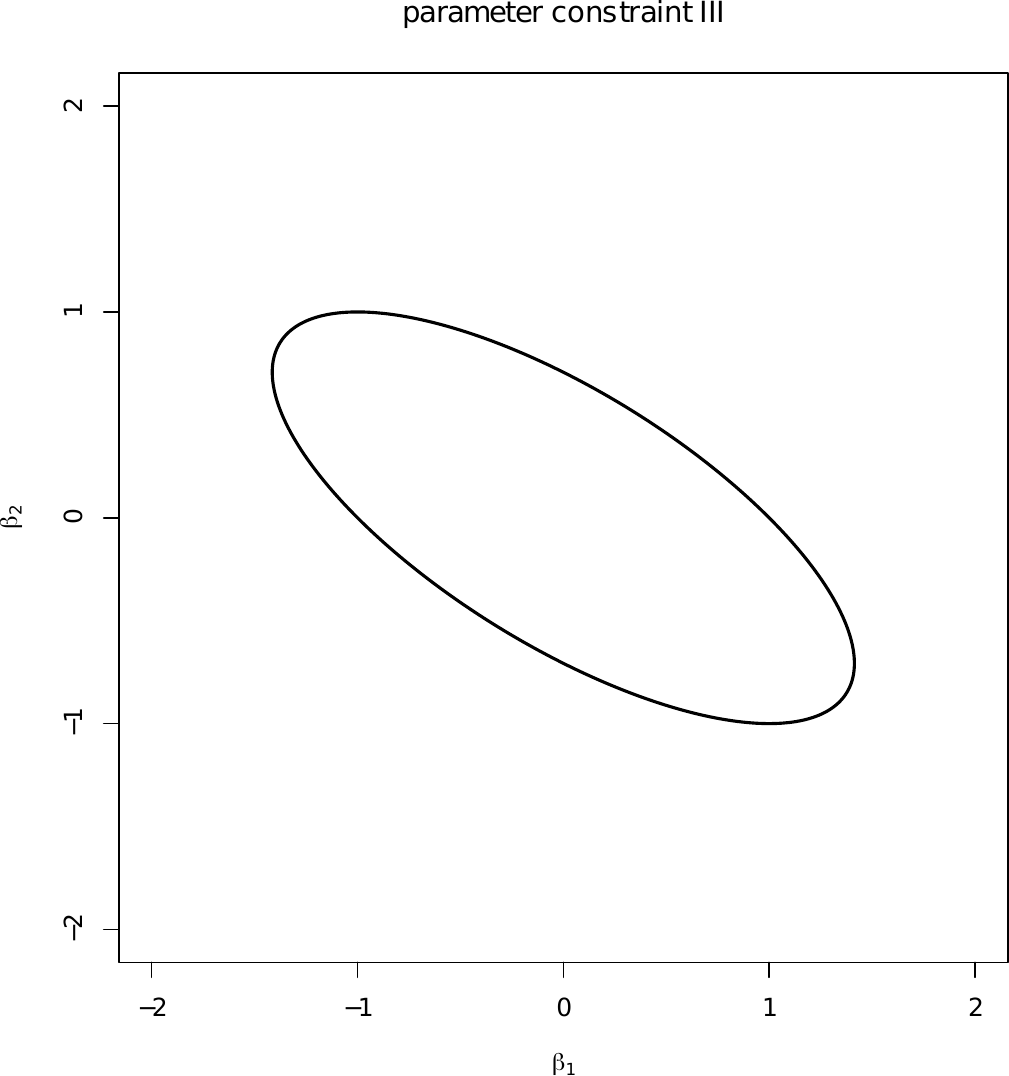}
\end{tabular}
\caption{Figure related to Question \ref{question:genRidgeConstraints}. 
\label{fig.constrainedEstExercise_genRidge}}
\end{figure}

\begin{question} \label{question:genRidgeConstraints} \mbox{ } \\
Consider the linear regression model $\mathbf{Y} = \mathbf{X} \bbeta + \vvarepsilon$ with $\vvarepsilon \sim \mathcal{N} ( \mathbf{0}_n, \sigma_{\varepsilon}^2 \mathbf{I}_{nn})$. The model is fitted using the generalized ridge regression estimator $\hat{\bbeta}(\lambda_2)  =  \arg \min\nolimits_{\bbeta \in \mathbb{R}^2} \| \mathbf{Y} - \mathbf{X} \bbeta \|_2^2 + \lambda \bbeta^{\top} \mathbf{\Delta} \bbeta$ with $\mathbf{\Delta}$ one of the below
\begin{eqnarray*}
\mathbf{\Delta}_1 \, \, \, = \, \, \, \left( \begin{array}{rr} 2 & 1 \\ 1 & 1 \end{array} \right), \, \, \mathbf{\Delta}_2 \, \, \, = \, \, \, \left( \begin{array}{rr} 1 & 1 \\ 1 & 2 \end{array} \right), \mbox{ and }  \mathbf{\Delta}_3 \, \, \, = \, \, \, \left( \begin{array}{rr} 2 & -1 \\ -1 & 1 \end{array} \right).
\end{eqnarray*}
Each of three penalty functions induces a constraint on the regression parameter. The boundary of these constraints is depicted in the panels of Figure \ref{fig.constrainedEstExercise_genRidge}. Which constraint corresponds to which penalty function? Motivate. 
\end{question}

\begin{question} \mbox{ }
\\
Consider the linear regression model $Y_i = \beta_1 X_{i,1} + \beta_2 X_{i,2} + \varepsilon_i$ for $i=1, \ldots, n$. Suppose estimates of the regression parameters $(\beta_1, \beta_2)$ of this model are obtained through the minimization of the sum-of-squares augmented with a ridge-type penalty:
\begin{eqnarray*}
\sum\nolimits_{i=1}^n (Y_i - \beta_1 X_{i,1} - \beta_2 X_{i,2})^2 + \lambda (\beta_1^2 + \beta_2^2 + 2 \nu \beta_1 \beta_2),
\end{eqnarray*}
with penalty parameters $\lambda \in \mathbb{R}_{> 0}$ and $\nu \in (-1, 1)$.

\begin{compactitem}
\item[\textit{a)}] Recall the equivalence between constrained and penalized estimation (cf. Section \ref{sect.constrainedEstimation}). Sketch (for both $\nu=0$ and $\nu=0.9$) the shape of the parameter constraint induced by the penalty above and describe in words the qualitative difference between both shapes.

\item[\textit{b)}] When $\nu  = -1$ and $\lambda \rightarrow \infty$ the estimates of $\beta_1$ and $\beta_2$ (resulting from minimization of the penalized loss function above) converge towards each other:
$\lim_{\lambda \rightarrow \infty} \hat{\beta}_1(\lambda, -1) = \lim_{\lambda \rightarrow \infty} \hat{\beta}_2(\lambda, -1)$. Motivated by this observation a data scientists incorporates the equality constraint $\beta_1 = \beta = \beta_2$ explicitly into the model, and s/he estimates the `joint regression parameter' $\beta$ through the minimization (with respect to $\beta$) of:
\begin{eqnarray*}
\sum\nolimits_{i=1}^n (Y_i - \beta  X_{i,1} - \beta X_{i,2})^2 + \delta \beta^2,
\end{eqnarray*}
with penalty parameter $\delta \in \mathbb{R}_{> 0}$. The data scientist is surprised to find that resulting estimate $\hat{\beta}(\delta)$ does not have the same limiting (in the penalty parameter) behavior as the $\hat{\beta}_1(\lambda, -1)$, i.e. $\lim_{\delta \rightarrow \infty} \hat{\beta} (\delta) \not= \lim_{\lambda \rightarrow \infty} \hat{\beta}_1(\lambda, -1)$. Explain the misconception of the data scientist.

\item[\textit{c)}] Assume that \textit{i)} $n \gg 2$, \textit{ii)} the unpenalized estimates $(\hat{\beta}_1(0, 0), \hat{\beta}_2(0, 0))^{\top}$ equal $(-2,2)$, and \textit{iii)} that the two covariates $X_1$ and $X_2$ are zero-centered, have equal variance, and are strongly negatively correlated. Consider $(\hat{\beta}_1(\lambda, \nu), \hat{\beta}_2(\lambda, \nu))^{\top}$ for both $\nu=-0.9$ and $\nu=0.9$. For which value of $\nu$ do you expect the sum of the absolute value of the estimates to be largest? \textit{Hint:} Distinguish between small and large values of $\lambda$ and think geometrically!
\end{compactitem}
\end{question}

\begin{question}\hspace{-0.05cm}$^\ast$ \mbox{ }
\\
Show that the genalized ridge regression estimator, $\hat{\bbeta}(\mathbf{\Delta}) = (\mathbf{X}^{\top} \mathbf{X} + \mathbf{\Delta})^{-1} \mathbf{X}^{\top} \mathbf{Y}$, too (as in Section \ref{sect:dataAugmentation}) can be obtained by ordinary least squares regression on an augmented data set. Hereto consider the Cholesky decomposition of the penalty matrix: $\mathbf{\Delta} = \mathbf{L}^{\top} \mathbf{L}$. Now augment the matrix $\mathbf{X}$ with $p$ additional rows comprising the matrix $\mathbf{L}$, and augment the response vector $\mathbf{Y}$ with $p$ zeros.
\end{question}

\begin{question}\hspace{-0.05cm}$^\ast$ \label{question.generalizedRidgeAndBayes} \mbox{ }
\\ 
Consider the linear regression model $\mathbf{Y} = \mathbf{X} \bbeta + \vvarepsilon$ with $\vvarepsilon \sim \mathcal{N}(\mathbf{0}_p, \sigma^2 \mathbf{I}_{pp})$. Assume $\bbeta \sim \mathcal{N}(\bbeta_0, \sigma^2 \mathbf{\Delta}^{-1})$ with $\bbeta_0 \in \mathbb{R}^p$ and $\mathbf{\Delta} \succ 0$ and a gamma prior on the error variance. Verify (i.e., work out the details of the derivation) that the posterior mean coincides with the generalized ridge estimator defined as: 
\begin{eqnarray*}
\hat{\bbeta} & = & (\mathbf{X}^{\top} \mathbf{X} + \mathbf{\Delta})^{-1} (\mathbf{X}^{\top} \mathbf{Y} + \mathbf{\Delta} \bbeta_0). 
\end{eqnarray*}
\end{question}

\begin{question} \label{question.consequencesOfHyperparameters} \mbox{ }
\\
Consider the Bayesian linear regression model $\mathbf{Y} = \mathbf{X} \bbeta + \vvarepsilon$ with $\vvarepsilon \sim \mathcal{N}(\mathbf{0}_n, \sigma^2 \mathbf{I}_{nn})$, a multivariate normal law as conditional prior distribution on the regression parameter: $\bbeta \, | \, \sigma^2 \sim \mathcal{N}(\bbeta_0, \sigma^2 \mathbf{\Delta}^{-1})$, and an inverse gamma prior on the error variance  $\sigma^2 \sim \mathcal{IG}(\gamma, \delta)$. The consequences of various choices for the hyper parameters of the prior distribution on $\bbeta$ are studied. 

\begin{compactitem}
\item[\textit{a)}] Consider the following conditional prior distributions on the regression parameters $\bbeta \, | \, \sigma^2 \sim \mathcal{N}(\bbeta_0, \sigma^2 \mathbf{\Delta}_a^{-1})$ and $\bbeta \, | \, \sigma^2 \sim \mathcal{N}(\bbeta_0, \sigma^2 \mathbf{\Delta}_b^{-1})$ with precision matrices $\mathbf{\Delta}_a, \mathbf{\Delta}_b \in \mathcal{S}_{++}^p$ such that $\mathbf{\Delta}_a \succeq \mathbf{\Delta}_b$, i.e. $\mathbf{\Delta}_a = \mathbf{\Delta}_b + \mathbf{D}$ for some positive semi-definite symmetric matrix of appropriate dimensions. Verify:
\begin{eqnarray*}
\mbox{Var}(\bbeta \, | \, \sigma^2, \mathbf{Y}, \mathbf{X}, \bbeta_0, \mathbf{\Delta}_a) & \preceq & \mbox{Var}(\bbeta \, | \, \sigma^2, \mathbf{Y}, \mathbf{X}, \bbeta_0, \mathbf{\Delta}_b),
\end{eqnarray*}
i.e. the smaller (in the positive definite ordering) the variance of the prior the smaller that of the posterior.

\item[\textit{b)}] In the remainder of this exercise assume $\mathbf{\Delta}_a =\mathbf{\Delta}  = \mathbf{\Delta}_b$. Let $\bbeta_t$ be the `true' or `ideal' value of the regression parameter, that has been used in the generation of the data, and show that a better initial guess yields a better posterior probability at $\bbeta_t$. That is, take two prior mean parameters $\bbeta_0 = \bbeta_0^{\mbox{{\tiny (a)}}}$ and $\bbeta_0 = \bbeta_0^{\mbox{{\tiny (b)}}}$ such that the former is closer to $\bbeta_t$ than the latter. Here close is defined in terms of the Mahalabonis distance, which  for, e.g. $\bbeta_t$ and $\bbeta_0^{\mbox{{\tiny (a)}}}$ is defined as $d_M(\bbeta_t, \bbeta_0^{\mbox{{\tiny (a)}}}; \mathbf{\Sigma}) =  [(\bbeta_t - \bbeta_0^{\mbox{{\tiny (a)}}})^{\top} \mathbf{\Sigma}^{-1} (\bbeta_t - \bbeta_0^{\mbox{{\tiny (a)}}})]^{1/2}$ with positive definite covariance matrix $\mathbf{\Sigma}$ with $\mathbf{\Sigma} = \sigma^2 \mathbf{\Delta}^{-1}$. Show that the posterior density $\pi_{\bbeta \, | \, \sigma^2} (\bbeta \, | \, \sigma^2, \mathbf{X}, \mathbf{Y}, \bbeta_0^{\mbox{{\tiny (a)}}}, \mathbf{\Delta})$ is larger at $\bbeta =\bbeta_t$ than with the other prior mean parameter. 

\item[\textit{c)}] Adopt the assumptions of part \textit{b)} and show that a better initial guess yields a better posterior mean. That is, show 
\begin{eqnarray*}
d_M[\bbeta_t, \mathbb{E}(\bbeta \, | \, \sigma^2, \mathbf{Y}, \mathbf{X}, \bbeta_0^{\mbox{{\tiny (a)}}}, \mathbf{\Delta}); \mathbf{\Sigma}] & \leq &  d_M[\bbeta_t, \mathbb{E}(\bbeta \, | \, \sigma^2, \mathbf{Y}, \mathbf{X}, \bbeta_0^{\mbox{{\tiny (b)}}}, \mathbf{\Delta}); \mathbf{\Sigma}],
\end{eqnarray*}
now with $\mathbf{\Sigma} = \sigma^2 (\mathbf{X}^{\top} \mathbf{X} + \mathbf{\Delta})^{-1}$.
\end{compactitem} 
\end{question}

\begin{question} \mbox{ }
\\
Revisit Exercise \ref{question:ridgeNumericalExerciseFindLambdaFromEstimator}. There the standard linear regression model $Y_i = \mathbf{X}_{i,\ast} \bbeta + \varepsilon_i$ for $i=1, \ldots, n$ and with $\varepsilon_i \sim_{i.i.d.}  \mathcal{N}(0, \sigma^2)$ is considered. The model comprises a single covariate and an intercept. Response and covariate data are: $\{(y_i, x_{i,1})\}_{i=1}^4 = \{ (1.4,  0.0),   (1.4, -2.0), (0.8,  0.0), (0.4,  2.0) \}$. 
\begin{compactitem}
\item[\textit{a)}] Evaluate the generalized ridge regression estimator of $\bbeta$ with target $\bbeta_0 = \mathbf{0}_2$ and penalty matrix $\mathbf{\Delta}$ given by $(\mathbf{\Delta})_{11} = \lambda = (\mathbf{\Delta})_{22}$ and $(\mathbf{\Delta})_{12} = \tfrac{1}{2} \lambda = (\mathbf{\Delta})_{21}$ in which $\lambda = 8$. 
\item[\textit{b)}] A data scientist wishes to leave the intercept unpenalized. Hereto s/he sets in part \textit{a)} $(\mathbf{\Delta})_{11} = 0$. Why does the resulting estimate not coincide with the answer to Exercise \ref{question:ridgeNumericalExerciseFindLambdaFromEstimator}? Motivate.
\end{compactitem}
\end{question}

\begin{question} \mbox{ } \\
Consider the linear regression model $\mathbf{Y} = \mathbf{X} \bbeta + \vvarepsilon$ with $\vvarepsilon \sim \mathcal{N} ( \mathbf{0}_n, \sigma_{\varepsilon}^2 \mathbf{I}_{nn})$. This model (without intercept) is fitted to data using the generalized regression estimator $\hat{\bbeta}(\lambda) = \arg \min_{\bbeta} \| \mathbf{Y} - \mathbf{X} \bbeta \|_2^2 + \lambda \bbeta^{\top} \mathbf{\Delta}  \bbeta$ with $\lambda > 0$. The penalty matrix $\mathbf{\Delta}$ and the data are:
\begin{eqnarray*}
\mathbf{\Delta} = \left( \begin{array}{rr} 4 & c \\ c & 1 \end{array} \right), \mathbf{X} = \left( \begin{array}{rr} -2 & 1 \end{array} \right), \mbox{ and } \, \mathbf{Y} = \left( \begin{array}{r} -2 \end{array} \right).
\end{eqnarray*}
for $c \in [-2, 2]$.
\begin{compactitem}
\item[\textit{a)}] Draw the parameter constraint on $\bbeta$ induced by the generalized ridge penalty for $c=0$ and $c=2$. Motivate the form of the drawn constraints.
\item[\textit{b)}] Evaluate the generalized  rige regression estimator $\hat{\bbeta} (\lambda, \mathbf{\Delta})$ for $c=2$.
\item[\textit{c)}] Why is the generalized ridge regression estimator $\hat{\bbeta} (\lambda, \mathbf{\Delta})$ well-defined for $c=2$ but not for $c=-2$? Motivate.
\item[\textit{d)}] Set $\lambda=1$ and $c=0$. Suppose $\bbeta = (2, 1)^{\top}$. Is the bias of of the generalized ridge regression estimator larger or smaller than that of its regular counterpart (i.e. with $\mathbf{\Delta} = \mathbf{I}_{22}$)? Motivate.
\end{compactitem}
\end{question}

\begin{question}\hspace{-0.05cm}$^\ast$ \mbox{ }
\\
Consider the linear regression model: $\mathbf{Y} = \mathbf{X} \bbeta + \vvarepsilon$ with $\vvarepsilon \sim \mathcal{N} ( \mathbf{0}_{n}, \sigma^2 \mathbf{I}_{nn})$. Let $\hat{\bbeta}(\lambda) = (\mathbf{X}^{\top} \mathbf{X} + \lambda \mathbf{I}_{pp})^{-1} \mathbf{X}^{\top} \mathbf{Y}$ be the ridge regression estimator with penalty parameter $\lambda$. The shrinkage of the ridge regression estimator propogates to the scale of the `ridge prediction' $\mathbf{X} \hat{\bbeta}(\lambda)$. To correct (a bit) for the shrinkage, \cite{deVlaming2015} propose the alternative ridge regression estimator:
$\hat{\bbeta}(\alpha)  = [ (1-\alpha) \mathbf{X}^{\top} \mathbf{X} + \alpha \mathbf{I}_{pp}]^{-1} \mathbf{X}^{\top} \mathbf{Y}$ with shrinkage parameter $\alpha \in [0,1]$.
\begin{compactitem}
\item[\textit{a)}] Let $\alpha = \lambda ( 1+ \lambda)^{-1}$. Show that $\hat{\bbeta}(\alpha) =  (1+\lambda) \hat{\bbeta}(\lambda)$ with $\hat{\bbeta}(\lambda)$ as in the introduction above. 

\item[\textit{b)}] Use part \textit{a)} and the parametrization of $\alpha$ provided there to
show that the some shrinkage has been undone. That is, show: $\mbox{Var}[ \mathbf{X} \hat{\bbeta}(\lambda)] < \mbox{Var}[ \mathbf{X} \hat{\bbeta}(\alpha)]$ for any $\lambda > 0$. 

\item[\textit{c)}] Use the singular value decomposition of $\mathbf{X}$ to show that $\lim_{\alpha \downarrow 0} \hat{\bbeta}(\alpha) = (\mathbf{X}^{\top} \mathbf{X})^{-1} \mathbf{X}^{\top} \mathbf{Y}$ (should it exist) and $\lim_{\alpha \uparrow 1} \hat{\bbeta}(\alpha) = \mathbf{X}^{\top} \mathbf{Y}$.

\item[\textit{d)}] Derive the expectation, variance and mean squared error of $\hat{\bbeta}(\alpha)$.

\item[\textit{e)}] Temporarily assume that $p=1$ and let $\mathbf{X}^{\top} \mathbf{X} = c$ for some $c > 0$. Then, $\mbox{MSE}[\hat{\bbeta}(\alpha)] = (c -1)^2 \beta^2 + \sigma^2 c [ (1-\alpha) c + \alpha ]^{-2}$. Does there exist an $\alpha \in (0,1)$ such that the mean squared error of $\hat{\bbeta}(\alpha)$ is smaller than that of its maximum likelihood counterpart? Motivate. 

\item[\textit{f)}] Now assume $p > 1$ and an orthonormal design matrix. Specify the regularization path of the alternative ridge regression estimator $\hat{\bbeta}(\alpha)$.
\end{compactitem} 
\end{question}

\begin{question}\hspace{-0.05cm}$^\ast$ \mbox{ } \\
Consider the linear regression model $\mathbf{Y} = \mathbf{X} \bbeta + \vvarepsilon$ with $\vvarepsilon \sim \mathcal{N}(\mathbf{0}_n, \sigma^2 \mathbf{I}_nn)$. \cite{goldstein1974ridge} proposed a novel generalized ridge estimator of its $p$-dimensional regression parameter:
\begin{eqnarray*}
\hat{\bbeta}_m(\lambda) & = & [ (\mathbf{X}^{\top} \mathbf{X})^m + \lambda \mathbf{I}_{pp} ]^{-1} (\mathbf{X}^{\top} \mathbf{X})^{m-1} \mathbf{X}^{\top} \mathbf{Y},
\end{eqnarray*}
with penalty parameter $\lambda > 0$ and `shape' or `rate' parameter $m$. 
\begin{compactitem}
\item[\textit{a)}] Assume, only for part \textit{a)}, that $n=p$ and the design matrix is orthonormal. Show that, irrespectively of the choice of $m$, this generalized ridge regression estimator coincides with the `regular' ridge regression estimator.

\item[\textit{b)}] Consider the generalized ridge loss function $\| \mathbf{Y} - \mathbf{X} \bbeta \|_2^2 + \bbeta^{\top} \mathbf{A} \bbeta$ with $\mathbf{A}$ a $p \times p$-dimensional symmetric matrix. For what $\mathbf{A}$, does $\hat{\bbeta}_m(\lambda)$ minimize this loss function?

\item[\textit{c)}] Let $d_j$ be the $j$-th singular value of $\mathbf{X}$. Show that in $\hat{\bbeta}_m(\lambda)$ the singular values are shrunken as $(d_j^{2m} + \lambda)^{-1} d_j^{2m-2}$. \textit{Hint:} use the singular value decomposition of $\mathbf{X}$.



\item[\textit{d)}] Do, for positive singular values, larger $m$ lead to more shrinkage? \textit{Hint:} Involve particulars of the singular value in your answer.

\item[\textit{e)}] Express $\mathbb{E}[\hat{\bbeta}_m(\lambda)]$ in terms of the design matrix, model and shrinkage parameters $\lambda$ and $m$.

\item[\textit{f)}] Express $\mbox{Var}[\hat{\bbeta}_m(\lambda)]$ in terms of the design matrix, model and shrinkage parameters $\lambda$ and $m$.


\end{compactitem}
\end{question}

\begin{question} \mbox{ } \\
Consider the linear regression model $\mathbf{Y} = \mathbf{X} \bbeta + \vvarepsilon$ with $\vvarepsilon \sim \mathcal{N} ( \mathbf{0}_n, \sigma_{\varepsilon}^2 \mathbf{I}_{nn})$. The model is fitted to the following (summaries of the) data:
\begin{eqnarray*}
\mathbf{X} = \left( \begin{array}{rr} 4 & -2 \end{array} \right), 
\mathbf{X}^{\top} \mathbf{X} = \left( \begin{array}{rr} 16 & -12 \\ -12 & 9 \end{array} \right), \mathbf{Y} = \left( \begin{array}{r} 0 \end{array} \right), \, \mbox{ and } \, \mathbf{X}^{\top} \mathbf{Y} = 
\left( \begin{array}{r} 0 \\ 0 \end{array} \right).
\end{eqnarray*}
by means of the generalized ridge regression estimator $\boldsymbol{\beta}(\lambda) = \arg \min_{\boldsymbol{\beta} \in \mathbb{R}^2} \| \mathbf{Y} - \mathbf{X} \boldsymbol{\beta} \|_2^2 + \lambda (\boldsymbol{\beta} - \boldsymbol{\beta}_0)^{\top} \mathbf{\Delta} (\boldsymbol{\beta} - \boldsymbol{\beta}_0)$ with $\boldsymbol{\beta}_0 = (0, 1)^{\top}$ and
\begin{eqnarray*}
\mathbf{\Delta} = \left( \begin{array}{rr} 2 & 1 \\ 1 & 2 \end{array} \right).
\end{eqnarray*}

\begin{compactitem}
\item[\textit{a)}] Draw the parameter constraint  of the generalized ridge regression estimator for an arbitrary but finite $\lambda > 0$. 

\item[\textit{b)}] Evaluate the ridge regression estimator for $\lambda = \tfrac{1}{3}$.

\item[\textit{c)}\hspace{-0.02cm}$^\ast$] Suppose that $\boldsymbol{\beta} = (1,1)^{\top}$ and $\sigma_{\varepsilon}^2 = 1$. Show that the Mean Squared Error (MSE) of the generalized estimator equals: $\mbox{MSE}[ \hat{\boldsymbol{\beta}} ( \lambda)] =  (3\lambda + 74)^{-2} ( 2721  + 180 \lambda + 9 \lambda^2 )$.

\item[\textit{d)}] Find the value of $\lambda$ that yields the best -- in the MSE sense -- generalized ridge regression estimator. Use part \textit{c)} in your answer.

\item[\textit{e)}] Now adopt the Bayesian formulation of the generalized ridge regression estimator, which assumes the prior $\boldsymbol{\beta} \, | \, \sigma^2 \sim \mathcal{N}(\boldsymbol{\beta}_0, \sigma^2 \lambda^{-1} \mathbf{\Delta})$ with $\lambda_2 = \tfrac{1}{3}$ and $\boldsymbol{\beta}_0$ and $\mathbf{\Delta}$ as above. Give the mean and variance of the conditional posterior distribution of $\boldsymbol{\bbeta} \, | \, \sigma^2 = 1, \mathbf{Y}, \mathbf{X}$.
\end{compactitem}
\end{question}

\begin{question} \mbox{ } \\
Consider the generalized ridge regression estimator of the regression parameter of the linear regression model. The employed generalized ridge penalty is:
\begin{eqnarray*}
\lambda \beta_1^2 + \lambda_f \sum\nolimits_{j=2}^p (\beta_{j} - \beta_1)^2.
\end{eqnarray*}
\begin{compactitem}
\item[\textit{a)}] 
Describe in words the shrinkage behavior of this generalized ridge regression estimator induced by either penalty parameter. 

\item[\textit{b)}] Assume $\lambda_f = \lambda$. The penalty can then be written in matrix notation $\lambda \bbeta^{\top} \mathbf{\Delta} \bbeta$ with $p \times p$-dimensional matrix $\mathbf{\Delta}$. Specify $\mathbf{\Delta}$. You may assume $p=4$.

\item[\textit{c)}] If $\lambda =0$ but $\lambda_f > 0$, $\mathbf{\Delta}$ is not positive definite. Is this generalized ridge regression estimator then uniquely defined? 
\end{compactitem}
\end{question}

\begin{question} \mbox{ } \label{exercise.mixturePrior} \\
Bayesian regression allows for other than normal priors. A normal prior on the regression parameter implicitly assumes that all elements of the parameter are (in some sense) comparable.   But when the linear regression model employs many explanatory covariates, it seems unlikely that all contribute roughly equally to the variation of the response variable. One may hypothesize that most of the regression coefficients are equal or close to zero with the remaining ones clearly deviating from zero. Such believe may be incorporated into Bayesian regression through the prior. 

The `\textit{sparse} regression parameter'-believe may be operationalized (as is done in \citealp{george1993variable}) by the following prior:
\begin{eqnarray*}
\beta_j \, | \, \sigma_\varepsilon^2 & \sim & \theta_{\mbox{{\tiny \texttt{<0}}}} \, \mathcal{N} (\beta_{\mbox{{\tiny \texttt{<0}}}}, \sigma_{\mbox{{\tiny \texttt{<0}}}}^2 \, \sigma_\varepsilon^2 \, \lambda^{-1}) + \theta_{\mbox{{\tiny \texttt{=0}}}} \, \mathcal{N}( \beta_{\mbox{{\tiny \texttt{=0}}}}, \sigma_{\mbox{{\tiny \texttt{=0}}}}^2 \, \sigma_\varepsilon^2 \, \lambda^{-1} ) + \theta_{\mbox{{\tiny \texttt{>0}}}} \, \mathcal{N}(\beta_{\mbox{{\tiny \texttt{>0}}}}, \sigma_{\mbox{{\tiny \texttt{>0}}}}^2 \, \sigma_\varepsilon^2 \, \lambda^{-1}),
\end{eqnarray*}
with mixing component probabilities $\theta_{\mbox{{\tiny \texttt{<0}}}},  \theta_{\mbox{{\tiny \texttt{=0}}}}, \theta_{\mbox{{\tiny \texttt{>0}}}} \in [0,1]$ such that $\theta_{\mbox{{\tiny \texttt{<0}}}} + \theta_{\mbox{{\tiny \texttt{=0}}}} + \theta_{\mbox{{\tiny \texttt{>0}}}} = 1$. To separate the locations of the mixture components, we assume $\beta_{\mbox{{\tiny \texttt{<0}}}} < 0$, $\beta_{\mbox{{\tiny \texttt{=0}}}} = 0$, $\beta_{\mbox{{\tiny \texttt{>0}}}} > 0$. Furthermore, we assume 
$\sigma_{\mbox{{\tiny \texttt{=0}}}}^2 \ll 1$. Moreover, $\lambda$ is introduced to align this set-up with that of generalized ridge regression and $\sigma_\varepsilon^2$  is only a convenient scaling. The second mixture component then represents the regression coefficients that are effectively indistinguishable from zero, whereas the first and third component represent the regression coefficients that are either negative and positive, respectively. If we introduce a latent indicator variable $Z_j$ for the mixture component, we can write, e.g. $\beta_j \, | \, Z_j = \mbox{`\texttt{<0}\mbox{}'}, \sigma_\varepsilon^2   \sim \mathcal{N} (\beta_{\mbox{{\tiny \texttt{<0}}}}, \sigma_{\mbox{{\tiny \texttt{<0}}}}^2 \, \sigma_\varepsilon^2 \, \lambda^{-1})$. In the remainder, we denote the vector of indicator variables $(Z_1, Z_2, \ldots, Z_p)$ by $\mathbf{Z}$. To wrap up the Bayesian set-up, we adopt a Dirichlet prior for the mixing components:
\begin{eqnarray*}
\boldsymbol{\theta} = (\theta_{\mbox{{\tiny \texttt{<0}}}}, \theta_{\mbox{{\tiny \texttt{=0}}}}, \theta_{\mbox{{\tiny \texttt{>0}}}}) & \sim & \mbox{Dir}(\alpha_{\mbox{{\tiny \texttt{<0}}}}, \alpha_{\mbox{{\tiny \texttt{=0}}}}, \alpha_{\mbox{{\tiny \texttt{>0}}}}) 
\end{eqnarray*}
which is conjugate for the multinomial distribution. Finally, for the error variance, we retain the inverse gamma prior $\sigma_\varepsilon^2 \sim \mathcal{IG}(a_0, b_0)$. 

\begin{compactitem}
\item[\textit{a)}] To construct a Gibbs sampler for the Bayesian regression specified above, derive all conditional posterior distributions one-by-one.
\begin{compactitem}
\item[\textit{i)}] Derive $\pi( \boldsymbol{\theta} \, | \, \mathbf{Z})$, the posterior of the mixing proportions given the mixture component indicators $\mathbf{Z}$. \textit{Hint}: use Bayes' rule.

\item[\textit{ii)}] Derive $\pi (Z_j \, | \, \beta_j, \boldsymbol{\theta})$, the posterior of indicator variables $Z_j$. \textit{Hint}: use Bayes' rule.

\item[\textit{iii)}] Derive $\pi(\boldsymbol{\beta} \, | \, \mathbf{Y}, \mathbf{X}, \mathbf{Z}, \sigma_{\varepsilon}^2)$, the posterior of regression parameter. 

\item[\textit{iv)}] Derive $\pi(\sigma^2_{\varepsilon} \, | \, \boldsymbol{\beta}, \mathbf{Y}, \mathbf{X}, \mathbf{Z})$, the posterior of the error variance.
\end{compactitem}

\item[\textit{b)}] Implement the Gibbs sampler in \texttt{R}. Use the computationally efficient sampler (Algorithm \ref{alg.highdimNormalGenRidge}) for drawing from a high-dimensional multivariate normal distribution.

\item[\textit{c)}] Revisit the microRNA data example of Section \ref{sect.ridgeRegressionDataIllustration}. Assume the expression levels of the MCM7 can be modelled by linear combination of those of the microRNAs, i.e. the linear regression model $\mathbf{Y} = \mathbf{X} \bbeta + \vvarepsilon$ serves as an reasonable description relating the two types of expression levels. Fit the Bayesian regression model to the data. Study the posteriors of $Z_j$ and $\beta_j$ and identify coefficients that clearly differ from zero. 

\item[\textit{d)}] Investigate the robustness of the posterior mean of the regression parameters against choices of the hyperparameters $
\beta_{\mbox{{\tiny \texttt{<0}}}}$, $\beta_{\mbox{{\tiny \texttt{>0}}}}$, $\sigma_{\mbox{{\tiny \texttt{<0}}}}^2$, $\sigma_{\mbox{{\tiny \texttt{?0}}}}^2$.

\item[\textit{e)}] Fit the Bayesian regression model with a normal prior and study the difference between the estimates of the two different priors.
\end{compactitem}
\end{question}

\pagestyle{fancy}

\chapter[Mixed model]{Mixed model} \label{chap.mixedModel}
Here the mixed model introduced by \cite{Henderson1953}, which generalizes the linear regression model, is studied and estimated in unpenalized (!) fashion. Nonetheless, it will turn out to have an interesting connection to ridge regression. This connection may be exploited to arrive at an informed choice of the ridge penalty parameter. 

The linear regression model, $\mathbf{Y} = \mathbf{X}  \bbeta + \vvarepsilon$, assumes the effect of each covariate to be fixed. In certain situations it may be desirable to relax this assumption. For instance, a study may be replicated. Conditions need not be exactly constant across replications. Among others this may be due to batch effects. These may be accounted for and are then incorporated in the linear regression model. But it is not the effects of these particular batches included in the study that are of interest. Would the study have been carried out at a later date, other batches may have been involved. Hence, the included batches are thus a random draw from the population of all batches. With each batch possibly having a different effect, these effects may also be viewed as random draws from some hypothetical `effect'-distribution. From this point of view the effects estimated by the linear regression model are realizations from the `effect'-distribution. But interest is not in the particular but the general. Hence, a model that enables a generalization to the distribution of batch effects would be more suited here.

Like the linear regression model the \textit{mixed model}, also called \textit{mixed effects model} or \textit{random effects model}, explains the variation in the response by a linear combination of the covariates. The key difference lies in the fact that the latter model distinguishes two sets of covariates, one with fixed effects and the other with random effects. In matrix notation mirroring that of the linear regression model, the mixed model can be written as:
\begin{eqnarray*}
\mathbf{Y} & = & \mathbf{X} \bbeta + \mathbf{Z} \ggamma + \vvarepsilon,
\end{eqnarray*}
where $\mathbf{Y}$ is the response vector of length $n$, $\mathbf{X}$ the $(n \times p)$-dimensional design matrix with the fixed vector $\bbeta$ with $p$ fixed effects, $\mathbf{Z}$ the $(n \times q)$- dimensional design matrix with an associated $q \times 1$ dimensional vector $\ggamma$ of random effects, and distributional assumptions $\vvarepsilon \sim \mathcal{N}( \mathbf{0}_{n}, \sigma_{\varepsilon}^2 \mathbf{I}_{nn})$, $\ggamma \sim \mathcal{N}(\mathbf{0}_{q}, \mathbf{R}_{\ttheta})$ and $\vvarepsilon$ and $\ggamma$ independent. In this $\mathbf{R}_{\ttheta}$ is symmetric, positive definite and parametrized by a low-dimensional parameter $\ttheta$.

The distribution of $\mathbf{Y}$ is fully defined by the mixed model and its accompanying assumptions. As $\mathbf{Y}$ is a linear combination of normally distributed random variables, it is itself normally distributed. Its mean is: 
\begin{eqnarray*}
\mathbb{E}(\mathbf{Y}) & = & \mathbb{E}(\mathbf{X} \bbeta + \mathbf{Z} \ggamma + \vvarepsilon) \, \, \, = \, \, \, \mathbb{E}(\mathbf{X} \bbeta) + \mathbb{E}(\mathbf{Z} \ggamma) + \mathbb{E}(\vvarepsilon) \, \, \, = \, \, \,  \mathbf{X} \bbeta + \mathbf{Z} \mathbb{E}(\ggamma)   \, \, \, = \, \, \,  \mathbf{X} \bbeta,
\end{eqnarray*}
while its variance is:
\begin{eqnarray*}
\mbox{Var}(\mathbf{Y}) & = & \mbox{Var}(\mathbf{X} \bbeta) + \mbox{Var}( \mathbf{Z} \ggamma) + \mbox{Var}(\vvarepsilon) \, \, \, = \, \, \, \mathbf{Z}  \mbox{Var}(\ggamma) \mathbf{Z}^{\top} + \sigma_\varepsilon^2 \mathbf{I}_{nn}  \, \, \, = \, \, \, \mathbf{Z}  \mathbf{R}_{\theta} \mathbf{Z}^{\top} + \sigma_\varepsilon^2 \mathbf{I}_{nn}
\end{eqnarray*}
in which the independence between $\vvarepsilon$ and $\ggamma$ 
and the standard algebra rules for the $\mbox{Var}(\cdot)$ and $\mbox{Cov}(\cdot)$ operators have been used. Put together, this yields: $\mathbf{Y} \sim \mathcal{N}(\mathbf{X} \bbeta, \mathbf{Z} \mathbf{R}_{\theta} \mathbf{Z}^{\top} + \sigma^2_{\varepsilon} \mathbf{I}_{nn})$. Hence, the random effects term $\mathbf{Z} \ggamma$ of the mixed model does not contribute to the explanation of the mean of $\mathbf{Y}$, but aids in the decomposition of its variance around the mean. From this formulation of the model it is obvious that the random part of two distinct observations of the response are -- in general -- not independent: their covariance is given by the corresponding element of $\mathbf{Z} \mathbf{R}_{\theta} \mathbf{Z}^{\top}$. Put differently, due to the independence assumption on the error two observations can only be (marginally) dependent through the random effect which is attenuated by the associated design matrix $\mathbf{Z}$. To illustrate this, temporarily set $\mathbf{R}_{\theta} = \sigma_{\gamma}^2 \mathbf{I}_{qq}$. Then, $\mbox{Var}(\mathbf{Y}) = \sigma_{\gamma}^2 \mathbf{Z} \mathbf{Z}^{\top} + \sigma_{\varepsilon}^2 \mathbf{I}_{nn}$. From this it is obvious that two variates of $\mathbf{Y}$ are now independent if and only if the corresponding rows of $\mathbf{Z}$ are orthogonal. Moreover, two pairs of variates have the same covariance if they have the same covariate information in $\mathbf{Z}$. Two distinct observations of the same individual have the same covariance as one of these observations with that of another individual with identical covariate information as the left-out observation on the former individual. In particular, their `between-covariance' equals their individual `within-covariance'.

The mixed model and the linear regression model are clearly closely related: they share a common mean, a normally distributed error, and both explain the response by a linear combination of the explanatory variables. Moreover, 
when $\ggamma$ is known, the mixed model reduces to a linear regression model. This is seen from the conditional distribution of $\mathbf{Y}$: $\mathbf{Y} \, | \, \ggamma \sim \mathcal{N}(\mathbf{X} \bbeta + \mathbf{Z} \ggamma, \sigma^2_{\varepsilon} \mathbf{I}_{nn})$. Conditioning on the random effect $\ggamma$ thus pulls in the term $\mathbf{Z} \ggamma$ to the systematic, non-random explanatory part of the model. In principle, the conditioned mixed model could now be rewritten as  a linear regression model by forming a new design matrix and parameter from
$\mathbf{X}$ and $\mathbf{Z}$ and $\bbeta$ and $\ggamma$, respectively.

\begin{example} \textit{(Mixed model for a longitudinal study)} \\
A longitudinal study looks into the growth rate of cells. At the beginning of the study cells are placed in $n$ petri dishes, with the same growth medium but at different concentrations. The initial number of cells in each petri dish is counted as is done at several subsequent time points. The change in cell count is believed to be -- at the log-scale -- a linear function of the concentration of the growth medium. The linear regression model may suffice. However, variation is omnipresent in biology. That is, apart from variation in the initial cell count, each cell -- even if they are from common decent -- will react (slightly?) differently to the stimulus of the growth medium. This intrinsic cell-to-cell variation in growth response may be accommodated for in the linear mixed model by the introduction of a random cell effect, both in off-set and slope. The (log) cell count of petri dish $i$ at time point $t$, denoted $Y_{it}$, is thus described by:
\begin{eqnarray*}
Y_{it} & = & \beta_0 + X_i \beta_1 + \mathbf{Z}_i \ggamma + \varepsilon_{it},
\end{eqnarray*}
with intercept $\beta_0$, growth medium concentration $X_i$ in petri dish $i$, and fixed growth medium effect $\beta_1$, and $\mathbf{Z}_i = (1, X_i)$, $\ggamma$ the $2$ dimensional random effect parameter bivariate normally distributed with zero mean and diagonal covariance matrix, and finally $\varepsilon_{it}\sim \mathcal{N}(0,\sigma_{\varepsilon}^2)$ the error in the cell count of petri dish $i$ at time $t$. In matrix notation the matrix $\mathbf{Z}$ would comprise of $2n$ columns: two columns for each cell, 
$\mathbf{e}_i$ and $X_i \mathbf{e}_i$ (with $\mathbf{e}_i$ the $n$-dimensional unit vector with a one at the $i$-th location and zeros elsewhere), corresponding to the random intercept and slope effect, respectively. 
\begin{figure}[!h]
\centering
\includegraphics[scale=0.30, angle=0]{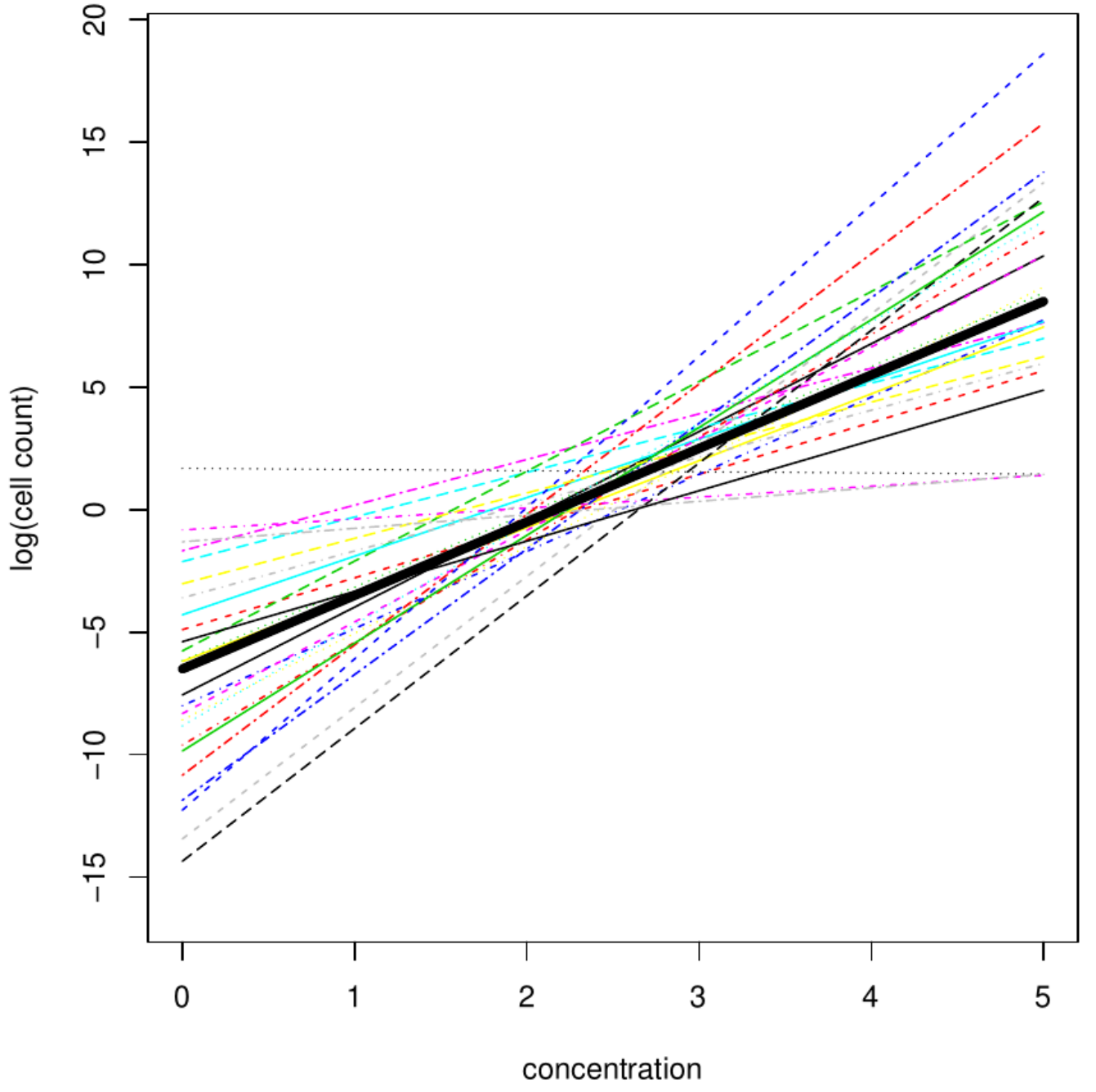}
\caption{Linear regession (thick black solid line) vs. mixed model fit (thin colored and patterned lines). } \label{fig:mixedModelExamples}
\end{figure}
The fact that the number columns of $\mathbf{Z}$, i.e. the explanatory random effects, equals $2n$ does not pose identifiability problems as per column only a single parameter is estimated. Finally, to illustrate the difference between the linear regression and the linear mixed model their fits on artifical data are plotted (top left panel, Figure \ref{fig:mixedModelExamples}). Where the linear regression fit shows the `grand mean relationship' between cell count and growth medium, the linear mixed model fit depicts the petri dish specific fits. 
\end{example}

The mixed model was motivated by its ability to generalize to instances not included in the study. From the examples above another advantage can be deduced. E.g., the cells' effects are modelled by a single parameter (rather than one per cell). More degrees of freedom are thus left to estimate the noise level. In particular, a test for the presence of a cell effect will have more power.
\\
\\
The parameters of the mixed model are estimated either by means of likelihood maximization or a related procedure known as restricted maximum likelihood. Both are presented, with the expos\'{e} loosely based on  \cite{bates2004linear}. First the maximum likelihood procedure is introduced, which requires the derivation of the likelihood. Hereto the assumption on the random effects is usually transformed. Let $\tilde{\mathbf{R}}_{\theta} = \sigma_{\varepsilon}^{-2} \mathbf{R}_{\theta}$, which is the covariance of the random effects parameter relevative to the error variance, and $\tilde{\mathbf{R}}_{\theta} = \mathbf{L}_{\theta} \mathbf{L}_{\theta}^{\top}$ its Cholesky decomposition.  Next define the change-of-variables $\ggamma = \mathbf{L}_{\theta} \tilde{\ggamma}$. This transforms the model to: $\mathbf{Y} =  \mathbf{X} \bbeta + \mathbf{Z} \mathbf{L}_{\ttheta} \tilde{\ggamma} + \vvarepsilon$ but now with the assumption $\tilde{\ggamma} \sim \mathcal{N}(\mathbf{0}_{q}, \sigma_{\varepsilon}^2 \mathbf{I}_{qq})$. Under this assumption the conditional likelihood, conditional on the random effects, is:
\begin{eqnarray*}
L(\mathbf{Y} \, | \, \tilde{\ggamma} = \mathbf{g}) & = & (2 \pi \sigma_{\varepsilon}^2)^{-n/2} \exp ( -\tfrac{1}{2} \sigma_{\varepsilon}^{-2} \| \mathbf{Y} - \mathbf{X} \bbeta - \mathbf{Z} \mathbf{L}_{\theta} \mathbf{g} \|_2^2 ).
\end{eqnarray*}
From this the unconditional likelihood is obtained through:
\begin{eqnarray*}
L(\mathbf{Y}) & = & \int_{\mathbb{R}^q} L(\mathbf{Y} \, | \, \tilde{\ggamma} = \mathbf{g}) \, f_{\tilde{\ggamma}}(\mathbf{g}) \, d \mathbf{g}
\\
& = & \int_{\mathbb{R}^q} (2 \pi \sigma_{\varepsilon}^2)^{-(n+q)/2} \exp [ -\tfrac{1}{2} \sigma_{\varepsilon}^{-2} (\| \mathbf{Y} - \mathbf{X} \bbeta - \mathbf{Z} \mathbf{L}_{\theta} \mathbf{g} \|_2^2 + \| \mathbf{g} \|_2^2 ) ] \, d \mathbf{g}.
\end{eqnarray*}
To evaluate the integral, the exponent needs rewriting. Hereto first note that:
\begin{eqnarray*}
\| \mathbf{Y} - \mathbf{X} \bbeta - \mathbf{Z} \mathbf{L}_{\theta} \mathbf{g} \|_2^2 + \| \mathbf{g} \|_2^2 & =  & (\mathbf{Y} - \mathbf{X} \bbeta - \mathbf{Z} \mathbf{L}_{\theta} \mathbf{g})^{\top} (\mathbf{Y} - \mathbf{X} \bbeta - \mathbf{Z} \mathbf{L}_{\theta} \mathbf{g})
+ \mathbf{g}^{\top} \mathbf{g}.
\end{eqnarray*}
Now expand the right-hand side as follows:
\begin{eqnarray*}
& & \hspace{-1cm} (\mathbf{Y} - \mathbf{X} \bbeta - \mathbf{Z} \mathbf{L}_{\theta} \mathbf{g})^{\top} (\mathbf{Y} - \mathbf{X} \bbeta - \mathbf{Z} \mathbf{L}_{\theta} \mathbf{g}) + \mathbf{g}^{\top} \mathbf{g}
\\
& = & \mathbf{Y}^{\top} \mathbf{Y} + \mathbf{g}^{\top}  
(\mathbf{L}_{\theta}^{\top} \mathbf{Z}^{\top} \mathbf{Z} \mathbf{L}_{\theta} + \mathbf{I}_{qq}) \mathbf{g} + \bbeta^{\top} \mathbf{X}^{\top} \mathbf{X} \bbeta - \mathbf{Y}^{\top} \mathbf{X} \bbeta - \bbeta^{\top} \mathbf{X}^{\top} \mathbf{Y}
\\
& & - (\mathbf{Y}^{\top} \mathbf{Z} \mathbf{L}_{\theta} - \bbeta^{\top} \mathbf{X}^{\top} \mathbf{Z} \mathbf{L}_{\theta}) \mathbf{g}
-  \mathbf{g}^{\top} (\mathbf{L}_{\theta}^{\top}  \mathbf{Z}^{\top} \mathbf{Y} - \mathbf{L}_{\theta}^{\top} \mathbf{Z}^{\top}
\mathbf{X} \bbeta)
\\
& = & (\mathbf{g} - \mmu_{\tilde{\ggamma} \, | \mathbf{Y}})^{\top} (\mathbf{L}_{\theta}^{\top} \mathbf{Z}^{\top} \mathbf{Z} \mathbf{L}_{\theta} + \mathbf{I}_{qq}) (\mathbf{g} - \mmu_{\tilde{\ggamma} \, | \, \mathbf{Y}})
\\
& & + ~ (\mathbf{Y} - \mathbf{X} \bbeta)^{\top} [\mathbf{I}_{nn} - \mathbf{Z} \mathbf{L}_{\theta} (\mathbf{L}_{\theta}^{\top} \mathbf{Z}^{\top} \mathbf{Z} \mathbf{L}_{\theta} + \mathbf{I}_{qq})^{-1} \mathbf{L}_{\theta}^{\top} \mathbf{Z}^{\top} ] (\mathbf{Y} - \mathbf{X} \bbeta)
\\
& = & (\mathbf{g} - \mmu_{\tilde{\ggamma} \, | \mathbf{Y}})^{\top} (\mathbf{L}_{\theta}^{\top} \mathbf{Z}^{\top} \mathbf{Z} \mathbf{L}_{\theta} + \mathbf{I}_{qq}) (\mathbf{g} - \mmu_{\tilde{\ggamma} \, | \, \mathbf{Y}})
\\
& & + ~ (\mathbf{Y} - \mathbf{X} \bbeta)^{\top} (\mathbf{I}_{nn} +  \mathbf{Z} \tilde{\mathbf{R}}_{\theta} \mathbf{Z}^{\top})^{-1} (\mathbf{Y} - \mathbf{X} \bbeta),
\end{eqnarray*}
where $\mmu_{\tilde{\ggamma} \, | \mathbf{Y}} = (\mathbf{L}_{\theta}^{\top} \mathbf{Z}^{\top} \mathbf{Z} \mathbf{L}_{\theta} + \mathbf{I}_{qq})^{-1} \mathbf{L}_{\theta}^{\top} \mathbf{Z}^{\top} (\mathbf{Y} - \mathbf{X} \bbeta)$ and the Woodbury identity has been used in the last step. As the notation suggests $\mmu_{\tilde{\ggamma} \, | \mathbf{Y}}$ is the conditional expectation of the random effect conditional on the data: $\mathbb{E}(\tilde{\ggamma} \, | \, \mathbf{Y})$. This may be verified from the conditional distribution $\tilde{\ggamma} \, | \, \mathbf{Y}$ when exploiting the equality derived in the preceeding display. Substitute the latter in the integral of the likelihood and use the change-of-variables:
$\mathbf{h} = (\mathbf{L}_{\theta}^{\top} \mathbf{Z}^{\top} \mathbf{Z} \mathbf{L}_{\theta} + \mathbf{I}_{qq})^{1/2} ( \mathbf{g} - \mmu_{\tilde{\ggamma} \, | \, \mathbf{Y}} )$ with Jacobian $| (\mathbf{L}_{\theta}^{\top} \mathbf{Z}^{\top} \mathbf{Z} \mathbf{L}_{\theta} + \mathbf{I}_{qq})^{1/2} |$:
\begin{eqnarray}
\nonumber
L(\mathbf{Y}) & = & \int_{\mathbb{R}^q} (2 \pi \sigma_{\varepsilon}^2)^{-(n+q)/2}  | \mathbf{L}_{\theta}^{\top} \mathbf{Z}^{\top} \mathbf{Z} \mathbf{L}_{\theta} + \mathbf{I}_{qq} |^{-1/2} \exp ( -\tfrac{1}{2} \sigma_{\varepsilon}^{-2}  \mathbf{h}^{\top} \mathbf{h})
\\
\nonumber
& & \qquad \qquad \exp [ -\tfrac{1}{2} \sigma_{\varepsilon}^{-2} (\mathbf{Y} - \mathbf{X} \bbeta)^{\top} (\mathbf{I}_{nn} +  \mathbf{Z} \tilde{\mathbf{R}}_{\theta} \mathbf{Z}^{\top})^{-1} (\mathbf{Y} - \mathbf{X} \bbeta) ] \, d \mathbf{g}
\\
\nonumber
& = & (2 \pi \sigma_{\varepsilon}^2)^{-n/2}  | \mathbf{I}_{nn} +  \mathbf{Z} \tilde{\mathbf{R}}_{\theta} \mathbf{Z}^{\top} |^{-1/2}
\\
\label{form.mixedModel_fullLikelihood}
& & \qquad \qquad \exp [ -\tfrac{1}{2} \sigma_{\varepsilon}^{-2} (\mathbf{Y} - \mathbf{X} \bbeta)^{\top} (\mathbf{I}_{nn} +  \mathbf{Z} \tilde{\mathbf{R}}_{\theta} \mathbf{Z}^{\top})^{-1} (\mathbf{Y} - \mathbf{X} \bbeta) ],
\end{eqnarray}
where in the last step Sylvester's determinant identity has been used.

The maximum likelihood estimators of the mixed model parameters $\bbeta$, $\sigma_{\varepsilon}^2$ and $\tilde{\mathbf{R}}_{\theta}$ are found through the maximization of the logarithm of the likelihood (\ref{form.mixedModel_fullLikelihood}). Find the roots of the partial derivatives of this log-likelihood with respect to the mixed model parameters. For $\bbeta$ and $\sigma_{\varepsilon}^2$ this yields:
\begin{eqnarray*} 
\hat{\bbeta} & = & [\mathbf{X}^{\top} (\mathbf{I}_{nn} + \mathbf{Z} \tilde{\mathbf{R}}_{\theta} \mathbf{Z}^{\top})^{-1} \mathbf{X}]^{-1}  \mathbf{X}^{\top} (\mathbf{I}_{nn} + \mathbf{Z} \tilde{\mathbf{R}}_{\theta} \mathbf{Z}^{\top} )^{-1} \mathbf{Y},
\\
\hat{\sigma}_{\varepsilon}^2 & = & \tfrac{1}{n}  (\mathbf{Y} - \mathbf{X} \bbeta)^{\top} (\mathbf{I}_{nn} +  \mathbf{Z} \tilde{\mathbf{R}}_{\theta} \mathbf{Z}^{\top})^{-1} (\mathbf{Y} - \mathbf{X} \bbeta).
\end{eqnarray*}
The former estimate can be substituted into the latter to remove its dependency on $\bbeta$. However, both estimators still depend on $\ttheta$. An estimator of $\ttheta$ may be found by substitution of $\hat{\bbeta}$ and $\hat{\sigma}_{\varepsilon}^2$ into the log-likelihood followed by its maximization. For general parametrizations of $\tilde{\mathbf{R}}_{\theta}$ by $\ttheta$ there are no explicit solutions. Then, resort to standard nonlinear solvers such as the Newton-Raphson algorithm and the like. With a maximum likelihood estimate of $\ttheta$ at hand, those of the other two mixed model parameters are readily obtained from the formula's above. As $\ttheta$ is unknown at the onset, it needs to be initiated followed by sequential updating of the parameter estimates until convergence.

Restricted maximum likelihood (REML) considers the fixed effect parameter $\bbeta$ as a `nuisance' parameter and concentrates on the estimation of the variance components. The nuisance parameter is integrated out of the likelihood, $\int_{\mathbb{R}^p} L(\mathbf{Y}) d\bbeta$, which is referred to as the restricted likelihood. Those values of $\ttheta$ (and thereby $\tilde{\mathbf{R}}_{\theta}$) and $\sigma_{\varepsilon}^2$ that maximize the restricted likelihood are the REML estimators. The restricted likelihood, by an argument similar to that used in the derivation of the likelihood, simplifies to:
\begin{eqnarray*}
\int_{\mathbb{R}^p} L(\mathbf{Y}) d\bbeta & = & (2 \pi \sigma_{\varepsilon}^2)^{-n/2} | \tilde{\mathbf{Q}} |^{-1/2} \exp \{ -\tfrac{1}{2} \sigma_{\varepsilon}^{-2} \mathbf{Y}^{\top}  [ \tilde{\mathbf{Q}}_{\theta}^{-1} -  \tilde{\mathbf{Q}}_{\theta}^{-1} \mathbf{X} (\mathbf{X}^{\top} \tilde{\mathbf{Q}}_{\theta}^{-1} \mathbf{X})^{-1} \mathbf{X}^{\top} \tilde{\mathbf{Q}}_{\theta}^{-1} ] \mathbf{Y} \} 
\\
& & \qquad \int_{\mathbb{R}^p} \exp \{ -\tfrac{1}{2} \sigma_{\varepsilon}^{-2} [\bbeta - (\mathbf{X}^{\top} \tilde{\mathbf{Q}}_{\theta}^{-1} \mathbf{X})^{-1} \mathbf{X}^{\top} \tilde{\mathbf{Q}}_{\theta}^{-1} \mathbf{Y}]^{\top} \mathbf{X}^{\top} \tilde{\mathbf{Q}}_{\theta}^{-1} \mathbf{X} 
\\
& & \qquad \qquad \qquad \qquad \qquad \qquad \qquad [\bbeta - (\mathbf{X}^{\top} \tilde{\mathbf{Q}}_{\theta}^{-1} \mathbf{X})^{-1} \mathbf{X}^{\top} \tilde{\mathbf{Q}}_{\theta}^{-1} \mathbf{Y} ] \} d \bbeta
\\
& = & (2 \pi \sigma_{\varepsilon}^2)^{-(n-p)/2} | \tilde{\mathbf{Q}}_{\theta} |^{-1/2}  | \mathbf{X}^{\top} \tilde{\mathbf{Q}}_{\theta}^{-1} \mathbf{X} |^{-1/2}
\\
&  & \qquad \qquad \qquad  \exp \{ -\tfrac{1}{2} \sigma_{\varepsilon}^{-2} \mathbf{Y}^{\top}  [ \tilde{\mathbf{Q}}_{\theta}^{-1} -  \tilde{\mathbf{Q}}_{\theta}^{-1} \mathbf{X} (\mathbf{X}^{\top} \tilde{\mathbf{Q}}_{\theta}^{-1} \mathbf{X})^{-1} \mathbf{X}^{\top} \tilde{\mathbf{Q}}_{\theta}^{-1} ] \mathbf{Y} \},
\end{eqnarray*}
where $\tilde{\mathbf{Q}}_{\theta} = \mathbf{I}_{nn} +  \mathbf{Z} \tilde{\mathbf{R}}_{\theta} \mathbf{Z}^{\top}$ is the relative covariance (relative to the error variance) of $\mathbf{Y}$. The REML estimators are now found by equating the partial derivatives of this restricted loglikelihood to zero and solving for $\sigma_{\varepsilon}^2$ and $\ttheta$. The former, given the latter, is:
\begin{eqnarray*}
\hat{\sigma}_{\varepsilon}^2 & = & \tfrac{1}{n-p}  \mathbf{Y}^{\top}  [ \tilde{\mathbf{Q}}_{\theta}^{-1} -  \tilde{\mathbf{Q}}_{\theta}^{-1} \mathbf{X} (\mathbf{X}^{\top} \tilde{\mathbf{Q}}_{\theta}^{-1} \mathbf{X})^{-1} \mathbf{X}^{\top} \tilde{\mathbf{Q}}_{\theta}^{-1} ] \mathbf{Y}
\\
& = & \tfrac{1}{n-p}  \mathbf{Y}^{\top} \tilde{\mathbf{Q}}_{\theta}^{-1/2} [ \mathbf{I}_{nn} -  \tilde{\mathbf{Q}}_{\theta}^{-1/2} \mathbf{X} (\mathbf{X}^{\top} \tilde{\mathbf{Q}}_{\theta}^{-1/2} \tilde{\mathbf{Q}}_{\theta}^{-1/2} \mathbf{X})^{-1} \mathbf{X}^{\top} \tilde{\mathbf{Q}}_{\theta}^{-1/2} ] \tilde{\mathbf{Q}}_{\theta}^{-1/2} \mathbf{Y},
\end{eqnarray*}
where the rewritten form reveals a projection matrix and, consequently, a residual sum of squares. Like the maximum likelihood estimator of $\ttheta$, its REML counterpart is generally unknown analytically and to be found numerically. Iterating between the estimation of both parameters until convergence yields the REML estimators. Obviously, REML estimation of the mixed model parameters does not produce an estimate of the fixed parameter $\bbeta$ (as it has been integrated out). Should however a point estimate be desired, then in practice the ML estimate of $\bbeta$ with the REML estimates of the other parameters is used. 
\\
\\
An alternative way to proceed (and insightful for the present purpose) follows the original approach of Henderson, who aimed to construct a linear predictor for $\mathbf{Y}$. 
\begin{definition} \mbox{ } \\
A predictand is the function of the parameters that is to be predicted. A predictor is a function of the data that predicts the predictand. When this latter function is linear in the observation it is said to be a linear predictor.
\end{definition}
\noindent In case of the mixed model the predictand is $\mathbf{X}_{{\mbox{{\tiny new}}}} \bbeta + \mathbf{Z}_{{\mbox{{\tiny new}}}} \ggamma$ for $(n_{{\mbox{{\tiny new}}}} \times p)$- and $(n_{{\mbox{{\tiny new}}}} \times q)$-dimensional design matrices $\mathbf{X}_{{\mbox{{\tiny new}}}}$ and $\mathbf{Z}_{{\mbox{{\tiny new}}}}$, respectively. Similarly, the predictor is some function of the data $\mathbf{Y}$. When it can be expressed as $\mathbf{A} \mathbf{Y}$ for some matrix $\mathbf{A}$ it is a linear predictor.

The construction of the aforementioned linear predictor requires estimates of $\bbeta$ and $\ggamma$. To obtain these estimates first derive the joint density of $(\ggamma, \mathbf{Y})$:
\begin{eqnarray*}
\left( 
\begin{array}{c}
\ggamma
\\
\mathbf{Y}
\end{array}
\right) & \sim & \mathcal{N}
\left(
\left( 
\begin{array}{l}
\mathbf{0}_q
\\
\mathbf{X} \bbeta
\end{array}
\right),
\left( 
\begin{array}{lr}
\mathbf{R}_{\theta} & \mathbf{R}_{\theta} \mathbf{Z}^{\top}
\\
\mathbf{Z} \mathbf{R}_{\theta} & \sigma_{\varepsilon}^2 \mathbf{I}_{nn} + \mathbf{Z} \mathbf{R}_{\theta} \mathbf{Z}^{\top}
\end{array}
\right) \right)
\end{eqnarray*}
From this the likelihood is obtained and after some manipulations the loglikelihood can be shown to be proportional to:
\begin{eqnarray} \label{mixedModel.penalizedLossFunction}
\sigma_{\varepsilon}^{-2} \|\mathbf{Y} - \mathbf{X} \bbeta - \mathbf{Z} \ggamma \|_2^2 +  \ggamma^{\top} \mathbf{R}_{\theta}^{-1} \ggamma,
\end{eqnarray}
in which -- following Henderson -- $\mathbf{R}_{\theta}$ and $\sigma_{\varepsilon}^2$ are assumed known (for instance by virtue of maximum likelihood or REML estimation). The estimators of $\bbeta$ and $\ggamma$ are now the minimizers of loss criterion (\ref{mixedModel.penalizedLossFunction}). Effectively, the random effect parameter $\ggamma$ is now temporarily assumed to be `fixed'. That is, it is temporarily treated as fixed in the derivations below that lead to the construction of the linear predictor. However, $\ggamma$ is a random variable and one therefore speaks of a linear predictor rather than linear estimator. 


To find the estimators of $\bbeta$ and $\ggamma$, defined as the minimizer of loss function, equate the partial derivatives of mixed model loss function (\ref{mixedModel.penalizedLossFunction}) with respect to $\bbeta$ and $\ggamma$ to zero. This yields the estimating equations (also referred to as Henderson's mixed model equations):
\begin{eqnarray*}
\mathbf{X}^{\top} \mathbf{Y} - \mathbf{X}^{\top} \mathbf{X} \bbeta - \mathbf{X}^{\top} \mathbf{Z} \ggamma & = & \mathbf{0}_{p},
\\
\sigma_{\varepsilon}^{-2} \mathbf{Z}^{\top} \mathbf{Y} - \sigma_{\varepsilon}^{-2}  \mathbf{Z}^{\top} \mathbf{Z} \ggamma - \sigma_{\varepsilon}^{-2}  \mathbf{Z}^{\top} \mathbf{X} \bbeta  - \mathbf{R}_{\theta}^{-1} \ggamma & = & \mathbf{0}_{q}.
\end{eqnarray*}
Solve each estimating equation for the parameters individually and find:
\begin{eqnarray} 
\label{form.mixedModel_penEstOfBeta}
\hat{\bbeta} & = & (\mathbf{X}^{\top} \mathbf{X})^{-1} \mathbf{X}^{\top} ( \mathbf{Y} - \mathbf{Z} \ggamma),
\\
\label{form.mixedModel_penEstOfGamma}
\hat{\ggamma} & = & (\mathbf{Z}^{\top} \mathbf{Z} +  \sigma_{\varepsilon}^{2}  \mathbf{R}_{\theta}^{-1})^{-1} \mathbf{Z}^{\top} ( \mathbf{Y} - \mathbf{X} \bbeta).
\end{eqnarray}
Note, using the Cholesky decomposition of $\tilde{\mathbf{R}}_{\theta}$ and applying the Woodbury identity twice (in both directions), that:
\begin{eqnarray*}
\hat{\ggamma} & = &  (\mathbf{Z}^{T} \mathbf{Z} + \tilde{\mathbf{R}}_{\theta}^{-1})^{-1} \mathbf{Z}^{\top} (\mathbf{Y} - \mathbf{X} \bbeta)
\\
& = & [\tilde{\mathbf{R}}_{\theta} -  \tilde{\mathbf{R}}_{\theta} \mathbf{Z}^{\top} (\mathbf{I}_{nn} + \mathbf{Z} \tilde{\mathbf{R}}_{\theta} \mathbf{Z}^{\top} )^{-1} \mathbf{Z} \tilde{\mathbf{R}}_{\theta}] \mathbf{Z}^{\top} (\mathbf{Y} - \mathbf{X} \bbeta)
\\
& = &  \mathbf{L}_{\theta} [ \mathbf{I}_{qq} - \mathbf{L}_{\theta}^{\top} \mathbf{Z}^{\top} (\mathbf{I}_{nn} + \mathbf{Z} \mathbf{L}_{\theta}  \mathbf{L}_{\theta}^{\top} \mathbf{Z}^{\top} )^{-1} \mathbf{Z} \mathbf{L}_{\theta} ] \mathbf{L}_{\theta}^{\top} \mathbf{Z}^{\top} (\mathbf{Y} - \mathbf{X} \bbeta)
\\
& = & \mathbf{L}_{\theta}  ( \mathbf{L}_{\theta}^{\top} \mathbf{Z}^{\top} \mathbf{Z} \mathbf{L}_{\theta} + \mathbf{I}_{qq})^{-1} \mathbf{L}_{\theta}^{\top} \mathbf{Z}^{\top} (\mathbf{Y} - \mathbf{X} \bbeta)
\\
& = & \mathbf{L}_{\theta} \, \mmu_{\tilde{\ggamma} \, | \mathbf{Y}}.
\end{eqnarray*}
It thus coincides with the conditional estimate of the $\ggamma$ found in the derivation of the maximum likelihood estimator of the mixed model. This expression could also have been found by conditioning with the multivariate normal above which would have given $\mathbb{E}(\ggamma \, | \, \mathbf{Y})$.

The estimator of both $\bbeta$ and $\ggamma$ can be expressed fully and explicitly in terms of $\mathbf{X}$, $\mathbf{Y}$, $\mathbf{Z}$ and $\mathbf{R}_{\theta}$. To obtain that of $\bbeta$ substitute the estimator of $\ggamma$ of equation (\ref{form.mixedModel_penEstOfGamma}) into that of $\bbeta$ given by equation \ref{form.mixedModel_penEstOfBeta}):
\begin{eqnarray*}
\bbeta & = & (\mathbf{X}^{\top} \mathbf{X})^{-1} \mathbf{X}^{\top} [ \mathbf{Y} - \mathbf{Z} (\mathbf{Z}^{\top} \mathbf{Z} +  \sigma_{\varepsilon}^{2} \mathbf{R}_{\theta}^{-1})^{-1} \mathbf{Z}^{\top} ( \mathbf{Y} - \mathbf{X} \bbeta)]
\\
& = & (\mathbf{X}^{\top} \mathbf{X})^{-1} \mathbf{X}^{\top} \{ \mathbf{Y} - [ \mathbf{I}_{nn} - (\sigma_{\varepsilon}^{-2} \mathbf{Z} \mathbf{R}_{\theta} \mathbf{Z}^{\top} + \mathbf{I}_{nn})^ {-1}] ( \mathbf{Y} - \mathbf{X} \bbeta) \},
\end{eqnarray*}
in which the Woodbury identity has been used. Now group terms and solve for $\bbeta$:
\begin{eqnarray} \label{form.mixedModel_penEstOfBeta_explicit}
\hat{\bbeta} & = & [\mathbf{X}^{\top} (\mathbf{Z} \mathbf{R}_{\theta} \mathbf{Z}^{\top} + \sigma_{\varepsilon}^2 \mathbf{I}_{nn})^{-1} \mathbf{X}]^{-1}  \mathbf{X}^{\top} (\mathbf{Z} \mathbf{R}_{\theta} \mathbf{Z}^{\top} + \sigma_{\varepsilon}^2 \mathbf{I}_{nn})^{-1} \mathbf{Y}.
\end{eqnarray}
This coincides with the maximum likelihood estimator of $\bbeta$ presented above (for known $\mathbf{R}_{\theta}$ and $\sigma_{\varepsilon}^2$). Moreover, in the preceeding display one recognizes a generalized least squares (GLS) estimator. The GLS regression estimator is BLUE (Best Linear Unbiased Estimator) when $\mathbf{R}_{\theta}$ and $\sigma_{\varepsilon}$ are known. To find an explicit expression for $\ggamma$ use $\mathbf{Q}_{\theta}$ as previously defined and substitute the explicit expression (\ref{form.mixedModel_penEstOfBeta_explicit}) for the estimator of $\bbeta$  in the estimator of $\ggamma$, shown in display (\ref{form.mixedModel_penEstOfGamma}) above. This gives:
\begin{eqnarray*}
\hat{\ggamma} & = & (\mathbf{Z}^{\top} \mathbf{Z} + \sigma_{\varepsilon}^2 \mathbf{R}_{\theta}^{-1})^{-1} \mathbf{Z}^{\top} [ \mathbf{I}_{nn}  - \mathbf{X} (\mathbf{X}^{\top} \mathbf{Q}_{\theta}^{-1} \mathbf{X})^{-1} \mathbf{X}^{\top} \mathbf{Q}_{\theta}^{-1}] \mathbf{Y},
\end{eqnarray*}
an explicit expression for the estimator  of $\ggamma$. 

The linear predictor constructed from these estimator can be shown (cf. Theorem \ref{theorem.BLUP}) to be optimal, in the BLUP sense. 

\begin{definition} \mbox{ } \\
A Best Linear Unbiased Predictor (BLUP) \textit{i)} is linear in the observations, \textit{ii)} is unbiased, and \textit{iii)} has a minimum (variance of its) predictor error, i.e. the difference among the predictor and predictand, among all unbiased linear predictors. 
\end{definition}

\begin{theorem} \label{theorem.BLUP} \mbox{ } \\
The predictor $\mathbf{X} \hat{\bbeta} + \mathbf{Z} \hat{\gamma}$ is the BLUP of $\tilde{\mathbf{Y}} = \mathbf{X} \bbeta + \mathbf{Z} \gamma$.
\end{theorem}

\begin{proof}
The predictor of $\mathbf{X} \hat{\bbeta} + \mathbf{Z} \hat{\ggamma}$ is:
\begin{eqnarray*}
\mathbf{X} \hat{\bbeta} + \mathbf{Z} \hat{\ggamma} & = & [ \mathbf{I}_{nn} - \sigma_{\varepsilon}^2 \mathbf{Q}^{-1}_{\theta} + \mathbf{Q}_{\theta}^{-1} \mathbf{X} (\mathbf{X}^{\top} \mathbf{Q}_{\theta}^{-1} \mathbf{X})^{-1} \mathbf{X}^{\top} \mathbf{Q}^{-1}_{\theta} ] \mathbf{Y} \, \, \, := \, \, \, \mathbf{B} \mathbf{Y}.
\end{eqnarray*}
Clearly, this is a linear function in $\mathbf{Y}$.

The expectation of the linear predictor is
\begin{eqnarray*}
\mathbb{E} (\mathbf{X} \hat{\bbeta} + \mathbf{Z} \hat{\ggamma} ) 
& = & \mathbb{E} [\mathbf{X} \hat{\bbeta} + \mathbf{Z} ( \mathbf{Z}^{\top} \mathbf{Z} + \sigma_{\varepsilon}^{2} \mathbf{R}_{\theta})^{-1} \mathbf{Z}^{\top} (\mathbf{Y} - \mathbf{X} \hat{\bbeta}) ]
\\
& = & \mathbf{X} \mathbb{E} (\hat{\bbeta}) + (\mathbf{I}_{nn} - \sigma_{\varepsilon}^2 \mathbf{Q}_{\theta}^{-1}) [\mathbb{E} (\mathbf{Y}) - \mathbb{E} (\mathbf{X} \hat{\bbeta})] \, \, \,  = \, \, \, \mathbf{X} \bbeta.
\end{eqnarray*}
This is also the expectation of the predictand $\mathbf{X} \bbeta + \mathbf{Z} \ggamma$. Hence, the predictor is unbiased.

To show the predictor $\mathbf{B} \mathbf{Y}$ has minimum prediction error variance within the class of unbiased linear predictors, assume the existence of another unbiased linear predictor $ \mathbf{A} \mathbf{Y}$ of $\mathbf{X} \bbeta + \mathbf{Z} \ggamma$. The predictor error variance of the latter predictor is:
\begin{eqnarray*}
\mbox{Var}(\mathbf{X} \bbeta + \mathbf{Z} \ggamma  - \mathbf{A} \mathbf{Y}) & = & \mbox{Var}(\mathbf{X} \bbeta + \mathbf{Z} \ggamma  - \mathbf{B} \mathbf{Y} - \mathbf{A} \mathbf{Y} + \mathbf{B} \mathbf{Y})
\\
& = & \mbox{Var}[ (\mathbf{A}  - \mathbf{B}) \mathbf{Y}]  + 
\mbox{Var}(\mathbf{X} \bbeta + \mathbf{Z} \ggamma  - \mathbf{B} \mathbf{Y})
\\
& & - 2 ~ \mbox{Cov}[ \mathbf{X} \bbeta + \mathbf{Z} \ggamma  - \mathbf{B} \mathbf{Y}, (\mathbf{A}  - \mathbf{B}) \mathbf{Y}].
\end{eqnarray*}
The last term vanishes as:
\begin{eqnarray*}
& & \hspace{-1.5cm} \mbox{Cov}[ \mathbf{X} \bbeta + \mathbf{Z} \ggamma  - \mathbf{B} \mathbf{Y}, (\mathbf{A}  - \mathbf{B}) \mathbf{Y}]  
\\
& = & 
[\mathbf{Z} \mbox{Cov}( \ggamma, \mathbf{Y}) - \mathbf{B} \mbox{Var}(\mathbf{Y}) ] (\mathbf{A}  - \mathbf{B})^{\top}
\\
& = & \{ \mathbf{Z} \mathbf{R}_{\theta} \mathbf{Z}^{\top} - 
[ \mathbf{I}_{nn} - \sigma_{\varepsilon}^2 \mathbf{Q}^{-1}_{\theta} + \mathbf{Q}_{\theta}^{-1} \mathbf{X} (\mathbf{X} \mathbf{Q}_{\theta}^{-1} \mathbf{X}^{\top})^{-1} \mathbf{X}^{\top} \mathbf{Q}^{-1}_{\theta} ]
\mathbf{Q}_{\theta} \} (\mathbf{A}  - \mathbf{B})^{\top}
\\
& = & \mathbf{Q}_{\theta}^{-1} \mathbf{X} (\mathbf{X} \mathbf{Q}_{\theta}^{-1} \mathbf{X}^{\top})^{-1} \mathbf{X}^{\top}  (\mathbf{A}  - \mathbf{B})^{\top}
\\
& = & \mathbf{Q}_{\theta}^{-1} \mathbf{X} (\mathbf{X} \mathbf{Q}_{\theta}^{-1} \mathbf{X}^{\top})^{-1} [ (\mathbf{A}  - \mathbf{B})
\mathbf{X}]^{\top} \, \, \, = \, \, \, \mathbf{0}_{nn},
\end{eqnarray*}
where the last step uses $\mathbf{A} \mathbf{X} = \mathbf{B} \mathbf{X}$, which follows from the fact that
\begin{eqnarray*}
\mathbf{A} \mathbf{X} \bbeta & = & \mathbb{E}(\mathbf{A} \mathbf{Y}) \, \, \, = \, \, \, \mathbb{E}(\mathbf{B} \mathbf{Y})  \, \, \, = \, \, \, 
\mathbf{B} \mathbf{X} \bbeta,
\end{eqnarray*}
for all $\bbeta \in \mathbb{R}^p$. Hence, 
\begin{eqnarray*}
\mbox{Var}(\mathbf{X} \bbeta + \mathbf{Z} \ggamma  - \mathbf{A} \mathbf{Y}) & = & \mbox{Var}[ (\mathbf{A}  - \mathbf{B}) \mathbf{Y}]  +  \mbox{Var}(\mathbf{X} \bbeta + \mathbf{Z} \ggamma  - \mathbf{B} \mathbf{Y}),
\end{eqnarray*}
from which the minimum variance follows as the first summand on the right-hand side is nonnegative and zero if and only if $\mathbf{A} = \mathbf{B}$.
\end{proof}

\section{Link to ridge regression} \label{sect:mixedModel2ridgeRegression}
The link with ridge regression, implicit in the expos\'{e} on the mixed model, is now explicated. Recall that ridge regression fits the linear regression model $\mathbf{Y} = \mathbf{X} \bbeta + \vvarepsilon$ by means of a penalized maximum likelihood procedure, which defines -- for given penalty parameter $\lambda$ -- the estimator as:
\begin{eqnarray*}
\hat{\bbeta} (\lambda) & = & \arg \min_{\bbeta \in \mathbb{R}^p} \| \mathbf{Y} - \mathbf{X} \bbeta \|_2^2 + \lambda \bbeta^{\top} \bbeta.
\end{eqnarray*}
Constrast this to a mixed model void of covariates with fixed effects and comprising only covariates with a random effects: $\mathbf{Y}  = \mathbf{Z} \ggamma + \vvarepsilon$ with distributional assumptions $\ggamma \sim \mathcal{N}( \mathbf{0}_q, \sigma_{\gamma}^2 \mathbf{I}_{qq})$ and
$\vvarepsilon \sim \mathcal{N}(\mathbf{0}_n, \sigma_{\varepsilon}^2 \mathbf{I}_{nn})$. This model, when temporarily considering $\ggamma$ as fixed, is fitted by the minimization of loss function (\ref{mixedModel.penalizedLossFunction}). The corresponding estimator of $\ggamma$ is then defined, with the current mixed model assumptions in place, as:
\begin{eqnarray*}
\hat{\ggamma} & = & \arg \min_{\ggamma \in \mathbb{R}^p}  \|\mathbf{Y}  - \mathbf{Z} \ggamma \|_2^2 + \sigma_{\gamma}^{-2} \ggamma^{\top} \ggamma.
\end{eqnarray*}
The estimators are -- up to a reparamatrization of the penalty parameter -- defined identically. This should not come as a surprise after the discussion of Bayesian regression (cf. Chapter \ref{chap:BayesianRegression}) and the alert reader would already have recognized a generalized ridge loss function in Equation (\ref{mixedModel.penalizedLossFunction}). The fact that we discarded the fixed effect part of the mixed model is irrelevant for the analogy as those would correspond to unpenalized covariates in the ridge regression problem. 

The link with ridge regression is also immenent from the linear predictor of the random effect. Recall: $\hat{\ggamma} = (\mathbf{Z}^{\top} \mathbf{Z} + \mathbf{R}_{\theta}^{-1})^{-1} \mathbf{Z}^{\top} (\mathbf{Y} - \mathbf{X} \bbeta)$. When we ignore $\mathbf{R}_{\theta}^{-1}$, the predictor reduces to a least squares estimator. But with a symmetric and positive definite matrix $\mathbf{R}_{\theta}^{-1}$, the predictor is actually of the shrinkage type as is the ridge regression estimator. This shrinkage estimator also reveals, through the term $(\mathbf{Z}^{\top} \mathbf{Z} + \mathbf{R}_{\theta}^{-1})^{-1}$, that a $q$ larger than $n$ does not cause identifiability problems as long as $\mathbf{R}_{\theta}$ is parametrized low-dimensionally enough.

The following mixed model result provide an alternative approach to choice of the penalty parameter in ridge regression. It assumes a mixed model comprising of the random effects part only. Or, put differently, it assume the linear regression model $\mathbf{Y} = \mathbf{X} \bbeta + \vvarepsilon$ with $\bbeta \sim \mathcal{N}(\mathbf{0}_p, \sigma_{\bbeta}^2 \mathbf{I}_{pp})$ and $\varepsilon \sim \mathcal{N}(\mathbf{0}_n, \sigma_{\varepsilon}^2 \mathbf{I}_{nn})$. 

\begin{theorem} (Theorem 2, Golub et al) \label{theorem.GCV2ridgePenalty} \\
The expected generalized cross-validation error $\mathbb{E}_{\bbeta} \{ \mathbb{E}_{\varepsilon} [ GCV(\lambda)] \}$ is minimized for $\lambda = \sigma_{\varepsilon}^2 / \sigma^2_{\beta}$.
\end{theorem}
\begin{proof}
The proof first finds an analytic expression of the expected $GCV(\lambda)$, then its minimum. Its expectation can be re-expressed as follows:
\begin{eqnarray*}
\mathbb{E}_{\bbeta} \{ \mathbb{E}_{\varepsilon} [ GCV(\lambda)] \} & = &  \mathbb{E}_{\bbeta} \big[ \mathbb{E}_{\varepsilon} \big( \tfrac{1}{n} 
\{\mbox{tr}[\mathbf{I}_{nn} - \mathbf{H}(\lambda)] / n \}^{-2} \| [\mathbf{I}_{nn} - \mathbf{H}(\lambda) ] \mathbf{Y} \big\|_2^2 \big) \big] 
\\
& = & n \{\mbox{tr}[\mathbf{I}_{nn} - \mathbf{H}(\lambda)] \}^{-2} \mathbb{E}_{\bbeta} \big( \mathbb{E}_{\varepsilon} \{ \mathbf{Y}^{\top} [\mathbf{I}_{nn} - \mathbf{H}(\lambda) ]^2 \mathbf{Y} \} \big)
\\
& = & n \{\mbox{tr}[\mathbf{I}_{nn} - \mathbf{H}(\lambda)] \}^{-2} \mathbb{E}_{\bbeta} \big[ \mathbb{E}_{\varepsilon} \big( \mbox{tr} \{ ( \mathbf{X} \bbeta + \vvarepsilon)^{\top} [\mathbf{I}_{nn} - \mathbf{H}(\lambda) ]^2  ( \mathbf{X} \bbeta + \vvarepsilon) \} \big) \big]
\\
& = & n \{\mbox{tr}[\mathbf{I}_{nn} - \mathbf{H}(\lambda)] \}^{-2}  \big( \mbox{tr} \{[\mathbf{I}_{nn} - \mathbf{H}(\lambda) ]^2
 \mathbf{X} \mathbf{X}^{\top} \mathbb{E}_{\bbeta} [ \mathbb{E}_{\varepsilon}  ( \bbeta \bbeta^{\top} ) ] \}
\\
& &  \qquad \qquad  \qquad \qquad \quad  + \, \mbox{tr}\{ [\mathbf{I}_{nn} - \mathbf{H}(\lambda) ]^2 \mathbb{E}_{\bbeta} [ \mathbb{E}_{\varepsilon}  (\vvarepsilon \vvarepsilon^{\top} ) ] \} \big)
\\
& = & n \big(  \sigma_{\beta}^2 \mbox{tr}\{ [\mathbf{I}_{nn} - \mathbf{H}(\lambda)]^{2} \mathbf{X} \mathbf{X}^{\top} \} + \sigma_{\varepsilon}^2  \mbox{tr}\{ [\mathbf{I}_{nn} - \mathbf{H}(\lambda)]^{2}\} \big) 
\{\mbox{tr}[\mathbf{I}_{nn} - \mathbf{H}(\lambda)] \}^{-2}.
\end{eqnarray*}
To get a handle on this expression, use $(\mathbf{X}^{\top} \mathbf{X} + \lambda \mathbf{I}_{pp})^{-1} \mathbf{X}^{\top} \mathbf{X}  =  \mathbf{I}_{pp} - \lambda (\mathbf{X}^{\top} \mathbf{X} + \lambda \mathbf{I}_{pp})^{-1} =  \mathbf{X}^{\top} \mathbf{X} (\mathbf{X}^{\top} \mathbf{X} + \lambda \mathbf{I}_{pp})^{-1}$, the cyclic property of the trace, and define 
$A(\lambda) = \sum\nolimits_{j=1}^p (d_{x,j}^2 + \lambda)^{-1}$, 
$B(\lambda) = \sum\nolimits_{j=1}^p (d_{x,j}^2 + \lambda)^{-2}$, and
$C(\lambda) = \sum\nolimits_{j=1}^p (d_{x,j}^2 + \lambda)^{-3}$. The traces in the expectation of $GCV(\lambda)$ can now be written as:
\begin{eqnarray*}
\mbox{tr}[\mathbf{I}_{nn} - \mathbf{H}(\lambda)] & = & \lambda ~  \mbox{tr}[ (\mathbf{X}^{\top} \mathbf{X} + \lambda \mathbf{I}_{pp})^{-1} ] \, \, \, \, \, = \, \, \,  \lambda A(\lambda),
\\
\mbox{tr} \{ [\mathbf{I}_{nn} - \mathbf{H}(\lambda)]^2 \} & = &  \lambda^2
\mbox{tr}[ (\mathbf{X}^{\top} \mathbf{X} + \lambda \mathbf{I}_{pp})^{-2} ] \, \, \, \, = \, \, \, \lambda^2 B(\lambda),
\\
\mbox{tr} \{ [\mathbf{I}_{nn} - \mathbf{H}(\lambda)]^2  \mathbf{X} \mathbf{X}^{\top} \} & = & \lambda^{2}  \mbox{tr}[ (\mathbf{X}^{\top} \mathbf{X} + \lambda \mathbf{I}_{pp})^{-1} ] - \lambda^{3} \mbox{tr}[ (\mathbf{X}^{\top} \mathbf{X} + \lambda \mathbf{I}_{pp})^{-2} ]
\\
& = & \lambda^2 A(\lambda) - \lambda^3 B(\lambda).
\end{eqnarray*}
The expectation of $GCV(\lambda)$ can then be reformulated as:
\begin{eqnarray*}
\mathbb{E}_{\bbeta} \{ \mathbb{E}_{\varepsilon} [ GCV(\lambda)] \} & = &  n \{ \sigma_{\beta}^2 [A(\lambda) - \lambda B(\lambda)]  + \sigma_{\varepsilon}^2  B(\lambda) \} [A(\lambda)]^{-2}.
\end{eqnarray*}
Equate the derivative of this expectation w.r.t. $\lambda$ to zero, which can be seen to be proportional to:
\begin{eqnarray*}
2 (\lambda \sigma^2_{\beta}  - \sigma_{\varepsilon}^2) [B(\lambda)]^2 + 2 (\lambda \sigma^2_{\beta} - \sigma_{\varepsilon}^2) A(\lambda) C(\lambda) & = & 0.
\end{eqnarray*}
Indeed, $\lambda = \sigma_{\varepsilon}^2 / \sigma^{2}_{\beta}$ is the root of this equation.
\end{proof}
\noindent Theorem \ref{theorem.GCV2ridgePenalty} can be extended to include unpenalized covariates. This leaves the result unaltered: the optimal (in the expected GCV sense) ridge penalty is the same signal-to-noise ratio. 

We have encountered the result of Theorem \ref{theorem.GCV2ridgePenalty} before. Revisit Example \ref{example:ridgeMSEorthonormal} which derived the mean squared error (MSE) of the ridge regression estimator when $\mathbf{X}$ is orthonormal. It was pointed out that this MSE is minized for $\lambda = p \sigma_{\varepsilon} / \bbeta^{\top} \bbeta$. As $\bbeta^{\top} \bbeta /p$ is an estimator for $\sigma_{\beta}^2$, this implies the same optimal choice of the penalty parameter.

To point out the relevance of Theorem \ref{theorem.GCV2ridgePenalty} for the choice of the ridge penalty parameter still assume the regression parameter random. The theorem then says that the optimal penalty parameter (in the GCV sense) equals the ratio of the error variance and that of the regression parameter. Both variances can be estimated by means of the mixed model machinery (provided for instance by the \texttt{lme4} package in \texttt{R}). These estimates may be plugged in the ratio to arrive at a choice of ridge penalty parameter (see Section \ref{sect:Psplines} for an illustration of this usage).

\section{REML consistency, high-dimensionally}
Here a result on the asymptotic quality of the REML estimators of the random effect and error variance parameters is presented and discussed. It is the ratio of these parameters that forms the optimal choice (in the expected GCV sense) of the penalty parameter of the ridge regression estimator. As in practice the parameters are replaced by estimates to arrive at a choice for the penalty parameter, the quality of these  estimators propogates to the chosen penalty parameter. 

Consider the standard linear mixed model $\mathbf{Y} = \mathbf{X} \bbeta + \mathbf{Z} \ggamma + \vvarepsilon$, now with equivariant and uncorrelated random effects:  $\ggamma \sim \mathcal{N}( \mathbf{0}_q, \sigma_{\gamma}^2 \mathbf{I}_{qq})$. Write $\theta =   \sigma_{\gamma}^2 / \sigma_{\varepsilon}^2$. The REML estimators of $\theta$ and $\sigma_{\varepsilon}^2$ are to be found from the estimating equations:
\begin{eqnarray*}
\mbox{tr}( \mathbf{P}_{\theta} \mathbf{Z} \mathbf{Z}^{\top}) & = & \sigma_{\varepsilon}^{-2} \mbox{tr}( \mathbf{Y}^{\top} \mathbf{P}_{\theta} \mathbf{Z} \mathbf{Z}^{\top} \mathbf{P}_{\theta} \mathbf{Y}),
\\
\sigma_{\varepsilon}^2 & = & (n-p)^{-1} \mathbf{Y}^{\top} \mathbf{P}_{\theta} \mathbf{Y},
\end{eqnarray*}
where $\mathbf{P}_{\theta} = \tilde{\mathbf{Q}}_{\theta}^{-1} - \tilde{\mathbf{Q}}_{\theta}^{-1} \mathbf{X} (\mathbf{X}^{\top} \tilde{\mathbf{Q}}_{\theta}^{-1} \mathbf{X})^{-1} \mathbf{X}^{\top} \tilde{\mathbf{Q}}_{\theta}^{-1}$ and  $\tilde{\mathbf{Q}}_{\theta} = \mathbf{I}_{nn} + \theta \mathbf{Z} \mathbf{Z}^{\top}$. To arrive at the REML estimators choose initial values for the parameters. Choose one of the estimating equations substitute the initial value of the one of the parameters and solve for the other. The found root is then substituted into the other estimating equation, which is subsequently solved for the remaining parameter. Iterate between these two steps until convergence. The discussion of the practical evaluation of a root for $\theta$ from these estimating equations in a high-dimensional context is postponed to the next section.

The employed linear mixed model assumes that each of the $q$ covariates included as a column in $\mathbf{Z}$ contributes to the variation of the response. However,  it may be that only a fraction of these covariates exerts any influence on the response. That is, the random effect parameter $\ggamma$ is sparse, which could be operationalized as $\ggamma$ having $q_0$ zero elements while the remaining $q_{c} = q - q_0$ elements are non-zero. Only for the latter $q_{c}$ elements of $\ggamma$ the normal assumption makes sense, but is invalid for the $q_0$ zeros in $\ggamma$. The posed mixed model is then misspecified. 

The next theorem states that the REML estimators of $\theta = \sigma_{\gamma}^2 / \sigma_{\varepsilon}^2$ and $\sigma_{\varepsilon}^2$ are consistent (possibly after adjustment, see the theorem), even under the above mentioned misspecification. 

\begin{theorem} (Theorem 3.1, \cite{Jiang2016}) \label{theorem.highDimconsistencyREML} \\
Let $\mathbf{Z}$ be standardized column-wise and with its unstandardized entries i.i.d. from a sub-Gaussian distribution. Furthermore, assume that $n, q,  q_{c}  \rightarrow \infty$ such that 
\begin{eqnarray*}
\frac{n}{q} \rightarrow \tau \qquad \mbox{ and } \qquad  \frac{ q_{c}}{q} \rightarrow \omega,
\end{eqnarray*}
where $\tau, \omega \in (0, 1]$. Finally, suppose that $\sigma_{\varepsilon}^2$ and $\sigma_{\gamma}^2$ are positive. Then:
\begin{compactitem}
\item[\textit{i)}] The `adjusted' REML estimator of the variance ratio $ \sigma_{\gamma}^2 / \sigma_{\varepsilon}^2$ is consistent: 
\begin{eqnarray*}
\frac{q}{ q_{c}} \widehat{(\sigma_{\gamma}^2 / \sigma_{\varepsilon}^2)} \stackrel{P}{\longrightarrow}  \sigma_{\gamma}^2 / \sigma_{\varepsilon}^2. 
\end{eqnarray*} 

\item[\textit{ii)}] The REML estimator of the error variance is consistent: $\hat{\sigma}_{\varepsilon}^2 \stackrel{P}{\longrightarrow} \sigma_{\varepsilon}^2$. 
\end{compactitem}
\end{theorem}
\begin{proof}
Confer \cite{Jiang2016}.
\end{proof}

\noindent
Before the interpretation and implication of Theorem  \ref{theorem.highDimconsistencyREML} are discussed, its conditions for the consistency result are reviewed:
\begin{compactitem}
\item The standardization and distribution assumption on the design matrix of the random effects has no direct practical interpretation. These conditions warrant the applicability of certain results from random matrix theory upon which the proof of the theorem hinges.

\item The positive variance assumption $\sigma_{\varepsilon}^2, \sigma_{\gamma}^2 > 0$, in particular that of the random effect parameter, effectively states that some -- possibly misspecified -- form of the mixed model applies.

\item Practically most relevant are the conditions on the sample size, random effect dimension, and sparsity. The $\tau$ and $\omega$ in Theorem \ref{theorem.highDimconsistencyREML} are the limiting ratio's of the sample size $n$ and non-zero random effects $ q_{c}$, respectively, to the total number of random effects $q$. The number of random effects thus exceeds the sample size, as long as the latter grows (in the limit) at some fixed rate with the former. Independently, the model may be misspecified. The sparsity condition only requires that (in the limit) a fraction of the random effects is nonzero. 
\end{compactitem}
Now discuss the interpretation and relevance of the theorem:
\begin{compactitem}
\item Theorem \ref{theorem.highDimconsistencyREML} complements the classical low-dimensional consistency results on the REML estimator. 

\item Theorem \ref{theorem.highDimconsistencyREML} shows that not all (i.e. consistency) is lost when the model is misspecified. 

\item The practical relevance of the part \textit{i)} of Theorem \ref{theorem.highDimconsistencyREML} is limited as the number of nonzero random effects $q_{c}$, or $\omega$ for that matter, is usually unknown. Consequently, the REML estimator of the variance ratio $\sigma_{\gamma}^2 / \sigma_{\varepsilon}^2$ cannot be adjusted correctly to achieve asymptotically unbiasedness and -- thereby -- consistency

\item Part \textit{ii)} in its own right may not seem very useful. But it is surprising that high-dimensionally (i.e. when the dimension of the random effect parameter exceeds the sample size) the standard (that is, derived for low-dimensional data) REML estimator of $\sigma_{\varepsilon}^2$ is consistent. Beyond this surprise, a good estimator of $\sigma_{\varepsilon}^2$ indicates how much of the variation in the response cannot be attributed to the covariates represented by the columns $\mathbf{Z}$. A good indication of the noise level in the data finds use at many place. In particular, it is helpful in deciding on the order of the penalty parameter. 

\item Theorem \ref{theorem.GCV2ridgePenalty} suggests to choose the ridge penalty parameter equal to the  ratio of the error variance and that of the random effects. Confronted with data the reciprocal of the REML estimator of $\theta = \sigma_{\gamma}^2 / \sigma_{\varepsilon}^2$ may be used as value for the penalty parameter. Without the adjustment for the fraction of nonzero random effects, this value is off.  But in the worst case this value is an over-estimation of the optimal (in the GCV sense) ridge penalty parameter. Consequently, too much penalization is applied and the ridge regression estimate of the regression parameter is conservative as it shrinks the elements too much to zero. 
\end{compactitem}

\section{Illustration: P-splines} \label{sect:Psplines}
An organism's internal circadian clock enables it to synchronize its activities to the earth's day-and-night cycle. The circadian clock maintains, due to environmental feedback, oscillations of approximately 24 hours. Molecularly, these oscillations reveal themselves in the fluctuation of the transcription levels of genes. The molecular core of the circadian clock is made up of $\pm 10$ genes. Their behaviour (in terms of their expression patterns) is described by a dynamical system with feedback mechanisms. Linked to this core are genes that tap into the clock's rythm and use it to regulate the molecular processes. As such many genes are expected to exhibit circadian rythms. This is investigated in a mouse experiment in which the expression levels of several transcipts have been measured during two days with a resolution of one hour, resulting in a time-series of 48 time points publicly available from the \texttt{R}-package \texttt{MetaCycle}. Circadian rythms may be identified simply by eye-balling the data. But to facilitate this identification the data are smoothed to emphasize the pattern present in these data.

\begin{figure}[!h]
\begin{tabular}{rcl}
\includegraphics[scale=0.30, angle=0]{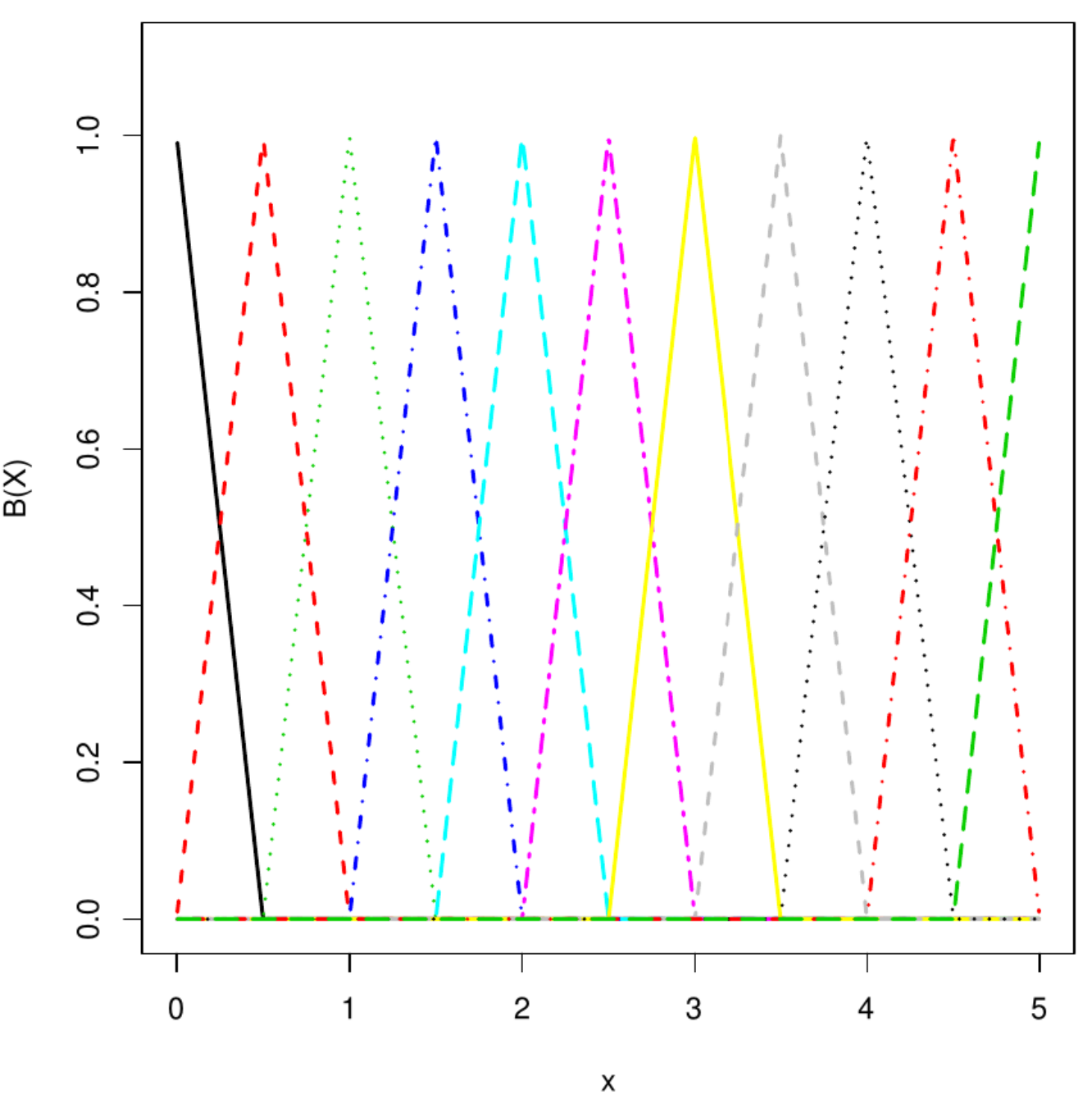} 
& & 
\includegraphics[scale=0.30, angle=0]{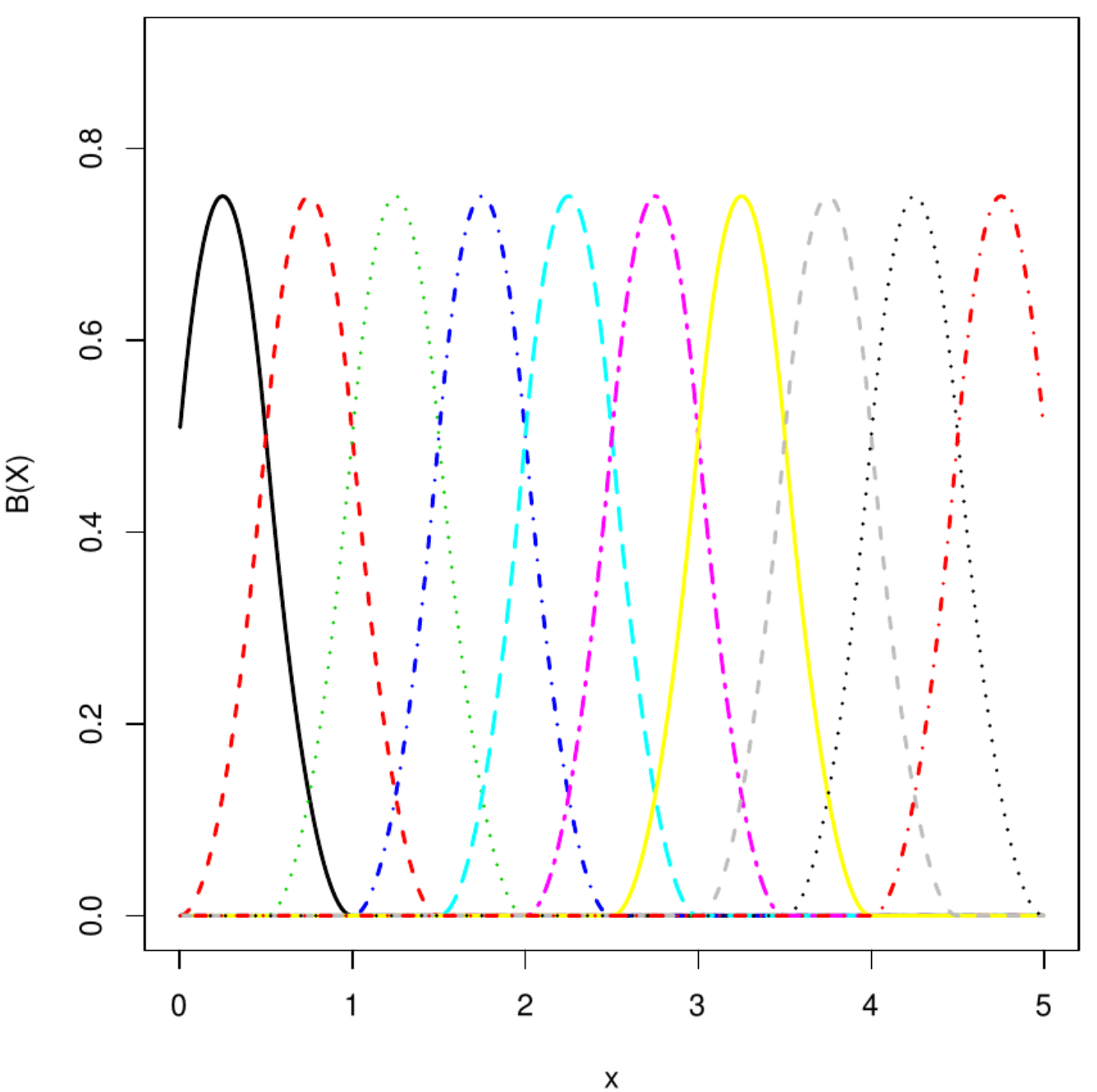}
\\
\includegraphics[scale=0.20, angle=0]{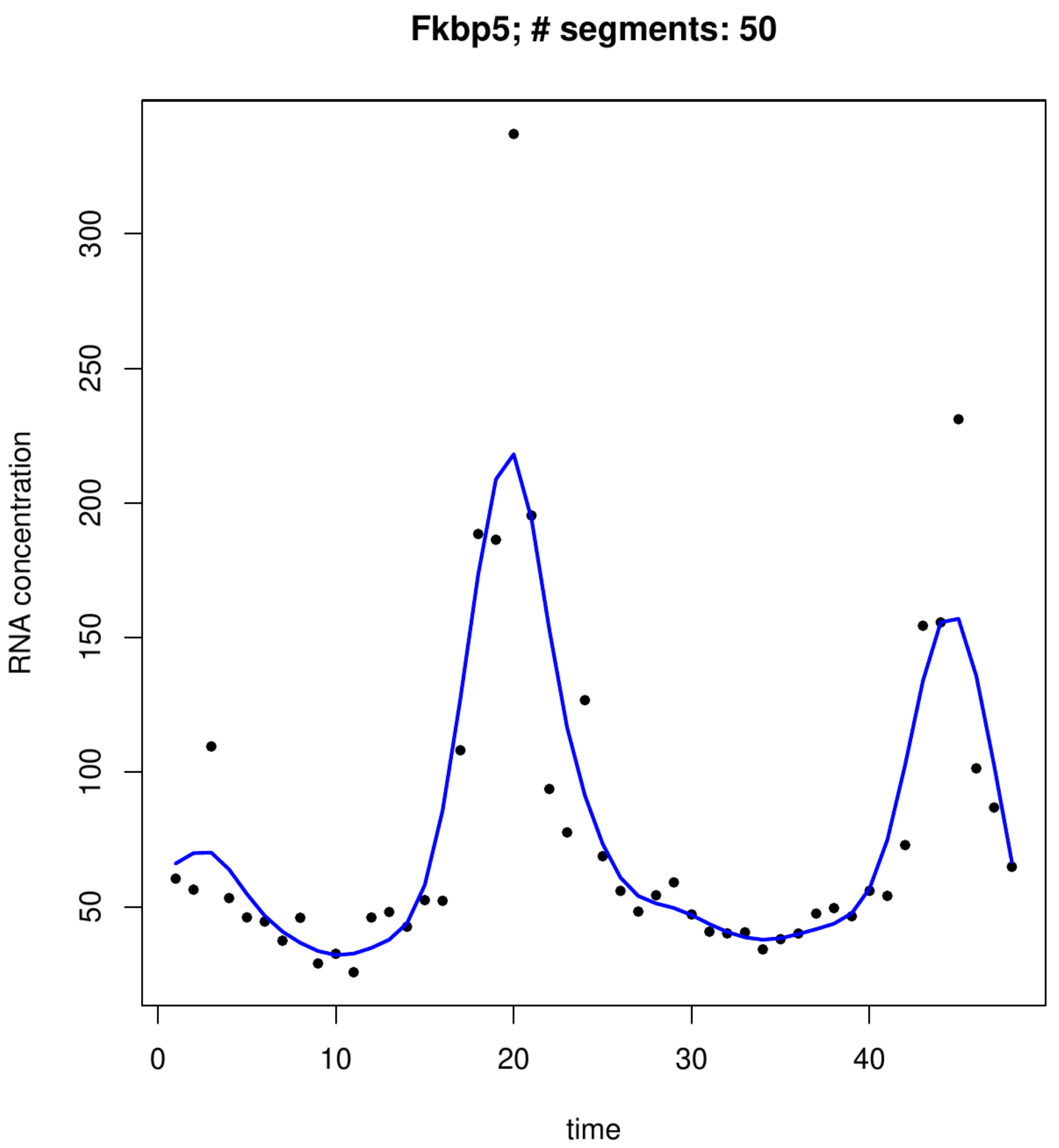}
& &
\includegraphics[scale=0.20, angle=0]{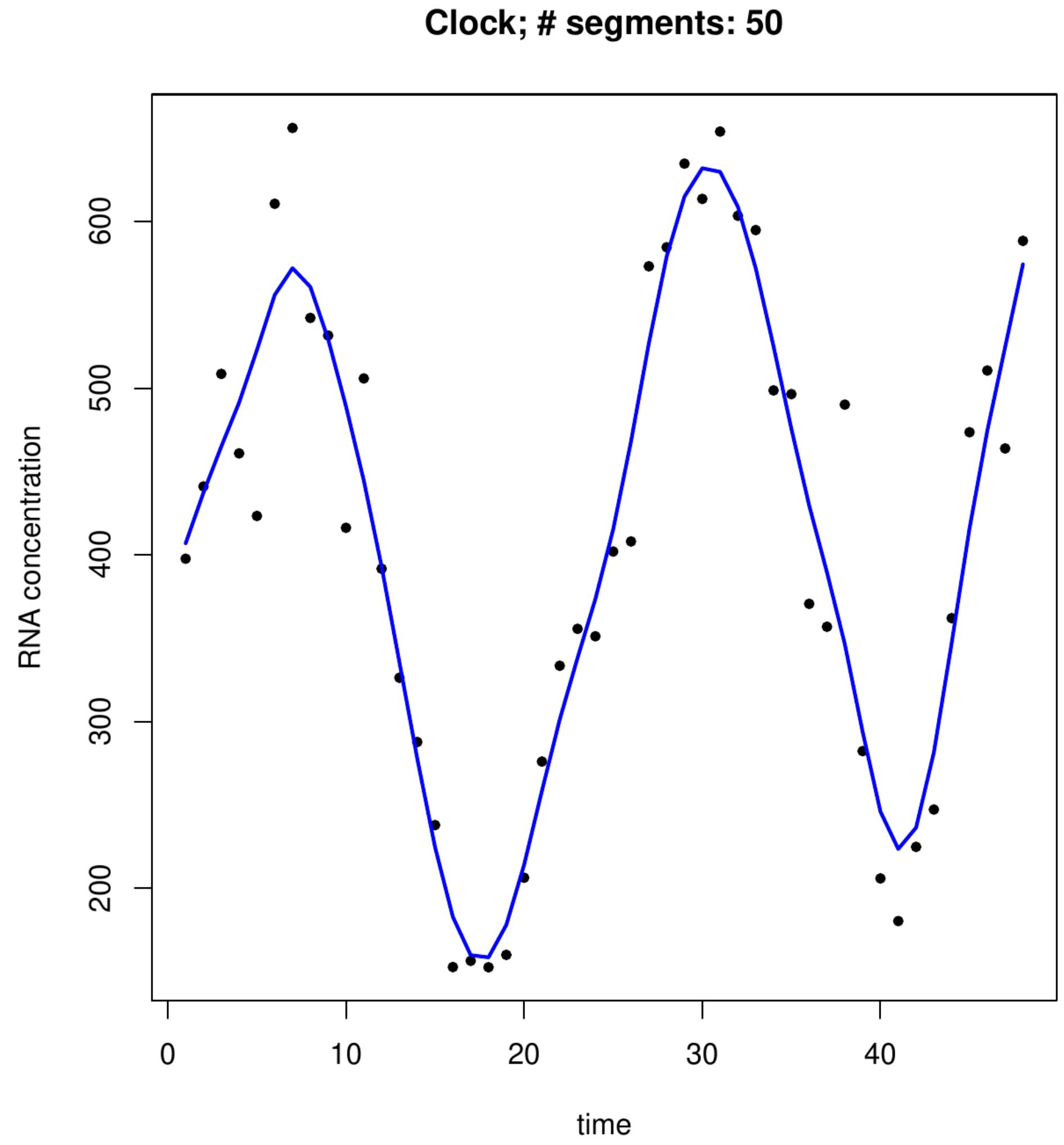}
\end{tabular}
\caption{Top left and right panels: B-spline basis functions of degree 1 and 2, respectively. Bottom left and right panel: P-spline fit to transcript levels of circadian clock experiment in mice.} \label{fig:PsplinesExamples}
\end{figure}

Smooothing refers to nonparametric -- in the sense that parameters have no tangible interpretation -- description of a curve. For instance, one may wish to learn some general functional relationship between two variable, $X$ and $Y$, from data. Statistically, the model $Y= f(X) + \varepsilon$, for unknown and general function $f(\cdot)$, is to be fitted to paired observations $\{ (y_i, x_i) \}_{i=1}^n$. Here we use P-splines, penalized B-splines with B for Basis \citep{eilers1996flexible}.

A B-spline is formed through a linear combination of (pieces of) polynomial basis functions of degree $r$. For their construction specify the interval $[x_{\mbox{{\tiny start}}}, x_{\mbox{{\tiny end}}}]$ on which the function is to be learned/approximated. Let $\{ t_j \}_{j=0}^{m+2r}$ be a grid, overlapping the interval, of equidistantly placed points called knots given by $t_j = x_{\mbox{{\tiny start}}} + (j- r) h$ for all $j=0, \ldots, m+2r$ with $h = \tfrac{1}{m}(x_{\mbox{{\tiny end}}} - x_{\mbox{{\tiny start}}})$. The B-spline base functions are then defined as:
\begin{eqnarray*}
B_{j}(x; r) & = & (-1)^{r+1} (h^r r!)^{-1} ~ \Delta^{r+1} [(x - t_j)^r \mathbbm{1}_{\{x \geq t_j \}} ]
\end{eqnarray*}
where $\Delta^r[f_j(\cdot)]$ is the $r$-th difference operator applied to $f_j(\cdot)$. For $r=1$: $\Delta[f_j(\cdot)] = f_j(\cdot) - f_{j-1}(\cdot)$, while $r=2$: $\Delta^2[f_j(\cdot)] = \Delta \{ \Delta [f_j(\cdot)] \} = \Delta[f_j(\cdot) - f_{j-1}(\cdot)] = f_j(\cdot) - 2f_{j-1}(\cdot) + f_{j-2}(\cdot)$, et cetera. The top right and bottom left panels of Figure \ref{fig:PsplinesExamples} show a $1^{\mbox{{\tiny st}}}$ and $2^{\mbox{{\tiny nd}}}$ degree B-spline basis functions. 
A P-spline is a curve of the form $\sum_{j=0}^{m+2r} \alpha_j B_j(x; r)$ fitted to the data by means of penalized least squares minimization. The least squares are $\| \mathbf{Y} - \mathbf{B} \aalpha \|_2^2$ where $\mathbf{B}$ is a $n \times (m + 2r)$-dimensional matrix with the $j$-th column equalling $(B_j(x_i; r), B_j(x_2; r), \ldots, B_j(x_{n}; r))^{\top}$. The employed penalty is of the ridge type: the sum of the squared difference among contiguous $\alpha_j$. Let $\mathbf{D}$ be the first order differencing matrix. The penalty can then be written as $\| \mathbf{D} \alpha \|_2^2 = \sum_{j=2}^{m+2r} (\alpha_j - \alpha_{j-1})^2$. A second order difference matrix would amount to $\| \mathbf{D} \alpha \|_2^2 = \sum_{j=3}^{m+2r} (\alpha_j - 2 \alpha_{j-1} + \alpha_{j-2})^2$. \cite{Eilers1999} points out how P-splines may be interpret as a mixed model. Hereto choose $\tilde{\mathbf{X}}$ such that its columns span the null space of $\mathbf{D}^{\top} \mathbf{D}$, which comprises a single column representing the intercept when $\mathbf{D}$ is a first order differencing matrix, and $\tilde{\mathbf{Z}} = \mathbf{D}^{\top} (\mathbf{D} \mathbf{D}^{\top})^{-1}$. Then, for any $\aalpha$:
\begin{eqnarray*}
\mathbf{B} \aalpha & = & \mathbf{B} (\tilde{\mathbf{X}} \bbeta + \tilde{\mathbf{Z}} \ggamma) \, \, \, := \, \, \, \mathbf{X} \bbeta + \mathbf{Z} \ggamma.
\end{eqnarray*}
This parametrization simplifies the employed penalty to:
\begin{eqnarray*}
\| \mathbf{D} \alpha \|_2^2 & = & 
\| \mathbf{D} (\tilde{\mathbf{X}} \bbeta + \tilde{\mathbf{Z}} \ggamma) \|_2^2 \, \, \, = \, \, \, \| \mathbf{D} \mathbf{D}^{\top} (\mathbf{D} \mathbf{D}^{\top})^{-1} \ggamma \|_2^2 \, \, \, = \, \, \, \| \ggamma \|_2^2,
\end{eqnarray*}
where $\mathbf{D} \tilde{\mathbf{X}} \bbeta$ has vanished by the construction of $\tilde{\mathbf{X}}$. Hence, the penalty only affects the random effect parameter, leaving the fixed effect parameter unshrunken. The resulting loss function, $\| \mathbf{Y} - \mathbf{X} \bbeta - \mathbf{Z} \ggamma \|_2^2 + \lambda \| \gamma \|_2^2$, coincides for suitably chosen $\lambda$ to that of the mixed model (as will become apparent later). The bottom panels of Figure \ref{fig:PsplinesExamples} shows the flexibility of this approach.

The following \texttt{R}-script fits a P-spline to a gene's transcript levels of the circadian clock study in mice. It uses a basis of $m=50$ truncated polynomial functions of degree $r=3$ (cubic), which is generated first alongside  several auxillary matrices. This basis forms, after post-multiplication with a projection matrix onto the space spanned by the columns of the difference matrix $\mathbf{D}$, the design matrix for the random coefficient of the mixed model $\mathbf{Y} = \mathbf{X} \bbeta + \mathbf{Z} \ggamma + \vvarepsilon$ with $\ggamma \sim \mathcal{N}(\mathbf{0}_q, \sigma_{\gamma}^2 \mathbf{I}_{qq})$ and  $\vvarepsilon \sim \mathcal{N}(\mathbf{0}_n, \sigma_{\varepsilon}^2 \mathbf{I}_{nn})$. The variance parameters of this model are then estimated by means of restricted maximum likelihood (REML). The final P-spline fit is obtained from the linear predictor using, in line with Theorem \ref{theorem.GCV2ridgePenalty},  $\lambda = \sigma_{\varepsilon}^2 / \sigma_{\gamma}^2$ in which the REML estimates of these variance parameters are substituted. The resulting P-spline fit of two transcripts is shown in the bottom panels of Figure \ref{fig:PsplinesExamples}.
\begin{lstlisting}
# load libraries
library(gridExtra)
library(MetaCycle)
library(MASS)

#------------------------------------------------------------------------------
# intermezzo: declaration of functions used analysis
#------------------------------------------------------------------------------

tpower <- function(x, knots, p) {
	# function for evaluation of truncated p-th
	# power functions positions x, given knots
	return((x - knots)^p * (x > knots))
}

bbase <- function(x, m, r) {
	# function for B-spline basis generation
	# evaluated at the extremes of x
	# with m segments and spline degree r
	h     <- (max(x) - min(x))/m
	knots <- min(x) + (c(0:(m+2*r)) - r) * h
	P     <- outer(x, knots, tpower, r)
	D     <- diff(diag(m+2*r+1), diff=r+1) / (gamma(r+1) * h^r)
	return((-1)^(r+1) * P %*% t(D))
}

thetaEstEqREML <- function(theta, Z, Y, X, sigma2e){
	# function for REML estimation: 
	# estimating equation of theta
	QthetaInv <- solve(diag(length(Y)) + theta * Z %*% t(Z))
	Ptheta    <- QthetaInv - 
	             QthetaInv %*% X %*% 
	             solve(t(X) %*% QthetaInv %*% X) %*% t(X) %*% QthetaInv
	return(sum(diag(Ptheta %*% Z %*% t(Z))) - 
	       as.numeric(t(Y) %*% Ptheta %*% Z %*% t(Z) %*% Ptheta %*% Y) / sigma2e)
}

#------------------------------------------------------------------------------

# load data
data(cycMouseLiverRNA)
id <- 14
Y  <- as.numeric(cycMouseLiverRNA[id,-1])
X  <- 1:length(Y)

# set P-spline parameters
m <- 50
r <- 3
B <- bbase(X, m=m, r=r)

# prepare some matrices
D <- diff(diag(m+r), diff = 2)
Z <- B %*% t(D) %*% solve(D %*% t(D))
X <- B %*% Null(t(D) %*% D)

# initiate
theta <- 1
for (k in 1:100){
	# for-loop, alternating between theta and error variance estimation
	thetaPrev <- theta
	QthetaInv <- solve(diag(length(Y)) + theta * Z %*% t(Z))
	Ptheta    <- QthetaInv - 
	             QthetaInv %*% X %*% 
	             solve(t(X) %*% QthetaInv %*% X) %*% t(X) %*% QthetaInv
	sigma2e   <- t(Y) %*% Ptheta %*% Y / (length(Y)-2)
	theta     <- uniroot(thetaEstEqREML, c(0, 100000), 
	                     Z=Z, Y=Y, X=X, sigma2e=sigma2e)$root
	if (abs(theta - thetaPrev) < 10^(-5)){ break }
}

# P-spline fit
bgHat <- solve(t(cbind(X, Z)) %*% cbind(X, Z) + 
         diag(c(rep(0, ncol(X)), rep(1/theta, ncol(Z))))) %*% 
         t(cbind(X, Z)) %*% Y

# plot fit
plot(Y, pch=20, xlab="time", ylab="RNA concentration", 
     main=paste(strsplit(cycMouseLiverRNA[id,1], "_")[[1]][1], 
     "; # segments: ", m, sep=""))
lines(cbind(X, Z) %*% bgHat, col="blue", lwd=2)



\end{lstlisting}
The fitted splines displayed Figure \ref{fig:PsplinesExamples} nicely match the data. From the circadian clock perspective it is especially the fit in the right-hand side bottom panel that displays the  archetypical sinusoidal behaviour associated by the layman with the sought-for rythm. Close inspection of the fits reveals some minor discontinuities in the derivative of the spline fit. These minor discontinuities are indicative of a little overfitting, due to too large an estimate of $\sigma_{\gamma}^2$. This appears to be due to numerical instability of the solution of the estimating equations of the REML estimators of the mixed model's variance parameter estimators when $m$ is large compared to the sample size $n$.

\pagestyle{fancy}

\chapter[Ridge logistic regression]{Ridge logistic regression} \label{chap.ridgeLogistic}
Ridge penalized estimation is not limited to the standard linear regression model, but can be used to estimate (virtually) any model. Here we illustrate how it may be used to fit the logistic regression model. To this end we first recap this model and the (unpenalized) maximum likelihood estimation of its parameters. After which the model is estimated by means of ridge penalized maximum likelihood, which will turn out to be a relatively straightforward modification of unpenalized estimation.

\section{Logistic regression} \label{sect:logisticRegression}
The logistic regression model explains a binary response variable (through some transformation) by a linear combination of a set of covariates (as in the linear regression model). Denote this response of the $i$-th sample by $Y_i$ with  $Y_i \in \{ 0, 1 \}$ for $i=1, \ldots, n$. The $n$-dimensional column vector $\mathbf{Y}$ stacks these $n$ responses. For each sample information on the $p$ explanatory variables $X_{i,1}, \ldots, X_{i,p}$ is available. In row vector form, this information is denoted $\mathbf{X}_{i,\ast} =  (X_{i,1}, \ldots, X_{i,p})$. The $n \times p$-dimensional matrix $\mathbf{X}$ aggregates these vectors, such that $\mathbf{X}_{i,\ast}$ is the $i$-th row vector.

The binary response cannot be modelled as in the linear model like $Y_i = \mathbf{X}_{i,\ast} \bbeta + \varepsilon_i$. With each element of $\mathbf{X}_{i,\ast}$ and $\bbeta$ assuming a value in $\mathbb{R}^p$, the linear predictor is not restricted to the domain of the response. This is resolved by modeling $p_i = P(Y_i = 1)$ instead. Still the linear predictor may exceed the domain of the response  as $p_i \in [0,1]$. Hence, a transformation is applied to map $p_i$ to $\mathbb{R}$, the range of the linear predictor.
\begin{figure}[!h]
\begin{tabular}{rcl}
\includegraphics[scale=0.40, angle=0]{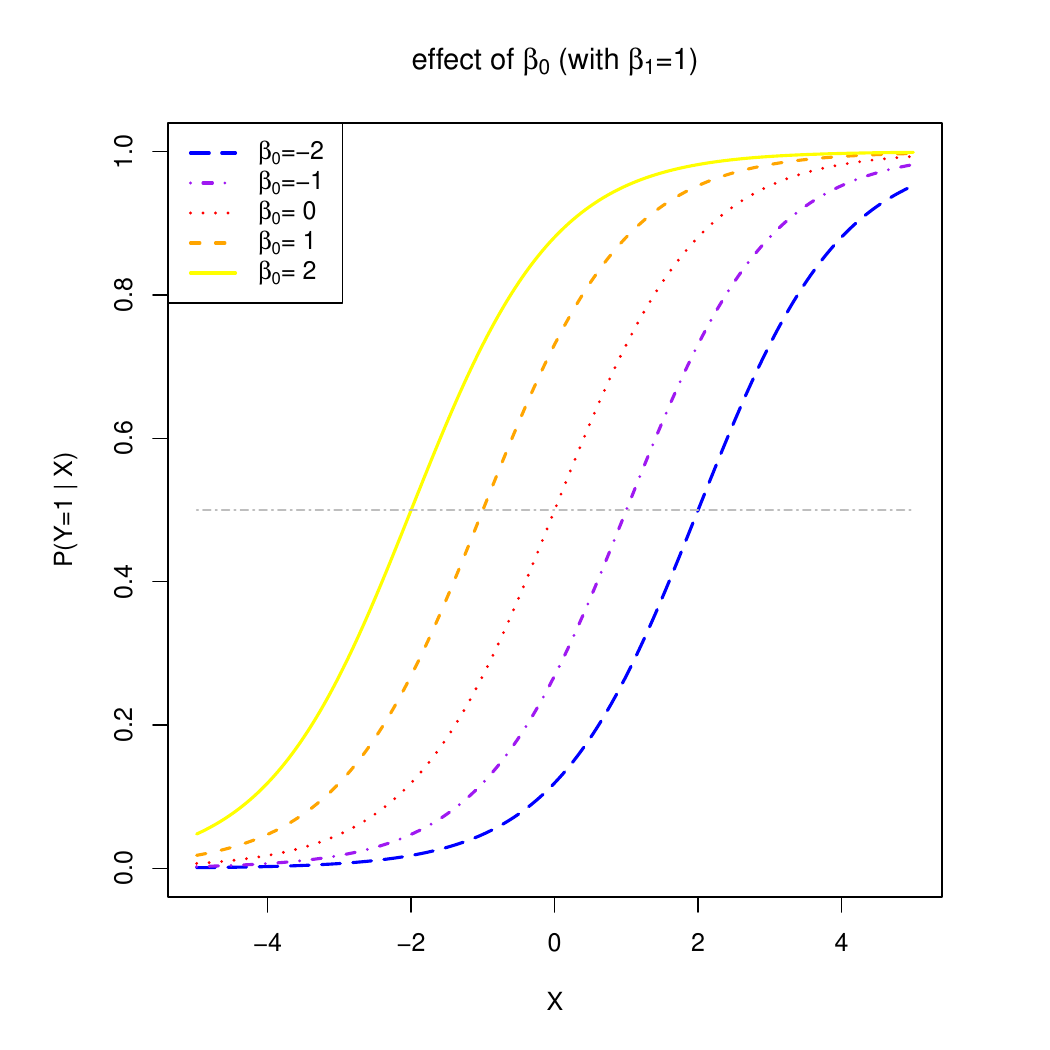}
&
\includegraphics[scale=0.40, angle=0]{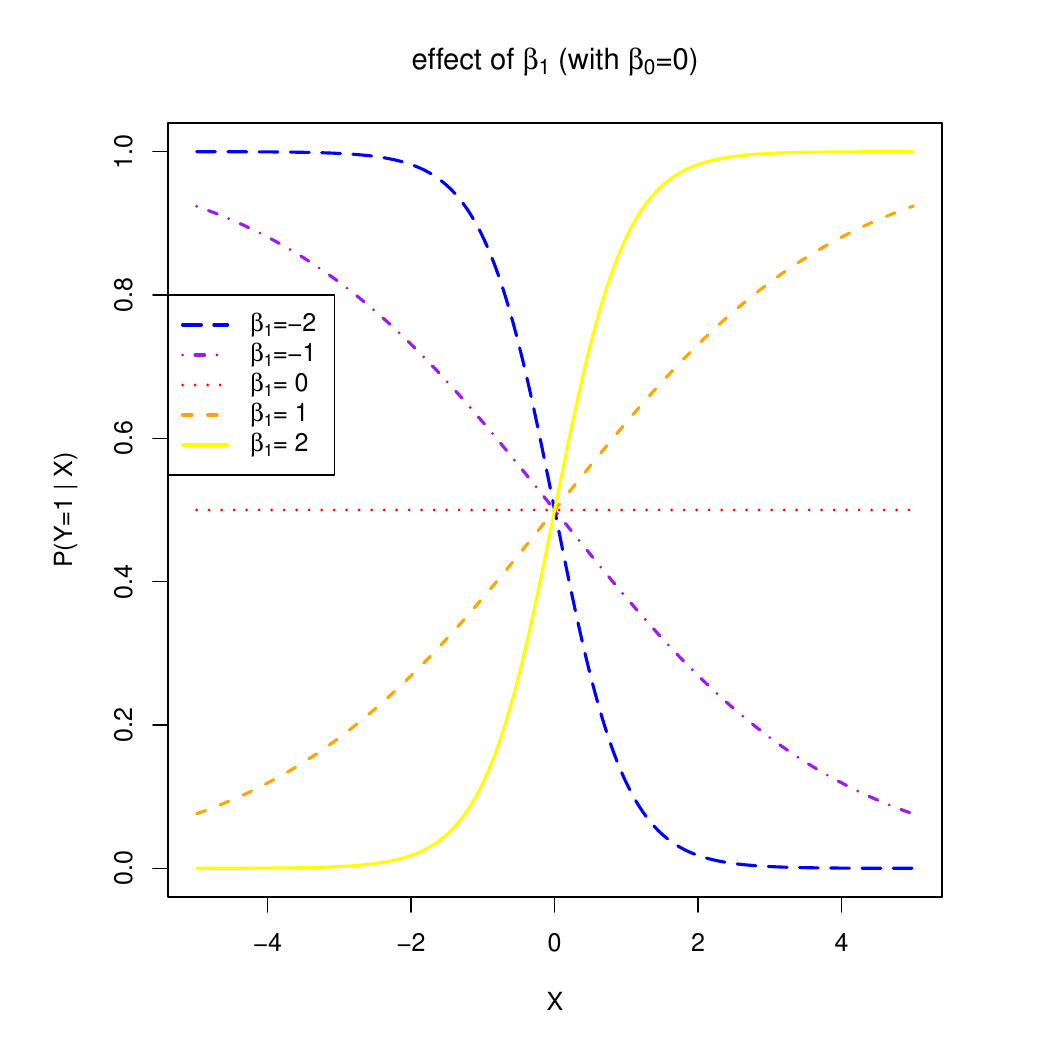}
\\
\includegraphics[scale=0.40, angle=0]{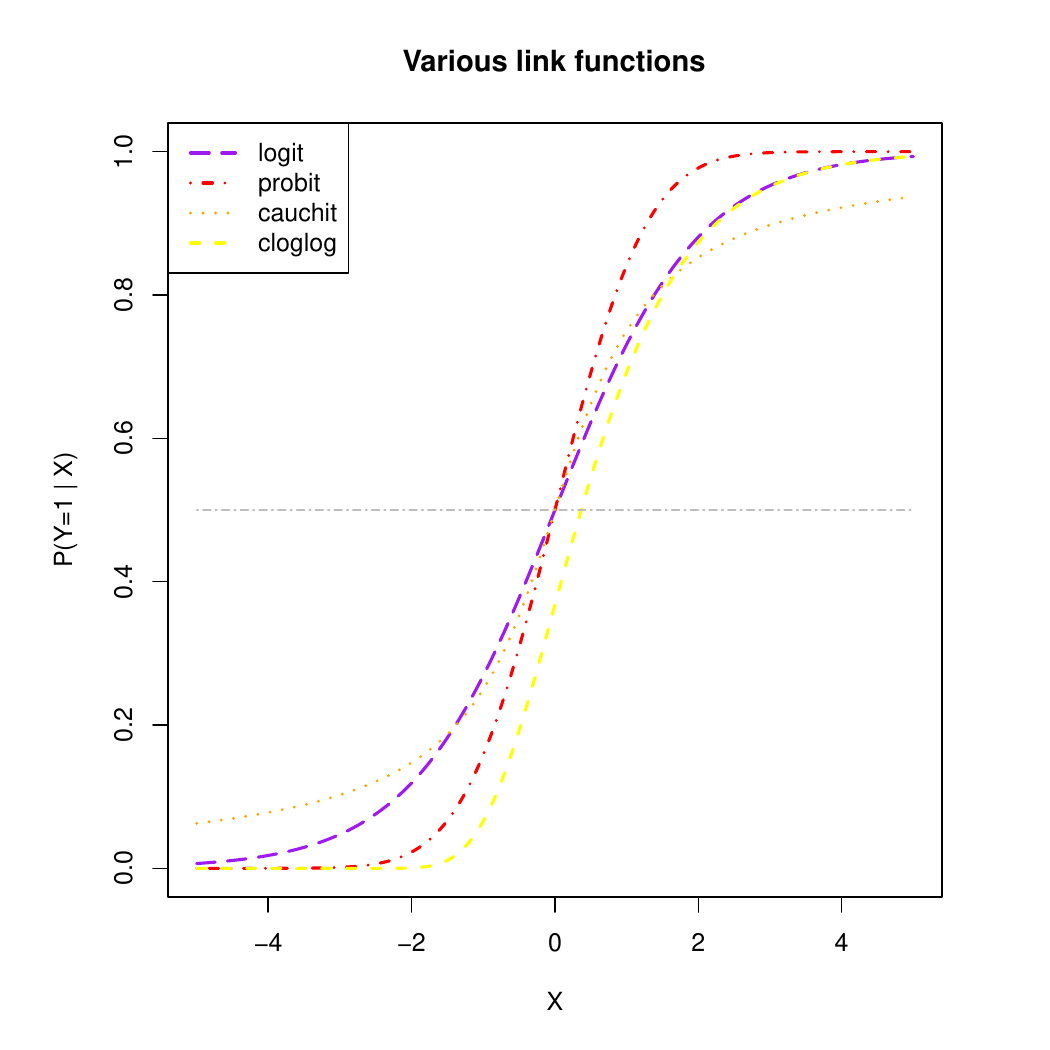}
&
\includegraphics[scale=0.40, angle=0]{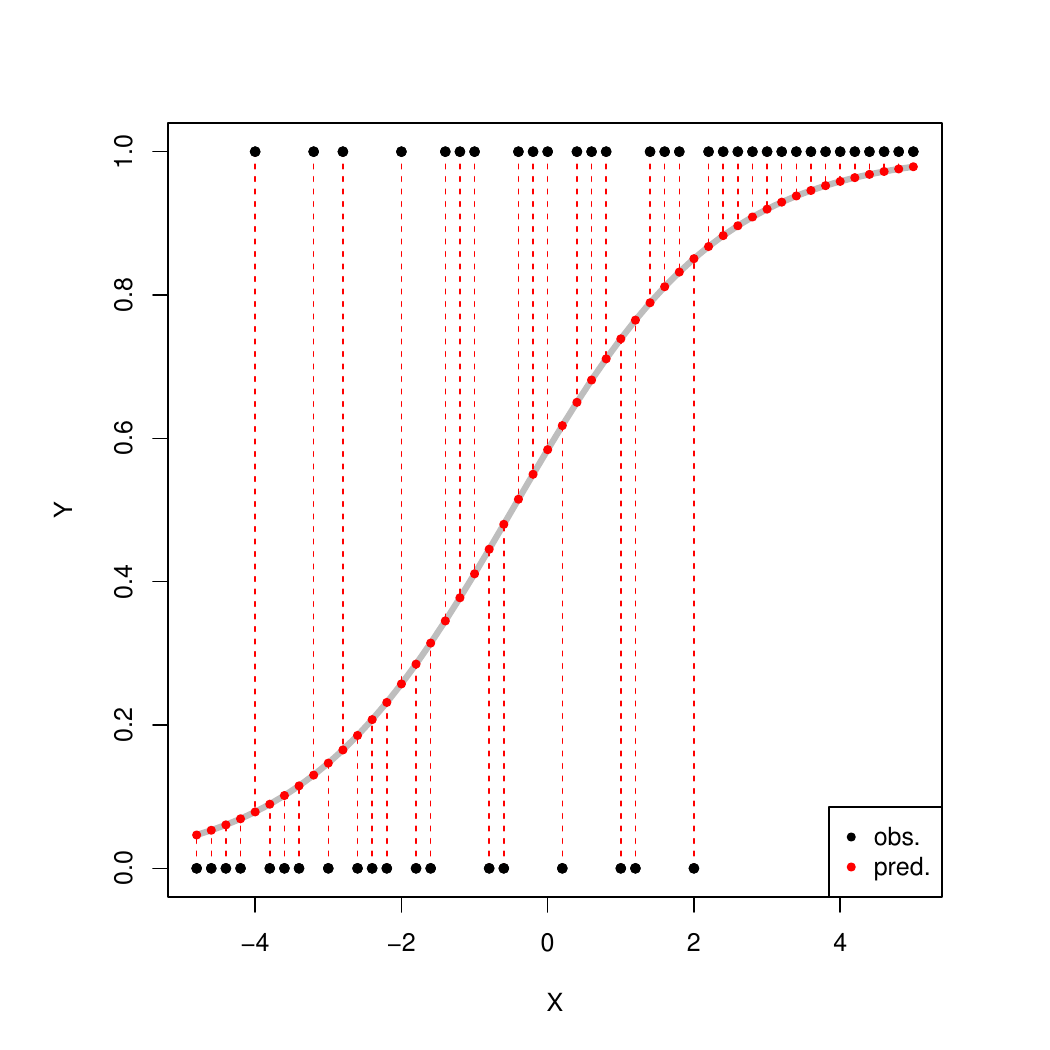}
\end{tabular}
\caption{Top row, left panel: the response curve for various choices of the intercept $\beta_0$. Top row, right panel: the response curve for various choices of the regression coefficent $\beta_1$. Bottom row, left panel: the responce curve for various choices of the link function. Bottom panel, right panel: observations, fits and their deviations. } \label{fig.logisticRidge_modelAndFitIllustration}
\end{figure}
The transformation associated with the logistic regression model is the logarithm of the odds, with the odds defined as: $\mbox{{\it odds}} = P(\mbox{succes}) / P(\mbox{failure})  = p_i/ (1-p_i)$. The logistic model is then written as $\log[ p_i / (1-p_i)] = \mathbf{X}_{i,\ast} \bbeta$ for all $i$. Or, expressed in terms of the response:
\begin{eqnarray*}
p_i & = & P(Y_i = 1) \, \, \, = \, \, \, g^{-1}(\mathbf{X}_{i,\ast}; \bbeta) \, \, \, = \, \, \,\exp(\mathbf{X}_{i,\ast} \bbeta) [1 + \exp(\mathbf{X}_{i,\ast} \bbeta) ]^{-1}.
\end{eqnarray*}
The function $g(\cdot; \cdot)$ is called the \textit{link function}. It links the response to the explanatory variables. The one above is called the logistic link function. Or short, logit. The regression parameters have tangible interpretations. When the first covariate represents the intercept, i.e. $X_{i,j} = 1$ for all $i$, then $\beta_1$ determines where the link function equals a half when all other covariates fail to contribute to the linear predictor (i.e. where $P (Y_i = 1 \, | \, \mathbf{X}_{i,\ast}) = 0.5$ when $\mathbf{X}_{i,\ast} \bbeta = \beta_1$). This is illustrated in the top-left panel of Figure \ref{fig.logisticRidge_modelAndFitIllustration} for various choices of the intercept. On the other hand, the regression parameters are directly related to the odds ratio: $\mbox{{\it odds ratio}} = \mbox{odds}(X_{i,j}+1) / \mbox{odds}(X_{i,j}) = \exp(\beta_j)$. Hence, the effect of a unit change in the $j$-th covariate on the odds ratio is $\exp(\beta_j)$ (see Figure \ref{fig.logisticRidge_modelAndFitIllustration}, top-right panel). Other link functions (depicted in Figure \ref{fig.logisticRidge_modelAndFitIllustration}, bottom-left panel) are common, e.g. the \textit{probit}:  $p_i = \Phi_{0,1}(\mathbf{X}_{i,\ast} \bbeta)$; the \textit{cloglog}: $p_i = \frac{1}{\pi} \arctan(\mathbf{X}_{i,\ast} \bbeta) + \frac{1}{2}$; the \textit{Cauchit}: $p_i = \exp[ - \exp(\mathbf{X}_{i,\ast} \bbeta)]$. All these link functions are invertible. Irrespective of the choice of the link function, the binary data are thus modelled as $Y_i \sim \mathcal{B}[g^{-1}(\mathbf{X}_{i,\ast}; \bbeta), 1]$. That is, as a single draw from the binomial distribution with success probability $g^{-1}(\mathbf{X}_{i,\ast}; \bbeta)$.

Let us now estimate the parameter of the logistic regression model by means of the maximum likelihood method. The likelihood of the experiment then, by the independence among the samples, is:
\begin{eqnarray*}
L(\mathbf{Y} \, | \, \mathbf{X}; \bbeta) & = & \prod_{i=1}^n \big[ P(Y_i = 1 \, | \, \mathbf{X}_{i,\ast}) \big]^{Y_i} \big[ P(Y_i = 0 \, | \, \mathbf{X}_{i,\ast}) \big]^{1-Y_i}.
\end{eqnarray*}
After taking the logarithm and some ready algebra, the log-likelihood is found to be:
\begin{eqnarray*}
\mathcal{L}(\mathbf{Y} \, | \, \mathbf{X}; \bbeta) & = & \sum\nolimits_{i=1}^n  \big\{ Y_i \mathbf{X}_{i,\ast} \bbeta - \log [ 1 + \exp(\mathbf{X}_{i,\ast} \bbeta) ] \big\}.
\end{eqnarray*}
Differentiate the log-likelihood with respect to $\bbeta$, equate it zero, and obtain the estimating equation for $\bbeta$:
\begin{eqnarray} \label{form:logisticRidge_estimatingEquationOfBeta}
\frac{\partial \mathcal{L}}{\partial \bbeta } & = & \sum_{i=1}^n \Big[ Y_i  - \frac{\exp(\mathbf{X}_{i,\ast} \bbeta)}{ 1 + \exp(\mathbf{X}_{i,\ast} \bbeta)} \Big] \mathbf{X}_{i,\ast}^{\top} \, \, \, = \, \, \, \mathbf{0}_p.
\end{eqnarray}
The maximum likelihood estimate of $\bbeta$ strikes a (weighted by the $\mathbf{X}_{i,\ast}$) balance between observation and model. Put differently (and illustrated in the bottom-right panel of Figure \ref{fig.logisticRidge_modelAndFitIllustration}), a curve is fit through data by minimizing the distance between them: at the maximum likelihood estimate of $\bbeta$ a weighted average of their deviations is zero.

The maximum likelihood estimate of $\bbeta$ is evaluated by solving Equation (\ref{form:logisticRidge_estimatingEquationOfBeta}) with respect to $\bbeta$ by means of the Newton-Raphson algorithm. The Newton-Raphson algorithm iteratively finds the zeros of a smooth enough function $f(\cdot)$. Let $x_0$ denote an initial guess of the zero. Then, approximate $f(\cdot)$ around $x_0$ by means of a first order Taylor series: $f(x) \approx f(x_0) + (x - x_0) \, (d f / d x) |_{x=x_0}$. Solve this for $x$ and obtain: $x = x_0 -  [ (d f / d x) |_{x=x_0} ]^{-1} f(x_0)$. Let $x_1$ be the solution for $x$, use this as the new guess and repeat the above until convergence. When the function $f(\cdot)$ has multiple arguments, is vector-valued and denoted by $\vec{\mathbf{f}}$, and the Taylor approximation becomes: $\vec{\mathbf{f}}(\mathbf{x}) \approx f(\mathbf{x}_0) + J \vec{\mathbf{f}} \big|_{\mathbf{x}=\mathbf{x}_0} (\mathbf{x} - \mathbf{x}_0)$ with
\begin{eqnarray*}
J \vec{\mathbf{f}}  = \left(
\begin{array}{llll}
\frac{\partial f_1}{\partial x_1} & \frac{\partial f_1}{\partial x_2} & \ldots & \frac{\partial f_1}{\partial x_p}
\\
\frac{\partial f_1}{\partial x_1} & \frac{\partial f_2}{\partial x_2} & \ldots & \frac{\partial f_2}{\partial x_p}
\\
\vdots & \vdots & \ddots & \vdots
\\
\frac{\partial f_q}{\partial x_1} & \frac{\partial f_q}{\partial x_2} & \ldots & \frac{\partial f_q}{\partial x_p}
\end{array}
\right),
\end{eqnarray*}
the Jacobi matrix. An update of $x_0$ is now readily constructed by solving (the approximation for) $\vec{\mathbf{f}}(\mathbf{x}) = \mathbf{0}$ for $\mathbf{x}$.

When applied here to the maximum likelihood estimation of the regression parameter $\bbeta$ of the logistic regression model, the Newton-Raphson update is:
\begin{eqnarray*}
\hat{\bbeta}^{\mbox{{\scriptsize new}}} & =  & \hat{\bbeta}^{\mbox{{\scriptsize old}}} -  \Big( \frac{\partial^2 \mathcal{L}}{\partial \bbeta \partial \bbeta^{\top}} \Big)^{-1} \Big|_{\bbeta =  \hat{\bbeta}^{\mbox{{\tiny old}}} } \, \, \frac{\partial \mathcal{L}}{\partial \bbeta } \Big|_{\bbeta =  \hat{\bbeta}^{\mbox{{\tiny old}}} }
\end{eqnarray*}
where the Hessian of the log-likelihood equals:
\begin{eqnarray*}
\frac{\partial^2 \mathcal{L}}{\partial \bbeta \partial \bbeta^{\top}} & = & - \sum\nolimits_{i=1}^n \frac{\exp(\mathbf{X}_{i,\ast} \bbeta)}{ [1 + \exp(\mathbf{X}_{i,\ast} \bbeta)]^2} \mathbf{X}_{i,\ast}^{\top} \mathbf{X}_{i,\ast}.
\end{eqnarray*}
Iterative application of this updating formula converges to the maximum likelihood estimate of $\bbeta$.

The Newton-Raphson algorithm is often reformulated as an iteratively re-weighted least squares (IRLS) algorithm. Hereto, first write the gradient and Hessian in matrix notation:
\begin{eqnarray*}
\frac{\partial \mathcal{L}}{\partial \bbeta } \, \, \, = \, \, \, \mathbf{X}^{\top} [\mathbf{Y} - \vec{\mathbf{g}}^{-1}( \mathbf{X}; \bbeta)]
& \mbox{ and } & \frac{\partial^2 \mathcal{L}}{\partial \bbeta \partial \bbeta^{\top}} \, \,  \, = \, \, \,  - \mathbf{X}^{\top} \mathbf{W} \mathbf{X},
\end{eqnarray*}
where $\vec{\mathbf{g}}^{-1}( \mathbf{X}; \bbeta)  = [g^{-1}( \mathbf{X}_{1, \ast}; \bbeta), \ldots, g^{-1}( \mathbf{X}_{n, \ast}; \bbeta)]^{\top}$ with $g^{-1}(\cdot; \cdot) = \exp(\cdot; \cdot) / [1 + \exp(\cdot; \cdot)]$ and $\mathbf{W}$ diagonal with $(\mathbf{W})_{ii} = \exp(\mathbf{X}_i \hat{\bbeta}^{\mbox{{\scriptsize old}}}  ) [ 1 + \exp(\mathbf{X}_i \hat{\bbeta}^{\mbox{{\scriptsize old}}}  ) ]^{-2}$. The notation $\mathbf{W}$ was already used in Chapter \ref{chap:genRidge} and, generally, refers to a (diagonal) weight matrix with the choice of the weights depending on the context. The updating formula of the estimate then becomes:
\begin{eqnarray*}
\hat{\bbeta}^{\mbox{{\scriptsize new}}} & =  &  \hat{\bbeta}^{\mbox{{\scriptsize old}}} + (\mathbf{X}^{\top} \mathbf{W} \mathbf{X})^{-1}  \mathbf{X}^{\top} [\mathbf{Y} - \vec{\mathbf{g}}^{-1}( \mathbf{X}; \bbeta^{\mbox{{\scriptsize old}}})]
\\
& = &  (\mathbf{X}^{\top} \mathbf{W} \mathbf{X})^{-1} \mathbf{X}^{\top} \mathbf{W} \{ \mathbf{X} \hat{\bbeta}^{\mbox{{\scriptsize old}}} +  \mathbf{W}^{-1} [\mathbf{Y} - \vec{\mathbf{g}}^{-1}( \mathbf{X}; \bbeta^{\mbox{{\scriptsize old}}})] \}
\\
& = & (\mathbf{X}^{\top} \mathbf{W} \mathbf{X})^{-1} \mathbf{X}^{\top} \mathbf{W} \mathbf{Z},
\end{eqnarray*}
where $\mathbf{Z} =  \mathbf{X} \hat{\bbeta}^{\mbox{{\scriptsize old}}} +  \mathbf{W}^{-1} [\mathbf{Y} - \vec{\mathbf{g}}^{-1}( \mathbf{X}; \bbeta^{\mbox{{\scriptsize old}}})]$. The Newton-Raphson update is thus the solution to the following weighted least squares problem:
\begin{eqnarray*}
\hat{\bbeta}^{\mbox{{\scriptsize new}}} & = &  \arg \min\nolimits_{\bbeta \in \mathbb{R}^p}  ~ (\mathbf{Z} - \mathbf{X} \bbeta)^{\top} \mathbf{W} (\mathbf{Z} - \mathbf{X} \bbeta).
\end{eqnarray*}
Effectively, at each iteration the \textit{adjusted response} $\mathbf{Z}$ is regressed on the covariates that comprise $\mathbf{X}$. For more on logistic regression confer, e.g. the monograph of \cite{Hosm2013}.

\section{Separable and high-dimensional data}  \label{sect.separableData}
The binary nature of the response may bring about another problem, called \textit{separable data}, that frustates the estimation of the logistic regression estimator. Separable data refer to the situation where the covariate space can be separated by a hyperplane such that the samples with indices in the set $\{i : Y_i =0\}$ fall on one side of this hyperplane, while those with an index in the set $\{i : Y_i =1\}$ on the other. More formally, the data are separable if there exist $\mathbf{w} \in \mathbb{R}^p$ and $b \in \mathbb{R}$ such that 
\begin{eqnarray*}
\left\{ 
\begin{array}{rrr}
Y_i = 0 & \mbox{if} & \mathbf{X}_{i,\ast} \mathbf{w} + b > 0,
\\
Y_i = 1 & \mbox{if} & \mathbf{X}_{i,\ast} \mathbf{w} + b < 0,
\end{array}
\right.
\end{eqnarray*}
for $i=1, \ldots, n$. Separable data mostly occur if either of the two response outcomes has a low prevalence.

The logistic regression parameter cannot be learned from separable data by means of the maximum likelihood method. For the existence of a separating hyperplane implies that the optimal fit is perfect and all samples have -- according to the fitted logistic regression model -- a probability of one of being assigned to the correct outcome, i.e. $P(Y_i = 1 \, | \, \mathbf{X}_i)$ equals either zero or one. Consequently, the loglikelihood vanishes as
\begin{eqnarray*}
\mathcal{L}(\mathbf{Y} \, | \, \mathbf{X}; \bbeta) & = & \sum\nolimits_{i=1}^n Y_i \log [ P(Y_i = 1 \, | \, \mathbf{X}_{i,\ast})] + (1-Y_i) \log [ P(Y_i = 0 \, | \, \mathbf{X}_{i,\ast}) \big] \, \, \, = \, \, \, 0,
\end{eqnarray*}
and does no longer involve the logistic regression parameter. The logistic regression parameter is then to be chosen such that $P(Y_i = 1 \, | \, \mathbf{X}_{i,\ast}) = \exp(\mathbf{X}_{i,\ast} \bbeta) [1 + \exp(\mathbf{X}_{i,\ast} \bbeta)]^{-1} \in \{ 0, 1 \}$ (depending on whether $Y_i$ is indeed of the `1' class). This only occurs when (some) elements of $\bbeta$ equal (minus) infinity. Hence, the maximum likelihood estimator is not well-defined.

The common workaround to learn the logistic regression parameter from separable data is a technique called Firth penalized estimation (\citealp{firth1993bias}; \citealp{heinze2002solution}). It amounts to the maximization of the loglikelihood augmented with the so-called Firth penalty:
\begin{eqnarray*}
\mathcal{L} (\mathbf{Y}, \mathbf{X}; \bbeta) + \tfrac{1}{2} \log \{ \mbox{det} [ \mathcal{I} (\bbeta)] \}, 
\end{eqnarray*}
where $\mathcal{I}(\bbeta) = \mathbf{X}^{\top} \mathbf{W}_{\hspace{-0.1cm}\bbeta} \mathbf{X}$ is the Fischer information matrix and the subscript of the weight matrix is to stress the dependency on the regression parameter. The Firth penalty corrects for the first order bias of the maximum likelihood estimator due to the unbalancedness in the class prevalences. The larger this unbalance, the more likely the data are separable. Alternatively, the Firth penalty can be motivated as a (very) weakly informative Jeffrey's prior.

High-dimensionally, a separable hyperplane can always be found, unless there is at least one pair of samples with a common variate vector, i.e. $\mathbf{X}_{i,\ast} = \mathbf{X}_{i',\ast}$ with $i \not= i'$ but different responses $Y_i \not= Y_{i'}$. Moreover, the maximum likelihood estimator of the logistic regression parameter is not well-defined high-dimensionally. To see this, assume $p > n$ and an estimate $\hat{\bbeta}$ to be available. Due to the high-dimensionality, the null space of $\mathbf{X}$ is non-trivial. Hence, let $\mathbf{v} \in \mbox{null}[\mbox{span}(\mathbf{X})]$. Then: $\mathbf{X} \hat{\bbeta} = \mathbf{X} \hat{\bbeta} + \mathbf{X} \mathbf{v} = \mathbf{X} (\hat{\bbeta} + \mathbf{v})$. As the null space is a $p-n$-dimensional subspace, $\mathbf{v}$ need not equal zero. Hence, an infinite number of estimators of the logistic regression parameter exists that yield the same loglikelihood.

Firth penalization may resolve the separable data issue low-dimensionally. It does not ensure the existence of the estimator high-dimensionally. This is illustrated by the next example. 

\begin{example} \mbox{ } \\
Let the covariates be mutually exclusive indicators. Then, for all $i=1,\ldots,n$, there is a $j \in \{ 1, \ldots, p\}$ such that $\mathbf{X}_{i,\ast} = \mathbf{e}_j^{\top}$, where $\mathbf{e}_j$ is the unit vector comprising all zero's except for a one at the $j$-th position. The Firth penalty then is
\begin{eqnarray*}
\sum\nolimits_{j=1}^p \sum\nolimits_{i=1}^n \mathbbm{1}_{\{ \mathbf{X}_{i,\ast} = \mathbf{e}_j^{\top} \}} \{ \tfrac{1}{2} \beta_j - \log[1+ \exp(\beta_j)] \}.
\end{eqnarray*} 
High-dimensionally, this Firth penalty is not strictly concave as its Hessian matrix is diagonal with diagonal elements $\mathbf{H}_{jj} = - \exp(\beta_j) [1+\exp(\beta_j) ]^{-2} \sum_{i=1}^n \mathbbm{1}_{\{ \mathbf{X}_{i,\ast} = \mathbf{e}_j^{\top} \}}$ for $j = 1, \ldots, p$. Hence, the maximizer of the Firth penalized loglikelihood is not well-defined high-dimensionally.  
\end{example}

\noindent 
We can circumvent the separability and high-dimensionality by resorting to early stopping. Hereto we choose an initial value for the regression parameter and update it via
\begin{eqnarray} \label{form.ridgeLogisticGradAscent}
\boldsymbol{\beta}^{(t+1)} & = & \boldsymbol{\beta}^{(t)} +  \delta \nabla_{\boldsymbol{\beta}} \mathcal{L} ( \mathbf{Y}, \mathbf{X}; \boldsymbol{\beta}) \hspace{0.04cm} |_{\boldsymbol{\beta} =  \boldsymbol{\beta}^{(t)}}  \, \, \, = \, \, \,   \boldsymbol{\beta}^{(t)} +  \delta  \mathbf{X}^{\top} [ \mathbf{Y} - \vec{\mathbf{g}}^{-1} ( \mathbf{X}; \boldsymbol{\beta}^{(t)})].
\end{eqnarray}
At each update we move closer to the optimum, but we stop after a finite number of steps prior to running into the convergence problems. For more on early stopping in the context of logistic regression, see Exercise \ref{question:logisticEarlyStopping}.

\section{Ridge estimation}  \label{sect.ridgeLogistic}
Ridge maximum likelihood estimates of the logistic model parameters are found by the maximization of the ridge penalized loglikelihood (cf. \citealt{Scha1984,LeCe1992}):
\begin{eqnarray*}
\mathcal{L}^{\mbox{{\tiny pen}}}(\mathbf{Y}, \mathbf{X}; \bbeta, \lambda) & = & \mathcal{L} (\mathbf{Y}, \mathbf{X}; \bbeta) - \tfrac{1}{2} \lambda \| \bbeta \|_2^2 \, \, \, = \, \, \, \sum\nolimits_{i=1}^n \big\{ Y_i \mathbf{X}_{i,\ast} \bbeta - \log [ 1 + \exp(\mathbf{X}_{i,\ast} \bbeta) ] \big\} - \tfrac{1}{2} \lambda \bbeta^{\top} \bbeta,
\end{eqnarray*}
where the second summand is the ridge penalty (the sum of the square of the elements of $\bbeta$) with $\lambda$ the penalty parameter.  Penalization of the loglikelihood now amounts to the substraction of the ridge penalty. This is due to the fact that the estimator is now defined as the maximizer (instead of a minimizer) of the loss function. Moreover, the $\tfrac{1}{2}$ in front of the penalty is only there to simplify derivations later, and could in principle be absorped in the penalty parameter. Finally, the augmentation of the loglikelihood with the ridge penalty ensures the existence of a unique estimator. The loglikelihood need not be strictly concave, but the ridge penalty, $-\tfrac{1}{2} \| \bbeta \|_2^2$, is. Together they form the ridge penalized loglikelihood above, which is thus also strictly concave. This warrants the existence of a unique maximizer, and a well-defined ridge logistic regression estimator.

\begin{figure}[!h]
\begin{tabular}{rcl}
\includegraphics[scale=0.40, angle=0]{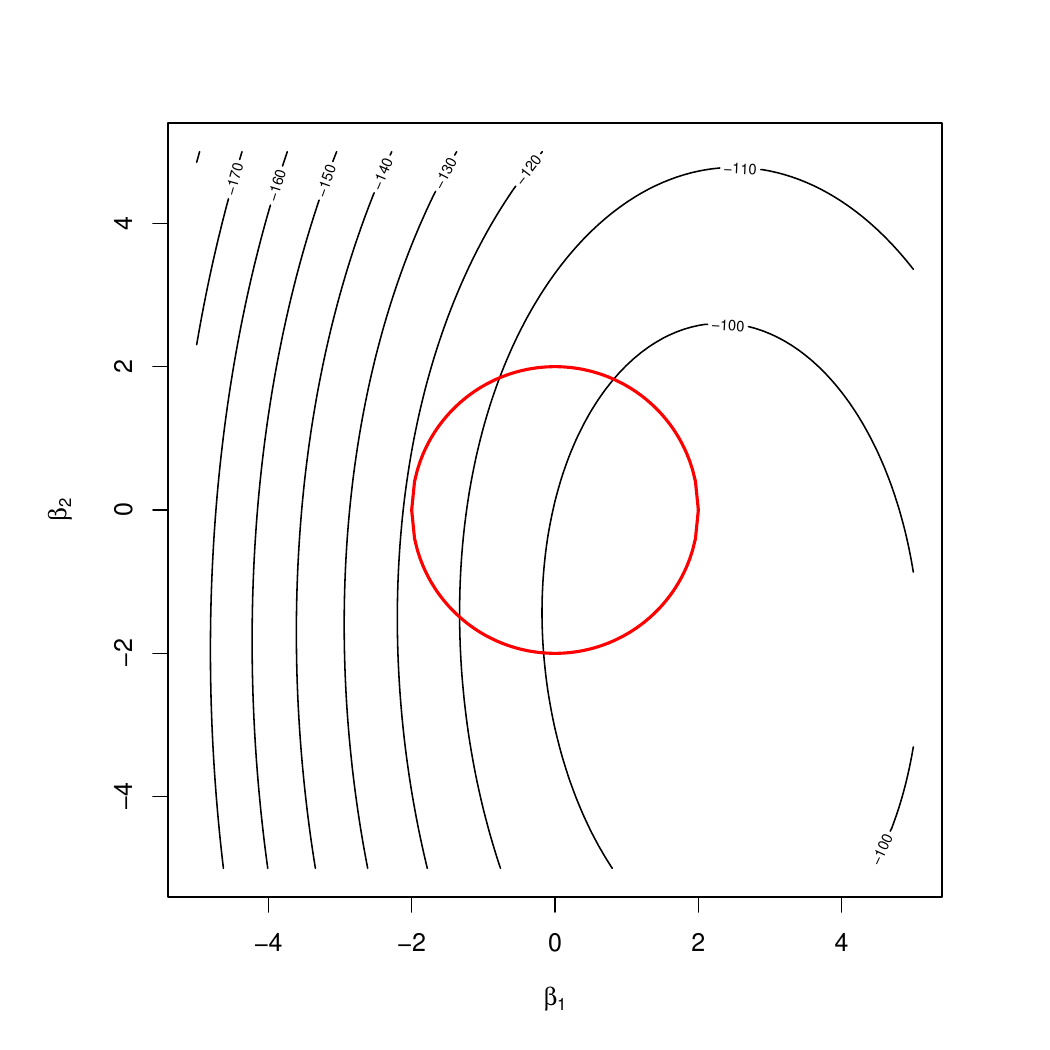}
&
\includegraphics[scale=0.40, angle=0]{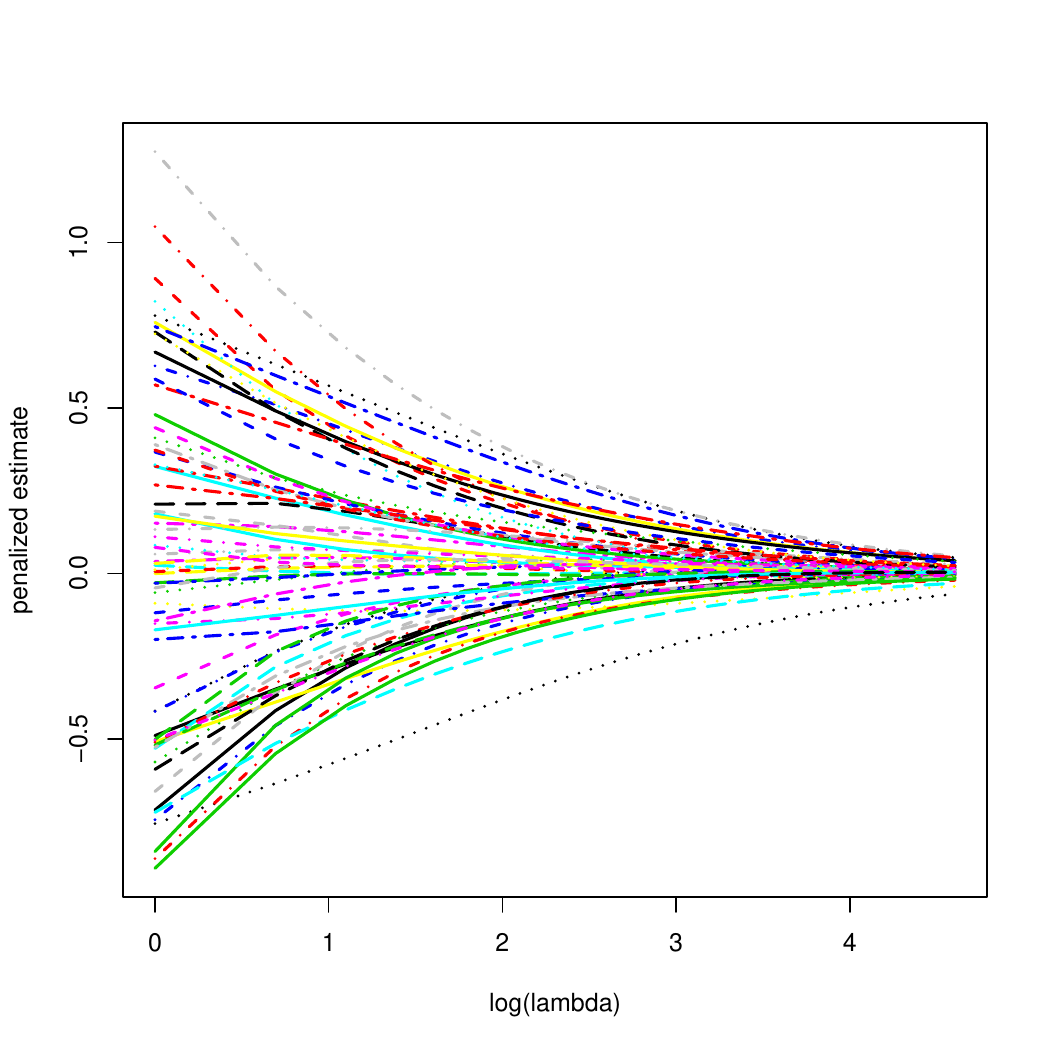}
\\
\includegraphics[scale=0.40, angle=0]{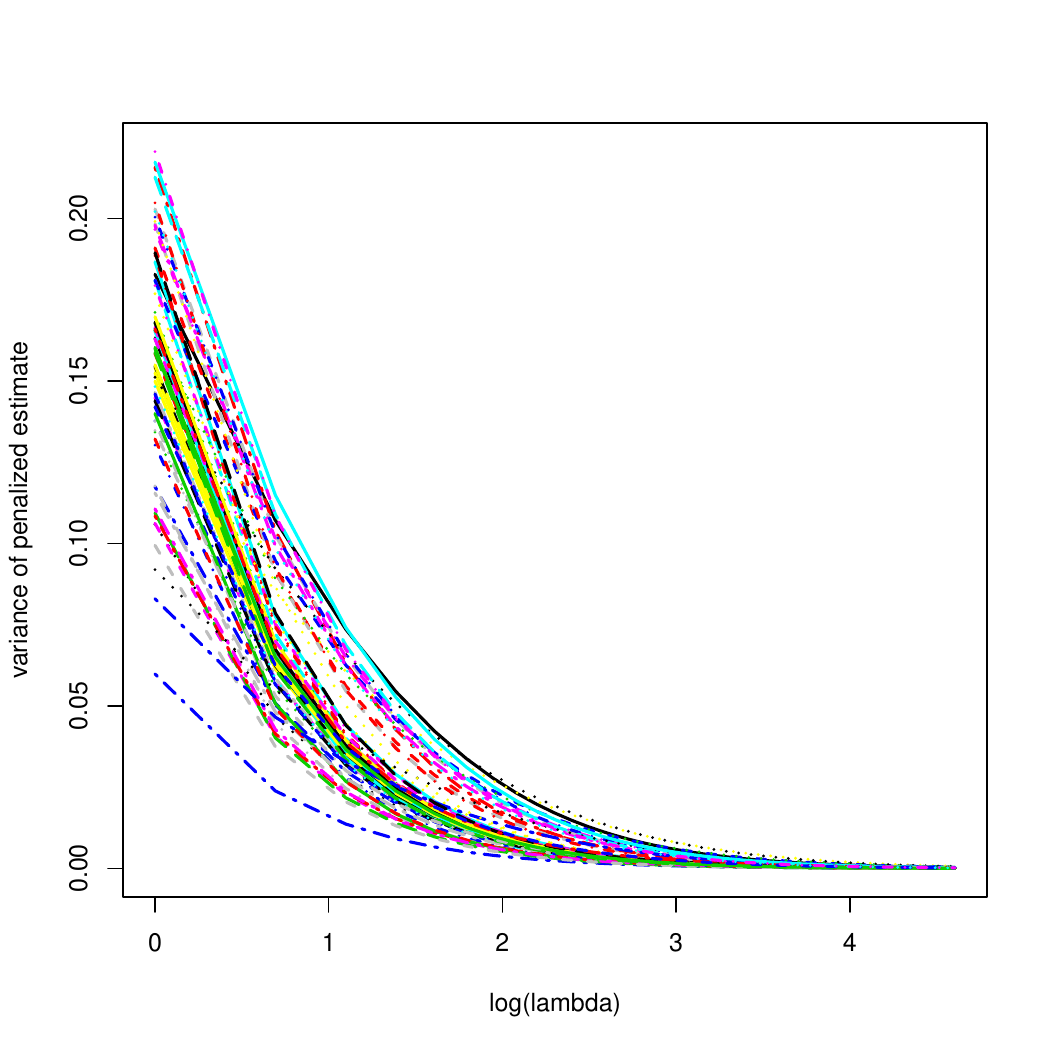}
&
\includegraphics[scale=0.40, angle=0]{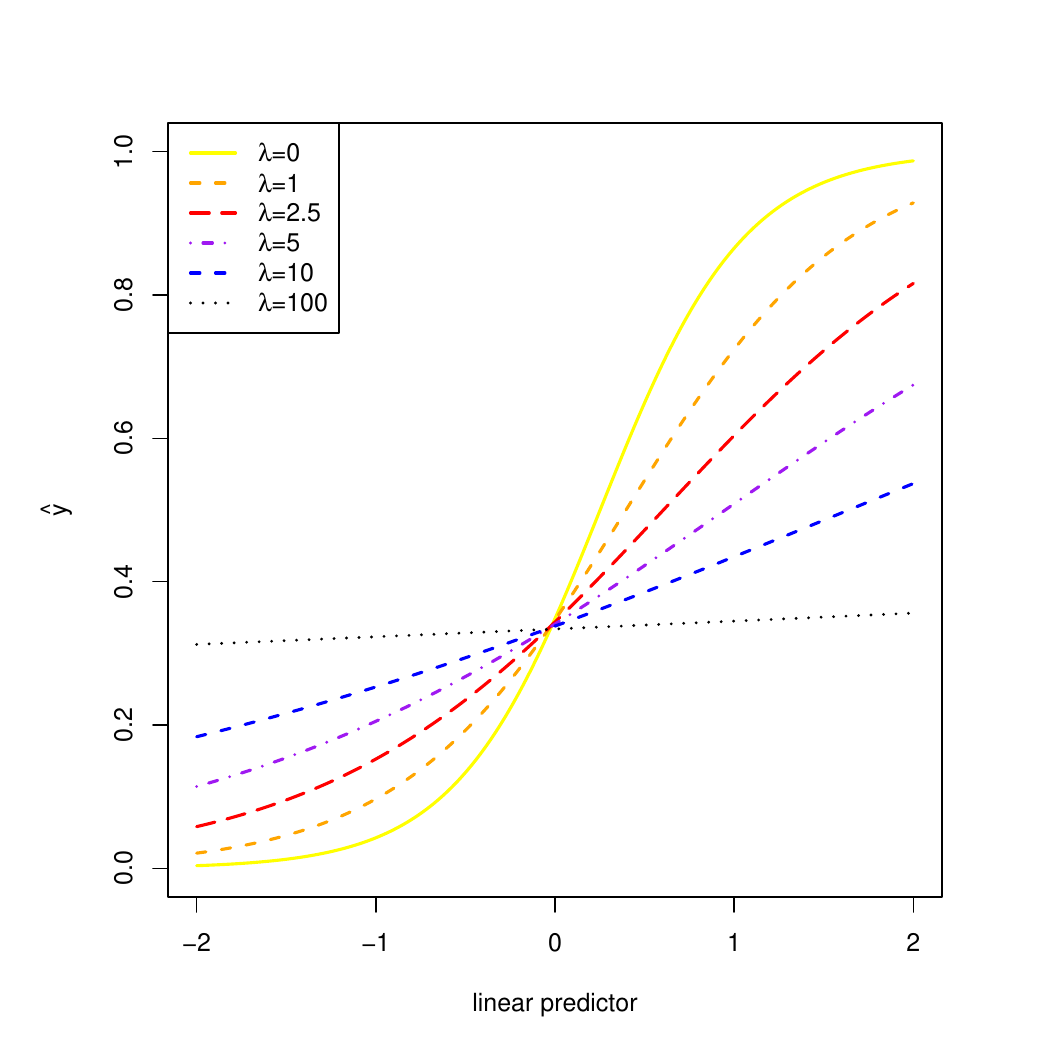}
\end{tabular}
\caption{Top row, left panel: contour plot of the penalized log-likelihood of a logistic regression model with the ridge constraint (red line). Top row, right panel: the regularization paths of the ridge estimator of the logistic regression parameter. Bottom row, left panel: variance of the ridge estimator of the logistic regression parameter against the logarithm of the penalty parameter. Bottom panel, right panel: the predicted success probability versus the linear predictor for various choices of the penalty parameter. } \label{fig.logisticRidge_effectOfPenalization}
\end{figure}


The optimization of the ridge penalized loglikelihood associated with the logistic regression model proceeds, due to the differentiability of the penalty, fully analogous to the unpenalized case and uses the Newton-Raphson algorithm for solving the (penalized) estimating equation. Hence, the unpenalized maximum likelihood estimation procedure is modified straightforwardly by replacing gradient and Hessian by their `penalized' counterparts:
\begin{eqnarray*}
\frac{\partial \mathcal{L}^{\mbox{{\tiny pen}}}}{\partial \bbeta } \, \, \,  = \, \, \, \frac{\partial \mathcal{L}}{\partial \bbeta } - \lambda \bbeta & \mbox{ and } & \frac{\partial^2 \mathcal{L}^{\mbox{{\tiny pen}}}}{\partial \bbeta \partial \bbeta^{\top}} \, \, \, = \, \, \, \frac{\partial^2 \mathcal{L}}{\partial \bbeta \partial \bbeta^{\top}} - \lambda \mathbf{I}_{pp}.
\end{eqnarray*}
With these at hand, the Newton-Raphson algorithm is (again) reformulated as an iteratively re-weighted least squares algorithm with the updating step changing accordingly to:
\begin{eqnarray*}
\hat{\bbeta}^{\mbox{{\scriptsize new}}} (\lambda) & =  & \hat{\bbeta}^{\mbox{{\scriptsize old}}}(\lambda) + \mathbf{V}^{-1}  \big(   \mathbf{X}^{\top} \{ \mathbf{Y} - \vec{\mathbf{g}}^{-1}[ \mathbf{X}; \bbeta^{\mbox{{\scriptsize old}}} (\lambda) ] \} - \lambda \bbeta^{\mbox{{\scriptsize old}}} (\lambda) \big)
\\
& = &  \mathbf{V}^{-1} \mathbf{V} \hat{\bbeta}^{\mbox{{\scriptsize old}}} -  \lambda \mathbf{V}^{-1}  \hat{\bbeta}^{\mbox{{\scriptsize old}}} (\lambda) + \mathbf{V}^{-1} \mathbf{X}^{\top} \mathbf{W}   \mathbf{W}^{-1} \{\mathbf{Y} - \vec{\mathbf{g}}^{-1}[ \mathbf{X}; \bbeta^{\mbox{{\scriptsize old}}}(\lambda)]\}
\\
& = & \mathbf{V}^{-1} \mathbf{X}^{\top} \mathbf{W}  \big( \mathbf{X} \hat{\bbeta}^{\mbox{{\scriptsize old}}} (\lambda) +  \mathbf{W}^{-1} \{ \mathbf{Y} - \vec{\mathbf{g}}^{-1}[ \mathbf{X}; \bbeta^{\mbox{{\scriptsize old}}}(\lambda)] \} \big)
\\
& = & (\mathbf{X}^{\top} \mathbf{W}  \mathbf{X} +  \lambda \mathbf{I}_{pp} )^{-1} \mathbf{X}^{\top} \mathbf{W}  \mathbf{Z},
\end{eqnarray*}
where $\mathbf{V} = \mathbf{X}^{\top} \mathbf{W}  \mathbf{X} + \lambda \mathbf{I}_{pp}$ and $\mathbf{W}$ and $\mathbf{Z}$ as before. Hence, use this to update the estimate of $\bbeta$ until convergence, which yields the desired ridge penalized estimate.

Obviously, the ridge estimate of the logistic regression parameter tends to zero as $\lambda \rightarrow \infty$. Now consider a linear predictor with an intercept that is left unpenalized. When $\lambda$ tends to infinity, all regression coefficients but the intercept vanish. The intercept is left to model the success probability. Hence, in this case $\lim_{\lambda \rightarrow \infty} \hat{\beta}_0 (\lambda) = \log [ \tfrac{1}{n} \sum_{i=1}^n Y_i / \tfrac{1}{n} \sum_{i=1}^n (1-Y_i)]$.

\begin{figure}[!h]
\begin{tabular}{rcl}
\includegraphics[scale=0.40, angle=0]{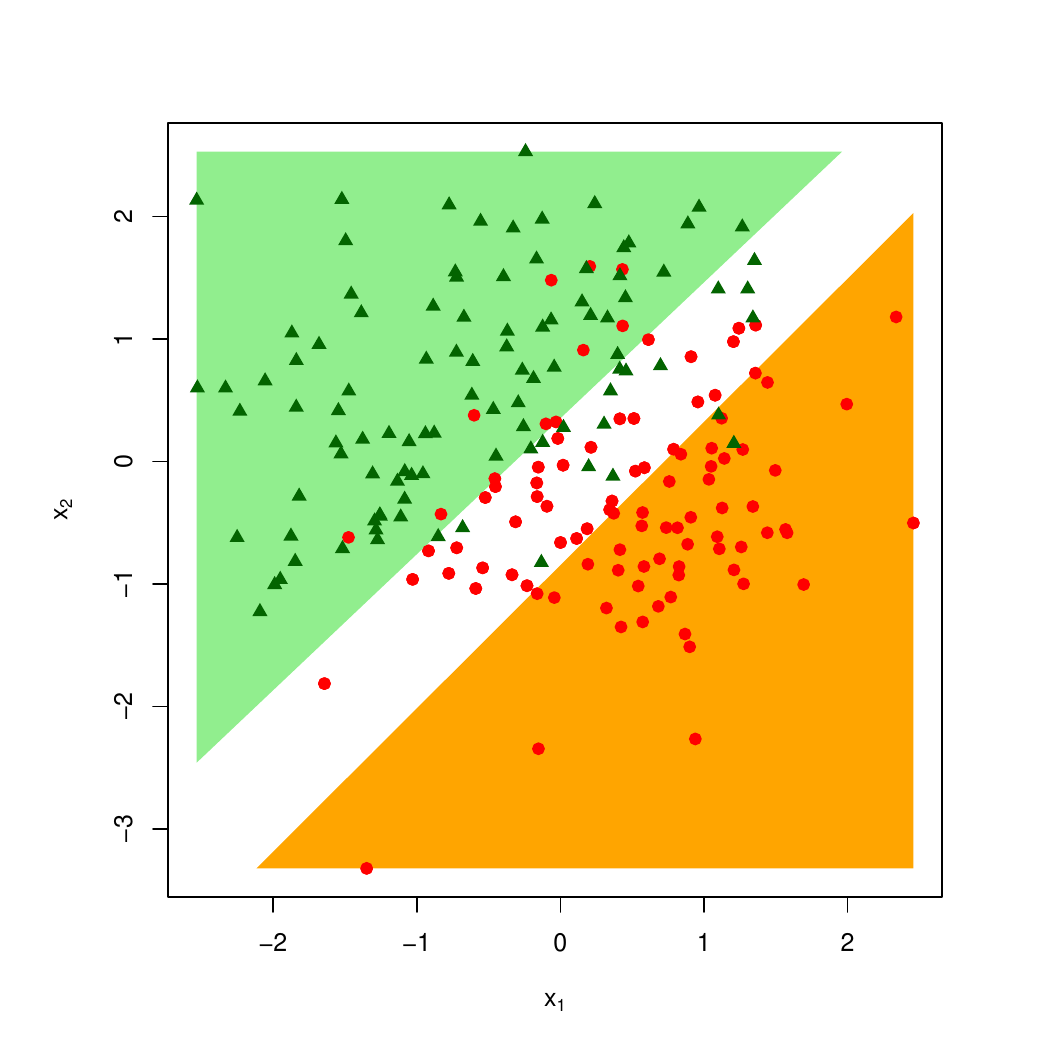}
&
\includegraphics[scale=0.40, angle=0]{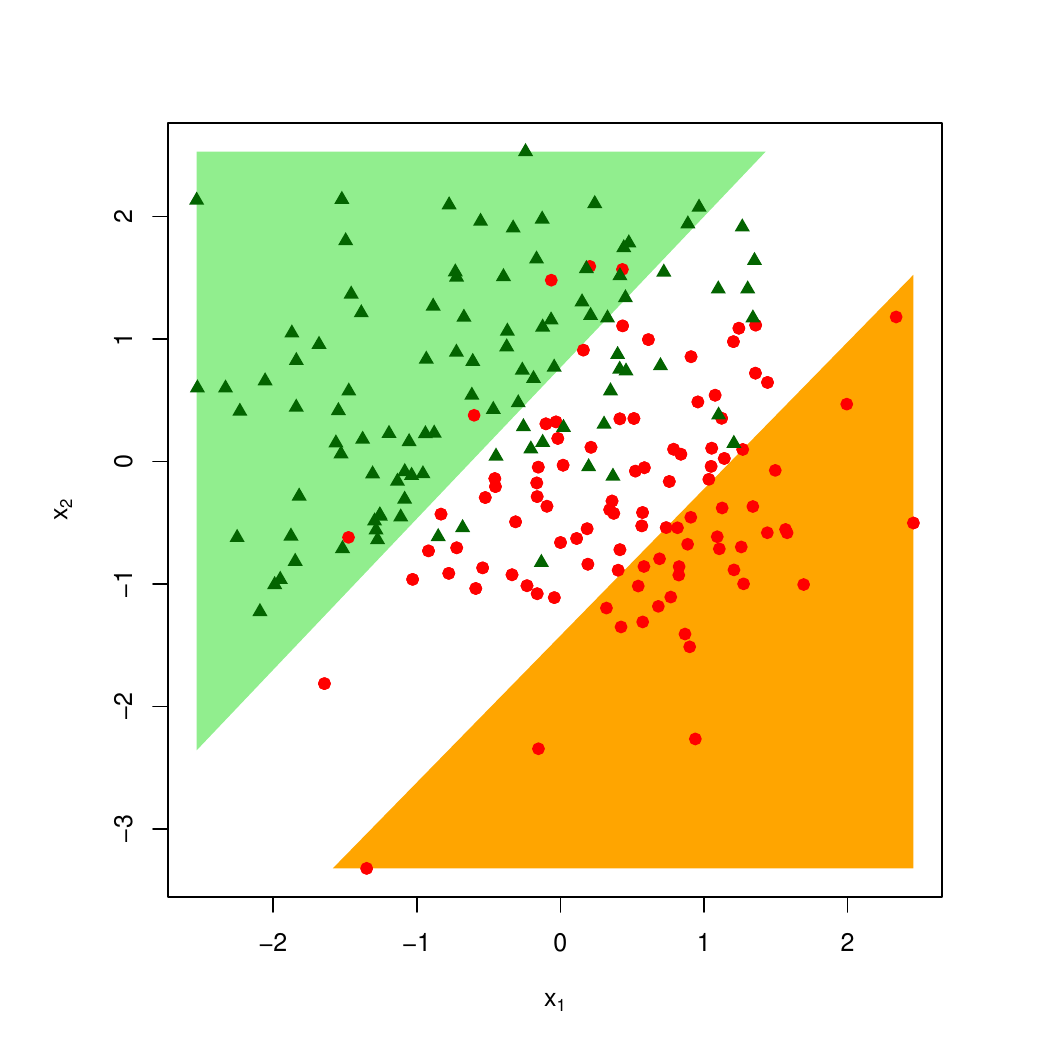}
\\
\includegraphics[scale=0.40, angle=0]{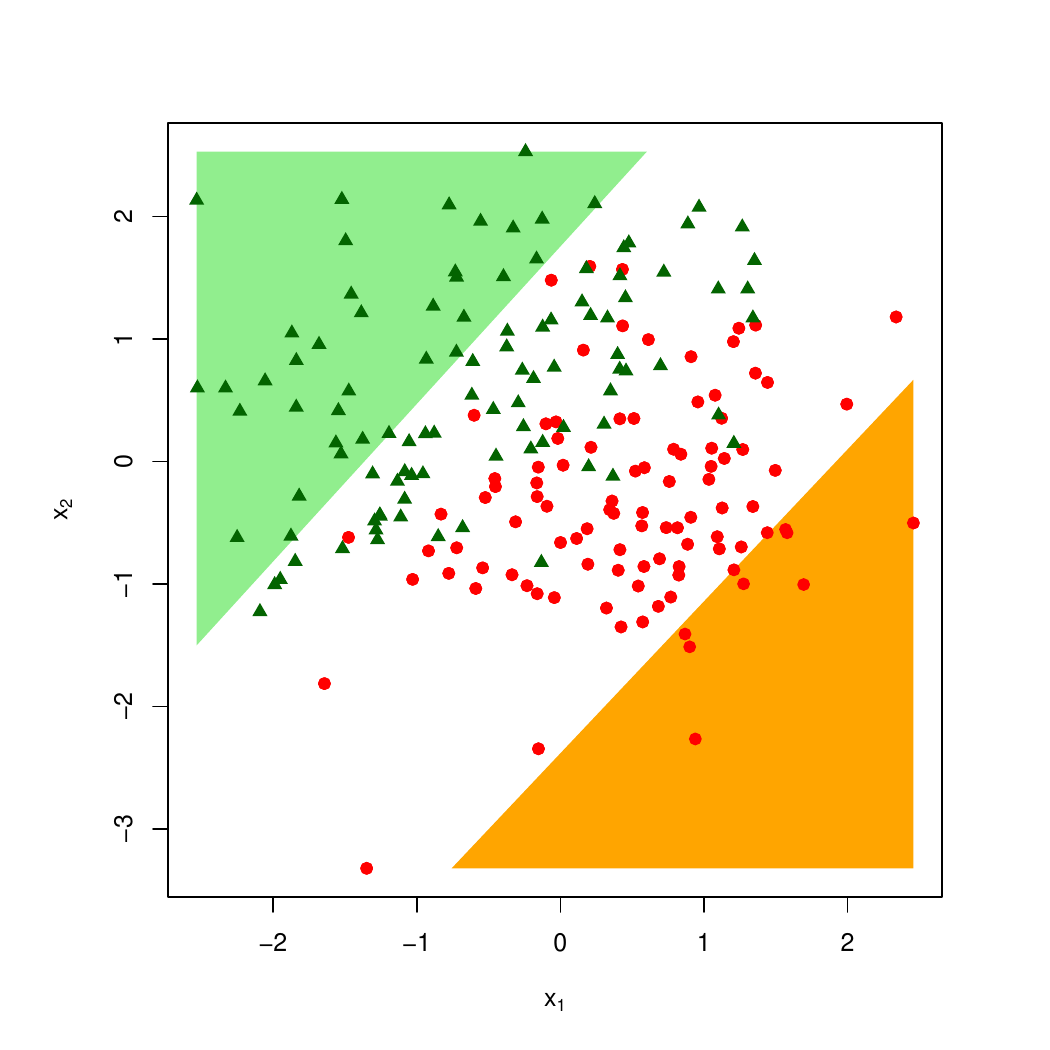}
&
\includegraphics[scale=0.40, angle=0]{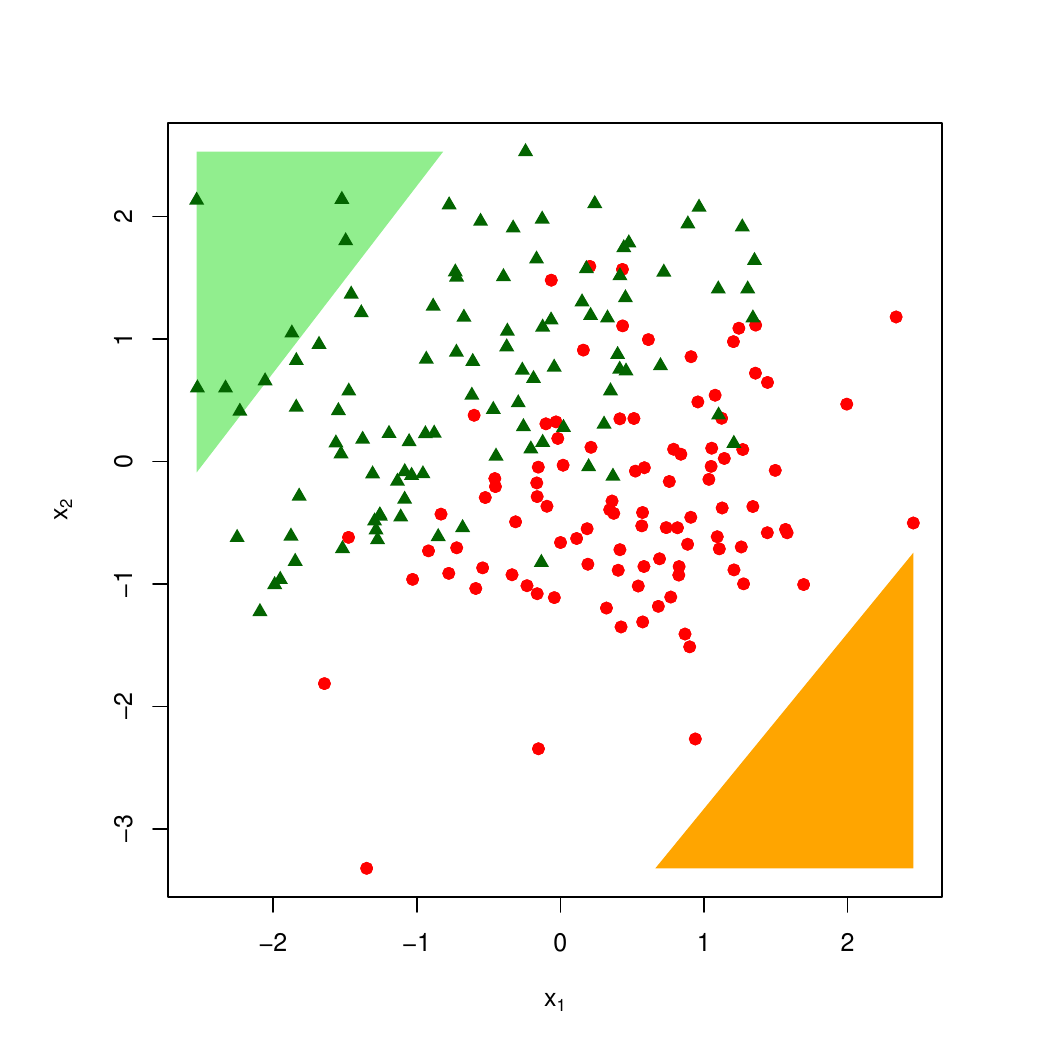}
\end{tabular}
\caption{The realized design as scatter plot ($X_1$ vs $X_2$ overlayed by the success (\textcolor{red}{red}) and failure regions (\textcolor{green}{green}) for various choices of the penalty parameter: $\lambda = 0$ (top row, left panel), $\lambda = 10$ (top row, right panel) $\lambda = 40$ (bottom row, left panel), $\lambda = 100$ (bottom row, right panel).} \label{fig.logisticRidge_effectOnPrediction}
\end{figure}

The effect of the ridge penalty on parameter estimates propagates to the predictor $\hat{p}_i$. The linear predictor of the linear regression model involving the ridge estimator $\mathbf{X}_{i, \ast} \hat{\bbeta}(\lambda)$ shrinks towards a common value for each $i$, leading to a scale difference between observation and predictor (as seen before in Section \ref{sect.ridgeRegressionDataIllustration}). This behaviour transfers to the ridge logistic regression predictor, as is illustrated in the next example.

\begin{example} \mbox{ }
\\
We simulate data to show the effect of regularization on the prediction. The dimension and sample size of these data are $p=2$ and $n=200$, respectively. The covariate data are drawn from the standard normal, while that of the response is sampled from a Bernoulli distribution with success probability $P(Y_i=1) = \exp(2 X_{i,1} - 2 X_{i,2}) / [ 1 + \exp(2 X_{i,1} - 2 X_{i,2})]$. The logistic regression model is estimated from these data by means of ridge penalized likelihood maximization with various choices of the penalty parameter. The bottom right plot in Figure \ref{fig.logisticRidge_effectOfPenalization} shows the predicted success probability versus the linear predictor for various choices of the penalty parameter. Larger values of the penalty parameter $\lambda$ flatten the slope of this curve. Consequently, for larger $\lambda$ more excessive values of the covariates are needed to achieve the same predicted success probability as those obtained with smaller $\lambda$ at more moderate covariate values. The implications for the resulting classification may become clearer when studying the effect of the penalty parameter on the `failure' and `success regions' respectively defined by:
\\
\indent $\{(x_1, x_2) : P({\color{green}{Y=0}} \, | \, X_1=x_1, X_2=x_2, \hat{\bbeta}(\lambda)) > 0.75 \}$,
\\
\indent $\{(x_1, x_2) : P({\color{red}{Y=1}} \, | \, X_1=x_1, X_2=x_2, \hat{\bbeta}(\lambda)) > 0.75 \}$.
\\
This separates the design space in a light red (`failure') and light green (`success') domain. The white bar between them is the domain where samples cannot be classified with high enough certainty. As $\lambda$ grows, so does the white area that separates the failure and success regions. Hence, as stronger penalization shrinks the logistic regression parameter estimate towards zero, it produces a predictor that is  less outspoken in its class assignments.
\end{example}

\begin{remark} \mbox{ } \label{remark:approximatedRidgeLogisticEstimator} \\
Historically, confer \cite{Scha1984}, the ridge logistic regression estimator is defined, analogously to the ridge regression estimator, as an ad-hoc fix to the collinearity among the covariates. The collinearity now causes $\mathbf{X}^{\top} \mathbf{W} \mathbf{X}$ to be ill-conditionedness, which is circumvented by the addition of the term $\lambda \mathbf{I}_{pp}$. This results in the following alternative definition of the ridge logistic regression estimator:
\begin{eqnarray*}
\hat{\bbeta}_a (\lambda) & = & ( \mathbf{X}^{\top} \mathbf{W}_{\mbox{\hspace{-0.1cm}{\tiny ml}}} \mathbf{X} + \lambda \mathbf{I}_{pp})^{-1} \mathbf{X}^{\top} \mathbf{W}_{\mbox{\hspace{-0.1cm}{\tiny ml}}} \mathbf{X} \hat{\bbeta}^{\mbox{{\tiny ml}}},
\end{eqnarray*}
where $\mathbf{W}_{\mbox{\hspace{-0.1cm}{\tiny ml}}}$ is defined as in the Iteratively Weighted Least Squares algorithm but with $\hat{\bbeta}^{\mbox{{\tiny ml}}}$ substituted for the previous update $\hat{\bbeta}^{\mbox{{\tiny old}}}(\lambda)$. This alternative definition assumes the availability of the maximum likelihood estimator and is thus not applicable to separable and, in particular, high-dimensional data. 

An alternative motivation of this approximate ridge logistic regression estimator follows from the following asymptotic argument. Hereto we assume that for large enough $n$ the maximum likelihood estimator of the logistic regression parameter $\hat{\bbeta}^{\mbox{{\tiny ml}}}$ exists. We can then develop a second order Taylor approximation of the penalized loglikelihood's summand $\log[ 1+ \exp( \mathbf{X} \bbeta) ]$ in $\bbeta$ around the point $\bbeta = \hat{\bbeta}^{\mbox{{\tiny ml}}}$:
\begin{eqnarray*}
\log[ 1+ \exp( \mathbf{X} \bbeta) ] & \approx & \log[ 1+ \exp( \mathbf{X} \hat{\bbeta}^{\mbox{{\tiny ml}}}) ] + [ \vec{\mathbf{g}} (\mathbf{X}; \hat{\bbeta}^{\mbox{{\tiny ml}}} )]^{\top} \mathbf{X} ( \bbeta - \hat{\bbeta}^{\mbox{{\tiny ml}}}) + \tfrac{1}{2} ( \bbeta - \hat{\bbeta}^{\mbox{{\tiny ml}}})^{\top} \mathbf{X}^{\top} \mathbf{W}_{\mbox{\hspace{-0.1cm}{\tiny ml}}} \mathbf{X} ( \bbeta - \hat{\bbeta}^{\mbox{{\tiny ml}}}),
\end{eqnarray*}
where $\mathbf{W}_{\mbox{\hspace{-0.1cm}{\tiny ml}}}$ defined as above. We substitute this Taylor approximation in the penalized loglikelihood to obtain an approximation of the latter:
\begin{eqnarray*}
\mathcal{L}^{\mbox{{\tiny pen}}}(\mathbf{Y}, \mathbf{X}; \bbeta, \lambda) & \approx & [\mathbf{Y} -  \vec{\mathbf{g}} (\mathbf{X}; \hat{\bbeta}^{\mbox{{\tiny ml}}} )]^{\top} \mathbf{X}  \bbeta - \tfrac{1}{2} \bbeta ^{\top} ( \mathbf{X}^{\top} \mathbf{W}_{\mbox{\hspace{-0.1cm}{\tiny ml}}} \mathbf{X} + \lambda \mathbf{I}_{pp}) \bbeta + \bbeta^{\top} \mathbf{X}^{\top} \mathbf{W}_{\mbox{\hspace{-0.1cm}{\tiny ml}}} \mathbf{X}  \hat{\bbeta}^{\mbox{{\tiny ml}}} + r_n,
\end{eqnarray*}
where the remainder term $r_n$ contains all remaining terms not involving $\bbeta$. The right-hand side is maximized at 
\begin{eqnarray*} 
( \mathbf{X}^{\top} \mathbf{W}_{\mbox{\hspace{-0.1cm}{\tiny ml}}} \mathbf{X} + \lambda \mathbf{I}_{pp})^{-1} \mathbf{X}^{\top} \mathbf{W}_{\mbox{\hspace{-0.1cm}{\tiny ml}}} \{  \mathbf{X}  \hat{\bbeta}^{\mbox{{\tiny ml}}} + \mathbf{W}_{\mbox{\hspace{-0.1cm}{\tiny ml}}}^{-1} [\mathbf{Y} -  \vec{\mathbf{g}} (\mathbf{X}; \hat{\bbeta}^{\mbox{{\tiny ml}}} )  ] \},
\end{eqnarray*}
which equals $\hat{\bbeta}_a(\lambda)$ for  $\mathbf{X}^{\top}   [\mathbf{Y} -  \vec{\mathbf{g}} (\mathbf{X}; \hat{\bbeta}^{\mbox{{\tiny ml}}} )  ] = \mathbf{0}_p$. 
\end{remark}

\section{Moments}
The $1^{\mbox{{\tiny st}}}$ and $2^{\mbox{{\tiny nd}}}$ order moments of the ridge maximum likelihood estimator of the logistic regression parameter is unknown analytically but typically approximated by that of the final update of the Newton-Raphson algorithm. This approximation assumes the one-to-last update $\hat{\bbeta}^{\mbox{{\scriptsize old}}}$ to be non-random and then proceeds as before with the regular ridge regression estimator of the linear regression model parameter to arrive at:
\begin{eqnarray*}
\mathbb{E} [ \hat{\bbeta}^{\mbox{{\scriptsize new}}} (\lambda) \, | \, \lambda, \mathbf{W}] & = & (\mathbf{X}^{\top} \mathbf{W}  \mathbf{X} + \lambda \mathbf{I}_{pp} )^{-1} \mathbf{X}^{\top} \mathbf{W} \mathbb{E}( \mathbf{Z}),
\\
\mbox{Var} [ \hat{\bbeta}^{\mbox{{\scriptsize new}}} (\lambda)\, | \, \lambda, \mathbf{W}] &  =  &  (\mathbf{X}^{\top} \mathbf{W}  \mathbf{X} + \lambda \mathbf{I}_{pp} )^{-1} \mathbf{X}^{\top} \mathbf{W} \big[  \mbox{Var} ( \mathbf{Z} ) \big] \mathbf{W}  \mathbf{X} ( \mathbf{X}^{\top} \mathbf{W}  \mathbf{X} + \lambda \mathbf{I}_{pp} )^{-1},
\end{eqnarray*}
with
\begin{eqnarray*}
\mathbb{E}(\mathbf{Z} \, | \, \lambda, \mathbf{W}) &  = & \{ \mathbf{X} \hat{\bbeta}^{\mbox{{\scriptsize old}}} +  \mathbf{W}^{-1} [\mathbb{E}(\mathbf{Y}) - \vec{\mathbf{g}}^{-1}( \mathbf{X}; \bbeta^{\mbox{{\scriptsize old}}})] \},
\\
\mbox{Var}(\mathbf{Z} \, | \, \lambda, \mathbf{W}) & = & \mathbf{W}^{-1} \mbox{Var}(\mathbf{Y}) \mathbf{W}^{-1} = \mathbf{W}^{-1},
\end{eqnarray*}
where the identity $\mbox{Var}(\mathbf{Y} \, | \, \lambda, \mathbf{W}) = \mathbf{W}$ follows from the variance of a binomial distributed random variable. From these expressions similar properties as for the ridge maximum likelihood estimator of the regression parameter of the linear model may be deduced. For instance, the ridge maximum likelihood estimator of the logistic regression parameter converges to zero as the penalty parameter tends to infinity (confer the top right panel of Figure \ref{fig.logisticRidge_effectOfPenalization}). Similarly, their variances vanish as $\lambda \rightarrow \infty$ (illustrated in the bottom left panel of Figure \ref{fig.logisticRidge_effectOfPenalization}).

The form of these moment approximation corroborates asymptotically with that of the approximated ridge logistic regression estimator $\hat{\bbeta}_a(\lambda)$ introduced in Remark \ref{remark:approximatedRidgeLogisticEstimator}.

\section{Constrained estimation}
The maximization of the penalized loglikelihood of the logistic regression model can be reformulated as a constrained estimation problem, as was done in Section \ref{sect.constrainedEstimation} for the linear regression model. The ridge logistic regression estimator is thus defined equivalently as:
\begin{eqnarray*}
\hat{\bbeta}(\lambda) & = & \arg \max\nolimits_{\{\bbeta \in \mathbb{R}^p \, : \, \| \bbeta \|_2^2 \leq c(\lambda) \} } \, \mathcal{L} ( \mathbf{Y}, \mathbf{X}; \bbeta).
\end{eqnarray*}
This is illustrated by the top left panel of Figure \ref{fig.logisticRidge_effectOfPenalization} for the `$p=2$'-case. It depicts the contours (black lines) of the loglikelihood and the spherical domain of the parameter constraint $\{ \bbeta \in \mathbb{R}^2 \, : \, \beta_1^2 + \beta_2^2 \leq c(\lambda) \}$ (red line). The parameter constraint can again be interpreted as a means to harness against overfitting, as it prevents the ridge logistic regression estimator from assuming very large values, thereby avoiding a perfect description of the data.

The parameter constraint of the logistic regression estimator exhibits behavior in $\lambda$ as that of the regular ridge estimator. Hereto we note that the squared radius of the parameter constraint is -- by identical argumentation as provided in Section \ref{sect.constrainedEstimation} -- equal to $c(\lambda) = \| \hat{\bbeta} (\lambda) \|_2^2$. However, no explicit expression for this squared radius exists, as none exists for the logistic ridge regression estimator, to verify the constraint's shrinkage behavior. Proposition \ref{prop:logisticConstraintBehavior} characterizes the essential properties of this behavior.

\begin{proposition} \label{prop:logisticConstraintBehavior} \mbox{ } \\
The squared norm of the ridge logistic regression estimator satisfies:
\begin{compactitem}
\item[\textit{i)}]
$d \| \hat{\bbeta} ( \lambda) \|_2^2 / d\lambda < 0$ for $\lambda > 0$,

\item[\textit{ii)}] $\lim_{\lambda \rightarrow \infty} \| \hat{\bbeta} ( \lambda) \|_2^2 = 0$.
\end{compactitem}
\end{proposition}

\begin{proof} For part \textit{i)} take the derivative of the estimating equation with respect to $\lambda$, solve for $\tfrac{d}{d\lambda} \hat{\bbeta} (\lambda)$, and find that
\begin{eqnarray*}
\frac {d}{d \, \lambda} \hat{\bbeta} (\lambda) & = &   - (\mathbf{X}^{\top} \mathbf{W} \mathbf{X} + \lambda \mathbf{I}_{pp})^{-1} \hat{\bbeta}(\lambda).
\end{eqnarray*}
Using this derivative and the chain rule, we obtain the derivative of the estimator's (squared) Euclidean length:
\begin{eqnarray*}
\frac{d}{d\lambda} \| \hat{\bbeta} (\lambda) \|_2^2 & = & \frac{d}{d\lambda} \{ [ \hat{\bbeta} (\lambda) ]^{\top}  \hat{\bbeta} (\lambda) \}
\, \, \, = \, \, \, 2 [ \hat{\bbeta} (\lambda) ]^{\top} \frac{d}{d\lambda}  \hat{\bbeta} (\lambda)  \, \, \, = \, \, \, - 2 [ \hat{\bbeta} (\lambda) ]^{\top}  (\mathbf{X}^{\top} \mathbf{W} \mathbf{X} + \lambda \mathbf{I}_{pp})^{-1} \hat{\bbeta}(\lambda).
\end{eqnarray*}
As the right-hand side is a quadratic form multiplied by a negative scalar, we conclude that $\tfrac{d}{d\lambda} \| \hat{\bbeta} (\lambda) \|_2^2 < 0$ for all $\lambda > 0$. 

To verify the claimed $\lambda$-limit of the squared length of the ridge
logistic regression estimator, note that it satisfies: $\hat{\bbeta}(\lambda)  = \lambda^{-1} \mathbf{X}^{\top} \{ \mathbf{Y} - \vec{\mathbf{g}}^{-1} [\mathbf{X}; \hat{\bbeta}(\lambda)]\}$, which can be derived from the estimating equation. But as all elements of the vector $\vec{\mathbf{g}}^{-1} [\mathbf{X}; \hat{\bbeta}(\lambda)]$ are in the interval $(0,1)$ and $Y_i \in \{ 0, 1 \}$, we obtain the inequality: 
\begin{eqnarray*}
\| \hat{\bbeta}(\lambda) \|_2^2 & = & \lambda^{-2} \{ \mathbf{Y} - \vec{\mathbf{g}}^{-1} [\mathbf{X}; \hat{\bbeta}(\lambda)]\}^{\top} \mathbf{X} \mathbf{X}^{\top}  \{ \mathbf{Y} - \vec{\mathbf{g}}^{-1} [\mathbf{X}; \hat{\bbeta}(\lambda)]\} \, \, \, \leq \, \, \, \lambda^{-2} \mathbf{1}_n^{\top}  \mathbf{X} \mathbf{X}^{\top}  \mathbf{1}_n.
\end{eqnarray*}
This bound can be sharpened by using the worst possible fit, where either $\lim_{\lambda \rightarrow \infty} \vec{\mathbf{g}}^{-1} [\mathbf{X}; \hat{\bbeta}(\lambda)] = \frac{1}{2}$ or $\lim_{\lambda \rightarrow \infty} \vec{\mathbf{g}}^{-1} [\mathbf{X}; \hat{\bbeta}(\lambda)] = \tfrac{1}{n} \sum_{i=1}^n Y_i$, depending on the presence of an unpenalized intercept in the model. Irrespectively, the derived bound vanishes as $\lambda \rightarrow \infty$. 
\end{proof}

\noindent
Part \textit{i)} of Proposition \ref{prop:logisticConstraintBehavior} tells us that the shrinkage behaviour of the ridge logistic regression estimator is monotone in $\lambda$. Monotone in the sense that the squared Euclidean length of the estimator is, as $\| \hat{\bbeta} (\lambda) \|_2^2  \geq 0$ for all $\lambda > 0$, is strictly decreasing in $\lambda$. Furthermore, part \textit{ii)} of Proposition \ref{prop:logisticConstraintBehavior} then ensures that ultimately the constraint collapses onto zero, and so does the estimator.

\section{Degrees of freedom}
The degrees of freedom consumed by the ridge logistic regression 
estimator can be derived from the definition (\ref{form.DOFdef}) previously introduced in Section \ref{sect.ridgeDOF}. It uses $\hat{\bbeta}(\lambda)$ as obtained from the last iteration of the IRLS algorithm, and assumes the weight matrix employed in this last iteration to be nonrandom. Then, but see also \cite{park2008penalized}, we can formulate the following lemma.
\begin{proposition} \label{prop:dofOfRidgeLogistic} \mbox{ } \\
The degrees of freedom, denoted $\mbox{df}(\lambda)$, of the ridge logistic regression estimator are, for known $\mathbf{W}$, approximately  $\mbox{tr} [ ( \mathbf{X}^{\top} \mathbf{W} \mathbf{X}^{\top} + \lambda \mathbf{I}_{pp})^{-1} \mathbf{X}^{\top} \mathbf{W} \mathbf{X}]$.
\end{proposition} 
\begin{proof}
We use the $1^{\mbox{{\tiny st}}}$ order Taylor approximation to the covariance of functions of a random variable, \\$\mbox{Cov}[ h_1(Y), h_2(Y)] \approx h_1'[\mathbb{E}(Y)] h_2'[\mathbb{E}(Y)] \mbox{Cov}(Y, Y)$, to obtain
\begin{eqnarray*}
\mbox{Cov} ( \hat{Y}_i, Y_i) & = & \mbox{Cov} \big( \exp[\mathbf{X}_{i,\ast} \hat{\bbeta}(\lambda)] \{ \exp[\mathbf{X}_{i,\ast} \hat{\bbeta}(\lambda)]\}^{-1}, Y_i \big)
\\
& \approx &  \left. \exp( \mu) [ 1+ \exp(\mu) ]^{-2} \right|_{\mu  = \mathbb{E}[ \mathbf{X}_{i,\ast} \hat{\bbeta}(\lambda)]} 
\mbox{Cov} [  \mathbf{X}_{i,\ast} \hat{\bbeta}(\lambda), Y_i \big]
\\
& \approx &  \exp[\mathbf{X}_{i,\ast} \hat{\bbeta}(\lambda)] \{1+ \exp[\mathbf{X}_{i,\ast} \hat{\bbeta}(\lambda)]\}^{-2} 
\, \mbox{Cov} [  \mathbf{X}_{i,\ast} \hat{\bbeta}(\lambda), Y_i \big]
\\
& = &  (\mathbf{W})_{ii}  \, \mbox{Cov} [  \mathbf{X}_{i,\ast} \hat{\bbeta}(\lambda), Y_i \big].
\end{eqnarray*}
In this, and for known $\mathbf{W}$, 
\begin{eqnarray*}
\mbox{Cov}[ \mathbf{X}_{i,\ast} \hat{\bbeta}(\lambda), Y_i ] & = & \mbox{Cov}[ \mathbf{X}_{i,\ast} (\mathbf{X}^{\top} \mathbf{W} \mathbf{X} + \lambda \mathbf{I}_{pp})^{-1} \mathbf{X}^{\top} \mathbf{W} \mathbf{Z}, Y_i]
\\
& = &   \mathbf{X}_{i,\ast} (\mathbf{X}^{\top} \mathbf{W} \mathbf{X} + \lambda \mathbf{I}_{pp})^{-1} \mathbf{X}^{\top} \mathbf{W} \, \mbox{Cov}( \mathbf{Z}, Y_i )
\\
& = & \mathbf{X}_{i,\ast} (\mathbf{X}^{\top} \mathbf{W} \mathbf{X} + \lambda \mathbf{I}_{pp})^{-1} \mathbf{X}^{\top} \mathbf{W} 
\\
& & \qquad  \quad \times \mbox{Cov}\{  \mathbf{X} \hat{\bbeta}^{\mbox{{\scriptsize old}}} +  \mathbf{W}^{-1} [\mathbf{Y} - \vec{\mathbf{g}}^{-1}( \mathbf{X}; \bbeta^{\mbox{{\scriptsize old}}})] , Y_i \}
\\
& = & \mathbf{X}_{i,\ast} (\mathbf{X}^{\top} \mathbf{W} \mathbf{X} + \lambda \mathbf{I}_{pp})^{-1} \mathbf{X}^{\top} \mathbf{W} \, \mbox{Cov}(   \mathbf{W}^{-1} \mathbf{Y} , Y_i )
\\
& = &   \mathbf{X}_{i,\ast} (\mathbf{X}^{\top} \mathbf{W} \mathbf{X} + \lambda \mathbf{I}_{pp})^{-1} \mathbf{X}^{\top} \mbox{Cov}(  \mathbf{Y} , Y_i)
\\
& = & \mathbf{X}_{i, \ast} ( \mathbf{X}^{\top} \mathbf{W}  \mathbf{X}^{\top} + \lambda \mathbf{I}_{pp})^{-1} \mathbf{X}^{\top} [ \mbox{Var}(Y_i) \, \mathbf{e}_i  ]
\\
& = & \mbox{Var}(Y_i) \, \mathbf{X}_{i, \ast} ( \mathbf{X}^{\top} \mathbf{W}  \mathbf{X}^{\top} + \lambda \mathbf{I}_{pp})^{-1} \mathbf{X}_{i, \ast}^{\top},
\end{eqnarray*}
in which we have used the independence among the individual observations. We now use the above together and find the claimed approximation of the ridge logistic regression estimator's degrees of freedom:
\begin{eqnarray*}
\mbox{df}(\lambda) & = &  \sum\nolimits_{i=1}^n [\mbox{Var}(Y_i)]^{-1} \mbox{Cov}(\widehat{Y}_i, Y_i)
\\
& \approx & \sum\nolimits_{i=1}^n [\mbox{Var}(Y_i)]^{-1} (\mathbf{W})_{ii} \mbox{Cov} [  \mathbf{X}_{i,\ast} \hat{\bbeta}(\lambda), Y_i \big]
\\
& = &  \sum\nolimits_{i=1}^n [\mbox{Var}(Y_i)]^{-1} (\mathbf{W})_{ii} \mbox{Var}(Y_i)  \mathbf{X}_{i, \ast} ( \mathbf{X}^{\top} \mathbf{W} \mathbf{X}^{\top} + \lambda \mathbf{I}_{pp})^{-1} \mathbf{X}_{i, \ast}^{\top}
\\
& = & \mbox{tr} [ ( \mathbf{X}^{\top} \mathbf{W}  \mathbf{X}^{\top} + \lambda \mathbf{I}_{pp})^{-1} \mathbf{X}^{\top}  \mathbf{W} \mathbf{X}].
\end{eqnarray*}
\end{proof}
From the approximation provided by Proposition \ref{prop:dofOfRidgeLogistic}, it is clear that the degrees of freedom of the ridge logistic regression estimator too decrease monotonically to zero as $\lambda$ increases, as 
\begin{eqnarray*}
\tfrac{d}{d\lambda} \mbox{df}(\lambda) & = &  - \mbox{tr} [ ( \mathbf{X}^{\top} \mathbf{W} \mathbf{X}^{\top} + \lambda \mathbf{I}_{pp})^{-1} \mathbf{X}^{\top} \mathbf{W} \mathbf{X}  ( \mathbf{X}^{\top} \mathbf{W} \mathbf{X}^{\top} + \lambda \mathbf{I}_{pp})^{-1}]
\end{eqnarray*}
with $\mathbf{W}$ still known.

\section{The Bayesian connection} \label{sect.BayesianLogistic}
All penalized estimators can be formulated as Bayesian estimators, including the ridge logistic estimator. In particular, ridge estimators correspond to Bayesian estimators with a multivariate normal prior on the regression coefficients. Thus, assume $\bbeta \sim \mathcal{N}(\mathbf{0}_p, \mathbf{\Delta}^{-1})$. The posterior distribution of $\bbeta$ then is:
\begin{eqnarray} \label{form:RidgeLogisticPosterior}
f_{\bbeta}(\bbeta \, | \, \mathbf{Y}, \mathbf{X}) & \propto & \Big\{ \prod_{i=1}^n \big[ P(Y_i = 1 \, | \, \mathbf{X}_{i,\ast}) \big]^{Y_i} \big[ P(Y_i = 0 \, | \, \mathbf{X}_{i,\ast}) \big]^{1-Y_i}  \Big\} \exp( - \tfrac{1}{2} \bbeta^{\top} \mathbf{\Delta} \bbeta).
\end{eqnarray}
This does not coincide with any standard distribution. But, under appropriate conditions, the posterior distribution is asymptotically normal. This invites a (multivariate) normal approximation to the posterior distribution above. The Laplace's method provides (cf. \citealp{Bish2006}).

\begin{figure}[!h]
\centering
\includegraphics[scale=0.40, angle=0]{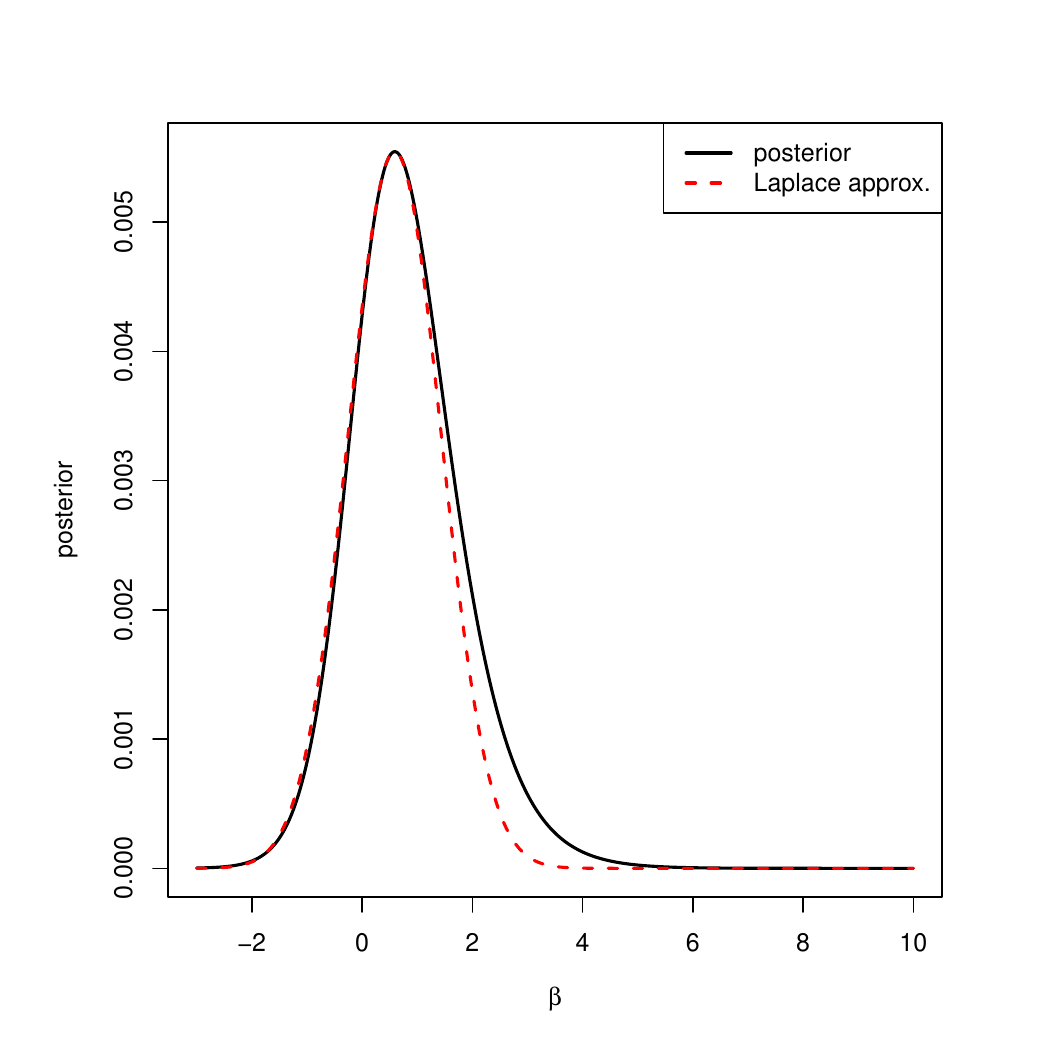}
\caption{
Laplace approximation to the posterior density of the Bayesian logistic regression parameter.} \label{fig.logisticRidge_MSEandLaplace2posterior}
\end{figure}

Laplace's method \textit{i)} centers the normal approximation at the mode of the posterior, and \textit{ii)} chooses the covariance to match the curvature of the posterior at the mode. The posterior mode is the location of the maximum of the posterior distribution. The location of this maximum coincides with that of the logarithm of the posterior. The latter is the log-likelihood augmented with a ridge penalty. Hence, the posterior mode, which is taken as the mean of the approximating Gaussian, coincides with the ridge logistic regression estimator. For the covariance of the approximating Gaussian, the logarithm of the posterior is approximated by a second order Taylor series around the posterior mode and limited to second order terms:
\begin{eqnarray*}
\log[f_{\bbeta}(\bbeta \, | \, \mathbf{Y}, \mathbf{X})] & \propto &
 \log[f_{\bbeta}(\bbeta \, | \, \mathbf{Y}, \mathbf{X})]
\big|_{\bbeta = \hat{\bbeta}_{\mbox{{\tiny MAP}}}}
\\
&  &  + \tfrac{1}{2} (\bbeta - \hat{\bbeta}_{\mbox{{\tiny MAP}}})^{\top} \left. \frac{\partial^2}{\partial \bbeta \partial \bbeta^{\top}}
\log[f_{\bbeta}(\bbeta \, | \, \mathbf{Y}, \mathbf{X})]
 \right|_{\bbeta = \hat{\bbeta}_{\mbox{{\tiny MAP}}}} (\bbeta - \hat{\bbeta}_{\mbox{{\tiny MAP}}})^{\top},
\end{eqnarray*}
in which the first order term cancels as the derivative of $f_{\bbeta}(\bbeta \, | \, \mathbf{Y}, \mathbf{X})$ with respect to $\bbeta$ vanishes at the posterior mode -- its maximum. Take the exponential of this approximation and match its arguments to that of a multivariate Gaussian $\exp[-\tfrac{1}{2} (\bbeta - \mmu_{\beta})^{\top} \mathbf{\Sigma}_{\bbeta}^{-1} (\bbeta - \mmu_{\beta})]$. The covariance of the sought Gaussian approximation is thus the inverse of the Hessian of the negative penalized log-likelihood. Put together the posterior is approximated by:
\begin{eqnarray} \label{form:ridgeLogisticPosteriorDistApprox}
\bbeta \,  | \, \mathbf{Y}, \mathbf{X} \sim \mathcal{N} [ \hat{\bbeta}_{\mbox{{\tiny MAP}}}, ( \mathbf{\Delta} + \mathbf{X}^{\top} \mathbf{W} \mathbf{X})^{-1} ].
\end{eqnarray}
The Gaussian approximation is convenient but need not be good. Fortunately, the Bernstein-Von Mises theorem \citep{VdVa2000} tells us that it is very accurate when the model is regular, the prior smooth, and the sample size sufficiently large. The quality of the approximation for an artificial example data set is shown in Figure \ref{fig.logisticRidge_MSEandLaplace2posterior}.
\\
\mbox{ }
\\
\cite{polson2013bayesian} present a Gibbs sampler to draw from the full posterior (\ref{form:RidgeLogisticPosterior}) of the Bayesian logistic regression parameter. We cannot sample from the approximate posterior as that involves the weight matrix $\mathbf{W}$, which depends on the to-be-drawn $\bbeta$. The Gibbs sampler of \cite{polson2013bayesian} thus alternates between the updating of the logistic regression parameter and its weights. \cite{polson2013bayesian} show that the distribution of $\bbeta$ conditional on data and weights is the form (\ref{form:ridgeLogisticPosteriorDistApprox}) and that of $\mathbf{W} \, | \, \bbeta, \mathbf{X}, \mathbf{Y}$ follows a P\'{o}lya-gamma distribution. A random variable $\omega$ is \textit{P\'{o}lya-gamma distributed} with parameters $b > 0$ and $c \in \mathbb{R}$, denoted by $\omega \sim \mathcal{PG}(b,c)$, if 
\begin{eqnarray*}
w & \sim & (2\pi^2)^{-1} \sum\nolimits_{k=1}^{\infty} [(k-\tfrac{1}{2})^2 + \tfrac{1}{4} c^2  \pi^{-2}]^{-1} \,  g_k,
\end{eqnarray*}
with independent and independently distributed $g_k \sim \mbox{Gamma}(b, 1)$.  Algorithm \ref{alg.gibbsSamplerLogisticRidgeRegression} then implements the Gibbs sampler for Bayesian logistic regression's posterior distribution (\ref{form:RidgeLogisticPosterior}) .
\\	
\mbox{ }
\\
\begin{minipage}[h]{\textwidth}
\begin{center}
\fbox{\parbox{14cm}{
\begin{algorithm}[H]
\SetKwInOut{Input}{input}
\SetKwInOut{Output}{output}
\RestyleAlgo{boxed}
\LinesNumbered
\SetKwFor{inpar}{for}{do in parallel}{\text{ end synchronize}}
\Input{sample size $T$;
\\
length of burn in period $T_{\mbox{{\tiny burn-in}}}$;
\\
thinning factor $f_{{\mbox{{\tiny thinning}}}}$;
\\
data $\{\mathbf{x}_i, \mathbf{y}_i \}_{i=1}^n$;
\\
prior mean $\bbeta_0$ and variance $\mathbf{\Delta}^{-1}$.
\\
}
\Output{draws from the posterior $\pi(\bbeta \, | \, \{\mathbf{x}_i, \mathbf{y}_i \}_{i=1}^n)$.}
\vspace{0.25cm}
{\bf initialize} {$\bbeta_1$ with the prior mean}.
\\
\vspace{0.25cm}
\For{ $t = 1$ to $T_{\mbox{{\tiny burn-in}}} + T f_{{\mbox{{\tiny thinning}}}}$  }{
\BlankLine
{sample the diagonal elements of $\mathbf{W}_{t+1}$ as $(\mathbf{W}_{t+1})_{ii} \, | \, \bbeta_t \sim \mathcal{PG}(1, \mathbf{X}_{i,\ast} \bbeta_t)$}.
\\
{sample $\bbeta_{t+1} \, | \, \mathbf{W}_{t+1} \sim \mathcal{N}\{(\mathbf{X}^{\top} \mathbf{W}_{t+1} \mathbf{X} + \mathbf{\Delta})^{-1} [\mathbf{X}^{\top} (\mathbf{Y} - \tfrac{1}{2} \mathbf{1}_p) + \mathbf{\Delta} \bbeta_0],$}
\\
\hspace{3.95cm} {$(\mathbf{X}^{\top} \mathbf{W}_{t+1} \mathbf{X} + \mathbf{\Delta})^{-1}\}$}.
\\
\BlankLine
}
\BlankLine
Remove the first $T_{\mbox{{\tiny burn-in}}}$ draws (representing the burn-in phase).
\\
Select every $f_{\mbox{{\tiny thinning}}}$-th sample (thinning).
\vspace{0.25cm}
\caption{{\small Pseudocode of the Gibbs sampler of the joint posterior of the
\\ \mbox{ } \hspace{2.08cm} Bayesian logistic regression parameter.}} \label{alg.gibbsSamplerLogisticRidgeRegression}
\end{algorithm}
}
}
\end{center}
\end{minipage}
\mbox{ }
\\
\cite{polson2013bayesian} provide a proof that $\bbeta$, if sampled in accordance with Algorithm \ref{alg.gibbsSamplerLogisticRidgeRegression}, is distributed as (\ref{form:RidgeLogisticPosterior}). They also describe an efficient sampler to draw from the P\'{o}lya-gamma distribution. Algorithm \ref{alg.gibbsSamplerLogisticRidgeRegression}  may be combined with Algorithm \ref{alg.highdimNormalGenRidge} of Chapter \ref{chap:genRidge} to draw efficiently from a high-dimensional multivariate normal distribution.

\section{Computationally efficient evaluation}
High-dimensionally, the IRLS algorithm of the ridge logistic regression estimator can be evaluated computationally efficiently as follows. Hereto, effectively, the problem of penalized likelihood maximization over all $\bbeta \in \mathbb{R}^p$ is converted into maximization of the same criterion but now over the weights $\mbox{diag}(\mathbf{W}) \in \mathbb{R}^n$, as those weights are all that are needed for the evaluation of the estimator. The key observations, that enable this conversion, are the following reformulations of the updated linear predictor and the penalty (see Question \ref{question.IRLSefficiently}):
\begin{eqnarray} \label{form.compEffRidgeLogistic1}
\mathbf{X} \hat{\bbeta} (\lambda) & = & \lambda^{-1} \mathbf{X} \mathbf{X}^{\top} (\lambda^{-1} \mathbf{X} \mathbf{X}^{\top} + \mathbf{W}^{-1})^{-1} \mathbf{Z},
\\
\label{form.compEffRidgeLogistic2}
\lambda \| \hat{\bbeta}(\lambda) \|_2^2 & = & \mathbf{Z}^{\top}  (\lambda^{-1} \mathbf{X} \mathbf{X}^{\top} + \mathbf{W}^{-1})^{-1}  ( \lambda^{-1}  \mathbf{X} \mathbf{X}^{\top}) (\lambda^{-1} \mathbf{X} \mathbf{X}^{\top} + \mathbf{W}^{-1})^{-1}  \mathbf{Z}
\end{eqnarray}
Both reformulations hinge upon the Woodbury matrix identity. On one hand, these reformulations avoid the inversion of a $p \times p$ dimensional matrix (as been shown for the regular ridge regression estimator, see Section \ref{sect.ridgeEfficientCalculation}). On the other, only the term $\lambda^{-1} \mathbf{X} \mathbf{X}^{\top}$ in the above expressions involves a matrix multiplication over the $p$ dimension. Exactly this term is not updated at each iteration of the IRLS algorithm. It can thus be evaluated and stored prior to the first iteration step, and at each iteration be called upon. Furthermore, the remaining quantities updated at each iteration are the weights and the penalized loglikelihood, but those are obtained straightforwardly from the linear predictor and the penalty. The pseudo-code of an efficient version of the IRLS algorithm of the ridge logistic estimator, as has been implemented in the \texttt{ridgeGLM}-function of the \texttt{porridge}-package \citep{vWie2021R}, thus comprises the following steps:
\begin{compactitem}
\item[1)] Evaluate and store $\lambda^{-1} \mathbf{X} \mathbf{X}^{\top}$.
\item[2)] Initiate the linear predictor. 
\item[3)] Update the weights, the adjusted response, the penalized loglikelihood, and the linear predictor.
\item[4)] Repeat the previous step until convergence.
\item[5)] Evaluate the estimator using the weights from the last iteration. 
\end{compactitem}
In the above, convergence refers to no further or a negligible improvement of the penalized loglikelihood. In the final step of this version of the IRLS algorithm, the ridge logistic regression estimator is obtained by:
\begin{eqnarray*}
\hat{\bbeta} (\lambda) & = & \lambda^{-1} \mathbf{X}^{\top} (\lambda^{-1} \mathbf{X} \mathbf{X}^{\top} + \mathbf{W}^{-1})^{-1} \mathbf{Z}.
\end{eqnarray*}
In the above display the weights $\mathbf{W}$ and adjusted response $\mathbf{Z}$ are obtained from the last iteration of the IRLS algorithm's third step. Alternatively, computational efficiency is obtained by the `SVD-trick' (see Question \ref{question.IRLSwithSVD}).


\section{Penalty parameter selection}
As before the penalty parameter may be chosen through $K$-fold cross-validation. For the $K=n$ case, \cite{Meij2013} describe a computationally efficient approximation of the leave-one-out cross-validated loglikelihood. It is based on the exact evaluation of the LOOCV loss, discussed in Section \ref{subsect.crossvalidation}, that avoided resampling. The approach of \cite{Meij2013} hinges upon the first-order Taylor expansion of the left-out penalized loglikelihood of the left-out estimate  $\hat{\bbeta}_{-i} (\lambda)$ around $\hat{\bbeta} (\lambda)$, which yields an approximation of the former:
\begin{eqnarray*}
\hat{\bbeta}_{-i} (\lambda) & \approx & \hat{\bbeta} (\lambda) -
\left( \left. \frac{\partial^2 \mathcal{L}_{-i}^{\mbox{{\tiny pen}}}}{\partial \bbeta \partial \bbeta^{\top}} \right|_{\bbeta = \hat{\bbeta}(\lambda)} \right)^{-1} \left. \frac{\partial \mathcal{L}_{-i}^{\mbox{{\tiny pen}}}}{\partial \bbeta } \right|_{\bbeta = \hat{\bbeta}(\lambda)}
\\
& = & \hat{\bbeta} (\lambda) + (\mathbf{X}_{- i, \ast}^{\top} \mathbf{W}_{-i, -i} \mathbf{X}_{- i, \ast} + \lambda \mathbf{I}_{pp})^{-1} \{ \mathbf{X}_{- i, \ast}^{\top} [\mathbf{Y}_{-i} - \vec{\mathbf{g}}^{-1}(\mathbf{X}_{-i, \ast}; \hat{\bbeta}(\lambda))] - \lambda \hat{\bbeta}(\lambda) \}.
\end{eqnarray*}
This approximation involves the inverse of a $p \times p$ dimensional matrix, which amounts to the evaluation of $n$ such inverses for the LOOCV loss. As in Section \ref{subsect.crossvalidation} this may be avoided. Rewrite both the gradient and the Hessian of the left-out loglikelihood in the approximation of the preceding display:
\begin{eqnarray*}
& & \hspace{-1.5cm} \mathbf{X}_{-i, \ast}^{\top} \{ \mathbf{Y}_{-i} - \vec{\mathbf{g}}^{-1}(\mathbf{X}_{-i, \ast}; \hat{\bbeta}(\lambda)]\} - \lambda \hat{\bbeta}(\lambda)
\\
& = & \mathbf{X}^{\top} \{ \mathbf{Y} - \vec{\mathbf{g}}^{-1}[\mathbf{X}; \hat{\bbeta}(\lambda)]\} - \lambda \hat{\bbeta}(\lambda)
- \mathbf{X}_{i, \ast}^{\top} \{Y_{i} - g^{-1}[\mathbf{X}_{i, \ast}; \hat{\bbeta}(\lambda)]\}
\\
& = & - \mathbf{X}_{i, \ast}^{\top} \{Y_{i} - g^{-1}[\mathbf{X}_{i, \ast}; \hat{\bbeta}(\lambda)]\}
\end{eqnarray*}
and
\begin{eqnarray*}
(\mathbf{X}_{- i, \ast}^{\top} \mathbf{W}_{-i, -i} \mathbf{X}_{- i, \ast} + \lambda \mathbf{I}_{pp})^{-1} & = & (\mathbf{X}^{\top} \mathbf{W} \mathbf{X} + \lambda \mathbf{I}_{pp})^{-1} + \mathbf{W}_{ii} (\mathbf{X}^{\top} \mathbf{W} \mathbf{X} + \lambda \mathbf{I}_{pp})^{-1} \mathbf{X}_{i, \ast}^{\top}
\\
& & \qquad \qquad \qquad \qquad \qquad \quad [ 1 - \mathbf{H}_{ii}(\lambda)]^{-1} \mathbf{X}_{i, \ast} (\mathbf{X}^{\top} \mathbf{W} \mathbf{X} + \lambda \mathbf{I}_{pp})^{-1},
\end{eqnarray*}
where the Woodbury identity has been used and now $\mathbf{H}_{ii}(\lambda) = \mathbf{W}_{ii} \mathbf{X}_{i, \ast}(\mathbf{X}^{\top} \mathbf{W} \mathbf{X} + \lambda \mathbf{I}_{pp})^{-1}  \mathbf{X}_{i, \ast}^{\top}$. Substitute both in the approximation of the left-out ridge logistic regression estimator and manipulate as in Section \ref{subsect.crossvalidation} to obtain:
\begin{eqnarray*}
\hat{\bbeta}_{- i}(\lambda) & \approx & \hat{\bbeta}(\lambda) - (\mathbf{X}^{\top} \mathbf{W} \mathbf{X} + \lambda \mathbf{I}_{pp})^{-1} \mathbf{X}_{i, \ast}^{\top} [ 1 - \mathbf{H}_{ii}(\lambda)]^{-1} [ Y_i - g^{-1}(\mathbf{X}_{i, \ast}; \hat{\bbeta}(\lambda)) ].
\end{eqnarray*}
Hence, the leave-one-out cross-validated loglikelihood $\sum_{i=1}^n \mathcal{L} [Y_i \, | \, \mathbf{X}_{i, \ast}, \hat{\bbeta}_{-i}(\lambda)]$ can now be evaluated by means of a single inverse of a $p \times p$ dimensional matrix and some matrix multiplications. And even this single inversion can be reduced to one of an $n \times n$-dimensional matrix when rewritten by means of the Woodbury matrix identity. For the performance of this approximation in terms of accuracy and speed, see \cite{Meij2013}.

For general $K$-fold cross-validation, computational efficiency is achieved through the same trick as used in Section \ref{subsect.crossvalidation} and reported by \citep{vdWiel2020fast}. It exploits the fact that the logistic regression parameter only appears in combination with the design matrix, forming the linear predictor, in the loglikelihood. For cross-validation there is thus no need to evaluate the estimator itself. To elaborate on the particulars of the logistic regression case, let $\mathcal{G}_1, \ldots, \mathcal{G}_K \subset \{1, \ldots, n\}$ be the mutually exclusive and exhaustive $K$-fold sample index sets. The linear predictor can then be written as:
\begin{eqnarray*}
\mathbf{X}_{\mathcal{G}_k, \ast} \hat{\bbeta}_{-\mathcal{G}_k}(\lambda) & = & 
\lambda^{-1} \mathbf{X}_{\mathcal{G}_k, \ast} \mathbf{X}_{-\mathcal{G}_k, \ast}^{\top} (\lambda^{-1} \mathbf{X}_{-\mathcal{G}_k, \ast} \mathbf{X}_{-\mathcal{G}_k, \ast}^{\top} + \mathbf{W}^{-1})^{-1} \mathbf{Z},
\end{eqnarray*}
which can be verified by means of the Woodbury matrix identity. Furthermore, all matrix products of submatrices of $\mathbf{X}$ are themselves submatrices of $\mathbf{X} \mathbf{X}^{\top}$. Further efficiency is gained if the latter is evaluated before the start of the cross-validation loop. It then only remains to subset $\mathbf{X} \mathbf{X}^{\top}$ inside this loop.

\section{Generalizing ridge logistic regression} \label{sect:genRidgeLogisticRegression}
The ridge logistic regression estimator may be generalized as the ridge regression counterpart. That is, the loglikelihood may be augmented by a nonzero centered quadratic form penalty $(\bbeta - \bbeta_0)^{\top} \mathbf{\Delta} (\bbeta - \bbeta_0)$, where $\bbeta_0 \in \mathbb{R}^p$ is the shrinkage target and $\mathbf{\Delta}$ is the nonnegative definite penalty matrix. The latter should be positive definite in high-dimensional settings.  Moreover, we may include covariates that we wish to leave unpenalized in order to estimate them as unbiasedly as possible. To accommodate that, we introduce $\mathbf{U}$, the $n \times q$-dimensional design matrix of the unpenalized covariates, and $\boldsymbol{\gamma} \in \mathbb{R}^q$, the associated regression parameter. The inclusion of the additional unpenalized covariates alters the distribution of the response variable. We now have:
\begin{eqnarray*}
P(Y_i = 1 \, | \, \mathbf{U}_{i,\ast}, \mathbf{X}_{i,\ast}) & = & \exp( \mathbf{U}_{i,\ast} \boldsymbol{\gamma} + \mathbf{X}_{i,\ast} \boldsymbol{\beta} ) [ 1+ \exp( \mathbf{U}_{i,\ast} \boldsymbol{\gamma} + \mathbf{X}_{i,\ast} \boldsymbol{\beta} )]^{-1}.
\end{eqnarray*} 
The derivation of the generalized ridge penalized loglikelihood of the logistic regression model is then analogous to that of the regular ridge penalized case. Its maximizer is the sought estimator:
\begin{eqnarray*}
\hat{\ggamma}(\cdot), \hat{\bbeta}(\cdot) & = & \arg \max_{\ggamma \in \mathbb{R}^q, \bbeta \in \mathbb{R}^p} \sum\nolimits_{i=1}^n Y_i (\mathbf{U}_{i,\ast} \ggamma +  \mathbf{X}_{i,\ast} \bbeta) - \log [ 1  + \exp(\mathbf{U}_{i,\ast} \ggamma +  \mathbf{X}_{i,\ast} \bbeta)] 
\\
& & \qquad \qquad \qquad \, - (\bbeta - \bbeta_0)^{\top} \mathbf{\Delta} (\bbeta - \bbeta_0).
\end{eqnarray*}
No analytic expression of this estimator is available. It is again evaluated numerically by means of the Iteratively Reweighted Least Squares (IRLS) algorithm that generates a sequence $\{ \hat{\ggamma}^{(t)}(\cdot), \hat{\bbeta}^{(t)}(\cdot) \}_{t=1}^\infty$ that converges to the generalized ridge logistic regression estimator. Given the $t$-th values of this sequence, the next are found by:
\begin{eqnarray*}
\hat{\ggamma}^{(t+1)}(\cdot) & = & \{ \mathbf{U}^{\top} (\mathbf{W}^{(t)})^{-1} [\mathbf{X} \mathbf{\Delta}^{-1} \mathbf{X}^{\top} + (\mathbf{W}^{(t)})^{-1}]^{-1} (\mathbf{W}^{(t)})^{-1} \mathbf{U}\}^{-1} 
\\
& & \qquad \quad \, \, \, \times \, \mathbf{U}^{\top} [ \mathbf{X}  \mathbf{\Delta}^{-1} \mathbf{X}^{\top} + (\mathbf{W}^{(t)})^{-1}]^{-1} ( \mathbf{Y}^{(\mbox{{\tiny adj}},t)} - \mathbf{X} \bbeta_0),
\\
\hat{\bbeta}^{(t+1)}(\cdot) & = & \bbeta_0 + (\mathbf{X}^{\top} \mathbf{W}^{(t)} \mathbf{X}  + \mathbf{\Delta})^{-1} \mathbf{X}^{\top} \mathbf{W}^{(t)} [ \mathbf{Y}^{(\mbox{{\tiny adj}},t)}  - \mathbf{U} \hat{\ggamma}^{(t+1)}(\cdot) - \mathbf{X} \bbeta_0].
\end{eqnarray*}
These expression have been obtained by application of the analytic expression of the inverse of a $2\times 2$ block matrix, the Woodbury matrix identity, and some linear algebraic manipulations (see \citealp{Lettink2023twoDimFusedRidge}, for details). The IRLS algorithm is terminated after a finite number of iterations, either when the loss does no longer substantially improve or subsequent iterations show little difference in their evaluation of the parameter estimates.

\section{Application}
The ridge logistic regression is used here to explain the status (dead or alive) of ovarian cancer samples at the close of the study from gene expression data at baseline. Data stem from the TCGA study \citep{TCGA2011ovarian}, which measured gene expression by means of sequencing technology. Available are 295 samples with both status and transcriptomic profiles. These profiles are composed of 19990 transcript reads. The sequencing data, being representative of the mRNA transcript count, is heavily skewed. \cite{Zwie2014} show that a simple transformation of the data prior to model building generally yields a better model than tailor-made approaches. Motivated by this observation the data were -- to accommodate the zero counts -- $\mbox{asinh}$-transformed. The logistic regression model is then fitted in ridge penalized fashion, leaving the intercept unpenalized. The ridge penalty parameter is chosen through  10-fold cross-validation minimizing the cross-validated error. R-code, and that for the sequel of this example, is to be found below.

\begin{lstlisting}
# load libraries
library(glmnet)
library(TCGA2STAT)

# load data
OVdata <- getTCGA(disease="OV", data.type="RNASeq", type="RPKM", clinical=TRUE)
Y      <- as.numeric(OVdata[[3]][,2])
X      <- asinh(data.matrix(OVdata[[3]][,-c(1:3)]))

# start fit
# optimize penalty parameter
cv.fit   <- cv.glmnet(X, Y, alpha=0, family=c("binomial"), 
                            nfolds=10, standardize=FALSE)
optL2    <- cv.fit$lambda.min

# estimate model
glmFit   <- glmnet(X, Y, alpha=0, family=c("binomial"), 
                         lambda=optL2, standardize=FALSE)

# construct linear predictor and predicted probabilities
linPred  <- as.numeric(glmFit$a0 + X %*% glmFit$beta)
predProb <- exp(linPred) / (1+exp(linPred))

# visualize fit
boxplot(linPred ~ Y, pch=20, border="lightblue", col="blue", 
                     ylab="linear predictor", xlab="response", 
                     main="fit")

# evaluate predictive performance
# generate k-folds balanced w.r.t. status
fold     <- 10
folds1   <- rep(1:fold, ceiling(sum(Y)/fold))[1:sum(Y)]
folds0   <- rep(1:fold, ceiling((length(Y)-length(folds1))
                                /fold))[1:(length(Y)-length(folds1))]
shuffle1 <- sample(1:length(folds1), length(folds1))
shuffle0 <- sample(1:length(folds0), length(folds0))
folds1   <- split(shuffle1, as.factor(folds1))
folds0   <- split(shuffle0, as.factor(folds0))
folds    <- list()
for (f in 1:fold){
	folds[[f]] <- c(which(Y==1)[folds1[[f]]], which(Y==0)[folds0[[f]]])
}
for (f in 1:fold){
	print(sum(Y[folds[[f]]]))
}

# build model
pred2obsL2 <- matrix(nrow=0, ncol=4)
colnames(pred2obsL2) <- c("optLambda", "linPred", "predProb", "obs")
for (f in 1:length(folds)){
	print(f)
	cv.fit     <- cv.glmnet(X[-folds[[f]],], Y[-folds[[f]]], alpha=0, 
	                        family=c("binomial"), nfolds=10, standardize=FALSE)
	optL2      <- cv.fit$lambda.min
	glmFit     <- glmnet(X[-folds[[f]],], Y[-folds[[f]]], alpha=0, 
	                     family=c("binomial"), lambda=optL2, standardize=FALSE)
	linPred    <- glmFit$a0 + X[folds[[f]],,drop=FALSE] %*% glmFit$beta
	predProb   <- exp(linPred) / (1+exp(linPred))
	pred2obsL2 <- rbind(pred2obsL2, cbind(optL2, linPred, predProb, Y[folds[[f]]]))
}

# visualize fit
boxplot(pred2obsL2[,3] ~ pred2obsL2[,4], pch=20, border="lightblue", col="blue", 
                                         ylab="linear predictor", xlab="response", 
                                         main="prediction")



\end{lstlisting}

\begin{figure}[!h]
\begin{tabular}{rcl}
\includegraphics[scale=0.30, angle=0]{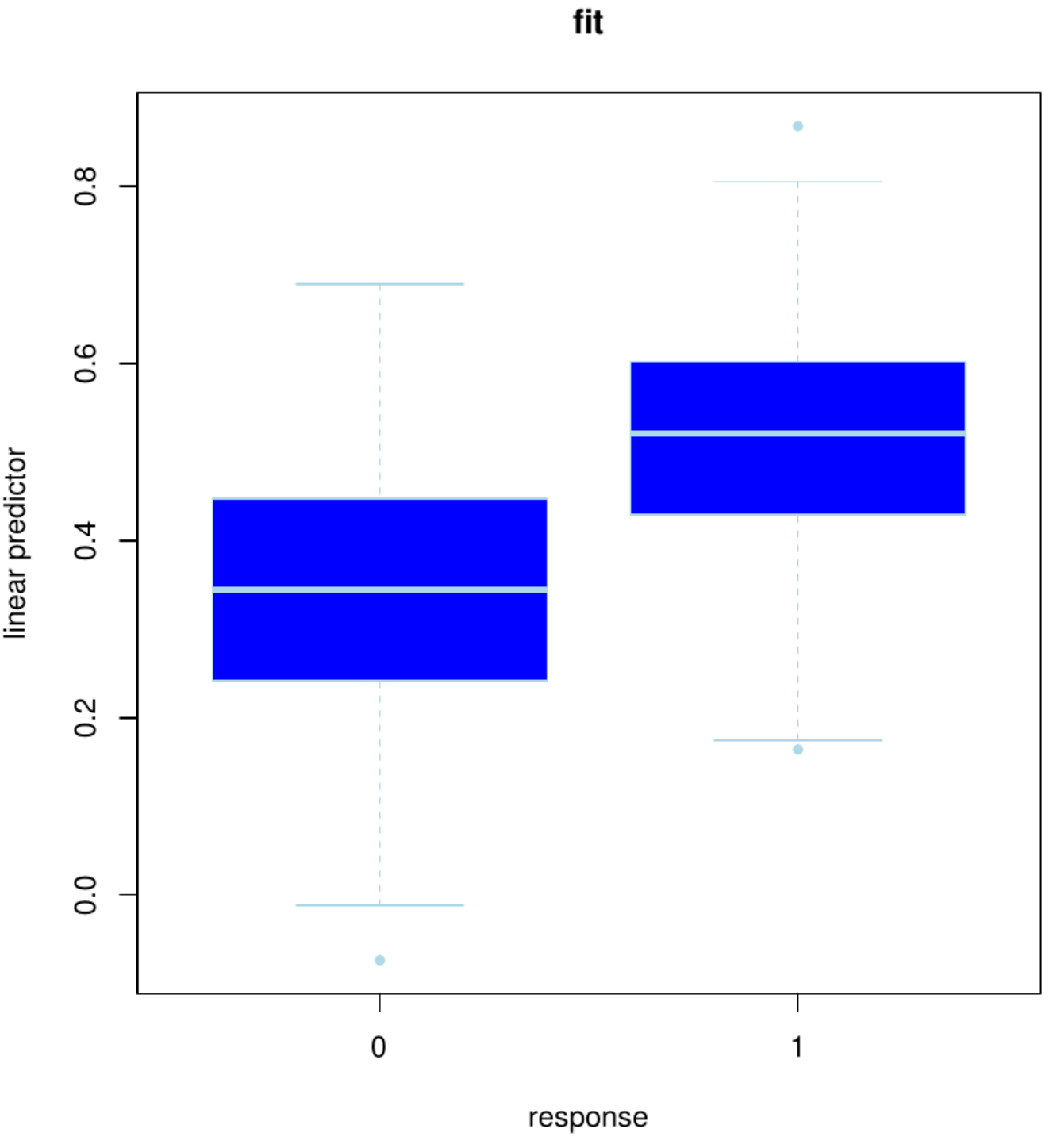}
&
\includegraphics[scale=0.30, angle=0]{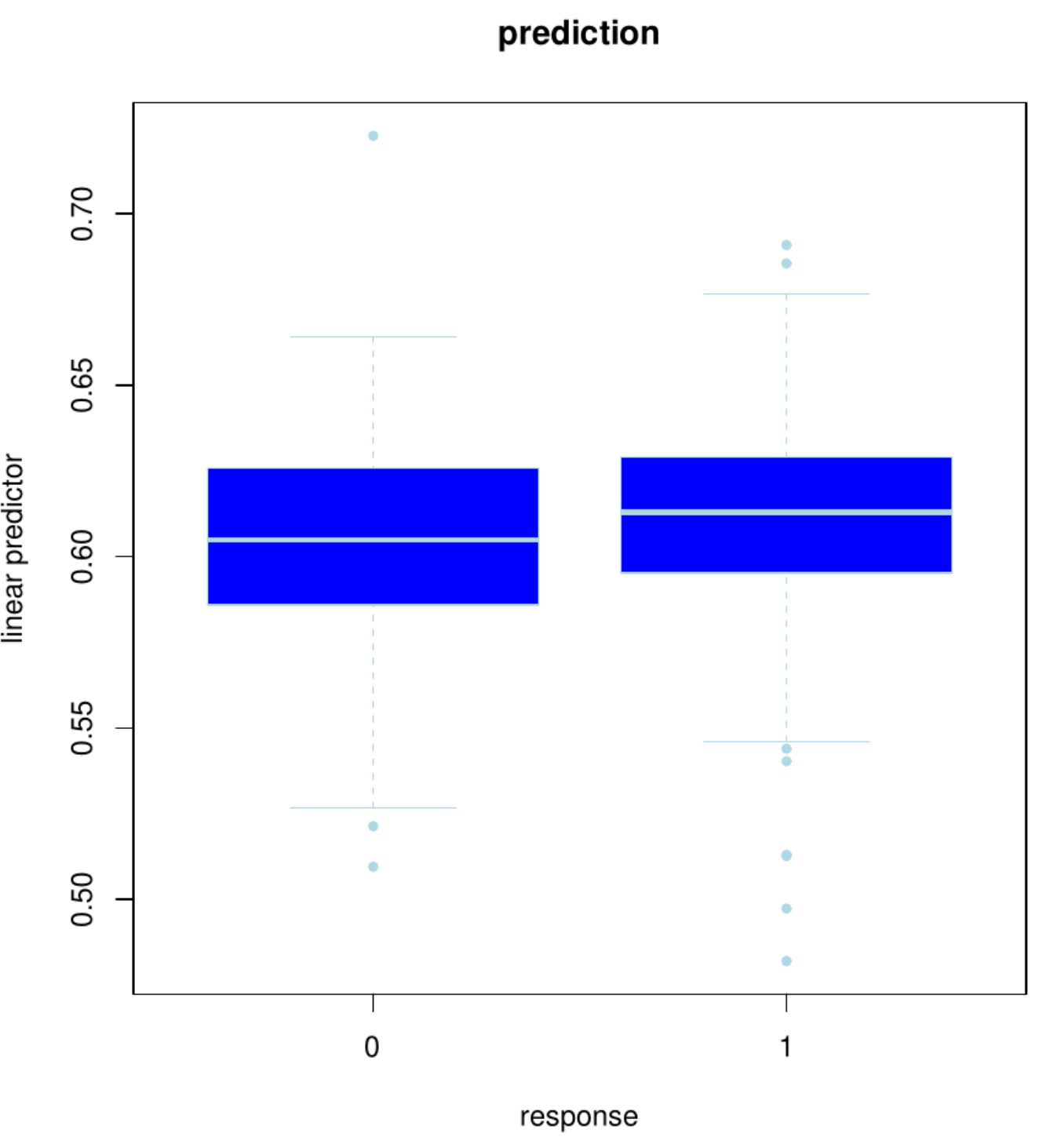}
\end{tabular}
\caption{Left panel: Box plot of the status vs. the fitted linear predictor using the full data set. Right panel: Box plot of the status vs. the linear prediction in the left-out samples of the 10 folds.} \label{fig:logisticRidge_ovarianExample}
\end{figure}

The fit of the resulting model is studied. Hereto the fitted linear predictor $\mathbf{X} \hat{\bbeta}(\lambda_{\mbox{\tiny opt}})$ is plotted against the status (Figure \ref{fig:logisticRidge_ovarianExample}, left panel). The plot shows some overlap between the boxes, but also a clear separation. The latter suggests gene expression at baseline thus enables us to distinguish surviving from the to-be-diseased ovarian cancer patients. Ideally, a decision rule based on the linear predictor can be formulated to predict an individual's outcome.

The fit, however, is evaluated on the samples that have been used to build the model. This gives no insight on the model's predictive performance on novel samples. A replication of the study is generally costly and comparable data sets need not be at hand. A common workaround is to evaluate the predictive performance on the same data \citep{Subr2010}. This requires to put several samples aside for performance evaluation while the remainder is used for model building. The left-out sample may accidently be chosen to yield an exaggerated (either dramatically poor or overly optimistic) performance. This is avoided through the repetition of this exercise, leaving (groups of) samples out one at the time. The left-out performance evaluations are then averaged and believed to be representative of the predictive performance of the model on novel samples. Note that, effectively, as the model building involves cross-validation and so does the performance evaluation, a double cross-validation loop is applied. This procedure is applied with a ten-fold split in both loops. Denote the outer folds by $f = 1, \ldots, 10$. Then, $\mathbf{X}_{f}$ and $\mathbf{X}_{-f}$ represent the design matrix of the samples comprising fold $f$ and that of the remaining samples, respectively. Define $\mathbf{Y}_{f}$ and $\mathbf{Y}_{-f}$ similarly. The linear prediction for the left-out fold $f$ is then $\mathbf{X}_{f} \hat{\bbeta}_{-f} (\lambda_{\mbox{{\tiny opt, -f}}})$. For reference to the fit, this is compared to $\mathbf{Y}_{f}$ visually by means of a boxplot as used above (see Figure \ref{fig:logisticRidge_ovarianExample}, right panel). The boxes overlap almost perfectly. Hence, little to nothing remains of the predictive power suggested by the boxplot of the fit. The fit may thus give a reasonable description of the data at hand, but it extrapolates poorly to new samples.

\section{Conclusion}
To deal with response variables other than continuous ones, ridge logistic regression was discussed. High-dimensionally, the empirical identifiability problem then persists. Again, penalization came to the rescue: the ridge penalty may be combined with other link functions than the identity. Properties of ridge regression were shown to carry over to its logistic equivalent.

\section{Exercises}
\begin{question} \mbox{ }
\\
The variation in a binary response $Y_i \in \{ 0, 1 \}$ due to two covariates $\mathbf{X}_{i,\ast} = (X_{i,1}, X_{i,2})$ is described by the logistic regression model: $P(Y_i = 1) = \exp(\mathbf{X}_{i,\ast} \bbeta) [1 +  \exp(\mathbf{X}_{i,\ast} \bbeta)]^{-1}$. The study design and the observed response are given by:
\begin{eqnarray*}
\mathbf{X} & = & \left( \begin{array}{rr} 1 & -1 \\ -1 &  1 \end{array} \right) \quad \mbox{and} \quad \mathbf{Y} \, \, \, = \, \, \, \left( \begin{array}{rr} 1 \\ 0 \end{array} \right).
\end{eqnarray*}

\begin{compactitem}
\item[\textit{a)}] Write down the loglikelihood and show that the maximum likelihood estimator $\hat{\bbeta}^{\mbox{{\tiny ml}}}$  is an element of $\{ (\beta_1, \beta_2)^{\top} \, : \, \beta_1 - \beta_2 = \infty \}$.

\item[\textit{b)}] Augment the loglikelihood with the ridge penalty $\tfrac{1}{2} \lambda \| \bbeta \|_2^2$ and show that the ridge logistic regression estimator satisfies $| \hat{\beta}_j (\lambda) | < \infty$ for $j=1, 2$.
\end{compactitem}
\end{question}

\begin{question} \mbox{ } \label{question:logisticEarlyStopping}
\\
Consider the maximization of the likelihood of the logistic regression model through gradient ascent with a finite number of iterations, i.e. stop early. Show -- by answering the following -- that the resulting parameter estimator is constrained.
\begin{compactitem}
\item[\textit{a)}] Use a first order Taylor expansion of the (inverse) link function around $\boldsymbol{\beta}^{(t-1)}$ in the gradient ascent's update rule (\ref{form.ridgeLogisticGradAscent}) to show that
\begin{eqnarray*}
\boldsymbol{\beta}^{(t+1)} & \approx & \boldsymbol{\beta}^{(t-1)}  +  \delta  \mathbf{X}^{\top} [ \mathbf{Y} - \vec{\mathbf{g}}^{-1}(  \mathbf{X}; \boldsymbol{\beta}^{(t-1)} )   ] + \delta (\mathbf{I}_{pp} - \delta \mathbf{X}^{\top} \mathbf{W}^{(t-1)} \mathbf{X})  \mathbf{X}^{\top} [ \mathbf{Y} - \vec{\mathbf{g}}^{-1}(  \mathbf{X}; \boldsymbol{\beta}^{(t-1)} )  ],
\end{eqnarray*}
where the weight matrix $\mathbf{W}^{(t)}$ is that corresponding to $\boldsymbol{\beta}^{(t)}$.

\item[\textit{b)}] Iterate the arguments used in part \textit{a)} to show that
\begin{eqnarray*}
\boldsymbol{\beta}^{(t+1)} & \approx & \boldsymbol{\beta}^{(0)}  +  \delta \big[\mathbf{I}_{pp} +  \sum\nolimits_{t'=0}^{t-1} \prod\nolimits_{t''=0}^{t'} (\mathbf{I}_{pp} - \delta \mathbf{X}^{\top} \mathbf{W}^{(t')} \mathbf{X})  \big] \mathbf{X}^{\top} [ \mathbf{Y} - \vec{\mathbf{g}}^{-1}(  \mathbf{X}; \boldsymbol{\beta}^{(0)} )  ].
\end{eqnarray*}

\item[\textit{c)}] Explain why $\mathbf{W}^{(t)} \preceq \tfrac{1}{4} \mathbf{I}_{pp}$ for all $t$. 

\item[\textit{d)}] Set $\boldsymbol{\beta}^{(0)} = \mathbf{0}_p$ and
suppose $\hat{\boldsymbol{\beta}}^{\mbox{{\tiny (ml)}}} = \lim_{t\rightarrow \infty} \hat{\boldsymbol{\beta}}^{(t)}$. Show, using parts \textit{b)} and \textit{c)}, that $\| \hat{\boldsymbol{\beta}}^{\mbox{{\tiny (ml)}}} - \hat{\boldsymbol{\beta}}^{(t+1)} \|$ can be bounded from below and away from zero, in a way that depends on $\delta$, $t$, and (properties of) $\mathbf{X}$ only. 
\end{compactitem}
\end{question}

\begin{question} \mbox{ }
\\
Load the leukemia data available via the {\tt multtest}-package (downloadable from BioConductor) through the following \texttt{R}-code:
\begin{lstlisting}
# activate library and load the data
library(multtest)
data(golub)

\end{lstlisting}
The objects {\tt golub} and {\tt golub.cl} are now available. The matrix-object {\tt golub} contains the expression profiles of 38 leukemia patients. Each profile comprises expression levels of 3051 genes. The numeric-object {\tt golub.cl} is an indicator variable for the leukemia type, either AML or ALL, of the patient.

\begin{compactitem}
\item[\textit{a)}] Relate the leukemia subtype and the gene expression levels by a logistic regression model. Fit this model by means of penalized maximum likelihood, employing the ridge penalty with penalty parameter $\lambda=1$. This is implemented in the {\tt penalized}-packages available from {\tt CRAN}. {\it Note:} center (gene-wise) the expression levels around zero.

\item[\textit{b)}] Obtain the fits from the regression model. The fit is almost perfect. Could this be due to overfitting the data? Alternatively, could it be that the biological information in the gene expression levels indeed determines the leukemia subtype almost perfectly?

\item[\textit{c)}]  To discern between the two explanations for the almost perfect fit, randomly shuffle the subtypes. Refit the logistic regression model and obtain the fits. On the basis of this and the previous fit, which explanation is more plausible?

\item[\textit{d)}]  Compare the fit of the logistic model with different penalty parameters, say $\lambda = 1$ and $\lambda = 1000$. How does $\lambda$ influence the possibility of overfitting the data?

\item[\textit{e)}]  Describe what you would do to prevent overfitting.
\end{compactitem}
\end{question}

\begin{question} \label{question.ridgeLogisticBreastData} \mbox{ }
\\
Load the breast cancer data available via the {\tt breastCancerNKI}-package (downloadable from BioConductor) through the following \texttt{R}-code:
\begin{lstlisting}
# activate library and load the data
library(breastCancerNKI)
data(nki)

# subset and center the data
X <- exprs(nki)
X <- X[-which(rowSums(is.na(X)) > 0),]
X <- apply(X[1:1000,], 1, function(X){ X - mean(X) })

# define ER as the response
Y <- pData(nki)[,8]

\end{lstlisting}
The \texttt{eSet}-object {\tt nki} is now available. It contains the expression profiles of 337 breast cancer patients. Each profile comprises expression levels of 24481 genes. The \texttt{R}-code above extracts the expression data from the object, removes all genes with missing values, centers the gene expression gene-wise around zero, and subsets the data set to the first thousand genes. The reduction of the gene dimensionality is only for computational speed. Furthermore, it extracts the estrogen receptor status (abbreviated to ER status), an important prognostic indicator for breast cancer, that is to be used as the response variable in the remainder of the exercise.

\begin{compactitem}
\item[\textit{a)}]  Relate the ER status and the gene expression levels by a logistic regression model, which is fitted by means of ridge penalized maximum likelihood. First, find the optimal value of the penalty parameter of $\lambda$ by means of cross-validation. This is implemented in {\tt optL2}-function of the {\tt penalized}-package available from {\tt CRAN}.

\item[\textit{b)}]  Evaluate whether the cross-validated likelihood indeed attains a maximum at the optimal value of $\lambda$. This can be done with the {\tt profL2}-function of the {\tt penalized}-package available from {\tt CRAN}.

\item[\textit{c)}]  Investigate the sensitivity of the penalty parameter selection with respect to the choice of the cross-validation fold.

\item[\textit{d)}]  Does the optimal lambda produce a reasonable fit?
\end{compactitem}
\end{question}

\begin{question} \label{question.IRLSwithSVD} \mbox{ }  \\
The iteratively reweighted least squares (IRLS) algorithm for the numerical evaluation of the ridge logistic regression estimator requires the inversion of a $p \times p$-dimensional matrix at each iteration. In Section \ref{sect.ridgeEfficientCalculation} the singular value decomposition (SVD) of the design matrix is exploited to avoid the inversion of such a matrix in the numerical evaluation of the ridge regression estimator. Use this trick to show that the computational burden of the IRLS algorithm may be reduced to one SVD prior to the iterations and the inversion of an $n \times n$ dimensional matrix at each iteration (as is done in \citealp{Eilers2001}).
\end{question}

\begin{question}  \label{question.IRLSefficiently}  \mbox{ } \\
Consider the generalized ridge logistic regression estimator of $\bbeta$ defined as: 
\begin{eqnarray*}
\mathcal{L}^{\mbox{{\tiny pen}}}(\mathbf{Y}, \mathbf{X}; \bbeta, \lambda) & = & \sum\nolimits_{i=1}^n \big\{ Y_i \mathbf{X}_i \bbeta - \log [ 1 + \exp(\mathbf{X}_i \bbeta) ] \big\} - \tfrac{1}{2} (\bbeta - \bbeta_0)^{\top} \mathbf{\Delta} (\bbeta - \bbeta_0),
\end{eqnarray*}
with nonrandom $\bbeta_0 \in \mathbb{R}^p$ and positive definite, symmetric $\mathbf{\Delta}$ (as in Chapter \ref{chap:genRidge}).
\begin{compactitem}
\item[\textit{a)}]  Modify the the iteratively reweighted least squares (IRLS) algorithm, which is outlined in Section \ref{sect.ridgeLogistic}, for the generalized ridge logistic regression estimator (as is done in \citealp{vanWieringen2022transfer}).

\item[\textit{b)}] Let $\mathbf{\Delta} = \lambda \mathbf{I}_{pp}$. Derive the equivalent of Equations (\ref{form.compEffRidgeLogistic1}) and (\ref{form.compEffRidgeLogistic2}) for computationally efficient evaluation of the targeted ridge logistic regression estimator.
\end{compactitem}
\end{question}


\begin{question} \label{question.logisticExerciseData} \mbox{ }
\\
Consider the logistic regression model to explain the variation of a binary random variable by a covariate. Let $Y_i \in \{ 0, 1\}$, $i=1, \ldots, n$, represent $n$ binary random variables that follow a Bernoulli distribution with parameter $P(Y_i = 1 \, | \, X_i) = \exp(X_i \beta) [1+ \exp(X_i \beta)]^{-1}$ and corresponding covariate $X_i \in \{ -1, 1 \}$. Data $\{ (y_i, X_i) \}_{i=1}^n$ are summarized as a contingency table (Table \ref{table.data4logisticExercise}):
\\
\mbox{ }
\begin{table}[h!]
\vspace{-0.45cm}
\centering
\begin{tabular}{rrr}
\hline \hline \vspace{-7pt} & & \\
\vspace{3pt} &         $y_i=\texttt{0}$ &    $y_i=\texttt{1}$ \\
\hline \vspace{-4pt} & & \\
$X_i=\texttt{-1}$ &  301 &  196 \\
$X_i= \texttt{\textcolor{white}{-}1}$ &  206 &  297 \\
\vspace{-9pt} & & \\
\hline \hline
\end{tabular}
\caption{Data for Exercise \ref{question.logisticExerciseData}.} \label{table.data4logisticExercise}
\end{table}
\vspace{-0.3cm}
\newline
The data of Table \ref{table.data4logisticExercise} is only to be used in parts $\textit{c)}$ and $\textit{g)}$.
\begin{compactitem}
\item[\textit{a)}] Show that the estimating equation of the ridge logistic regression estimator of $\bbeta$ is of the form:
\begin{eqnarray*}
c - n \exp (\beta) [ 1 + \exp (  \beta)]^{-1} -  \lambda \beta & = & 0,
\end{eqnarray*}
and specify $c$. 
\item[\textit{b)}] Show that $\hat{\beta} (\lambda) \in  (\lambda^{-1} (c-n), \lambda^{-1} c)$ for all $\lambda > 0$. Ensure that this is a meaningful interval, i.e. that it is non-empty, and verify that $c > 0$. Moreover, conclude from the interval that $\lim_{\lambda \rightarrow \infty} \hat{\beta} (\lambda) = 0$.
\item[\textit{c)}] The Taylor expansion of $\exp (\beta) [ 1 + \exp (  \beta)]^{-1}$ around $\beta=0$ is:
\begin{eqnarray*}
\exp (\beta) [ 1 + \exp (  \beta)]^{-1}  & = & \tfrac{1}{2} + \tfrac{1}{4} \beta - \tfrac{1}{48} \beta^3 + \tfrac{1}{480} \beta^5 + \mathcal{O} (\beta^6).
\end{eqnarray*}
Substitute the $3^{\mbox{{\tiny rd}}}$ order Taylor approximation into the estimating equations and use the data of Table \ref{table.data4logisticExercise} to evaluate its roots using the \texttt{polyroot}-function (of the \texttt{base}-package). 
Do the same for the $1^{\mbox{{\tiny st}}}$ and $2^{\mbox{{\tiny nd}}}$ order Taylor approximations of $\exp (\beta) [ 1 + \exp (  \beta)]^{-1}$. Compare these approximate estimates to the one provided by the \texttt{ridgeGLM}-function (of the \texttt{porridge}-package, \citealp{vWie2021R}).
\end{compactitem}
In the remainder consider the $1^{\mbox{{\tiny st}}}$ order Taylor approximated ridge logistic regression estimator: $\hat{\beta}_{1^{\mbox{{\tiny st}}}} ( \lambda ) = (\lambda + \tfrac{1}{4} n)^{-1} (c - \tfrac{1}{2} n)$.
\begin{compactitem}
\item[\textit{d)}] Find an expression for $\mathbb{E} [ \hat{\beta}_{1^{\mbox{{\tiny st}}}} (\lambda) ]$.
\item[\textit{e)}] Find an expression for $\mbox{Var} [ \hat{\beta}_{1^{\mbox{{\tiny st}}}} (\lambda) ]$.
\item[\textit{f)}] Combine the answers of parts \textit{d)} and \textit{e)} to find an expression for $\mbox{MSE} [ \hat{\beta}_{1^{\mbox{{\tiny st}}}} (\lambda) ]$. 
\item[\textit{g)}] Plot this MSE against $\lambda$ for the data provided in Table \ref{table.data4logisticExercise} and reflect on an analogue of Theorem \ref{theo.Theobald2} for the ridge logistic regression estimator.

\end{compactitem}
\end{question}

\begin{question} \mbox{ } \\
Consider the logistic regression model to explain the variation of a binary random variable by a covariate. Let $Y_i \in \{ 0, 1\}$, $i=1, \ldots, n$, represent $n$ binary random variables that follow a Bernoulli distribution with parameter $P(Y_i = 1 \, | \, \mathbf{X}_{i, \ast}) = \exp(\mathbf{X}_{i,\ast} \bbeta) [1+ \exp(\mathbf{X}_{i,\ast} \bbeta)]^{-1}$ with regression paramater $\bbeta$ and the $i$-th sample's covariate information vector $\mathbf{X}_{i,\ast}^{\top} \in \mathbb{R}^p$. Adopt the prior $\beta \sim \mathcal{N}(0, \lambda^{-1})$ and consider fitting the model to the following data $\mathbf{X}^{\top} = ( -1, 2)^{\top}$ and $\mathbf{Y} = ( 0,1 )^{\top}$. 

\begin{compactitem}
\item[\textit{a)}] Denote the Maximum A Posteriori estimator by $\hat{\beta}_{\mbox{{\tiny MAP}}}$. Verify, using \texttt{R}, that  $\hat{\beta}_{\mbox{{\tiny MAP}}} \approx 0.7148331$.
\item[\textit{b)}] Verify, using part \textit{a)} and possibly \texttt{R}, that the posterior variance $\mbox{Var}(\beta \, | \, \mathbf{Y}, \mathbf{X}, \lambda) \approx 0.5423074$. 
\end{compactitem}

\end{question}

\begin{question} \mbox{ } \\
Consider the logistic regression model $Y_i \sim \mbox{Bernoulli} \{ \exp(\mathbf{X}_{i,\ast} \bbeta)  [\exp(\mathbf{X}_{i,\ast} \bbeta)]^{-1} \}$ for $i =1, \ldots, n$ and regression parameter $\bbeta \in \mathbb{R}^p$, response variable $Y_i$ and the $p$-dimensional row vector $\mathbf{X}_{i,\ast}$ with the covariate information. Suppose the samples can be divided into two groups, one of size $n_0$ as the covariate information and the other of size $n_1$ such that $n=n_0 + n_1$, with data  $Y_i  = 0$ and $\mathbf{X}_{i,\ast} = \mathbf{X}_{1,\ast}$ for $i=1, \ldots, n_0$ and $Y_i  = 1$ and $\mathbf{X}_{i,\ast} = -\mathbf{X}_{1,\ast}$ for $i=n_0+1, \ldots, n_0+n_1$. From these data, the regression parameter is estimated with the ridge logistic regression parameter which maximizes the loglikelihood augmented with the ridge penalty $\tfrac{1}{2} \lambda \| \bbeta \|_2^2$ with penalty parameter $\lambda_2 > 0$. Show that the ridge logistic regression estimator $\hat{\bbeta}(\lambda)$ is of the form $c (\lambda_2) \mathbf{X}_{1, \ast}^{\top}$ with $c (\lambda) \in \mathbb{R}$.
\end{question}

\begin{question} \mbox{ } \\
Suppose data on a binary response variable $Y_i \in \{ 0, 1\}$ and covariate information $\mathbf{X}_{i,\ast}^{\top} \in \mathbb{R}^p$ of sample $i=1,\ldots,n$ are available. Use the logistic regression model to explain the variation of a binary random variable by that of a set covariates: the  $Y_i$ follow a Bernoulli distribution with parameter $P(Y_i = 1 \, | \, \mathbf{X}_{i, \ast}) = \exp(\mathbf{X}_{i,\ast} \bbeta) [1+ \exp(\mathbf{X}_{i,\ast} \bbeta)]^{-1}$ with regression parameter $\bbeta$.  Throughout assume $\mathbf{X}_{1,\ast} = \ldots = \mathbf{X}_{n,\ast}$ with all $X_{i,j}$ nonzero and finite and consider the ridge logistic regression estimator $\hat{\bbeta}(\lambda)$ of $\bbeta$.
\begin{compactitem}
\item[\textit{a)}] Show that $X_{1,1}^{-1} \hat{\beta}_1(\lambda)  = X_{1,2}^{-1} \hat{\beta}_2(\lambda) = \ldots = X_{1,p}^{-1} \hat{\beta}_p(\lambda)$.

\item[\textit{b)}] Show that the linear prediction of the $i$-th sample can be bounded as:
\begin{eqnarray*}
- \lambda^{-1}  \| \mathbf{X}_{i,\ast}^{\top} \|_2^2 \big| n - \big( \sum\nolimits_{i'=1}^n Y_{i'} \big) \big|  & \leq \, \, \,  \mathbf{X}_{i,\ast} \hat{\bbeta}(\lambda)  \, \, \, \leq &   \lambda^{-1}  \| \mathbf{X}_{i,\ast}^{\top} \|_2^2 \big| n - \big( \sum\nolimits_{i'=1}^n Y_{i'} \big) \big|.
\end{eqnarray*} 

\item[\textit{c)}] Keep $n$ fixed. How should the regularization parameter $\lambda$ scale with the dimension $p$ to ensure that each $\mathbb{E} [ \hat{\beta}_j (\lambda) ]$ is finite as $p \rightarrow \infty$?

\item[\textit{d)}] Keep $n$ fixed.  How should the regularization parameter $\lambda$ scale with the dimension $p$ to ensure that $\mathbb{E} [ \| \hat{\bbeta} (\lambda) \|_2^2]$ is finite as $p \rightarrow \infty$?

\item[\textit{e)}] Part \textit{b)} implies that we can bound the size of the estimator as $\| \hat{\bbeta}(\lambda_2) \|_2^2  \leq   n^2 \lambda^{-2}  \| \mathbf{X}_{i,\ast}^{\top} \|_2^2$. For fixed $n$ and $\lambda$, this bound becomes weaker as the right-hand side increases monotonically with the dimension $p$. Weaker as, still for fixed $n$ and $\lambda$, the size $\| \hat{\bbeta}(\lambda) \|_2^2$ decreases monotonically with the dimension $p$. Explain the latter. 
\end{compactitem}
\end{question}






\pagestyle{fancy}

\chapter[Lasso regression]{Lasso regression} \label{chap:lassoRegression}
In this chapter, we return to the linear regression model, which is still fitted in penalized fashion, but this time with a so-called lasso penalty. Yet another penalty? Yes, but the resulting estimator will turn out to have interesting properties. The outline of this chapter loosely follows that of its counterpart on ridge regression (Chapter \ref{chap:ridgeRegression}). The chapter can -- at least partially -- be seen as an elaborated version of the original work on lasso regression, i.e. \cite{Tibs1996}, with most topics covered and visualized more extensively and incorporating results and examples published since. 

Recall that ridge regression finds an estimator of the parameter of the linear regression model through the minimization of:
\begin{eqnarray} \label{form.penLeastSquares}
\| \mathbf{Y} - \mathbf{X} \bbeta \|_2^2 + f_{\mbox{{\tiny pen}}}(\bbeta, \lambda),
\end{eqnarray}
with $f_{\mbox{{\tiny pen}}}(\bbeta, \lambda) = \lambda \| \bbeta \|_2^2$. The particular choice of the penalty function originated in a post-hoc motivation of the ad-hoc fix to the singularity of the matrix $\mathbf{X}^{\top} \mathbf{X}$, stemming from the design matrix $\mathbf{X}$ not being of full rank, i.e. $\mbox{rank}(\mathbf{X}) < p$. The ad-hoc nature of the fix suggests that the choice for the squared Euclidean norm of $\bbeta$ as a penalty is arbitrary and other choices may be considered, some of which were already encountered in Chapter \ref{chap:genRidge}.

One such choice is the so-called lasso penalty, giving rise to lasso regression, as introduced by \cite{Tibs1996}. Like ridge regression, lasso regression fits the linear regression model $\mathbf{Y} = \mathbf{X} \bbeta + \vvarepsilon$ with the standard assumption on the error $\vvarepsilon$. Like ridge regression, it does so by minimization of the sum of squares augmented with a penalty. Hence, lasso regression too minimizes loss function (\ref{form.penLeastSquares}). The difference with ridge regression is in the penalty function. Instead of the squared Euclidean norm, lasso regression uses the $\ell_1$-norm: $f_{\mbox{{\tiny pen}}}(\bbeta, \lambda_1) = \lambda_1 \| \bbeta \|_1$, the sum of the absolute values of the regression parameters multiplied by the lasso penalty parameter $\lambda_1$. To distinguish the ridge and lasso penalty parameters, they are henceforth denoted $\lambda_2$ and $\lambda_1$, respectively, with the subscript referring to the norm used in the penalty. The lasso regression loss function is thus:
\begin{eqnarray}
\label{form.lassoLossFunction}
\mathcal{L}_{\mbox{{\tiny lasso}}}(\bbeta; \lambda) & = & \| \mathbf{Y} - \mathbf{X} \, \bbeta \|^2_2 + \lambda_1 \| \bbeta \|_1 \, \, \, = \, \, \, \sum\nolimits_{i=1}^n (Y_i - \mathbf{X}_{i\ast} \, \bbeta)^2 + \lambda_1 \sum\nolimits_{j=1}^p | \beta_j |.
\end{eqnarray}
The lasso regression estimator is then defined as the minimizer of this loss function. As with the ridge regression loss function, the maximum likelihood estimator (should it exists) of $\bbeta$ minimizes the first part, while the second part is minimized by setting $\bbeta$ equal to the $p$ dimensional zero vector. For $\lambda_1$ close to zero, the lasso estimate is close to the maximum likelihood estimate. Whereas for large $\lambda_1$, the penalty term overshadows the sum-of-squares, and the lasso estimate is small (in some sense). Intermediate choices of $\lambda_1$ mold a compromise between those two extremes, with the penalty parameter determining the contribution of each part to this compromise. The lasso regression estimator thus is not one but a whole sequence of estimators of $\bbeta$, one for every $\lambda_1 \in \mathbb{R}_{>0}$. This sequence is the lasso regularization path, defined as $\{ \hat{\bbeta}(\lambda_1) \, : \, \lambda_1 \in \mathbb{R}_{>0} \}$. To arrive at a final lasso estimator of $\bbeta$, like its ridge counterpart, the lasso penalty parameter $\lambda_1$ needs to be chosen (see Section \ref{section:penaltyParameterSelectionL1}).

\begin{figure}[!h]
\begin{tabular}{rcl}
\includegraphics[scale=0.32, angle=0]{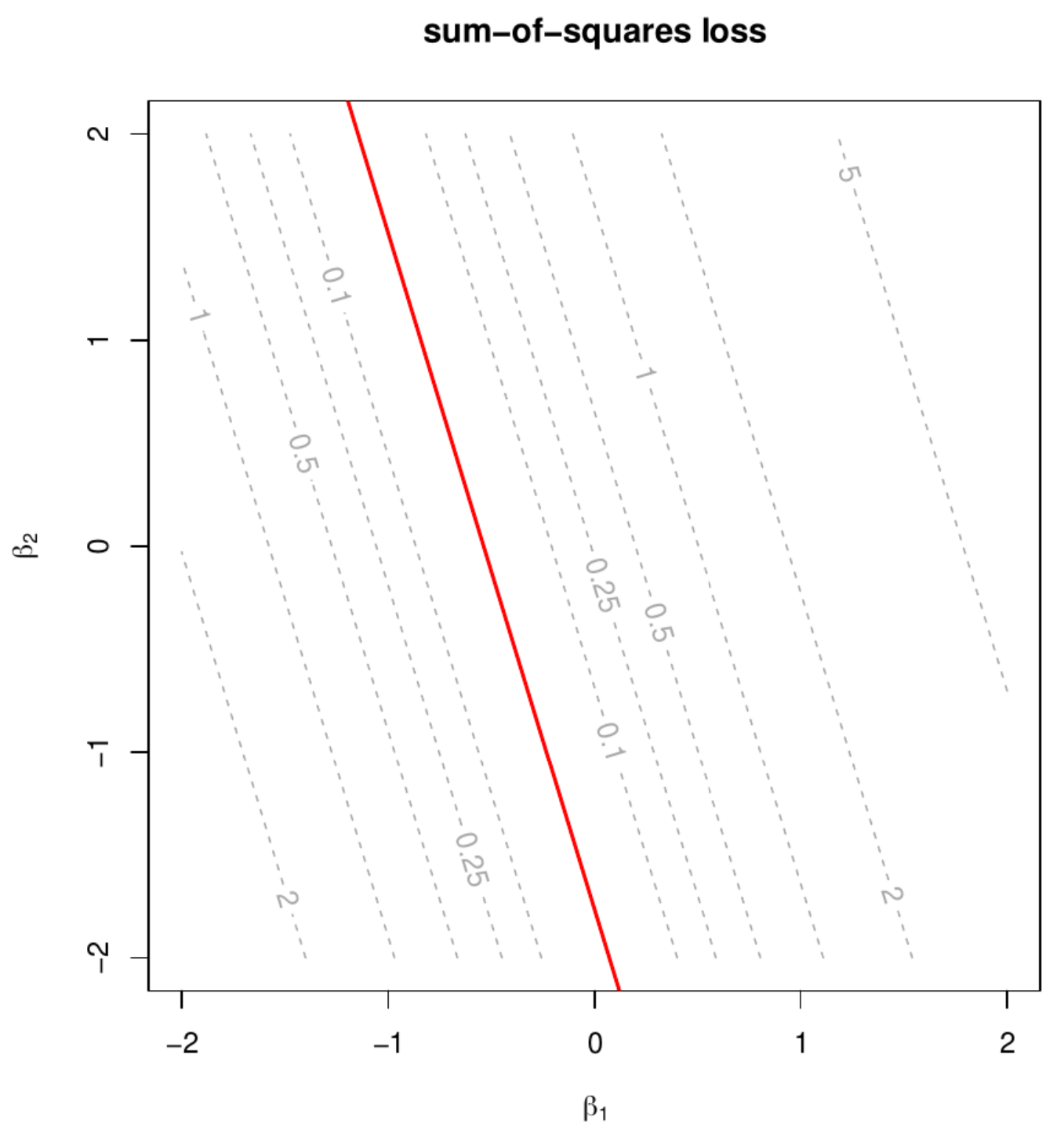}
& &
\includegraphics[scale=0.23, angle=0]{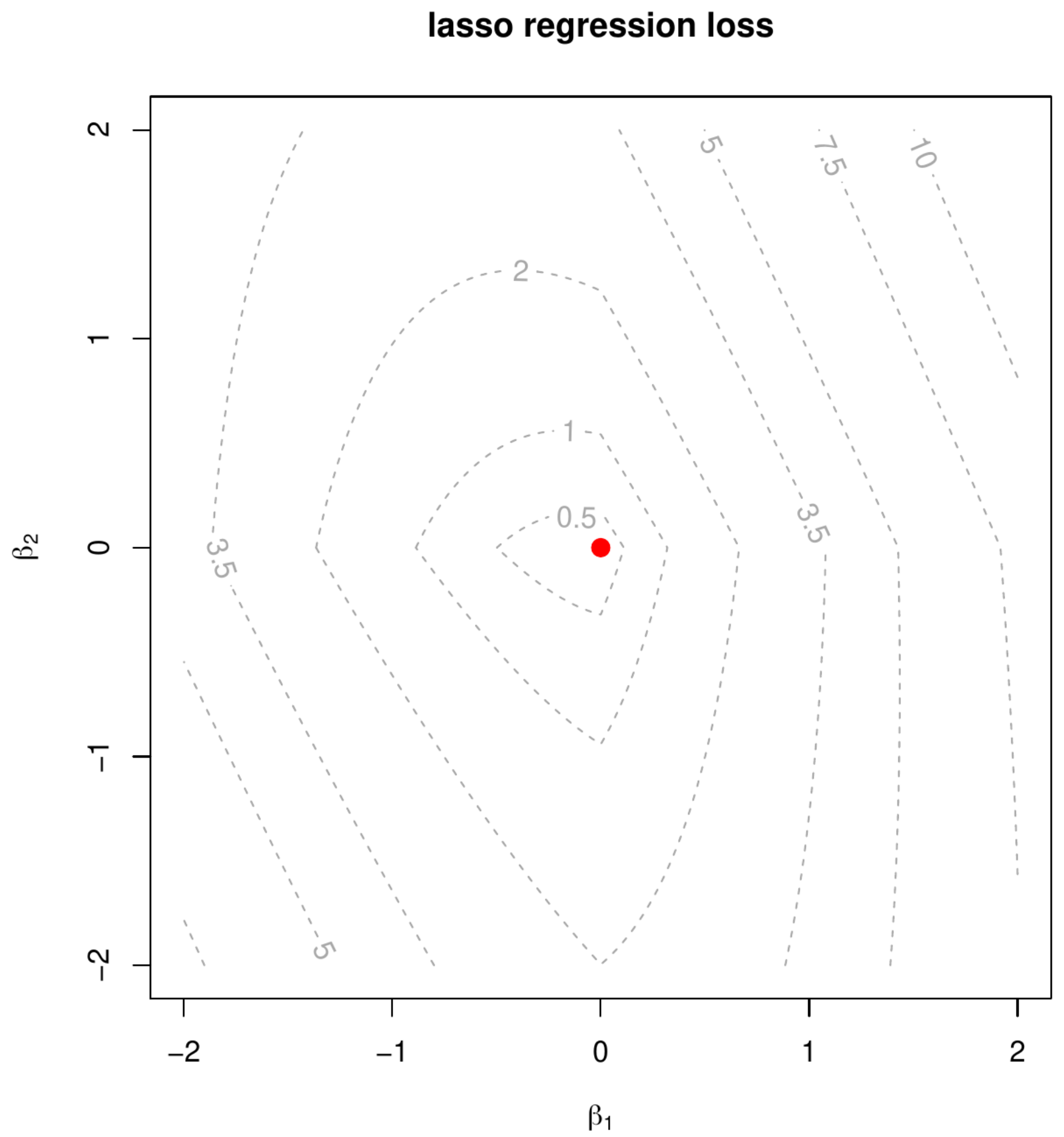}
\end{tabular}
\caption{Contour plots of the sum-of-squares and the lasso regression loss (left and right panel, respectively). The dotted grey line represent level sets. The red line and dot represent the location of minimum in both panels.} \label{fig.contourLoss}
\end{figure}

The $\ell_1$ penalty of lasso regression is equally arbitrary as the $\ell_2^2$-penalty of ridge regression. The latter ensured the existence of a well-defined estimator of the regression parameter $\bbeta$ in the presence of super-collinearity in the design matrix $\mathbf{X}$, in particular when the dimension $p$ exceeds the sample size $n$. The augmentation of the sum-of-squares with the lasso penalty achieves the same. This is illustrated in Figure \ref{fig.contourLoss}. For the high-dimensional setting with $p=2$ and $n=1$ and arbitrary data the level sets of the sum-of-squares $\| \mathbf{Y} - \mathbf{X} \, \bbeta \|^2_2$ and the lasso regression loss $\| \mathbf{Y} - \mathbf{X} \, \bbeta \|^2_2 + \lambda_1 \| \bbeta \|_1$ are plotted (left and right panel, respectively). In both panels the minimum is indicated in red. For the sum-of-squares the minimum is a line. As pointed out before in Section \ref{sect.ridgeRegression} of Chapter \ref{chap:ridgeRegression} on ridge regression, this minimum is determined up to an element of the null set of the design matrix $\mathbf{X}$, which due to the rank deficiency of $\mathbf{X}$ is non-trivial. In contrast, the lasso regression loss exhibits a unique well-defined minimum. Hence, the augmentation of the sum-of-squares with the lasso penalty yields a well-defined estimator of the regression parameter. (This needs some attenuation: in general the minimum of the lasso regression loss need not be unique, confer Section \ref{sect:lassoUniqueness}).
\\
\\
The mathematics involved in the derivation in this chapter tends to be more intricate than for ridge regression. This is due to the non-differentiability of the lasso penalty at zero. This has consequences on all aspects of the lasso regression estimator as is already obvious in the right-hand panel of Figure \ref{fig.contourLoss}: confer the non-differentiable points of the lasso regression loss level sets.

\section{Uniqueness}  \label{sect:lassoUniqueness}
The lasso regression estimator need not be unique. This can loosely be concluded from the lasso regression loss function, which is the sum of the sum-of-squares criterion and a sum of absolute value functions. Both are convex in $\bbeta$: the former is not strictly convex due to the high-dimensionality and the fact that the absolute value function is convex but not strictly convex due to its piece-wise linearity. Thereby the lasso loss function too is convex but not necessarily strictly convex. Consequently, its minimum need not be uniquely defined. But the set of solutions of a convex minimization problem is convex (Theorem 9.4.1, Fletcher, 1987). Hence, would there exist multiple minimizers of the lasso loss function, they can be used to construct a convex set of minimizers. Thus, if $\hat{\bbeta}_a (\lambda_1)$ and  $\hat{\bbeta}_b (\lambda_1)$ are lasso estimators, then so are $(1-\theta) \hat{\bbeta}_a (\lambda_1) + \theta \hat{\bbeta}_b (\lambda_1)$ for $\theta \in (0,1)$. This is illustrated in Example \ref{example.nonuniqueLasso}.

\begin{example} \textit{(Super-collinear covariates)} \label{example.nonuniqueLasso}
\\
Consider the standard linear regression model $Y_i = \mathbf{X}_{i,\ast} \bbeta + \varepsilon_i$ for $i=1, \ldots, n$ and with the $\varepsilon_i$ i.i.d. normally distributed with zero mean and a common variance. The rows of the design matrix $\mathbf{X}$ are of length two, neither column represents the intercept, but $\mathbf{X}_{\ast, 1} = \mathbf{X}_{\ast, 2}$. Suppose an estimate of the regression parameter $\bbeta$ of this model is obtained through the minimization of the sum-of-squares augmented with a lasso penalty, $\| \mathbf{Y} - \mathbf{X} \bbeta \|_2^2 + \lambda_1 \| \bbeta \|_1$ with penalty parameter $\lambda_1 > 0$. To find the minimizer define $u = \beta_1 + \beta_2$ and $v = \beta_1 - \beta_2$ and rewrite the lasso loss criterion to:
\begin{eqnarray*}
\| \mathbf{Y} - \mathbf{X} \bbeta \|_2^2 + \lambda_1 \| \bbeta \|_1
& = &  \| \mathbf{Y} - \mathbf{X}_{\ast,1} u \|_2^2 + \tfrac{1}{2} \lambda_1 ( | u + v |+ | u - v |).
\end{eqnarray*}
The function $| u + v |+ | u - v |$ is minimized with respect to $v$ for any $v$ such that $|v| < |u|$ and the corresponding minimum equals $2| u|$. The  estimator of $u$ thus minimizes:
\begin{eqnarray*}
\| \mathbf{Y} - \mathbf{X}_{\ast,1} u \|_2^2 + \lambda_1  | u|.
\end{eqnarray*}
For sufficiently small values of $\lambda_1$, the estimate of $u$ will be unequal to zero. Then, any $v$ such that $|v| < |u|$ will yield the same minimum of the lasso loss function. Consequently,
$\hat{\bbeta}(\lambda_1)$ is not uniquely defined as
$\hat{\beta}_1(\lambda_1) = \tfrac{1}{2} [\hat{u}(\lambda_1) + \hat{v}(\lambda_1)]$ need not equal $\hat{\beta}_2(\lambda_1) = \tfrac{1}{2} [\hat{u}(\lambda_1) - \hat{v}(\lambda_1)]$ for any $\hat{v}(\lambda_1)$ such that $0 < | \hat{v}(\lambda_1) | < |\hat{u}(\lambda_1)|$.
\end{example}

\noindent
The non-uniquess of the lasso regression estimator may not be its most appealing property, it is unproblematic in many practical settings. The following lemma states that the estimator is, with high probability and under a mild condition on the design matrix, unique. 

\begin{lemma} \label{lemma.lassoUniqueness} (Lemma 4, \citealp{Tibs2013}) 
\\
If the elements of $\mathbf{X} \in \mathcal{M}^{n,p}$ are drawn from  a continuous probability distribution on $\mathbb{R}^{n \times p}$, 
then for any $\mathbf{Y}$ and $\lambda  > 0$, the lasso solution is unique with probability one.
\end{lemma}
\begin{proof}
See \cite{Tibs2013}.
\end{proof}

\noindent
Lemma \ref{lemma.lassoUniqueness} rules out two important features of Example \ref{example.nonuniqueLasso} that lead to the non-uniqueness of the lasso regression estimator. First, the randomness of a low-dimensional design matrix' elements ensures that the probability of supercollinearity among the columns $\mathbf{X}$ is zero. Secondly, the length of these columns is unlikely to be equal, i.e. $P(\| \mathbf{X}_{\ast,j} \|_2 \not= \| \mathbf{X}_{\ast,j'} \|_2 ) = 1$ for $j, j' = 1, \ldots, p$ such that $j \not= j'$. 

While the lasso regression estimator $\hat{\bbeta}(\lambda_1)$ need not be unique, its linear predictor  $\mathbf{X} \hat{\bbeta}(\lambda_1)$ is. This suffices as prediction is typically of primary interest leaving the non-uniqueness of the estimator a lesser concern. 

\begin{theorem} \label{theorem:uniqueLassoPredictor} \citep{Tibs2013} \\
The lasso linear predictor, $\mathbf{X} \hat{\bbeta}(\lambda_1)$, is unique.
\end{theorem}
\begin{proof}
The proof is by contradiction. Suppose there exists two lasso regression estimators of $\bbeta$, denoted $\hat{\bbeta}_a (\lambda_1)$ and  $\hat{\bbeta}_b (\lambda_1)$, such that $\mathbf{X} \hat{\bbeta}_a (\lambda_1) \not= \mathbf{X} \hat{\bbeta}_b (\lambda_1)$. Define $c$ to be the minimum of the lasso loss function. Then, by definition of the lasso regression estimators $\hat{\bbeta}_a (\lambda_1)$ and $\hat{\bbeta}_b (\lambda_1)$ satisfy:
\begin{eqnarray*}
\| \mathbf{Y} - \mathbf{X} \hat{\bbeta}_a (\lambda_1)  \|_2^2
+ \lambda_1 \| \hat{\bbeta}_a (\lambda_1) \|_1 & = & c \, \, \, = \, \, \,  \| \mathbf{Y} - \mathbf{X} \hat{\bbeta}_b (\lambda_1)  \|_2^2 + \lambda_1 \| \hat{\bbeta}_b (\lambda_1) \|_1.
\end{eqnarray*}
For $\theta \in (0,1)$, we then have:
\begin{eqnarray*}
& & \hspace{-1cm} \| \mathbf{Y} - \mathbf{X} [(1-\theta) \hat{\bbeta}_a (\lambda_1) + \theta \hat{\bbeta}_b (\lambda_1)] \|_2^2
+ \lambda_1 \| (1-\theta) \hat{\bbeta}_a (\lambda_1) + \theta \hat{\bbeta}_b (\lambda_1) \|_1
\\
& = & \| (1-\theta)[ \mathbf{Y} - \mathbf{X} \hat{\bbeta}_a (\lambda_1) ] + \theta [ \mathbf{Y} - \mathbf{X} \hat{\bbeta}_b (\lambda_1)] \|_2^2
+ \lambda_1 \| (1-\theta) \hat{\bbeta}_a (\lambda_1) + \theta \hat{\bbeta}_b (\lambda_1) \|_1
\\
& < & (1-\theta)
\| \mathbf{Y} - \mathbf{X} \hat{\bbeta}_a (\lambda_1) \|_2^2  + \theta \| \mathbf{Y} - \mathbf{X} \hat{\bbeta}_b (\lambda_1) \|_2^2
+ (1-\theta) \lambda_1 \| \hat{\bbeta}_a (\lambda_1) \|_1 + \theta \lambda_1 \| \hat{\bbeta}_b (\lambda_1) \|_1
\\
& = & (1-\theta) c + \theta c \, \, \, = \, \, \, c,
\end{eqnarray*}
by the strict convexity of $\| \mathbf{Y} - \mathbf{X} \bbeta \|_2^2$ in $\mathbf{X} \bbeta$ and the convexity of $\| \bbeta \|_1$ on $\theta \in (0,1)$. This implies that $(1-\theta) \hat{\bbeta}_a (\lambda_1) + \theta \hat{\bbeta}_b (\lambda_1)$ yields a lower minimum of the lasso loss function and contradicts our assumption that $\hat{\bbeta}_a (\lambda_1)$ and $\hat{\bbeta}_b (\lambda_1)$ are lasso regression estimators. Hence, the lasso linear predictor is unique.
\end{proof}
We illustrate Lemma \ref{theorem:uniqueLassoPredictor} on Example \ref{example.nonuniqueLasso}.
\begin{example} \textit{(Perfectly super-collinear covariates, revisited)}
\\
Revisit the setting of Example \ref{example.nonuniqueLasso}, where a linear regression model without intercept and only two but perfectly correlated covariates is fitted to data. The example revealed that the lasso estimator need not be unique. The lasso predictor, however, is
\begin{eqnarray*}
\widehat{\mathbf{Y}}(\lambda_1) \, \, \, = \, \, \,\mathbf{X} \hat{\bbeta}(\lambda_1) \, \, \, = \, \, \, \mathbf{X}_{\ast,1} \hat{\beta}_1(\lambda_1) + \mathbf{X}_{\ast,2} \hat{\beta}_2(\lambda_1) & = & \mathbf{X}_{\ast,1} [ \hat{\beta}_1(\lambda_1) + \hat{\beta}_2(\lambda_1)] \, \, \, = \, \, \,
\mathbf{X}_{\ast,1} \hat{u}(\lambda_1),
\end{eqnarray*}
with $u$ defined and (uniquely) estimated as in Example \ref{example.nonuniqueLasso} and $v$ dropping from the predictor.
\end{example}

\begin{example} \mbox{ }
\\
The issues, non- and uniqueness of the lasso-estimator and predictor, respectively, raised above are illustrated in a numerical setting. Hereto data are generated in accordance with the linear regression model $\mathbf{Y} = \mathbf{X} \bbeta + \vvarepsilon$ where the $n=5$ rows of $\mathbf{X}$ are sampled from $\mathcal{N}[\mathbf{0}_p, (1-\rho) \mathbf{I}_{pp} + \rho \mathbf{1}_{pp}]$ with $p=10$, $\rho=0.99$, $\bbeta = (\mathbf{1}_3^{\top}, \mathbf{0}_{p-3}^{\top})^{\top}$ and $\vvarepsilon \sim \mathcal{N}(\mathbf{0}_p, \tfrac{1}{10} \mathbf{I}_{nn})$. With these data the lasso estimator of the regression parameter $\bbeta$ for $\lambda_1=1$ is evaluated using two different algorithms (see Section \ref{sect:lassoEstimation}). Employed implementations of the algorithms are those available through the \texttt{R}-packages \texttt{penalized} and \texttt{glmnet}. Both estimates, denoted $\hat{\bbeta}_{\texttt{p}} (\lambda_1)$ and $\hat{\bbeta}_{\texttt{g}} (\lambda_1)$ (the subscript refers to the first letter of the package), are given in Table \ref{table:lassoNonunique}.
\begin{table}[h!]
\begin{center}
\begin{tabular}{rrrrrrrrrrrr}
\hline
& & & & & & & & & & &
\vspace{-7pt}
\\
& & $\beta_1$ & $\beta_2$ & $\beta_3$ & $\beta_4$ & $\beta_5$ & $\beta_6$ & $\beta_7$ & $\beta_8$ & $\beta_9$ & $\beta_{10}$
\\
\hline
& & & & & & & & & & &
\vspace{-7pt}
\\
\texttt{penalized} & $\hat{\bbeta}_{\texttt{p}} (\lambda_1)$ & 0.267 & 0.000 & 1.649 & 0.093 & 0.000 & 0.000 & 0.000 & 0.571 & 0.000 & 0.269
\\
\texttt{glmnet} & $\hat{\bbeta}_{\texttt{g}} (\lambda_1)$	& 0.269 & 0.000 & 1.776 & 0.282 & 0.195 & 0.000 & 0.000 & 0.325 & 0.000 & 0.000
\\
\hline
\end{tabular}
\label{table:lassoNonunique}
\caption{Lasso estimates of the linear regression $\bbeta$ for both algorithms.}
\end{center}
\end{table}
The table reveals that the estimates differ, in particular in their support (i.e. the set of nonzero values of the estimate of $\bbeta$). This is troublesome when it comes to communication of the optimal model. From a different perspective the realized loss $\| \mathbf{Y} - \mathbf{X} \bbeta \|_2^2 + \lambda_1 \| \bbeta \|_1$ for each estimate is approximately equal to $2.99$, with the difference possibly due to convergence criteria of the algorithms. On another note, their corresponding predictors, $\mathbf{X} \hat{\bbeta}_{\texttt{p}} (\lambda_1)$ and $\mathbf{X} \hat{\bbeta}_{\texttt{g}} (\lambda_1)$, correlate almost perfectly:
$\mbox{cor}[\mathbf{X} \hat{\bbeta}_{\texttt{p}} (\lambda_1), \mathbf{X} \hat{\bbeta}_{\texttt{g}} (\lambda_1)] = 0.999$. These results thus corroborate the non-uniqueness of the estimator and the uniqueness of the predictor.
\\
\\
The \texttt{R}-script provides the code to reproduce the analysis.
\begin{lstlisting}
# set the random seed
set.seed(4)

# load libraries
library(penalized)
library(glmnet)
library(mvtnorm)

# set sample size
p <- 10

# create covariance matrix
Sigma <- matrix(0.99, p, p)
diag(Sigma) <- 1

# sample the design matrix
n <- 5
X <- rmvnorm(10, sigma=Sigma)

# create a sparse beta vector
betas <- c(rep(1, 3), rep(0, p-3))

# sample response
Y <- X %*% betas + rnorm(n, sd=0.1)

# evaluate lasso estimator with two methods
Bhat1 <- matrix(as.numeric(coef(penalized(Y, X, lambda1=1, unpenalized=~0), 
                                "all")), ncol=1)
Bhat2 <- matrix(as.numeric(coef(glmnet(X, Y, lambda=1/(2*n), standardize=FALSE, 
                                intercept=FALSE)))[-1], ncol=1)

# compare estimates
cbind(Bhat1, Bhat2)

# compare the loss
sum((Y - X %*% Bhat1)^2) + sum(abs(Bhat1))
sum((Y - X %*% Bhat2)^2) + sum(abs(Bhat2))

# compare predictor
cor(X %*% Bhat1, X %*% Bhat2)

\end{lstlisting}
Note that in the code above the evaluation of the lasso estimator appears to employ a different lasso penalty parameter $\lambda_1$ for each package. This is due to the fact that internally (after removal of standardization of $\mathbf{X}$ and $\mathbf{Y}$) the loss functions optimized are $\| \mathbf{Y} - \mathbf{X} \bbeta \|_2^2 + \lambda_1 \| \bbeta \|_1$ vs. $(2n)^{-1} \| \mathbf{Y} - \mathbf{X} \bbeta \|_2^2 + \lambda_1 \| \bbeta \|_1$. Rescaling of $\lambda_1$ resolves this issue.
\end{example}

\section{Analytic solutions} \label{sect:lassoAnalytic}
In general, no explicit expression for the lasso regression estimator exists. There are exceptions, as illustrated in Examples \ref{example.orthonormalDesignLasso} and \ref{example:lassoPequals}. Nonetheless, it is possible to show properties of the lasso estimator, amongst others of the smoothness of its regularization path (Theorem \ref{theo.lassoPiecewiseLinear}) and the limiting behaviour as $\lambda_1 \rightarrow \infty$ (see the end of this section).

\begin{theorem} (Theorem 2, \citealp{Ross2007}) \label{theo.lassoPiecewiseLinear}
\\
The lasso regression loss function (\ref{form.lassoLossFunction}) yields a piecewise linear, in $\lambda_1$, regularization path $\{ \hat{\bbeta}(\lambda_1) : \lambda_1 \in \mathbb{R}_{>0} \}$.
\end{theorem}
\begin{proof}
Confer \cite{Ross2007}.
\end{proof}

\begin{figure}[h!]
\begin{tabular}{rcl}
\includegraphics[scale=0.23, angle=0]{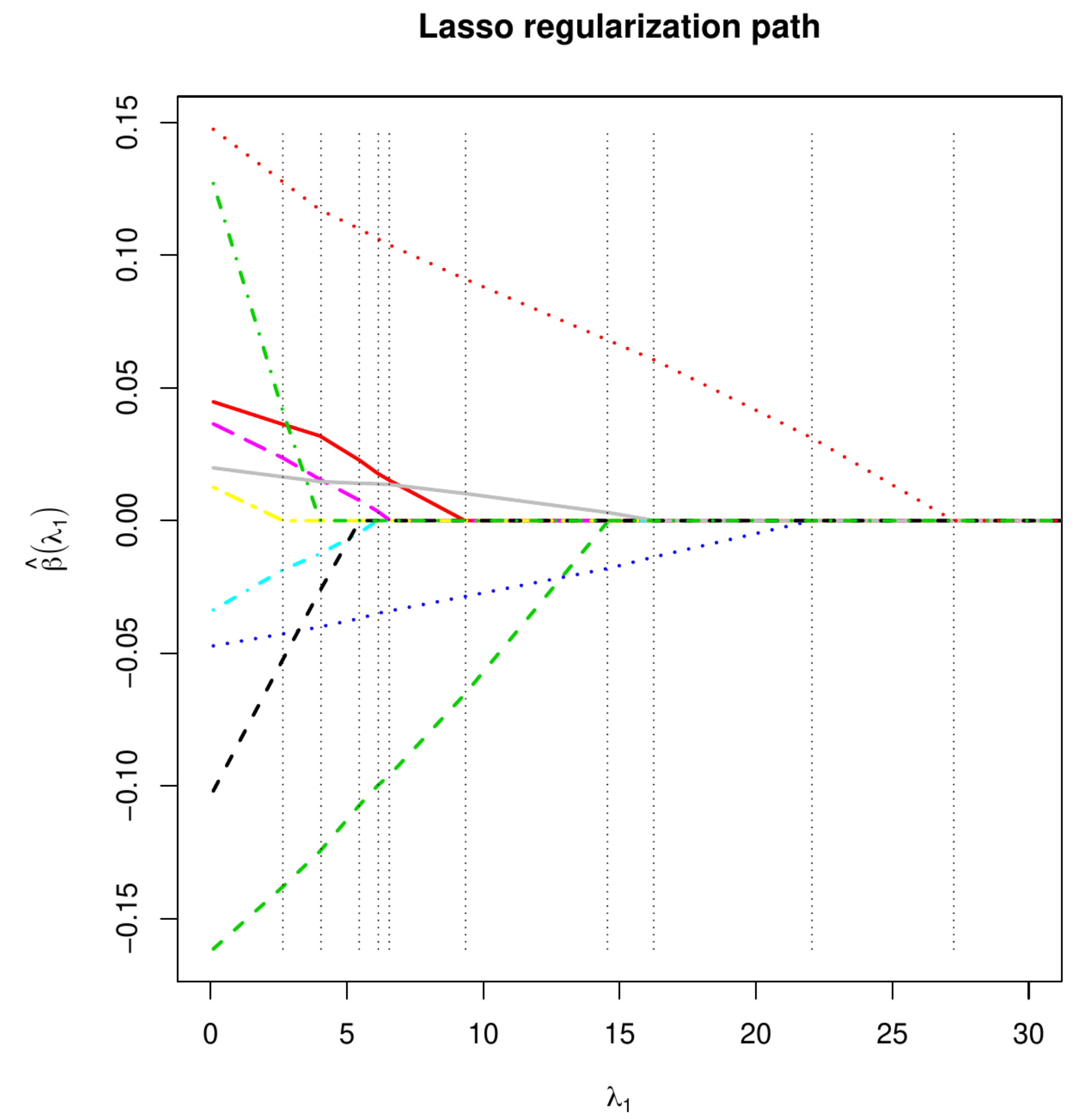}
& &
\includegraphics[scale=0.23, angle=0]{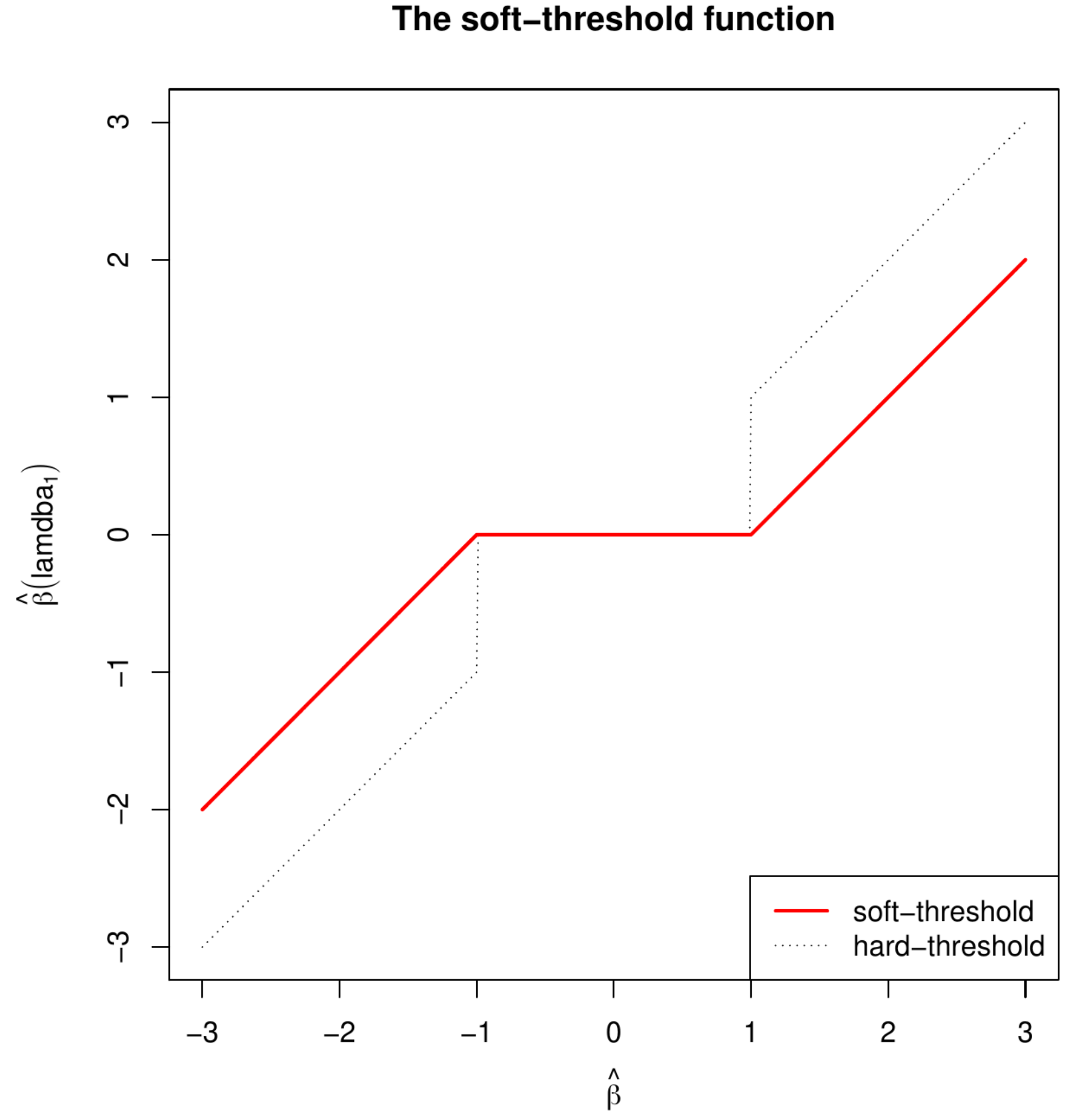}
\end{tabular}
\caption{The left panel shows the regularization path of the lasso regression estimator for simulated data. The vertical grey dotted lines indicate the values of $\lambda_1$ at which there is a discontinuity in the derivative (with respect to $\lambda_1$) of the lasso regularization path of one the regression estimates. The right panel displays the soft (solid, red) and hard (grey, dotted) threshold functions.
\label{fig.lassoSolution}}
\end{figure}

\noindent
This piecewise linear nature of the lasso solution path is illustrated in the left-hand panel of Figure \ref{fig.lassoSolution} of an arbitrary data set. At each vertical dotted line a discontinuity in the derivative with respect to $\lambda_1$ of the regularization path of a lasso estimate of an element of $\bbeta$ may occur. The plot also foreshadows the $\lambda_1 \rightarrow \infty$ limiting behaviour of the lasso regression estimator: the estimator tends to zero. This is no surprise knowing that the ridge regression estimator exhibits the same behaviour and the lasso regression loss function is of similar form as that of ridge regression: a sum-of-squares plus a penalty term (which is linear in the penalty parameter).
\\
\\
For particular cases, an orthonormal design (Example \ref{example.orthonormalDesignLasso}) and $p=2$ (Example \ref{example:lassoPequals}), an analytic expression for the lasso regression estimator exists. While the latter is of limited use, the former is exemplary and will come of use later in the numerical evaluation of the lasso regression estimator in the general case (see Section \ref{sect:lassoEstimation}).

\begin{example} \textit{(Orthonormal design matrix)} \label{example.orthonormalDesignLasso}
\\
Consider an orthonormal design matrix $\mathbf{X}$, i.e. $\mathbf{X}^{\top} \mathbf{X} = \mathbf{I}_{pp}  =  (\mathbf{X}^{\top} \mathbf{X})^{-1}$. The lasso estimator then is:
\begin{eqnarray*}
\hat{\beta}_j (\lambda_1) & = & \mbox{sign} (\hat{\beta}_j ) (|\hat{\beta}_j| - \tfrac{1}{2} \lambda_1)_+,
\end{eqnarray*}
where $\hat{\bbeta} = (\mathbf{X}^{\top} \mathbf{X})^{-1} \mathbf{X}^{\top} \mathbf{Y} = \mathbf{X}^{\top} \mathbf{Y}$ is the maximum likelihood estimator of $\bbeta$ and $\hat{\beta}_j$ its $j$-th element and $(x)_+ = \max\{x, 0\}$. This expression for the lasso regression estimator can be obtained as follows.  Rewrite the lasso regression loss criterion:
\begin{eqnarray*}
\min_{\bbeta} \, \big( \| \mathbf{Y} - \mathbf{X} \, \bbeta \|^2_2 + \lambda_1 \| \bbeta \|_1  \big)  & = & \min_{\bbeta}  \, \big( \mathbf{Y}^{\top} \mathbf{Y} - \mathbf{Y}^{\top} \mathbf{X} \, \bbeta - \bbeta^{\top} \mathbf{X}^{\top} \mathbf{Y} + \bbeta^{\top} \mathbf{X}^{\top}  \mathbf{X} \, \bbeta + \lambda_1 \sum\nolimits_{j=1}^p | \beta_j | \big)
\\
& = & \min_{\bbeta} \, \big( - \hat{\bbeta}^{\top} \, \bbeta - \bbeta^{\top} \hat{\bbeta} + \bbeta^{\top} \, \bbeta + \lambda_1 \sum\nolimits_{j=1}^p | \beta_j | \big)
\\
& = & \min_{\beta_1, \ldots, \beta_p} \, \Big[ \sum\nolimits_{j=1}^p \big( - 2 \hat{\beta}_j^{\mbox{\tiny ml}} \, \beta_j + \beta_j^2 + \lambda_1 | \beta_j | \big) \Big]
\\
& = & \sum\nolimits_{j=1}^p \, \big[ \min_{\beta_j} \, ( - 2 \hat{\beta}_j \, \beta_j + \beta_j^2 + \lambda_1 | \beta_j | ) \big]. 
\end{eqnarray*}
The minimization problem can thus be solved per regression coefficient. This gives:
\begin{eqnarray*}
\min_{\beta_j} \, \big(  - 2 \hat{\beta}_j \, \beta_j + \beta_j^2 + \lambda_1 | \beta_j | \big) & = &
\left\{
\begin{array}{lcr}
\min_{\beta_j} \, \big( - 2 \hat{\beta}_j \, \beta_j + \beta_j^2 + \lambda_1  \beta_j \big) & \mbox{if} & \beta_j > 0,
\\
\min_{\beta_j} \, \big(  - 2 \hat{\beta}_j \, \beta_j + \beta_j^2 - \lambda_1  \beta_j \big) & \mbox{if} & \beta_j < 0.
\end{array}
\right.
\end{eqnarray*}
The minimization within the sum over the covariates is with respect to each  element of the regression parameter separately. Optimization with respect to the $j$-th one gives:
\begin{eqnarray*}
\hat{\beta}_j (\lambda_1) & = &
\left\{
\begin{array}{lcr}
\hat{\beta}_j  - \frac{1}{2} \lambda_1  & \mbox{if} & \hat{\beta}_j (\lambda_1) > 0
\\
\hat{\beta}_j + \frac{1}{2} \lambda_1  & \mbox{if} & \hat{\beta}_j (\lambda_1) < 0
\\
0 & \multicolumn{2}{l}{\mbox{otherwise}}
\end{array}
\right.
\end{eqnarray*}
This case-wise solution can be compressed to the form of the lasso regression estimator above.

The analytic expression for the lasso regression estimator above provides insight in how it relates to the maximum likelihood estimator of $\bbeta$. The right-hand side panel of Figure \ref{fig.lassoSolution} depicts this relationship. Effectively, the lasso regression estimator thresholds (after a translation) its maximum likelihood counterpart. The function is also referred to as the \textit{soft-threshold function} (for contrast the hard-threshold function is also plotted -- dotted line -- in Figure  \ref{fig.lassoSolution}).
\end{example}

\begin{example} \textit{(Orthogonal design matrix)} \label{example.orthogonalDesignLasso}
\\
The analytic solution of the lasso regression estimator for experiments with an orthonormal design matrix applies to those with an orthogonal design matrix. This is illustrated by a numerical example. Use the lasso estimator with $\lambda_1=10$ to fit the linear regression model to the response data and the design matrix:
\begin{eqnarray*}
\mathbf{Y}^{\top} & = &
\left(
\begin{array}{rrrrrr}
-4.9 & -0.8 & -8.9 & 4.9 & 1.1 &  -2.0
\end{array}
\right),
\\
\mathbf{X}^{\top} & = & \left(
\begin{array}{rrrrrr}
1 &  -1 &   3 &  -3 &   1 &   1
\\
-3 &  -3 &  -1 &   0 &   3 &   0
\end{array}
\right).
\end{eqnarray*}
Note that the design matrix is orthogonal, i.e. its columns are orthogonal (but not normalized to one). The orthogonality of $\mathbf{X}$ yields a diagonal $\mathbf{X}^{\top} \mathbf{X}$, and so is its inverse $(\mathbf{X}^{\top} \mathbf{X})^{-1}$. Here $\mbox{diag}(\mathbf{X}^{\top} \mathbf{X}) = (22, 28)$. Rescale $\mathbf{X}$ to an orthonormal design matrix, denoted $\tilde{\mathbf{X}}$, and rewrite the lasso regression loss function to:
\begin{eqnarray*}
\| \mathbf{Y} - \mathbf{X} \bbeta \|_2^2 + \lambda_1 \| \bbeta \|_1 & = &
\Bigg\| \, \mathbf{Y} - \mathbf{X}
\left(
\begin{array}{rrrrrr}
\sqrt{22} &  0
\\
0 &  \sqrt{28} &
\end{array}
\right)^{-1}
\left(
\begin{array}{rrrrrr}
\sqrt{22} &  0
\\
0 &  \sqrt{28} &
\end{array}
\right)
\bbeta \, \Bigg\|_2^2 + \lambda_1 \| \bbeta \|_1
\\
& = & \| \mathbf{Y} - \tilde{\mathbf{X}}  \ggamma \|_2^2 +
(\lambda_1 / \sqrt{22}) | \gamma_1 |  + (\lambda_1 / \sqrt{28}) | \gamma_2 |,
\end{eqnarray*}
where $\ggamma = (\sqrt{22} \beta_1, \sqrt{28} \beta_2)^{\top}$. By the same argument this loss can be minimized with respect to each element of $\ggamma$ separately. In particular, the soft-threshold function provides an analytic expression for the estimates of $\ggamma$:
\begin{eqnarray*}
\hat{\gamma}_1 (\lambda_1 / \sqrt{22}) & = &  \mbox{sign} (\hat{\gamma}_1 ) [|\hat{\gamma}_1| - \tfrac{1}{2} (\lambda_1 / \sqrt{22})]_+ \, \, \, = \, \, - [ 9.892513 - \tfrac{1}{2} (10 / \sqrt{22})]_+ \, \, \, = \, \, -8.826509,
\\
\hat{\gamma}_2 (\lambda_1 / \sqrt{28}) & = &  \mbox{sign} (\hat{\gamma}_2 ) [|\hat{\gamma}_2| - \tfrac{1}{2} (\lambda_1 / \sqrt{28})]_+ \, \, \, = \, \, \, \, \, \, [5.537180 - \tfrac{1}{2} (10 / \sqrt{28})]_+ \, \, \, = \, \, \, \, \, \, \, 4.592269,
\end{eqnarray*}
where $\hat{\gamma}_1$ and $\hat{\gamma}_2$ are the ordinary least square estimates of $\gamma_1$ and $\gamma_2$ obtained from regressing $\mathbf{Y}$ on  the corresponding column of $\tilde{\mathbf{X}}$. Rescale back and obtain the lasso regression estimate: $\hat{\bbeta}(10) = (-1.881818, 0.8678572)^{\top}$.
\end{example}

\begin{example} \textit{($p=2$ with equivariant covariates, \citealp{Leng2006})} \label{example:lassoPequals} \\
Let $p=2$ and suppose the design matrix $\mathbf{X}$ has equivariant covariates. Without loss of generality they are assumed to have unit variance. We may thus write
\begin{eqnarray*}
\mathbf{X}^{\top} \mathbf{X} & = & \left( \begin{array}{rr} 1 & \rho \\ \rho & 1 \end{array} \right),
\end{eqnarray*}
for some $\rho \in (-1, 1)$. The lasso regression estimator is then of similar form as in the orthonormal case but the soft-threshold function now depends on $\lambda_1$, $\rho$ and the maximum likelihood estimate $\hat{\bbeta}$ (see Exercise \ref{exercise:lassoAnalyticSolutionp=2}).
\end{example}

\noindent
Apart from the specific cases outlined in the two examples above no other explicit solution for the minimizer of the lasso regression loss function appears to be known. Locally though, for large enough values of $\lambda_1$, an analytic expression for solution can also be derived. 
\begin{lemma} \mbox{ } \\
If $\lambda_1 > 2 \| \mathbf{X}^{\top} \mathbf{Y} \|_{\infty}$, 
where $\| \mathbf{a} \|_{\infty}$ is the supremum norm of vector $\mathbf{a}$ defined as $\| \mathbf{a} \|_{\infty} = \max \{ |a_1|, |a_2|, \ldots, |a_p | \}$, then $\hat{\bbeta}(\lambda_1) = \mathbf{0}_p$.  
\end{lemma}
\begin{proof}
We first observe that (\textbf{[details to be included later]}) the lasso regression estimator satisfies the following estimating equation:
\begin{eqnarray*}
\mathbf{X}^{\top} \mathbf{X} \hat{\bbeta} (\lambda_1) & = & \mathbf{X}^{\top} \mathbf{Y} - \tfrac{1}{2} \lambda_1 \hat{\mathbf{z}}
\end{eqnarray*}
for some $\hat{\mathbf{z}} \in \mathbb{R}^p$ with $(\hat{\mathbf{z}})_j = \mbox{sign}\{[\hat{\bbeta}(\lambda_1)]_j\}$ whenever $[\hat{\bbeta}_j(\lambda_1)]_j \not= 0$ and $(\hat{\mathbf{z}})_j \in [-1,1]$ if $[\hat{\bbeta}_j(\lambda_1)]_j = 0$. Then:
\begin{eqnarray*}
0 \, \, \, \leq \, \, \, [ \hat{\bbeta} (\lambda_1)]^{\top} \mathbf{X}^{\top} \mathbf{X} \hat{\bbeta} (\lambda_1) & = &  [ \hat{\bbeta} (\lambda_1)]^{\top} (\mathbf{X}^{\top} \mathbf{Y} - \tfrac{1}{2} \lambda_1 \hat{\mathbf{z}} )
\, \, \, = \, \, \, \sum\nolimits_{j=1}^p [\hat{\bbeta} (\lambda_1)]_j (\mathbf{X}^{\top} \mathbf{Y} - \tfrac{1}{2} \lambda_1 \hat{\mathbf{z}} )_j.
\end{eqnarray*}
For $\lambda_1 > 2 \| \mathbf{X}^{\top} \mathbf{Y} \|_{\infty}$ the summands on the right-hand side satisfy:
\begin{eqnarray*}
\begin{array}{rclcl}
\, [\hat{\bbeta} (\lambda_1)]_j (\mathbf{X}^{\top} \mathbf{Y} - \tfrac{1}{2} \lambda_1 \hat{\mathbf{z}} )_j & < & 0 & \mbox{if} &  [\hat{\bbeta} (\lambda_1)]_j > 0,
\\
\, [\hat{\bbeta} (\lambda_1)]_j (\mathbf{X}^{\top} \mathbf{Y} - \tfrac{1}{2} \lambda_1 \hat{\mathbf{z}} )_j & = & 0 & \mbox{if} &  [\hat{\bbeta} (\lambda_1)]_j = 0,
\\
\, [\hat{\bbeta} (\lambda_1)]_j (\mathbf{X}^{\top} \mathbf{Y} - \tfrac{1}{2} \lambda_1 \hat{\mathbf{z}} )_j & < & 0 & \mbox{if} &  [\hat{\bbeta} (\lambda_1)]_j < 0.
\end{array}
\end{eqnarray*}
This implies that $\hat{\bbeta} (\lambda_1) = \mathbf{0}_p$ if $\lambda_1 > 2 \| \mathbf{X}^{\top} \mathbf{Y} \|_{\infty}$.
\end{proof}

\section{Sparsity} \label{sect:sparsity}
The change from the $\ell_2^2$-norm to the $\ell_1$-norm in the penalty may seem only a detail. Indeed, both ridge and lasso regression fit the same linear regression model. But the attractiveness of the lasso lies not in {\it what} it fits, but in a {\it consequence of how} it fits the linear regression model. The lasso estimator of the vector of regression parameters may contain some or many zero's. In contrast, ridge regression yields an estimator of $\bbeta$ with elements (possibly) close to zero, but unlikely to be equal to zero. Hence, lasso penalization results in $\hat{\beta}_j (\lambda_1) = 0$ for some $j$ (in particular for large values of $\lambda_1$, see Section \ref{sect:lassoUniqueness}), while ridge penalization yields an estimate of the $j$-th element of the regression parameter $\hat{\beta}_j (\lambda_2) \not= 0$. A zero estimate of a regression coefficient means that the corresponding covariate has no effect on the response and can be excluded from the model. Effectively, this amounts to variable selection. Where traditionally the linear regression model is fitted by means of maximum likelihood followed by a testing step to weed out the covariates with effects indistinguishable from zero, lasso regression is a one-step-go procedure that simulatenously estimates and selects.

\begin{figure}[!h]
\begin{tabular}{rcl}
\includegraphics[scale=0.40, angle=0]{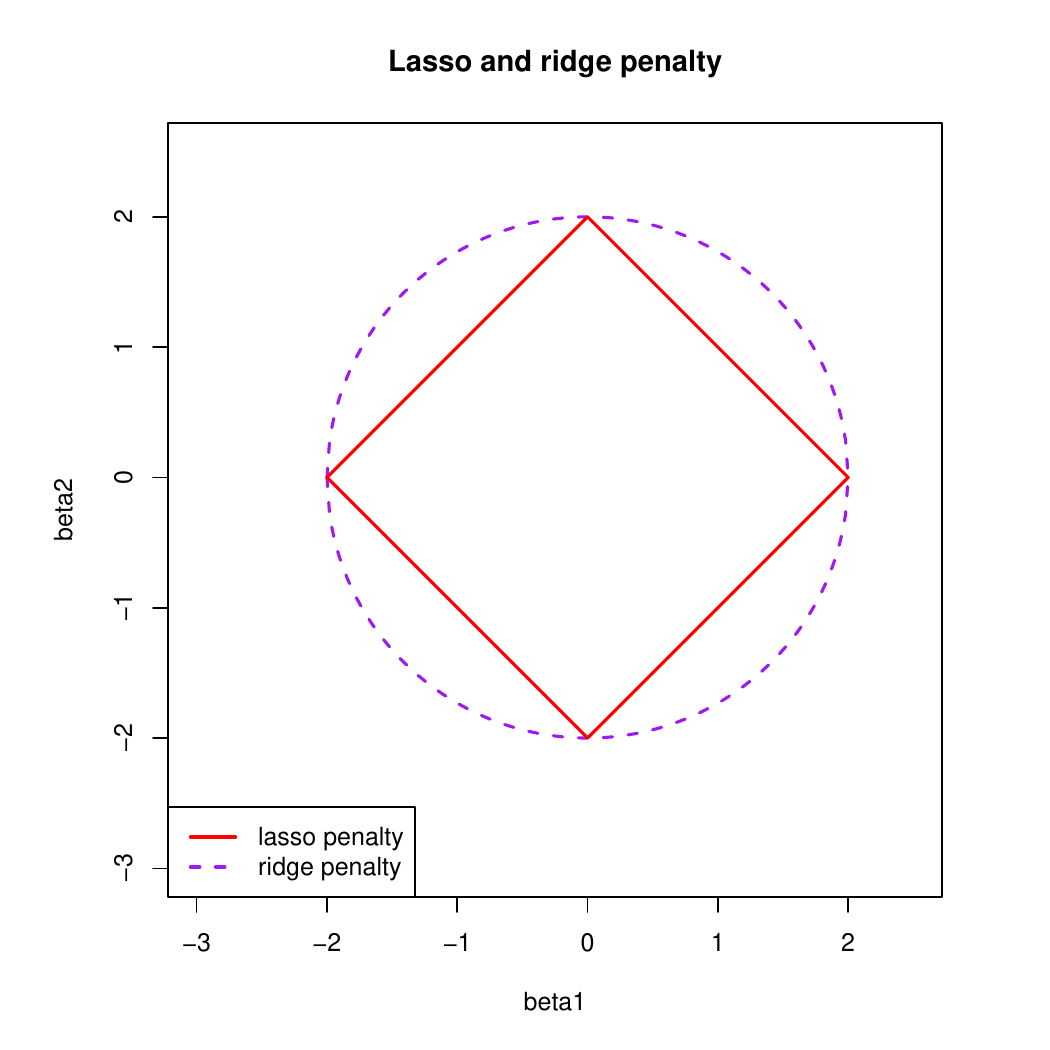}
& &
\includegraphics[scale=0.40, angle=0]{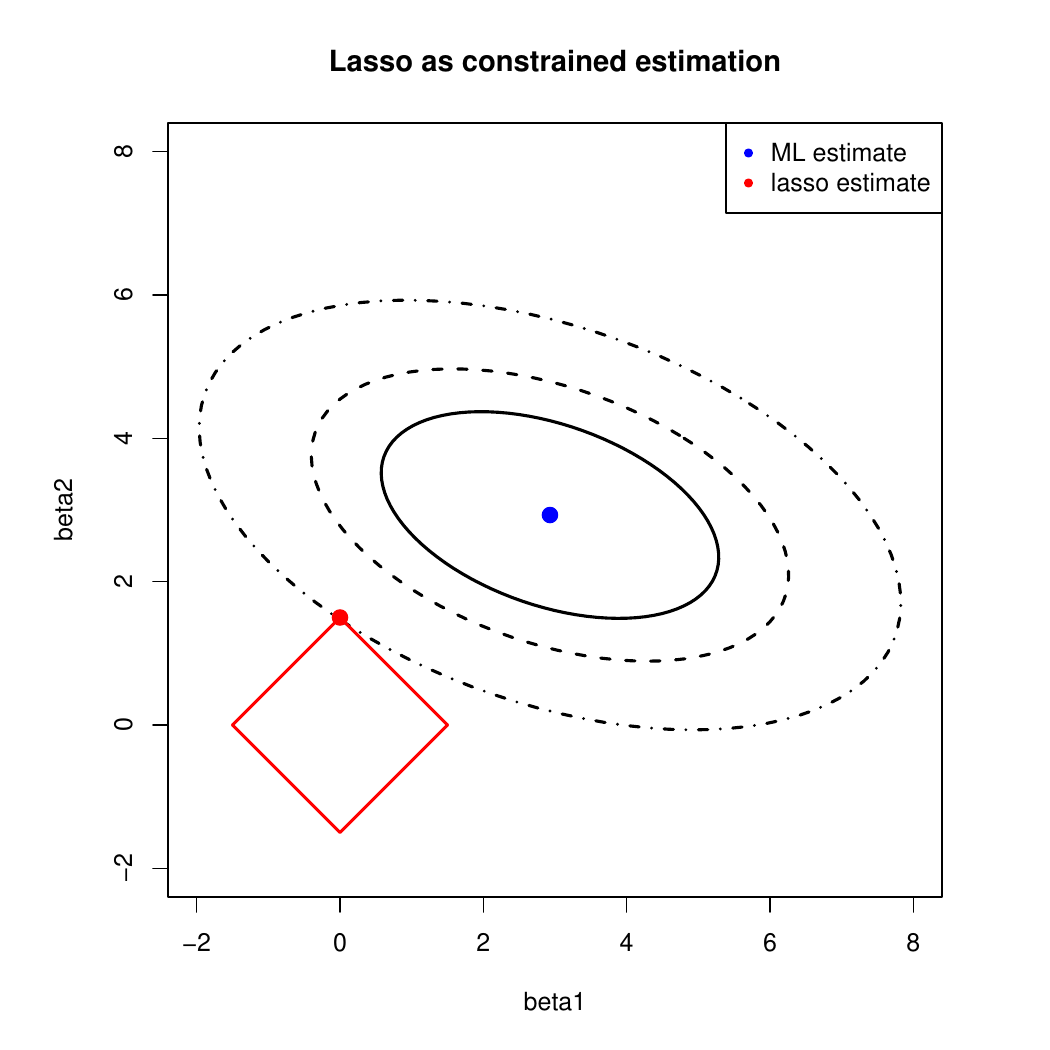}
\end{tabular}
\caption{Left panel: The lasso parameter constraint ($|\beta_1| + |\beta_2| \leq 2$) and its ridge counterpart ($\beta_1^2 + \beta_2^2 \leq 2$). Solution path of the ridge estimator and its variance. Right panel: the lasso regression estimator as a constrained least squares estimator.} \label{fig.lassoPenalty}
\end{figure}

The in-built variable selection of the lasso regression estimator is a geometric accident. To understand how it comes about, the lasso regression loss optimization problem (\ref{form.lassoLossFunction}) is reformulated as a constrained estimation problem (using the same argumentation as previously employed for ridge regression, see Section \ref{sect.constrainedEstimation}):
\begin{eqnarray*}
\hat{\bbeta} (\lambda_1) & = & \arg \min\nolimits_{ \{\bbeta \in \mathbb{R}^p \, : \, \|\bbeta \|_1 \leq c(\lambda_1) \} } \| \mathbf{Y} - \mathbf{X} \bbeta \|_2^2.
\end{eqnarray*}
where $c(\lambda_1) = \| \hat{\bbeta}(\lambda_1)\|_1$. Again, this is the standard least squares problem, with the only difference that the sum of the (absolute value of the) regression parameters $\beta_1, \beta_2, \ldots, \beta_p$ is required to be smaller than $c(\lambda_1)$.  The effect of this requirement is that the lasso estimator of the regression parameter coefficients can no longer assume any value (from $-\infty$ to $\infty$, as is the case in standard linear regression), but are limited to a certain range of values. With the lasso and ridge regression estimators minimizing the same sum-of-squares, the key difference with the constrained estimation formulation of ridge regression is not in the explicit form of $c(\lambda_1)$ (and is set to some arbitrary convenient value in the remainder of this section) but in what is bounded by $c(\lambda_1)$ and the domain of acceptable values for $\bbeta$ that it implies. For the lasso regression estimator the domain is specified by a bound on the $\ell_1$-norm of the regression parameter, while for its ridge counterpart the bound is applied to the squared $\ell_2$-norm of $\bbeta$. The parameter constraints implied by the lasso and ridge norms result in balls in different norms:
\begin{eqnarray*}
& & \{ \bbeta \in \mathbb{R}^p \, : \, | \beta_1 | + | \beta_2 | + \ldots + | \beta_p |  \leq c_1(\lambda_1) \},
\\
& & \{ \bbeta \in \mathbb{R}^p \, : \, \, \beta_1^2 \, \, + \, \beta_2^2 \, \, + \ldots + \, \beta_p^2  \, \leq c_2(\lambda_2) \},
\end{eqnarray*}
respectively, and where $c(\cdot)$ is now equipped with a subscript referring to the norm to stress that it is different for lasso and ridge. The left-hand panel of Figure \ref{fig.lassoPenalty} visualizes these parameter constraints for $p=2$ and $c_1(\lambda_1) = 2 = c_2(\lambda_2)$. In the Euclidean space ridge yields a spherical constraint for $\bbeta$, while a diamond-like shape for the lasso. The lasso regression estimator is then that $\bbeta$ inside this diamond domain that yields the smallest sum-of-squares (as is visualized by right-hand panel of Figure \ref{fig.lassoPenalty}).

The right-hand panel of Figure \ref{fig.lassoPenalty}) illustrates 
Lemma \ref{lemma.lassoUniqueness}. The lasso solution is non-unique if the levels sets of the sum-of-squares are parallel to one of the sides of the diamond parameter constraint. This occurs with probability zero. 

\begin{figure}[h!]
\begin{center}
\includegraphics[scale=0.28, angle=0]{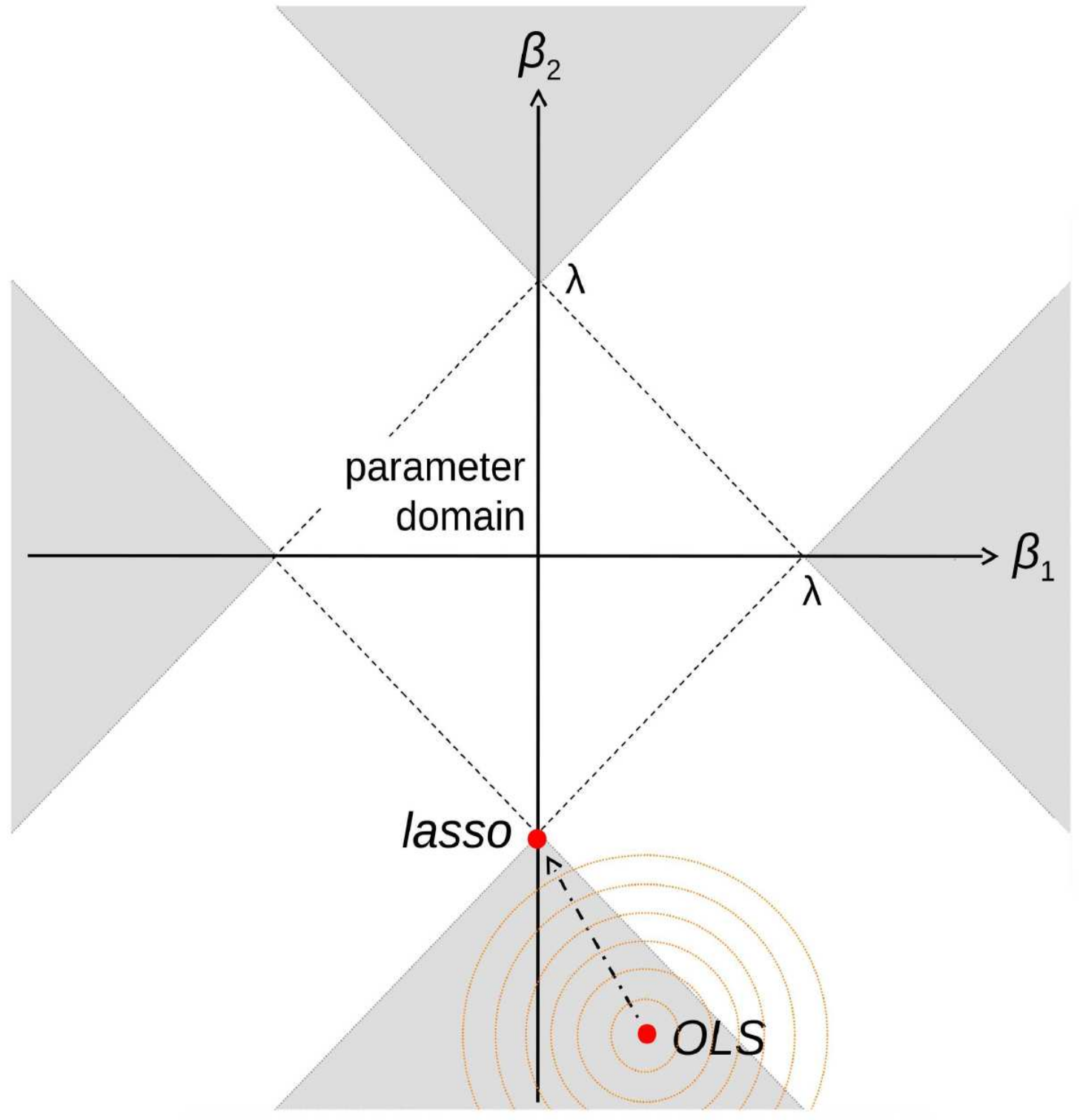}
\end{center}
\caption{Shrinkage with the lasso. The range of possible lasso estimates is demarcated by the diamond around the origin. The grey areas contain all points that are closest to one of the diamond's corners than to any other point inside the diamond. If the maximum likelihood estimate falls inside any of these grey areas, the lasso shrinks it to the closest diamond tip (which corresponds to a sparse solution). For example, let the red dot in the fourth quadrant be an maximum likelihood estimate. It is in a grey area. Hence, its lasso estimate is the red dot at the lowest tip of the diamond.
\label{fig.lassoParameterDomain}}
\end{figure}

The selection property of the lasso is due to the fact that the diamond-shaped parameter constraint has its corners falling on the axes. For a point to lie on an axis, one coordinate needs to equal zero. The lasso regression estimator coincides with the point inside the diamond closest to the maximum likelihood estimator. This point may correspond to a corner of the diamond, in which case one of the coordinates (regression parameters) equals zero and, consequently, the lasso regression estimator does not select this element of $\bbeta$. Figure \ref{fig.lassoParameterDomain} illustrates the selection property for the case with $p=2$ and an orthonormal design matrix. An orthornormal design matrix yields level sets (orange dotted circles in Figure \ref{fig.lassoParameterDomain}) of the sum-of-squares that are spherical and centered around the maximum likelihood estimate (red dot in Figure \ref{fig.lassoParameterDomain}). For maximum likelihood estimates inside the grey areas the closest point in the diamond-shaped parameter domain will be on one of its corners. Hence, for these maximum likelihood estimates the corresponding lasso regression estimate will include on a single covariate in the model. The geometrical explanation of the selection property of the lasso regression estimator also applies to non-orthonormal design matrices and in dimensions larger than two. In particular, high-dimensionally, the sum-of-squares may be a degenerated ellipsoid, that can and will still hit a corner of the diamond-shaped parameter domain.  Finally, note that a zero value of the lasso regression estimate does imply neither that the parameter is indeed zero nor that it will be statistically indistinguishable from zero.

\begin{figure}[!h]
\begin{tabular}{rcl}
\includegraphics[scale=0.23, angle=0]{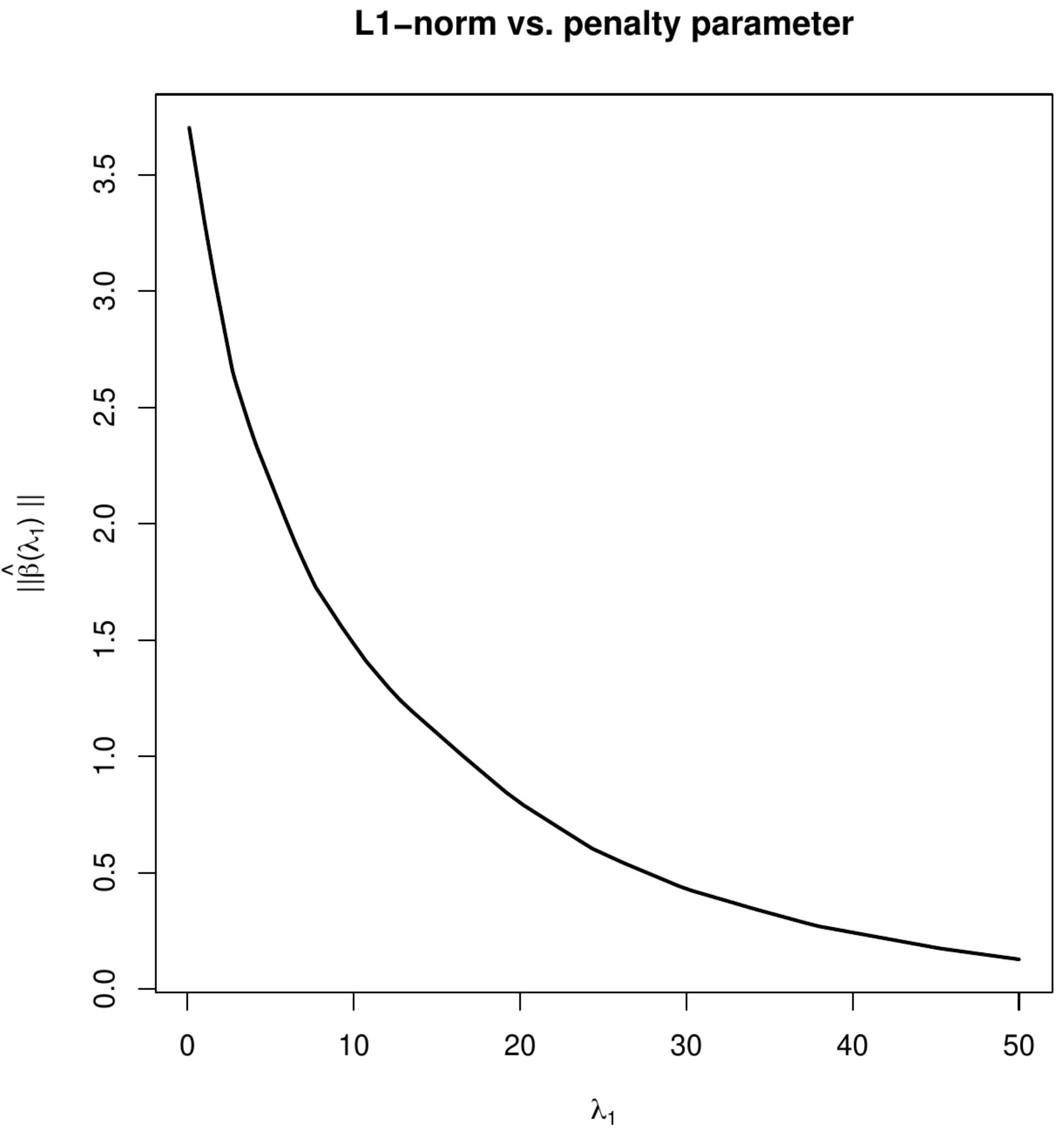}
& &
\includegraphics[scale=0.23, angle=0]{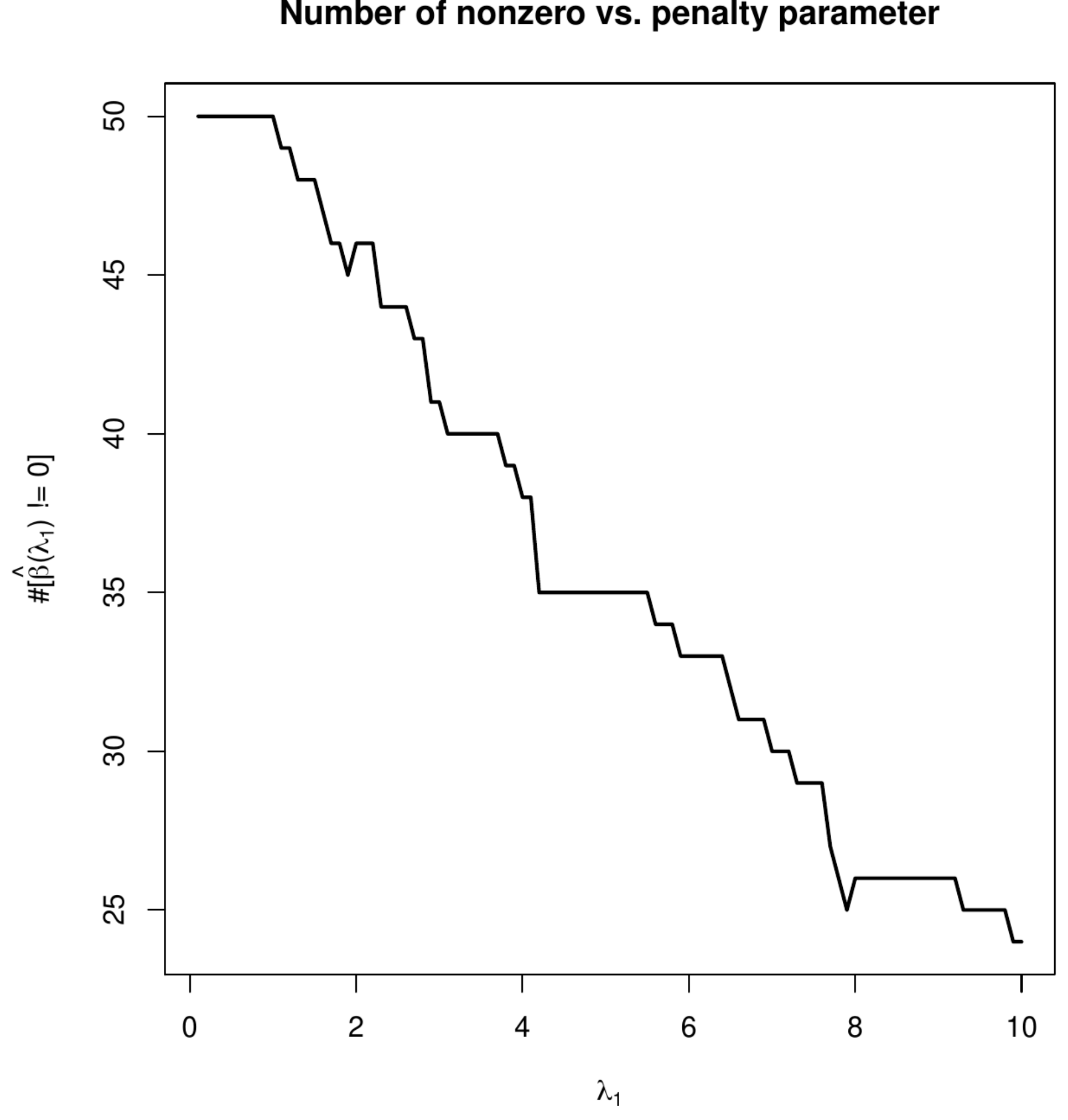}
\end{tabular}
\caption{Left panel: the $\ell_1$ norm of the lasso regression estimator against its penalty parameter. Right panel: number of nonzero elements of the lasso regression estimator against its penalty parameter.
} \label{fig:monotonePenalty}
\end{figure}

Larger values of the lasso penalty parameter $\lambda_1$ induce tighter parameter constraints. Consequently, the number of zero elements in the lasso regression estimator of $\bbeta$ increases as $\lambda_1$ increases. However, where $\| \hat{\bbeta}(\lambda_1) \|_1$ decreases monotonically as $\lambda_1$ increases (left panel of Figure \ref{fig:monotonePenalty} for an example and Exercise \ref{exercise:lassoMonotoneBetaNorm}), the number of non-zero coefficients does not. Locally, at some finite $\lambda_1$, the number of non-zero elements in $\hat{\bbeta}(\lambda_1)$ may temporarily increase with $\lambda_1$, to only go down again as $\lambda_1$ is sufficiently increased (as in the $\lambda_1 \rightarrow \infty$ limit the number of non-zero elements is zero, see the argumentation at the end of Section \ref{sect:lassoAnalytic}). The right panel of Figure \ref{fig:monotonePenalty} illustrates this behavior for an arbitrary data set.

The attractiveness of the lasso regression estimator is in its simultaneous estimation and selection of parameters. For large enough values of the penalty parameter $\lambda_1$, the estimated regression model comprises only a subset of the supplied covariates. High-dimensionally, typically demanding a large penalty parameter, the number of selected parameters by the lasso regression estimator is usually small relative to the total number of parameters. It thus producing a so-called sparse model. Would one adhere to the parsimony principle, such a sparse and thus simpler model is preferable over a full model. Simpler may be better, but too simple is worse. The phenomenon or system that is to be described by the model need not be sparse. For instance, in molecular biology the regulatory network of the cell is no longer believed to be sparse \citep{Boyle2017}. Similarly, when analyzing brain image data, the connectivity of the brain is not believed to be sparse.


\subsection{Maximum number of selected covariates}
The number of parameter/covariates selected by the lasso regression estimator is bounded non-trivially. The cardinality (i.e. the number of included covariates) of every lasso estimated linear regression model is smaller than or equal to $\min\{n, p\}$ \citep{Buhlmann2011}. As \cite{Buhlmann2011} point out this is obvious from the analysis of the LARS algorithm of \cite{Efro2004lars} (which is to be discussed in Section \ref{sect:LARS}). For now we just provide an \texttt{R}-script that generates the regularization paths using the \texttt{lars}-package for the \texttt{diabetes} data included in the package for a random number of samples $n$ not exceeding the number of covariates $p$.
\begin{lstlisting}
# activate library
library(lars)

# load data
data(diabetes)
X <- diabetes$x
Y <- diabetes$y

# set sample size
n  <- sample(1:ncol(X), 1)
id <- sample(1:length(Y), n)

# plot regularization paths
plot(lars(X[id,], Y[id], intercept=FALSE))
\end{lstlisting}
Irrespective of the drawn sample size $n$ the plotted regularization paths all terminate before the $n+1$-th variate enters the model. This could of course be circumstantial evidence at best, or even be labelled a bug in the software.

But even without the LARS algorithm the nontrivial part of the inequality, that the number of selected variates $p$ does not exceed the sample size $n$,  can be proven \citep{Osbo2000}.

\begin{theorem} (Theorem 6, \citealp{Osbo2000}) \label{theo.maximumNonzeros}
\\
If $p > n$ and $\hat{\bbeta}(\lambda_1)$ is a minimizer of the lasso regresssion loss function (\ref{form.lassoLossFunction}), then $\hat{\bbeta}(\lambda_1)$ has at most $n$ non-zero entries.
\end{theorem}
\begin{proof}
Confer \cite{Osbo2000}.
\end{proof}
In the high-dimensional setting, when $p$ is large compared to $n$ small, this implies a considerable dimension reduction. It is, however, somewhat unsatisfactory that it is the study design, i.e. the inclusion of the number of samples, that determines the upperbound of the model size.

\section{Estimation} \label{sect:lassoEstimation}
In the absence of an analytic expression for the optimum of the lasso loss function (\ref{form.lassoLossFunction}), much attention is devoted to numerical procedures to find it.

\subsection{Quadratic programming} \label{sect:lassoQuadProg}
In the original lasso paper \cite{Tibs1996} reformulates the lasso optimization problem to a quadratic program. A quadratic problem optimizes a quadratic form subject to linear constraints. This is a well-studied optimization problem for which many readily available implementations exist (e.g., the \texttt{quadprog}-package in \texttt{R}). The quadratic program that is equivalent to the lasso regression problem, which minimizes the least squares criterion, $\| \mathbf{Y} - \mathbf{X} \bbeta \|_2^2$ subject to the constraint $\bbeta \in \mathbb{R}^p$ such that $\| \bbeta \|_1 < c(\lambda_1)$, is:
\begin{eqnarray} \label{form:quadProgam}
\min\nolimits_{\{\bbeta \in \mathbb{R}^p \, : \, \mathbf{R} \bbeta \geq \mathbf{0}_q \}} \, \tfrac{1}{2} (\mathbf{Y} - \mathbf{X} \bbeta )^{\top}
(\mathbf{Y} - \mathbf{X} \bbeta ),
\end{eqnarray}
where $\mathbf{R}$ is a suitably chosen $q \times p$ dimensional linear constraint matrix that specifies the linear constraints on the parameter $\bbeta$. For $p=2$ the domain implied by lasso parameter constraint $\{ \bbeta \in \mathbb{R}^2 : \| \bbeta \|_1 < c(\lambda_1) \mathbf{1}_4 \}$ is equal to:
\begin{eqnarray*}
& & \{ \bbeta \in \mathbb{R}^2 :  \beta_1 + \beta_2 \leq c(\lambda_1)  \} \cap \{ \bbeta \in \mathbb{R}^2 :  \beta_1 - \beta_2 \geq -c(\lambda_1)  \} \cap \{ \bbeta \in \mathbb{R}^2 :  \beta_1 - \beta_2 \leq c(\lambda_1)  \}
\\
& & \qquad \qquad \qquad \qquad \qquad \qquad
\cap \, \{ \bbeta \in \mathbb{R}^2 :  \beta_1 + \beta_2 \geq - c(\lambda_1)  \}.
\end{eqnarray*}
This collection of linear parameter constraints can be reformulated, using:
\begin{eqnarray*}
\mathbf{R}^{\top} & = & \left(
\begin{array}{rrrr}
1 & -1 & 1 & -1
\\
1 & -1 & -1 & 1
\end{array}
\right),
\end{eqnarray*}
into $\{ \bbeta \in \mathbb{R}^2 \, : \, \mathbf{R} \bbeta \geq -c(\lambda_1) \mathbf{1}_4 \}$.

To solve the quadratic program (\ref{form:quadProgam}) it is usually reformulated in terms of its dual. Hereto we introduce the Lagrangian:
\begin{eqnarray} \label{form.lagrangian}
L(\bbeta, \nnu) & = & \tfrac{1}{2}  (\mathbf{Y} - \mathbf{X} \bbeta )^{\top}  (\mathbf{Y} - \mathbf{X} \bbeta ) - \nnu^{\top} [ \mathbf{R} \bbeta + c(\lambda_1) \mathbf{1}_q],
\end{eqnarray}
where $\nnu = (\nu_1, \ldots, \nu_{q})^{\top}$ is the vector of non-negative multipliers. The dual function is now defined as $\inf_{\bbeta} L(\bbeta, \nnu)$. This infimum is attained at:
\begin{eqnarray} \label{form.solution.primal.problem}
\bbeta^* & = & (\mathbf{X}^{\top} \mathbf{X})^{-1}(  \mathbf{X}^{\top} \mathbf{Y} + \mathbf{R}^{\top} \nnu),
\end{eqnarray}
which can be verified by equating the first order partial derivative with respect to $\bbeta$ of the Lagrangian to zero and solving for $\bbeta$. Substitution of $\bbeta = \bbeta^*$ into the dual function gives, after changing the minus sign:
\begin{eqnarray*}
\tfrac{1}{2} \nnu^{\top}  \mathbf{R}  (\mathbf{X}^{\top} \mathbf{X} )^{-1} \mathbf{R}^{\top} \nnu  + \nnu^{\top} [ \mathbf{R} (\mathbf{X}^{\top} \mathbf{X} )^{-1} \mathbf{X}^{\top} \mathbf{Y} + c(\lambda_1) \mathbf{1}_q ] - \tfrac{1}{2} \mathbf{Y}^{\top} [ \mathbf{I}_{nn} - \mathbf{X} (\mathbf{X}^{\top}  \mathbf{X} )^{-1}  \mathbf{X}^{\top} ] \mathbf{Y}.
\end{eqnarray*}
The dual problem minimizes this expression (from which the last term is dropped as is does not involve $\nnu$) with respect to $\nnu$, subject to $\nnu \geq \mathbf{0}$. Although also a quadratic programming problem, the dual problem \textit{i)} has a simpler formulation of the linear constraints and \textit{ii)} is defined on a lower dimensional space  than the primal problem (should the number of columns of $\mathbf{R}$ exceeds its number of rows). If $\tilde{\nnu}$ is the solution of the dual problem, the solution of the primal problem is obtained from Equation (\ref{form.solution.primal.problem}). Refer to, e.g., \cite{Bert2014} for more on quadratic programming.

\begin{contexample}\textbf{\ref{example.orthogonalDesignLasso}} \hspace{3pt}\textit{(Orthogonal design matrix,  continued)}
\\
The evaluation of the lasso regression estimator by means of quadratic programming is illustrated using the data from the numerical Example \ref{example.orthogonalDesignLasso}. The \texttt{R}-script below solves,
the implementation of the \texttt{quadprog}-package, the quadratic program associated with the lasso regression problem of the aforementioned example.
\begin{lstlisting}
# load library
library(quadprog)

# data
Y <- matrix(c(-4.9, -0.8, -8.9, 4.9, 1.1, -2.0), ncol=1)
X <- t(matrix(c(1, -1, 3, -3, 1, 1, -3, -3, -1, 0, 3, 0), nrow=2, byrow=TRUE))

# constraint radius
L1norm <- 1.881818 + 0.8678572

# solve the quadratic program
solve.QP(t(X) %*% X, t(X) %*% Y, 
         t(matrix(c(1, 1, -1, -1, 1, -1, -1, 1), ncol=2, byrow=TRUE)), 
         L1norm*c(-1, -1, -1, -1))$solution
\end{lstlisting}
The resulting estimates coincide with those found earlier.
\end{contexample}

\noindent
For relatively small $p$ quadratic programming is a viable option to find the lasso regression estimator. For large $p$ it is practically not feasible. Above the linear constraint matrix $\mathbf{R}$ is $4 \times 2$ dimensional for $p=2$. When $p =3$, it requires a linear constraint matrix $\mathbf{R}$ with six rows corresponding to the six side of a 3-dimensional cube. In general, $2 p$ linear constraints are required to fully specify the parameter constraint of the lasso regression estimator. If $p$ large, the specification of only the linear constraint matrix $\mathbf{R}$ will take endlessly, as it is a $2p \times p$ matrix, leave alone solving the corresponding quadratic program.

\subsection{Iterative ridge} \label{sect:iterativeRidge}
\begin{figure}[!h]
\begin{tabular}{rcl}
\includegraphics[scale=0.45, angle=0]{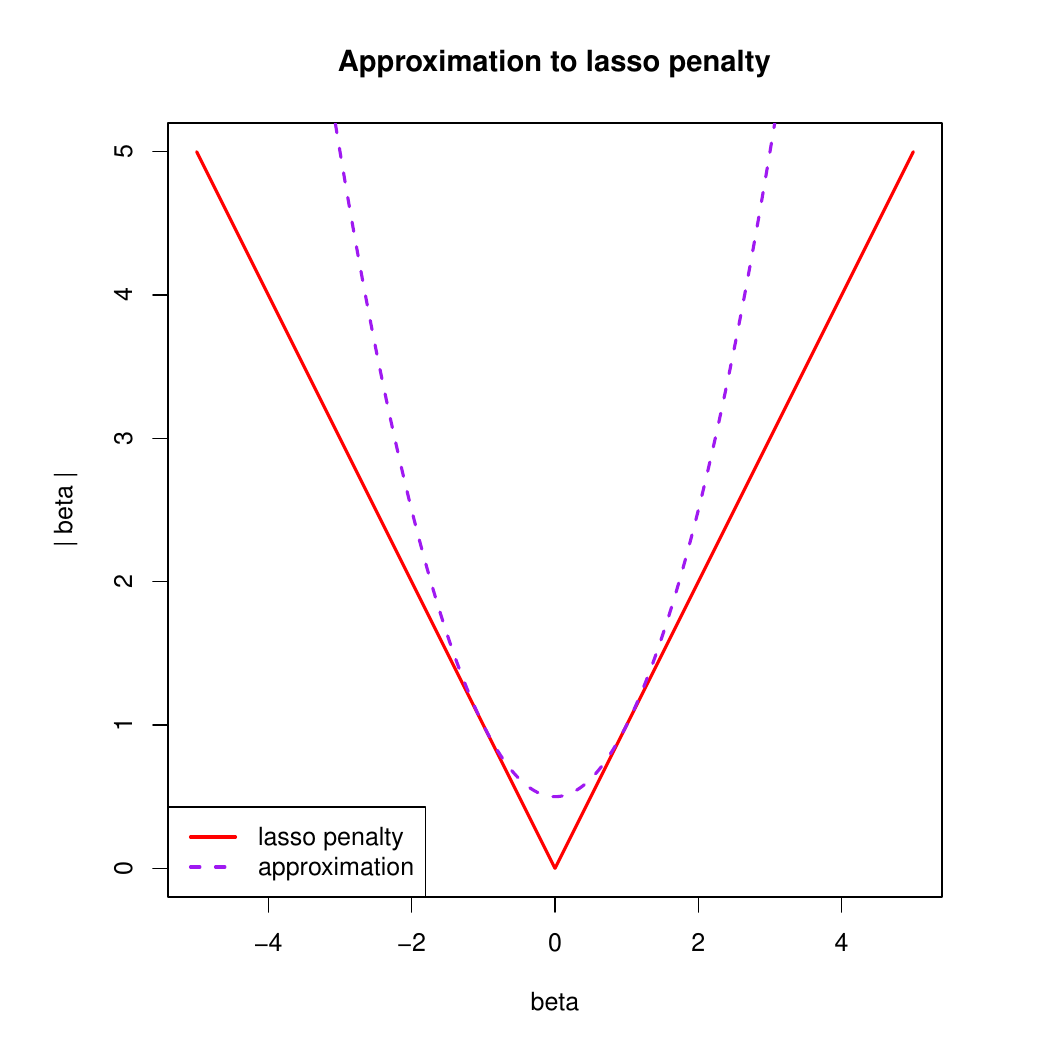}
& &
\includegraphics[scale=0.45, angle=0]{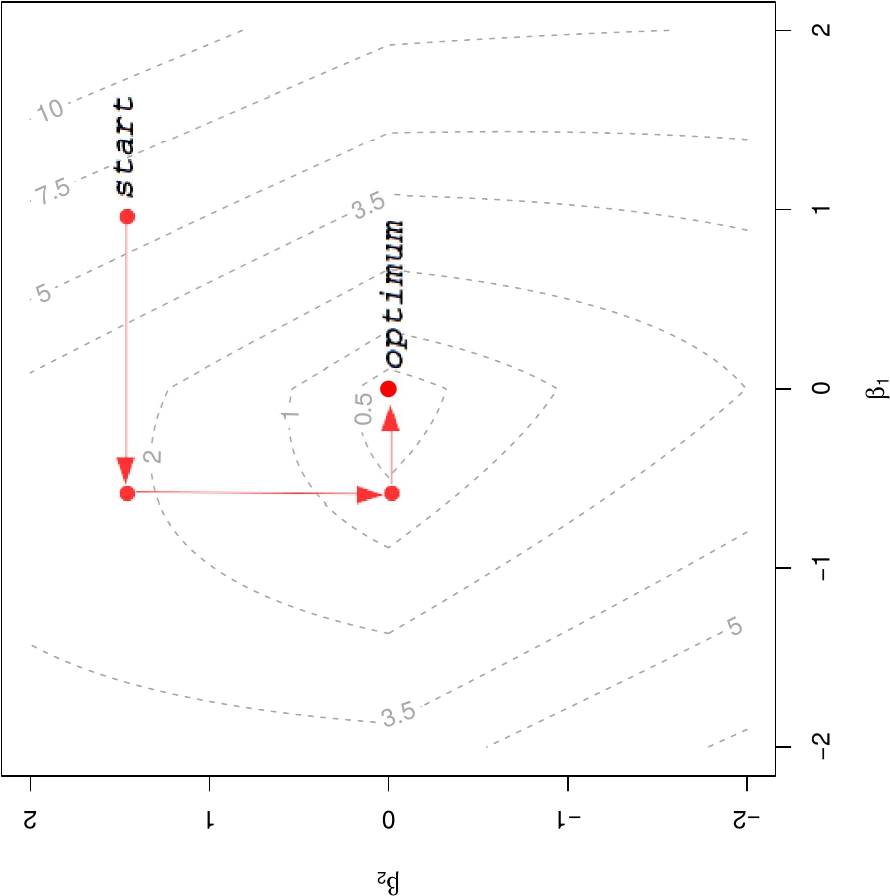}
\end{tabular}
\caption{Left panel: quadratic approximation  (i.e. the ridge penalty) to the absolute value function (i.e. the lasso penalty). Right panel: Illustration of the coordinate descent algorithm. The dashed grey lines are the level sets of the lasso regression loss function. The red arrows depict the parameter updates. These arrows are parallel to either the $\beta_1$ or the $\beta_2$ parameter axis, thus indicating that the regression parameter $\bbeta$ is updated coordinate-wise. \label{fig:ridgeApproxAndcoordinateDescent}}
\end{figure}

Why develop something new, when one can also make do with existing tools? The loss function of the lasso regression estimator can be optimized by iterative application of ridge regression (as pointed out in \citealp{Fan2001}). It requires an approximation of the lasso penalty, or the absolute value function. Set $p=1$ and let $\beta_0$ be an initial parameter value for $\beta$ around which the absolute value function $| \beta |$ is to be approximated. Its quadratic approximation then is:
\begin{eqnarray*}
|\beta | & \approx & |\beta_0 | + \tfrac{1}{2} | \beta_0 |^{-1}  (\beta^2 - \beta_0^2 ).
\end{eqnarray*}
An illustration of this approximation is provided in the left panel of Figure \ref{fig:ridgeApproxAndcoordinateDescent}.

The lasso regression estimator is evaluated through iterative application of the ridge regression estimator. This iterative procedure needs initiation by some guess $\bbeta^{(0)}$ for $\bbeta$. For example, the ridge estimator itself may serve as such. Then, at the $k+1$-th iteration an update $\bbeta^{(k+1)}$ of the lasso regression estimator of $\bbeta$ is to be found. Application of the quadratic approximation to the absolute value functions of the elements of $\bbeta$ (around the $k$-th update $\bbeta^ {(k)}$) in the lasso penalty yields an approximation to the lasso regression loss function:
\begin{eqnarray*}
\| \mathbf{Y} - \mathbf{X} \, \bbeta^{(k+1)} \|^2_2 + \lambda_1 \| \bbeta^{(k+1)} \|_1 & \approx & \| \mathbf{Y} - \mathbf{X} \, \bbeta^{(k+1)} \|^2_2 + \lambda_1 \| \bbeta^{(k)} \|_1
\\
&  &   + \tfrac{1}{2} \lambda_1 \sum\nolimits_{j=1}^p |\beta_j^{(k)}|^{-1} [ \beta_j^{(k+1)} ]^2 - \tfrac{1}{2} \lambda_1 \sum\nolimits_{j=1}^p |\beta_j^{(k)}|^{-1} [ \beta_j^{(k)} ]^2
\\
& \propto & \| \mathbf{Y} - \mathbf{X} \, \bbeta^{(k+1)} \|^2_2 + \tfrac{1}{2} \lambda_1 \sum\nolimits_{j=1}^p |\beta_j^{(k)}|^{-1} [ \beta_j^{(k+1)} ]^2.
\end{eqnarray*}
The loss function now contains a weighted ridge penalty. In this we recognize a generalized ridge regression loss function (see Chapter \ref{chap:genRidge}). As its minimizer is known, the approximated lasso regression loss function is optimized by:
\begin{eqnarray*}
\bbeta^{(k+1)}(\lambda_1) & = & \{ \mathbf{X}^{\top} \mathbf{X} + \lambda_1 \PPsi[ \bbeta^{(k)}(\lambda_1) ] \}^{-1} \mathbf{X}^{\top} \mathbf{Y}
\end{eqnarray*}
where
\begin{eqnarray*}
\mbox{diag}\{ \PPsi[ \bbeta^{(k)}(\lambda_1) ] \} & = & (1/|\beta_1^{(k)}|, 1/|\beta_2^{(k)}|, \ldots, 1/|\beta_p^{(k)}|).
\end{eqnarray*}
The thus generated sequence of updates $\{ \bbeta^{(k)}(\lambda_1) \}_{k=0}^{\infty}$ converges (under `nice' conditions) to the lasso regression estimator $\hat{\bbeta}(\lambda_1)$.

A note of caution. The in-built variable selection property of the lasso regression estimator may -- for large enough choices of the penalty parameter $\lambda_1$ -- cause elements of $\bbeta^{(k)}(\lambda_1)$ to become arbitrary close to zero (or, in \texttt{R} exceed machine precision and thereby being effectively zero) after enough updates. Consequently, the ridge penalty parameter for the $j$-th element of regression parameter may approach infinity, as the $j$-th element of $\PPsi[ \bbeta^{(k)}(\lambda_1)]$ equals $|\beta_j^{(k)}|^{-1}$. To accommodate this, the iterative ridge regression algorithm for the evaluation of the lasso regression estimator requires a modification. Effectively, that amounts to the removal of $j$-th covariate from the model all together (for its estimated regression coefficient is indistinguishable from zero). After removal, this covariate does not return to the set of covariates. This may be problematic if two covariates are (close to) super-collinear.

\subsection{Gradient ascent} \label{sect:gradientAscent}
\begin{figure}[!h]
\begin{tabular}{c}
\includegraphics[scale=0.50, angle=0]{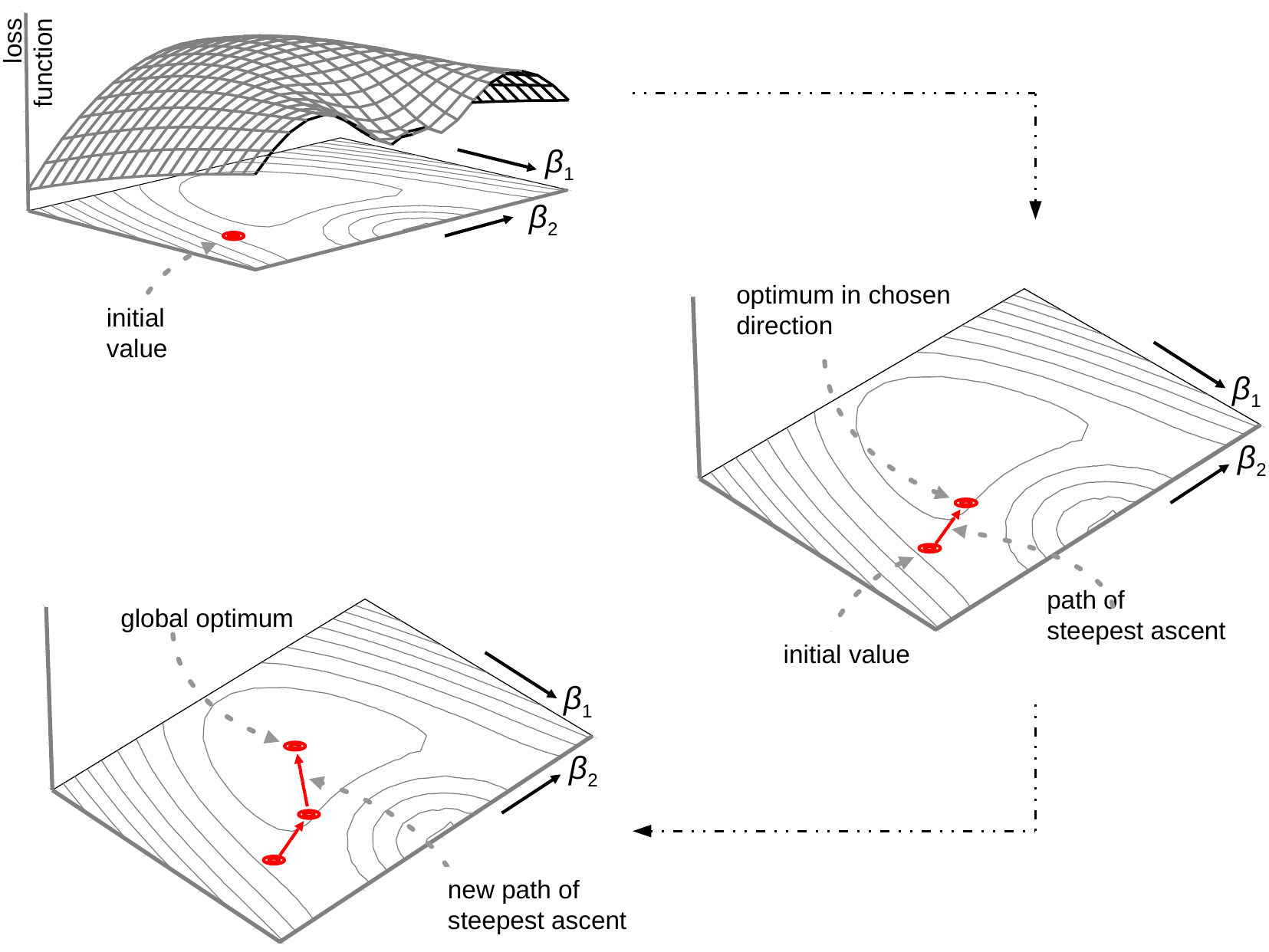}
\end{tabular}
\caption{Illustration of the gradient ascent procedure.} \label{fig:gradientAscent}
\end{figure}

Another method of finding the lasso regression estimator and implemented in the \texttt{penalized}-package \citep{Goem2010} makes use of gradient ascent.
Gradient ascent/descent is an maximization/minization method that finds the optimum of a smooth function by iteratively updating a first-order local approximation to this function. Gradient ascent runs through the following sequence of steps repetitively until convergence:
\begin{compactitem}
\item Choose a starting value.
\item Calculate the derivative of the function, and determine the direction in which the function increases most. This direction is the path of steepest ascent.
\item Proceed in this direction, until the function no longer increases.
\item Recalculate at this point the gradient to determine a new path of steepest ascent.
\item Repeat the above until the (region around the) optimum is found.
\end{compactitem}
The procedure above is illustrated in Figure \ref{fig:gradientAscent}. The top panel shows the choice of the initial value. From this point the path of the steepest ascent is followed until the function no longer increases (right panel of Figure \ref{fig:gradientAscent}). Here the path of steepest ascent is updated along which the search for the optimum is proceeded (bottom panel of Figure \ref{fig:gradientAscent}).

The application of gradient ascent to find the lasso regression estimator is frustrated by the non-differentiability (with respect to any of the regression parameters) of the lasso penalty function at zero. In \cite{Goem2010}, this is overcome by the use of a generalized derivative. Define the \textit{directional} or \textit{G\^{a}teaux} derivative of the function $f : \mathbb{R}^p \rightarrow \mathbb{R}$ at $\mathbf{x} \in \mathbb{R}^p$ in the direction of $\mathbf{v} \in \mathbb{R}^p$ as:
\begin{eqnarray*}
f'(\mathbf{x}) & = & \lim_{\tau \downarrow 0} \tau^{-1} \big[ f(\mathbf{x} + \tau \mathbf{v}) - f(\mathbf{x}) \big],
\end{eqnarray*}
assuming this limit exists. The G\^{a}teaux derivative thus gives the infinitesimal change in $f(\cdot)$ at $\mathbf{x}$ in the direction of $\mathbf{v}$. As such $f'(\mathbf{x})$ is a scalar (as is immediate from the definition after noting that $f(\cdot) \in \mathbb{R}$) and should not be confused with the gradient (the vector of partial derivatives). Furthermore, at each point $\mathbf{x}$ there are infinitely many G\^{a}teaux differentials (as there are infinitely many choices for $\mathbf{v} \in \mathbb{R}^p$). In the particular case when $\mathbf{v} = \mathbf{e}_j$, $\mathbf{e}_j$ the unit vector along the axis of the $j$-th coordinate, the directional derivative coincides with the partial derivative of $f$ in the direction of $x_j$. Relevant for the case at hand is the absolute value function $f(x) = | x|$ with $x \in \mathbb{R}$. Evaluation of the limits in its G\^{a}teaux derivative yields:
\begin{eqnarray*}
f'(x) & = & \left\{ \begin{array}{lcl}
\mbox{v}  \frac{x}{ | x| }  & \mbox{if} & x \not= 0,
\\
 \mbox{v}  & \mbox{if} &   x=0,
\end{array}
\right.
\end{eqnarray*}
for any $\mbox{v} \in \mathbb{R} \setminus \{ 0 \}$. Hence, the G\^{a}teaux derivative of $| x|$ does exits at $x=0$. In general, the  G\^{a}teaux differential may be uniquely defined by limiting the directional vectors $\mathbf{v}$ to \textit{i)} those with unit length (i.e. $\| \mathbf{v} \| = 1$) and \textit{ii)} the direction of steepest ascent. Using the
G\^{a}teaux derivative, a gradient of $f(\cdot)$ at $\mathbf{x} \in \mathbb{R}^p$ can then be defined as:
\begin{eqnarray} \label{def:GateauxDerivative}
\nabla f(\mathbf{x}) & = & \left\{ \begin{array}{lcl}
f'(\mathbf{x}) \circ \mathbf{v}_{\mbox{{\tiny opt}}} & \mbox{if} & f'(\mathbf{x}) \geq 0,
\\
\mathbf{0}_{p} & \mbox{if} &   f'(\mathbf{x}) < 0,
\end{array}
\right.
\end{eqnarray}
in which $\circ$ is the Hadamard (i.e,. element-wise) product and $\mathbf{v}_{\mbox{{\tiny opt}}} = \arg \max_{\{ \mathbf{v} \, : \, \| \mathbf{v} \| = 1 \} } f'(\mathbf{x})$. This is the direction of steepest ascent, $\mathbf{v}_{\mbox{{\tiny opt}}}$, scaled by G\^{a}teaux derivative, $f'(\mathbf{x})$, in the direction of $\mathbf{v}_{\mbox{{\tiny opt}}}$.

\cite{Goem2010} applies the definition of the G\^{a}teaux gradient to the lasso penalized likelihood (\ref{form.lassoLossFunction}) using the direction of steepest ascent as $\mathbf{v}_{\mbox{{\tiny opt}}}$. The resulting partial G\^{a}teaux derivative with respect to the $j$-th element of the regression parameter $\bbeta$ is:
\begin{eqnarray*}
\frac{\partial}{ \partial \beta_j } \mathcal{L}_{\mbox{{\tiny lasso}}}(\mathbf{Y}, \mathbf{X}; \bbeta) & = & \left\{ \begin{array}{lcl}
\frac{\partial}{ \partial \beta_j } \mathcal{L}(\mathbf{Y}, \mathbf{X}; \bbeta) - \lambda_1 \mbox{sign} (\beta_j)
& \mbox{if} & \beta_j \not= 0
\\
\frac{\partial}{ \partial \beta_j } \mathcal{L}(\mathbf{Y}, \mathbf{X}; \bbeta) - \lambda_1 \mbox{sign} \big[ \frac{\partial}{ \partial \beta_j } \mathcal{L}(Y, X; \bbeta)
\big] & \mbox{if} &  \beta_j = 0 \mbox{ and } | \partial \mathcal{L}/ \partial \beta_j  | > \lambda_1
\\
0 & \multicolumn{2}{l}{\mbox{otherwise}},
\end{array}
\right.
\end{eqnarray*}
where $\partial \mathcal{L}/ \partial \beta_j = \sum_{j'=1}^p (\mathbf{X}^\top \mathbf{X})_{j', j} \beta_j - (\mathbf{X}^\top \mathbf{Y})_{j}$. This can be understood through a case-by-case study. The partial derivative above is assumed to be clear for the $\beta_j \not= 0$ and the `otherwise' cases. That leaves the clarification of the middle case. When $\beta_j = 0$, the direction of steepest ascent of the penalized loglikelihood points either into $\{ \bbeta \in \mathbb{R}^p \, : \, \beta_j > 0 \}$, or $\{ \bbeta \in \mathbb{R}^p \, : \, \beta_j < 0 \}$, or stays in $\{ \bbeta \in \mathbb{R}^p \, : \, \beta_j = 0 \}$. When the direction of steepest ascent points into the positive or negative half-hyperplanes, the contribution of $\lambda_1 | \beta_j|$ to the partial G\^{a}teaux derivative is simply $\lambda_1$ or $-\lambda_1$, respectively. Then, only when the partial derivative of the log-likelihood 
together with this contribution is larger then zero, the penalized loglikelihood improves and the direction is of steepest ascent. Similarly, the direction of steepest ascent may be restricted to $\{ \bbeta \in \mathbb{R}^p \, | \, \beta_j = 0 \}$ and the contribution of $\lambda_1 | \beta_j|$ to the partial G\^{a}teaux derivative vanishes. Then, only if the partial derivative of the loglikelihood is positive, this direction is to be pursued for the improvement of the penalized loglikelihood.

Convergence of gradient ascent can be slow close to the optimum. This is due to its linear approximation of the function. Close to the optimum the linear term of the Taylor expansion vanishes and is dominated by the second-order quadratic term. To speed-up convergence close to the optimum the gradient ascent implementation offered by the \texttt{penalized}-package switches to a Newton-Raphson procedure.

\subsection{Coordinate descent} \label{sect:coordinateDescent}
Coordinate descent is another optimization algorithm that may be used to evaluate the lasso regression estimator numerically, as is done by the implemention offered via the \texttt{glmnet}-package. Coordinate descent, instead of following the gradient of steepest descent (as in Section \ref{sect:gradientAscent}),  minimizes the loss function along the coordinates one-at-the-time. For the $j$-th regression parameter, this amounts to finding:
\begin{eqnarray*}
\arg \min_{\beta_j} \| \mathbf{Y} - \mathbf{X} \bbeta \|_2^2 + \lambda_1 \| \bbeta \|_1 & = & \arg \min_{\beta_j} \| \mathbf{Y} - \mathbf{X}_{\ast, \setminus j} \bbeta_{\setminus j} - \mathbf{X}_{\ast, j} \beta_{j} \|_2^2 + \lambda_1 | \beta_j |_1
\\
& = & \arg \min_{\beta_j} \| \tilde{\mathbf{Y}} - \mathbf{X}_{\ast, j} \beta_{j} \|_2^2 + \lambda_1 | \beta_j |_1,
\end{eqnarray*}
where $\tilde{\mathbf{Y}} = \mathbf{Y} - \mathbf{X}_{\ast, \setminus j} \bbeta_{\setminus j}$. After a simple rescaling of both $\mathbf{X}_{\ast, j}$ and $\beta_j$, the minimization of the lasso regression loss function  with respect to $\beta_j$ is equivalent to one with an orthonormal design matrix. From Example \ref{example.orthonormalDesignLasso}, it is known that the minimizer is obtained by application of the soft-threshold function to the corresponding maximum likelihood estimator (now derived from $\tilde{\mathbf{Y}}$ and $\mathbf{X}_{\ast,j}$). The coordinate descent algorithm iteratively runs over the $p$ elements until convergence. The right panel of Figure \ref{fig:ridgeApproxAndcoordinateDescent} provides an illustration of the coordinate descent algorithm.

Convergence of the coordinate descent algorithm to the minimum of the lasso regression loss function  (\ref{form.lassoLossFunction}) is warranted by the convexity of this function. At each minimization step the coordinate descent algorithm yields an update of the parameter estimate that corresponds to an equal or smaller value of the loss function. It, together with the compactness of diamond-shaped parameter domain and the boundedness (from below) of the lasso regression loss function, implies that the coordinate descent algorithm converges to the minimum of this lasso regression loss function. Although convergence is assured, it may take many steps for it to be reached. In particular, when \textit{i)} two covariates are strongly collinear, \textit{ii)} one of the two covariates contributes only slightly more to the response, and \textit{iii)} the algorithm is initiated with the weaker explanatory covariate. The coordinate descent algorithm will then take many iterations to replace the latter covariate by the preferred one. In such cases simultaneous updating, as is done by the gradient ascent algorithm (Section \ref{sect:gradientAscent}), may be preferable.

\section{Moments} \label{sect.lassoMoments}
In general the moments of the lasso regression estimator appear to be unknown. In certain cases an approximation can be given. This is pointed out here. Use the quadratic approximation to the absolute value function of Section \ref{sect:iterativeRidge} and approximate the lasso regression loss function around the lasso regression estimate:
\begin{eqnarray*}
\| \mathbf{Y} - \mathbf{X} \bbeta \|_2^2 +  \lambda_1 \| \bbeta \|_1 & \approx & \| \mathbf{Y} - \mathbf{X} \bbeta \|_2^2 +  \tfrac{1}{2} \lambda_1 \sum\nolimits_{j=1}^p |\hat{\beta}_j (\lambda_1)|^{-1} \, \beta_j^2.
\end{eqnarray*}
Optimization of the right-hand side of the preceeding display with respect to $\bbeta$ gives a `ridge approximation' to the lasso estimator:
\begin{eqnarray*}
\hat{\bbeta}(\lambda_1) & \approx & \{\mathbf{X}^\top \mathbf{X} + \lambda_1 \mathbf{\Psi}[\hat{\bbeta}(\lambda_1)] \}^{-1} \mathbf{X}^\top \mathbf{Y},
\end{eqnarray*}
with $(\mathbf{\Psi}[\hat{\bbeta}(\lambda_1)])_{jj} = | \hat{\beta}_j(\lambda_1) |^{-1}$ if $\hat{\beta}_j(\lambda_1) \not=0$. We now use this `ridge approximation' to obtain the approximation to the moments of the lasso regression estimator:
\begin{eqnarray*}
\mathbb{E} [\hat{\bbeta}(\lambda_1)] & \approx & \mathbb{E} \big( \{\mathbf{X}^\top \mathbf{X} + \lambda_1 \mathbf{\Psi}[\hat{\bbeta}(\lambda_1)] \}^{-1} \mathbf{X}^\top \mathbf{Y} \big) \, \, \, = \, \, \, \{\mathbf{X}^\top \mathbf{X} + \lambda_1 \mathbf{\Psi}[\hat{\bbeta}(\lambda_1)] \}^{-1} \mathbf{X}^\top \mathbf{X} \bbeta
\end{eqnarray*}
and
\begin{eqnarray*}
\mbox{Var} [\hat{\bbeta}(\lambda_1)] & \approx & \mbox{Var}\big( \{\mathbf{X}^\top \mathbf{X} + \lambda_1 \mathbf{\Psi}[\hat{\bbeta}(\lambda_1)] \}^{-1} \mathbf{X}^\top \mathbf{Y} \big)
\\
& = & \sigma^2 \{\mathbf{X}^\top \mathbf{X} + \lambda_1 \mathbf{\Psi}[\hat{\bbeta}(\lambda_1)] \}^{-1} \mathbf{X}^\top \mathbf{X} \{\mathbf{X}^\top \mathbf{X} + \lambda_1 \mathbf{\Psi}[\hat{\bbeta}(\lambda_1)] \}^{-1},
\end{eqnarray*}
where we have assumed $\mathbf{\Psi}[\hat{\bbeta}(\lambda_1)]$ to be nonrandom. The approximations above can only be used if the lasso regression estimate is not sparse, which is at odds with its selection property. A better approximation of the variance of the lasso regression estimator can be found in \cite{Osbo2000}, but even this becomes poor when many elements of $\bbeta$ are estimated as zero.


Although the above approximations are only crude, they indicate that the moments of the lasso regression estimator exhibit similar behaviour as those of its ridge counterpart. The (approximation of the) mean $\mathbb{E}[\hat{\bbeta}(\lambda_1)]$ tends to zero as $\lambda_1 \rightarrow \infty$. This was intuitively already expected from the form of the lasso regression loss function (\ref{form.lassoLossFunction}), in which the penalty term dominates for large $\lambda_1$ and is minimized for $\hat{\bbeta}(\lambda_1) = \mathbf{0}_p$. This may also be understood geometrically when appealing to the equivalent constrained estimation formulation of the lasso regression estimator. The parameter constraint shrinks to zero with increasing $\lambda_1$. Hence, so must the estimator. Similarly, the (approximation of the) variance of the lasso regression estimator vanishes as the penalty parameter $\lambda_1$ grows. Again, its loss function (\ref{form.lassoLossFunction}) provides the intuition: for large $\lambda_1$ the penalty term, which does not depend on data, dominates. Or, from the perspective of the constrained estimation formulation, the parameter constraint shrinks to zero as $\lambda_1 \rightarrow \infty$. Hence, so must the variance of the estimator as less and less room is left for it to fluctuate.

The behaviour of the mean squared error, bias squared plus variance, of the lasso regression estimator in terms of $\lambda_1$ is hard to characterize exactly  without knowledge of the quality of the approximations. In particular, does there exist a $\lambda_1$ such that the MSE of the lasso regression estimator outperforms that of its maximum likelihood counterpart? Nonetheless, a first observation may be obtained from reasoning in extremis. Suppose $\bbeta = \mathbf{0}_p$, which corresponds to an empty or maximally sparse model. A large value of $\lambda_1$ then yields a zero estimate of the regression parameter:  $\hat{\bbeta}(\lambda_1) = \mathbf{0}_p$. The bias squared is thus minimized as $\| \hat{\bbeta}(\lambda_1) - \bbeta \|_2^2 = 0$. With the bias vanished and the (approximation of the) variance decreasing in $\lambda_1$, so must the MSE decrease for $\lambda_1$ larger than some value. So, for an empty model the lasso regression estimator with a sufficiently large penalty parameter yields a better MSE than the maximum likelihood estimator. For very sparse models this property may be expected to uphold, but for non-sparse models the bias squared will have a substantial contribution to the MSE, and it is thus not obvious whether a $\lambda_1$ exists that yields a favourable MSE for the lasso regression estimator. This is investigated \textit{in silico} in \cite{Hans2015}. The simulations presented there indicate that the MSE of the lasso regression estimator is particularly sensitive to the actual $\bbeta$. Moreover, for a large part of the parameter space $\bbeta \in \mathbb{R}^p$ the MSE of $\hat{\bbeta}(\lambda_1)$ is behind that of the maximum likelihood estimator.

\section{Degrees of freedom}
The degrees of freedom consumed the lasso regression estimator can be estimated simply by its number of nonzero elements (as follows from the following theorem).  

\begin{theorem} (\citealp{Zou2007degrees}, \citealp{tibshirani2012degrees}) \\
Let $\mathbf{X} \in \mathcal{M}^{n,p}$ be any design matrix and $\mathbf{Y} = \mathbf{X} \boldsymbol{\beta} + \boldsymbol{\varepsilon}$ with $\boldsymbol{\beta} \in \mathbb{R}^p$ and $\boldsymbol{\varepsilon} \sim \mathcal{N}(\mathbf{0}_n, \sigma^2 \mathbf{I}_{nn})$. Then,
\begin{eqnarray*}
\mbox{df}(\lambda_1) & = & \mathbb{E} \big( | \{ j \, : \, [\hat{\boldsymbol{\beta}}(\lambda_1)]_j \not= 0 \} | \big),
\end{eqnarray*}
for $\lambda_1 > 0$. If $\mbox{rank}(\mathbf{X}) = p$, the lasso regression estimator's number of nonzero's is even a consistent estimator of the degrees of freedom. 
\end{theorem}
\begin{proof}
The full proof is beyond the scope of these notes, and here we limit ourselves to show the unbiasedness of the degrees of freedom estimator for orthonormal $\mathbf{X}$. The proof makes use of Lemma 2 of \cite{stein1981estimation}, which states that if $Y \sim \mathcal{N}(\mu, \sigma^2)$ and $g(\cdot)$ is an absolute continuous function with $\mathbb{E}[ g'(Y)] < \infty$, then
$\mathbb{E}[ g(Y) (Y - \mu)] =   \sigma^2 \mathbb{E}[ g'(Y)]$. Starting from the degrees of freedom definition introduced in Section \ref{sect.ridgeDOF}, we then manipulate (with minor abuse of notation) as follows:
\begin{eqnarray*}
\mbox{df}(\lambda_1) & = &  \sum\nolimits_{i=1}^n [\mbox{Var}(Y_i)]^{-1} \mbox{Cov}(\widehat{Y}_i, Y_i)
\\
& = & \sum\nolimits_{i=1}^n \sigma^{-2} \mathbb{E} \{ [ \widehat{Y}_i - \mathbb{E}(\widehat{Y}_i)] [ Y_i - \mathbb{E}(Y_i)] \}
\\
& = & \sum\nolimits_{i=1}^n \sigma^{-2} \mathbb{E} \{  \mathbf{X}_{i,\ast} \hat{\boldsymbol{\beta}}(\lambda_1) [ Y_i - \mathbb{E}(Y_i)] \}
\\
& = & \sum\nolimits_{i=1}^n \sigma^{-2}  \sum\nolimits_{j=1}^p X_{i,j} \mathbb{E} \{   [\hat{\boldsymbol{\beta}}(\lambda_1)]_j [ Y_i - \mathbb{E}(Y_i)] \}
\\
& = & \sum\nolimits_{i=1}^n \sigma^{-2}  \sum\nolimits_{j=1}^p X_{i,j} \mathbb{E} \{   \mbox{sign} (\hat{\beta}^{\mbox{{\tiny ml}}}_j) ( | \hat{\beta}^{\mbox{{\tiny ml}}}_j| - \tfrac{1}{2} \lambda_1 )_+  [ Y_i - \mathbb{E}(Y_i)] \}
\\
& = & \sum\nolimits_{i=1}^n \sum\nolimits_{j=1}^p X_{i,j} \mathbb{E} \Big\{  \frac{d}{dY_i}  \mbox{sign} \Big( \sum\nolimits_{i'=1}^n X_{i',j} Y_{i'} \Big) \Big( \Big| \sum\nolimits_{i'=1}^n X_{i',j} Y_{i'} \Big| - \tfrac{1}{2} \lambda_1 \Big)_+  \Big\} 
\\
& = & \sum\nolimits_{i=1}^n \sum\nolimits_{j=1}^p X_{i,j}^2 \mathbb{E} \big[ \mathbbm{1}_{ \{ | \sum\nolimits_{i'=1}^n X_{i',j} Y_{i'} | > \frac{1}{2} \lambda_1 \} } \big]
\\
& = & \mathbb{E} \Big[ \sum\nolimits_{j=1}^p \mathbbm{1}_{ \{ | \sum\nolimits_{i=1}^n X_{i,j} Y_{i} | > \frac{1}{2} \lambda_1 \} } \Big],
\end{eqnarray*}
where we have used the definition of the covariance, the analytic expression of the lasso regression estimator in the orthonormal case, and the independence among the samples.
\end{proof}
Would one select the lasso regression parameter's penalty parameter on the basis of an information criterion, this simple unbiased estimator of its degrees of freedom is rather convenient.

\section{The Bayesian connection} \label{sect.BayesianLasso}
\begin{figure}[!h]
\begin{tabular}{rcl}
\includegraphics[scale=0.31, angle=0]{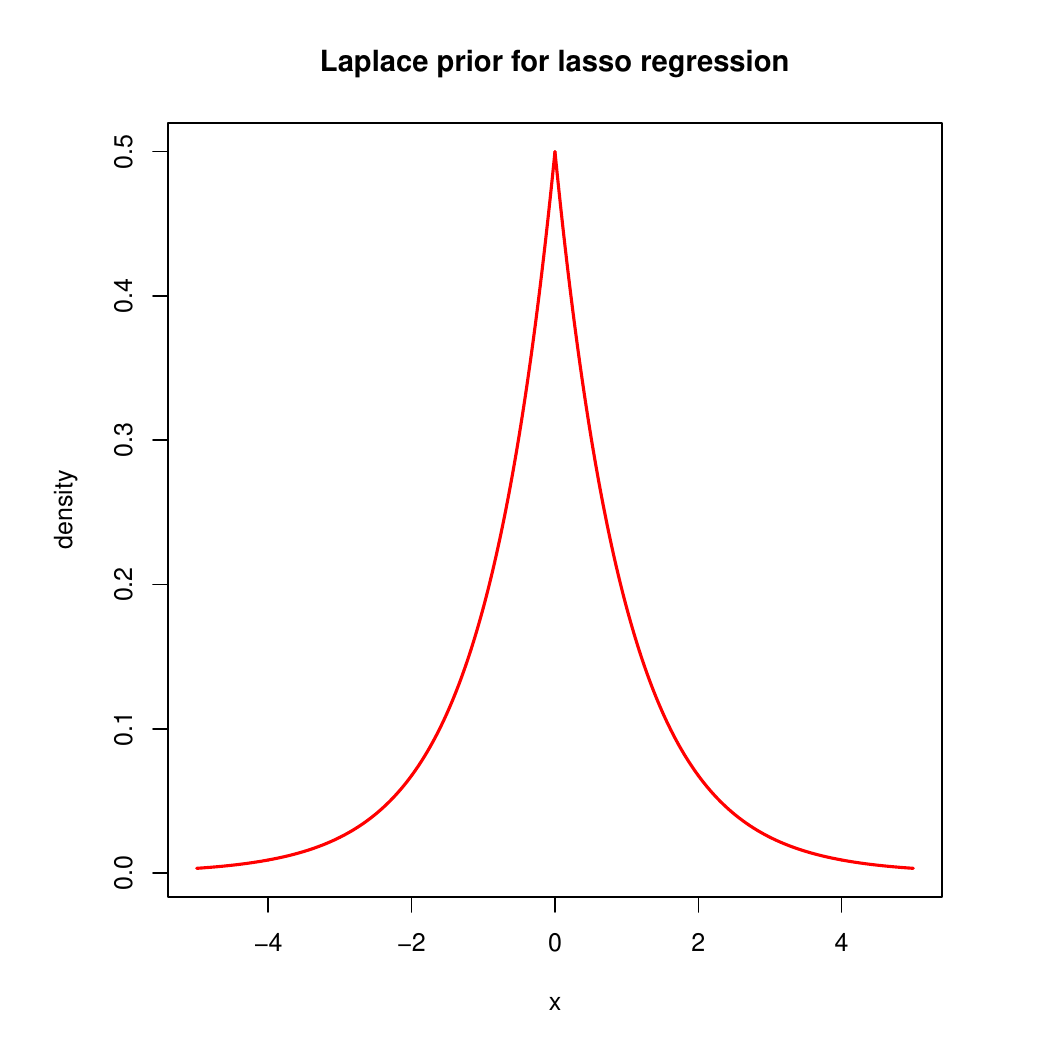}
&
\includegraphics[scale=0.15, angle=0]{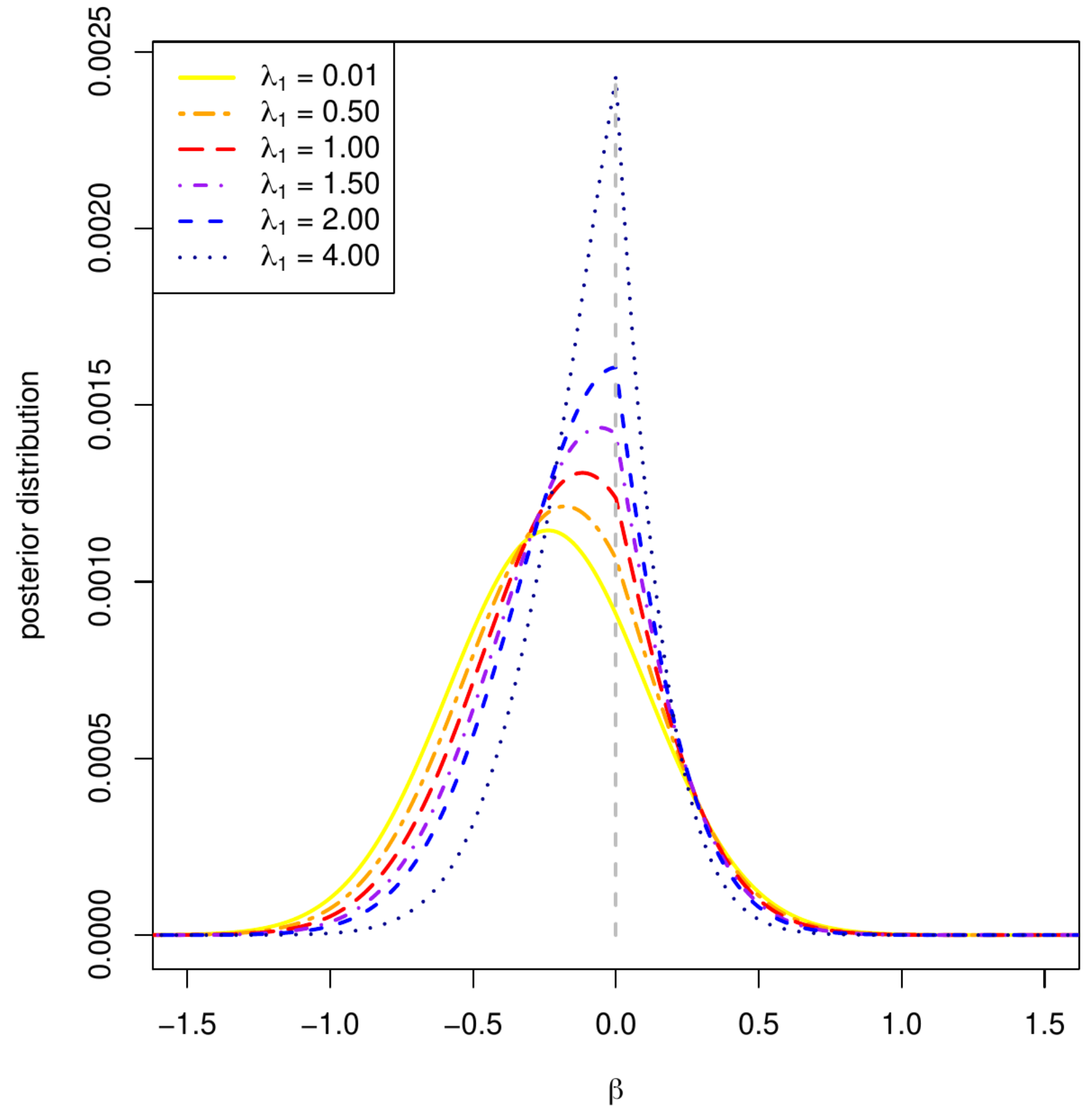} 
& 
\includegraphics[scale=0.21, angle=0]{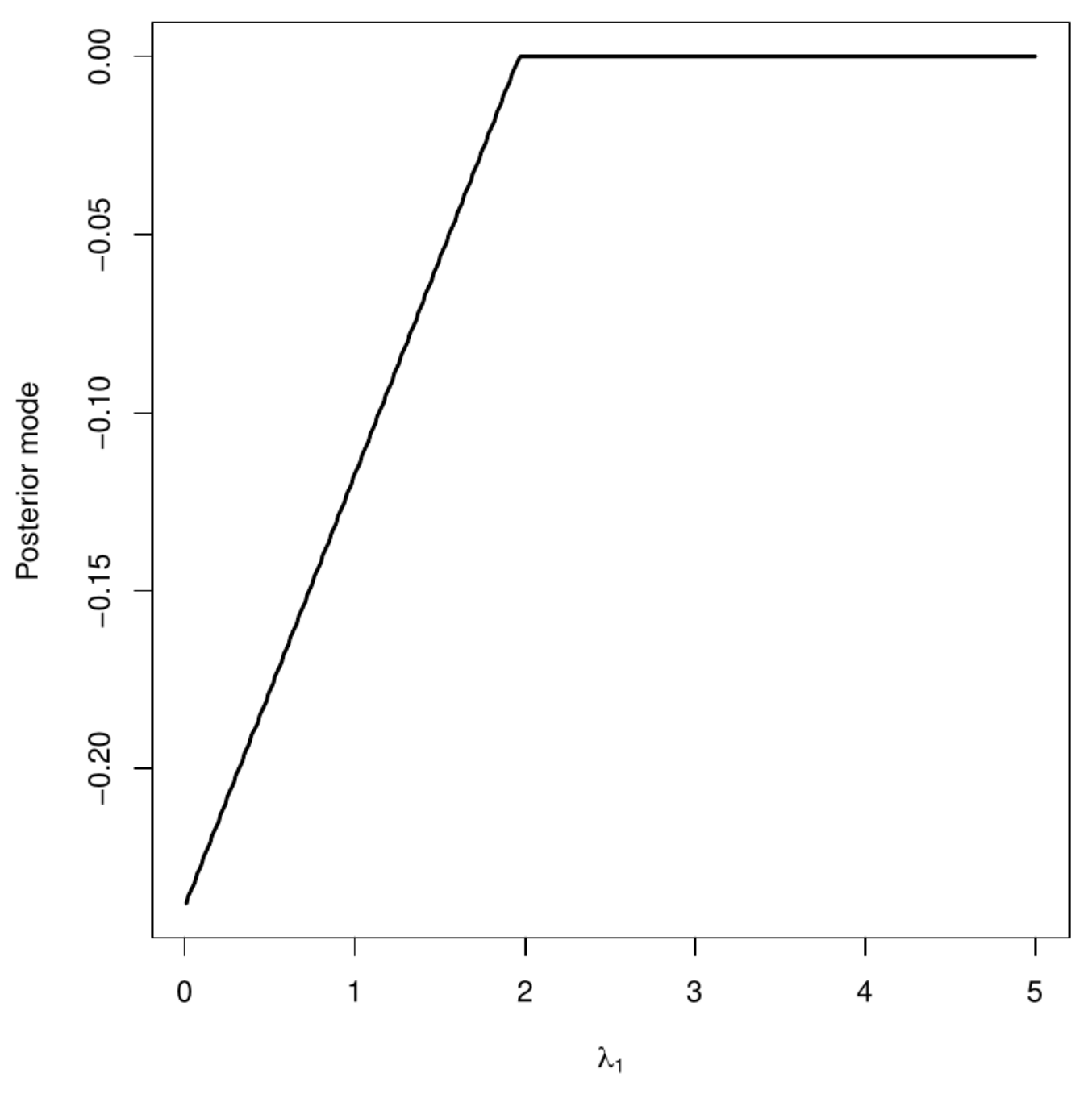}
\end{tabular}
\caption{Left panel: the Laplace prior associated with the Bayesian counterpart of the lasso regression estimator. Middle panel: the posterior distribution of the regression parameter for various Laplace priors. Right panel: posterior mode vs. the penalty parameter $\lambda_1$.
} \label{fig.lassoPrior}
\end{figure}

The lasso regression estimator knows a Bayesian counterpart, much like the (generalized) ridge regression estimator could be viewed as a Bayesian estimator if a Gaussian prior is imposed on the regression parameter (cf. Chapter \ref{chap:BayesianRegression} and Section \ref{sect:genRidgeBayes}). Here we replace the normal prior by a zero-centered Laplacian a.k.a. a double exponential prior. A zero-centered Laplace distributed random variable $X$ has density $f_X(x) = \tfrac{1}{2} b^{-1} \exp(-|x| /b)$ with scale parameter $b > 0$. The left panel of Figure \ref{fig.lassoPrior} shows the Laplace prior  and -- for reference -- the normal prior. This figure reveals that the `lasso prior' puts more mass close to zero and in the tails than the Gaussian `ridge prior'. This corroborates with the tendency of the lasso regression estimator to produce either zero or large (compared to ridge) estimates.

The lasso regression estimator corresponds to the maximum a posteriori (MAP) estimator of $\bbeta$, when the prior is a Laplace distribution. The posterior distribution is then proportional to:
\begin{eqnarray*}
\prod_{i=1}^n (2 \pi \sigma^2)^{-1/2} \exp [ - (2\sigma^2)^{-1} (Y_i - \mathbf{X}_{i,\ast} \bbeta)^2] \times \prod_{j=1}^p (2b)^{-1} \exp( -|\beta_j |/b),
\end{eqnarray*}
with the Laplace' scale parameter $b = \sqrt{\sigma^2} / \lambda_{1}^{\mbox{{\tiny (bayes)}}}$. The super- and subscript of $\lambda_{1}^{\mbox{{\tiny (bayes)}}}$ have been chosen to contrast and match, respectively, it to the frequentist setting. The posterior is not a well-known and characterized distribution. This need not be as interest often concentrates on its maximum. The location of the posterior mode coincides with the location of the maximum of logarithm of the posterior. The log-posterior is proportional to: $- (2\sigma^2)^{-1} \| \mathbf{Y} - \mathbf{X}\bbeta \|_2^2   -  \lambda_{1}^{\mbox{{\tiny (bayes)}}}  \, \| \bbeta \|_1 / \sqrt{\sigma^2}$, with its maximizer minimizing  $\| \mathbf{Y} - \mathbf{X}\bbeta \|_2^2   +  \lambda_{1}^{\mbox{{\tiny (freq)}}} \| \bbeta \|_1$ where $\lambda_{1}^{\mbox{{\tiny (freq)}}} = 2 \sqrt{\sigma^2} \lambda_{1}^{\mbox{{\tiny (bayes)}}}$. In this, one recognizes the form of the lasso regression loss function (\ref{form.lassoLossFunction}). It is thus clear that the scale parameter of the Laplace distribution reciprocally relates to lasso penalty parameter, which is similar to the relation of the ridge penalty parameter and the variance of the Gaussian prior.

The posterior may not be a standard distribution, in the univariate case ($p=1$) it can be visualized. Specifically, the behaviour of the MAP can then be illustrated, which -- as the MAP estimator corresponds to the lasso regression estimator -- should also exhibit the selection property (see Exercise \ref{question:lassoMAP}). The middle panel of Figure \ref{fig.lassoPrior} shows the posterior distribution for various choices of the Laplace scale parameter (i.e. lasso penalty parameter). Clearly, the mode shifts towards zero as the scale parameter decreases / lasso penalty parameter increases. In particular, the posterior obtained from the Laplace prior with the smallest scale parameter (i.e. largest penalty parameter), although skewed to the left, has a mode placed exactly at zero. The Laplace prior may thus produce MAP estimators that select. However, for smaller values of the lasso penalty parameter the Laplace prior is not concentrated enough around zero and the contribution of the likelihood in the posterior outweighs that of the prior. The mode is then not located at zero and the parameter is `selected' by the MAP estimator. The bottom right panel of Figure \ref{fig.lassoPrior} plots the mode of the normal-Laplace posterior vs. the Laplace scale parameter. In line with Theorem \ref{theo.lassoPiecewiseLinear} it is piece-wise linear.
\\
\\
\cite{Park2008bayesian} present a Gibbs sampler for the Bayesian regression with a Laplace prior on the regression parameter. Their sampler circumvents the sampling from the conditional posterior of the regression parameter. This conditional posterior is not a standard distribution and would require its own Metropolis-Hastings sampler, rendering the overall sampler slow. This workaround comprises the rewriting of the Laplace distribution as a scale mixture of normal distributions with an exponential mixing density. To see this, we adopt the priors $\beta_j \, | \, \sigma^2, \tau^2 \sim \mathcal{N}(0, \sigma^2 \tau^2)$ and $\tau^2 \sim \mbox{Exp}[\tfrac{1}{2}( \lambda_1^{\mbox{{\tiny (bayes)}}})^2]$. The unconditional prior of $\beta_j$ can, after substitution of the analytic distributions and grouping of terms, then be written as the convolution:
\begin{eqnarray*}
\pi ( \beta_j) & \propto & \int_0^{\infty} \pi (\beta_j \, | \, \tau^2 ) \, \pi ( \tau^2 ) \, d \tau^2
\\
& = &  \tfrac{1}{2} ( \lambda_1^{\mbox{{\tiny (bayes)}}})^2  (2\pi \sigma^2 )^{-1/2} \int_{0}^{\infty} (\tau^2)^{-1/2} \exp \{ -\tfrac{1}{2} [  \beta_j^2 \sigma^{-2}
 \tau^{-2} + ( \lambda_1^{\mbox{{\tiny (bayes)}}})^2 \tau^2 ] \} \, d\tau^2.
\end{eqnarray*}
The integrand is the (relative) density of the generalized inverse Gaussian distribution (with a specific choice for one of its parameters). Hence, the integral, mathematically known as a Bessel function, is the normalization constant of this distribution. Bessel functions are notoriously difficult, but for the specific parameter choice here, it simplifies to the exponential function. This gives:
\begin{eqnarray*} 
\pi ( \beta_j) & \propto & \tfrac{1}{2} ( \lambda_1^{\mbox{{\tiny (bayes)}}} / \sqrt{\sigma^2}) \exp ( - \lambda_1^{\mbox{{\tiny (bayes)}}} \hspace{0.5pt} | \beta_j | / \sqrt{\sigma^2}),
\end{eqnarray*}
which is the Laplace distribution. \cite{Park2008bayesian} use the latent variable $\tau^2$ explicitly in the construction of a Gibbs sampler. The relevant conditional distributions then are:
\begin{eqnarray*}
\boldsymbol{\beta} \, | \, \mathbf{Y}, \mathbf{X},  \sigma^2, \tau^2 \hspace{0.2cm} & \sim & \mathcal{N} [ (\mathbf{X}^{\top} \mathbf{X} + \mathbf{D}_\tau^{-1})^{-1} \mathbf{X}^{\top} \mathbf{Y},  \sigma^2 (\mathbf{X}^{\top} \mathbf{X} +  \mathbf{D}_\tau^{-1})^{-1}],
\\
\sigma^2 \, | \, \mathbf{Y}, \mathbf{X},  \boldsymbol{\beta} \hspace{0.83cm} & \sim &  \mathcal{IG} [a_0 + \tfrac{1}{2} (n + p), b_0 + \tfrac{1}{2} \| \mathbf{Y} - \mathbf{X} \boldsymbol{\beta} \|_2^2 +  \tfrac{1}{2}  \boldsymbol{\beta}^{\top} \mathbf{D}_\tau^{-1} \boldsymbol{\beta}],
\\
\tau_j^{-2} \, | \, \beta_j, \sigma^2, \lambda_1^{\mbox{{\tiny (bayes)}}} & \sim &  \mbox{Wald} [ (\sigma^2 / \beta_j^2 )^{1/2}  \lambda_1^{\mbox{{\tiny (bayes)}}}, (\lambda_1^{\mbox{{\tiny (bayes)}}})^2 ],
\end{eqnarray*}
where $\mathbf{D}_\tau$ is a diagonal matrix with $\mbox{diag}(\mathbf{D}_\tau) = (\tau_1^2, \ldots, \tau_p^2)$. The first two conditional distributions are derived as in the case of a normal prior, while we omit the derivation of the third one. Note that the latter is the conditional distribution of $\tau^{-2}$, instead of $\tau^{2}$. This is motivated from computational convenience: the Wald distribution, also known as the inverse Gaussian distribution, has been implemented in most software packages. Finally, the conditional distribution of $\tau^{-2}$ is also conditional on $\lambda_1^{\mbox{{\tiny (bayes)}}}$, for \cite{Park2008bayesian} endow it with a gamma prior. We can now construct a Gibbs sampler for the Bayesian lasso regression by iteratively sampling from the conditional distributions of the preceding display. This Gibbs sampler enables the construction of credible sets to express the uncertainty of the MAP estimator. However, this MAP estimator may be coincide with the lasso regression estimator, its  posterior distribution cannot be blindly used for uncertainty quantification. In high-dimensional sparse settings, the posterior distribution of $\bbeta$ need not concentrate around the true parameter, even though its mode is a good estimator of the regression parameter (cf. Theorem 7 of \citealp{Castillo2015bayesian}).


\section{Penalty parameter selection} \label{section:penaltyParameterSelectionL1}
The informed choice of the lasso penalty parameter can be made by similar means as for the ridge penalty parameter, e.g. cross-validation, information criteria (Section  \ref{sect:penaltyParameterSelectionL2}). Here we describe an alternative procedure for the selection of the lasso penalty parameter based on another heuristic, which aims to ensure a reliable covariate selection by the estimator.

\subsection{Stability selection} \label{sect:stabilitySelection}
Stability selection is an alternative procedure to choose the penalty parameter of the lasso regression estimator \citep{Meinshausen2010stability}. Central to the procedure is a map of $\lambda_1$  from its unitless scale of the positive real numbers to one with a tangible interpretation. This requires the generation of many perturbed versions of the data set. The lasso regression estimator is fitted on each of these version. The number of times a covariate is selected by the estimator over all versions is called its \textit{selection frequency}. The selection frequency is a quantity that is directly related to $\lambda_1$ as $\lambda_1$ controls which covariates enter the model. Moreover, by their stability covariates with a high selection frequency are preferred over those with a low one. The stability selection procedure then chooses $\lambda_1$ such that the resulting model includes only `stable' covariates. In particular, the procedure bounds the number of falsely selected covariates using a cut-off on the selection frequency, in other words, what is considered a stable covariate. This guides an informed choice on the amount of penalization.

Let us detail the stability selection procedure. Define the sets $\mathcal{S} = \{ j \, : \, (\boldsymbol{\beta})_j \not= 0 \}$ and $\mathcal{N} = \{ j \, : \, (\boldsymbol{\beta})_j = 0 \}$ that indicate at which elements the regression parameter $\bbeta$ has support and where it does not, respectively. An estimate of the set $\mathcal{S}$ is the set of nonzero coefficient of the lasso regression estimate $\hat{\mathcal{S}}_{\lambda_1} = \{ j \, : \, [ \hat{\boldsymbol{\beta}} (\lambda_1)]_j \not= 0 \}$, and its complement is an estimate of $\mathcal{N}$. Ideally, we choose $\lambda_1$ such that $\hat{\mathcal{S}}_{\lambda_1} \cap \mathcal{S} = \mathcal{S}$ and $\hat{\mathcal{S}}_{\lambda_1} \cap \mathcal{N} = \emptyset$. 

The stability selection procedure chooses $\lambda_1$ on the basis of the covariates' stability. \cite{Meinshausen2010stability} assess this stability over perturbed versions of the data created by subsampling. Each subsample yields an evaluation of the lasso regression estimator and, thus, an estimate of $\mathcal{S}$. Selection behavior over the subsamples is described by the concept of selection frequency/probability. 

\begin{definition} \mbox{ } \\
Let $\mathcal{I}$ be random drawn without replacement from $\{ 1, \ldots, n \}$ such that $| \mathcal{I} | = \lfloor n/2 \rfloor$. For the $j$-th covariate, the probability of being in the selected set $\hat{\mathcal{S}}_{\lambda_1} (\mathcal{I})$ is $\hat{\Pi}_{\lambda_1} (j) = P [ j \in \hat{\mathcal{S}}_{\lambda_1} (\mathcal{I}) ]$, where the dependence on the subsample is explicated in the notation. 
\end{definition}

\noindent
The probability definition above is with respect to all possible subsamples of the same size and for the particular value of $\lambda_1$. This probability is studied over a set of penalty parameters denoted $\Lambda$.  Hereto we introduce the \textit{stability path} of the $j$-th covariate, which is the set of the selection probabilities $\hat{\Pi}_{\lambda_1} (j)$ obtained by running the regularization parameter $\lambda_1$ over the elements of $\Lambda$ for subsample $\mathcal{I}$. Furthermore, we define $\hat{\mathcal{S}}_{\Lambda} (\mathcal{I}) = \bigcup_{\lambda_1 \in \Lambda} \hat{\mathcal{S}}_{\lambda_1} (\mathcal{I})$, the set of selected covariates at any point of their stability path. For a covariate to be in $\hat{\mathcal{S}}_{\Lambda} (\mathcal{I})$ for a particular subsample $\mathcal{I}$ is not of interest. But if it is for many subsamples, it is considered stable.  

\begin{definition} \mbox{ } \\
The set of stable covariates is:
\begin{eqnarray*}
\hat{\mathcal{S}}^{\mbox{{\tiny stable}}} & = & \{ j \, : \, \max\nolimits_{\lambda_1 \in \Lambda}   \hat{\Pi}_{\lambda_1} (j)  \geq  \pi_{\mbox{{\tiny thr}}} \},
\end{eqnarray*} 
for threshold $\pi_{\mbox{{\tiny thr}}} \in (0, 1)$ and the set of regularization parameters $\Lambda$.
\end{definition}

\noindent
A covariate is thus considered stable if at some point on its stability path it is selected in more than $100 \pi_{\mbox{{\tiny thr}}} \%$ of the subsamples. \cite{Meinshausen2010stability} claim that selection on the basis of stability is relatively insensitive to either the choice of $\pi_{\mbox{{\tiny thr}}}$ or that of $\lambda_1$ and $\Lambda$. \cite{Meinshausen2010stability} then prove that error control of selection based on stability is possible.

\begin{theorem} (Theorem 1, \citealp{Meinshausen2010stability}) \label{theorem:selectionStability} \\
Assume the distribution of $\mathbbm{1}_{\{j \in \hat{\mathcal{S}}_{\lambda_1} \} }$ to be exchangeable over the random subsamples for all $\lambda_1 \in \Lambda$ and $j \in \mathcal{N}$. Also, assume the original procedure is not worse than random guessing, i.e. 
\begin{eqnarray*}
\frac{\mathbb{E} ( |\mathcal{S} \cap \hat{\mathcal{S}}_{\Lambda} |) }
{\mathbb{E} ( |\mathcal{N} \cap \hat{\mathcal{N}}_{\Lambda} |) }
& = & \frac{ | \mathcal{S} | }{ |\mathcal{N} | }.
\end{eqnarray*}
The expected number $V = | \mathcal{N} \cap \hat{\mathcal{S}}^{\mbox{{\tiny stable}}} |$ of falsely selected but stable variables is then bounded for $\pi_{\mbox{{\tiny thr}}} \in (\tfrac{1}{2}, 1)$ by $V \leq (2 \pi_{\mbox{{\tiny thr}}} - 1)^{-1} p^{-1} q_{\Lambda}^2$, 
where $q_{\Lambda} = \mathbb{E}_{\mathcal{I}} [  | \hat{\mathcal{S}}_{\Lambda} (\mathcal{I}) | ]$ is the expected number of selected covariates over all subsamples of equal size. 
\end{theorem}
\begin{proof}
See  \cite{Meinshausen2010stability}.
\end{proof}

\noindent
Exchangeability thus requires that for the covariates with a zero coefficient the probability of being selected is invariant under subsampling over the full range of penalty parameters. The validity of the exchangeability may be hard to assess in practice. The random guessing assumption, however, seems unproblematic for any minimally sophisticated method.


Theorem \ref{theorem:selectionStability} can be put to practical use and guide the decision on the optimal value(s) of $\lambda_1$ as follows. Specify the stability threshold, i.e. the bound on the covariates' selection frequency beyond which they are considered stable, $\pi_{\mbox{{\tiny thr}}}$. Then, to ensure (say) $\mathbb{E}(V) \leq  1$ choose the penalty parameter $\lambda_1$ such that $q_{\Lambda} \leq \sqrt{(2 \pi_{\mbox{{\tiny thr}}} - 1) p}$. While $q_{\Lambda}$ is in principle unknown, it can be estimated by means of the resampling.


\section{Comparison to ridge}
Here an inventory of the similarities and differences between the lasso and ridge regression estimators is presented. To recap what we have seen so far: both estimators optimize a loss function of the form (\ref{form.penLeastSquares}) and can be viewed as Bayesian estimators. But in various respects the lasso regression estimator exhibited differences from its ridge counterpart: \textit{i)} the former need not be uniquely defined (for a given value of the penalty parameter) whereas the latter is, \textit{ii)} an analytic form of the lasso regression estimator does in general not exists, but \textit{iii)} it is sparse (for large enough values of the lasso penalty parameter). The remainder of this section expands this inventory.

\subsection{Linearity}
The ridge regression estimator is a linear -- in the observations -- estimator, while the lasso regression estimator is not. This is immediate from the analytic expression of the ridge regression estimator, $\hat{\bbeta}(\lambda_2) = (\mathbf{X}^{\top} \mathbf{X} + \lambda_2 \mathbf{I}_{pp})^{-1} \mathbf{X}^{\top} \mathbf{Y}$, which is a linear combination of the observations $\mathbf{Y}$. To show the non-linearity of the lasso regression estimator available, it suffices to study the analytic expression of $j$-th element of $\hat{\bbeta}(\lambda_1)$ in the orthonormal case: $\hat{\beta}_j(\lambda_1) = \mbox{sign}(\hat{\beta}_j) ( |\hat{\beta}_j| - \tfrac{1}{2} \lambda_1)_+ = \mbox{sign}(\mathbf{X}_{\ast,j}^{\top} \mathbf{Y} ) ( | \mathbf{X}_{\ast,j}^{\top} \mathbf{Y} | - \tfrac{1}{2} \lambda_1)_+$. This clearly is not linear in $\mathbf{Y}$. Consequently, the response $\mathbf{Y}$ may be scaled by some constant $c$, denoted $\tilde{\mathbf{Y}} = c \mathbf{Y}$, and the corresponding ridge regression estimators are one-to-one related by this same factor $\hat{\bbeta} (\lambda_2) = c \tilde{\bbeta} (\lambda_2)$. The lasso regression estimator based on the unscaled data is not so easily recovered from its counterpart obtained from the scaled data.

\subsection{Shrinkage}
\begin{figure}[!h]
\begin{tabular}{rcl}
\includegraphics[scale=0.20, angle=0]{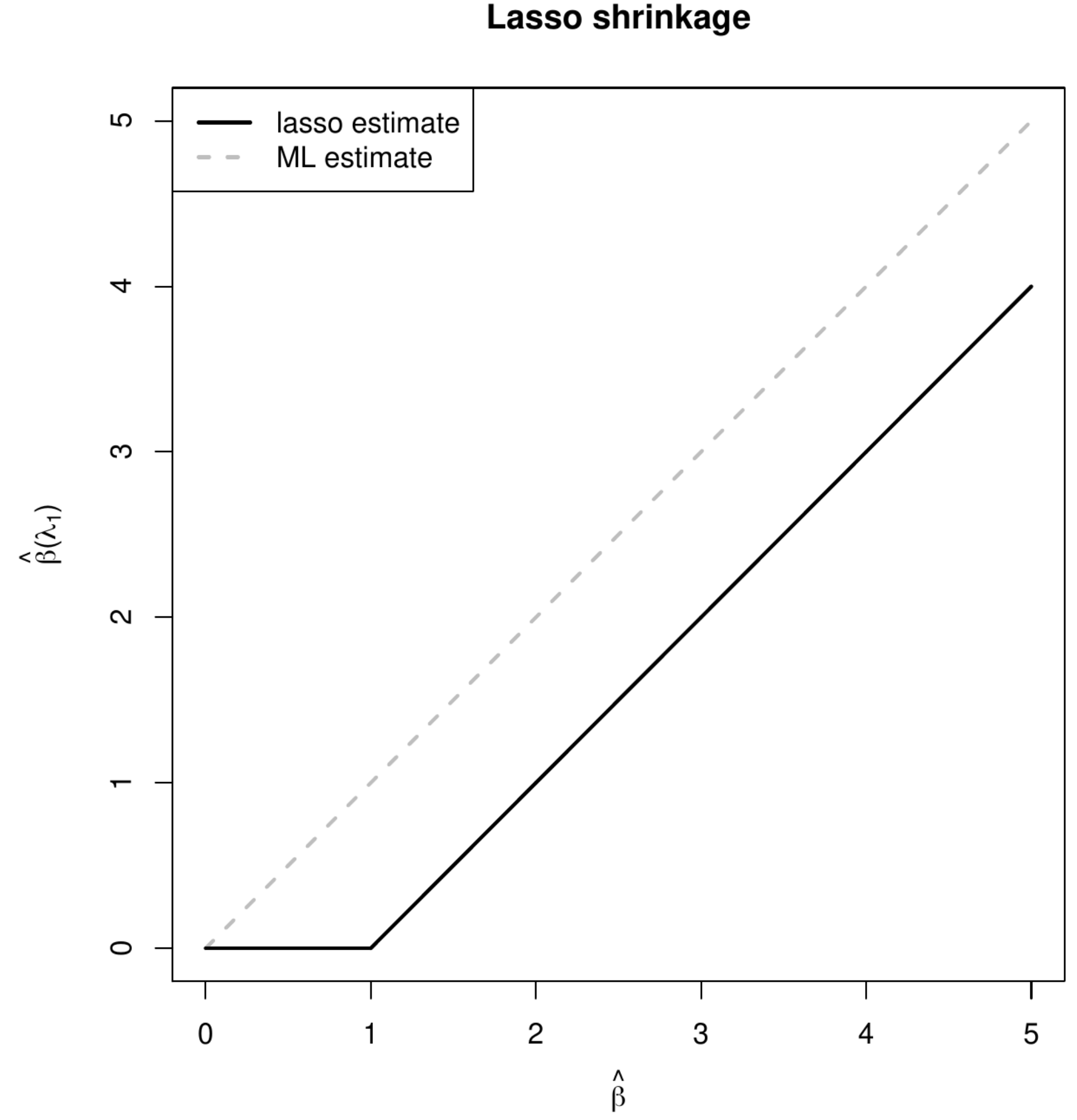}
& &
\includegraphics[scale=0.20, angle=0]{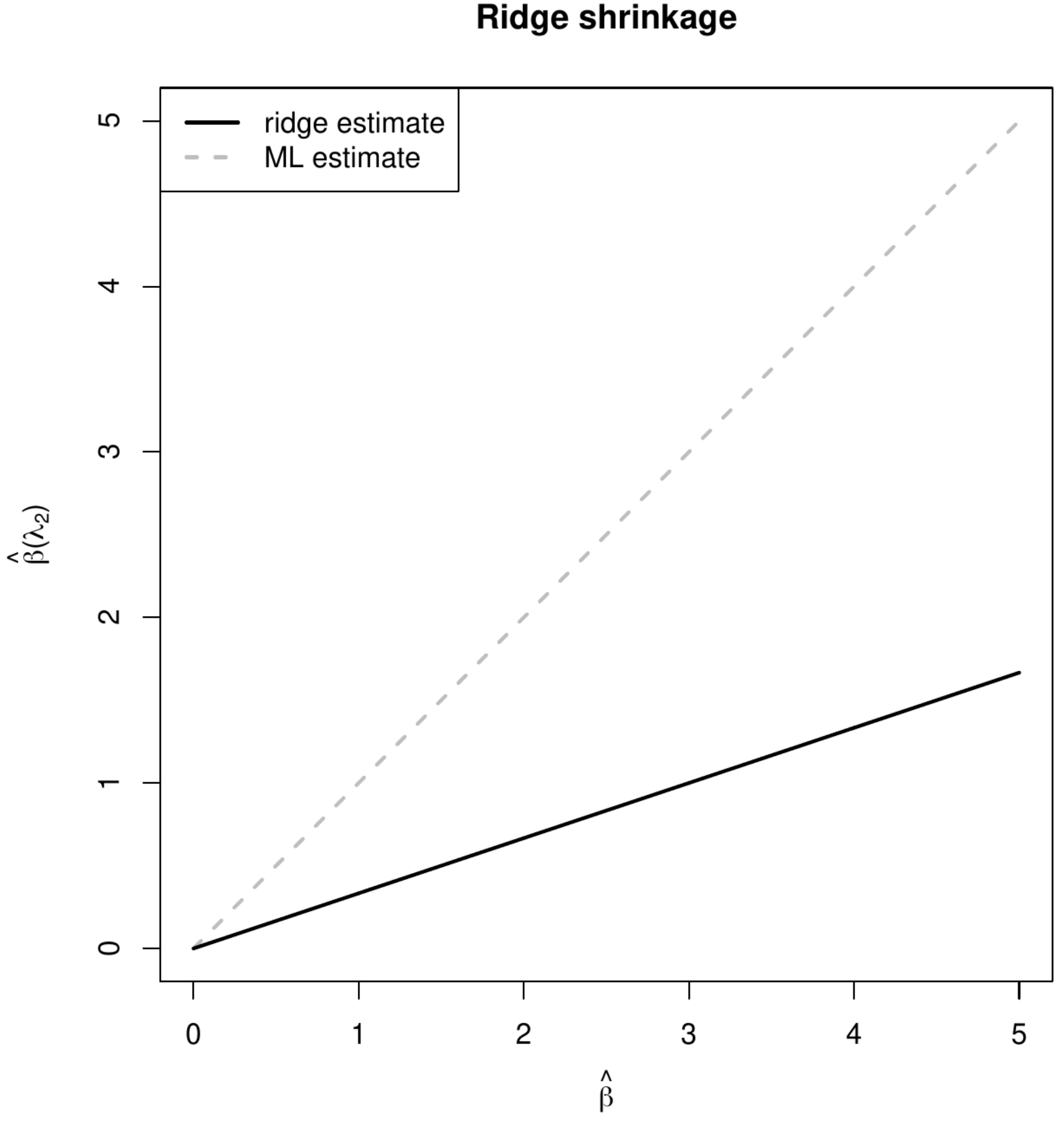}
\end{tabular}
\caption{Illustration of the shrinkage effect of the lasso and ridge regression estimators, left- and right-hand side panel, respectively, for data with an orthonormal design matrix. The maxmimum likelihood estimator is on $x$-axis, while its shrunken counterpart (lasso or ridge), is on the $y$-axis.} \label{fig.lassoVsRidge_linear2proportionalShrinkage}
\end{figure}

Both lasso and ridge regression estimation minimize the sum-of-squares plus a penalty. The latter encourages the estimator to be small, in particular closer to zero. This behavior is called shrinkage. The particular form of the penalty yields different types of this shrinkage behavior. This is best grasped in the case of an orthonormal design matrix. The $j$-the element of the ridge regression estimator then is: $\hat{\beta}_j (\lambda_2) = (1+\lambda_2)^{-1} \hat{\beta}_j$, while that of the lasso regression estimator is: $\hat{\beta}_j (\lambda_1) = \mbox{sign}(\hat{\beta}_j) ( |\hat{\beta}_j| - \tfrac{1}{2} \lambda_1)_+$. In Figure \ref{fig.lassoVsRidge_linear2proportionalShrinkage} these two estimators $\hat{\beta}_j (\lambda_2)$ and $\hat{\beta}_j (\lambda_1)$ are plotted as a function of the maximum likelihood estimator $\hat{\beta}_j$. Figure \ref{fig.lassoVsRidge_linear2proportionalShrinkage} shows that lasso and ridge regression estimator translate and scale, respectively, the maximum likelihood estimator, which could also have been concluded from the analytic expression of both estimators. The scaling of the ridge regression estimator amounts to substantial and little shrinkage (in an absolute sense) for elements of the regression parameter $\bbeta$ with a large and small maximum likelihood estimate, respectively. In contrast, the lasso regression estimator applies an equal amount of shrinkage to each element of $\bbeta$, irrespective of the coefficients' sizes.

\subsection{Simulation I: covariate selection} \label{example:lassoEffectOfCovariateVariance}
Here it is investigated whether lasso regression exhibits the same behaviour as ridge regression in the presence of covariates with differing variances. Recall: the simulation of Section \ref{ridge:covariateVariances} showed that ridge regression shrinks the estimates of covariates with a large spread less than those with a small spread. That simulation has been repeated, with the exact same parameter choices and sample size, but now with the ridge regression estimator replaced by the lasso regression estimator. To refresh the memory: in the simulation of Section \ref{ridge:covariateVariances} the linear regression model is fitted, now with the lasso regression estimator. The $(n=1000) \times (p=50)$ dimensional design matrix $\mathbf{X}$ is sampled from a multivariate normal distribution: $\mathbf{X}_{i,\ast}^{\top} \sim \mathcal{N}(\mathbf{0}_{50}, \mathbf{\Sigma})$ with $\mathbf{\Sigma}$ diagonal and $(\mathbf{\Sigma})_{jj} = j / 10$ for $j=1, \ldots, p$. The response $\mathbf{Y}$ is generated through $\mathbf{Y} = \mathbf{X} \bbeta + \vvarepsilon$ with $\bbeta$ a vector of all ones and $\vvarepsilon$ sampled from the multivariate standard normal distribution. Hence, all covariates contribute equally to the response.

\begin{figure}[!h]
\begin{tabular}{rcl}
\includegraphics[scale=0.22, angle=0]{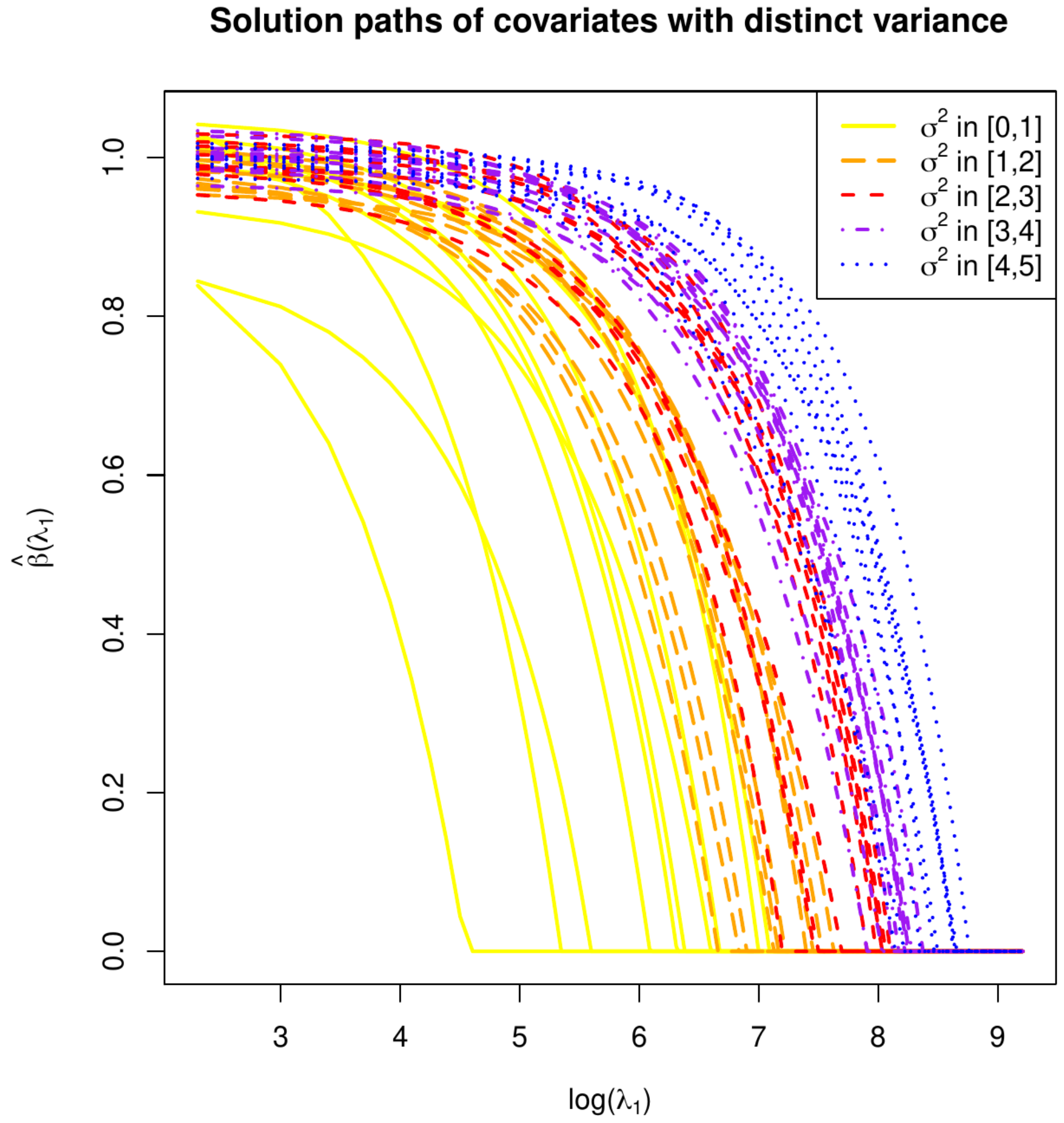}
& &
\includegraphics[scale=0.22, angle=0]{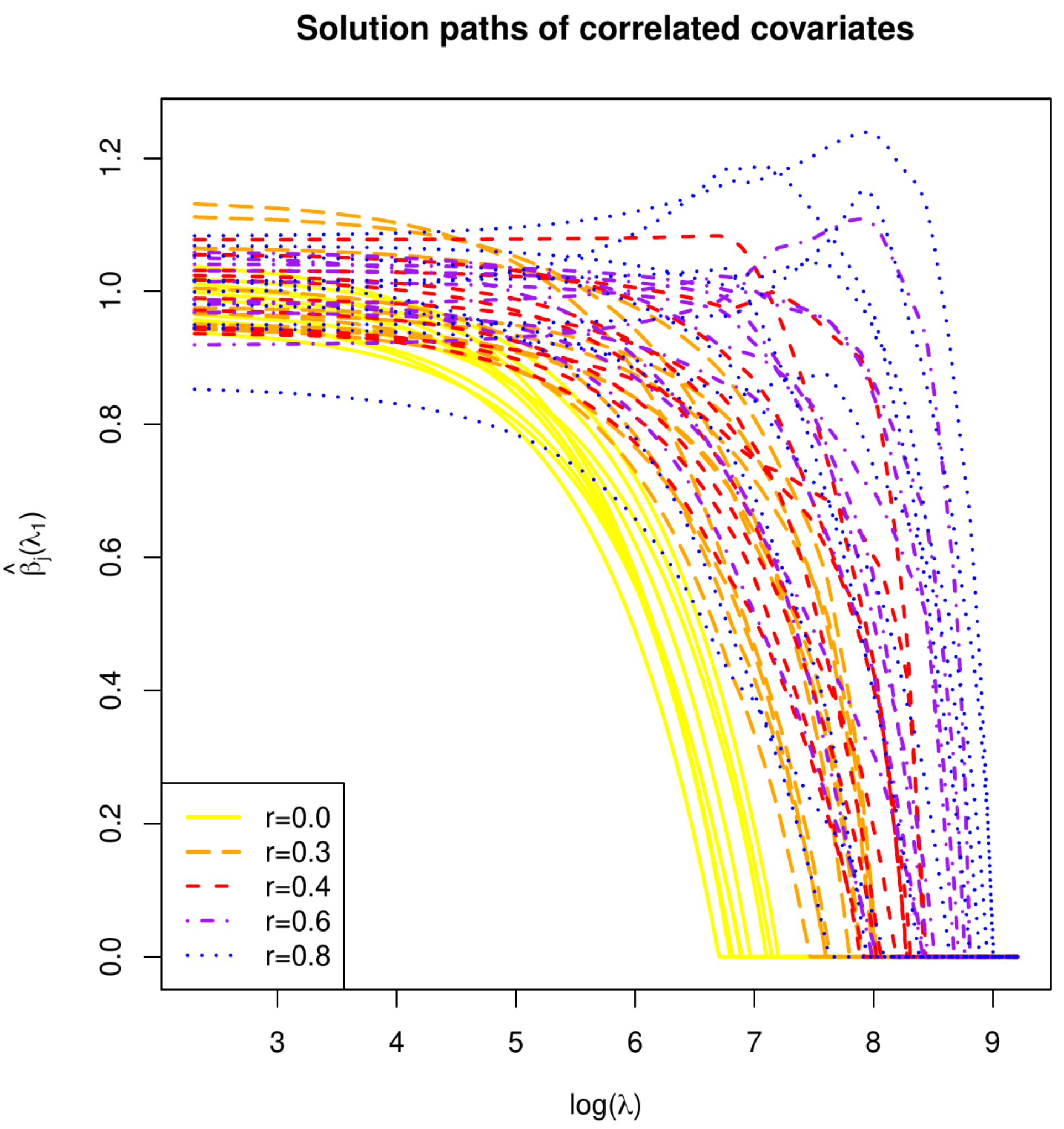}
\end{tabular}
\caption{Regularization paths of the lasso regression estimator related to the simulations presented in Sections \ref{example:lassoEffectOfCovariateVariance} and \ref{example.lassoCorrelatedVariables} in the left and right panel, respectively.} \label{fig.lassoCorrelatedCovariates}
\end{figure}

The results of the simulation are displayed in Figure \ref{fig.lassoCorrelatedCovariates}, which shows the regularization paths of the $p=50$ covariates. The regularization paths are demarcated by color and style to indicate the size of the spread of the corresponding covariate.  These regularization paths show that -- like its ridge counterpart -- the lasso regression estimator shrinks most the regression coefficients of the covariates with the smallest range/variance. For the lasso regression, this translates (for sufficiently large values of the penalty parameter) into a preference for the selection of covariates with largest range/variance.

Intuition for this behavior of the lasso regression estimator may be obtained through geometrical arguments analogous to that provided for the similar behaviour of the ridge regression estimator in Section \ref{ridge:covariateVariances}. Algebraically, it is easily seen when assuming an orthogonal design with $\mbox{Var}(X_1) \gg \mbox{Var}(X_2)$. The lasso regression loss function can then be rewritten, as in Example \ref{example.orthogonalDesignLasso}, to:
\begin{eqnarray*}
\| \mathbf{Y} - \mathbf{X} \bbeta \|_2^2 + \lambda_1 \| \bbeta \|_1 & = &
\| \mathbf{Y} - \tilde{\mathbf{X}}  \ggamma \|_2^2 +
\lambda_1 [\mbox{Var}(X_1)]^{-1/2} | \gamma_1 |  + \lambda_1 [\mbox{Var}(X_2)]^{-1/2} | \gamma_2 |,
\end{eqnarray*}
where $\gamma_1 = [\mbox{Var}(X_1)]^{1/2} \beta_1$ and $\gamma_2 = [\mbox{Var}(X_2)]^{1/2} \beta_2$. The rescaled design matrix $\tilde{\mathbf{X}}$ is now orthonormal and analytic expressions of estimators of $\gamma_1$ and $\gamma_2$ are available. The former parameter is penalized substantially less than the latter as $\lambda_1 [\mbox{Var}(X_1)]^{-1/2} \ll \lambda_1 [\mbox{Var}(X_2)]^{-1/2}$. As a result, if for large enough values of $\lambda_1$ one variable is selected, it is more  likely to be $\gamma_1$.

\subsection{Simulation II: correlated covariates}  \label{example.lassoCorrelatedVariables}
The behaviour of the lasso regression estimator is now studied in the presence of collinearity among the covariates. Previously, in simulation, Section \ref{sect:collinearCovariates}, the ridge regression estimator was shown to exhibit the joint shrinkage of strongly collinear covariates. This simulation is repeated for the lasso regression estimator. The details of the simulation are recapped. The linear regression model is fitted by means of the lasso regression estimator. The $(n=1000) \times (p=50)$ dimensional design matrix $\mathbf{X}$ is samples from a multivariate normal distribution: $\mathbf{X}_{i,\ast}^{\top} \sim \mathcal{N}(\mathbf{0}_{50}, \mathbf{\Sigma})$ with a block-diagonal $\mathbf{\Sigma}$. The $k$-th, $k=1,\ldots,5$, diagonal block, denoted $\mathbf{\Sigma}_{kk}$ comprises ten covariates and equals $\frac{k-1}{5} \, \mathbf{1}_{10 \times 10} + \frac{6-k}{5} \, \mathbf{I}_{10 \times 10}$ for $k=1, \ldots, 5$. The response vector $\mathbf{Y}$ is then generated by $\mathbf{Y}  = \mathbf{X} \bbeta + \vvarepsilon$, with $\vvarepsilon$ sampled from the multivariate standard normal distribution and $\bbeta$ containing only ones. Again, all covariates contribute equally to the response.

The results of the above simulation results are captured in Figure \ref{fig.lassoCorrelatedCovariates}. It shows the lasso regularization paths for all elements of the regression parameter $\bbeta$. The regularization paths of covariates corresponding to the same block of $\mathbf{\Sigma}$ (indicative of the degree of collinearity) are now marcated by different colors and styles. Whereas the ridge regularization paths are nicely grouped per block, their lasso counterparts do not. The selection property spoils the party. Instead of shrinking the regression parameter estimates of collinear covariates together, the lasso regression estimator (for sufficiently large values of its penalty parameter $\lambda_1$) tends to pick one covariate to enter the model while forcing the others out (by setting their estimates to zero).

\subsection{Simulation III: sparsity}  \label{example.lasso2ridge_sparsity}
Our next simulation investigates the effect of sparsity on the predictive performance of the lasso and ridge regression estimators. Hereto we draw again from the linear model $\mathbf{Y} = \mathbf{X} \boldsymbol{\beta} + \boldsymbol{\varepsilon}$ with $\boldsymbol{\varepsilon} \sim \mathcal{N}(\mathbf{0}_n, \sigma^2 \mathbf{I}_{nn})$. Throughout we set the sample size to $n=100$ and adopt a unit error variance $\sigma^2 = 1$. Furthermore, to see the effect of the dimension, we vary it such that $p \in \{ 50, 500, 5000 \}$. An $n\times p$-dimensional design matrix $\mathbf{X}$ is then formed by subsetting the  \texttt{breastCancerVDX}-package's gene expression data \citep{breastCancerVDX2022}, with covariates and samples randomly drawn. Subsequently, we center the covariates at zero. This data-driven choice of the design matrix harbors strongly collinear covariates that have different variances. Furthermore, with respect to the regression parameter we assume either a sparse or dense structure:
\begin{compactitem}
\item[$\circ$] \textit{sparse} : i.e. $\beta_j = j$ for $j=1, \ldots, 5$ and $\beta_j = 0$ otherwise, or

\item[$\circ$] \textit{dense}\, : i.e. $\beta_j = 10 p^{-1}$ for all $j=1, \ldots, p$.
\end{compactitem}
In the dense case, the size of the elements of $\bbeta$ depends on $p$. This aims to keep the variance of the response approximately constant over the various employed dimensions. With the design matrix and regression parameter at hand, we draw the response in accordance with the distributional assumption of the linear regression model. Next we choose the penalty parameter of both estimators by means of leave-one-out cross-validation. The estimators are then evaluated with the optimal penalty parameter, and subsequently, the prediction of the left-out samples are obtained. These predictions are compared to the corresponding observations by means of Spearman's rank correlation. This performance measure is hardly affected by the shrunken scale of the predictions due the regularization. For the lasso regression estimate we also register which elements of the parameter are nonzero. The aforementioned has been repeated a thousand times for each setting. 

\begin{lstlisting}
# activate libraries
library(penalized)
library(breastCancerVDX)
library(Biobase)

# set parameters
structure <- c("dense", "sparse")[2]
p         <- c(50, 500, 5000)[1]
n         <- 100

# load data
data(vdx)

# define storage variables
selFreqsL1     <- matrix(0, nrow=p, ncol=1)
cors           <- matrix(nrow=0, ncol=4)
colnames(cors) <- c("n", "p", "corL1", "corL2")
for (u in 1:1000){
	# create design matrix, regression parameter, and response 
	X            <- exprs(vdx)
	idCovariates <- sample(1:nrow(X), p)
	idSamples    <- sample(1:ncol(X), n)
	X            <- t(X[idCovariates, idSamples])
	X            <- sweep(X, 2, apply(X, 2, mean), "-")
	if (structure == "dense"){
		betas <- matrix(rep(10/p, p), ncol=1) 
	} 
	if (structure == "sparse"){
 		betas      <- matrix(rep(0, p), ncol=1)
		betas[1:5] <- c(1:5)
	}
	Y <- X %*% betas + rnorm(n)

	# lasso: optimal lambda, estimate model, extract prediction and selection
	optLasso   <- optL1(Y, X, trace=FALSE)
	YpredL1    <- optLasso$prediction[,1]
	selFreqsL1 <- selFreqsL1 + 1*(coef(optLasso$fullfit, "all")[-1] != 0)

	# ridge: optimal lambda, estimate model and extract prediction
	optRidge <- optL2(Y, X, trace=FALSE)
	YpredL2  <- optRidge$prediction[,1]
		
	# results	
	cors <- rbind(cors, c(n, p, cor(Y, YpredL1, m="s"), 
	                            cor(Y, YpredL2, m="s")))
}




\end{lstlisting}

\begin{figure}[h!]
\begin{center}
\centering
\begin{tabular}{ll}
\hspace{-2.3cm}
\includegraphics[angle=0, scale=0.40]{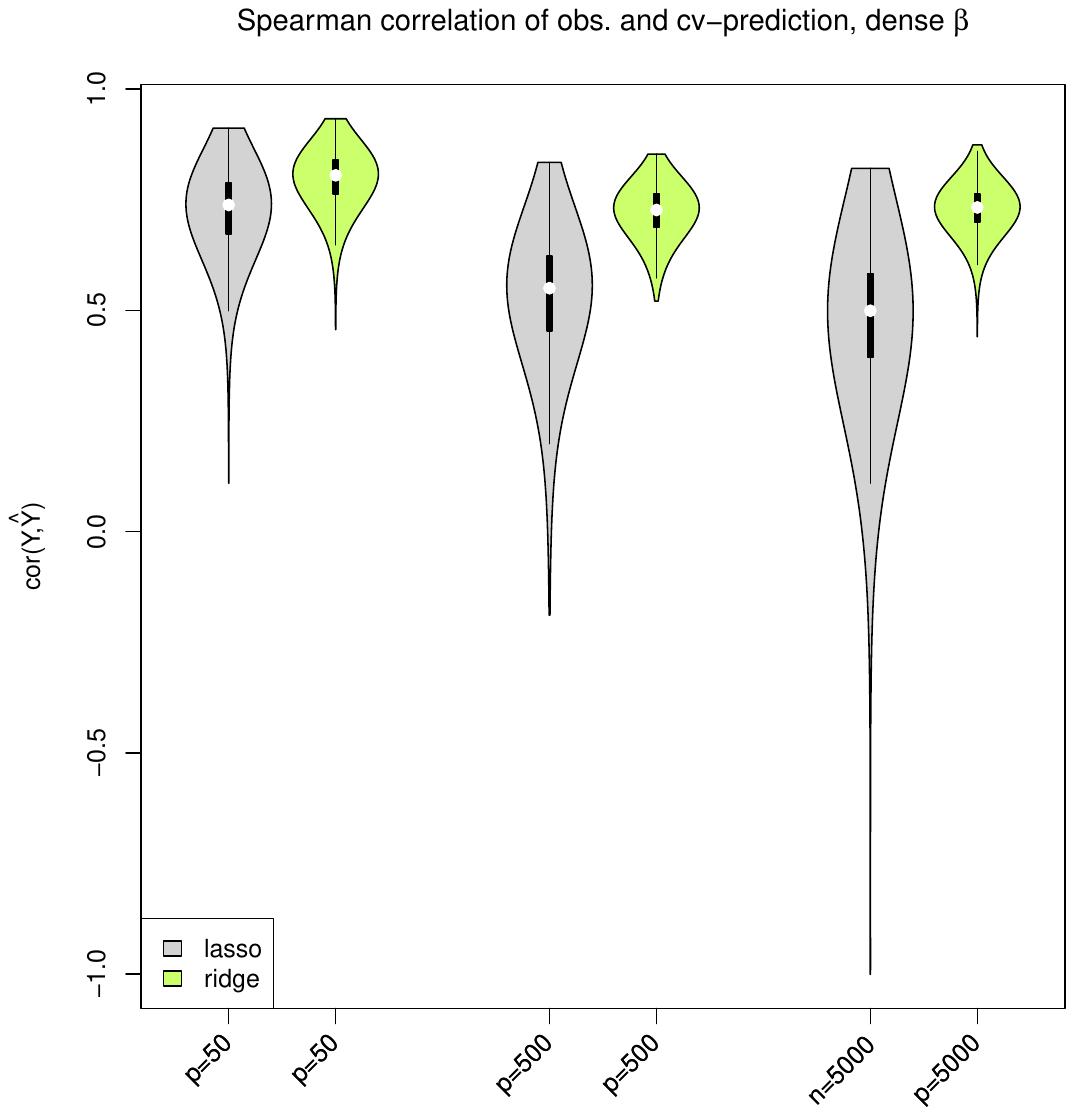}
&
\hspace{-3.5cm}
\includegraphics[angle=0, scale=0.40]{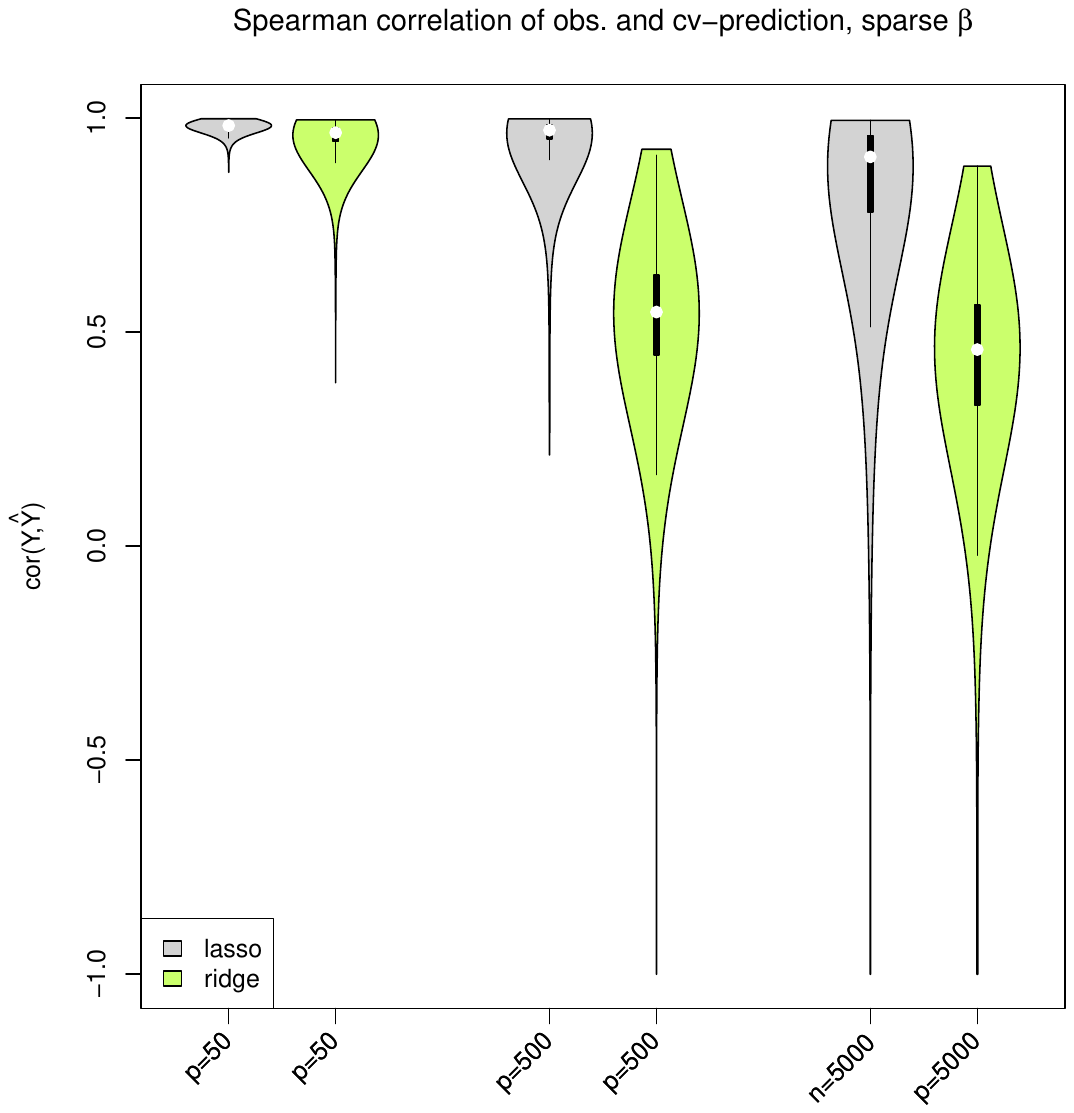}
\end{tabular}
\end{center}
\caption{Violinplots of the predictive performance, operationalized as Spearman's rank correlation among observation and prediction, of the lasso and ridge regression estimates with fixed absolute sparsity at 10\% for various increasing $p$ and a dense (\textit{left}) and sparse (\textit{right}) regression parameter. \label{fig.lasso2ridge_perfComp_dense2sparse}}
\end{figure}

\noindent
The predictive performances as measured by Spearman's rank correlation over all iterations of the simulation is depicted by violinplots in Figure \ref{fig.lasso2ridge_perfComp_dense2sparse}. In the setting with a dense regression parameter, the ridge regression estimates show a better performance than its lasso counterpart. This becomes more pronounced for larger dimensions: the former's performance is reasonably constant over the dimensions, while the latter's performance dilutes slightly and, additionally, becomes more unstable, i.e. the spread among the thousand correlations increases. In the sparse case, the roles are reversed and the lasso regression estimates exhibit a better performance. But with larger dimensions the spread among both estimates' performance measures increases, although most for the ridge regression estimator. Overall, the ridge and lasso regression estimators are prefered if the true regression parameter is dense and sparse, respectively.

\begin{figure}[h!]
\begin{center}
\centering
\hspace{-1.3cm}
\includegraphics[angle=0, scale=0.45]{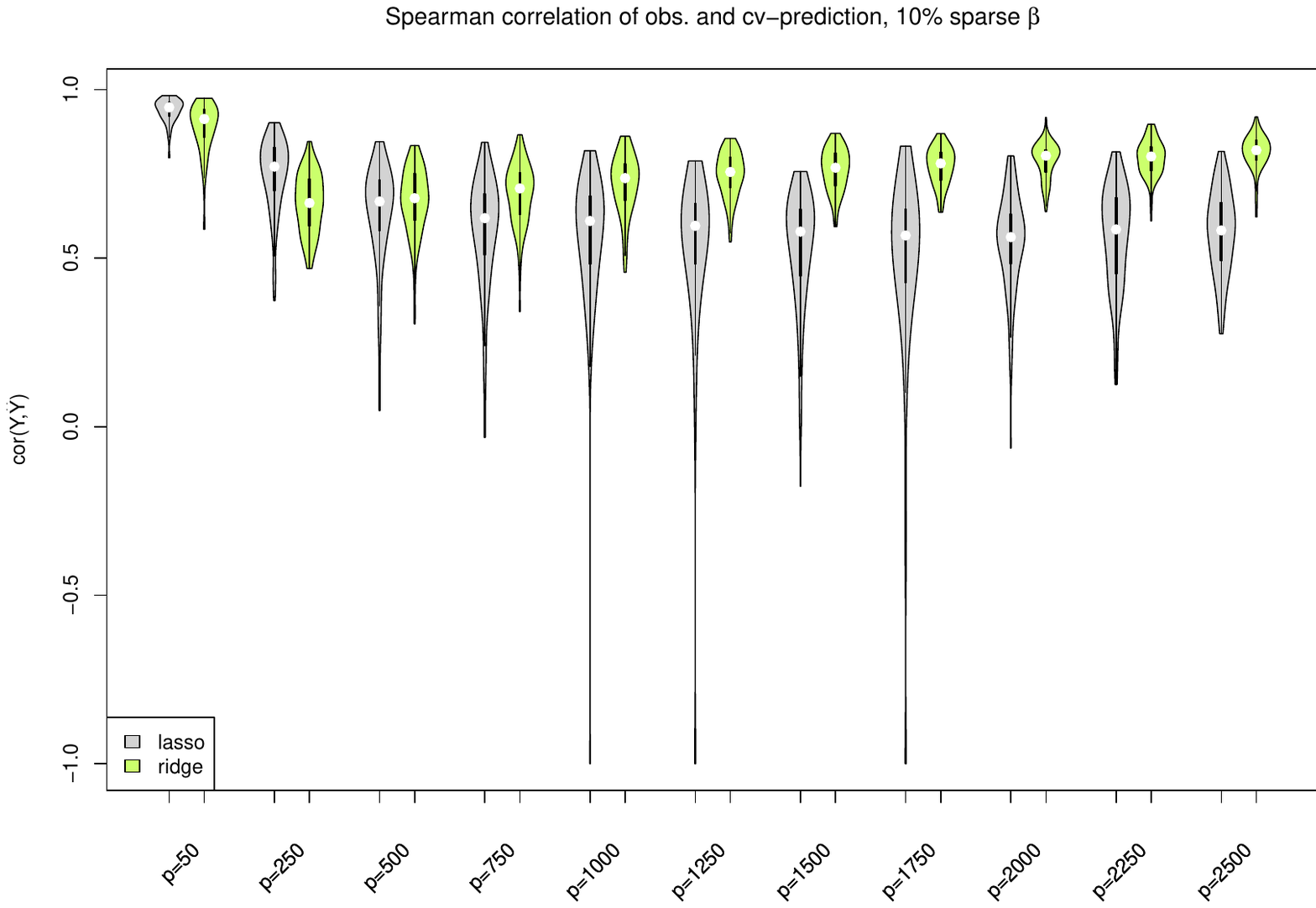}
\end{center}
\caption{Violinplots of the predictive performance, operationalized as Spearman's rank correlation among observation and prediction, of the lasso and ridge regression estimates with fixed relative sparsity at 10\% and increasing $p$. \label{fig.lasso2ridge_perfComp_dense2sparsity}}
\end{figure}

The selection frequency of the lasso regression estimator (not shown) reveals that larger (in absolute sense) elements of the parameter are selected more often than smaller ones. The selection frequency of either waters down as the dimension increases. In principle, we could do a similar excercise for the ridge estimator by studying the ranks of the estimate's absolute value. This indicates similar behavior, i.e. the ranks are concordant with the true parameter values, but a formal selection procedure is absent, although we could formulate some thresholding-type procedure.
\\
\\
We have repeated the simulation but fix the sparsity in a relative sense instead of an absolute one. That is, the relative sparsity is fixed at 10\%, i.e. $\lceil p/10 \rceil$ elements of the regression parameter are non-zero, instead of the five non-zero elements irrespective of the dimension. The elements of the regression parameter are set to $\beta_j = 500 \, j \,  p^{-7/4}$ for $j = 1, \ldots, \lceil p/10 \rceil$ and $\beta_j = 0$ for $j= \lceil p/10 \rceil +1, \ldots, p$. The particular dependence of the nonzero elements of the regression parameter on the dimension is a clumsy attempt to fix the variance of the response over the employed range of dimensions. The latter now ranges over $p \in \{50, 250, 500, 750, \ldots, 2500 \}$. Figure \ref{fig.lasso2ridge_perfComp_dense2sparsity} shows the violinplots of the resulting thousand Spearman's rank correlations between the predictions and observations of the lasso and ridge estimates for the various dimensions. For small dimensions, $p=50$ and $p=250$, the predictive performance of the ridge regression estimates falls behind that of its lasso counterpart. But as the dimension grows beyond $p = 500$, the roles reverse and the ridge regression estimate performs better than its lasso counterpart. 

Let us close with some intuition why we see those results. The lasso regression estimator selects, and thereby performs dimension reduction. This may be unstable and thereby less reproducible in high-dimensional settings. The ridge regression estimator can be conceived as doing some form of averaging, which is relatively robust and reproducible. 

In all, these simulations may suggest that overall the `ridge predictor' tends to have a better predictive performance than its lasso counterpart. Should we prefer thus the ridge over the lasso regression estimator? No, that would be shortsighted. The simulations are far too limited in nature, they are only meant to illustrate the dependence of the behavior of the estimators under various choices of the regression parameter. One may interpret the simulations as suggesting that a choice for either estimator comes with a believe on the structure of the regression parameter. E.g., the lasso regression estimator implicitly assumes the system under study is sparse (and the nonzero elements are clearly distinguishable -- in some sense -- from zero). The validity of this assumption is not directly obvious in, e.g., biological phenomena. Similarly, the ridge regression estimator implicity assumes all -- or at least many -- covariates contribute in comparable manner to the explanation of the variation of the response. Such an assumption too is difficult to verify.

\section{Application}
A seminal high-dimensional data set, that has often been re-analyzed in many articles to illustrate novel (regularization) methods, is presented in \cite{vanTVeer2002gene,vanDeVijver2002gene}. It is part of a study into breast cancer and comprises the overall survival information and expression levels of 24158 genes of 291 breast cancer samples. Alongside the high-throughput omics data, clinical information like age and sex is available but discarded here. The data are provided via the \texttt{breastCancerNKI}-package \citep{breastCancerNKI2022}.

In the original work of \cite{vanTVeer2002gene,vanDeVijver2002gene} the data are used to build a linear predictor of overall survival. This resulted in a so-called `signature' of 70 genes -- quite a dimension reduction! -- that together are predictive of overall survival. Upon this work, a commercial enterprise has been founded that introduced a diagonostic test called the Mammaprint (\url{https://en.wikipedia.org/wiki/MammaPrint}). In a nutshell, the workings of the test (essentially) amount to the evaluation of the linear predictor, which is subsequently compared to a reference value to decide on the expected survival outcome. Within the context of our data set, the reference value will be the median of all linear predictions:
\vspace{-0.2cm}
\begin{eqnarray*}
\left\{
\begin{array}{rcl}
\texttt{poor prognosis} & \mbox{if} & \mathbf{X}_{i, \ast} \hat{\boldsymbol{\beta}} > \mbox{median} \{ \mathbf{X}_{1, \ast} \hat{\boldsymbol{\beta}}, \ldots, \mathbf{X}_{n, \ast} \hat{\boldsymbol{\beta}} \},
\\
\texttt{good prognosis} & \mbox{if} & \mathbf{X}_{i, \ast} \hat{\boldsymbol{\beta}} < \mbox{median} \{ \mathbf{X}_{1, \ast} \hat{\boldsymbol{\beta}}, \ldots, \mathbf{X}_{n, \ast} \hat{\boldsymbol{\beta}} \}.
\end{array} \right.
\end{eqnarray*}
\vspace{-0.4cm}
\\
The prognosis determines the individual's follow-up treatment. For instance, should the test indicate a `\texttt{good prognosis}', the individual may be spared chemo-therapy without a substantial overall survival reduction but a considerable gain in the person's quality of life.

\begin{figure}[h!]
\begin{center}
\centering
\begin{tabular}{ccc}
\includegraphics[angle=0, scale=0.18]{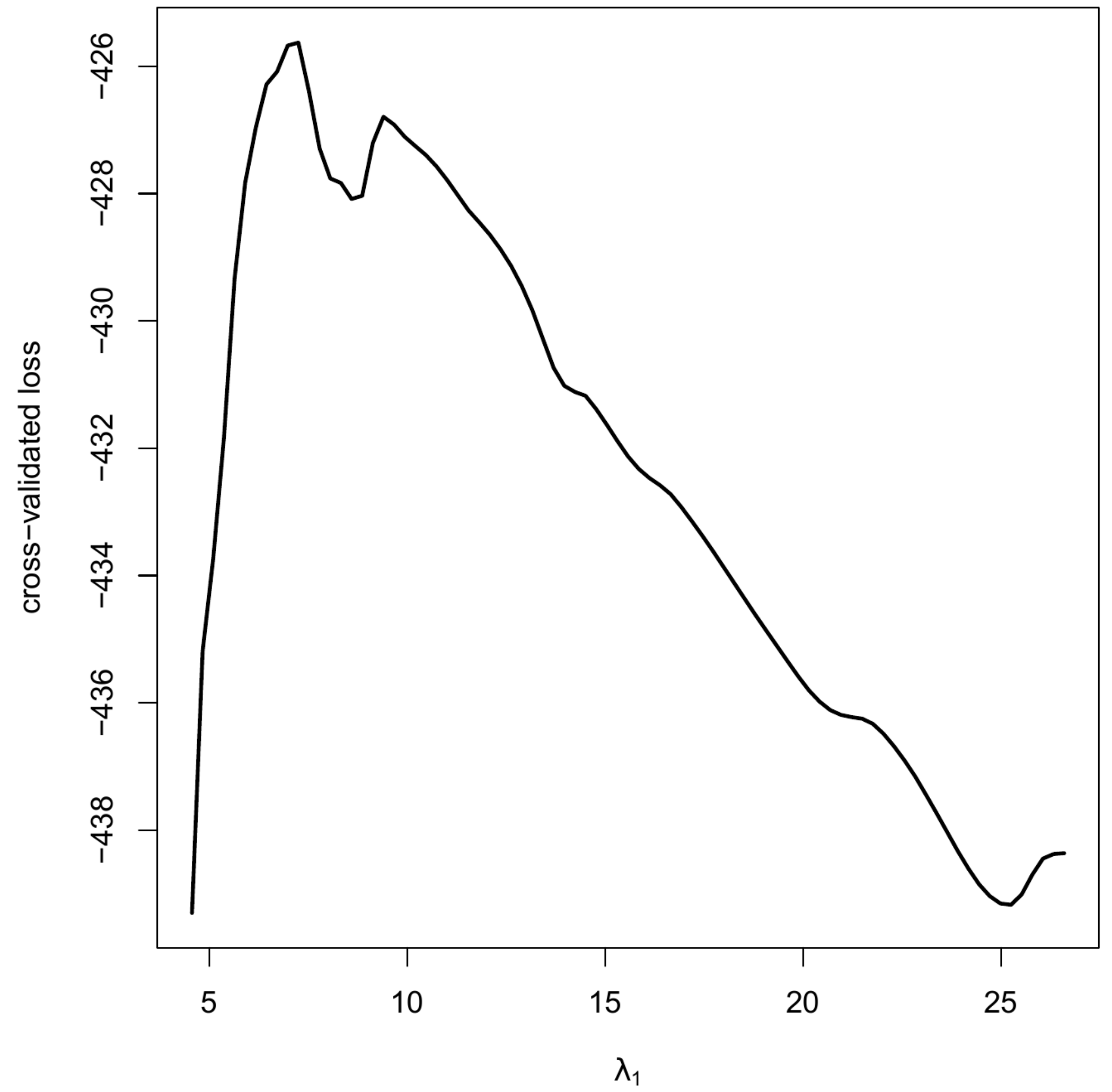}
&  \mbox{ } \hspace{-2.5cm} \mbox{ }   &
\includegraphics[angle=0, scale=0.16]{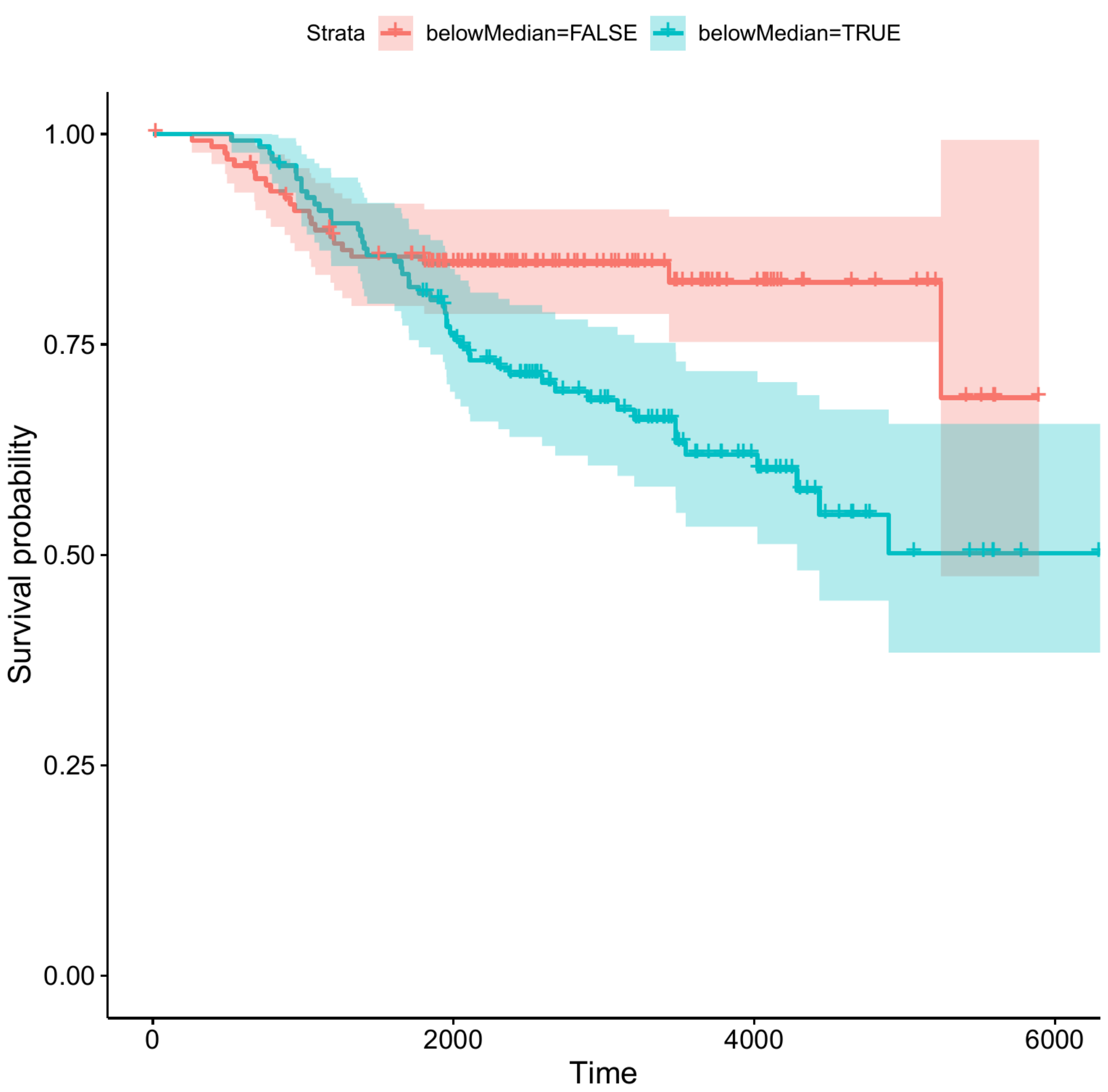}
\\
\includegraphics[angle=0, scale=0.35]{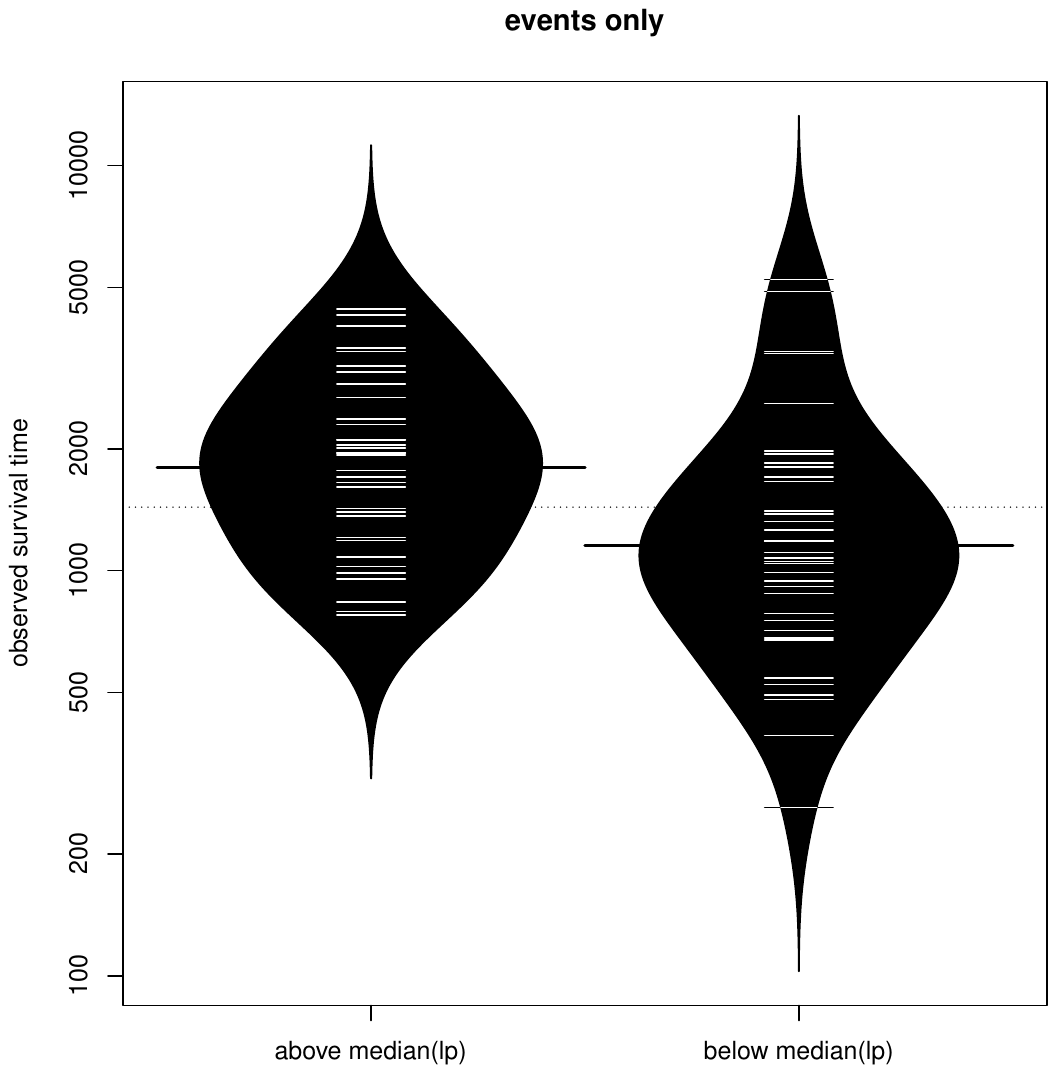}
& \mbox{ } \hspace{-2.5cm} \mbox{ } &
\includegraphics[angle=0, scale=0.35]{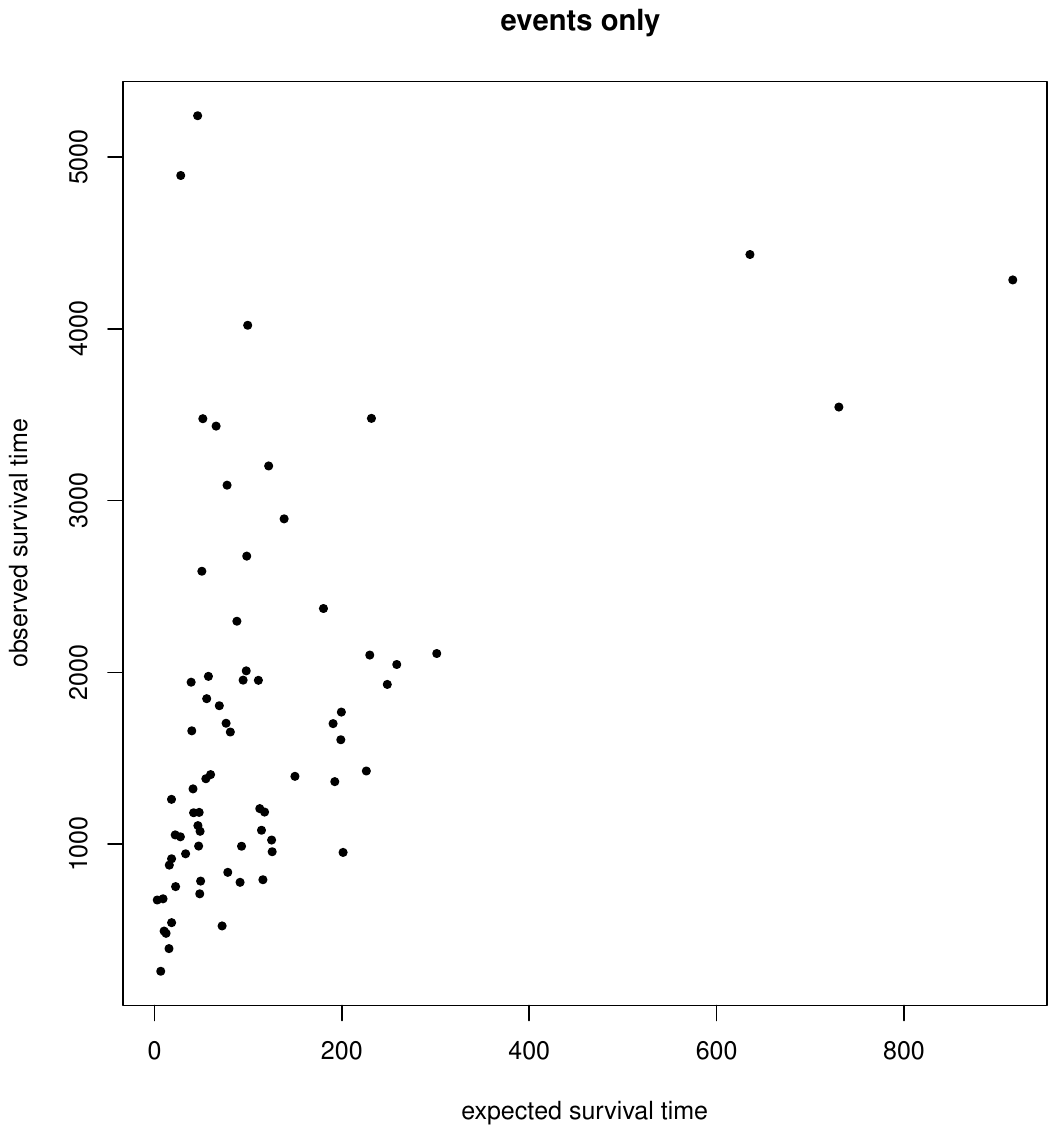}
\end{tabular}
\vspace{-15pt}
\end{center}
\caption{State diagrams of two Markov chains with corresponding transition matrices. The `$\ast$' in the transition matrices indicate unknown values of the transition probabilities.} \label{fig:lassoApplication}
\end{figure}

Here we re-analyse the data set of \cite{vanTVeer2002gene,vanDeVijver2002gene} and predict the breast cancer overall survival by gene expression levels. To this end we adopt the Cox proportional hazards model that describes the instantaneous rate of death at time $t$ conditional on survival until then, or later. Mathematically, the hazard of individual $i$ is then $h_0(t) \exp(\mathbf{X}_{i,\ast} \bbeta)$, where $h_0(t)$ is the baseline hazard function common to all. The covariates affect the hazard in a multiplicative manner, that is independent of time. The regression parameter is estimated by maximization of the (partial) likelihood. Due to the high-dimensionality, the latter is  augmented with a penalty. To mimick the parameter selection used in the contruction of the gene signature, we employ the lasso penalty. The penalty parameter $\lambda_1$ is chosen through leave-one-out cross-validation. The following \texttt{R}-code loads the data, trains the model, and performs some simple diagnostics of the model's fit. 

\begin{lstlisting}
# activate libraries
library(breastCancerNKI)
library(Biobase)
library(penalized)
library(impute)
library(beanplot)
library(survival)
library(survminer)

# load data to memory
data(nki)
X      <- t(exprs(nki))
time   <- pData(nki)[,20]
status <- pData(nki)[,21]

# remove samples without outcome
X      <- X[     -which(is.na(time)),]
status <- status[-which(is.na(time))]
time   <- time[  -which(is.na(time))]

# remove genes with too many missings
X <- X[, -which(colSums(is.na(X)) > 25)]

# remove samples with too many missings
time   <- time[  -which(rowSums(is.na(X)) > 100)]
status <- status[-which(rowSums(is.na(X)) > 100)]
X      <- X[     -which(rowSums(is.na(X)) > 100),]

# impute missing values
X <- t(impute.knn(t(X))$data)

# make Surv object
Y <- Surv(time, status)

# cross-validated plot
profL1list <- profL1(Y ~ X, model="cox")
plot(profL1list[[3]] ~ profL1list[[1]], type="l", lwd=2, 
     xlab=expression(lambda[1]), ylab="cross-validated loss", main="")

# find optimal lambda + estimate model for this lambda
optLasso <- optL1(Y, X, model="cox")

# generate Kaplan-Meier plot
linPredL1   <- X %*% coef(optLasso$fullfit, "all") 
belowMedian <- (linePredL1 > median(linPredL1))
ggsurvplot(survfit(Y ~ belowMedian, 
                   data=as.data.frame(belowMedian)), conf.int=TRUE)

# obtain fit  
coxFit <- coxph(Y ~ linPredL1)

# compare expectation to observation
expTime <- predict(coxFit, type="expected")[status==1]*365
plot(expTime, time[status==1], pch=20, xlab="expected survival time", 
     ylab="observed survival time", main="events only ")

# compare grouping to observation
belowMedian         <- 1*(expTime > median(expTime))
slh                 <- rep("below median(lp)", length(belowMedian))
slh[belowMedian==1] <- "above median(lp)"
beanplot(time[status==1] ~ as.factor(slh), 
         main="events only", ylab="observed survival time")




\end{lstlisting}

The upper left panel of Figure \ref{fig:lassoApplication} shows the cross-validated loglikelihood versus the penalty parameter. The plot reveals that it has multiple local maxima. This is frequently seen for the lasso-type estimators when trained through cross-validation. Hence, care should be exercised when adopting the outcome of a search algorithm for the optimal penalty parameter: it may correspond to a local maximum.

The estimated linear predictor with the optimal cross-validated penalty parameter comprises 20 genes, an even stronger dimension reduction than in the original analysis of \cite{vanTVeer2002gene} (but they have used different statistical machinery to do so). The resulting fit is depicted in various ways by three panels of Figure \ref{fig:lassoApplication}. The upper right panel shows the Kaplan-Meier curves of the groups of individuals falling below and above the linear predictor's median. The bottom left panels compares the observed non-censored survival times of these groups, while the bottom right panel plots these survival times against their expected values. The diagnostic plots reveal some value on the formed linear predictor.

One may now ask oneself whether the 20 genes selected by our lasso estimator are well-established (or novel) contributors to breast cancer. Let us first answer this anecdotally. A few years after the work of \cite{vanTVeer2002gene,vanDeVijver2002gene}, the results of a similar study were published. It too presented a gene signature for overall survival of breast cancer patients based on expression levels. This signature comprises a similar amount of genes, but the two gene signatures, that became known as the `Amsterdam' and `Rotterdam' signatures, showed little overlap. Clinicians did not know which signature to prefer (football fans may see a potential rivalry along familiar lines here), as neither set of signature genes was clearly implicated in breast cancer. The signatures' minimal overlap and their lack of a clear relation to breast cancer was a cause for concern. To investigate this, \cite{eindor2005outcome} conducted a simple \textit{in silico} experiment. They took the original Amsterdam signature of 70 genes. Subsequently, they removed these 70 genes from the set of covariates, and built another predictor comprising 70 covariates/genes. In turn the latter 70 covariates were removed too, and again a novel predictor of 70 covariates was built. And so on, until ten predictors of equal size were obtained. These predictors differed little in terms of their performance. Hence, predictors with non-overlapping gene sets may perform equally well. The line of thought behind this experiment was pushed \textit{in extremis} by \cite{venet2011most}. The authors skipped the training of a model and simply formed a signature from a randomly selected gene set of size $p \approx 100$. The title of their article captures the essence of their conclusion: ``Most random gene expression signatures are significantly associated with breast cancer outcome''. To conclude, \cite{eindor2006thousands} showed by simulation that a training set of thousands of samples is needed to produce a predictor with a stable gene set. The lasso estimator selects, but it does so to explain the response  best, not to facilitate a convenient contextual interpretation. Moreover, as supercollinearity abounds in high-dimensional settings, there typically is an alternative parameter selection with equal performance. Hence, we better restrain ourselves when assigning (too?) much interpretational weight to the selected set of covariates.

\section{Exercises}
\begin{question} \mbox{ } \\
Consider the linear regression model $\mathbf{Y} = \mathbf{X} \bbeta + \vvarepsilon$ with $\vvarepsilon \sim \mathcal{N} ( \mathbf{0}_n, \sigma_{\varepsilon}^2 \mathbf{I}_{nn})$. This model (without intercept) is fitted to data using the ridge regression estimator $\hat{\bbeta}(\lambda_1) = \arg \min_{\bbeta} \| \mathbf{Y} - \mathbf{X} \bbeta \|_2^2 + \lambda_1 \| \bbeta\|_1$ with $\lambda_1 > 0$. The data  are:
\begin{eqnarray*}
\mathbf{X}^{\top} = \left( \begin{array}{rrrr} 
1/2 & -1/2 &  1/2 &  -1/2 \\
\end{array} \right) \, \mbox{ and } \,
\mathbf{Y}^{\top} 
= \left( \begin{array}{rrrr} 
-1.5 &  2.9 & -3.5 & 0.7 
\end{array} \right).
\end{eqnarray*}
\begin{compactitem}
\item[\textit{a)}] Verify that the design matrix is orthonormal.

\item[\textit{b)}] Evaluate the maximum likelihood estimator of the regression parameter, i.e. $\hat{\beta}(\lambda)$ for $\lambda=0$.

\item[\textit{c)}] Evaluate the lasso regression estimator for $\lambda=1$.  

\item[\textit{d)}] Verify that the lasso regression estimator $\hat{\beta}(\lambda_1)$ shrinks as $\lambda_1$ increases. Hereto combine your answer to parts \textit{b)} and \textit{c)} and the evaluation of the ridge regression estimator for $\lambda_1=4$ and $\lambda_1=8$. Is the order of the employed choices of $\lambda_1$ and the absolute value of the corresponding estimates concordant?  
\end{compactitem}
\end{question}

\begin{question} \mbox{ } \\
Find the lasso regression solution for the data below for a general value of $\lambda$ and for the straight line model $Y = \beta_0 + \beta_1 X + \varepsilon$ (only apply the lasso penalty to the slope parameter, not to the intercept). Show that when $\lambda_1$ is chosen as 14, the lasso solution fit is $\hat{Y} = 40 + 1.75 X$. Data: $\mathbf{X}^{\top} = (X_1, X_2, \ldots, X_{8})^{\top} = (-2, -1, -1, -1, 0, 1, 2, 2)^{\top}$, and $\mathbf{Y}^{\top} = (Y_1, Y_2, \ldots, Y_{8})^{\top} = (35, 40, 36, 38, 40, 43, 45, 43)^{\top}$.
\end{question}

\begin{question} \mbox{ } \\
Consider the standard linear regression model $Y_i = \mathbf{X}_{i,\ast} \bbeta + \varepsilon_i$ for $i=1, \ldots, n$ and with $\varepsilon_i \sim_{i.i.d.}  \mathcal{N}(0, \sigma^2)$. The model comprises a single covariate and, depending on the subquestion, an intercept. Data on the response and the covariate are: $\{(y_i, x_{i,1})\}_{i=1}^4 = \{ (1.4,  0.0),   (1.4, -2.0), (0.8,  0.0), (0.4,  2.0) \}$.

\begin{compactitem}
\item[\textit{a)}] Evaluate the lasso regression estimator of the model without intercept for the data at hand with $\lambda_1 = 0.2$.

\item[\textit{b)}] Evaluate the lasso regression estimator of the model with intercept for the data at hand with $\lambda_1 = 0.2$ that does not apply to the intercept (which is left unpenalized).
\end{compactitem}
\end{question}

\begin{question}  \mbox{ }
\\
Plot the regularization path of the lasso regression estimator over the range $\lambda_1 \in (0, 160]$ using the data of Example \ref{example.collinearyFlotin}. 
\end{question}

\begin{question} \mbox{ } \\
Consider the standard linear regression model $Y_i = X_{i,1} \beta_1 + X_{i,2} \beta_2 + \varepsilon_i$ for $i=1, \ldots, n$ and with the $\varepsilon_i$ i.i.d. normally distributed with zero mean and some known common variance. In the estimation of the regression parameter $(\beta_1, \beta_2)^{\top}$ a lasso penalty is used: $\lambda_{1,1} | \beta_1 | + \lambda_{1,2} | \beta_2 |$ with penalty parameters $\lambda_{1,1}, \lambda_{1,2} > 0$.
\begin{compactitem}
\item[\textit{a)}] Let $\lambda_{1,1} = \lambda_{1,2}$ and assume the covariates are orthogonal with the spread of the first covariate being much larger than that of the second. Draw a plot with $\beta_1$ and $\beta_2$ on the $x$- and $y$-axis, repectively. Sketch the parameter constraint as implied by the lasso penalty. Add the levels sets of the sum-of-squares, $\| \mathbf{Y} - \mathbf{X} \bbeta \|_2^2$, loss criterion. Use the plot to explain why the lasso tends to select covariates with larger spread.

\item[\textit{b)}] Assume the covariates to be orthonormal. Let $\lambda_{1,2} \gg \lambda_{1,1}$. Redraw the plot of part a of this exercise. Use the plot to explain the effect of differening $\lambda_{1,1}$ and $\lambda_{1,2}$ on the resulting lasso estimate.

\item[\textit{c)}] Show that the two cases (i.e. the assumptions on the covariates and penalty parameters) of part a and b of this exercise are equivalent, in the sense that their loss functions can be rewritten in terms of the other.
\end{compactitem}
\end{question}

\begin{question} \mbox{ } \\
Investigate the effect of the variance of the covariates on variable selection by the lasso. Hereto consider the toy model: $Y_i = X_{i,1} + X_{i,2} + \varepsilon_i$, where $\varepsilon_i \sim \mathcal{N}(0, 1)$, $X_{i,1} \sim \mathcal{N}(0, 1)$, and $X_{i,2} = a \, X_{i,1}$ with $a \in [0, 2]$. Draw a hundred samples for both $X_{i,1}$ and $\varepsilon_i$ and construct both $X_{i,2}$ and $Y_i$ for a grid of $a$'s. Fit the model by means of the lasso regression estimator with $\lambda_1=1$ for each choice of $a$. Plot e.g. in one figure {\it a)} the variance of $X_{i,1}$, {\it b)} the variance of $X_{i,2}$, and {\it c)} the indicator of the selection of $X_{i,2}$. Which covariate is selected for which values of scale parameter $a$?
\end{question}

\begin{question}\hspace{-0.02cm}$^\ast$  \label{exercise:lassoAnalyticSolutionp=2} \mbox{ } \\
Consider the linear regression model $\mathbf{Y} = \mathbf{X} \bbeta + \vvarepsilon$ with $\vvarepsilon \sim \mathcal{N}(0,\sigma^2)$ and an $n \times 2$-dimensional design matrix with zero-centered and standardized but collinear columns, i.e.:
\begin{eqnarray*}
\mathbf{X}^{\top} \mathbf{X} & = &  \left( \begin{array}{ll} 1 & \rho \\ \rho & 1 \end{array} \right)
\end{eqnarray*}
with $\rho \in (-1, 1)$. Then, an analytic expression for the lasso regression estimator exists. Show that:
\begin{eqnarray*}
\hat{\beta}_j (\lambda_1)  & = & \left\{ 
\begin{array}{lcl}
\mbox{sgn}(\hat{\beta}_j^{\mbox{{\tiny (ml)}}}) [| \hat{\beta}_j^{\mbox{{\tiny (ml)}}} | - \tfrac{1}{2} \lambda_1 (1+\rho)^{-1}]_+ & \mbox{ if } & \mbox{sgn}[\hat{\beta}_1 (\lambda_1)] = \mbox{sgn}[\hat{\beta}_2 (\lambda_1)],
\\
& & \hat{\beta}_j (\lambda_1) \not= 0 \not= \hat{\beta}_2 (\lambda_1),
\\
\mbox{sgn}(\hat{\beta}_j^{\mbox{{\tiny (ml)}}}) [| \hat{\beta}_j^{\mbox{{\tiny (ml)}}} | - \tfrac{1}{2} \lambda_1 (1-\rho)^{-1}]_+ & \mbox{ if } & \mbox{sgn}[\hat{\beta}_1 (\lambda_1)] \not= \mbox{sgn}[\hat{\beta}_2 (\lambda_1)], 
\\
& & \hat{\beta}_1 (\lambda_1) \not= 0 \not= \hat{\beta}_2 (\lambda_1),
\\
\left\{
\begin{array}{lcl}
0 & \mbox{ if } & j \not= \arg \max_{j'} \{ | \hat{\beta}_{j'}^{\mbox{{\tiny (ml)}}} | \} 
\\
\mbox{sgn}(\tilde{\beta}_j^{\mbox{{\tiny (ml)}}}) ( | \tilde{\beta}_j^{\mbox{{\tiny (ml)}}} | - \tfrac{1}{2} \lambda_1)_+ & \mbox{ if } & j = \arg \max_{j'} \{ | \hat{\beta}_{j'}^{\mbox{{\tiny (ml)}}} | \} 
\end{array}
\right.
&
\multicolumn{2}{l}{\mbox{ otherwise }}.
\end{array} 
\right.
\end{eqnarray*}
where $\tilde{\beta}_j^{\mbox{{\tiny (ml)}}} = (\mathbf{X}_{\ast,j}^{\top} \mathbf{X}_{\ast,j})^{-1} \mathbf{X}_{\ast,j}^{\top} \mathbf{Y}$.
\end{question}

\begin{question}  \mbox{ } \\
Consider the linear regression model $\mathbf{Y} = \mathbf{X} \bbeta + \vvarepsilon$ with $\vvarepsilon \sim \mathcal{N} ( \mathbf{0}_n, \sigma_{\varepsilon}^2 \mathbf{I}_{nn})$. This model (without intercept) is fitted to data using the lasso regression estimator $\hat{\bbeta}(\lambda_1) = \arg \min_{\bbeta} \| \mathbf{Y} - \mathbf{X} \bbeta \|_2^2 + \lambda_1 \| \bbeta \|_1$. The relevant summary statistics of the data are:
\begin{eqnarray*}
\mathbf{X}^{\top} \mathbf{X} = \left( \begin{array}{rr} 1 & -\tfrac{1}{5} \\ -\tfrac{1}{5} & 1 \end{array} \right), \mbox{ and } \, \mathbf{X}^{\top} \mathbf{Y} = \left( \begin{array}{r} -7 \\ 5 \end{array} \right).
\end{eqnarray*}

\begin{compactitem}
\item[\textit{a)}] Evaluate for $\lambda_1 = 6$ the lasso regression estimator.

\item[\textit{b)}] For which $\lambda_1$ does the lasso regression estimator have a single non-zero element?
\end{compactitem}
\end{question}

\begin{question} \mbox{ } \\
Consider the linear regression model $\mathbf{Y} = \mathbf{X} \bbeta + \vvarepsilon$ with $\bbeta \in \mathbb{R}^p$ and $\vvarepsilon \sim \mathcal{N} ( \mathbf{0}_n, \sigma_{\varepsilon}^2 \mathbf{I}_{nn})$. This model is fitted to data using the lasso regression estimator $\hat{\bbeta}(\lambda_1) = \arg \min_{\bbeta} \| \mathbf{Y} - \mathbf{X} \bbeta \|_2^2 + \lambda_1 \| \bbeta \|_1$. 

\begin{compactitem}
\item[\textit{a)}] Suppose $n=2$ and $p=3$. Could it be that $\hat{\bbeta}(\lambda_1) = (-1.3, 2.7, 0.9)^{\top}$ for some $\lambda_1 > 0$? Motivate.

\item[\textit{b)}] Suppose $n=3$ and $p=2$. Could it be that $\hat{\bbeta}(\lambda_1) = (-1.3, 0.9)^{\top}$ for $\lambda_1 =1$ and 
$\hat{\bbeta}(\lambda_1) = (-1.5, 0.8)^{\top}$ for $\lambda_1 =4$? Motivate.
 
\item[\textit{c)}] Suppose $n=3$ and $p=2$. Could it be that $\hat{\bbeta}(\lambda_1) = (-4, 2)^{\top}$ for $\lambda_1 =1$, $\hat{\bbeta}(\lambda_1) = (-2,  1.5)^{\top}$ for $\lambda_1 =2$, and $\hat{\bbeta}(\lambda_1) = (-1,  1)^{\top}$ for $\lambda_1 =3$? Motivate.
\end{compactitem}
\end{question}

\begin{question} \mbox{ } 
\\
Consider the standard linear regression model $Y_i = \mathbf{X}_{i,\ast} \bbeta + \varepsilon_i$ for $i=1, \ldots, n$ and with the $\varepsilon_i$ i.i.d. normally distributed with zero mean and a common variance. Moreover, $\mathbf{X}_{\ast,j} = \mathbf{X}_{\ast,j'}$ for all $j, j'=1, \ldots, p$ and $\sum_{i=1}^n X_{i,j}^2 = 1$. Question \ref{question.RidgeEstimatorWithIdenticalCovariates} revealed that in this case all elements of the ridge regression estimator are equal, irrespective of the choice of the penalty parameter $\lambda_2$. Does this hold for the lasso regression estimator? Motivate your answer.
\end{question}

\begin{question} \mbox{ } \\
Consider the linear regression model $Y_i = \mathbf{X}_{i,\ast} \bbeta + \varepsilon_i$ for $i=1, \ldots, n$ and with the $\varepsilon_i$ i.i.d. normally distributed with zero mean and a common variance. Relevant information on the response and design matrix are summarized as:
\begin{eqnarray*}
\mathbf{X}^{\top} \mathbf{X} = \left( \begin{array}{rr} 3 & -2 \\ -2 & 2 \end{array} \right), \qquad \mathbf{X}^{\top} \mathbf{Y} = \left( \begin{array}{r} 3 \\ -1 \end{array} \right).
\end{eqnarray*}
The lasso regression estimator is used to learn parameter $\bbeta$.
\begin{compactitem}
\item[\textit{a)}] Show that the lasso regression estimator is given by:
\begin{eqnarray*}
\hat{\bbeta}(\lambda_1) & = & \arg \min_{\bbeta \in \mathbb{R}^2} 3 \beta_1^2 + 2 \beta_2^2 - 4 \beta_1 \beta_2 - 6 \beta_1 + 2 \beta_2 + \lambda_1 | \beta_1 | + \lambda_1 | \beta_2|.
\end{eqnarray*}

\item[\textit{b)}] For $\lambda_{1} = 0.2$ the lasso estimate of the second element of $\bbeta$ is $\hat{\beta}_2(\lambda_1) = 1.25$. Determine the corresponding value of $\hat{\beta}_1(\lambda_1)$.

\item[\textit{c)}] Determine the smallest $\lambda_1$ for which it is guaranteed that $\hat{\bbeta}(\lambda_1) = \mathbf{0}_2$.
\end{compactitem}
\end{question}

\begin{question} \label{exercise:lassoMonotoneBetaNorm} \mbox{ } \\
Show $\| \hat{\bbeta}(\lambda_1)\|_1$ is monotone decreasing in $\lambda_1$. In this assume orthonormality of the design matrix $\mathbf{X}$.
\end{question}

\begin{question} \mbox{ } \\
Consider the linear regression model $\mathbf{Y} = \mathbf{X} \bbeta + \vvarepsilon$ with $\vvarepsilon \sim \mathcal{N} ( \mathbf{0}_n, \sigma^2 \mathbf{I}_{nn})$. This model is fitted to data, $\mathbf{X}_{1,\ast} = (4, -2)$ and $Y_1 = 10$, using the lasso regression estimator $\hat{\bbeta}(\lambda_1) = \arg \min_{\bbeta} \| Y_1 - \mathbf{X}_{1,\ast} \bbeta \|_2^2 + \lambda_1 \| \bbeta \|_1$.
\begin{compactitem}
\item[\textit{a)}] How many nonzero elements does the lasso regression estimator with an arbitrary $\lambda_1 > 0$ have for these data?

\item[\textit{b)}] Ignore the second covariate and evaluate the lasso regression estimator for $\lambda_1=8$.

\item[\textit{c)}] Suppose that, when regressing the response on each covariate separately, the corresponding lasso regression estimates with $\lambda_1=8$ are $\hat{\beta}_1 (\lambda_1) = 2 \tfrac{1}{4}$ and $\hat{\beta}_2 (\lambda_1) = -4$. Now consider the regression problem with both covariates in the model. Does the lasso regression estimate with $\lambda_1=8$ then equal $\hat{\bbeta} (\lambda_1) = (2 \tfrac{1}{4}, 0)^{\top}$, $\hat{\bbeta} (\lambda_1) = (0, -4)^{\top}$, or some other value? Motivate!
\end{compactitem}
\end{question}

\begin{question} \label{question:lassoMAP} \mbox{ } \\
Consider a single draw, denoted $\mathbf{Y}$, from the normal means model $\mathbf{Y} \sim \mathcal{N} ( \boldsymbol{\mu}, \sigma^2 \mathbf{I}_{pp})$ with $\boldsymbol{\mu} \in \mathbb{R}^p$ and $\sigma^2 =1$. Assume the $\mu_i$ to be i.i.d. distributed following the double exponential distribution with density $f(x) = (2b)^{-1} \exp(- b^{-1} | x| )$. Show that, if $b = \lambda_1^{-1}$, the MAP (Maximum A Posteriori) estimator  $\hat{\mu}_i^{\mbox{{\tiny map}}}$ equals $\mbox{sign}(Y_j) (|Y_j| - \lambda_1)_+$ for $j=1, \ldots, p$.
\end{question}




\begin{question} \mbox{ } \\
Consider the standard linear regression model $Y_i = \mathbf{X}_{i,\ast} \bbeta + \varepsilon_i$ for $i=1, \ldots, n$ and with the $\varepsilon_i$ i.i.d. normally distributed with zero mean and a common variance. Let the first covariate correspond to the intercept. The model is fitted to data by means of the minimization of the sum-of-squares augmented with a lasso penalty in which the intercept is left unpenalized: $\lambda_1 \sum_{j=2}^p | \beta_j |$ with penalty parameter $\lambda_1 > 0$. The penalty parameter is chosen through leave-one-out cross-validation (LOOCV). The predictive performance of the model is evaluated, again by means of LOOCV. Thus, creating a double cross-validation loop. At each inner loop the optimal $\lambda_1$ yields an empty intercept-only model, from which a prediction for the left-out sample is obtained. The vector of these prediction is compared to the corresponding observation vector through their Spearman correlation (which measures the monotonicity of a relatonship and -- as a correlation measure -- assumed values on the $[-1,1]$ interval with an analogous interpretation to the `ordinary' correlation). The latter equals $-1$. Why?
\end{question}

\begin{question} \mbox{ } \\
Load the breast cancer data available via the {\tt breastCancerNKI}-package (downloadable from BioConductor) through the following \texttt{R}-code:
\begin{lstlisting}
# activate library and load the data
library(breastCancerNKI)
data(nki)

# subset and center the data
X <- exprs(nki)
X <- X[-which(rowSums(is.na(X)) > 0),]
X <- apply(X[1:1000,], 1, function(X){ X - mean(X) })

# define ER as the response
Y <- pData(nki)[,8]

\end{lstlisting}
The \texttt{eSet}-object {\tt nki} is now available. It contains the expression profiles of 337 breast cancer patients. Each profile comprises expression levels of 24481 genes. The \texttt{R}-code above extracts the expression data from the object, removes all genes with missing values, centers the gene expression gene-wise around zero, and subsets the data set to the first thousand genes. The reduction of the gene dimensionality is only for computational speed. Furthermore, it extracts the estrogen receptor status (short: ER status), an important prognostic indicator for breast cancer, that is to be used as the response variable in the remainder of the exercise.

\begin{compactitem}
\item[\textit{a)}] Relate the ER status and the gene expression levels by a logistic regression model, which is fitted by means of the lasso penalized maximum likelihood method. First, find the optimal value of the penalty parameter of $\lambda_1$ by means of cross-validation. This is implemented in {\tt optL1}-function of the {\tt penalized}-package available from {\tt CRAN}.

\item[\textit{b)}]  Evaluate whether the cross-validated likelihood indeed attains a maximum at the optimal value of $\lambda_1$. This can be done with the {\tt profL1}-function of the {\tt penalized}-package available from {\tt CRAN}.

\item[\textit{c)}]  Investigate the sensitivity of the penalty parameter selection with respect to the choice of the cross-validation fold.

\item[\textit{d)}]  Does the optimal lambda produce a reasonable fit? And how does it compare to the `ridge fit'?
\end{compactitem}
\end{question}

\pagestyle{fancy}

\chapter[Generalizing lasso regression]{Generalizing lasso regression}
Many variants of penalized regression, in particular of lasso regression, have been presented in the literature. Here we give an overview of some of the more current ones. Not a full account is given, but rather a brief introduction with emphasis on their motivation and use.

\section{Elastic net}
The elastic net regression estimator is a modication of the lasso regression estimator that preserves its strength and harnesses its weaknesses. The biggest appeal of the lasso regression estimator is clearly its ability to perform selection. Less pleasing are \textit{i)} the non-uniquess of the lasso regression estimator due to the non-strict convexity of its loss function, \textit{ii)} the bound on the number of selected variables, i.e. maximally $\min \{ n, p \}$ can be selected, and \textit{iii)} the observation that strongly (positively) collinear covariates are not shrunken together: the lasso regression estimator selects among them while it is hard to distinguish their contributions to the variation of the response. While it does not select, the ridge regression estimator does not exhibit these less pleasing features. These considerations led \cite{Zou2005} to combine the strengths of the lasso and ridge regression estimators and form a `best-of-both-worlds' estimator, called the \textit{elastic net} regression estimator, defined as:
\begin{eqnarray*}
\hat{\bbeta}(\lambda_1, \lambda_2) & = & \arg \min_{\bbeta \in \mathbb{R}^p} \| \mathbf{Y} - \mathbf{X} \bbeta \|_2^2  + \lambda_1 \| \bbeta \|_1 + \tfrac{1}{2} \lambda_2 \| \bbeta \|_2^2.
\end{eqnarray*}
The elastic net penalty -- defined implicitly in the preceeding display -- is thus simply a linear combination of the lasso and ridge penalties. Consequently, the elastic net regression estimator encompasses its lasso and ridge counterparts. Hereto just set $\lambda_2=0$ or $\lambda_1=0$, respectively. A novel estimator is defined if both penalties act simultaneously, i.e. if their corresponding penalty parameters are both nonzero.

\begin{figure}[!h]
\begin{tabular}{rcl}
\includegraphics[scale=0.40, angle=0]{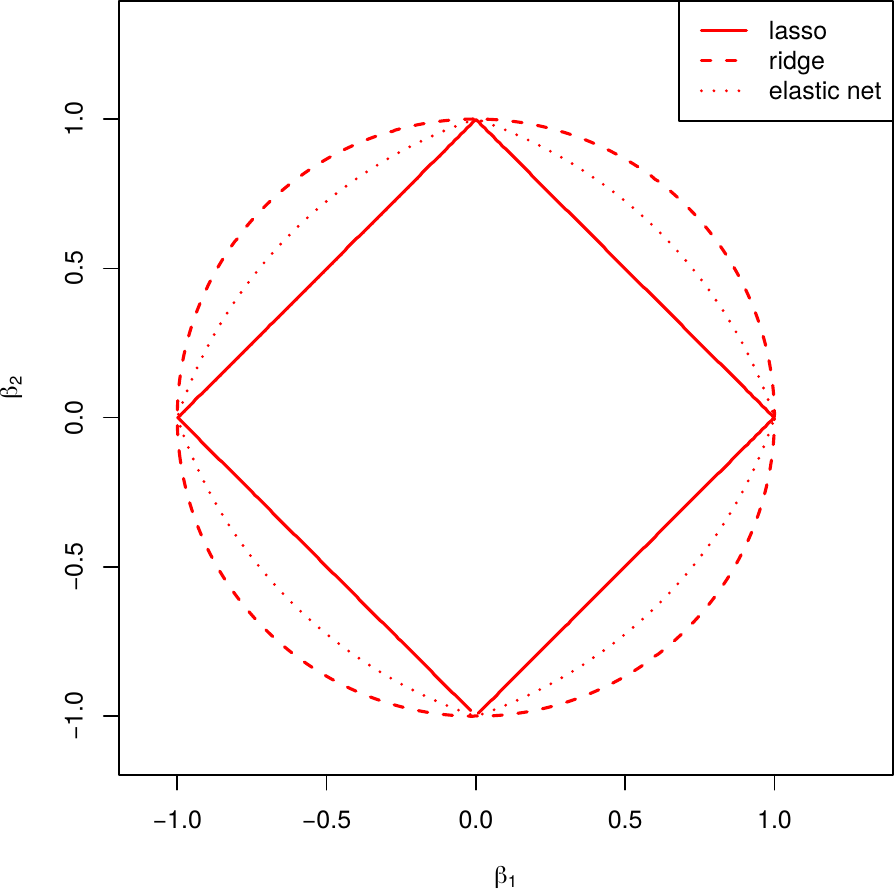}
& &
\includegraphics[scale=0.45, angle=0]{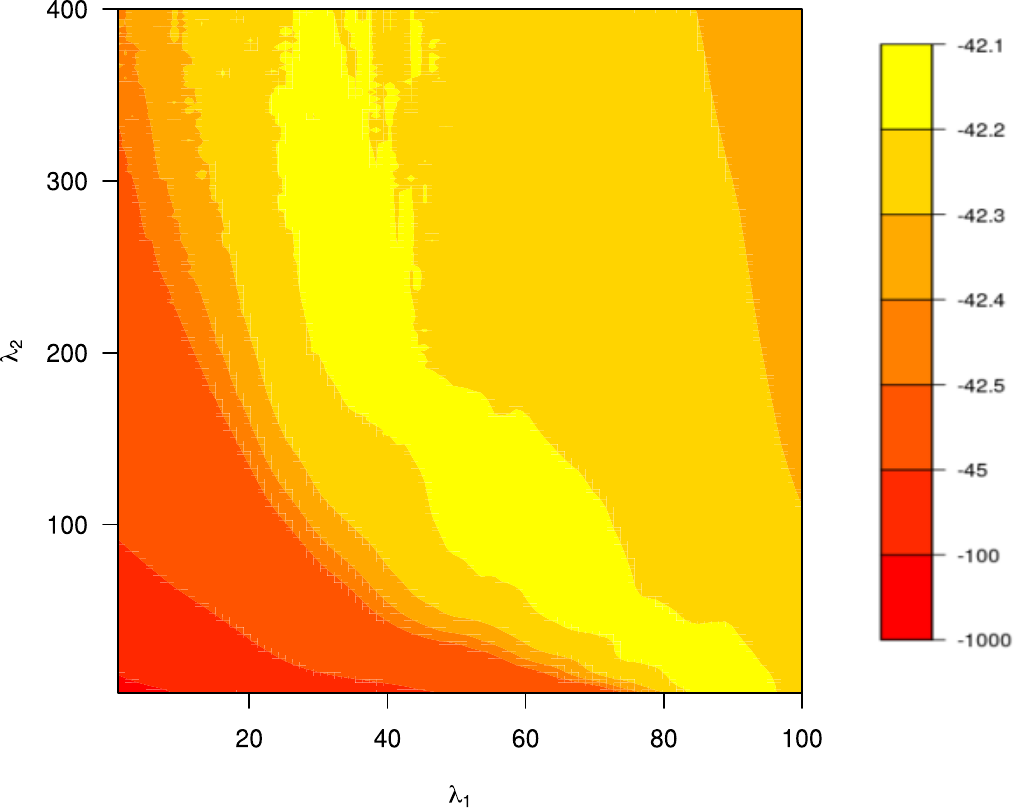}
\end{tabular}
\caption{The left panel depicts the parameter constraint induced by the elastic net penalty and, for reference, those of the lasso and ridge are added. The right panel shows the contour plot of the cross-validated loglikelihood vs. the two penalty parameters of the elastic net estimator.  \label{fig.elasticNetConstraintAndCVcontour}}
\end{figure}

Does this novel elastic net estimator indeed inherit the strengths of the lasso and ridge regression estimators? Let us turn to the aforementioned motivation behind the elastic net estimator. Starting with the uniqueness, the strict convexity of the ridge penalty renders the elastic net loss function strictly convex, as it is a combination of the ridge penalty and the lasso loss function -- notably non-strict convex when the dimension $p$ exceeds the sample size $n$. This warrants the existence of a unique minimizer of the elastic net loss function. To assess the preservation of the selection property, now without the bound on the maximum number of selectable variables, exploit the equivalent constraint estimation formulation of the elastic net estimator. Figure \ref{fig.elasticNetConstraintAndCVcontour} shows the parameter constraint of the elastic net estimator for the `$p=2$'-case, which is defined by the set:
\begin{eqnarray*}
\{(\beta_1, \beta_2) \in \mathbb{R}^2 \, : \, \lambda_1 (| \beta_1 | + | \beta_2 |) + \tfrac{1}{2} \lambda_2 (\beta_1^2 + \beta_2^2) \leq c(\lambda_1, \lambda_2) \}.
\end{eqnarray*}
Visually, the `elastic net parameter constraint' is a compromise between the circle and the diamond shaped constraints of the ridge and lasso regression estimators. This compromise inherits exactly the right geometrical features:  the strict convexity of the `ridge circle' and the `corners' (referring to points at  which the constraint's boundary is non-smootness/non-differentiability) falling at the axes of the `lasso constraint'. The latter feature, by the same argumentation as presented in Section \ref{sect:sparsity}, endows the elastic net estimator with the selection property. Moreover, it can -- in principle -- select $p$ features as the point in the parameter space where the smallest level set of the unpenalized loss hits the elastic net parameter constraint need not fall on any axis. For example, in the `$p=2, n=1$'-case the level sets of the sum-of-squares loss are straight lines that, if running almost parallel to the edges of the `lasso diamond', are unlikely to first hit the elastic net parameter constraint at one of its corners. Finally, the largest penalty parameter relates (reciprocally) to the volume of the elastic net parameter constraint, while the ratio between $\lambda_1$ and $\lambda_2$ determines whether it is closer to the `ridge circle' or to the `lasso diamond'.

Whether the elastic net regression estimator also delivers on the joint shrinkage property is assessed by simulation (not shown). The impression given by these simulations is that the elastic net has joint shrinkage potential. This, however, usually requires a large ridge penalty, which then dominates the elastic net penalty.

The elastic net regression estimator can be found with procedures similar to those that evaluate the lasso regression estimator (see Section \ref{sect:lassoEstimation}) as the elastic net loss can be reformulated as a lasso loss. Hereto the ridge part of the elastic net penalty is absorbed into the sum of squares using the data augmentation trick of Exercise \ref{question:ridgeAugmentation}, which showed that the ridge regression estimator is the ML regression estimator of the related regression model with $p$ zeros and rows added to the response and design matrix, respectively. That is, write $\tilde{\mathbf{Y}} = (\mathbf{Y}^{\top}, \mathbf{0}_{p}^{\top})^{\top}$ and $\tilde{\mathbf{X}} = (\mathbf{X}^{\top}, \sqrt{\lambda_2} \, \mathbf{I}_{pp})^{\top}$. Then:
\begin{eqnarray*}
\| \tilde{\mathbf{Y}} - \tilde{\mathbf{X}} \bbeta \|_2^2 & = & \| \mathbf{Y} - \mathbf{X} \beta \|_2^2 + \lambda_2 \| \bbeta \|_2^2.
\end{eqnarray*}
Hence, the elastic net loss function can be rewritten to $\| \tilde{\mathbf{Y}} - \tilde{\mathbf{X}} \bbeta \|_2^2 + \lambda_1 \| \bbeta \|_1$. This is familiar territory and the lasso algorithms of Section \ref{sect:lassoEstimation} can be used. \cite{Zou2005} present a different algorithm for the evaluation of the elastic net estimator that is faster and numerically more stable (see also Exercise \ref{question:elasticNet}). Irrespectively, the reformulation of the elastic net loss in terms of augmented data also reveals that the elastic net regression estimator can select $p$ variables. Even for $p > n$, which is immediate from the observation that $\mbox{rank}(\tilde{\mathbf{X}})=p$.
\\
\\
The penalty parameters need tuning, e.g. by cross-validation. They are subject to some empirically indeterminancy. That is, often a range of $(\lambda_1, \lambda_2)$-combinations will yield a similar cross-validated performance as can be witnessed from Figure \ref{fig.elasticNetConstraintAndCVcontour}. It shows the contourplot of the penalty parameters vs. this performance. There is a yellow `banana-shaped' area that corresponds to the same and optimal performance. Hence, no single $(\lambda_1, \lambda_2)$-combination can be distinguished as all yield the best performance. This behaviour may be understood intuitively. Reasoning loosely, while the lasso penalty $\lambda_1 \| \bbeta \|_1$ ensures the selection property and the ridge penalty $\tfrac{1}{2} \lambda_2 \| \bbeta \|_2^2$ warrants the uniqueness and joint shrinkage of coefficients of collinear covariates,  they have a similar effect on the size of the estimator. Both shrink, although in different norms. But a reduction in the size of $\hat{\bbeta}(\lambda_1, \lambda_2)$ in one norm implies a reduction in another. An increase in either the lasso and ridge penalty parameter will have a similar effect on the elastic net estimator: it shrinks. The selection and `joint shrinkage' properties are only consequences of the employed penalty and are not criteria in the optimization of the elastic net loss function. There, only size matters. The size of $\bbeta$ refers to $\lambda_1 \| \bbeta \|_1 + \tfrac{1}{2} \lambda_2 \| \bbeta \|_2^2$. As in the size the lasso and ridge penalties appear as a linear combination in the elastic net loss function and have a similar effect on the elastic net estimator: there are many positive $(\lambda_1, \lambda_2)$-combinations that constrain the size of the elastic net estimator equally. In contrast, for both the lasso and ridge regression estimators, different penalty parameters yield estimators of different sizes (defined accordingly). Moreover, it is mainly the size that determines the cross-validated performance as the size determines the shrinkage of the estimator and, consequently, the size of the errors. But only a fixed size leaves enough freedom to distribute this size over the $p$ elements of the regression parameter estimator $\hat{\bbeta}(\lambda_1, \lambda_2)$ and, due to the collinearity, among them many that yield a comparable performance. Hence, if a particularly sized elastic net estimator $\hat{\bbeta}(\lambda_1, \lambda_2)$ optimizes the cross-validated performance, then high-dimensionally there are likely many others with a different $(\lambda_1, \lambda_2)$-combination but of equal size and similar performance.

The empirical indeterminancy of penalty parameters touches upon another issues. In principle, the elastic net regression estimator can decide whether a sparse or non-sparse solution is most appropriate. The indeterminancy indicates that for any sparse elastic net regression estimator a less sparse one can be found with comparable performance, and vice versa. Care should be exercised when concluding on the sparsity of the linear relation under study from the chosen elastic net regression estimator.

A solution to the indeterminancy of the optimal penalty parameter combination is to fix their ratio. For interpretation purposes this is done through the introduction of a `mixing parameter' $\alpha \in [0,1]$. The elastic net penalty is then written as $\lambda [ \alpha \| \bbeta \|_1 + \tfrac{1}{2} (1-\alpha) \| \bbeta \|_2^2 ]$. The mixing parameter is set by the user while $\lambda > 0$ is typically found through cross-validation (cf. the implementation in the \texttt{glmnet}-package) \citep{Frie2009}. Generally, no guidance on the choice of mixing parameter $\alpha$ can be given. In fact, it is a tuning parameter and as such needs tuning rather then setting out of the blue.

\section{Fused lasso}
\begin{figure}[!h]
\begin{tabular}{rcl}
\includegraphics[scale=0.20, angle=0]{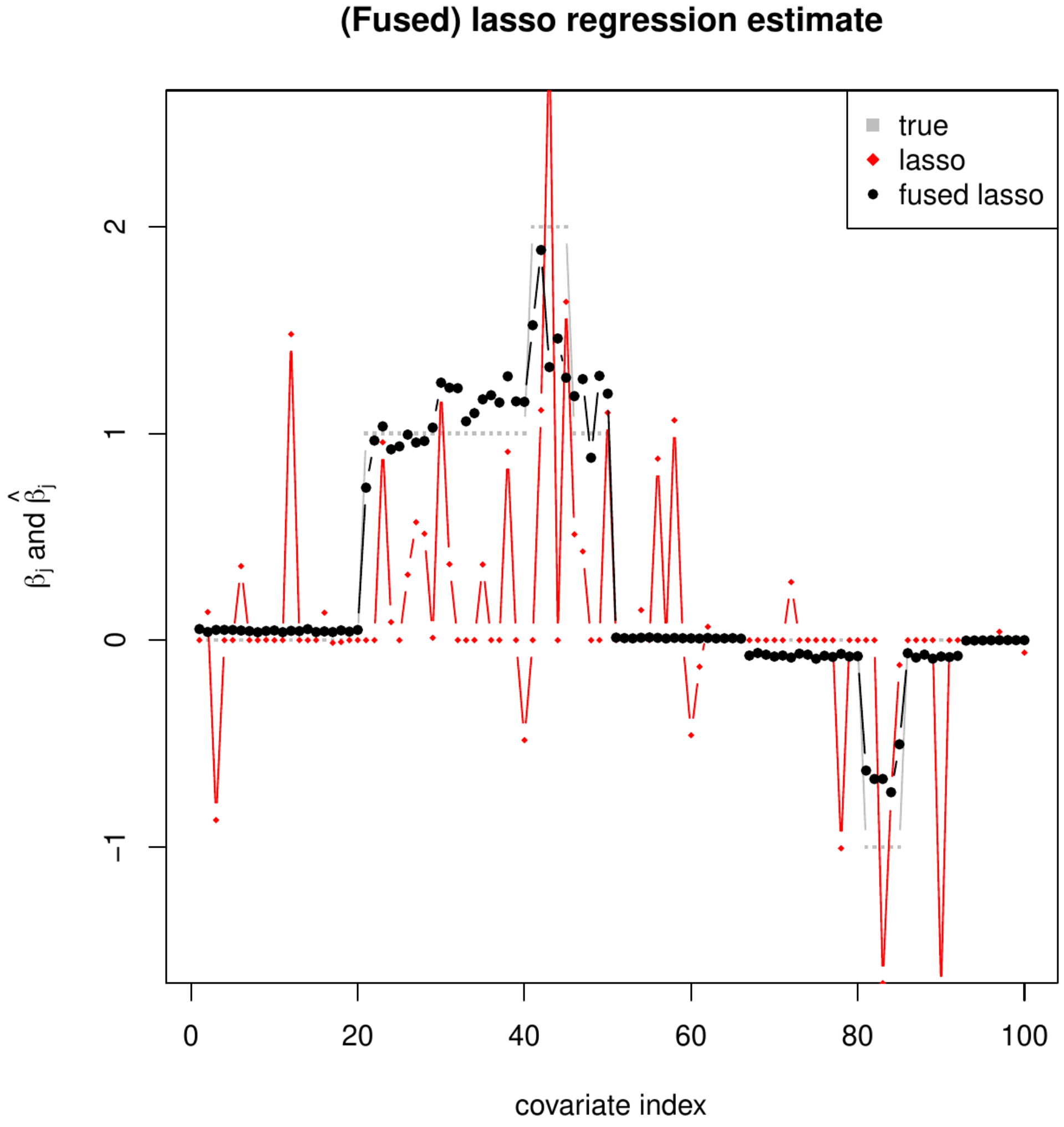}
& &
\includegraphics[scale=0.20, angle=0]{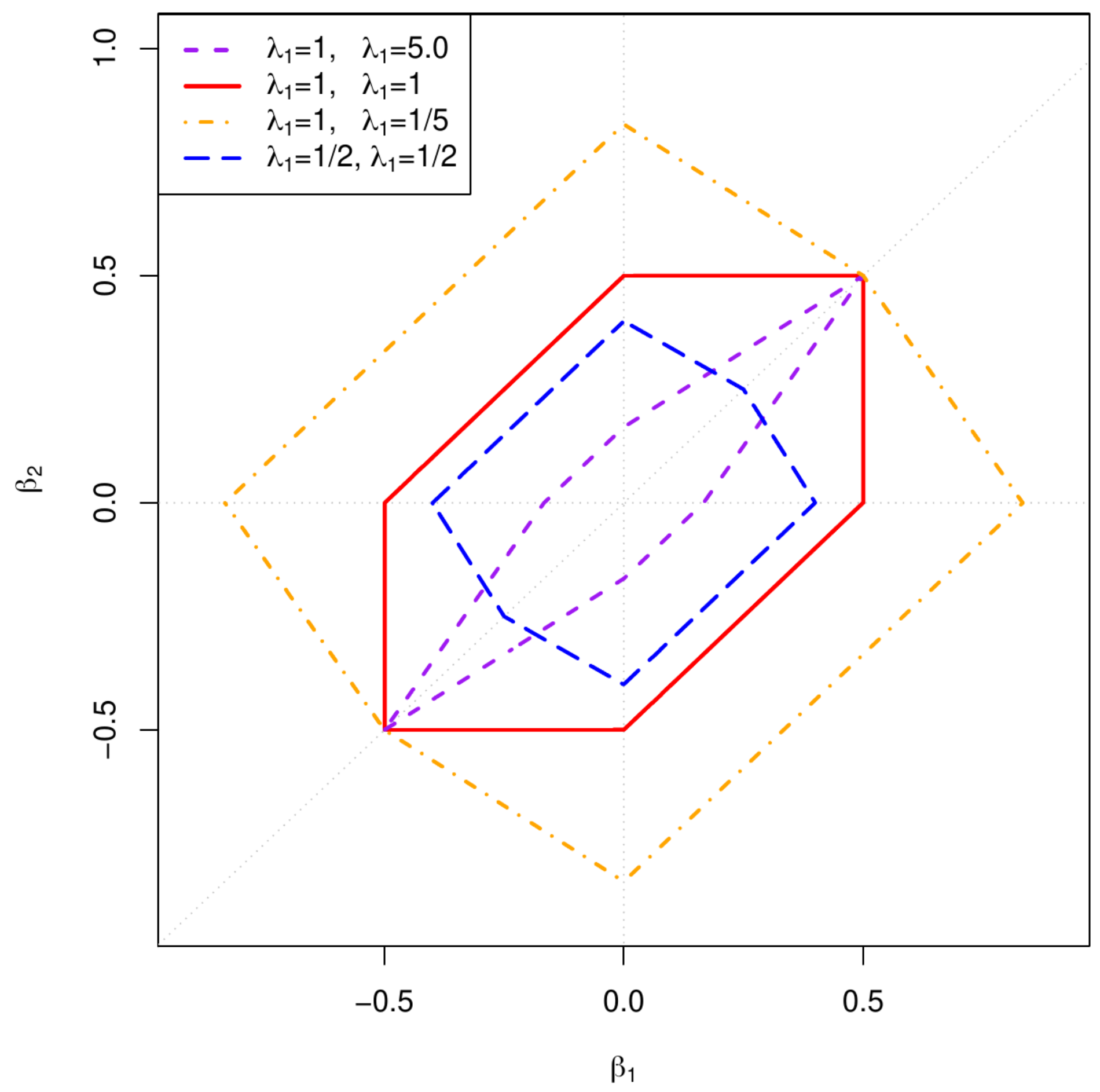}
\end{tabular}
\caption{The left panel shows the lasso (red, diamonds) and fused lasso (black circles) regression parameter estimates and the true parameter values (grey circles) plotted against their index $j$. The $(\beta_1, \beta_2)$-parameter constraint induced by the fused lasso penalty for various combinations of the lasso and fusion penalty parameters $\lambda_1$ and $\lambda_{1,f}$, respectively. The grey dotted lines corresponds to the `$\beta_1 = 0$'-, `$\beta_2 = 0$'- and `$\beta_1 = \beta_2$'-lines where selection takes place. \label{fig.fusedLassoConstraintAndEstimate}}
\end{figure}

The fused lasso regression estimator proposed by \cite{Tibs2005} is the counterpart of the fused ridge regression estimator encountered in Example \ref{sect:fusedRidgeEstimation}. It is a generalization of the lasso regression estimator for situations where the order of index $j$, $j=1, \ldots, p$, of the covariates has a certain meaning such as a spatial or temporal one. The fused lasso regression estimator minimizes the sum-of-squares augmented with the lasso penalty, the sum of the absolute values of the elements of the regression parameter, and the $\ell_1$-\textit{fusion} (or simply \textit{fusion} if clear from the context) penalty, the sum of the first order differences of the regression parameter. Formally, the fused lasso estimator is defined as:
\begin{eqnarray*}
\hat{\bbeta} (\lambda_1, \lambda_{1,f}) & =& \arg \min_{\bbeta \in \mathbb{R}^p} \| \mathbf{Y} - \mathbf{X} \bbeta \|_2^2 + \lambda_1 \| \bbeta \|_{1}  + \lambda_{1,f} \sum\nolimits_{j=2}^{p} | \beta_{j} - \beta_{j-1} |,
\end{eqnarray*}
which involves two penalty parameters $\lambda_{1}$ and $\lambda_{1,f}$ for the lasso and fusion penalties, respectively. As a result of adding the fusion penalty the fused lasso regression estimator not only shrinks elements of $\bbeta$ towards zero but also the difference of neighboring elements of $\bbeta$. In particular, for large enough values of the penalty parameters the estimator selects elements and differences of neighboring elements of $\bbeta$. This corresponds to a sparse estimate of $\bbeta$ while the vector of its first order differences too is dominated by zeros. That is many elements of $\hat{\bbeta} (\lambda_1, \lambda_{1,f})$ equal zero with few changes in $\hat{\bbeta} (\lambda_1, \lambda_{1,f})$ when running over $j$. Hence, the fused lasso regression penalty encourages large sets of neighboring elements of $j$ to have a common (or at least a comparable) regression parameter estimate. This is visualized -- using simulated data from a simple toy model with details considered irrelevant for the illustration -- in Figure \ref{fig.fusedLassoConstraintAndEstimate} where the elements of the fused lasso regression estimate $\hat{\bbeta} (\lambda_1, \lambda_{1,f})$ are plotted against the index of the covariates. For reference the true $\bbeta$ and the lasso regression estimate with the same $\lambda_1$ are added to the plot. Ideally, for a large enough fusion penalty parameter, the elements of $\hat{\bbeta} (\lambda_1, \lambda_{1,f})$ would form a step-wise function in the index $j$, with many steps equalling zero and exhibiting few changes, as the elements of $\bbeta$ do. While this is not the case, it is close, especially in comparison to the elements of its lasso cousin $\hat{\bbeta}(\lambda_1)$, thus showing the effect of the inclusion of the fusion penalty.

It is insightful to view the fused lasso regression problem as a constrained estimation problem. The fused lasso penalty induces a parameter constraint: $\{ \bbeta \in \mathbb{R}^p \, : \, \lambda_1 \| \bbeta \|_1 + \lambda_{1,f} \sum_{j=2}^p | \beta_{j} - \beta_{j-1} | \leq c(\lambda_1, \lambda_{1,f}) \}$. This constraint is plotted for $p=2$ in the right panel of Figure \ref{fig.fusedLassoConstraintAndEstimate} (clearly, it is not the intersection of the constraints induced by the lasso and fusion penalty separately as one might accidently conclude from Figure 2 in \citealp{Tibs2005}). The constraint is convex, although not strict, which is convenient for optimization purposes. Morever, the geometry of this fused lasso constraint reveals why the fused lasso regression estimator selects the elements of $\bbeta$ as well as its first order difference. Its boundary, while continuous and generally smooth, has six points at which it is non-differentiable. These all fall on the grey dotted lines in the right panel of Figure \ref{fig.fusedLassoConstraintAndEstimate} that correspond to the axes and the diagonal, put differently, on either the `$\beta_1 = 0$', `$\beta_2 = 0$' , or `$\beta_1 = \beta_2$'-lines. The fused lasso regression estimate is the point where the smallest level set of the sum-of-squares criterion $\| \mathbf{Y} - \mathbf{X} \bbeta \|_2^2$, be it an ellipsoid or hyper-plane, hits the fused lasso constraint. For an element or the first order difference to be zero it must fall on one of the dotted greys lines of the right panel of Figure \ref{fig.fusedLassoConstraintAndEstimate}. Exactly this happens on one of the aforementioned six point of the constraint. Finally, the fused lasso regression estimator has, when -- for reasonably comparable penalty parameters $\lambda_1$ and $\lambda_{1,f}$ -- it shrinks the first order difference to zero, a tendency to also estimate the corresponding individual elements as zero. In part, this is due to the fact that $| \beta_1 | = 0 = | \beta_2|$ implies that $| \beta_1 - \beta_2 | = 0$, while the reverse does not necessary hold. Moreover, if $|\beta_1| = 0$, then $|\beta_1 - \beta_2|= | \beta_2|$. The fusion penalty thus converts to a lasso penalty of the remaining nonzero element of this first order difference, i.e. $(\lambda_1 + \lambda_{1,f}) | \beta_2|$, thus furthering the shrinkage of this element to zero.




The evaluation of the fused lasso regression estimator is more complicated than that of the `ordinary' lasso regression estimator. For `moderately sized' problems, \cite{Tibs2005} suggest to use a variant of the quadratic programming method (see also Section \ref{sect:lassoQuadProg}) that is computationally efficient when many linear constraints are active, i.e. if many elements and first order difference of $\bbeta$ are zero. \cite{Chat2014} extend the gradient ascent approach discussed in Section \ref{sect:gradientAscent} to solve the minimization of the fused lasso loss function. For the limiting `$\lambda_1=0$'-case the fused lasso loss function can be reformulated as a lasso loss function (see Exercise \ref{question:fusedLasso}). Then, the algorithms of Section \ref{sect:lassoEstimation} may be applied to find the estimate $\hat{\bbeta} (0, \lambda_{1,f})$.

\section{Group lasso}
The lasso regression estimator selects covariates, irrespectively of the relation among them. However, groups of covariates may be discerned. For instance, a group of covariates may be dummy variables representing levels of a categorical factor. Or, within the context of gene expression studies such groups may be formed by so-called pathways, i.e. sets of genes that work in concert to fulfill a certain function in the cell. In such cases a group-structure can be overlayed on the covariates  and it may be desirable to select the whole group, i.e. all covariates together, rather than an individual covariate of the group. To achieve this \cite{Yuan2006} proposed the group lasso regression estimator. It minimizes the sum-of-squares now augmented with the \textit{group lasso} penalty, i.e.:
\begin{eqnarray*}
\lambda_{1,G} \sum\nolimits_{g = 1}^G \sqrt{|\mathcal{J}_g|} \,  \| \bbeta_g \|_2 & = & \lambda_{1,G} \sum\nolimits_{g = 1}^G \sqrt{|\mathcal{J}_g|} \sqrt{\sum\nolimits_{j \in \mathcal{J}_g} \beta_j^2},
\end{eqnarray*}
where $\lambda_{1, G}$ is the group lasso penalty parameter (with subscript $G$ for Group), $G$ is the total number of groups, $\mathcal{J}_g \subset \{1, \ldots, p \}$ is covariate index set of the $g$-th group such that the $\mathcal{J}_g$ are mutually exclusive and exhaustive, i.e. $\mathcal{J}_{g_1} \cap \mathcal{J}_{g_2} = \emptyset$ for all $g_1 \not= g_2$ and $\cup_{g=1}^G \mathcal{J}_g = \{1, \ldots, p \}$, and $|\mathcal{J}_g|$ denotes the cardinality (the number of elements) of $\mathcal{J}_g$.

The group lasso estimator performs covariate selection at the group level but does not result in a sparse within-group estimate. This may be achieved through employment of the \textit{sparse group lasso} regression estimator \citep{Simon2013}:
\begin{eqnarray*}
\hat{\bbeta}(\lambda_1, \lambda_{1,G}) & = & \arg \min_{\bbeta \in \mathbb{R}^p} \| \mathbf{Y} - \mathbf{X} \bbeta \|_2^2 + \lambda_1 \| \bbeta \|_1 + \lambda_{1,G} \sum\nolimits_{g = 1}^G \sqrt{|J_g|} \,  \| \bbeta_g \|_2,
\end{eqnarray*}
which combines the lasso with the group lasso penalty. The inclusion of the former encourages within-group sparsity, while the latter performs selection at the group level. The sparse group lasso penalty resembles the elastic net penalty with the $\| \bbeta \|_2$-term replacing the $\| \bbeta \|_2^2$-term of the latter.

\begin{figure}[!h]
\begin{tabular}{rcl}
\includegraphics[scale=0.20, angle=0]{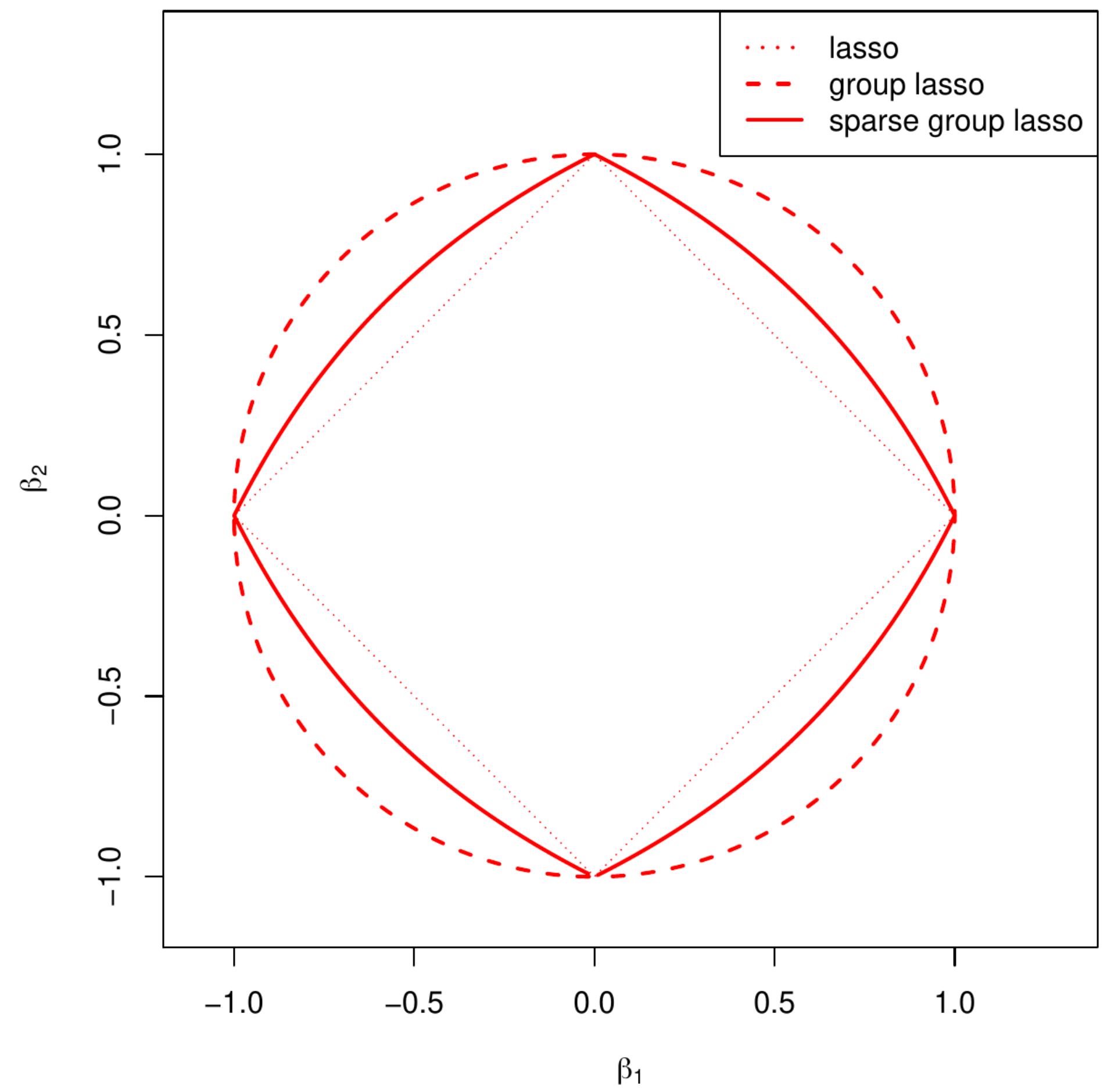}
& &
\includegraphics[scale=0.20, angle=0]{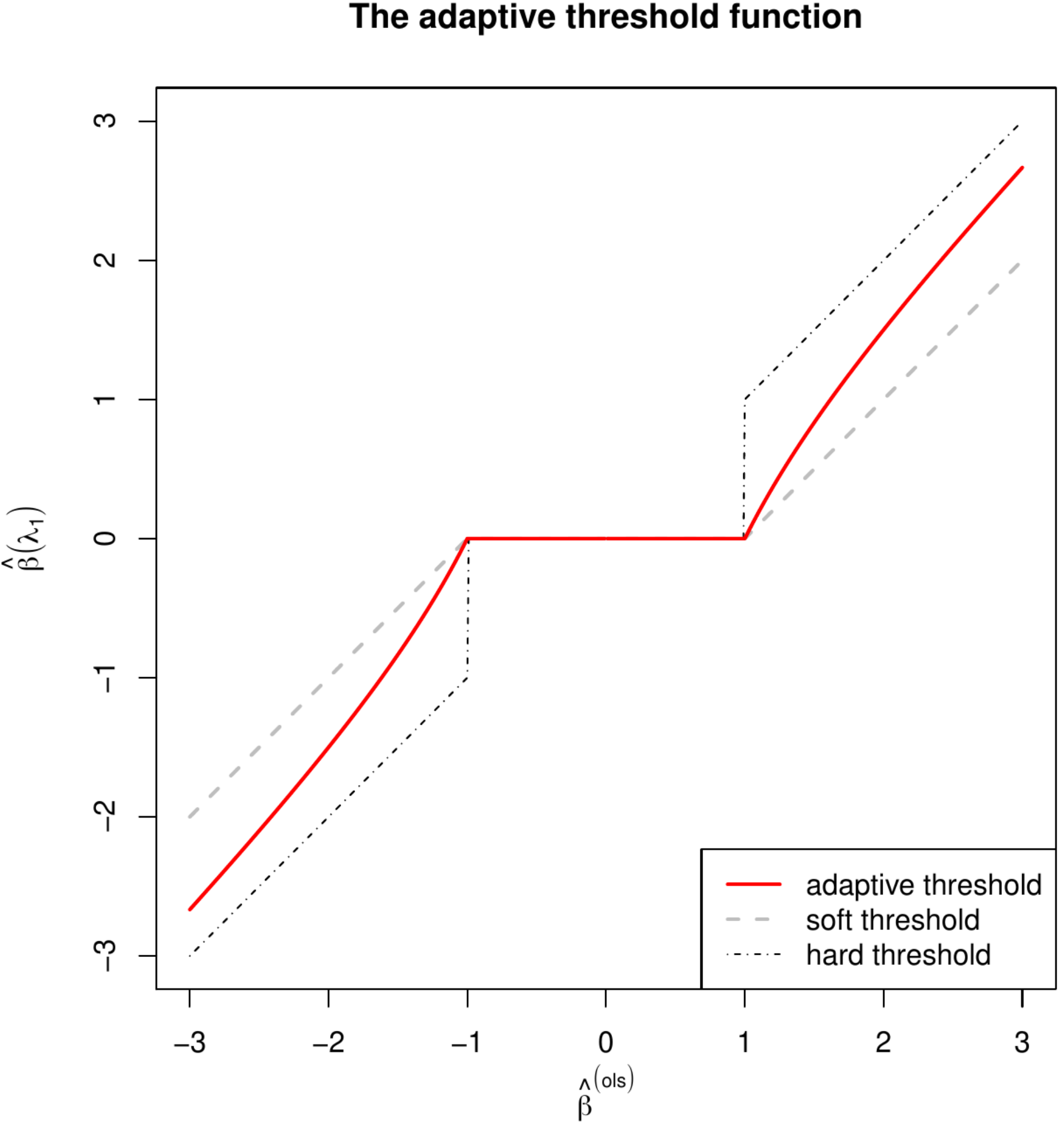}
\\
\includegraphics[scale=0.20, angle=0]{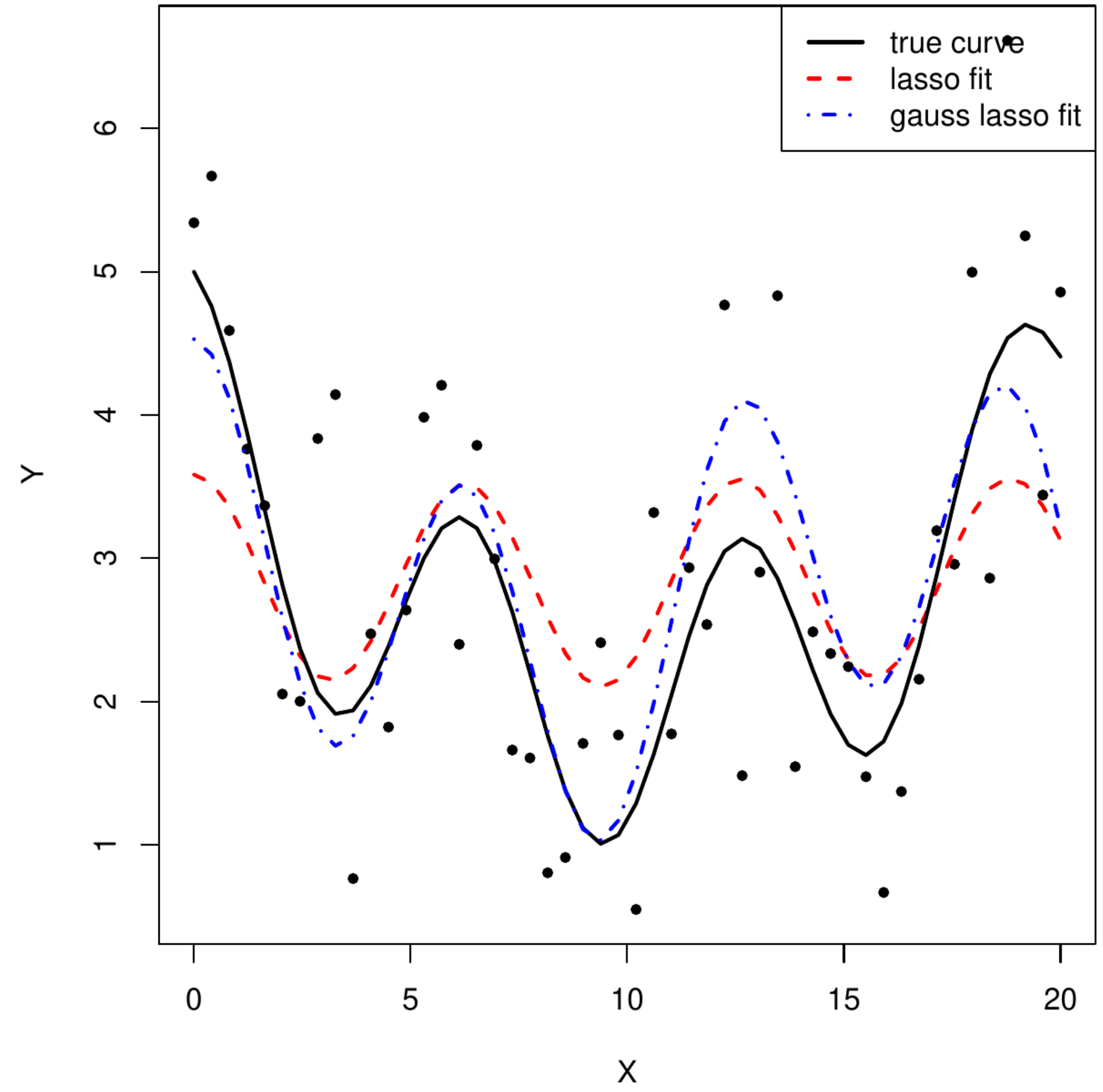}
& &
\includegraphics[scale=0.20, angle=0]{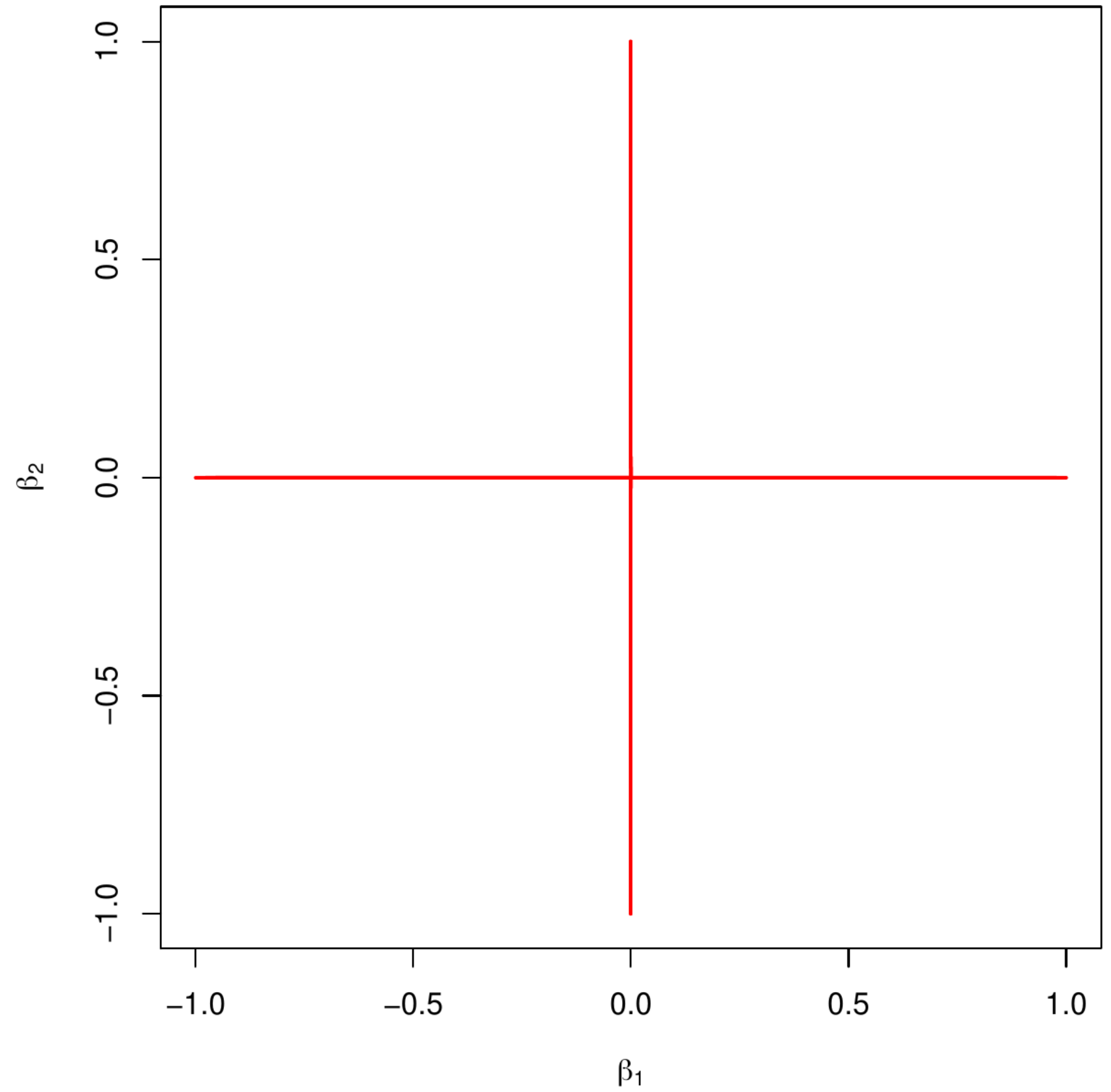}
\end{tabular}
\caption{Top left panel: The $(\beta_1, \beta_2)$-parameter constraint induced by the lasso, group lasso and sparse group lasso penalty. Top right panel: the adaptive threshold function associated with the adaptive lasso regression estimator for orthonormal design matrices. Bottom left panel: illustration of the lasso and adaptive lasso fit and the true curve. Bottom right panel: the parameter constraint induced by the $\ell_0$-penalty. \label{fig.groupLassoConstraint}}
\end{figure}

The (sparse) group lasso regression estimation problems can be reformulated as constrained estimation problems. Their parameter constraints are depicted in Figure \ref{fig.groupLassoConstraint}. By now the reader will be familiar with the charactistic feature, i.e. the non-differentiability of the boundary at the axes, of the constraint that endows the estimator with the potential to select. This feature is clearly present for the sparse group lasso regression estimator, covariate-wise. Although both the group lasso and the sparse group lasso regression estimator select group-wise, illustration of the associated geometrical feature requires plotting in dimensions larger than two and is not attempted. However, if all groups are singletons, the (sparse) group lasso penalties are equivalent to the regular lasso penalty.

The sparse group lasso regression estimator is found through exploitation of the convexity of the loss function \citep{Simon2013}. It alternates between group-wise and within-group optimization. The resemblance of the sparse group lasso and elastic net penalties propagates to the optimality conditions of both estimators. In fact, the within-group optimization amounts to the evaluation of an elastic net regression estimator \citep{Simon2013}. When within each group the design matrix is orthonormal and  $\lambda_{1,G} = 0$, the group lasso regression estimator can be found by a group-wise coordinate descent procedure for the evaluation of the estimator (cf. Exercise \ref{question:groupLasso}).

Both the sparse group lasso and the elastic net regression estimators have two penalty parameters that need tuning. In both cases the corresponding penalties have a similar effect: shrinkage towards zero. If the $g$-th group's contribution to the group lasso penalty has vanished, then so has the contribution of all covariates to the regular lasso penalty. And vice versa. This complicates the tuning of the penalty parameters as it is hard to distinguish which shrinkage effect is most beneficial for the estimator. \cite{Simon2013} resolve this by setting the ratio of the two penalty parameters $\lambda_1$ and $\lambda_{1,G}$ to some arbitrary but fixed value, thereby simplifying the tuning.

\section{Adaptive lasso}
The lasso regression estimator does not exhibit some key and desirable asymptotic properties. \cite{Zou2006} proposed the \text{adaptive lasso} regression estimator to achieve these properties. The adaptive lasso regression estimator is a two-step estimation procedure. First,  an initial estimator of the regression parameter $\bbeta$, denoted $\hat{\bbeta}
^{\mbox{{\tiny init}}}$, is to be obtained. The adaptive lasso regression estimator is then defined as:
\begin{eqnarray*}
\hat{\bbeta}^{\mbox{{\tiny adapt}}} (\lambda_1) & = & \arg \min_{\bbeta \in \mathcal{R}^p}
\| \mathbf{Y} - \mathbf{X} \bbeta \|_2^2 + \lambda_1 \sum\nolimits_{j=1}^p  | \hat{\beta}^{\mbox{{\tiny init}}}_j |^{-1}    | \beta_j |.
\end{eqnarray*}
Hence, it is a generalization of the lasso penalty with covariate-specific weighing. The weight of the $j$-th covariate is reciprocal to the $j$-th element of the initial regression parameter estimate $\hat{\bbeta}^{\mbox{{\tiny init}}}$. If the initial estimate of $\beta_j$ is large or small, the corresponding element in the adaptive lasso estimator will be penalized less or more and thereby determine the amount of shrinkage, which may now vary between the estimates of the elements of $\bbeta$. In particular, if $\hat{\bbeta}_j^{\mbox{{\tiny init}}} = 0$, the adaptive lasso penalty parameter corresponding to the $j$-th element is infinite and yields $\hat{\bbeta}^{\mbox{{\tiny adapt}}} = 0$.

The adaptive lasso regression estimator, given an initial regression estimate $\hat{\bbeta}^{\mbox{{\tiny init}}}$, can be found numerically by minor changes of the algorithms presented in Section \ref{sect:lassoEstimation}. In case of an orthonormal design an analytic expression of the adaptive lasso estimator exists (see Exercise \ref{question:adaptiveLasso}):
\begin{eqnarray*}
\hat{\beta}_j^{\mbox{{\tiny adapt}}} (\lambda_1) & = & \mbox{sign}(\hat{\beta}_j^{\mbox{{\tiny ols}}}) ( | \hat{\beta}_j^{\mbox{{\tiny ols}}} | - \tfrac{1}{2} \lambda_1 / | \hat{\beta}_j^{\mbox{{\tiny ols}}} |)_+.
\end{eqnarray*}
This adaptive lasso estimator can be viewed as a compromise between the soft thresholding function, associated with the lasso regression estimator for orthonormal design matrices (Section \ref{sect:lassoAnalytic}), and the hard thresholding function, associated with truncation of the ML regression estimator (see the top right panel of Figure \ref{fig.groupLassoConstraint} for an illustration of these thresholding functions).

How is the initial regression parameter estimate $\hat{\bbeta}^{\mbox{{\tiny init}}}$ to be chosen? Low-dimensionally the maximum likelihood regression estimator one may used. The resulting adaptive lasso regression estimator is sometimes referred to as the \textit{Gauss-Lasso} regression estimator. High-dimensionally, the lasso or ridge regression estimators will do. Any other estimator may in principle be used. But not all yield the desirable asymptotic properties.

A different motivation for the adaptive lasso is found in its ability to undo some or all of the shrinkage of the lasso regression estimator due to penalization. This is illustrated in the left bottom panel of Figure \ref{fig.groupLassoConstraint}. It shows the lasso and adaptive lasso regression fits. The latter clearly undoes some of the bias of the former.

\section{The $\ell_0$ penalty}
An alternative estimators, considered to be superior to both the lasso and ridge regression estimators, is the $\ell_0$-penalized regression estimator. It too minimizes the sum-of-squares now with the $\ell_0$-penalty. The $\ell_0$-penalty, denoted $\lambda_0 \| \bbeta \|_0$, is defined as $\lambda_0 \| \bbeta \|_0 = \lambda_0 \sum_{j=1}^p \mathbbm{1}_{ \{ \beta_j \not= 0 \} }$ with penalty parameter $\lambda_0$ and $\mathbbm{1}_{\{ \cdot \}}$ the indicator function.  The parameter constraint associated with this penalty is shown in the right panel of Figure \ref{fig.groupLassoConstraint}. From the form of the penalty it is clear that the $\ell_0$-penalty penalizes only for the presence of a covariate in the model. As such it is concerned with  the number of covariates in the model, and not the size of their regression coefficients (or a derived quantity thereof). The latter is considered only  a surrogate of the number of covariates in the model. As such the lasso and ridge estimators are proxies to the $\ell_0$-penalized regression estimator.

The $\ell_0$-penalized regression estimator is not used for large dimensional problems as its evaluation is computationally too demanding. It requires a search over all possible subsets of the $p$ covariates to find the optimal model. As each covariate can either be in or out of the model, in total $2^p$ models need to be considered. This is not feasible with present-day computers.

The adaptive lasso regression estimator may be viewed as an approximation of the $\ell_0$-penalized regression estimator. When the covariate-wise weighing employed in the penalization of the adaptive lasso regression estimation is equal to $\lambda_{1,j} = \lambda / |\beta_j|$ with $\beta_j$ known true value of the $j$-th regression coefficient, the $j$-th covariate's estimated regression coefficient contributes only to the penalty if it is nonzero. In practice, this weighing involves an initial estimate of $\bbeta$ and is therefore an approximation at best, which the quality of the approximation hinging upon that of the weighing.

\section{Conclusion}
Finally, a note of caution in similar spirit as that which concludes Chapter \ref{chap:genRidge}. It is a joy to play with the penalty and see how it encourages the regression parameter estimate to exhibit hypothesized behaviour. Nonetheless, for a deeper and more profound understanding of the data generating mechanism, such knowledge is ideally incorporated explicitly in the model itself.

\section{Exercises}
\begin{question} \mbox{ } \label{question:fusedLasso} \\
Augment the lasso penalty with the sum of the absolute differences all pairs of successive regression coefficients:
\begin{eqnarray*}
\lambda_1 \sum\nolimits_{j=1}^p | \beta_j | + \lambda_{f} \sum\nolimits_{j=2}^p | \beta_j  - \beta_{j-1} |.
\end{eqnarray*}
This augmented lasso penalty is referred to as the {\it fused lasso penalty}.
\begin{compactitem}
\item[\textit{a)}] Consider the standard multiple linear regression model:
$Y_i = \sum_{j=1}^p X_{ij} \, \beta_j + \varepsilon_i$. Estimation of the regression parameters takes place via minimization of penalized sum of squares, in which the fused lasso penalty is used with $\lambda_1 =0$. Rewrite the corresponding loss function to the standard lasso problem by application of the following change-of-variables: $\gamma_1 = \beta_1$  and $\gamma_{j} = \beta_j - \beta_{j-1}$.

\item[\textit{b)}] Investigate on simulated data the effect of the second summand of the fused lasso penalty on the parameter estimates. In this, temporarily set $\lambda_1 = 0$.

\item[\textit{c)}] Let $\lambda_1$ equal zero still. Compare the regression estimates of part \textit{b)} to the ridge estimates with a first-order autoregressive prior. What is qualitatively the difference in the behavior of the two estimates? {\it Hint:} plot the full solution path for the penalized estimates of both estimation procedures.

\item[\textit{d)}] How do the estimates of part \textit{b)} of this question change if we allow $\lambda_1 >0$?
\end{compactitem}
\end{question}

\begin{question} \mbox{ } \\
Consider the standard linear regression model $Y_i = \mathbf{X}_{i,\ast} \bbeta + \varepsilon_i$ for $i=1, \ldots, n$ and with $\varepsilon_i \sim_{i.i.d.}  \mathcal{N}(0, \sigma^2)$. The rows of the design matrix $\mathbf{X}$ are of length 2, neither column represents the intercept. Relevant summary statistics from the data on the response $\mathbf{Y}$ and the covariates are: 
\begin{eqnarray*}
\mathbf{X}^{\top} \mathbf{X} & = & \left( \begin{array}{rr} 40 & -20 \\ -20 & 10 \end{array} \right) \qquad \mbox{ and } \qquad \mathbf{X}^{\top} \mathbf{Y} \, \, \, = \, \, \, \left( \begin{array}{rr} 26 \\ -13
\end{array} \right).
\end{eqnarray*}

\begin{compactitem}
\item[\textit{a)}] Use lasso regression estimator to fit the linear regression model without intercept and only the first covariate. Draw (i.e. not sketch!) the regularization path of lasso regression estimator.


\item[\textit{b)}] The two covariates are perfectly collinear. However, their regularization paths do not coincide. Why? 

\item[\textit{c)}] Fit the linear regression model with both covariates (and still without intercept) by means of the fused lasso, i.e. $\hat{\bbeta} (\lambda_1, \lambda_f) = \arg \min_{\bbeta} \| \mathbf{Y} - \mathbf{X} \bbeta \|_2^2 + \lambda_1  \| \bbeta \|_1 + \lambda_f | \beta_1 - \beta_2 |$ with $\lambda_1 = 10$ and $\lambda_f = \infty$. \textit{Hint:} at some point in your answer you may wish to write $\mathbf{X} = ( \mathbf{X}_{\ast,1} \, \, \, c \mathbf{X}_{\ast,1})$ and deduce $c$ from $\mathbf{X}^{\top} \mathbf{X}$.
\end{compactitem}
\end{question}

\begin{question} \mbox{ } \\
Consider the linear regression model $\mathbf{Y} = \mathbf{X} \bbeta + \vvarepsilon$ with $\vvarepsilon \sim \mathcal{N} ( \mathbf{0}_n, \sigma^2 \mathbf{I}_{nn})$. This model (without intercept) is fitted to data using the lasso regression estimator $\hat{\bbeta}(\lambda_1) = \arg \min_{\bbeta} \| \mathbf{Y} - \mathbf{X} \bbeta \|_2^2 + \lambda_1 \| \bbeta \|_1$. The relevant summary statistics of the data are:
\begin{eqnarray*}
\mathbf{X}  = \left( \begin{array}{rr} 1 & -1 \\ -1 & 1 \end{array} \right), \, 
\mathbf{Y}  = \left( \begin{array}{r} -5 \\ 4 \end{array} \right), \,
\mathbf{X}^{\top} \mathbf{X} = \left( \begin{array}{rr} 2 & -2 \\ -2 & 2 \end{array} \right), \mbox{ and } \, \mathbf{X}^{\top} \mathbf{Y} = \left( \begin{array}{r} -9 \\ 9 \end{array} \right).
\end{eqnarray*}

\begin{compactitem}
\item[\textit{a)}] Specify the full set of lasso regression estimates with $\lambda_1 = 2$ that minimize the lasso loss function for these data.  

\item[\textit{b)}] Now consider fitting the linear regression model with the fused lasso estimator $\hat{\bbeta}(\lambda_f) = \arg \min_{\bbeta} \| \mathbf{Y} - \mathbf{X} \bbeta \|_2^2 + \lambda_f | \beta_1 - \beta_2|$. Determine $\lambda_{f}^{(0)} > 0$ such that $\hat{\bbeta}(\lambda_f) = (0, 0)^{\top}$ for all $\lambda_f > \lambda_{f}^{(0)}$.
\end{compactitem}
\end{question}

\begin{question} \mbox{ } \\
Consider the linear regression model $\mathbf{Y} = \mathbf{X} \bbeta + \vvarepsilon$ with $\vvarepsilon \sim \mathcal{N} ( \mathbf{0}_n, \sigma_{\varepsilon}^2 \mathbf{I}_{nn})$. This model (without intercept) is fitted to data using the lasso regression estimator $\hat{\bbeta}(\lambda_1, \lambda_f) = \arg \min_{\bbeta} \| \mathbf{Y} - \mathbf{X} \bbeta \|_2^2 + \lambda_1 \| \bbeta \|_1 + \lambda_f \sum_{j=2}^p | \beta_{j} - \beta_{j-1} |$. The data are:
\begin{eqnarray*}
\mathbf{X} = \left( \begin{array}{rrr} 3 & -1 \\ 1 & 1  \end{array} \right), \mbox{ and } \mathbf{Y} = \left( \begin{array}{r} -2 \\ 3 \end{array} \right).
\end{eqnarray*}

\begin{compactitem}
\item[\textit{a)}] Evaluate the fused lasso regression estimator $\hat{\bbeta}(\lambda_1, \lambda_f)$ for  $\lambda_1 = 1$  and $\lambda_f = \infty$.

\item[\textit{b)}] How many covariates can the fused lasso regression estimator select? Motivate.
\end{compactitem}
\end{question}

\begin{question} \mbox{ } \\
Consider the linear regression model $\mathbf{Y} = \mathbf{X} \bbeta + \vvarepsilon$ with $\vvarepsilon \sim \mathcal{N} ( \mathbf{0}_n, \sigma_{\varepsilon}^2 \mathbf{I}_{nn})$. The model is fitted to the following (summaries of the) data:
\begin{eqnarray*}
\mathbf{X} = \left( \begin{array}{rr} -1 & 2 \\ 1 & -1 \end{array} \right),  \,
\mathbf{X}^{\top} \mathbf{X} = \left( \begin{array}{rr} 2 & -3 \\ -3 & 5 \end{array} \right), \mathbf{Y} = \left( \begin{array}{r} 1 \\ -2 \end{array} \right), \, \mbox{ and } \, \mathbf{X}^{\top} \mathbf{Y} = 
\left( \begin{array}{r} -3 \\ 4 \end{array} \right),
\end{eqnarray*}
using the sparse fused lasso regression estimator: $\hat{\bbeta}(\lambda_1, \lambda_f) = \arg \min_{\bbeta \in \mathbb{R}^2} \| \mathbf{Y} - \mathbf{X} \bbeta \|_2^2 + \lambda_1 \| \bbeta \|_1 + \lambda_f | \beta_1 - \beta_2 |$. 

\begin{compactitem}
\item[\textit{a)}] Evaluate the sparse fused lasso regression estimator for $\lambda_1 = 1$ and $\lambda_f = \infty$. 

\item[\textit{b)}] Drop the first observation and set $\lambda_1 = 0$ while $\lambda_f$ is some finite positive value. Is $\hat{\bbeta}(\lambda_1, \lambda_f)$ uniquely defined? Motivate! 
 
\item[\textit{c)}] Should $\| \hat{\bbeta}(\lambda_1, \lambda_f) \|_1 > \| \hat{\bbeta}(\lambda_1, \lambda_f') \|_1$ if $\lambda_f' > \lambda_f$? Motivate.
\end{compactitem}
\end{question}

\begin{question} \textit{(The elastic net regression estimator)} \label{question:elasticNet} \\
Consider fitting the linear regression model by means of the elastic net regression estimator.
\begin{compactitem}
\item[{\it a)}] Recall the data augmentation trick of Question \ref{question:ridgeAugmentation} of the ridge regression exercises. Use the same trick to show that the elastic net least squares loss function can be reformulated to the form of the traditional lasso function. {\it Hint}: absorb the ridge part of the elastic net penalty into the sum of squares.

\item[{\it b)}] The elastic net regression estimator can be evaluated by a coordinate descent procedure outlined in Section \ref{sect:coordinateDescent}. Show that in such a procedure at each step the $j$-th element of the elastic net regression estimate is updated according to:
\begin{eqnarray*}
\hat{\beta}_j (\lambda_1, \lambda_2)  & = & (\| \mathbf{X}_{\ast, j} \|_2^2 + \lambda_2)^{-1}  \mbox{sign}(\mathbf{X}_{\ast, j}^{\top} \tilde{\mathbf{Y}}) \big[  | \mathbf{X}_{\ast, j}^{\top} \tilde{\mathbf{Y}} | - \tfrac{1}{2} \lambda_1 \big]_+.
\end{eqnarray*}
with $\tilde{\mathbf{Y}} = \mathbf{Y} - \mathbf{X}_{\ast, \setminus j} \bbeta_{\setminus j}$.
\end{compactitem}
\end{question}

\begin{question} \textit{(The elastic net regression estimator)}
\\
Consider the linear regression model $\mathbf{Y} = \mathbf{X} \bbeta + \vvarepsilon$ with $\vvarepsilon \sim \mathcal{N} ( \mathbf{0}_n, \sigma_{\varepsilon}^2 \mathbf{I}_{nn})$. This model (without intercept) is fitted to data using the elastic net estimator $\hat{\bbeta}(\lambda_1, \lambda_2) = \arg \min_{\bbeta} \| \mathbf{Y} - \mathbf{X} \bbeta \|_2^2 + \lambda_1 \| \bbeta \|_1 + \tfrac{1}{2} \lambda_2 \| \bbeta \|_2^2$. The relevant summary statistics of the data are:
\begin{eqnarray*}
\mathbf{X}  = \left( \begin{array}{r} 1 \\ -1 \\ -1 \end{array} \right), \,  \mathbf{Y}  = \left( \begin{array}{r} -5 \\ 4 \\ 1 \end{array} \right), \, \mathbf{X}^{\top} \mathbf{X} = \left( \begin{array}{r} 3  \end{array} \right), \mbox{ and } \, \mathbf{X}^{\top} \mathbf{Y} = \left( \begin{array}{r} -10 \end{array} \right).
\end{eqnarray*}

\begin{compactitem}
\item[\textit{a)}] Evaluate for $(\lambda_1, \lambda_2) = (3,2)$ the elastic net regression estimator of the linear regression model.

\item[\textit{b)}] Now consider the evaluation the elastic net regression estimator of the linear regression model for the same penalty parameters, $(\lambda_1, \lambda_2) = (3,2)$, but this time involving two covariates. The first covariate is as in part \textit{a)}, the second is orthogonal to that one. Do you expect the resulting elastic net estimate of the first regression coefficient $\hat{\beta}_1 ( \lambda_1, \lambda_2)$ to be larger, equal or smaller (in an absolute sense) than your answer to part \textit{a)}? Motivate. 

\item[\textit{c)}] Now take in part \textit{b)} the second covariate equal to the first one. Show that the first coefficient of elastic net estimate $\hat{\beta}_1 ( \lambda_1, 2 \lambda_2)$ is half that of part \textit{a)}. \textit{Note:} there is no need to know the exact answer to part \textit{a)}.
\end{compactitem}
\end{question}

\begin{question} \mbox{ } \\
Consider fitting the linear regression model $\mathbf{Y} = \mathbf{X}\bbeta + \vvarepsilon$ with $\bbeta \in \mathbb{R}^p$ with $\vvarepsilon \sim \mathcal{N} (\mathbf{0}_n , \sigma_{\varepsilon}^2 \mathbf{I}_{nn} )$ to data using an elastic net-type regression estimator $\hat{\bbeta}(\lambda_1 ) = \arg \min_{\bbeta \in \mathbb{R}^2} \| \mathbf{Y} - \mathbf{X} \bbeta \|_2^2 + \lambda_1 (|\beta_1 | + \beta_2^2 )$. The (summaries of the) data are
\begin{eqnarray*}
\mathbf{X}  = \left( \begin{array}{rr} -1 & 0 \\ -2 & 1 \end{array} \right), \,  \mathbf{Y}  = \left( \begin{array}{r} 5 \\ -2 \end{array} \right), \, \mathbf{X}^{\top} \mathbf{X} = \left( \begin{array}{rr} 5 & -2 \\ -2 & 1 \end{array} \right),  \mbox{ and } \, \mathbf{X}^{\top} \mathbf{Y} = \left( \begin{array}{r} -1 \\ -2 \end{array} \right).
\end{eqnarray*}

\begin{compactitem}
\item[\textit{a)}]
Draw the parameter constraint on $\bbeta$ induced by the elastic net-type penalty for an arbitrary but finite $\lambda_1$. Motivate the form of the drawn constraints from the archetypical lasso and ridge parameter constraints.

\item[\textit{b)}] Evaluate the estimator for $\lambda_1 = 4$. \textit{Hint}: Use the data augmentation trick of Exercise \ref{question:ridgeAugmentation} and optimize the elastic net-type loss function for $\beta_2$.
\end{compactitem}
\end{question}

\begin{question}\footnote{This question is freely copied from \cite{Buhlmann2011}: Problem 2.5a, page 43.} \mbox{ } \label{question:adaptiveLasso} \\
Consider the linear regression model $\mathbf{Y} = \mathbf{X} \bbeta + \vvarepsilon$. It is fitted to data from a study with an orthonormal design matrix by means of the adaptive lasso regression estimator initiated by the OLS/ML regression estimator. Show that the $j$-th element of the resulting adaptive lasso regression estimator equals:
\begin{eqnarray*}
\hat{\beta}_j^{\mbox{{\tiny adapt}}} (\lambda_1) & = & \mbox{sign}(\hat{\beta}_j^{\mbox{{\tiny ols}}}) ( | \hat{\beta}_j^{\mbox{{\tiny ols}}} | - \tfrac{1}{2} \lambda_1 / | \hat{\beta}_j^{\mbox{{\tiny ols}}} |)_+.
\end{eqnarray*}
\end{question}

\begin{question} \label{question:lassoPenalty2constraint} \mbox{ } \\
Regularized estimation is equivalent to constrained estimation, for corresponding penalty and constraint. 
\begin{compactitem}
\item[\textit{a)}] Consider the following three penalty functions:
\begin{compactitem}
\item[\textit{i)}] $f^{\mbox{{\tiny (pen)}}} (\beta_1, \beta_2; \lambda_1, \lambda_{1,f}) = \lambda_1 ( |\beta_1 | + | \beta_2 |) + \lambda_{1,f} |\beta_1 - 2 \beta_2 |$,
\item[\textit{ii)}]  $f^{\mbox{{\tiny (pen)}}} (\beta_1, \beta_2; \lambda_1, \lambda_{1,f}) = \lambda_1 | \beta_2 | + \lambda_{1,f} | 2\beta_1 -  \beta_2 |$, and
\item[\textit{iii)}]  $f^{\mbox{{\tiny (pen)}}} (\beta_1, \beta_2; \lambda_1, \lambda_{1,f}) = \lambda_1 |\beta_1 | + \lambda_{1,f} |\beta_1 - 2 \beta_2 |$.
\end{compactitem}
Each of three penalty functions induces a constraint on the regression parameter. The top panels in Figure \ref{fig.constrainedEstExercise} depict three such constraints. Which constraint corresponds to which penalty function? Motivate! \textit{Note}: the values of $\lambda_1$ and $\lambda_{1,f}$ employed to plot the constraints may differ between the panels.

\item[\textit{b)}] Consider the following three penalty functions:
\begin{compactitem}
\item[\textit{i)}] $f^{\mbox{{\tiny (pen)}}} (\beta_1, \beta_2) = \lambda \beta_1^2 + \lambda_f |\beta_1 - \beta_2 |$,
\item[\textit{ii)}]  $f^{\mbox{{\tiny (pen)}}} (\beta_1, \beta_2) = \lambda |\beta_1 | + \lambda_f (\beta_1 - \beta_2 )^2$, and
\item[\textit{iii)}]  $f^{\mbox{{\tiny (pen)}}} (\beta_1, \beta_2) = \lambda (\beta_1^2 + \beta_2^2) + \lambda_f |\beta_1 - \beta_2 |$.
\end{compactitem}
Each of three penalty functions induces a constraint on the regression parameter. The bottom panels in Figure \ref{fig.constrainedEstExercise} depict three parameter constraints. Which constraint corresponds to which penalty function? Motivate! \textit{Note}: The values of $\lambda$ and $\lambda_f$ employed to plot depicted parameter constraints may differ between the panels.

\begin{figure}[!h]
\begin{tabular}{rcl}
\hspace{-1cm}
\includegraphics[scale=0.31, angle=0]{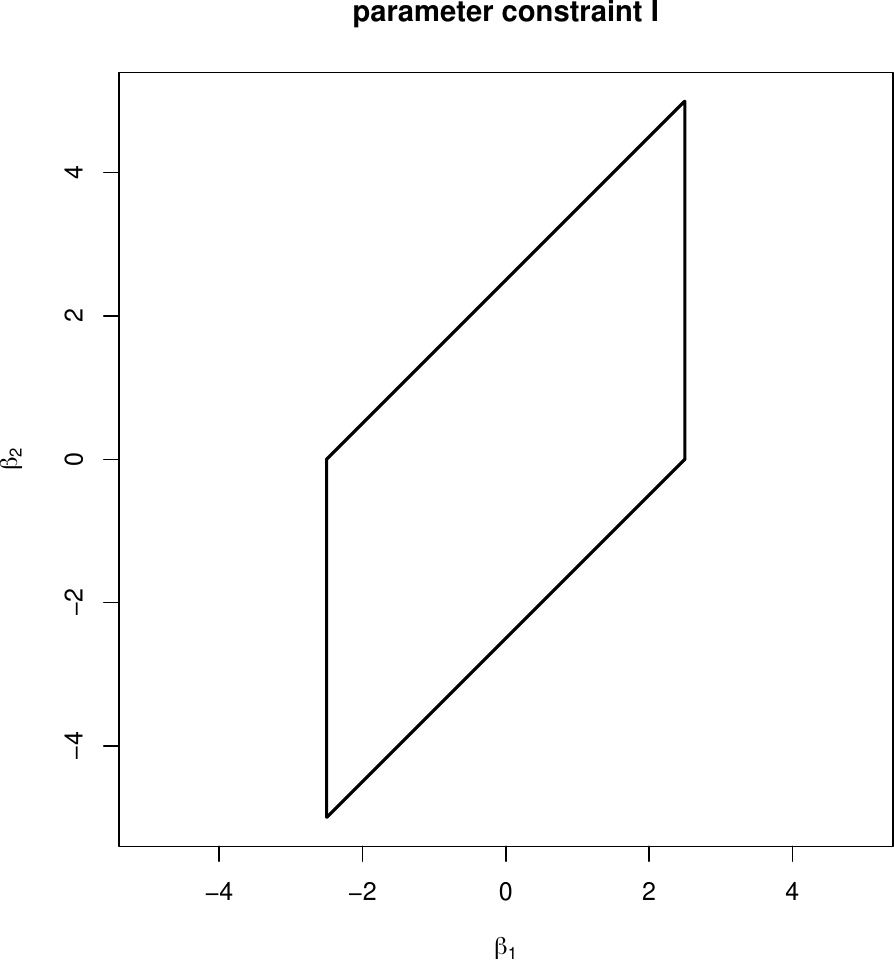}
&
\hspace{-0.7cm}
\includegraphics[scale=0.31, angle=0]{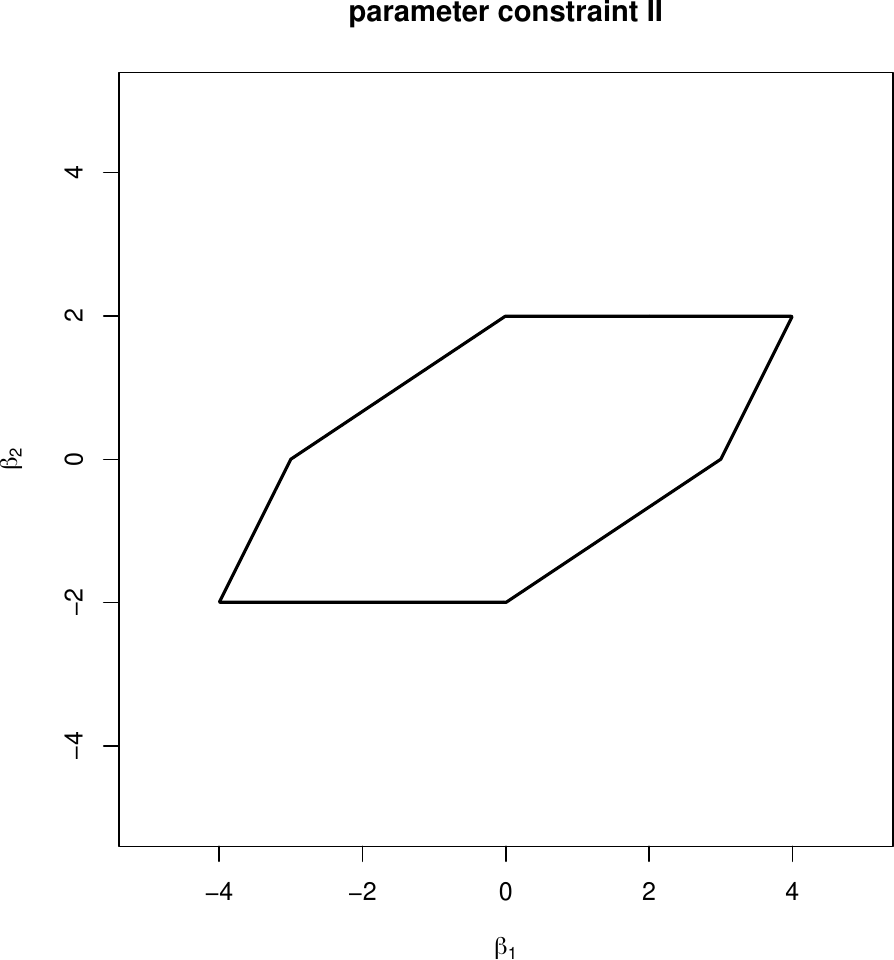}
&
\hspace{-0.7cm}
\includegraphics[scale=0.31, angle=0]{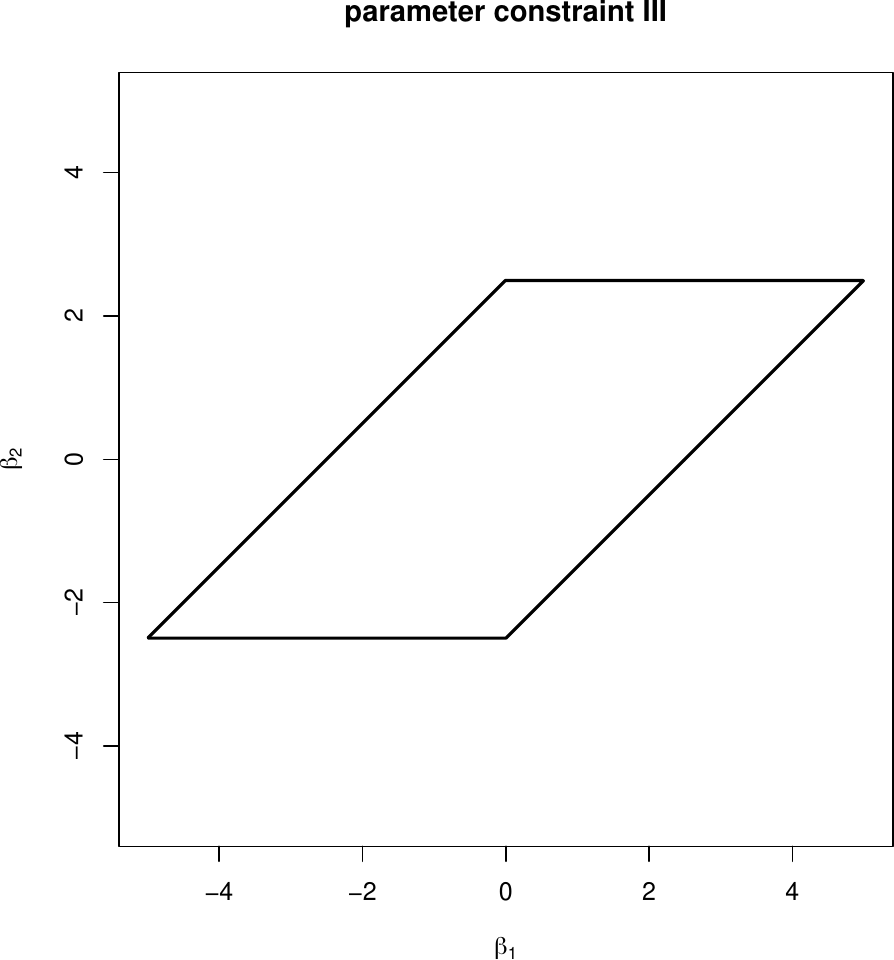}
\\
\hspace{-1cm}
\includegraphics[scale=0.30, angle=0]{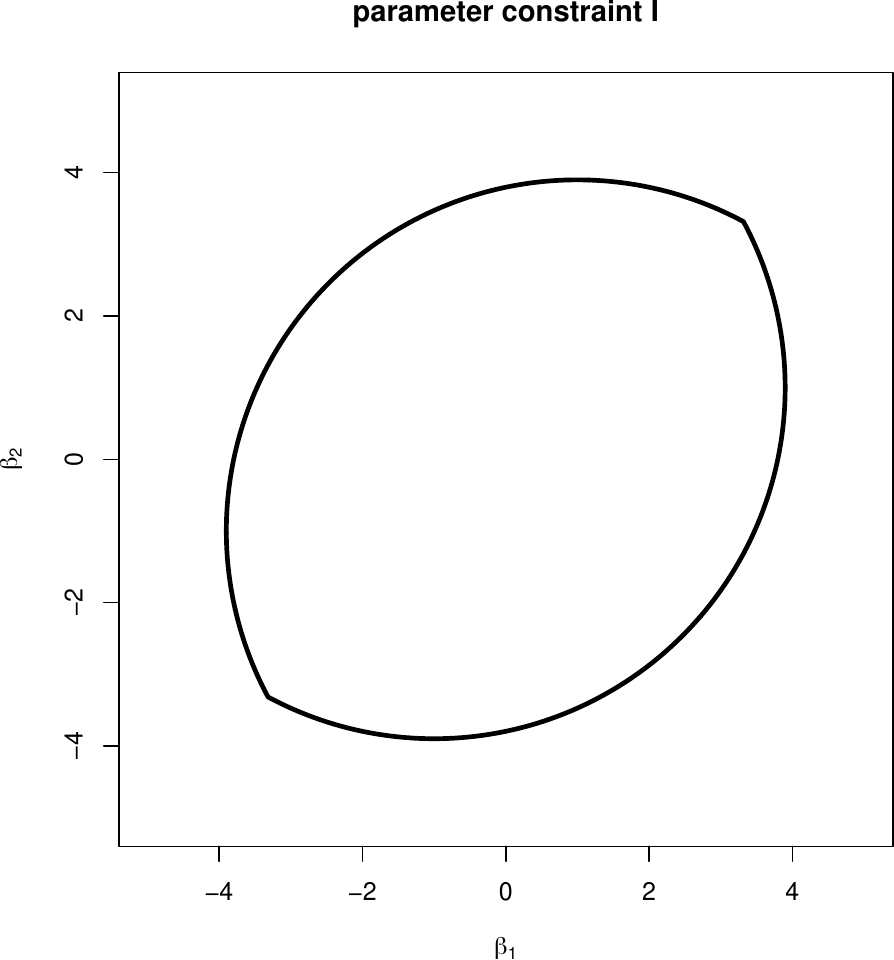}
&
\hspace{-0.7cm}
\includegraphics[scale=0.30, angle=0]{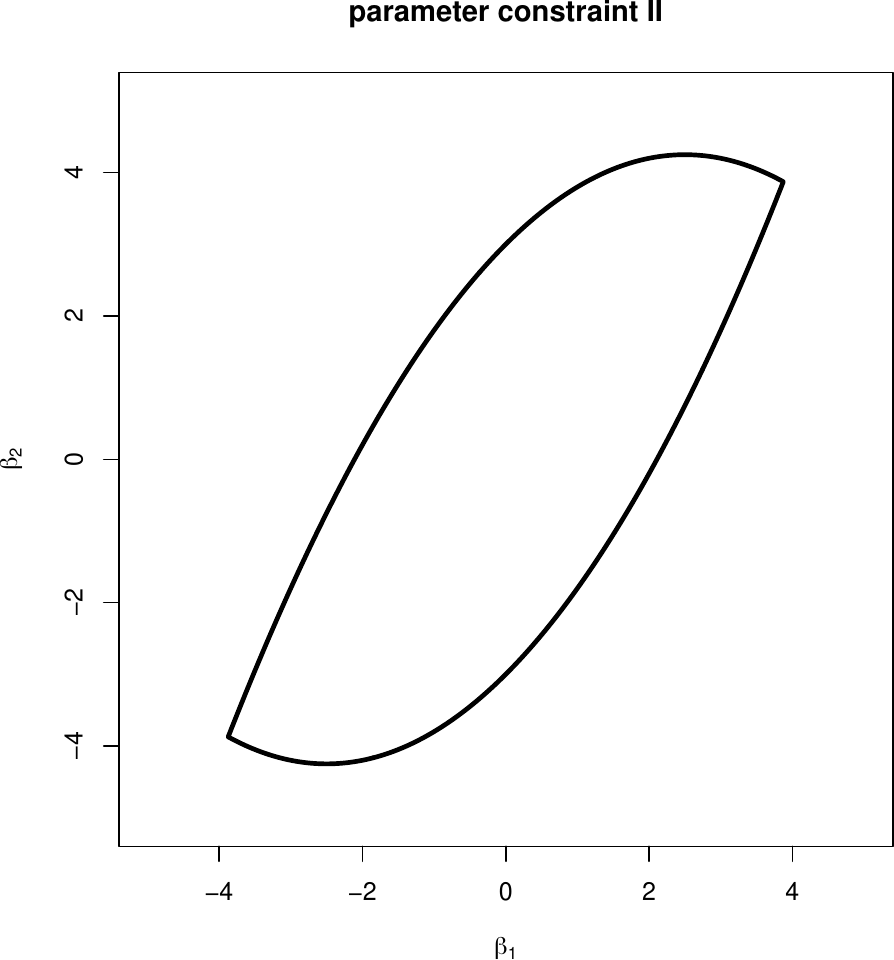}
&
\hspace{-0.7cm}
\includegraphics[scale=0.30, angle=0]{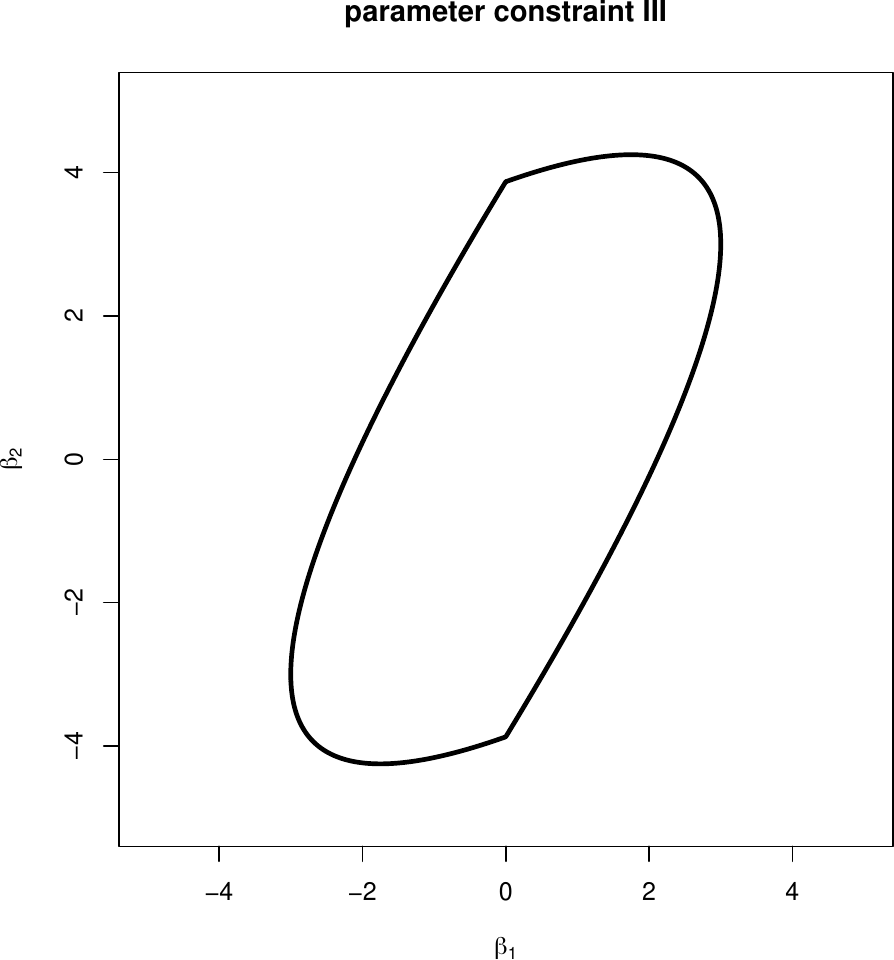}
\end{tabular}
\caption{Figure related to Question \ref{question:lassoPenalty2constraint}, with the top panels corresponding to part \textit{a)} and the bottom ones to part \textit{b)}. 
\label{fig.constrainedEstExercise}}
\end{figure}
\end{compactitem}
\end{question}

\setlength{\bibsep}{2pt}
\bibliographystyle{natbib}
\bibliography{genomics_literature}

\end{document}